\documentclass[10pt, a4paper, fleqn]{book}

\usepackage{bookmark}

\usepackage{etex}
\usepackage[utf8]{inputenc}
\usepackage[TS1,T1]{fontenc}
\usepackage{xspace}
\usepackage{array}
\usepackage{tikz-cd}
\usetikzlibrary{external}
\usepackage{smartdiagram}

\usetikzlibrary{decorations.markings}
\usepackage{layout} 
\usetikzlibrary{arrows.meta} 
\usepackage{pdfpages}
\usepackage[a4paper,tmargin=4truecm,bmargin=3truecm,hmargin=2.5truecm,marginparwidth=10mm]{geometry}

\usetikzlibrary{arrows,matrix,positioning,shapes,arrows}
\usetikzlibrary{shapes.geometric, arrows, calc, intersections}
\newcommand{\tikznode}[2]{\relax
	\ifmmode%
	\tikz[remember picture,baseline=(#1.base),inner sep=0pt] \node (#1) {$#2$};
	\else
	\tikz[remember picture,baseline=(#1.base),inner sep=0pt] \node (#1) {#2};%
	\fi}

\usepackage[english]{babel}
\usepackage{blindtext} 
\usepackage{epigraph} 
\usepackage[]{libertinus-type1}
\usepackage{amsfonts, mathtools,slashed,mathdots,bm}
\usepackage{libertinust1math}
\usepackage[bb=ams]{mathalpha}
\makeatletter
\newcommand{\addbar@}[3]{%
	\makebox[0pt][l]{%
		\raisebox{#1}[0pt][0pt]{%
			\kern#2
			\scalebox{#3}[0.8]{$\m@th\mathchar"84$}%
		}%
	}%
}
\DeclareRobustCommand{\lambdabar}{\text{\addbar@{0.1ex}{0.18em}{1}}\lambda}
\makeatother
\usepackage{slashed}
\usepackage{enumitem}
\newlist{enum-hypothesis}{enumerate}{1}
\setlist[enum-hypothesis]{label=(\arabic*),itemsep=0pt, parsep=0pt}

\setlist[enumerate,1]{label=\arabic*., ref=\arabic*, topsep=1pt, itemsep=2pt, parsep=0pt, leftmargin=1.5em, itemindent=0em, labelsep=0.2em, labelwidth=1.3em}
\setlist[enumerate,2]{label=\alph*., ref=\theenumi.\alph*, topsep=1pt, itemsep=2pt, parsep=0pt, leftmargin=0.5em, itemindent=0em, labelsep=0.2em, labelwidth=1.5em}
\setlist[enumerate,3]{label=\roman*., ref=\theenumii.\roman*, topsep=1pt, itemsep=2pt, parsep=0pt, leftmargin=0.5em, itemindent=0em, labelsep=0.2em, labelwidth=1.2em}

\usepackage{booktabs, dcolumn}
\usepackage{graphicx, subfig}

\usepackage{pbox}
 
\usepackage[round,authoryear,sort]{natbib}

\let\cite\citep 

\usepackage[thmmarks,amsmath]{ntheorem}

\newtheorem{theorem}{Theorem}[section]
\newtheorem{proposition}[theorem]{Proposition}
\newtheorem{lemma}[theorem]{Lemma}
\newtheorem{corollary}[theorem]{Corollary}

\newtheorem{definition}[theorem]{Definition}

\theoremsymbol{$\square$}
\theoremstyle{plain}
\theorembodyfont{\upshape}
\newtheorem{remark}[theorem]{Remark}
\theoremsymbol{$\Diamond$}
\newtheorem{example}[theorem]{Example}

\theoremsymbol{}
\theoremstyle{break}
\theorembodyfont{\slshape}
\newtheorem{hypothesis}[theorem]{Hypothesis}

\theoremheaderfont{\scshape}
\theorembodyfont{\upshape} 
\theoremstyle{nonumberplain}
\theoremseparator{} 
\theoremsymbol{\rule{1.3ex}{1.3ex}} 
\newtheorem{proof}{Proof}

\usepackage{hyperref}

\hypersetup{
plainpages=false,
colorlinks=true,
linkcolor=blue, 
anchorcolor=black, 
citecolor=blue, 
urlcolor=blue, 
menucolor=black, 
filecolor=black, 
bookmarksopen=true,
bookmarksnumbered=true}

\hypersetup{
pdfauthor={Gaston Nieuviarts},
pdftitle={Titre},
pdfsubject={},
pdfcreator={pdflatex},
pdfproducer={pdflatex},
pdfkeywords={}
}

\newcommand{\bbC}{\mathbb{C}}

\newcommand{\bbH}{\mathbb{H}}

\newcommand{\bbN}{\mathbb{N}}
\newcommand{\bbR}{\mathbb{R}}

\newcommand\bbZ{\mathbb{Z}}

\newcommand{\Man}{M}
\newcommand{\SpinBun}{S} 
\newcommand{\TanBun}{T(\Man)} 
\newcommand{\bbbone}{{\text{\usefont{U}{bbold}{m}{n}\char49}}} 
\newcommand{\bbbzero}{{\text{\usefont{U}{bbold}{m}{n}\char48}}} 
\newcommand{\hbbbone}{\widehat{\bbbone}}
\newcommand{\boldone}{\mathbf{1}}
\newcommand{\boldzero}{\mathbf{0}}

\newcommand{\calA}{\mathcal{A}}
\newcommand{\calB}{\mathcal{B}}
\newcommand{\calC}{\mathcal{C}}
\newcommand{\calD}{\mathcal{D}}

\newcommand{\calG}{\mathcal{G}}
\newcommand{\calH}{\mathcal{H}}

\newcommand{\calL}{\mathcal{L}}
\newcommand{\calM}{\mathcal{M}}
\newcommand{\calN}{\mathcal{N}}
\newcommand{\calO}{\mathcal{O}}
\newcommand{\calP}{\mathcal{P}}

\newcommand{\calS}{\mathcal{S}}
\newcommand{\calT}{\mathcal{T}}
\newcommand{\calU}{\mathcal{U}}
\newcommand{\calV}{\mathcal{V}}

\newcommand{\calZ}{\mathcal{Z}}

\newcommand{\algA}{\calA}
\newcommand{\algB}{\calB}
\newcommand{\hspm}{\mathbf{T}} 
\newcommand{\hs}{\calH}
\newcommand{\inj}{\iota}
\newcommand{\injhs}{\inj_{\hs}}
\newcommand{\injA}{\inj_{\algA}}

\newcommand{\injHA}{\inj_{\hs_\algA}}

\newcommand{\Fin}{F}
 
\newcommand{\BTU}{\overline{\widetilde{U}}}
\newcommand{\BTu}{\bar{\tilde{u}}}
\newcommand{\TU}{\widetilde{U}}
\newcommand{\Tu}{\tilde{u}}
\newcommand{\Mu}{u} 
\newcommand\hw{\hat{w}}
\newcommand{\proj}{\pi}
\newcommand{\projhs}{\proj^{\hs}}
\newcommand{\projA}{\proj^{\algA}}
\newcommand{\projB}{\proj^{\algB}}
\newcommand{\projHA}{\proj^{\hs_\algA}}
\newcommand{\projHB}{\proj^{\hs_\algB}}
\newcommand{\spe}{\mathbf{t}} 
\newcommand{\vm}{\sigma} 
\newcommand{\Bvm}{\overline{\sigma}} 
\newcommand{\wm}{\tau} 
\newcommand{\DA}[1][]{D_{\algA #1}}
\newcommand{\DB}[1][]{D_{\algB #1}}
\newcommand{\bbboneA}{\bbbone_\algA}
\newcommand{\bbboneB}{\bbbone_\algB}
\newcommand{\uA}[1][]{{u_{\algA #1}}}
\newcommand{\uB}[1][]{{u_{\algB#1}}}
\newcommand{\UA}{{U_\algA}}
\newcommand{\UB}{{U_\algB}}
\newcommand{\Bpsi}{\overline{\psi}}
\newcommand{\Blambda}{\overline{\lambda}}

\newcommand{\BK}{\overline{K}}
\newcommand{\BL}{\overline{L}}
\newcommand{\BA}{\overline{A}}
\newcommand{\Bxi}{\bar{\xi}}
\newcommand{\Beta}{\bar{\eta}}
\newcommand{\Be}{\bar{e}}
\newcommand{\Bv}{\bar{v}}
 \newcommand{\Bell}{\bar{\ell}}
 \newcommand{\Bp}{\bar{p}}
 \newcommand{\Bq}{\bar{q}}
 \newcommand{\BU}{\bar{U}}
 \newcommand{\Tv}{\tilde{v}}
 \newcommand{\Oset}{\mathbb{O}}
 
\newcommand\hD{\widehat{D}}
\newcommand{\Dst}{\calD_{st}}
\newcommand{\piD}{\pi_{D}}

\newcommand{\hA}{\hat{A}}
\newcommand{\hB}{\hat{B}}
\newcommand{\kX}{\mathfrak{X}}
\newcommand{\kY}{\mathfrak{Y}}

\newcommand{\ksl}{\mathfrak{sl}}
\newcommand{\ksu}{\mathfrak{su}}
\newcommand{\kT}{\mathfrak{T}}
\newcommand{\kS}{\mathfrak{S}}

\newcommand{\spm}{\mathbf{T}} 

\newcommand{\modM}{\calM}
\newcommand{\modN}{\calN}

\newcommand{\hstar}{\star} 
\newcommand{\grast}{\bullet} 
\newcommand{\omi}[1]{\buildrel { \buildrel{#1}\over{\vee} } \over .} 
\newcommand\exter{{\textstyle\bigwedge}} 

\newcommand\cdotaction{\mathord{\cdot}}
\newcommand\halgA{\widehat{\algA}}
\newcommand\halgB{\widehat{\algB}}

\newcommand\hphi{\widehat{\phi}}
\newcommand\tphi{\widetilde{\phi}}

\newcommand{\Tpsi}{\widetilde{\psi}}

\newcommand{\defeq}{\vcentcolon=} 

\DeclareMathOperator{\tsum}{\textstyle\sum}
\DeclareMathOperator{\toplus}{\textstyle\oplus}

\DeclareMathOperator{\ad}{ad}
\DeclareMathOperator{\Aut}{Aut}
\DeclareMathOperator{\card}{card}
\DeclareMathOperator{\Der}{Der}
\DeclareMathOperator{\diag}{diag}
\DeclareMathOperator{\Diff}{Diff} 

\DeclareMathOperator{\End}{End} 
\DeclareMathOperator{\Hom}{Hom} 
\DeclareMathOperator{\Id}{Id}
\DeclareMathOperator{\Inn}{Inn}
\DeclareMathOperator{\Int}{Int}
\DeclareMathOperator{\Ker}{Ker}	 

\DeclareMathOperator{\Out}{Out}
\DeclareMathOperator{\Ran}{Ran}	 
\DeclareMathOperator{\rank}{rank}	 
\DeclareMathOperator{\Span}{Span}

\DeclareMathOperator{\tr}{tr}	   
\DeclareMathOperator{\Tr}{Tr}	   

\DeclarePairedDelimiter\abs{\lvert}{\rvert}
\DeclarePairedDelimiter\norm{\lVert}{\rVert}

\newcommand{\dd}{\text{\textup{d}}}
\newcommand{\ddU}{\text{\textup{d}}_U}
\newcommand{\dddR}{\dd_{\text{dR}}}
\newcommand{\vol}{\text{\textup{vol}}}

\newcommand{\iotaDer}{\iota^{\Der}}
\newcommand{\iotaMod}{\iota_{\text{\textup{Mod}}}}
\newcommand{\piMod}{\pi^{\text{\textup{Mod}}}}
\newcommand{\phiMod}[1][]{\phi_{\text{\textup{Mod}}#1}}

\newcommand{\OmegaDer}{\Omega_{\Der}}
\newcommand{\PsiDer}{\Psi_{\Der}}
\newcommand{\PsiMod}{\Psi_{\text{\textup{Mod}}}}

\newcommand{\mrnabla}{\mathring{\nabla}}
\newcommand{\mromega}{\mathring{\omega}}
\newcommand{\bi}{\mathbf{i}}
\newcommand{\bn}{\mathbf{n}}
\newcommand{\bem}{{\mathbf{m}}}
\newcommand{\bOmega}{\mathbf{\Omega}}

\newcommand{\af}{\calA_\Fin}
\newcommand{\hshA}[1][]{\hs_{\halgA #1}} 

\newcommand{\gammaM}{\gamma_\Man}
\newcommand{\gammahA}[1][]{\gamma_{\halgA #1}}
\newcommand{\DM}{D_\Man}
\newcommand{\JM}{J_\Man}
\newcommand{\algAp}{\calA'} 
\newcommand{\algBp}{\calB'}

\newcommand{\DAp}[1][]{D_{\algAp #1}}
\newcommand{\DBp}[1][]{D_{\algBp #1}}
\newcommand{\JA}[1][]{J_{\algA #1}}
\newcommand{\JB}[1][]{J_{\algB #1}}
\newcommand{\JAp}[1][]{J_{\algAp #1}}
\newcommand{\JBp}[1][]{J_{\algBp #1}}
\newcommand{\JhA}[1][]{J_{\halgA #1}}
\newcommand{\JhB}[1][]{J_{\halgB #1}}
\newcommand{\gammaA}[1][]{\gamma_{\algA #1}}
\newcommand{\gammaB}[1][]{\gamma_{\algB #1}}
\newcommand{\DAF}[1][]{D_{\af #1}} 
\newcommand{\hsAF}[1][]{\hs_{\af #1}} 
\newcommand{\JAF}[1][]{J_{\af #1}} 
\newcommand{\gammaAF}[1][]{\gamma_{\af #1}}

\newcommand{\gammahB}[1][]{\gamma_{\halgB #1}}
\newcommand{\gammaAp}[1][]{\gamma_{\algAp #1}}
\newcommand{\gammaBp}[1][]{\gamma_{\algBp #1}}
\newcommand{\GammaA}[1][]{\Gamma_{\algA #1}}
\newcommand{\TGamma}{\widetilde{\Gamma}}
\newcommand{\TGammaA}[1][]{\TGamma_{\algA #1}}
\newcommand{\GammaB}[1][]{\Gamma_{\algB #1}} 
\newcommand{\TGammasub}[1][]{\TGamma_{#1}}
\newcommand{\Gammasub}[1][]{\Gamma_{ #1} }

 \newcommand{\hsA}[1][]{\hs_{\algA #1}} 
 \newcommand{\hsB}[1][]{\hs_{\algB #1}} 
 \newcommand{\hsAp}[1][]{\hs_{\algAp #1}} 
 \newcommand{\hsBp}[1][]{\hs_{\algBp #1}} 
 \newcommand{\hsiA}[1][]{\hs_{\algA #1}} 
 \newcommand{\hsiB}[1][]{\hs_{\algB #1}} 
 \newcommand{\phiH}[1][]{\phi_{\hs #1}} 
 
 \newcommand{\hphiH}[1][]{\hphi_{\hs #1}}
 
 \newcommand{\DhA}[1][]{D_{\halgA #1}}
 \newcommand{\DhB}[1][]{D_{\halgB #1}}
 \newcommand{\piDhA}{\pi_{\DhA}}
 \newcommand{\piDhB}{\pi_{\DhB}}
 \newcommand{\hshB}[1][]{\hs_{\halgB #1}} 
 
  \newcommand{\Ths}[1][]{\widetilde{\hs}}

 \newcommand{\ThshA}[1][]{\widetilde{\hs}_{\halgA #1}}
 \newcommand{\ThshB}[1][]{\widetilde{\hs}_{\halgB #1}}
  \newcommand{\hMan}{\widehat{\Man}}
 \newcommand{\phiA}{\phi_\algA}
 \newcommand{\phiB}{\phi_\algB}

 \newcommand{\piA}[1][]{\pi_{\algA #1}}
 \newcommand{\piAp}[1][]{\pi_{\algAp #1}}
 \newcommand{\piDA}{\pi_{\DA}}
 \newcommand{\piB}[1][]{\pi_{\algB #1}}
 \newcommand{\piBp}[1][]{\pi_{\algBp #1}}
 \newcommand{\piDB}{\pi_{\DB}}

 \newcommand{\epsilonA}{\epsilon_\algA}
 \newcommand{\epsilonB}{\epsilon_\algB}
 \newcommand{\epsilonhA}{\epsilon_{\halgA}}
 \newcommand{\epsilonhB}{\epsilon_{\halgB}}

 \newcommand{\hGammaB}{\widehat{\Gamma}_\algB}

\newcommand{\Jim}{\kappa} 
\newcommand{\JimA}[1][]{\Jim_{\algA #1}} %
\newcommand{\JimB}[1][]{\Jim_{\algB #1}} %
\newcommand{\hJim}{\widehat{\Jim}} %
\newcommand{\hJimA}[1][]{\widehat{\Jim}_{\algA #1}} %

\newcommand{\TNIC}{{\small TNIC}}
\newcommand{\lig}{\calG} 
\newcommand{\lag}{\calL} 
\newcommand{\act}{\calS} 

\newcommand{\EspPhas}{S_{ps}} 
\newcommand{\EspConf}{S_{cs}} 
\newcommand{\StrMath}{\mathfrak{M}} 
\newcommand{\hhs}{\widehat{\hs}} 

\newcounter{mnotecount}[section]
\renewcommand{\themnotecount}{\thesection.\arabic{mnotecount}}
\newcommand{\mnote}[1]%
{\protect{\stepcounter{mnotecount}}${}^{\text{\footnotesize$\bullet$\themnotecount}}$%
\reversemarginpar%
\marginpar{\raggedleft\footnotesize$\bullet$\themnotecount: #1}}

\newlength{\mnotewidth}
\newcommand{\GN}[1]{{ } }

\newcommand{\smallpmatrix}[1]{\left(\begin{smallmatrix}#1\end{smallmatrix}\right)}
\newcommand{\lieAlg}{\mathfrak{g}}
\newcommand{\bmu}{\bm{\mu}}
\newcommand{\bddU}{\text{\textbf{d}}_U}
\newcommand{\bomega}{\bm{\omega}}
\definecolor{blueamu}{RGB}{0, 101, 189}
\definecolor{cyanamu}{RGB}{61, 183, 228}
\newcommand{\dhorline}[3][0]{%
	\tikz[baseline=-2pt]{\path[decoration={markings, 
			mark=between positions 0 and 1 step 2*#3
			with {\node[color=blueamu, fill, circle, minimum width=#3, inner sep=0pt, anchor=south west] {};}},postaction={decorate}]  (0,#1) -- ++(#2,0);}}
\newcommand{\dvertline}[3][0]{%
	\tikz[baseline=2em]{\path[decoration={markings,
			mark=between positions 0 and 1 step 2*#2
			with {\node[color=black!50, fill, circle, minimum width=#2, inner sep=0pt, anchor=south west] {};}},postaction={decorate}] (0, #1) -- ++(0,#3);}}

\newcommand\titl[1]{{\usefont{T1}{tit}{l}{n} #1 }}

\newcommand\titb[1]{{\usefont{T1}{tit}{b}{n} #1 }}
\makeatletter\newcommand\HUGE{\@setfontsize\Huge{28}{0}}\makeatother		
\numberwithin{equation}{section}

\allowdisplaybreaks
\begin{document}
	
\pdfbookmark[0]{Page de titre}{titre}
\thispagestyle{empty}
\newgeometry{margin=2em}

\begin{center}
	\begin{minipage}[c]{.5\linewidth}
		\raggedright\includegraphics[height=7em]{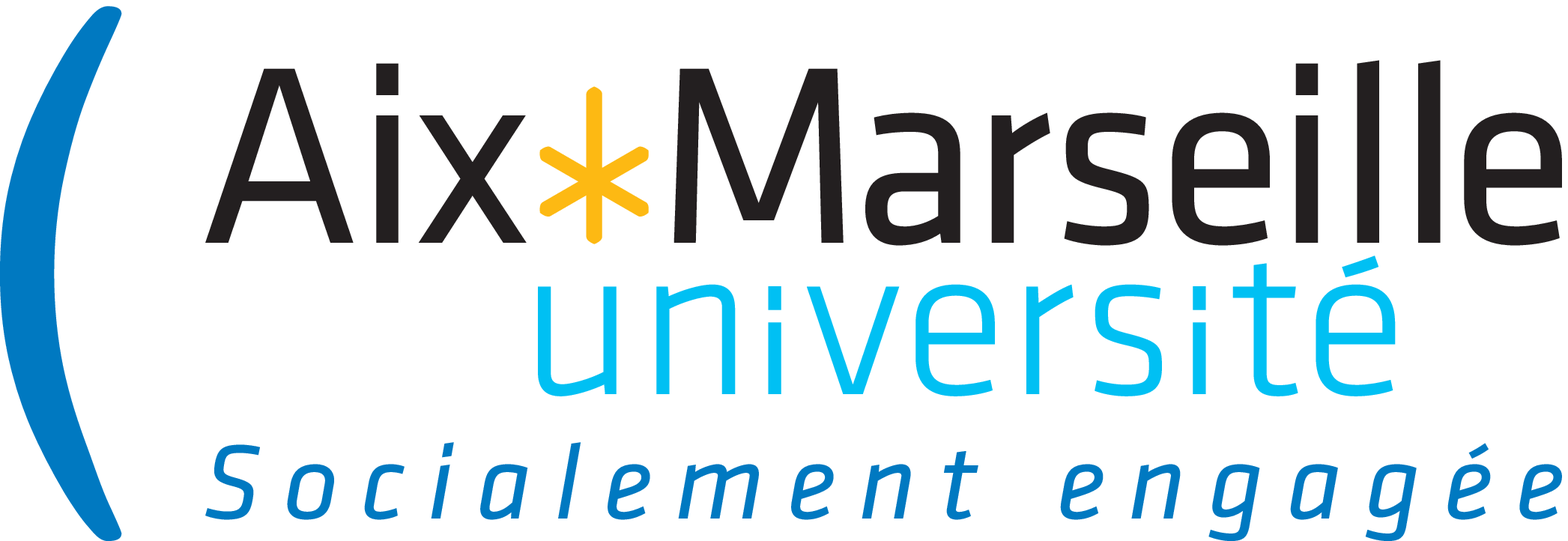}
 
	\end{minipage}\hfill
	\begin{minipage}[c]{.5\linewidth}
	\end{minipage}\hfill
\end{center}

\vspace{1em}

\begin{center}
	\begin{minipage}[c]{.63\linewidth}
		\dhorline{\textwidth}{4pt}
	\end{minipage}\hfill
	\begin{minipage}[c]{.35\linewidth}
		\raggedleft\titl{NNT/NL : 2020AIXM0001/001ED000}
	\end{minipage}\hfill
\end{center}


\begin{flushleft}
    \titb{\HUGE\textcolor{cyanamu}{THÈSE DE DOCTORAT}}\\
	\Large Soutenue à Aix-Marseille Université\\
	\Large le 18 novembre 2022 par\\
\end{flushleft}
\vspace{2em}
\begin{center}
	 \Huge \textbf{Gaston Nieuviarts} \\
    \vspace{1em}
	 \LARGE \textbf{Géométrie non commutative\\
	 	et théories de jauge sur les algèbres $AF$.} \\
\end{center}

\vspace{4em}

\begin{center}
	\begin{minipage}[t]{.45\linewidth}
    	    \vspace{.5em}
        	 \textbf{Discipline} 
        	
        	 Physique et Sciences de la Matière 
        	
    	    \vspace{1em}
        	 \textbf{Spécialité} 
        	
        	Physique Théorique et Mathématique
        		
    	    \vspace{2em}
        	\textbf{École doctorale}
        	
        	ED 352
        	
    	    \vspace{1em}
        	\textbf{Laboratoire/Partenaires de recherche}
        	
        	Centre de physique théorique (CPT), Luminy
        	
	\end{minipage}\hfill
	\begin{minipage}[t]{.03\linewidth}
	    \dvertline{4pt}{-16em}
	\end{minipage}\hfill
	\begin{minipage}[t]{.52\linewidth}
	    \vspace{.5em}
    	 \small \textbf{Composition du jury }

	    \vspace{1em}
    	 
        \begin{tabular}{p{12em} p{9.5em}}
        	Walter van Suijlekom & Rapporteur \\
        	Radboud University \\
        	\\
        	Patrizia Vitale & Rapporteure \\
        	Frederic II University \\
            \\
            Pierre Martinetti & Examinateur \\
        	Genova University \\
            \\
             Roberta Iseppi & Examinatrice \\
            Göttingen University \\
            \\
             Christoph Stephan & Examinateur\\
            Postdam University \\
            \\
        	Serge Lazzarini & Président du jury \\
        	Aix-Marseille University \\
            \\
        	Thierry Masson & Directeur de thèse \\
        	Aix-Marseille University \\
        \end{tabular}
         
	\end{minipage}\hfill
\end{center}       

\vspace{2em}

\begin{center} 
	\begin{minipage}[c]{.25\linewidth}
		\centering\includegraphics[height=5em]{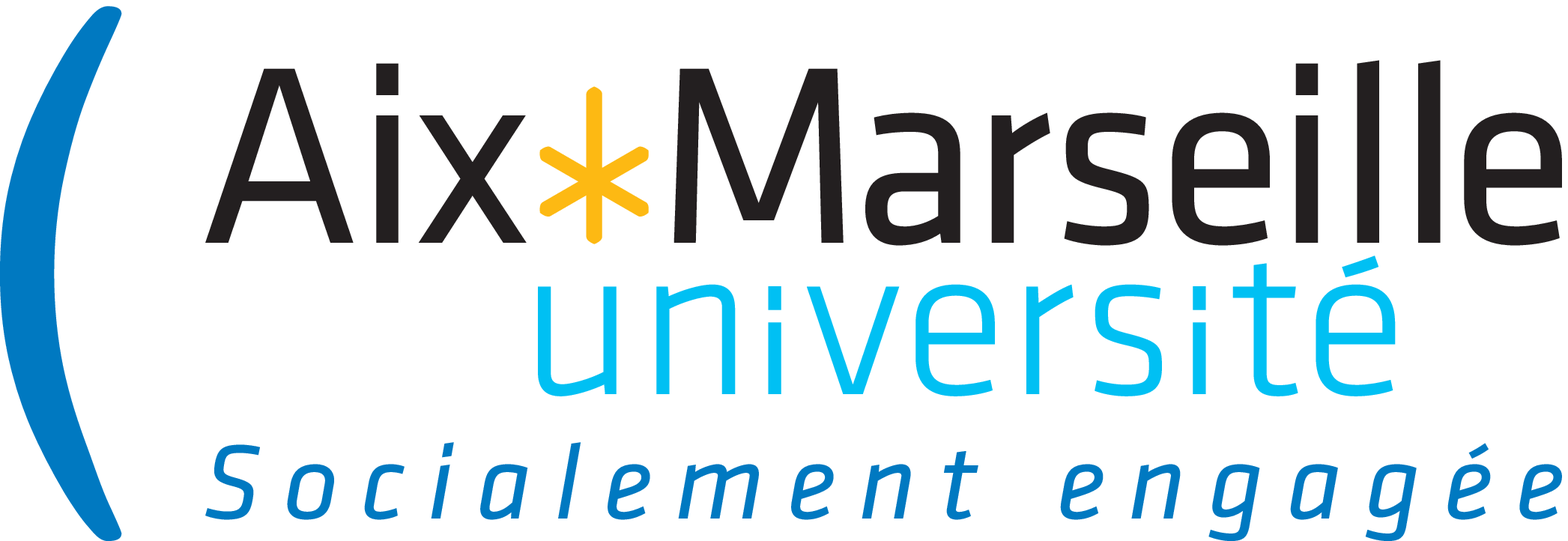} 
	\end{minipage}\hfill
	\begin{minipage}[c]{.25\linewidth}
		\centering\includegraphics[height=5em]{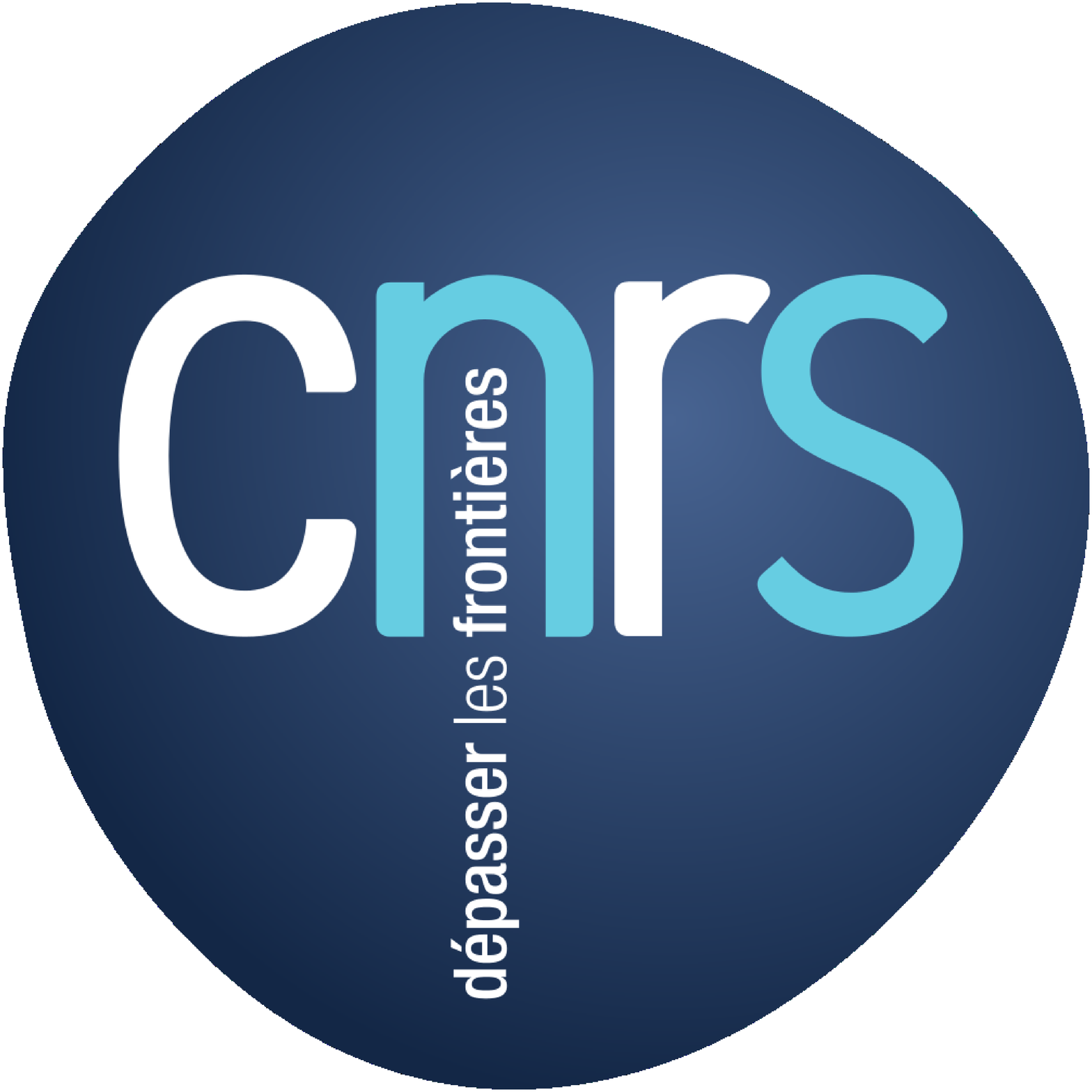} 
	\end{minipage}\hfill
	\begin{minipage}[c]{.25\linewidth}
		\centering\includegraphics[height=5em]{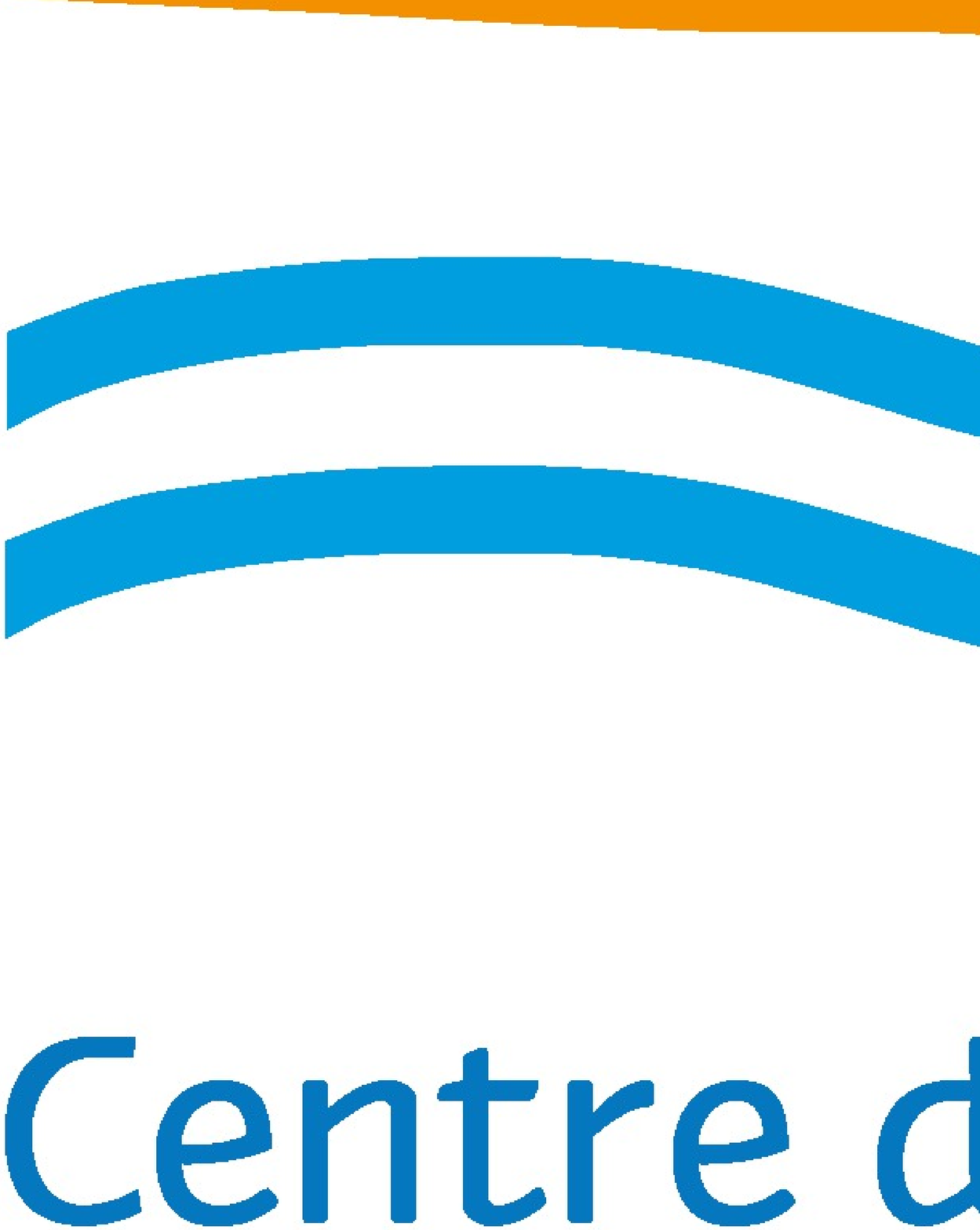}
	\end{minipage}\hfill
\end{center}
\thispagestyle{empty}
\restoregeometry

\renewcommand\figurename{Fig.}

{
\makeatletter\def\@fnsymbol{\@arabic}\makeatother 

\title{Noncommutative Geometry\\
and Gauge theories on $AF$ algebras}
\author{ G. Nieuviarts\\
{\small Centre de Physique Théorique}%
\\
\small{Aix Marseille Univ, CPT, Marseille, France}\\[2ex]
}
\date{}
\maketitle
}
  
\setcounter{tocdepth}{1}

\newpage

\newpage
 
\chapter*{Liste de publications et participation aux conférences}
 
\subsection*{Liste des publications réalisées dans le cadre du projet de thèse:}
\begin{enumerate}
\item Derivation-based Noncommutative Field Theories on AF algebras, Authors: Thierry Masson, Gaston Nieuviarts (arXiv:2106.08358)
\item Lifting Bratteli Diagrams between Krajewski Diagrams: Spectral Triples, Spectral Actions, and AF algebras:  Thierry Masson, Gaston Nieuviarts (arXiv: 2207.04466)
\end{enumerate}

\subsection*{Participation aux conférences et écoles d’été au cours de la période de thèse:}
\begin{enumerate}
\item Young Researchers
	Symposium (YRS)
	29 - 31 July 2021 Geneva,
	Switzerland
\item International Congress
on Mathematical Physics (ICMP)
2 - 7 August 2021 Geneva,
Switzerland
\item The 34th International Colloquium on Group Theoretical Methods in Physics (ICGTMP) 18 - 22 July 2022 Strasbourg, France
\item Noncommutative geometry: metric and spectral aspects 28 - 30 september 2022 Crakovia, Poland
\end{enumerate}

\chapter*{Résumé}
 
 \vspace{0.5cm}
\textbf{Mots clefs:} Géométrie non-commutative, théorie de Jauge, théories Grand-Unifiés (GUT), AF-algebres. 

\vspace{0.5cm}
La géométrie non-commutative (GNC) est une discipline mathématique développée dans les années 90 par Alain Connes. Elle se présente comme la nouvelle généralisation de la géométrie usuelle, englobant et dépassant le cadre Riemannien, et ce dans un formalisme purement algébrique. À l'instar de la géométrie Riemannienne, la GNC possède elle aussi des liens avec la physique. En effet, la GNC a fourni un cadre puissant pour reformuler le Modèle Standard de la Physique des Particules (SMPP) en tenant compte de la relativité générale, et ce, en une seule représentation "géométrique", basée sur les Théories de Jauge Non-Commutatives (NCGFT). De plus, cette réalisation offre un cadre propice à l'étude des diverses possibilités pour aller au-delà du SMPP, comme les Théories Grands Unifiées (GUT). Cette thèse cherche à montrer une méthode élégante, récemment développée par Thierry Masson et moi-même proposant un schéma général pour élaborer des GUTs dans le cadre des NCGFT. Ceci concerne l'étude de NCGFTs basées sur les $C^*$-algebres Approximativement Finies (AF-algèbres), en utilisant soit les dérivations de l'algèbre, soit des triplets spectraux afin de mettre en place la structure différentielle sous-jacente à la théorie de Jauge. La séquence inductive définissant l'algèbre AF est utilisée pour permettre la construction d'une séquence de NCGFTs de types Yang-Mills Higgs, le rang $n+1$ pouvant ainsi représenter la théorie grande unifiée de celle du rang $n$. Le principal avantage de ce cadre est de contrôler, à l'aide de conditions adéquates, l'interaction des degrés de liberté le long de la séquence inductive sur l'algèbre AF, et de suggérer un moyen d'obtenir des modèles de type GUT, tout en offrant de nombreuses voies d'exploration théorique pour aller au-delà du SMPP.

\chapter*{Abstract}
 
\selectlanguage{english}

\vspace{0.5cm}
\textbf{Keywords:} Non-commutative geometry, Gauge theory, Grand Unified Theories (GUT), AF-algebras. 

\vspace{0.5cm}
Non-commutative geometry (NCG) is a mathematical discipline developed in the 1990s by Alain Connes. It is presented as a new generalization of usual geometry, both encompassing and going beyond the Riemannian framework, within a purely algebraic formalism. Like Riemannian geometry, NCG also has links with physics. Indeed, NCG provided a powerful framework for the reformulation of the Standard Model of Particle Physics (SMPP), taking into account General Relativity, in a single "geometric" representation, based on Non-Commutative Gauge Theories (NCGFT). Moreover, this accomplishment provides a convenient framework to study various possibilities to go beyond the SMPP, such as Grand Unified Theories (GUTs). This thesis intends to show an elegant method recently developed by Thierry Masson and myself, which proposes a general scheme to elaborate GUTs in the framework of NCGFTs. This concerns the study of NCGFTs based on approximately finite $C^*$-algebras (AF-algebras), using either derivations of the algebra or spectral triples to build up the underlying differential structure of the Gauge Theory. The inductive sequence defining the AF-algebra is used to allow the construction of a sequence of NCGFTs of Yang-Mills Higgs types, so that the rank $n+1$ can represent a grand unified theory of the rank $n$. The main advantage of this framework is that it controls, using appropriate conditions, the interaction of the degrees of freedom along the inductive sequence on the AF algebra. This suggests a way to obtain GUT-like models while offering many directions of theoretical investigation to go beyond the SMPP.

\chapter*{Remerciements}
\thispagestyle{empty}

First of all, I would like to thank Thierry Masson for having accepted to supervise this work. He introduced me to the NCG's landscapes, communicating to me (as much as he could), with pedagogy and patience his taste for the organization of ideas and works as well as for simplicity and clarity in formulating things. I am very grateful for his presence, his support, his kindness, and his ability to guide me in the best possible ways according to my personality. I appreciated his disponibility, which went far beyond the scope of this thesis, as we spent hours talking about many things, concerning all kinds of topics. 
\medskip 
\par 
I would like to thank Thierry Martin, Head of the Center of Theoretical physics, for having received me in this laboratory, where I enjoyed the working atmosphere during my thesis. I would like to thank very specially Serge Lazzarini who took the time to read my entire thesis and to make a number of important suggestions for improvement. I would also like to thank Thomas Krajewski, Thomas Schücker, and Laurent Raymond who were always present during my thesis, for discussions, advice, and to teach me things.
\medskip 
\par 
I would also thank my colleagues and friends of the CPT, in particular (in an order that does not matter) Loïc Marsot San, Sophie Mutzel San, Aurélien Cordonnier, Antoine Rignon Bret, Kenza Zeghari, Jordan François, and many others, for all their discussions, their help, or their friendship.    
\medskip 
\par 
I would also like to thank Patrizia Vitale and Walter van Suijlekom for agreeing to be the referees of my thesis, Serge Lazzarini for being the president of the jury, and Pierre Martinetti, Roberta Iseppi for agreeing to be the examiners and Christoph Stephan to be guest.
\medskip 
\par
I don't know how to thank my parents for always being available during my studies, and for creating the atmosphere I needed.
\medskip 
\par
And finally, I would like to thank the Nature, for reasons that seem obvious.

\tableofcontents
 
\chapter*{Symbols and Acronyms}
The symbols table is organized with a logic that respects a given view of algebraic structures and the order in which they can be constructively considered. First, the various algebras used here are presented. Then the module and Hilbert space structures on which these algebras can act. The third block concerns other structures that can be extracted from the algebra, such as the character space and the manifold. Then we consider the various constructions on these last structures, tangent spaces, and others. The 5th block concerns automorphisms and groups acting on the algebraic structures. The 6th block concerns metric and scalar product structures. The 7th block deals with the operators used to study differential structures, and the connections that can be associated with these structures in the 8th one. Finally, the 9th block contains the various operators that are useful in NCG, while the last block contains all the other symbols used in this thesis.

\medskip
\par 
 
\begin{flalign*}
	&\algA&  &\text{an algebra}\qquad\qquad\qquad\qquad\qquad\qquad\qquad\qquad\qquad\qquad\qquad&\\
	&\af&  &\text{a finite algebra}&\\
	&\calC^\infty(\Man)&  &\text{the algebra of differentiable functions over $\Man$}&\\
	&\algA_{SM}&  &\text{the finite algebra of the SMPP}&\\
	&\halgA \defeq \calC^\infty(\Man)\otimes \af&  &\text{the algebra of the AC-manifold}&\\
	& &  &\text{ }&\\
	&\modM&  &\text{a module (over an algebra)}&\\
	&\calH&  &\text{the Hilbert space}&\\
	&L^2(\SpinBun)&  &\text{the Hilbert space of square integrable sections of a spinor bundle $\SpinBun$}&\\
	&\calH_{SM}&  &\text{the Hilbert space/fermionic representation of the SMPP}&\\
	&\hshA&  &\text{the Hilbert space of the AC-manifold}&\\
	& &  &\text{ }&\\
	&\mathbf{\Man}(\calA)&  &\text{the set of all characters of $\algA$}&\\
	&\Man&  &\text{a manifold}&\\
	&\Fin&   &\text{a finite set/space}&\\
	&\hMan=\Man\times\Fin&  &\text{the AC-manifold}&\\
	& &  &\text{ }&\\
	&\Gamma^\infty(\Man)&   &\text{the Lie algebra of smooth vector fields on $\Man$}&\\
	&\TanBun&   &\text{the tangent bundle over $\Man$}&\\
	&\Gamma(E)&   &\text{the space of all sections on the total space $E$}&\\
	& &  &\text{ }&\\
	&\Aut(\algA)&  &\text{the automorphisms space of $\algA$}&\\
	&\Out(\algA)&  &\text{the Outer automorphisms space of $\algA$}&\\
	&\Inn(\algA)&  &\text{the Inner automorphisms space of $\algA$}&\\
	&\Diff(\Man)&   &\text{the space of diffeomorphism of $\Man$}&\\
	&\Pi,\Lambda&  &\text{the module and algebra automorphism}&\\
	&G&  &\text{the finite dimensional structure group}&\\
	&\lig=Map(\Man, G)&  &\text{the Gauge group (locally) parameterized by $\Man$}&\\
	&\lieAlg&  &\text{the Lie algebra of $\lig$}&\\
	&\Aut(\halgA)=\Diff(\Man)\ltimes\lig &  &\text{the automorphisms of $\halgA$}&\\
	& &  &\text{ }&\\
	&\tilde{g}&   &\text{the metric on $\Man$}&\\
	&g&   &\text{the metric on $\algA$ or $\calH$}&\\
	& \langle \,\, |\,\,\rangle&  &\text{the local Hodge inner product induced by $\tilde{g}$}&\\
	&(\,\, |\,\,)&  &\text{the Hodge inner product}&\\
	& \langle \,\, ,\,\,\rangle&  &\text{the local scalar product}&\\
	&(\,\, ,\,\,)&  &\text{the scalar product}&\\
	& &  &\text{ }&\\
	&\Der(\algA)&  &\text{the space of all derivations on $\algA$}&\\
	&\kX&  &\text{a derivation over an algebra}&\\
	&\dd&  &\text{the differential}&\\
	&D&  &\text{the Dirac operator}&\\
	&\DM&  &\text{the Dirac operator on the spin manifold}&\\
	&D_{SM} &  &\text{the Dirac operator of the SMPP}&\\
	&\DhA&  &\text{the Dirac operator of the AC-manifold}&\\
	& &  &\text{ }&\\
	&\omega&  &\text{a (sometime connection) one form}&\\
	&\DhA[,\omega]&  &\text{the fluctuated Dirac operator }&\\
	&\Phi&  &\text{the Higgs field}&\\
	&A_\mu&  &\text{a Gauge potential field} &\\
	&\nabla_{\kX}&  &\text{a non-commutative connection along the derivation $\kX$}&\\
	& &  &\text{ }&\\
	&J&  &\text{the reality operator}&\\
	&\JM&  &\text{the reality operator of the spin manifold}&\\
	&J_{SM}&  &\text{the reality operator of the SMPP}&\\
	&\JhA&  &\text{the reality operator of the AC-manifold}&\\
	&\gamma&  &\text{the grading}&\\
	&\gammaM&  &\text{the grading over the spin manifold}&\\
	&\gamma_{SM}&  &\text{the grading of the SMPP}&\\
	&\gammahA&  &\text{the grading of the AC-manifold}&\\
	& &  &\text{ }&\\
	&SMPP&  &\text{standard model of particles physics}&\\
	&QM&  &\text{quantum mechanics}&\\
	&GR&  &\text{general relativity}&\\
	&GFT&  &\text{gauge field theory}&\\
	&GUT&  &\text{grand unified theory}&\\
	&C&  &\text{commutative}&\\
	&NC&  &\text{noncommutative}&\\
	&NCG&  &\text{noncommutative geometry}&\\
	&NCSMPP&  &\text{noncommutative standard model of particles physics}&\\
	&AF&  &\text{almost finite}&\\
	&\EspPhas&  &\text{a state phase space}&\\
	&\EspConf&  &\text{a state configuration space}&\\
	&\phi&  &\text{an injective morphism}&\\
	&\psi&  &\text{a state/matter field element in a Hilbert space $\calH$}&\\
	&\lag&  &\text{the Lagrangian}&\\
	&\act&  &\text{the action}&\\
	&\Dst(\psi)&  &\text{a space-time domain over which $\psi\psi^*\neq 0$}&\\
	&\Oset&  &\text{the collection data of observation changes}&\\
	&ER&  &\text{an element of (physical) reality}&\\
	&\StrMath&  &\text{an "abstract" mathematical structure}&\\
	&\pi_\StrMath(ER)&  &\text{the $\StrMath$-representation of the $ER$}&\\
	&Mes(ER)&  &\text{the measurement process of the $ER$}&\\
	&O(ER)&  &\text{observable deduced from the $ER$}&\\
	&F(ER)&  &\text{the resulting $ER$ after the measurement process}&\\
	&UP&  &\text{uncertainty principle}&\\
	&CP&  &\text{complementarity principle}&\\
	&RC&  &\text{repeatability criterion}&\\
	&KC&  &\text{kickability criterion}&\\
	&OR&  &\text{observable realism}&\\
	&ONR&  &\text{observable non realism}&\\
	&SR&  &\text{state realism}&\\
	&SNR&  &\text{state non realism}&\\
	&\psi\text{-OSR}&  &\text{$\psi$-ontic theory admitting SR}&\\
	&\psi\text{-ESR}&  &\text{$\psi$-epistemic theory admitting SR}&\\
	&\psi\text{-ESNR}&  &\text{$\psi$-epistemic theory admitting only SNR}&\\ 
\end{flalign*}

\thispagestyle{empty} 

\chapter*{Introduction}
\addcontentsline{toc}{chapter}{Introduction}
\label{sec Intro}

\setcounter{page}{1}
\medskip
\par 
In this last century, theoretical physics has known two important paradigm changes, each of them going in its way beyond the thought framework offered by Newton. These are General Relativity (GR) and Quantum Mechanics (QM). The first is written in the framework of differential geometry, and the second is in a pure algebraic framework. The first describes more precisely what space-time is, the second does the same for the matter and light.
\medskip
\par 
The formalism in which the physical processes obeying the quantum behavior are described is that of operators acting on Hilbert spaces. This mathematical structure did not exist before QM, it was progressively set up while the quantum theory was being built. It can therefore be considered as physically motivated. The functions of the classical mechanic's framework (CM) are thus replaced by operators acting on the states described as elements in Hilbert spaces. Several important novelties appear as simple consequences of this. Operators become free not to commute, superposed states can coexist and non-factorizable or entangled states (which are not tensor products states) combining more than two different particles appear. These three aspects alone induce almost all the non-classical aspects of quantum. This is the Non-Commutativity (NC) of observables (which are available data, extracted from the world through experiments performed on it) which is the main origin of the need for such a formalism. The CM could also be described in this framework (as we will see in \ref{KvN}), but it would be too complicated because the latter only requires simple phase spaces (with commuting functions) to describe the observables and the mechanics of the concerned processes. 
\medskip
\par 
This was W. Heisenberg who first discover the NC properties of QM observables, giving a formulation of this fact in terms of matrix mechanics. Then von Neumann reformulated QM using operators acting on Hilbert spaces as elements of $C^*$-algebras. He was the first to mention the potential relevance of an extended geometry for which functions algebras are replaced by NC ones. Then P.A.M. Dirac followed him in this presentiment of the need to introduce such concepts for the QM, making the analogy between Poisson's bracket in CM which is linked to differentiable structures, and the QM's commutator. But these were feelings, analogies, and incomplete conjectures since no such things as geometry were drawn. In 1943, I. Gelfand and M. Naimark proved the so-called Gelfand-Naimark theorem which will be presented in \ref{GNMath}. This theorem offers a strong link between states on such algebras and topological properties, offering a path to extract topological data such as the notion of points of an underlying space from these algebras. This was an important step to understand how to “make speak” these algebras about the “geometry” of the space-time in which they live.
\medskip 
\par 
It is only in the 1990's that A. Connes, set up the complete conceptual framework of NC differential geometry, concluding satisfactorily on these preconceptions. By revealing the structures extending the usual geometrical framework, and showing how to study these structures, he offered to the world of physicists and mathematicians a new, more general, and clearly defined geometrical framework: Non-Commutative Geometry (NCG).     	
\medskip
\par 
NCG has since then inspired many fields within pure mathematics and theoretical physics. It is not only a generalization of the Riemannian geometry, but a reformulation of this one (for its commutative restriction) in a completely new and purely algebraic framework, being free to receive the NC of operators and to go beyond the notion of point. It is all the more interesting to work on the algebraic side that we can notice a very simple fact concerning our understanding in the form of a geometrical representation of what is space-time, it is through the observables that we receive all the information that makes us determine space-time. It seems, therefore, more logical to have a theory that starts from this algebra of observables as fundamental objects, rather than from a hypothetical geometrical spacetime, more arguments in this sense will be given in chapter \ref{GeomAndSpacetime}. 
\medskip
\par
The general way of doing, taken by many researchers for applying NCG to various domains was to take objects belonging to the usual commutative framework (function algebra, CM, Riemannian geometry, groups...) which we will designate as belonging to Commutative Geometry (CG), then find its algebraic re-formulation, then try to extend this to NCG framework dropping some commutativity conditions or changing the product:
\begin{align*}
	\text{CG} \to \text{NCG}.
\end{align*}
This offers numerous generalizations of common concepts. In physics, this way of thinking is often linked to the quantization process of observables, especially when the relaxed commutation relation concerns the latter.
\medskip
\par
There is another way to see things, in the reverse order: $\text{NCG} \to \text{CG}$. This way is less intuitive since our intuition is based on classical representations and then starts from CG. In physics, this may be called dequantization. I nevertheless think that this is a more fundamental and potentially fruitful way to see things, in mathematics and physics. For the mathematician who is familiar with NCG, the emerging feeling can be that it is a more natural conceptual framework since “commutative mathematics” appear as restrictions, sometimes losing nice properties and then looking like to be secondary. Non-commutativity appears as the negation of commutativity, but it is a negation in the direction of the enlargement, and in my opinion, the more fundamental level of mathematical structures are most of the time in this direction. Then taking the continuation of classical (or commons) concepts to go to their NC generalization seems not to be totally justified for the one who wants to find “good” NC generalizations. 
\medskip
\par
In physics, colossal efforts have been made to make NC, \textit{i.e.} to quantify classical concepts, starting from classical theories. It gives results as for math but brings many problems as well. The fact that no general quantization scheme has been found is for me an indicator of the limitation of this first way of doing. For those who believe that the classical world must be an emerging picture from pure quantum phenomena (which seems to be reasonable since it is composed of such objects with such behaviors), the correct path, again, is to find a pure quantum description of Nature and then see how the classical picture emerges. QM is actually a hybrid theory, parameterized by the classical world, for its outcomes specifically, but intrinsically not reducible to CM. My hope is then that a pure NC-inspired mechanics will replace QM, giving asymptotically the classical picture, for a large number of particles and/or denser environments. Decoherence is a good candidate in this direction.
\medskip
\par 
Strong mathematical reasons have motivated the creation of NCG. An example is provided by the Gelfand-Naimark theorem, which is at the origin of the study of NC topology. By providing a link between algebra and topology, it opened the door to the exploration of NC topology (induced by NC algebras), that the formalism of QM invited us to cross. We thus have various equivalences between algebraic and topological properties, like the notion of point and pure state on a commutative algebra, compactness, and unitarity, or the property of connectedness equivalent to the one of the non-existence of (non-trivial) projections for $C^*$-algebras (see section \ref{GNMath} for more details).
\medskip
\par 
A second important aspect concerning the study of topological properties via algebraic properties is provided by the invariants of K-theory which we will address in section \ref{Ktheory}. This is one of the main “tools” for the study of non-commutative spaces.
\medskip
\par 
Another aspect related to K-theory concerns the work of A. Connes on the continuous cyclic cohomology which appears to be related to the de Rham homology in the case of commutative algebras. This notion does not yet introduce a differentiable structure that would generalize the de Rham complex of differential forms. However, these results, in addition to their links with the Atiyah–Singer index theorem, offer strong motivation for the introduction of spectral triples with Dirac operators to define a differential structure.
\medskip
\par 
Again linked to K-theory, another motivation is provided by the Serre-Swan theorem. The latter provides an explicit link between fiber bundles (at the basis of the formulation of gauge theories (GFT)) and projective modules of finite type on the algebra (see section \ref{SerSwann}). It is this double link with K-theory and gauge theories that provides one of the main mathematical motivations for the construction of gauge theories on $AF$ $C^*$-algebras, which is the subject of this thesis.
\medskip
\par    
The strong points of this new “geometrical” framework are that it allows to treat both the discrete and the continuous within the same formalism (operators and Hilbert spaces), to find all the known results in Riemannian geometry, while allowing to geometrize the other forces (see chapter \ref{NCSMPP}), and to establish more links with other mathematical fields. A nice review on NCG's landscape can be found in \cite{connes2019noncommutativeSpectralStandpoint}, for reference books, see \cite{Connes94noncommutativegeometry} and \cite{gracia2013elements}.
\medskip
\par
From the physicist's point of view, a strong motivation for NCG comes from the simple fact that this mathematical field was initially developed for the pure physical purpose and that mathematics tells us clearly that NC algebras like those of observables in QM cannot be thought of as living on the points of a manifold. In theoretical physics, the quantization of a theory means to find a way to link the commutative operators underlying the classical observables to the NC operators associated with the quantum equivalent of these observables. A natural way to introduce a NC algebra is to deform a commutative algebra of functions. For example, one can consider Moyal quantization, which deforms a function algebra with a Poisson structure. However, there are non-commutative algebras that are not deformations of function algebras, like for example the algebra of matrices. This algebra being one of the most famous, it will be used as a toy model to study NC structures throughout this thesis (see Chapters \ref{FiniteAlg} and \ref{ChapDiffMatrixAlg}). There is no theory explaining in a systematic way the quantization process, which seems to be problematic from the point of view of conceptual clarity. One of the great challenges in physics today is to find a quantum theory of gravitation, and thus to “quantize space-time”. If the motivation for such an approach is well founded, it seems clear now that we lack theoretical and conceptual arguments to know how to carry it out.
\medskip
\par 
In physics, Lie algebras and groups are omnipresent, mainly in gauge theory. NCG proposes a natural generalization of these structures, Hopf algebras, and quantum groups. Initially introduced to study the cohomology of Lie groups, a Hopf algebra is a bialgebra with an operation called antipode, generalizing the notion of the inverse operation in groups. A particular type of Hopf algebras are the quantum groups. They are often deformed Hopf algebras, generally NC. These objects are at the heart of NCG and are considered as the natural generalization of function algebras on a Lie group manifold, when this became a NCG. Indeed, the deformed algebra can be considered as the algebra of functions on a NCG space. Some well-known books on quantum groups are for instance \cite{CharPres94a} and \cite{KlimSchm97a}.
\medskip
\par 
In physics, example of NC spaces using NC coordinates are given by the $\kappa$-Minkowski space \cite{Mass39,Mass40,pachol2015kappa} (whose coordinate algebra contains the bicrossproduct of the Hopf algebra of the group of $\kappa$-Poincaré, describing its symmetries) and the Moyal space \cite{gracia1988algebras,Mass32,martinetti2013noncommutative,wallet2008noncommutative}. One of the main motivations behind such an introduction was to reconcile gravity with QM and to explore the intuition that such spaces can help to solve the re-normalization problems. However, nothing in NCG or in the current description of QM implies that the coordinates are NC. NCG is primarily concerned with NC algebras and is not just about making coordinates noncommutative. These attempts should therefore not be considered as the only possible ones, and it would be interesting to study generalizations of spacetime of another kind, without necessarily making the coordinates NC. 
\medskip
\par 
One of the greatest progresses made with NCG is to have succeeded in going beyond the equivalence with topology and measure theory, to include and extend differential geometry (differentiable manifolds, Lie groups, fiber bundles, connections...). These structures are of major importance in mathematics and particularly in theoretical physics. Indeed, they are of primary interest in field theory, as they are the basis for the formulation of all known forces, whether it is gravitation in the framework of GR or the forces of the Standard Model of Particles Physics (SMPP) in the framework of Gauge theories. We will explore the two main ways to extend the framework of differential geometry, the one based on derivations of the algebra in section \ref{DeDS}, and the one based on spectral triplets in section \ref{DiDS}.
\medskip
\par 
The application of the NCG in field theory has been done following two main ways. The first one concerns the re-foundation of the structure of field theories, by changing the background space-time on which they are defined. One of the main attempts concerns the replacement of the background space-time with a Moyal space. However, theories on Moyal space are not renormalizable and suffer from the phenomenon called
Ultraviolet/Infrared (UV/IR) mixing. A significant advance was made in 2004, the proposal of an additional term to the action (harmonic oscillator term) leads to a fully renormalizable NCFT. More details can be found in \cite{grosse2005power,grosse2005renormalisation}, and a nice review on the topic is given by \cite{wallet2008noncommutative}.
\medskip
\par
The second path concerns the reformulation of field theories within the framework of NCG, but retaining the nature of the field and background space-time theories. This is the path followed throughout this thesis, the NCSMPP (presented in chapter \ref{NCSMPP}) being the main realization of this field of applications of the NCG.
Indeed, NCG has been used to develop GTF (hereafter mentioned as NCGFT) in which scalar fields are part of the generalized notion of connections. Then, the naturally constructed Lagrangians produce quadratic potentials for these fields, providing Spontaneous Symmetry Breaking Mechanisms (SSBM) in these models. Our works with T. Masson can be located in this last field of applications of NCG \textit{i.e.} the elaboration of NCGFTs.    
\medskip
\par
GFTs are an essential ingredient to model high energy particle physics. From the pioneer work by Yang and Mills to the Standard Model of Particle Physics (SMPP), the gauge principle has shown how insightful it is both technically and conceptually, and the search for the “right gauge group” has stimulated physicists to construct Grand Unified Theories (GUT). Unfortunately, none of these theories has been retained until now as a convincing model beyond the SMPP. 
\medskip
\par 
The way these GUT are constructed relies on the classical mathematics of fiber bundles and connections: the gauge groups (infinite dimensional spaces) are the groups of vertical automorphisms of principal fibers over space-time with some convenient structure groups (finite dimensional Lie groups). The choice of the structure group is then the choice of the gauge group of the theory. GUT rely on finite dimensional Lie groups which are “big enough”, for instance $SU(5)$, to contain the group of the SMPP, $U(1)\times SU(2) \times SU(3)$. This unifying approach (for interactions) is then controlled by the choice of possible “not too large” finite dimensional Lie groups. This group has to be “not too large” because one has to reduce it to the group of the SMPP that we actually see in experiments, and the larger the original group, the more hypothesis it requires to perform this reduction, usually using some successive SSBM.
\medskip
\par 
Since the 90's, noncommutative geometry (NCG) has shown that one can construct more general gauge field theories in a natural way (see \cite{ConnLott90a, DuboKernMado90a, DuboKernMado90b} for the seminal papers and \cite{Suij15a} for a review and references for more recent developments). NCG has permitted to include in a natural way the scalar fields used in the SMPP to make manifest the SSBM which gives masses to fermions and some of the gauge particles. In this approach, the gauge group is the group of inner automorphisms of an algebraic structure, which in general is an associative algebra (it could be the group of automorphisms of a module in some cases). According to the NC differential structure we choose, two main frameworks can be used to construct NCGFT. The first one (from a historical perspective), initially proposed by M. Dubois-Violette, R. Kerner, and J. Madore in \cite{DuboKernMado90a, DuboKernMado90b} concerns the use of derivation-based differential structure as fundamental structure. The second one, mainly developed by A. Connes, J. Lott, A. Chamseddine, and M. Marcolli in \cite{ConnLott90a,chamseddine1997spectral,ChamConnMarc07a} concerns the use of spectral triples based differential structure. These ways to build NCGFT will respectively be called derivation and spectral triple-based NCGFT.
\medskip
\par 
With T. Masson, we start the investigation of a new natural approach to “unifying” noncommutative gauge field theories (NCGFT). This approach is based on approximately finite-dimensional ($AF$) $C^*$-algebras, a very natural class of algebras in NCG (see \cite{Blac06a, Davi96a, RordLarsLaus00a} for instance). By definition, $AF$ $C^*$-algebras are inductive limits of sequences of finite-dimensional $C^*$-algebras. Let us recall the following two important points (see Sect.~\ref{AFalg} and Sect.~\ref{sec AFA} for more details):
\begin{enumerate}
	\item A finite-dimensional $C^*$-algebra is, up to isomorphism, a finite sum of matrix algebras: $\algA = M_{n_1} \toplus \cdots \toplus M_{n_r}$ where $M_n \defeq M_n(\bbC)$ is the space of $n \times n$ matrices over $\bbC$. For a manifold $M$, NCGFT have been investigated on the algebras $C^\infty(M) \otimes \algA$ (referred to in the literature as “almost commutative” algebras) and these NCGFT are of Yang-Mills-Higgs types.
	
	\item An $AF$ $C^*$-algebra is constructed in such a way that we get a control of the approximation of this algebra by the successive finite-dimensional $C^*$-algebras in its defining inductive sequence. This control (in terms of $C^*$-norms) can be used to approximate some structures defined on the $AF$ $C^*$-algebra. The best example is the one of the $K_0$-group that we briefly recall in Sect.~\ref{sec AFKT} for sake of illustration.
\end{enumerate}
\medskip
\par 
The motivation for the present approach can be summarized in the following way: point~1 suggests to use the defining inductive sequence of an $AF$ $C^*$-algebra to construct a sequence of NCGFT of Yang-Mills-Higgs types and  point~2 could be used to get some control on this sequence of NCGFT in a meaningful way as successive approximations of a “unifying” NCGFT on the full $AF$ $C^*$-algebra. This would be a way to implement inclusions of (finite dimensional) “gauge groups” as successive approximations of an (infinite dimensional) “unifying gauge group”. More details on the mathematical and physical reasons will be given (respectively) in sections \ref{SerSwann} and \ref{GUTandAF}. The setting up of the underlying general mathematical structure was done using derivation-based NCGFT in \cite{MassNieu21q} and using spectral triple-based NCGFT in \cite{masson2022lifting}. This thesis is based on these two works.
\medskip
\par 
\GN{ccl ?}Notice that the NCGFT that we can define on the full (maybe infinite dimensional) $AF$ $C^*$-algebra can be quite unusual from a physical point of view since it can involve an infinite number of degrees of freedom in the gauge sector. But if the control of approximations by “finite dimensional“ NCGFT (as suggested by point~2) is possible, then the content of this NCGFT could be understood as a limit of usual and manageable Yang-Mills-Higgs theories constructed gradually, for instance using empirical data.
\medskip
\par 
In the GUT approach, the “big enough” gauge group must contain all the empirical phenomenology of present particle physics. But, in our approach, thanks to the approximation procedure, we only require any step in the inductive limit of NCGFT to \emph{approximate} the present empirical phenomenology and the future possible discoveries in particle physics. As the probing energy increases, a better approximating NCGFT in the sequence has to be taken at a farther position. In our bottom-up approximation, the number of degrees of freedom (in the gauge sector) can increase along the sequence, contrary to the usual SSBM, which is a top-down procedure that relies on reduction of degrees of freedom.
\medskip
\par 
Notice that one might encounter a stationary sequence starting at some point, so that the full $AF$ $C^*$-algebra would be of finite dimension. In that case, the “final” NCGFT would make appear only a finite number of degrees of freedom in the gauge sector, very much like ordinary GUT. But even in that situation, some interesting features might be depending on how the successive NCGFT in the sequence (here finite) of NCGFT are connected to each other, in particular concerning the mass spectra of the gauge bosons, see Sect.~\ref{sect direct limit NCGFT}. At present time, we are not aware of any empirical fact suggesting that such a radical new approach could be relevant. But we hope that our new way to construct unifying gauge field theories beyond the SMPP could reveal new empirical content that would be suggestive to answer open questions in particle physics. Works in this direction will be done after this PhD thesis.  
\vspace{0.2cm}
\medskip
\par
\medskip
\par
\vspace{0.2cm}
This thesis has been written with the intention of synthesizing and organizing my knowledge and perspectives on this topic, and in a more general way on the meaning of NCG. This work is far from covering the whole field of NCG. It represents a selection, according to my tastes, my interests, and the perspective I have chosen to put forward to present NCG. I have preferred an approach and writing rather oriented on physical principles (observables, symmetries, degrees of freedom...) for the first three parts because I think that other researchers having more taste and competence to deal with the most mathematically abstract aspects of NCG had already depicted in a sufficiently complete way these fundamental points. I hereby express my hopes for future applications of NCG that will be enlightening for the physicist's mind, ideally as much as the application of Riemannian geometry to describe the structure of space was. I have written this thesis intending to address it to different types of readers on a continuous spectrum between two idealized ones. The first one would be a reader having a mathematical or theoretical physics background and who would be curious to understand some points concerning NCG for their culture or to develop new physics. They will thus be mainly interested in reading the first three parts (\ref{PartCalToNCG},\ref{part2ThinkingNCG},\ref{TowardNCGFT}). In particular, they could gain intuition by carefully following the applications of the NCG to matrix algebras in chapter \ref{ChapDiffMatrixAlg} and then to the extensions of the framework of the usual Gauge theories via the NCG. I will try throughout this text to show how the usual structures, whether topological or geometrical, such as the differential structures, the notions of point and space, and the gauge theories become when moving to their NC generalizations, in the purely algebraic framework of $C^*$-algebras. I hope to bring some light to their understanding of this subject which remains difficult to access.  At the other extreme of the spectrum of my intended readers, my alternative ideal reader is a researcher closer to the themes of NCG and Gauge theories, more able to appreciate the technical details, the direction followed during this work as well as the contribution provided to the pre-existing works, while perceiving the perspectives of research on the structure developed in the work on which this thesis is based. This reader will be more concerned by the parts \ref{part2ThinkingNCG} and \ref{partNCGFTAF}. 
\medskip
\par 
\textbf{Outline.} This thesis is divided into four parts. Part \ref{PartCalToNCG} is composed of four chapters. In chapter \ref{sec C algebras}, I recall the essentials of the theory of $C^*$-algebras, putting ahead the properties which allow to consider these algebras as noncommutative generalizations of locally compact topological spaces. In chapter \ref{FiniteAlg}, I introduce the notions of finite and approximately finite algebras (AF) which will be the basic structures on which the works of this thesis are based. These algebras, which will always be matrix algebras, are the prototype of the non-commutative algebra and offer one of the most interesting, rich, and simple examples of NC structure, thus offering a pedagogical illustration of noncommutative spaces, which will be followed throughout this thesis. Chapter \ref{DifCalc} proposes to highlight the path from usual differential geometry to its NC generalizations, these structures forming the skeleton of the NCGFT which will be developed later on. And finally, for pedagogical and practical reasons, I will present in chapter \ref{ChapDiffMatrixAlg}, how the two main methods of studying differential structures put forward earlier allow us to study such matrix algebras. This will, among other things, introduce the notations that will be used throughout this thesis. This first part can be considered as a general (and incomplete) presentation of the technical path to NCG.
\medskip
\par
Now that some technical grounds have been set up, I propose in Part \ref{part2ThinkingNCG} a general discussion around NC in QM, on the meaning of geometry in physics, and of a potential interpretation we can have of NCG.
\medskip
\par
When people asked me to explain what NCG is intuitively, I was blocked, I could answer that it is a geometry without points, that it cannot be visualized. In short, I was saying what it is not, without really understanding what it is. When I went into the literature, I realized that no author was offering such an intuitive view and that most of them stood at the mathematical level for the understanding of NCG. This part can be seen as a personal and partial answer to this question. Its claims are of particular importance for me because it seems clear today that we lack new physical ideas to find relevant and innovative ways to progress. Specially concerning relevant applications of the mathematical framework offered by NCG in physics. This is why I thought it was essential to understand the origin of NC in QM and to understand what this meant for the potential NCG that would be deduced from these observables (if it must be).
\medskip
\par
I think that it is physics that should drive our mathematical constructions. Then, without an intuitive representation of what NCG is, the physicist's intuition cannot express itself and allow them to guide themselves in the elaboration of mathematical theories and structures able to grasp the underlying structures of reality, and potentially express them in the beautiful structures revealed in the NC framework. Part \ref{part2ThinkingNCG} is then composed of five chapters. In chapter \ref{GeneralOutline}, I present the general aim and structure of the argumentation of this part. In chapter \ref{NCShadow}, I explain my view on the general misunderstanding in which NCG lives, particularly in the scope of its applications to physics. In chapter \ref{PhysTh}, I try to show what is the meaning of non-commutativity of observables in QM, what it induces, and how we can interpret it. In chapter \ref{GeomAndSpacetime}, I discuss the notion of geometry to describe space-time, its meaning, and its progressive algebraization, giving arguments to show that the algebraic view is more fundamental than the geometrical one, with the points as primary objects. Then in chapter \ref{ObjectiveNCGFromQM}, using all the arguments presented in this part and arguing that the formalism of QM must be taken seriously concerning its geometrical consequences, I present some nice potential geometrical implications coming from such a formalism. Part \ref{part2ThinkingNCG} corresponds to insights that I want to submit to an editorial review, once I have made more progress on clarifying the suggested ideas.
\medskip
\par
Part \ref{TowardNCGFT} is composed of three chapters. In chapter \ref{SMPPGFT}, I propose a presentation of the gauge principle, and its mathematical formulation in fibers bundles while presenting the SMPP, its links to gravitation, and the ways to go beyond this theory. Then in chapter \ref{NCGFTGener}, I show how these structures can be generalized in the NC framework, and then show how this is done when using derivations-based and then spectral triplets-based differential structures. I conclude this part by chapter \ref{NCSMPP} with a very general presentation of the NCSMPP, explaining how it offers a unification of GR and SMPP at the classical level while providing an adequate mathematical framework for thinking about formulations of theories going beyond SMPP. 
\medskip
\par
Part \ref{partNCGFTAF} is composed of four chapters. In chapter	\ref{sec AFA}, I present the general motivation behind the works on which this thesis is based \textit{i.e.} why building NCGFT based on AF-algebras can offer an interesting framework to go beyond the SMPP. Then, I highlight how the embedding structure of $AF$-algebras works, and how we can connect structures along its inductive sequence using the so-called $\phi$-compatibility condition, which is an essential condition to relate NCGFTs at successive steps. Then I show how to build and relate NCGFTs along the sequence, first in the derivation-based framework in chapter \ref{sec DBA} then with spectral triples one in chapter \ref{sec STA}. I conclude part \ref{partNCGFTAF} with chapter \ref{CompMethAFNCGFT} where I summarize the results obtained so far with the derivations and spectral triples frameworks for the building of NCGFT on AF-algebras, putting into perspective where they have common features and where they differ. This part contains the main results of this thesis.
\medskip
\par
This thesis can also be partially read vertically. This may allow a reader more interested in technical details to get faster to the point. Indeed, two main guidelines have been followed throughout this thesis. The first one concerns the study of NCG via differential structures based on derivations presented in section \ref{DeDS}. The use of this method on matrix algebras is then done in section \ref{ResMatrDer}, the construction of NCGFTs based on derivations is then presented in section \ref{DBNCGFT} to finish with chapter \ref{sec DBA} by showing what it gives when the finite algebra on which the gauge theory is constructed is an algebra of type AF. The second line in this vertical reading grid concerns exactly the same four points (made in \ref{DiDS},\ref{ResMatrDirac},\ref{STNCGFT},\ref{sec STA}), but using spectral triples to construct the basic underlying differential structure. This two lines are almost verbatim transcriptions of \cite{MassNieu21q} and \cite{masson2022lifting} and are mainly addressed to the second type of readers. These to lines correspond to the main work done during this thesis.

\selectlanguage{english}
\part{\texorpdfstring{From $C^*$-Algebras to Noncommutative Geometry}{From C*-algebras to noncommutative geometry}}
\label{PartCalToNCG}
\chapter{\texorpdfstring{$C^*$-Algebras}{C*-algebras}}
\label{sec C algebras}
\epigraph{Algebra is the intellectual instrument which has been created for rendering clear the quantitative aspects of the world.}{\textit{A.N. Whitehead}}

We will try in the following to give a very general presentation of what algebras are, then $C^*$-algebras, and to explore some nice structures in this context. An algebra $\calA$ is a set $A$ equipped with some operations on it's elements $\calA=\{A,\star(1),\star(2),\dots\}$. $A$ can be a set of numbers, symbols, vectors, or functions... And the more common operations can be $+,\times, [\,,\,]$... Therefore, an algebra is a pattern of relations, a structure with respect to abstract operations between its elements. Thus, algebras are very general objects, and therefore useful in math and physics, where associative, Lie, and $C^*$-algebras are very common, with additional structures such as norm, states, spectrum... 

Fundamental concepts of quantum physics belong to a mathematical framework that cannot be reduced to the classical one; they form a generalization of the framework used in classical physics, as we will see later in section \ref{KvN}. The mathematical framework of $C^*$-algebras was mainly developed to formulate quantum physics. It is the result of a long work that can be considered as initiated by Werner Heisenberg's matrix mechanics in 1925, then transmitted in more mathematical language by Pascual Jordan around 1933, and finally by John von Neumann's work setting up a general framework for these algebras. The $C^*$-algebras thus offer a formalism that generalizes that of classical and quantum theories, while preserving as much as possible of the structures they share. In the context of physics, the elements of these $C^*$-algebras are operators called observables, they are the generalization of the notion of classical observable (functions) in the quantum world with operators, susceptible to not commute. More details can be found in \cite{landsman2006between}. We propose here to highlight the main features of the structure of $C^*$-algebras, being motivated by the fact that they form the ground where non-commutative geometry was born through the Gelfand Naimark theorem presented in section \ref{GNMath}.

\section{\texorpdfstring{General Features of $C^*$-Algebras}{General features of C*-algebras}}

\epigraph{I would like to make a confession which may seem immoral: I do not believe absolutely in Hilbert space any more.}{\textit{J. von Neumann}}

\textbf{An algebra $\calA$} on $\mathbb{C}$ is a complex vector space on which an internal law, multiplication, is defined. This multiplication can be compatible with addition: $a(b + c) = ab + ac$ $\forall a, b, c \in \calA$. $\calA$ is said to be associative if $a(bc) = a(bc)$; and unital if it contains the identity element $\bbbone$ such that $\bbbone a = a\bbbone = a$.
\medskip
\par
The \textbf{center} of an algebra $\calA$ is the subset of $\algA$ such that:
\begin{align*}
	\calZ(\algA) = \{ a \in \algA \ / \ ab = ba,\, \forall \, b \in \algA \}
\end{align*}
An \textbf{involution} on an algebra $\calA$ is a real linear application of $\calA\to \calA^*$ such that for all $a,b\in \calA$ and $\lambda\in\bbC$ we have :
\begin{align*}
	a^{**}=a\qquad\qquad (ab)^*=b^*a^*\qquad\qquad (\lambda a)^*=\overline{\lambda}a^*.
\end{align*}
\par
The pair $(V,\|\|)$ of a vector space with a norm is said to be complete for this norm if any Cauchy sequence converges.
\medskip
\par
Given a vector space $V$, a norm on $V$ defines a metric $d$ by $d(v,w):=\|v-w\|$ with $v$ and $w$ $\in V$. A vector space with a complete norm in the sense of the associated metric is called \textbf{Banach space} and will be denoted $B$. This notion of complete distance makes possible to set up topological structures on the vector space, which makes Banach spaces very useful in functional analysis.
\medskip
\par
An example of a Banach space is the Cartesian space $\mathbb{R}^n$ with norm :
\begin{align*}
	\|(x_1, x_2, ... , x_n)\|_p:=\sqrt[p]{\sum_{i=1}^n|x_i|^p}
\end{align*}
with $1\leq p\leq\infty$ taking the limit for $p$ going to infinity, we get : $\|(x_1, x_2, ... , x_n)\|_\infty=\max_i|x_i|$.

\medskip
\par
A \textbf{Banach algebra $\calA$} is a Banach space which is furthermore an algebra such that for any $a,b\in \calA$ we have :
\begin{align*}
	\|ab\|\leq \|a\|\ \|b\|
\end{align*}
A \textbf{unit element} in a Banach space $\calA$ is an element $\bbbone$ satisfying $\bbbone a=a\bbbone=a$ for all $a\in\calA$, in particular, we have $\|\bbbone\|=1$.
\medskip
\par
A \textbf{functional on a Banach space $B$} is a linear map $\rho : B\to\mathbb{C}$ which is continuous; such that $|\rho(v)|\leq C\|v\|$ for a given constant $C$ and all $v\in B$. The smallest $C$ satisfying this condition is called the norm of $\rho$ : 
\begin{align*}
	\|\rho\|:=\sup\{|\rho(v)|\quad / \quad  v\in B,\ \|v\|=1\}.
\end{align*}
The space of all functionals on $B$ is the dual space of $B$ and will be denoted $B^*$, itself being a Banach space.
\medskip
\par
A \textbf{bounded operator} on a Banach space $B$ is a linear map $a\ :\ B\to B$ for which :
\begin{align*}
	\|a\|:=\sup\{\|av\|\quad /\quad v\in B,\  \|v\|=1\}\leq \infty\qquad\text{with}\qquad \|av\|\leq \|a\|\ \|v\|
\end{align*}
The space of all bounded operators on $B$ will be denoted $\calB(B)$, it is a Banach space for the norm on operators.
\medskip
\par
\textbf{A sesquilinear product} on a vector space $V$ is a map denoted by ( , ) : $V\times V\to \mathbb{C}$ such that for all $v,w,w_1,w_2\in V$ et $\lambda_1$ and $\lambda_2\in \mathbb{C}$ :\\
\begin{itemize}
	\item $\overline{(v,w)}=(w,v)$
	\item $(v, \lambda_1w_1+\lambda_2w_2)=\lambda_1(v,w_1)+\lambda_2(v,w_2)$
	\item $(v,v)\geq 0$ and such that $(v,v)=0$ if $v=0$
\end{itemize}
This is also called inner product.
\medskip
\par
The \textbf{norm} of an element $v$ is defined using the sesquilinear product by $\sqrt{(v,v)}$.
\medskip
\par
A \textbf{Hilbert space $\calH$} is a vector space with a complete sesquilinear product in the sense of its associated norm.
\medskip
\par
All Hilbert spaces that we will consider will be assumed to be separable, \textit{i.e.} to have a dense countable subset. An \textbf{operator} on this space is a continuous linear map from $\calH$ to $\calH$, we will denote $\calB(\calH)$ the set of bounded operators on $\calH$; and we will most of the time consider bounded operators in the following.\\

The simplest example of a Hilbert space is $\bbC^n$ with $(x,y)=\sum_{i=1}^n\overline{x}_iy_i$.
Observe that if the Hilbert space is finite dimensional $\calH = \bbC^n$ then both the algebra of bounded and compact operators coincide and are isomorphic to $M_n(\bbC)$.
\medskip
\par
It was von Neumann who first defined the abstract notion of a Hilbert space and who later developed the theory of operators (bounded and unbounded) on these spaces \cite{von2018mathematical}. He also proved the equivalence between the matrix formalism of Heisenberg's approach and the wave formalism developed by Schrödinger to describe quantum phenomenon. Similarly, all Hilbert spaces of a given dimension are isomorphic so that a Hilbert space can be completely characterised by its dimension. Real Hilbert spaces thus offer a potentially infinite dimensional generalisation of usual Euclidean spaces, in which the geometric notions of distance and angle still make sense, thanks to a generalisation of the usual product, made by the scalar product.
\medskip
\par
\textbf{The adjoint operator $a\to a^*$} on a Hilbert space is defined by the following property, for all $\psi_1,\psi_2\in \calH$: 
\begin{align*}
	(a^*\psi_1,\psi_2):=(\psi_1,a\psi_2).
\end{align*}
It defines an involution on $\calB(\calH)$. An operator $a$ is self-adjoint if $a=a^*$; it is called \textbf{normal} if it commutes with its adjoint.
\medskip
\par
The set of self-adjoint operators will be denoted $\mathcal{U}_\mathbb{R}$ and we can decompose any operator $a$ into two self-adjoint operators as follows:
\begin{align*}
	A=A'+iA'':=\frac{A+A^*}{2}+i\frac{A-A^*}{2i}
\end{align*}
with $A'$ et $A''$ belonging to $\mathcal{U}_\mathbb{R}$.
\medskip
\par
A \textbf{$C^*$-algebra $\calA$} is a complex Banach space which is at the same time a $^*$-algebra (the $^*$ denoting an algebra endowed with an involution) such that for all $a,b\in \calA$ we have :
\begin{align*}
	\|ab\|\leq \|a\|\ \|b\|\qquad \text{and}\qquad \|a\|^2=\|a^*\|\|a\|=\|a^*a\|
\end{align*}
The important the property $\|a^*a\|=\|a\|^2$ for $a\in \calB(\calH)$ is the origin of the definition of $C^*$-algebras; it is the property that makes the described norm and involution match together and renders the structure much more workable. Initially, the “$C$” of $C^*$-algebras stood for closed. For example, any $M_n(A)$-algebra of matrices with coefficients in a $C^*$-algebra is also a $C^*$-algebra. In particular $M_n(\mathbb{C})$ is a $C^*$-algebra for all $n\in\mathbb{N}$. 
\medskip
\par 
An \textbf{homomorphism of $C^*$-algebras} is a map preserving all the structures of the $C^*$-algebra. The $C^*$-algebras with these homomorphisms form the category of $C^*$-algebras.
\medskip
\par
A \textbf{representation of $C^*$-algebra} $\calA$ on a Hilbert space $\calH$ is a $^*$-homomorphism from $\calA$ to the algebra $\calB(\calH)$ of bounded operators on $\calH$.
\medskip
\par

\begin{theorem}[Gelfand-Naimark-Segal]
	Every abstract $C^*$-algebra $\calA$ is isometrically $*$-isomorphic to a concrete $C^*$-algebras of operators on a Hilbert space $\calH$. If the algebra $\calA$ is separable then we can take $\calH$ to be separable. For a recent reference see \cite{FriendlyGNC}. 
\end{theorem}
Thus, any abstract $C^*$-algebra can be made “concrete”, \textit{i.e.} representable as an algebra of bounded operators on a Hilbert space. When this Hilbert space is of finite dimension, this algebra is finite and can be represented as a matrix algebra. In the following, every algebra will be of finite type.
$$\text{$C^*$-algebras}\leftrightarrow\text{bounded operators in $\calB(\calH)$}$$
This establishes the possibility of considering $C^*$-algebras as abstract algebraic entities without reference to particular realizations as an operator algebra acting on Hilbert spaces.
\medskip
\par
$C^*$-algebras are of deep interest because of these three properties:
\begin{itemize}
	\item They abstract the properties of bounded operations on Hilbert spaces: they do not require any particular realization as acting on Hilbert spaces. This allows us to think of structures from the simple point of view of the algebraic relations living on them.
	\item As we will see in section \ref{GNMath}, they generalize the properties of locally Compact Hausdorff topological spaces.
	\item They provide a successful mathematical framework for QM, but also of classical mechanics as we will see in section \ref{KvN}.
\end{itemize}

\medskip
\par
\begin{definition}[Spectrum of $a\in\calA$]
	For $a\in\calA$ an element of a $C^*$-algebra, we call $Sp(a)$ the spectrum of $a$ which is the set given by 
	\begin{align}
		Sp(a):=\{\lambda \in \mathbb{C} \mid a-\lambda \bbbone \text { is not invertible }\}
	\end{align}
	The collection of these sets (over all $a\in\calA)$ is $Sp(\calA)$. It is also the set of unitary equivalences of irreducible $^*$-representations of $\calA$
\end{definition}
\begin{definition}[Eigenspaces of an operator] For $a\in\calA$ a $C^*$-algebra, we call $V_\lambda$ the eigenspace associated to the element $\lambda\in Sp(a)$ such that
	\begin{align*}
		V_\lambda\defeq \{\,\psi\, \in\, \calH \, |\, a\psi = \lambda\psi\,\}
	\end{align*}
\end{definition}
\begin{definition}[Spectral decomposition of normal operator with discrete spectrum] For $a\in\calA$ a normal operator in a $C^*$-algebra, taking $P_\lambda$ to be the orthogonal projection onto the associated eigenspace $V_\lambda$, and $\{\lambda_i\}_{i\in S_a}$ the set of all eigenvalues, with the corresponding index set $S_a$. Then we have the decomposition
	\begin{align}
		\label{spectralDec}
		a=\sum_{i\in S_a}\lambda_i P_{\lambda_i}
	\end{align}
\end{definition}
For any normal element $a$ in a $C^*$-algebra, the norm of $a$ is equal to its spectral radius:
\begin{align*}
	\|a\|=\sup\{|\lambda|\quad / \quad \lambda\in Sp(a)\}
\end{align*}
We define here three notions of topologies on operators, based on the previously constructed norm:
\begin{itemize}
	\item The \textbf{normed topology} on $\calB(\calH)$ is defined by the convergence criterion:\\ $A_n\to A$ if $\|A_n-A\|\to 0$.
	\item The \textbf{strong topology} on $\calB(\calH)$ is defined by the convergence criterion:\\ $A_n\to A$ if $\|(A_n-A)\psi\|\to 0$ for all $\psi\in\calH$.
	\item The \textbf{weak topology} on $\calB(\calH)$ is defined by the convergence criterion:\\ $A_n\to A$ if $\|(\psi_1,(A_n-A)\psi_2)\|\to 0$ for all $\psi_{1,2}\in\calH$.\\
\end{itemize}
It can be shown that normed convergence implies strong convergence, which in turn implies weak convergence. A special kind of $C^*$-algebras are the von Neumann algebras:
\medskip
\par
A \textbf{von Neumann algebra} is an *-algebra of bounded operators on a Hilbert space which is unital and closed under the weak operator topology. These are of deep interest in QM and axiomatic approach to quantum field theory.
\medskip
\par
A \textbf{factor} is a von Neumann algebra with trivial center (the  center consist only of scalar operators).
\medskip
\par
An element $a$ of a $C^*$-algebra is said to be \textbf{positive} if $a^*=a$ and if its spectrum is positive $Sp(a)\subseteq\mathbb{R}^+$. The space of positive elements will be denoted $\calA^+$.

\medskip
\par
A bounded operator $A\in\calB(\calH)$ on a Hilbert space $\calH$ is said to be positive if for all $\psi\in\calH$ we have $(\psi, A\psi)\geq 0$. This property is equivalent to $A^*=A$ and $Sp(A)\subseteq\mathbb{R}^+$ and applies to closed subalgebras of $\calB(\calH)$. In QM, positive self-adjoint operators represent an extension of the notion of positive real functions, whose values are often associated with observables in classical physics. It is possible to prove that an element $a\in\calA$ is positive if and only if it can be written as $a=b^*b$ for $b\in\calA$.
\medskip
\par
The fact that the positivity of a bounded operator $A$ coincides with its self-adjointness and that $Sp(A)\subseteq\mathbb{R}^+$ induces that the expected value of an observable $A$ in QM will always be real and positive. The characterisation of quantum mechanical systems is thus done in this framework via algebraic conditions on the observables, the Hilbert space allowing to organise the so-called “pure” (non-decomposable) states of quantum systems. 
\medskip
\par
Behind the notion of observable, attached to the real world, a more abstract and underlying notion (from the physicist's point of view) is that of state:
\medskip
\par
\begin{definition}[Positive linear functional]
	A linear functional $\rho : \calA\to\bbC$  is called positive if $\rho(a)\geq0$ for all positive element $a\in\calA$. (equivalently $\rho(b^*b)\geq 0\quad \forall b\in\calA$)
\end{definition}
\begin{definition}[State on an algebra]
	We call state on a $C^*$-algebra $\calA$, a positive linear functional $\rho$ of  norm one, ie  $||\rho||=1$. A state is called a trace if  $\forall a,b\in\calA$ we have $\rho(ab)=\rho(ba)$
\end{definition}
More details can be found in \cite{sergeevIntroToNCG}.
\medskip
\par
The \textbf{Gelfand-Naimark-Segal-construction (GNS-construction)} is a procedure which permits from any state $\rho$ on a $C^*$-algebra $\calA$ to construct a representation $\pi_\rho$ of $\calA$ in a Hilbert space $\calH$. A simplified view of the procedure can be seen as follows, if the vector $\psi\in\calH$ is such that for all $a\in\calA$ we have $\rho(a)=\langle\psi|a|\psi\rangle$ (in Dirac Bra–ket notation), then we define $\pi_\rho:\calA\,\to\, End(\calH)$ the representation of $\calA$ by the relation $\rho(a)=\langle\psi|a|\psi\rangle:=\langle\psi,\pi_\rho(a)\psi\rangle$.
\medskip
\par
Note the importance of the nature of the product (which is the inner product) in this procedure. The analogy with QM is that $\rho(a)$ provides the
expectation value of the observable $a$ applied to the state $\psi$. As $\psi$ becomes the main parameter of the procedure, it can be summarized as:
\begin{align*}
	\calA\to \psi \to \rho \to \pi_\rho(\calA)
\end{align*}
More details can be found in \cite{landsman2012mathematical,Blac06a,landsman2017foundations}. 

\section{Gelfand-Naimark Theorem}
\label{GNMath}

One of the main theorems at the origin of NCG is the Gelfand-Naimark theorem, induced by the categorie equivalence between commutative $C^*$-algebras and Hausdorff-types topological spaces  as we will see.
\medskip
\par
\begin{definition}[Character of $\calA$]
	A character over an algebra $\calA$ is a surjective non-zero homomorphism\\ $\mu : \calA\to \bbC$ which respects the multiplicativity property $\mu(ab)=\mu(a)\mu(b)$ $\forall a,b\in \calA$. The set of all characters is denoted by $\mathbf{\Man}(\calA)$.
\end{definition}
As $\mu(a-\mu(a))=0$ we have that $a-\mu(a)\bbbone$ is not invertible, then $\mu(a) \in S p(a)$.
\medskip
\par
\begin{example}
	\label{exemple1}
	An example can be given by $\calA=\calC^0(\Man)$ the algebra of continuous functions on a locally compact space $\Man$, vanishing at infinity, with $p_x:f\to f(x)$ the evaluation map on $f\in\calA$ at $x\in M$.
\end{example}

\begin{lemma}
	\label{lem Commut}
	For a commutative Banach algebra $\calA$, $\mathbf{\Man}(\calA)$ endowed with the Gelfand topology is a locally compact space.
	
\end{lemma}

\begin{definition}[Gelfand transform]
	For $\calA$ a commutative Banach algebra, we call Gelfand transform (also called Gelfand representation) of $a\in\calA$   the function $\hat{a}:\mathbf{\Man}(\calA)\to \bbC$ given by the evaluation of $a$ at $\mu$ :
	\begin{align}
		\label{GelfTf}
		\hat{a}(\mu):=\mu(a) \text {. }
	\end{align}
	The \textbf{Gelfand Transformation} is a map $ \Pi : a \to \hat{a}$ from $\calA$ to $C(\mathbf{\Man}(\calA))$.
\end{definition}
\begin{lemma}
	\label{lem VpReel}
	Let $a\in\calA$ be a self-adjoint element in $C^*$-algebra. Then $\mu(a)\in\bbR$ for
	all $\mu\in \mathbf{\Man}(\calA)$. 
	
\end{lemma}

\begin{theorem}[Gelfand-Naimark theorem for commutative $C^*$-algebras]
	If a $C^*$-algebra is commutative then it is an algebra of continuous functions on some (locally compact, Hausdorff) topological space.
\end{theorem}

\medskip
\par 
Thus, the Gelfand-Naimark theorem tells us that if $\calA$ is a commutative $C^*$-algebra, then the Gelfand transformation between $\calA$ and the algebra $C^0(X)$ of continuous functions on the spectrum $X$ of $\calA$ is an isomorphism. The Gelfand transform allows us to define a functor that makes the categories of topological spaces and $C^*$-algebras equivalent. More generally, the Gelfand-Naimark theorem provides a method for representing commutative algebras as algebras of continuous functions on a topological space, it tells us that a space $\Man$ is equal to the set of characters on $C(\Man)$, that is $\Man \equiv\mathbf{\Man}(C(\Man))$, which clarifies the choice of notation made for the set of all characters.
\medskip
\par
Using the example \ref{exemple1}, the relationship \ref{GelfTf} becomes $\hat{f}(x)=x(f)$, This relation is quite surprising, because it shows that we can equally well perceive the points of topological spaces as functions on an algebra of observables, as we can perceive this algebra as a function on a topological space. This remark is of great importance and philosophical depth. We will return to it in sections \ref{BackToObs} and \ref{AlgeGeom}.
\medskip
\par
The above mentioned category equivalence implies that we can speak of topology by referring to the algebra of functions living on that topology. Thus, many characteristic properties of topological spaces have their algebraic counterparts. For example, the group of homeomorphisms of a compact space $\Man$ is isomorphic to the group of automorphisms of the $C^*$-algebra $\calC(\Man)$. A more detailed list of these equivalences can be found in table \ref{figEquivTopoAlg}.
\begin{figure}[h]
	\begin{center}
		{\setlength{\doublerulesep}{0pt}
			\begin{tabular}{|c|c|}
				\hline
				Algebra &Topology\\\hline\hline\hline\hline
				Commutative $C^*$-algebra $\calA$ &Topological space $\Man$ \\\hline
				Projectionless&Connectedness\\\hline
				Projection/Pure state &Point\\\hline
				Finite projective module &Vector bundle \\\hline
				Automorphism&Homeomorphism\\\hline
				Unital&Compact\\\hline
				Separable&Metrisable\\\hline
				Ideal&Open subset\\\hline
				Quotient algebra&Closed subset\\\hline
				tensor product&Cartesian product\\\hline
				
			\end{tabular}
		}
	\end{center}
	\caption{Equivalences between algebraic and topological properties.}
	\label{figEquivTopoAlg}
\end{figure}\\
We can therefore talk about topological spaces without any mention of them. Hence the manifold $\Man$ no longer needs to be taken as input to express the algebra, since it can be derived from it.
\medskip
\par 
An important remark is that nothing in this category equivalence requires the commutativity of the algebra. It is by placing us on the side of the algebra, and by relaxing the commutativity condition of the latter that the NCG came into being, as the extension of the geometric framework that accompanied that of commutative algebras. This opens a new door for mathematics, where we can consider a non-commutative $C^*$-algebra, and try to make sense of a corresponding generalization of topological space. This was the beginning of NCG. A schematic view of the underlying idea of NCG is given in figure \ref{figGN}.
\begin{center}
	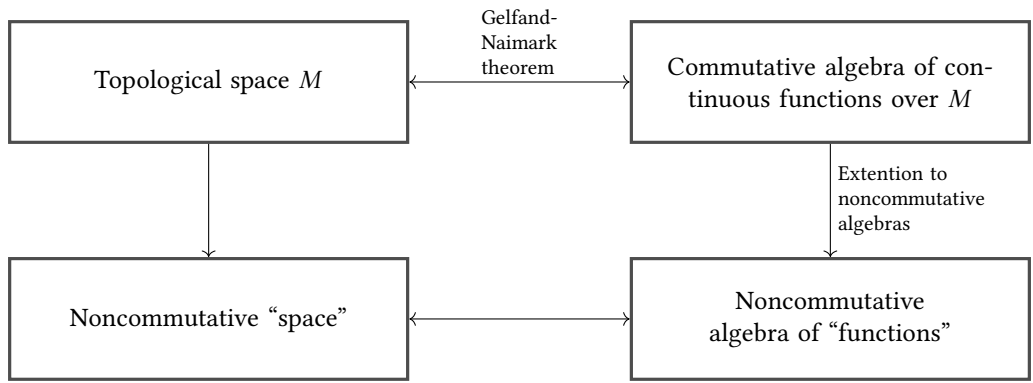
\begin{figure}[h]
		\centering
		\begin{tikzpicture}
			[
			squarednode/.style={%
				rectangle,
				draw=black!70,
				fill=white,
				very thick,
				minimum size=16mm,
				text centered,
				text width=5cm,
			}
			]
			\node[squarednode]      (bg)                              {Noncommutative “space”};
			\node[squarednode, name=f]      (hg)       [above=1.5cm of bg] { Topological space $\Man$};
			
			\node[squarednode]      (bd)       [right=2.9cm of bg] {Noncommutative\\ algebra of “functions”};
			
			\node[squarednode, name=g]      (hd)   [right=of hg]    [above=1.5cm of bd] {Commutative algebra of continuous functions over $\Man$};
			
			\footnotesize
			\draw[->] (hg.south) -- node[anchor=east] {} (bg.north);
			\draw[<->] (bg.east) -- node[anchor=south] {} (bd.west);
			\draw[<-] (bd.north) -- node[anchor=south, right, text width=1cm] {\footnotesize Extention$\,\, $to\\ noncommutative\\ algebras\normalsize} (hd.south);
			\draw[<->] (hg.east) -- node[sloped, anchor=center, above, text width=1cm] {\footnotesize Gelfand-Naimark\\ theorem\normalsize} (hd.west);

		\end{tikzpicture}
		\caption{Extending topology through Gelfand-Naimark theorem.}
		\label{figGN}
	\end{figure}
\end{center} 
Thus, the concept of point being algebraically characterized, we are now free to explore what happens in the non-commutative context. In non-commutative “spaces”, what was only a point in commutative geometry now acquires a structure. However, a NC algebra cannot be considered as an algebra of functions on a space thought as a set of points. NC geometry therefore consists of conveying the properties of spaces (in the commutative setting) into an algebraic form and then exploring what happens to these properties when moving to the NC world. NC spaces do not exist as spaces, only their algebras have existence.
\begin{remark}
	The notion of point is not lost, it is just no longer fundamental and is therefore destitute of its general significance.
\end{remark}
Then, coordinates are replaced by the generators of an algebra. Because they do not commute, they cannot be simultaneously diagonalized and the usual space picture can no longer stand.
\begin{remark}
	It is important to note that insofar as this theorem can be considered as one of the foundations of NCG, there is no mention of NC coordinates, only of NC observables.
\end{remark}
More details can be found in \cite{FriendlyGNC} and   \cite{coquereaux1998espaces}.

\section{\texorpdfstring{K-Theory of $C^*$-Algebras}{K-theory of C*-algebras}}

\label{Ktheory}
K-theory is a branch of mathematics related to geometry, topology, number theory, and ring theory. The objects of geometry, algebra, or arithmetic are associated with an object called K-group which contain important information about them. It can be seen as the study of additive or abelian invariants over matrix algebras, simple examples of abelian invariants being traces and determinants \cite{atiyah2000k}. Examples of results collected from K-theory approach include the Grothendieck–Riemann–Roch theorem, Bott periodicity, the Atiyah–Singer index theorem, this last one being one of the big results with the Gelfand-Naimark theorem that motivated the development of the NCG. The link made with topological K-theory provides another equivalence between topology and algebra, through the invariants of the K-theory.
\medskip
\par
Take $\calA$ an unital $C^*$-algebra, we consider the projectors of $\calA$; \textit{i.e.} elements that are both self-adjoint and idempotent $P^*=P=P^2$. We define $M_\infty(\algA) \defeq \cup_{p\geq 1} M_p(\algA)$ where $M_p(\algA) \defeq M_p \otimes \algA$ is the $C^*$-algebra of $p\times p$ matrices with entries in $\algA$. For all $n\geq 1$, we consider the set $\calP_n(\calA)$ of projections onto $M_n(\calA)$, and we introduce the object $\calP_\infty(\calA)$ : 
\begin{align*}
	\calP_\infty(\calA)=\bigcup^\infty_{n=1}\calP_n(\calA).
\end{align*}
We then construct an equivalence class $\sim_0$ on $\calP_\infty(\calA)$; that is $P$ and $Q$ belonging to $\calP_\infty(\calA)$ are equivalent if there is $S\in M_\infty(\algA)$ such that $P=SS^*$ and $Q=S^*S$. We thus construct $\calD(\calA)$ as the quotient of $\calP_\infty(\calA)$ by $\sim_0$ :
\begin{align*}
	\calD(\calA)=\calP_\infty(\calA)/\sim_0.	
\end{align*}
$\calD(\calA)$ can be equipped with an additive abelian law denoted $+$, moreover,  $\calD(\calA)$ is not a group because there is no inverse element. We then create another equivalence relation $\sim$ on $\calD(\calA)\times \calD(\calA)$, take $(x_1,y_1)$ and $(x_2,y_2)$ two elements of $\calD(\calA)\times \calD(\calA)$; we have $(x_1,y_1)\sim (x_2,y_2)$ if and only if there is $z\in \calD(\calA)$ such that $x_1+y_2+z=x_2+y_1+z$. We consider $K_0(\calA)=\calD(\calA)\times \calD(\calA)/\sim$, which is an abelian group, $K_0(\calA)$ is the Grothendieck group $G(\calD(\calA))$:
\begin{align*}
	K_0(\calA)=G(\calD(\calA))=\calD(\calA)\times \calD(\calA)/\sim
\end{align*}
This functor will thus allow us to transfer the study of algebras to that of abelian groups. In the case of inductive sequences of algebras: 
\begin{align*}
	\calA_1\lhook\joinrel\xrightarrow{\phi_1}\calA_2\lhook\joinrel\xrightarrow{\phi_2}\calA_3 \quad ... \ \rightsquigarrow \calA_{AF}
\end{align*}
we will be able to build the sequence:
\begin{align*}
	K_0(\calA_1)\lhook\joinrel\xrightarrow{K_0(\phi_1)}K_0(\calA_2)\lhook\joinrel\xrightarrow{K_0(\phi_2)}K_0(\calA_3) \quad ... \ \rightsquigarrow K_0(\calA_{AF})	
\end{align*}
with $K_0(\phi_i)$ defined in
\cite{rordam2000introduction}. In the subsection \ref{sec AFKT} we will see how K-theory can be used to classify $AF$ algebras, and in \ref{SerSwann}, that the projections used to build the construction that leads to the invariants of K-theory can also be used to create projective modules. As we shall see, this provides a nice link with the topological constructions such as the vector bundles in Gauge theory, more precisely by connecting these projective modules to the sections of the bundle. As for the Gelfand-Naimark theorem, this result is the consequence of an equivalence of category.

\chapter{\texorpdfstring{Finite and Approximately Finite ($AF$) Dimensional Algebras}{Finite and Approximately Finite (AF) dimensional algebras}}
\label{FiniteAlg}
In the following, as said before, we will only consider algebras of finite dimension, these are the ones with which the quantum physicist is the more familiar with since she/he deals with experiments with finite or countable outcome results. In this section, we will offer a more intuitive view of these algebras, adopting the natural matrix representation. This will give a view of what the commutative limit represents in this unifying scheme. Then, I will show how any $C^*$-algebra admits a natural direct sum decomposition into sub-algebras. In section \ref{AFalg}, I will describe an important kind of algebra which is based on this direct sum decomposition \textit{i.e.} the $AF$-algebras, on which the work of this thesis is based. Then we will show in section \ref{sec AFKT} how K-theory offers a nice classification of these algebras. 

\section{Finite Algebras}
\label{secFinite}
A first observation is that through the Gelfand-Naimark-Segal presented in chapter \ref{sec C algebras}, any finite $C^*$-algebra can be represented by a matrix algebra  acting on a finite-dimensional Hilbert space. These are isomorphic to the space $\bbC^n$ equipped with the inner product $(\xi,\eta)=\sum_{i=1}^n\overline{\xi_i}\eta_i$ with $\xi,\eta\in \bbC^n$, where $i$ is the index of an orthonormal basis of $\bbC^n$. More generally, these can be extended by associating a multiplicity space to them: $\bbC^n\otimes\bbC^\mu$ where $\mu$ is the multiplicity of this representation. In the context of NCG, these indices $i$ will be associated (in a one to one way) with elements of a finite space $\Fin$.
\medskip
\par 
To proceed pedagogically, let us start with the simplest example, that of commutative algebra. A first observation is that any commutative (finite and acting on $\calH=\bbC^n$) algebra $\calA_C$ is of the form $\calA_C=\bbC^n$, where $n$ is its dimension, which coincides with that of $\calZ(\calA_C)$. An algebraically equivalent writing is therefore the following, $\forall f\in\calA_C$:
\begin{align*}
	f=\begin{pmatrix}
		f(1) & 0 & \cdots & 0 \\
		0 & f(2) & \cdots & 0 \\
		\vdots & \vdots & \ddots & \vdots \\
		0 & 0 & \cdots & f(n) \\
	\end{pmatrix}
\end{align*}
where the indices $1, \cdots, n$ refer to the elements of the Hilbert space on which each $f(i)$ acts, and thus to elements of $\Fin$. In the commutative case, the space $\Fin$ can be considered as a topological space with $n$ points. We can see perfectly well that nothing can give the differential structure taking this form alone, so an additional structure is needed to link the values of $f(i)$ together. We can also provide this topological space with a metric, giving a notion of distance between the points, at the level of the algebra. This is common in NCG and notably done in \cite[p~19]{Suij15a}. We will see in chapter \ref{DifCalc} different procedures to build such metric and differential structures.
\medskip
\par 
A first step in the NC generalization of this algebra is to replace $f(i)$ by algebras of NC matrices, and thus obtain the algebra $\calA_{NC}$ such that $\forall a \in\calA_{NC}$:
\begin{align*}
	a=\begin{pmatrix}
		a_1 & 0 & \cdots & 0 \\
		0 & a_2 & \cdots & 0 \\
		\vdots & \vdots & \ddots & \vdots \\
		0 & 0 & \cdots & a_n \\
	\end{pmatrix}
\end{align*}
This expression is algebraically equivalent to $a=\bigoplus_{i=1}^na_i=\bigoplus_{i=1}^nM_{n_i}(\bbC)$, $M_{n_i}(\bbC)$ being matrix algebras of dimension $n_i$, as we will see, every finite $C^*$-algebra admits such a writing. Natural Hilbert spaces for this kind of algebras will be written $\bigoplus_{i=1}^n\bbC^{n_i}\otimes\bbC^{\mu_i}$. This example is interesting because the only definable point notions are associated with each of these matrix algebras independently, there is no more refined point notion, the projectors used to create these points being the elements $\bbbone_{n_i}$, identity matrices reduced to algebras $M_{n_i}(\bbC)$. One way of representing what happens is to consider that these $n$ points are now endowed with an inner structure, whose degrees of freedom can be called NC degrees of freedom. Go even further by filling in the non-diagonal terms, we obtain an algebra $\calA_{NC}^\prime$ “even more NC”, with $\forall a\in \calA_{NC}^\prime$:
\begin{align*}
	a=\begin{pmatrix}
		a_{11} & a_{12} & \cdots & a_{1n} \\
		a_{21}& a_{22} & \cdots & a_{2n} \\
		\vdots & \vdots & \ddots & \vdots \\
		a_{n1} & a_{n2} & \cdots & a_{nn} \\
	\end{pmatrix}\qquad\qquad a_{i_1i_2}\in M_{n_1\times n_2}(\bbC)
\end{align*}
In this case, the space $F$ is the space reduced to a point, none of its elements having the status of a point for the algebra (through-Gelfand Naimark theorem).
\medskip
\par 
These finite spaces are the basis of the NC standard model which is “built” on the algebra $\calC^\infty(\Man)\otimes \calA$ (with $\Man$ a topological space) which can be considered as the one acting on the Cartesian product space $\Man\times\Fin$.
\medskip
\par 
Finally, an interesting analogy can be made with QM. If we take the observable defined by the density matrix $\rho$, then we can see how the set of these density matrices forms a NC algebra. The decoherence process can be understood as a process making this algebra commutative according to a given basis, that of the classical observables:
\begin{align*}
	\rho=\begin{pmatrix}
		\rho_{11} & \rho_{12} & \cdots & \rho_{1n} \\
		\rho_{21}& \rho_{22} & \cdots & \rho_{2n} \\
		\vdots & \vdots & \ddots & \vdots \\
		\rho_{n1} & \rho_{n2} & \cdots & \rho_{nn} \\
	\end{pmatrix}
	\qquad \to \qquad\rho^\prime=
	\begin{pmatrix}
		\rho_{11} & 0 & \cdots & 0 \\
		0& \rho_{22} & \cdots & 0 \\
		\vdots & \vdots & \ddots & \vdots \\
		0 & 0 & \cdots & \rho_{nn} \\
	\end{pmatrix}
\end{align*}

\medskip
\par
It is interesting to note that any finite $C^*$-algebra admits a natural decomposition into direct sum. Indeed, by the Artin–Wedderburn Theorem any finite-dimensional $C^*$-algebra can be represented as a direct sum of matrix algebras:  
\begin{align*}
	\calA\simeq \bigoplus_{i=1}^{r}M_{n_i}(\bbC)
\end{align*}
With the $n_i$'s and $r$ being positive finite integers. This can be easily seen in this example of reconstruction, taking any finite $*$-algebra $\algA$ of complex matrices, there exists a minimal set of mutually orthogonal projectors $\{p_i\}_{i=1}^r$ spanning the center $\calZ(\algA)$ and forming a resolution of the identity ($p_i=\bbbone_{n_i}$ and $0$ outside for example). Then for any $a\in \algA$ we have: 
\begin{align*}
	a=\bbbone_{\algA}a=\left(\sum_i^rp_i\right)a=\sum_i^rp_ip_ia=\sum_i^rp_iap_i\simeq a_1\toplus a_2\toplus\dots\toplus a_r\toplus 0	
\end{align*}
because the projections are in the center and mutually orthogonal. As the projectors depend only on $\algA$, this decomposition holds for any $a\in \algA$, then we get :
\begin{align*}
	\algA\simeq\mathcal{U}\algA\mathcal{U}^\dag=\algA_1\toplus \algA_2\toplus\dots\toplus \algA_r\toplus 0 
\end{align*}
with $\mathcal{U}$ the unitary matrix coming from the set of projections, the center of each algebra is trivial $\calZ(\algA_i)=\bbC\bbbone_{\algA_i}$. 
A complete derivation can be found in \cite{beny2015algebraic}.
\medskip
\par
This is why we will often study what happens on such representations later on, as they are more convenient to use, have many links with physics, and because the parameters of the embedding structure of $AF$-algebras act specially on the elements of the direct sum.
\medskip
\par
The concept of direct sum can be generalized to direct integral. It was mostly developed for von Neumann algebras which are at the heart of the axiomatic approaches to quantum field theory. A deep and important result goes as follows:
If one assumes that von Neumann algebras have the right properties to represent the general algebra of observables in physics and that one considers $\calA$ as a potential candidate of interest for physics (without assuming any parametrization of this algebra with respect to points in space-time, and assuming that this algebra is finite dimensional), then it is possible to obtain a central decomposition of $\calA$, such that, as above for the Artin-Wedderburn theorem, (via the collection of mutually orthogonal projections of the center forming a resolution of the identity) given the projections set $\{p_x\}_{x\in M}$ of such a decomposition, with $\Man$ the set of indices of these projections, one has:
\begin{align*}
	\calA\equiv\int _{M}^{\oplus }\calA_x d\mu(x)
\end{align*}
$d\mu(x)$ being a measure over $\Man$. So in other words we have a diagonalization of $\calA$ according to its center. Each $\calA_x$ is obtained from a $p_x$, it turns out that (almost) each $\calA_x$ is a factor, which is analogous to full matrix algebras over a field. We have taken $x$ as a subscript to emphasize the potential link with points of space-time. More details can be found in \cite{kadison1986fundamentals}.
\medskip
\par 
This is of capital significance because it proves that given any algebra (potentially NC) having reasonable properties concerning physics, it can naturally be rewritten as defined on “a field” parametrized by its center, where matrix algebras “live” above each point, which is exactly the description made in gauge theories as we will see in part \ref{TowardNCGFT}. One must therefore understand that starting from such a NC algebra, without any notion of point, one can rewrite it as a matrix field over a set of points, which one can imagine to be those of space-time. This theorem thus provides a clue that NCG is a natural framework for gauge theories, since as we will see, considering a general NC algebra lead naturally to structure like the ones described by fiber bundles. We will see in section \ref{ACManifold} how this fact is implemented with AC-manifolds.
\medskip
\par
Moreover, this construction is not without reminding the Gelfand-Naimark theorem, because these projections of the center can be taken to construct pure states defined on the algebra (using the trace), and thus define a notion of disjoint points (mutually orthogonal) for the algebra since they belong to the center of this one. It is interesting to note the contribution of this approach compared to that of the Gelfand-Naimark theorem which is more general. Indeed, the notion of point can here be defined on an NC algebra from its own elements, \textit{i.e.} without external structure!

\section{\texorpdfstring{Approximately Finite ($AF$) Algebras}{Approximately Finite (AF) algebras}}
\label{AFalg}
\par 
In this subsection, we would like to recall the necessary structures involved in the definition of $AF$ $C^*$-algebras that will be used in the following. We would like also to illustrate in section \ref{sec AFKT}, with the $K_0$-group example, the powerful approximation procedure by finite dimensional structures that we inherit in this framework.
\medskip
\par 
A $C^*$-algebra $\algA$ is said to be $AF$ (approximately finite-dimensional) if it is the closure of an increasing union of finite dimensional subalgebras $\algA_n$ \textit{i.e.} $\algA = \overline{\cup_{n\geq 0} \algA_n}$. We will always suppose that $\algA$ is unital and that $\algA_0 \simeq \bbC \bbbone_{\algA}$ \cite{Davi96a}. The equivalent of $AF$ C$^*$-algebras in von Neumann algebras are hyperfinite factors. The $AF$ $C^*$-algebras are non-commutative generalizations of $C^0(\Man)$, with $\Man$ a totally disconnected space.
\medskip
\par 
It is convenient to describe $\algA$ as the direct limit $\algA_\infty = \varinjlim \algA_n$ of the inductive sequence of the finite dimensional (sub)algebras $\{ (\algA_n, \phi_{n,m}) \, / \,  0 \leq n < m \}$ where $\phi_{n,m} : \algA_n \to \algA_{m}$ are one-to-one unital $*$-homomorphisms such that $\phi_{m,p} \circ \phi_{n,m} = \phi_{n,p}$ for any $0 \leq n < m < p$. From this composition property, one needs only to describe the homomorphisms $\phi_{n,n+1} : \algA_n \to \algA_{n+1}$. This can be done in two steps.
\medskip
\par 
Firstly, any finite dimensional $C^*$-algebra $\algA$ is $*$-isomorphic to the direct sum of matrix algebras, $\algA \simeq \toplus_{i=1}^{r} M_{n_i}$ \cite[Thm.~III.1.1]{Davi96a}. Secondly, any unital $*$-homomorphism $\phi : \algA = \toplus_{i=1}^{r} M_{n_i} \to \algB = \toplus_{k=1}^{s} M_{m_k}$ is determined up to unitary equivalence in $\algB$ by a $s \times r$ matrix $A = ( \alpha_{ki} )$ where $\alpha_{ki} \in \bbN$ (non-negative integers) is the multiplicity of the inclusion of $M_{n_i}$ into $M_{m_k}$ \cite[Lemma~III.2.1]{Davi96a}. The multiplicity matrix $A$ is such that $\sum_{i=1}^{r} \alpha_{ki} n_i = m_k$. The action of $A$ can be seen on the index sets as follow:
\begin{align*}
	A\left(\begin{smallmatrix} n_1\\ 
		n_2\\ 
		.\\  .\\ n_r\\ 
	\end{smallmatrix}\right)\to\left(\begin{smallmatrix} m_1\\ 
		m_2\\ 
		.\\ .\\ m_s\\ 
	\end{smallmatrix}\right)
\end{align*}
The Bratteli diagram is constructed in such a way that each $n_i$ and $m_k$ corresponds to points in the diagram, the $n_i$'s being all on the same line, and the $m_k$'s on another, the $\alpha_{ki}$ are then used to define arrows between these points, so that if $\alpha_{ki}=0$ there is no arrow, and if $\alpha_{ki}\neq 0$ there is an arrow of multiplicity $\alpha_{ki}$: 
\begin{align*}
	\overset{n_i}{\bullet} \overset{\alpha_{ki}}{\longrightarrow}  \overset{m_k}{\bullet}
\end{align*} 
An important fact is that if  $\mathcal{A}=\overline{\cup_{n\geq 0}\mathcal{A}_n}$ and $\mathcal{B}=\overline{\cup_{n\geq 0}\mathcal{B}_n}$ have the same Bratteli diagram, then they are isomorphic. Furthermore if $\mathcal{A}_n$ and $\mathcal{B}_n$ are isomorphic $\forall n$, the algebras are not necessarily isomorphic, since the nature of the embedding structure defined by the $\alpha_{ki}$ contain crucial properties. 
This fact is very important since as mentionned in \cite{christensen2006spectral}, $\phi$ contains many important geometric and topological data. A natural question will be how to characterize the additional geometric and topological data coming from the structure of the $AF$ algebra, when we will have constructed differential structures based on such algebras. The characterization of $\phi$ up to unitary equivalence in $\algB$ permits to take a convenient presentation of the inclusions of the $M_{n_i}$'s into the $M_{m_k}$'s, for instance by increasing order of the $n_i$'s along the diagonal of $M_{m_k}$. 
\medskip
\par 
In what follows, we will not be interested in the $C^*$ aspect of $AF$ $C^*$-algebras since we will focus on the differentiable structures compatible with the increasing sequence of $\{ (\algA_n, \phi_{n,n+1}) \, / \,  n \geq 0 \}$. In our point of view, we will consider $\algA_\infty \defeq \cup_{n\geq 0} \algA_n$ as the dense subalgebra of $\algA = \overline{\cup_{n\geq 0} \algA_n}$ of “smooth” elements. $\algA_\infty$, as the direct limit of $\{ (\algA_n, \phi_{n,n+1}) \, / \,  n \geq 0 \}$ in the category of associative (unital) algebras, inherits some algebraic structures of the algebras $\algA_n$. 
\medskip
\par 
Let us notice that one key result for the study of $AF$ algebras is given by \cite[Lemma~III.2.1]{Davi96a}, which describes the possible unital $*$-homomorphisms $\phi : \algA = \toplus_{i=1}^{r} M_{n_i} \to \algB = \toplus_{k=1}^{s} M_{m_k}$. For reasons that will be explained below (see Sect.~\ref{sect direct limit NCGFT}), we will consider non unital $*$-homomorphisms $\phi : \algA = \toplus_{i=1}^{r} M_{n_i} \to \algB = \toplus_{k=1}^{s} M_{m_k}$. In that case, we can use \cite[Lemma~III.2.2]{Davi96a} to describe $\phi$ up to unitary equivalence in $\algB$ with a matrix $A = ( \alpha_{ki} )$, with $\alpha_{ki} \in \bbN$, such that $\sum_{i=1}^{r} \alpha_{ki} n_i \leq m_k$.
\medskip
\par 
Another important point to notice is that in the mathematical considerations described before, the $*$-homomorphisms $\phi : \algA = \toplus_{i=1}^{r} M_{n_i} \to \algB = \toplus_{k=1}^{s} M_{m_k}$ need only be characterized up to unitary equivalence in $\algB$. This is a consequence of the fact that we need only consider “classes” (modulo isomorphisms for instance) for the purpose of classifying the structures. \textit{A priori}, in physics, we may need to consider two $*$-homomorphisms $\phi$ as different even if they are related by a unitary equivalence. This is related to the fact mentioned above that we consider the algebraic structure $\algA_\infty \defeq \cup_{n\geq 0} \algA_n$ instead of its completion, and that its presentation (the sequence of $*$-homomorphisms $\phi_{n,n+1}$) may contain some phenomenological information. But, as will be shown, see Examples~\ref{example Mn} and \ref{example C(\Man) otimes Mn},  the action of (unitary) inner automorphisms is not relevant from a physical point of view since it consists to a transport of structures. These inner automorphisms are similar to gauge transformations in the sense that one can chose a particular representative in the class of equivalent structures to describe a physical situation. This explains why the analysis in this thesis relies on a chosen “standard form” for these $*$-homomorphisms which simplifies the presentation.

\section{\texorpdfstring{The Classification of $AF$ $C^*$-Algebras by K-Theory:}{The classification of AF C*-algebras by K-theory:}}
\label{sec AFKT}

The complete classification of $AF$ $C^*$-algebras was made by G.A. Elliott in the framework of K-theory thanks to the $K_0$-functor producing a category equivalence between the theory of $C^*$-unital algebras and that of abelian groups. This is one of the main arguments motivating the development of NC gauge theories based on $AF$-algebras. Because it makes feel their structural importance (from the point of view of topology/algebra equivalence), as well as the link with a classification of the gauge theories built on it, with respect to the link between this classification, and the gauge theories, via the idempotents of the algebra, as we will see in section \ref{SerSwann} with the Serre-Swan theorem. Indeed, for any $AF$ algebra, at each rank of the inductive process, we can construct a K-theory and obtain the corresponding invariants. The fact that this K-theory is linked to the one of the next rank induces the possibility to compute the K-theory of the $AF$ algebra at the limit of the inductive process. The interest is therefore to use this fact to make gauge theories in the same way, \textit{i.e.} by constructing a gauge theory at each given rank, starting from the previous rank. The fact that K-theory allows us to fully classify $AF$ algebras will thus allow us to characterize the limit gauge theory associated with the induced $AF$ algebra. We propose here to show the construction of the functor $K_0$ from an $AF$ $C^*$-algebra to the associated abelian group.
\cite{rordam2000introduction}
\medskip
\par
Let us now illustrate how the defining sequence $\{ (\algA_n, \phi_{n,n+1}) \, / \,  n \geq 0 \}$ can be used to construct “approximations” of the $K_0$ group of $\algA$, using the same notations as in section \ref{Ktheory}. The definition (see \cite{RordLarsLaus00a,Davi96a} for details) of the $K_0$ group of a unital $C^*$-algebra $\algA$ starts with an equivalence class of projections in $M_\infty(\algA)$. The space $\calD(\algA) \defeq \calP_\infty(\algA)/\sim_0$ of equivalence classes $[P]$ of projections in $M_\infty(\algA)$ is an Abelian semigroup for the additive law $[P] + [Q] \defeq [P \oplus Q]$ where $P \oplus Q \defeq \begin{psmallmatrix} P & 0 \\ 0 & Q \end{psmallmatrix} \in M_\infty(\algA)$. Then $K_0(\algA)$ is the Grothendieck group of $\calD(\algA)$.\footnote{The Grothendieck group $G(S)$ of an Abelian semigroup $(S,+)$ is the unique Abelian group $K$ which satisfies the following universal property: there is a morphism of semigroups $\iota : S \to G(S)$ such that for any morphism of semigroups $\varphi : S \to T$ for any Abelian group $T$, there is a morphism of groups $\widetilde{\varphi} : G(S) \to T$ with $\varphi = \widetilde{\varphi} \circ \iota$.} We denote by $\iota : \calD(\algA) \to K_0(\algA)$ the map defining the universal property of $K_0(\algA)$. Then, for any $P, Q \in \calP_\infty(\algA)$, one has \cite[Prop.~3.1.7]{RordLarsLaus00a} $\iota([P \oplus Q]) = \iota([P]) + \iota([Q])$; $\iota([P]) = \iota([Q])$ if and only if there exists $R \in \calP_\infty(\algA)$ such that $P \oplus R \sim_0 Q \oplus R$; and $K_0(\algA) = \{ \iota([P]) - \iota([Q]) \, / \, P, Q \in \calP_\infty(\algA) \}$. So, describing all the $\iota([P])$ is sufficient to get $K_0(\algA)$.
\medskip
\par 
For a matrix algebra $\algA = M_n$, one has $M_p(M_n) = M_p \otimes M_n = M_{pn}$ and,  for any $P, Q \in \calP_\infty(M_n)$, $P \sim_0 Q$ iff $\rank(P) = \rank(Q)$. So $\calD(M_n) \simeq \bbN$, $K_0(M_n) = \bbZ$ and $\iota$ is the inclusion $\iota : \bbN \hookrightarrow \bbZ$ and so it can be omitted in that case. For a finite dimensional algebra $\algA = \toplus_{i=1}^{r} M_{n_i}$, this result generalizes as $P = \toplus_{i=1}^{r} P_i \sim_0 Q = \toplus_{i=1}^{r} Q_i$ iff $\rank(P_i) = \rank(Q_i)$ for any $i$, so that $\calD(\toplus_{i=1}^{r} M_{n_i}) \simeq \bbN^r$ and $K_0(\toplus_{i=1}^{r} M_{n_i}) = \bbZ^r$ \cite[Ex.~IV.2.1]{Davi96a}. Here again $\iota$ is the natural inclusion and it can be omitted.
\medskip
\par
Any morphism of $C^*$-algebras $\phi : \algA \to \algB$ induces a canonical morphism of groups $\phi_* : K_0(\algA) \to K_0(\algB)$ by $\phi_* \circ \iota_\algA([P]) = \iota_\algB([\phi(P)])$ where $\phi(P) \in \calP_\infty(\algB) \subset M_\infty(\algB)$ is defined by applying $\phi$ to the entries of the matrix $P \in M_p(\algA) \subset M_\infty(\algA)$. So, from the defining inductive sequence $\{ (\algA_n, \phi_{n,m}) \, / \,  0 \leq n < m \}$ of an $AF$ $C^*$-algebra $\algA = \varinjlim \algA_n$, we get an inductive sequence $\{ (K_0(\algA_n), \phi_{n,m\, *}) \, / \,  0 \leq n < m \}$. Then one has $K_0(\algA) = \varinjlim K_0(\algA_n)$ \cite[Thm~6.3.2]{RordLarsLaus00a}.
\medskip
\par
To compute $\varinjlim K_0(\algA_n)$, one has to describe the morphisms $\phi_{n,n+1\, *} : K_0(\algA_n) \to K_0(\algA_{n+1})$. This can be done easily in terms of the multiplicity matrices $A_{n, n+1}$ associated to the morphisms $\phi_{n, n+1} : \algA_n \to \algA_{n+1}$. In order to do that, we switch from projections to finitely generated projective modules. 
\medskip
\par
Let $P \in M_p(M_n)$ be a projection with $\rank(P) = \alpha \ (\leq pn)$. Then $P$ can be diagonalized as $P = U^* E_\alpha U$ for a unitary $U \in M_{pn}$ and $E_\alpha = \begin{psmallmatrix} \bbbone_\alpha & 0 \\ 0 & 0 \end{psmallmatrix} \in M_{pn}$ where $\bbbone_\alpha$ is the unit $\alpha \times \alpha$ matrix. Then $S \defeq E_\alpha U$ satisfies $S^* S = (E_\alpha U)^* (E_\alpha U) = U^* E_\alpha^* E_\alpha U = U^* E_\alpha U = P$ and $S S^* = (E_\alpha U) (E_\alpha U)^* = E_\alpha U U^* E_\alpha^* = E_\alpha$ so that $[P] = [E_\alpha]$. Consider the free left $M_n$-module $(M_n)^p \simeq M_{pn, n}$ (row of $p$ copies of $M_n$ or rectangular matrices $pn \times n$). Then up to the unitary equivalence by $U$ (acting on the right on rectangular matrices), $P$ defines the submodule $\modM_P \simeq M_{pn, n} E_\alpha \simeq M_{\alpha, n} \simeq \bbC^n \otimes \bbC^\alpha$ (a finitely generated projective module over $M_n$). 
\medskip
\par
In the same way, a class $[P = \toplus_{i=1}^{r} P_r] \in \calD(\toplus_{i=1}^{r} M_{n_i})$ defines a class (modulo isomorphisms) of left (finitely generated projective) modules $[\modM_P]$ with $\modM_P = \bbC^{n_1} \otimes \bbC^{\alpha_1} \toplus \cdots \toplus \bbC^{n_r} \otimes \bbC^{\alpha_r}$ where $\alpha_i = \rank(P_i)$. Indeed, if $P \in M_p(\toplus_{i=1}^{r} M_{n_i}) = \toplus_{i=1}^{r} M_p \otimes M_{n_i}$ then $P_i \in M_p \otimes M_{n_i} = M_p(M_{n_i})$ and then we are in the previous situation for every $P_i$. The map $\phi_* : \calD(\toplus_{i=1}^{r} M_{n_i}) \to \calD(\toplus_{k=1}^{s} M_{m_i})$ induced by $\phi : \toplus_{i=1}^{r} M_{n_i} \to \toplus_{k=1}^{s} M_{m_k}$ with multiplicity matrix $A = ( \alpha_{ki} )$ sends $[P]$ to $[Q = \toplus_{k=1}^{s} Q_k]$ where every entry along $M_p$ in $Q_k \in M_{p}(M_{m_k})$ contains $\alpha_{ki}$ copies, distributed along the diagonal of $M_{m_k}$, of the entry at the same position along $M_p$ in $P_i$. Since the rank of a matrix projection is its trace, one gets $\beta_k \defeq \rank(Q_k) = \sum_{i=1}^{r} \alpha_{ki} \alpha_i$ and the associated module is then $\modN_Q = \bbC^{m_1} \otimes \bbC^{\beta_1} \toplus \cdots \toplus \bbC^{m_s} \otimes \bbC^{\beta_s}$ by the previous construction. So, in terms of modules, $\phi_*$ sends the class of $\bbC^{n_1} \otimes \bbC^{\alpha_1} \toplus \cdots \toplus \bbC^{n_r} \otimes \bbC^{\alpha_r}$ to the class of $\bbC^{m_1} \otimes \bbC^{\beta_1} \toplus \cdots \toplus \bbC^{m_s} \otimes \bbC^{\beta_s}$, where $\bbC^{n_i} \otimes \bbC^{\alpha_i} = M_{\alpha_i, n_i}$ is repeated $\alpha_{ki}$ times on the diagonal of $\bbC^{m_k} \otimes \bbC^{\beta_k} = M_{\beta_k, m_k}$. The diagonals are filled thanks to the relations $m_k = \sum_{i=1}^{r} \alpha_{ki} n_i$ and $\beta_k = \sum_{i=1}^{r} \alpha_{ki} \alpha_i$. See \cite[Ex.~IV.3.1]{Davi96a} where the identification of $\phi_*$ with $A$ is also presented using projections.
\medskip
\par
This describes the maps $\phi_{n,n+1\, *} : \calD(\algA_n) \to \calD(\algA_{n+1})$ in terms of (finitely generated projective) modules. For $AF$ $C^*$-algebras,  $\calD(\algA)$ equals the space $K_0^+(\algA)$ of stable equivalence classes of projections in $\calP_\infty(\algA)$,\footnote{$P,Q \in \calP_\infty(\algA)$ are stably equivalent, $P \approx Q$, if there is a projection $R \in \calP_\infty(\algA)$ such that $P \oplus R \sim_0 Q \oplus R$.} and this is a cone in $K_0(\algA)$ such that $K_0(\algA) = K_0^+(\algA) - K_0^+(\algA)$ \cite[Thm~IV.1.6, Thm~IV.2.3, Thm~IV.2.4]{Davi96a}. So, for $AF$ $C^*$-algebras, it can be of practical importance to know what it means to approximate elements of $\calD(\algA) = K_0^+(\algA)$. A class $[P] \in \calD(\algA)$ can be looked at as a sequence of classes $[P_n] \in \calD(\algA_n)$ for $n \geq n_0$, related step-by-step by the maps $\phi_{n,n+1\, *}$. The sequence $\{  [P_n] \}_{n \geq n_0}$ corresponds then to a sequence $\{ [\modM_n] \}_{n \geq n_0}$ of equivalence classes of isomorphisms of finitely generated projective modules on the algebras $\algA_n$. Notice that it is only the sequence in the whole that permits to reconstruct the target element $[P] \in \calD(\algA)$. We could say that, for some $n \geq n_0$, the module $\modM_n$ “approximates” (as a representative element in $[\modM_n]$) the class $[P]$, but some information are encoded in the embedding maps $\phi_{n,n+1\, *} : \modM_n \to \modM_{n+1}$ which then participate to this notion of approximation. As seen before, concretely, the maps $\phi_{n,n+1\, *}$ are written in terms of the multiplicity matrices $A_{n,n+1}$ associated to the $\phi_{n,n+1}$.\footnote{The full sequence of multiplicity matrices $A_{n,n+1}$ is provided by the $AF$ $C^*$-algebra. It can be represented graphically by a Bratteli diagram, and it is known that two $AF$ $C^*$-algebras with the same Bratteli diagram are isomorphic \cite[Prop.~III.2.7]{Davi96a}}
\medskip
\par
It is well-known (Elliott's Theorem, see for instance \cite[Thm~IV.4.3]{Davi96a}) that the $K_0$-group, supplemented with a structure of scaled dimension group, is sufficient to classify $AF$ $C^*$-algebras. So there is no more information outside of the one encoded in the sequence of multiplicity matrices $A_{n,n+1}$ to be expected in the constructions described before since it determines a unique $AF$ $C^*$-algebra and it permits to construct its scaled dimension group.\footnote{Keep in mind that an $AF$ $C^*$-algebra can be obtained from different sequences of multiplicity matrices.}, and so to go on from the sequence:
\begin{align*}
	\calA_1\lhook\joinrel\xrightarrow{\phi_1}\calA_2\lhook\joinrel\xrightarrow{\phi_2}\calA_3 \quad ... \ \rightsquigarrow \calA_{AF}
\end{align*}
to the sequence:
\begin{align*}
	K_0(\calA_1)\lhook\joinrel\xrightarrow{K_0(\phi_1)}K_0(\calA_2)\lhook\joinrel\xrightarrow{K_0(\phi_2)}K_0(\calA_3) \quad ... \ \rightsquigarrow K_0(\calA_{AF}).
\end{align*}

In the part \ref{partNCGFTAF}, we will use again these $\phi_n$ morphisms to link the essential structures of the gauge theories which will be associated to each $\calA_i$, in order to build inductive sequences of NCGFTs on top AF-algebras.

\chapter{Noncommutative Differential Structures}
\label{DifCalc}
We have seen with the Gelfand Naimark theorem that properties of an algebra can be mapped to topological properties of an underlying space in the commutative case. Therefore, we know how to reconstruct the topological manifold $\Man$ from $\calA$, and then obtain the algebra $\calC^0(\Man)$ of continuous functions on $\Man$. It should be noted that this connection provides only incomplete information about the geometry of this underlying “space”, and thus for example does not allow one to distinguish between two equivalent topological spaces such as the ones of torus and teacup for example. The additional information required to fully characterize the geometry would be to introduce an object enabling the elaboration of a differential structure (on the algebraic side), and from which one could extract a metric structure (geometric side). It is therefore very important to complete the algebraic/geometric equivalence by finding techniques to define such differential and metric structures, and thus construct the algebra $\calC^\infty(\Man)$ of differentiable functions. Having done this with tools that do not depend on the commutativity of the algebra, the generalization to NC algebras can be started.
\medskip
\par
The study of differential structure can be seen as the study of variations of what is observed. This can be achieved according to any abstract or concrete parameter like time, space, another observable, or numbers. This can be done infinitesimally or not for the variation of this parameters. In QM, the observational outcomes are given by
$O_a(\psi)=\langle\psi|a|\psi\rangle$ (in the equivalent bra-ket notation). Then if we denote by $\delta O_a(\psi)$ its variation, they are two equivalent ways to see its origin:
\begin{itemize}
	\item The 1st one is to consider that it comes from a variation of the state:  $\psi^\prime=\psi+\epsilon\delta\psi$ which gives at first order $O_a(\psi^\prime)=\langle\psi|a|\psi\rangle+\epsilon(\langle\delta\psi|a|\psi\rangle+\langle\psi|a|\delta\psi\rangle)=O_a(\psi)+\epsilon\delta^\psi O_a(\psi)$ if we neglect the terms of order two in $\epsilon$. We can call this the Schrödinger picture view of variations.
	\item The 2nd one is to consider that it comes from a variation at the level of the algebra: $a^\prime=a+\tilde{\epsilon}\delta a$ which give $O_{a^\prime}(\psi)=O_{a}(\psi)+\tilde{\epsilon}\langle\psi|\delta a|\psi\rangle=O_{a}(\psi)+\tilde{\epsilon}\delta^aO_{a}(\psi)$ which can be called in the same way the Heisenberg picture view of variations..
\end{itemize}
Taking these two variations to be the same: $O_a(\psi^\prime)=O_{a^\prime}(\psi)$ (via the equivalence between Schrödinger and Heisenberg pictures) which mean that $\delta^\psi O_a(\psi)=\delta^aO_{a}(\psi)=\delta O_a(\psi)$, we are left with two (in principle) equivalent ways to study differential structures in this context. As we will see in sections \ref{UsualSD} and \ref{KvN}, this works also in the classical framework since it can also be describe in the framework of Hilbert spaces an operator algebras.
\medskip
\par
Different techniques for elaborating differential calculus have been developed, the most general being the universal differential calculus. Differential calculi based on derivations of the algebra, as well as those based on the Dirac operator can be obtained as quotients of this universal structure in favorable cases. Thus the derivation technique can be said to study differential structure directly on $\calA$ while the spectral triple technique is at the level of operators on $\calH$. Then, the two main approaches explored to build NC differential structures in NCG belong respectively to the two categories of differential structures which will be presented below. Indeed in this view, the derivation approach to algebra can be seen as a kind of Heisenberg picture (where only variations of the observables are taken into account), whereas the spectral triplet approach belongs to the Schrödinger picture (variation of the states).
\medskip
\par  
From a mathematical point of view, the derivation approach seems to be more economic since it has the good feature that it can be defined from the algebra itself, without added structures (like the Dirac), and that it considers only the algebra, which is more fundamental than the Hilbert space according to GNS construction. However, strong mathematical motivations have also motivated the spectral triplet approach, whose mathematical and physical developments are quite remarkable. Giving the usual Riemannian geometrical framework and having nice realizations in physics as we will see with the NCSMPP in section \ref{NCSMPP}.

\section{Usual Differential Structures, open door to Noncommutative Extension}
\label{UsualSD}
In what follows I will present these three ways of proceeding. But let us first clarify what we mean by differential structure, in order to feel as much as possible the potential meaning of it's NC generalization.
\medskip
\par
\textbf{Differential calculus} is the study of the rate at which quantities like functions evolve. The main object of differential calculus is the \textbf{derivative}, which measures this rate according to the evolution of its argument, which can be a coordinate for example. Using spatial coordinate, the derivative is defined as follows:
\begin{align}
	\label{DiffLimit}
	\frac{df(x)}{dx}= \lim_{dx \to 0\atop dx\ne 0}\frac{f(x+dx)-f(x)}{dx}
\end{align}
The identification of a derivative is called \textbf{differentiation}.
\begin{remark}
	In practice, the study of the variation of an observable can only be done by correlating it with another observable, known as the reference observable, such as the pointer of a clock or the graduations of a ruler, it is not intrinsic!
\end{remark}
Derivatives are of great importance in physics, and in almost all scientific fields studying quantities and their evolution. For example, velocity is the derivative corresponding to the displacement of a moving body with respect to time, the derivative of velocity is the acceleration, and the derivative of momentum (function of velocity) is equal to the sum of forces acting on the body. Moreover, most of the fundamental equations used to describe phenomenon are differential equations, whose constituents are once again derivatives. As soon as it is a question of the evolution of a quantity, or position, according to certain degrees of freedom of the associated space, derivations associated with these infinitesimal degrees of freedom come into play, this is why differential geometry is of such great importance and universality in physics. More general definitions of derivative, not involving the notion of a limit as in eq.( \ref{DiffLimit}) (which becomes limited in the NC framework) can be defined, using the so-called $q$ or $h$-calculus, taking $q=\exp(ih)$:
\begin{align*}
	d_q(f(x))=f(qx)-f(x)\qquad \text{or}\qquad d_h(f(x))=f(x+h)-f(x)
\end{align*}
When $h\to 0$ or equivalently $q\to 1$, these formulas become equivalent to the usual derivative \ref{DiffLimit}.
\medskip
\par
Taking a derivative, we can construct the corresponding \textbf{differential}:
\begin{align}
	\label{FormDiff}
	df(x,dx)=\frac{df(x)}{dx}dx
\end{align}
If $f$ is a real differentiable function, its differential $df$ is a differential 1-form (called exact) which at each point $x$ is the linear form $df(x)$. Locally, the differential 1-forms are expressed as combinations of function differentials.
\medskip
\par
As mentioned before, we can express all these results in the world of operator algebras and Hilbert spaces. Taking the differential over $\calC^\infty(\Man)$, the Hilbert space can be defined as the completion in the $L^2$ norm of the module of 1-forms (considered as vector space here). The derivative $d$ can then be considered as an operator on this Hilbert space (densely defined) such that $\forall\psi\in\calH$ and $f\in\calC^\infty (\Man)$:
\begin{align*}
	&(df)\psi=[d,f]\psi 	=d(f\cdot \psi)-f\cdot (d\psi)\qquad\qquad \text{Heisenberg picture}&\\
	&f. (d\psi)=d(f\cdot \psi)-(df)\psi\qquad\qquad\qquad\qquad \text{Schrodinger picture}&
\end{align*}
With this way of expressing things, with commutators in the framework of operators algebras, we are therefore able to explore the NC framework extension, adopting more general notions, when needed.
\medskip
\par
A \textbf{derivation} is a linear map over an algebra that generalizes the derivative operator, but this time on the algebraic side. Taking an algebra $\calA$, a derivation is defined to be a linear map $\kX: \calA\to\calA$ which respects the Leibniz's rule: 
\begin{align*}
	\kX(ab)=a\kX(b)+\kX(a)b
\end{align*}
For example, partial and Lie derivatives with respect to a vector field are derivations over algebras on $\bbR^n$ or a differentiable manifold. This formula can be deduced from this simple argument, if we take $\kX(a)$ and $\kX(b)$ as representing small variations of $a$ and $b$, then the variation $\kX(ab)$ can be written:
\begin{align*}
	\kX(ab)=(a+\kX(a))(b+\kX(b))-ab=\kX(a)b+a\kX(b)
\end{align*}
because $\kX(a)\kX(b)$ is negligible since we consider small variations. As we will see, this notion is of great importance to extend the usual derivative to the NCG framework, partly because Leibniz rule still makes sense on NC algebras the novelty being that because of now $[a,\kX(b)]\neq 0$, then:
\begin{align*}
	\kX([a,b])=a\kX(b)+\kX(a)b-\kX(b)a-b\kX(a)\neq 0 
\end{align*}  
Differential calculus can therefore also be constructed on the basis of such derivations.
\medskip
\par
It is important to realize that derivations on the algebraic side can be viewed to be equivalent to tangent spaces on the geometric side. Indeed, directional derivatives can be defined as follow: if we take a point $x\in \Man$, and a smooth curve $\gamma:\bbR\,\to\, \Man$ parameterized by $t$ such that $\gamma(0)=x$, therefore, taking any $f\in\calC^\infty(\Man)$ we have 
\begin{align*}
	\frac{d f(\gamma(t))}{dt}=\frac{d\gamma(t)}{dt}\frac{d f(\gamma(t))}{d\gamma(t)}\qquad\qquad \text{at} \qquad t=0
\end{align*}
Where, by comparison to eq.(\ref{FormDiff}), the derivation $\frac{df(x)}{dx}$ can be identified to $\frac{d f(\gamma(t))}{d\gamma(t)}$ and the infinitesimal displacement $dx$ to $\frac{d\gamma(t)}{dt}$. Therefore, for a given $x$, taking the set of all derivations $\frac{d f(\gamma(t))}{dt}$ (at $t=0$) corresponding to all curves $\gamma$ gives the tangent space $T_x\Man$. This is interesting since, we can see that in this way, the differential structure on the algebraic side contains the tangent space which is an essential geometric structure. These tangent spaces being at the hearth of most of the natural geometric constructions. This is why derivations can offer the algebraic equivalent of tangent spaces, “styling” the points provided by pure states through the Gelfand Naimark theorem, and fulfilling the algebraic to geometric correspondence. In a more general context than commutative algebras, some derivations will not correspond to directional derivatives in the usual meaning, but to derivations along inner degrees of freedom.
\medskip
\par
Furthermore, derivations can be viewed as infinitesimal generators of the automorphisms in the algebra, as discussed in \cite{sakai1991operator}, and later in section \ref{DeDS}, taking an algebra $\calA$, and $\phi_t\, :\, \calA\to\calA$ an automorphism with parameter $t$ such that $\phi_0=\bbbone_\calA$ and naturally $\phi_t(ab)=\phi_t(a)\phi_t(b)$, then the map $a\, \to \, \frac{d}{dt}|_{t=0}\phi_t(a)$ is a derivation. You can therefore think of derivations as a kind of vector fields, linked to the potential infinitesimal displacements which are allowed in the geometry. In the context of differential geometry (then general relativity), it is possible to show that diffeomorphisms of $\Man$ are equivalent to automorphisms of $\calC^\infty(\Man)$, therefore derivations can be seen to be in relation with the set of infinitesimal displacements generating these diffeomorphisms. As we will see in chapter \ref{NCGFTGener} automorphisms of the algebra can be seen as linked with the symmetries of a theory, and are of great importance in the framework of Gauge field theories since they will correspond to gauge transformations.
\medskip
\par
An important structure can be built from the algebra and its derivations, that of the differential calculus. Let $\calA$ be an unital associative algebra, a differential calculus on $\calA$ is a graded differential algebra $(\Omega^\grast(\calA),\dd)$, with $\Omega^\grast(\calA)=\toplus_{n\geq 0}\Omega^n(\calA)$, $\Omega^0=\calA$ and by definition $\dd:\Omega^\grast\,\to\, \Omega^{\grast+1}$. The derivation must satisfy $\dd^2=0$ together with the graded Leibniz rule:
\begin{align*}
	\dd(\omega\eta)=(\dd\omega)\eta+(-1)^n\omega(\dd\eta)
\end{align*}
for all $\omega\in\Omega^n$ and $\eta\in\Omega^m$. A direct consequence is that $\dd\bbbone=0$. The space $\Omega^n$ is called the space of non-commutative $n$-forms, which is a $\calA$-bimodule. In general, every element of $\Omega^n$ can be expressed as a combination of elements $a_0\dd a_1\dots\dd a_n$, and we have:
\begin{align*}
	\dd(a_0\dd a_1\dots\dd a_n)=\dd a_0\dd a_1\dots\dd a_n.
\end{align*} \GN{pas toujour, à discuter}
There are many possibilities to make such a differential calculus on $\calA$, these possibilities are distinguished by the nature of the chosen derivation. In section \ref{UDS}, I will present the most general of them, based on the differential algebra of universal forms $\Omega_U(\calA)$ \cite{coquereaux1998espaces}.  Then in sections \ref{DeDS} and \ref{DiDS} I will present the methods which use the derivations of the algebra $\OmegaDer(\algA)$, and the one which uses the spectral triples $\Omega_{D}(\calA)$, both being obtained as a quotient of $\Omega_U(\calA)$. More details can be found in \cite{muller1998introduction},  \cite{jordan2014gauge}, \cite{gracia2013elements} and \cite{masson2001geometrie}.
\GN{exemple de Rham}

\section{Universal Differential Structures}
\label{UDS}
The universal differential calculus can be seen as the more general and abstract way to look at differential structures. In the following, we will need a convenient presentation of the differential graded algebra $(\Omega^\grast_U(\algA), \ddU)$. We follow the presentation in \cite{Mass95a} and \cite{gracia2013elements}.\footnote{We owe this presentation to Thierry Masson who owed this presentation to Michel~Dubois-Violette.}
\medskip
\par 
Taking an associative algebra $\calA$, we define the tensor algebra of degree $n$ on $\calA$:
\begin{align*}
	\calT^n \algA \defeq  \underbrace{\mathcal{A} \otimes \cdots \otimes \mathcal{A}}_{n+1 \text { times }}= \algA^{\otimes^{n+1}}
\end{align*}
with $\calT^0\algA=\algA$. We define $\calT^\grast$ by $\calT^\grast \algA = \toplus_{n\geq 0} \calT^n \algA$. These are bimodules on $\calA$ and a graded algebra for the product $\calT^n \algA \otimes \calT^{n'} \algA \to \calT^{n+n'} \algA$ defined by $(a^0 \otimes \cdots \otimes a^n) (a'^0 \otimes \cdots \otimes a'^{n'}) \defeq a^0 \otimes \cdots \otimes a^n a'^0 \otimes \cdots \otimes a'^{n'}$. In particular, $\calT^\grast \algA$ is a bimodule over $\algA = \calT^0 \algA$. Define $\ddU : \calT^n \algA \to \calT^{n+1} \algA$ as 
\begin{align*}
	\ddU (a^0 \otimes \cdots \otimes a^n) =
	& \bbbone \otimes a^0 \otimes \cdots \otimes a^n
	\\
	& + \tsum_{p=1}^{n} (-1)^p a^0 \otimes \cdots \otimes a^{p-1} \otimes \bbbone \otimes a^{p} \otimes \cdots \otimes a^n
	\\
	& + (-1)^{n+1} a^0 \otimes \cdots \otimes a^n \otimes \bbbone.
\end{align*}
Then $\ddU$ is a derivation of degree $1$ on the graded algebra $\calT^\grast \algA$ such that $\ddU^2 = 0$. Notice that $\ddU(a) = \bbbone \otimes a - a \otimes \bbbone$ on $\calT^0 \algA$. It satisfies the Leibniz rule since: 
\begin{align*}
	\ddU(ab)=\bbbone\otimes ab-ab\otimes \bbbone=\bbbone\otimes ab-a\otimes b+a\otimes b-ab\otimes \bbbone=\ddU(a)b+a\ddU(b).
\end{align*}
Let first see how this relates to derivations on the algebra. A derivation $\kX$ of an algebra $\calA$ into a bimodule $\calM$, is a linear map satisfying the Leibniz rule.
\begin{proposition}
	For any derivation $\kX$ of $\algA$, in a bimodule $\calM$, there is a unique homomorphism of bimodules:
	\begin{align*}
		h_\kX\, :\, \Omega^1_U(\algA)\,\to\, \calM
	\end{align*}
	such that $\kX=h_\kX \circ \ddU$.
\end{proposition}
\begin{proof}
	Taking $h_\kX(\ddU a)=\kX(a)$, as $\Omega^1_U(\algA)$ is generated by the $\ddU a$ taking all $a\in\calA$, we see that
	\begin{align*}
		\kX(ab)=h_\kX(\ddU(ab))=h_\kX(\ddU(a))h_\kX(b)+h_\kX(a)h_\kX(\ddU(b))=\kX(a)h_\kX(b)+h_\kX(a)\kX(b).
	\end{align*}
	are compatible structures iff $\forall a\in\calA$ we have $h_\kX(a)=a$, this requirement inducing the unicity of this homomorphism.
\end{proof}

It is convenient to introduce the maps $i^p_{\bbbone}(a^0 \otimes \cdots \otimes a^n) \defeq a^0 \otimes \cdots \otimes a^{p-1} \otimes \bbbone \otimes a^p \otimes \cdots \otimes a^n$ for any $p = 0, \dots, n+1$, with the convention that for $p=0$, the tensor factor $\bbbone$ is added before $a^0$ (for $p=n+1$, it is added after $a^n$). Then $\ddU = \tsum_{p=0}^{n+1} (-1)^p \,  i^p_{\bbbone} : \calT^n \algA \to \calT^{n+1} \algA$.
\medskip
\par 
Let $\mu : \calT^1 \algA \to \calT^0 \algA$ be the multiplication map $a^0 \otimes a^1 \mapsto a^0 a^1$, and define $\Omega^1_U(\algA) \defeq \ker \mu \subset \calT^1 \algA$. Then $\ddU$ maps $\calT^0 \algA$ into $\Omega^1_U(\algA)$ and $\Omega^1_U(\algA)$ is generated, as a bimodule on $\algA$, by the $\ddU a$'s for $a \in \algA$:\footnote{If $\tsum_{i} a_i^0 \otimes a_i^1 \in \calT^1 \algA$ is such that $\mu(\tsum_{i} a_i^0 \otimes a_i^1) = \tsum_{i} a_i^0 a_i^1 = 0$, then $\tsum_{i} a_i^0 \otimes a_i^1 = \tsum_{i} a_i^0 (\bbbone \otimes a_i^1 -  a_i^1 \otimes \bbbone) = \tsum_{i} a_i^0 \ddU a_i^1$.}
\begin{align*}
	a\ddU b=a\otimes b-ab\otimes \bbbone\in \Omega^1_U(\algA)\subset \algA\otimes \algA=\algA^{\otimes^{2}}
\end{align*}
We can check that this belongs to $\ker\mu$: $\mu(a\ddU b)=ab-ab=0$. Then, elements in $\Omega^n_U(\algA)$ will take the form:
\begin{align*}
	a\ddU b_1\ddU b_2\dots \ddU b_n\,\in\,\Omega^n_U(\algA)\subset \algA^{\otimes^{n}} .
\end{align*}
These definitions are very abstract and give the feeling that many intuitions that we usually associate with derivations are lost. Let's see what this can represent in the case of commutative algebras, in order to get an idea of what this universal differential calculus actually means.
Let's take $\algA$ to be a commutative algebra. Since the elements of $\algA$ can be represented as functions on $\Man$, the element of $\algA\otimes \algA$ are of the form $c(x)\otimes d(y)$. Then $a\ddU b$ can be written as:
\begin{align*}
	[a\ddU b](x,y)=a(x)\otimes b(y)-a(x)b(x)\otimes\bbbone(y)
\end{align*}
We see that $\mu([a\ddU b](x,y))=a(x)(b(x)-b(y))$ is equal to $0$ if $x=y$. When $\Man$ is a differentiable manifold, we recognize the usual differential $[\ddU b](x,y)=b(x)-b(y)$, taking the limit $y\to x$. we recover the usual differential one-form of equation \eqref{FormDiff}:
\begin{align*}
	\dd b(x,dx)=\frac{\dd b(x)}{dx}dx\qquad\qquad \text{with}\qquad\qquad \dd x=y-x
\end{align*}
As mentioned in \cite{coquereaux1998espaces}, when considering general $\dd x$, this can be seen as a non local differential calculus.
\medskip
\par 
Let come back to general associative algebras, taking $\Omega^0_U(\algA) \defeq \algA$ and $\Omega^n_U(\algA) \defeq \Omega^1_U(\algA) \otimes_{\algA} \cdots \otimes_{\algA} \Omega^1_U(\algA)$ ($n$ times tensor product over $\algA$) for any $n \geq 2$ and $\Omega^\grast_U(\algA) \defeq \toplus_{n \geq 0} \Omega^n_U(\algA)$. Equivalently, $\Omega^\grast_U(\algA)$ is the graded sub-algebra of $\calT^\grast \algA$ generated by $\Omega^0_U(\algA)$ and $\Omega^1_U(\algA)$. One can then check that $\Omega^n_U(\algA) \subset \calT^n \algA$ is generated by the $a^0 \ddU a^1 \cdots \ddU a^n$ for $a^0, \dots, a^n \in \algA$, so that $\ddU$ restricts to maps $\Omega^n_U(\algA) \to \Omega^{n+1}_U(\algA)$, and then $(\Omega^\grast_U(\algA), \ddU)$ is a graded differential sub-algebra of $(\calT^\grast \algA, \ddU)$.
\medskip
\par 
Let us consider the case $\algA = \toplus_{i=1}^{r} \algA_i$, where $\algA_i$ are unital algebras with units $\bbbone_{\algA_i}$. It will be useful for future discussions to use explicit presentations of $(\calT^\grast \algA, \ddU)$ and $(\Omega^\grast_U(\algA), \ddU)$ constructed as follows. Let 
\begin{align*}
	\kT^0\algA \defeq \left\{
	\smallpmatrix{ a_1 & 0 & \cdots & 0 \\
		0 & a_2 & \cdots & 0 \\
		\vdots & & \ddots &  \\
		0 & 0 & \cdots & a_r}
	\mid 
	a = \toplus_{i=1}^{r} a_i \in \algA \right\}.
\end{align*}
For any $n \geq 1$ and any $1 \leq i_0, \dots, i_n \leq r$, let us introduce the notation $\algA^\otimes_{i_0, \dots, i_{n}} \defeq \algA_{i_0} \otimes \cdots \otimes \algA_{i_n}$. Now, let $\kT^n_{i_1, \dots, i_{n-1}}\algA$ be the set of matrices with entries in $\algA^\otimes_{i, i_1, \dots, i_{n-1}, j}$ at row $i$ and column $j$. This can be schematically visualized as
\begin{align*}
	\begin{pmatrix} 
		\algA^\otimes_{1, i_1, \dots, i_{n-1}, 1} 
		& \algA^\otimes_{1, i_1, \dots, i_{n-1}, 2} 
		& \cdots 
		& \algA^\otimes_{1, i_1, \dots, i_{n-1}, r}
		\\
		\algA^\otimes_{2, i_1, \dots, i_{n-1}, 1} 
		& \algA^\otimes_{2, i_1, \dots, i_{n-1}, 2} 
		& \cdots 
		& \algA^\otimes_{2, i_1, \dots, i_{n-1}, r}
		\\
		\vdots & \vdots &  & \vdots 
		\\
		\algA^\otimes_{r, i_1, \dots, i_{n-1}, 1} 
		& \algA^\otimes_{r, i_1, \dots, i_{n-1}, 2} 
		& \cdots 
		& \algA^\otimes_{r, i_1, \dots, i_{n-1}, r}
	\end{pmatrix}
\end{align*}
where the first and last algebras in the tensor products will play a crucial role in the following. Combining the products 
\begin{align*}
	\algA^\otimes_{i, i_1, \dots, i_{n-1}, k}
	\otimes 
	\algA^\otimes_{k, j_1, \dots, j_{n'-1}, j}  
	\to 
	\algA^\otimes_{i, i_1, \dots, i_{n-1}, k, j_1, \dots, j_{n'-1}, j},
\end{align*} 
defined by the product in $\algA_{k}$, and the usual rules for matrix multiplications, one gets products
\begin{align*}
	\kT^n_{i_1, \dots, i_{n-1}}\algA \otimes \kT^{n'}_{j_1, \dots, j_{n'-1}}\algA
	\to \toplus_{k=1}^{r} \kT^{n+n'}_{i_1, \dots, i_{n-1}, k, j_1, \dots, j_{n'-1}}\algA .
\end{align*} 
Let us introduce
\begin{align*}
	\kT^n \algA \defeq \toplus_{i_1, \dots, i_{n-1} = 1}^{r} \kT^n_{i_1, \dots, i_{n-1}}\algA
	\quad \text{and} \quad 
	\kT^\grast \algA \defeq \toplus_{n \geq 0} \kT^n \algA
\end{align*}
then $\kT^\grast \algA$ is a graded algebra for the global product induced by the products defined above. Explicitly, for $\toplus_{i_1, \dots, i_{n-1} =1}^{r} \big( a^0_{i_0} \otimes a^1_{i_1} \otimes \cdots \otimes a^{n-1}_{i_{n-1}} \otimes a^n_{i_n} \big)_{i_0, i_n = 1}^{r} \in \kT^n \algA$ and $\toplus_{j_1, \dots, j_{n'-1} =1}^{r} \big( b^0_{j_0} \otimes b^1_{j_1} \otimes \cdots \otimes b^{n'-1}_{j_{n'-1}} \otimes b^{n'}_{j_{n'}} \big)_{j_0, j_{n'} = 1}^{r} \in \kT^{n'} \algA$, their product in $\kT^{n+n'} \algA$ is
\begin{align}
	\label{eq product kT n np}
	\toplus_{\substack{i_1, \dots, i_{n-1}, i_n,\\ j_1, \dots, j_{n'-1} =1}}^{r}
	\big(
	a^0_{i} \otimes a^1_{i_1} \otimes \cdots \otimes a^{n-1}_{i_{n-1}} \otimes a^n_{i_n} 
	b^0_{i_n} \otimes b^1_{j_1} \otimes \cdots \otimes b^{n'-1}_{j_{n'-1}} \otimes b^{n'}_{j}
	\big)_{i, j=1}^{r}
\end{align}

Let $\bmu$ be the component-wise product on $\kT^1 \algA$. Since multiplications by elements in $\algA_i$ and $\algA_j$ are zero for $i \neq j$, the resulting matrix is diagonal, and so one gets a natural map $\bmu : \kT^1 \algA \to \kT^0 \algA$. Let $\bOmega^1_U(\algA) \defeq \ker \bmu \subset \kT^1 \algA$ and $\bOmega^\grast_U(\algA) \subset \kT^\grast \algA$ be the graded sub-algebra generated by $\bOmega^0_U(\algA) \defeq \kT^0\algA$ and $\bOmega^1_U(\algA)$. For any $p = 0, \dots, n+1$, define $\bi^p_{\bbbone} : \kT^n_{i_1, \dots, i_{n-1}}\algA \to \toplus_{k=0}^{r} \kT^{n+1}_{i_1, \dots, i_{p-1}, k, i_{p}, \dots, i_{n-1}}\algA$ by inserting $\bbbone = \toplus_{k=1}^{r} \bbbone_{\algA_k}$ component-wise, \textit{i.e.} $\bi^p_{\bbbone} = \toplus_{k=1}^{r} \bi^p_{\bbbone_{\algA_k}}$ with obvious notations. Then one can define $\bddU \defeq \tsum_{p=0}^{n+1} (-1)^p \,  \bi^p_{\bbbone} : \kT^n \algA  \to \kT^{n+1} \algA$.

\begin{proposition}
	The map $\bddU$ is a differential on $\kT^\grast \algA$ and there is an isomorphism $t : \calT^\grast \algA \to \kT^\grast \algA$ of graded differential algebras which induces an isomorphism of the graded differential (sub)algebras $\Omega^\grast_U(\algA)$ and $\bOmega^\grast_U(\algA)$.
\end{proposition}

\begin{proof}
	For $n=0$, one defines $t(\toplus_{i=1}^{r} a_i) = \smallpmatrix{ a_1 & \cdots & 0 \\
		& \ddots &  \\
		0 & \cdots & a_r} \in \kT^0\algA$ for any $\toplus_{i=1}^{r} a_i \in \algA$. For $n \geq 1$, consider any $a^{0} \otimes \cdots \otimes a^{n} \in \calT^n \algA$ with $a^{p} = \toplus_{i=1}^{r} a^{p}_i$ where $a^{p}_i \in \algA_i$. Expanding the tensor products along these direct sums, one gets a sum of terms of the form $a^{0}_{i_0} \otimes \cdots \otimes a^{n}_{i_n} \in \algA^\otimes_{i_0, \dots, i_{n}}$ that we assemble as elements in $\kT^n_{i_1, \dots, i_{n-1}}\algA$. This defines the map $t : \calT^n \algA \to \kT^n \algA$, which, for any $n \geq 0$, is by construction an isomorphism of vector spaces. A straightforward computation shows that the product on $\kT^\grast \algA$ is such that $t$ is a morphism of graded algebras.
	\medskip
	\par 	
	By construction of $\bi^p_{\bbbone}$, one has $t \circ i^p_\bbbone = \bi^p_{\bbbone} \circ t$, so that $\bddU$ is a differential on $\kT^\grast \algA$ and $t$ is an isomorphism of differential algebras.
	\medskip
	\par 	
	Finally, the map $\bmu$ has been defined such that $t \circ \mu = \bmu \circ t$ so that $t$ identifies $\Omega^1_U(\algA)$ with $\bOmega^1_U(\algA)$, and so $\Omega^\grast_U(\algA)$ with $\bOmega^\grast_U(\algA)$.
\end{proof}

Notice that, with $\hbbbone \defeq t(\bbbone \otimes \bbbone) \in \kT^1 \algA$, one has $\bddU t(a) = [ \hbbbone, t(a)]$ for any $a \in \algA$.

\section{Derivation-based Differential Structures}
\label{DeDS}
The derivation-based differential calculus was defined by Dubois-Violette in \cite{Dubo88a} and studied for various algebras, by Dubois-Violette, Kerner, Madore, Masson, Michor, Vitale, de Goursac, Wallet, Wulkenhaar, see for instance \cite{DuboKernMado90a, DuboKernMado90b, Mass96a, DuboMass98a, Mass99a, DuboMich94a, DuboMich96a, DuboMich97a, CagnMassWall11a,de2007noncommutative,de2008vacuum, martinetti2013noncommutative}. Some reviews can also be found in \cite{Dubo01a, Mass08c, Mass08b, masson2008introduction}. The main ingredient is the space of derivations on an associative algebra on which a natural differential calculus can be based.
\medskip
\par 
Let $\algA$ be an associative algebra with unit $\bbbone$, and let $\calZ(\algA)$ be its center. The space of derivations of $\algA$ is
\begin{equation*}
	\Der(\algA) = \{ \kX : \algA \rightarrow \algA \ / \ \kX \text{ linear}, \kX\cdotaction(ab) = (\kX\cdotaction a) b + a (\kX\cdotaction b), \forall a,b\in \algA\}.
\end{equation*}
This vector space is a Lie algebra for the bracket $[\kX, \kY ]a = \kX  \kY\cdotaction a - \kY \kX\cdotaction a$ for all $\kX,\kY \in \Der(\algA)$, and a $\calZ(\algA)$-module for the product $(f \kX )\cdotaction a = f ( \kX \cdotaction a)$ for all $f \in \calZ(\algA)$ and $\kX \in \Der(\algA)$. The subspace 
\begin{equation*}
	\Int(\algA) = \{ \ad_a : b \mapsto [a,b]\ / \ a \in \algA\} \subset \Der(\algA)
\end{equation*}
is called the vector space of inner derivations: it is a Lie ideal and a $\calZ(\algA)$-submodule. The fact that the map $\algA\,\to\,\Int(\algA)$ defined by $a\,\to\,\ad_a$ is surjective with kernel $\calZ(\algA)$ implies that $\Int(\algA)$ can take the corresponding form:
\begin{align}
	\label{InQuotCenter}
	\Int(\algA)\simeq \algA/\calZ(\algA).
\end{align}
Taking any $\kX\in \Der(\algA)$ we have $[\kX, \ad_a]=\ad_{\kX \cdotaction a}$, then $\Int(\algA)$ is an ideal of $\Der(\algA)$ and so, $\Der(\algA)/\Int(\algA)$ is also a Lie algebra. This define the subspace of Outer derivations with the quotient $\Out(\algA)=\Der(\algA)/\Int(\algA)$. Then there is a short exact sequence of Lie algebras and $\calZ(\algA)$-modules:
\begin{align}
	\label{ExactSeqDer}
	0\, \longrightarrow \, \Int(\calA)\,\longrightarrow\, \Der(\calA)\,\longrightarrow\, \Out(\calA)\,\longrightarrow\, 0
\end{align}
Commutative algebras only have outer derivations.
\medskip
\par 
Suppose that $\algA$ has an involution $a \mapsto a^*$. Then a real derivation on $\algA$ is a derivation $\kX$ such that $(\kX \cdotaction a)^* = \kX \cdotaction a^*$ for any $a \in \algA$.
\medskip
\par 
Let $\OmegaDer^p(\algA)$ be the vector space of $\calZ(\algA)$-multilinear antisymmetric maps from $\Der(\algA)^p$ to $\algA$, with $\OmegaDer^0(\algA) = \algA$. Then the total space
\begin{equation*}
	\OmegaDer^\grast(\algA) = \toplus_{p \geq 0} \OmegaDer^p(\algA)
\end{equation*}
gets a structure of $\bbN$-graded differential algebra for the product
\begin{align*}
	\label{eq def form product}
	(\omega \wedge \eta)(\kX_1, \dots, \kX_{p+q}) 
	\defeq
	\frac{1}{p!q!} \sum_{\sigma\in \kS_{p+q}} (-1)^{\abs{\sigma}} \omega(\kX_{\sigma(1)}, \dots, \kX_{\sigma(p)}) \eta(\kX_{\sigma(p+1)}, \dots, \kX_{\sigma(p+q)})
\end{align*}
for any $\omega \in \OmegaDer^p(\algA)$, any $\eta \in \OmegaDer^q(\algA)$ and any $\kX_i \in \Der(\algA)$ where $\kS_{n}$ is the group of permutations of $n$ elements. A differential $\dd$ is defined by the so-called Koszul formula
\begin{align*}
	\label{eq def differential}
	\dd \omega (\kX_1, \dots , \kX_{p+1}) 
	\defeq
	\sum_{i=1}^{p+1} (-1)^{i+1} \kX_i \cdotaction \omega( \kX_1, \dots \omi{i} \dots, \kX_{p+1}) 
	+ \sum_{1 \leq i < j \leq p+1} (-1)^{i+j} \omega( [\kX_i, \kX_j], \dots \omi{i} \dots \omi{j} \dots, \kX_{p+1}). 
\end{align*}
This makes $(\OmegaDer^\grast(\algA), \dd)$ a graded differential algebra.
\medskip
\par 
Let us take the more intuitive commutative algebra $\calC^\infty(\Man)$, we have that $\calZ(\calC^\infty(\Man))=\calC^\infty(\Man)$. We can see that the algebra of derivations correspond to $\Gamma(\Man)$ the Lie algebra of smooth vector fields on $\Man$ such that $\Der(\calC^\infty(\Man))=\Gamma(\Man)=\Out(\calC^\infty(\Man))$ since $\Int(\calC^\infty(\Man))=0$. In this case, we recover the graded differential algebra of de Rham forms on $\Man$:
\begin{align*}
	\OmegaDer^\grast(\calC^\infty(\Man))=\Omega^\grast(\Man)
\end{align*}	
This shows how natural is this construction.
\medskip
\par 
\begin{proposition}[Transport of forms by automorphisms]
	\label{prop transport forms}
	Let $\Psi : \algA \to \algA$ be an algebra automorphism. Then $\Psi$ induces an automorphism on $\calZ(\algA)$.
	\medskip
	\par 	
	The map $\PsiDer : \Der(\algA) \to \Der(\algA)$ defined by $\PsiDer(\kX) \cdotaction a \defeq \Psi( \kX \cdotaction \Psi^{-1}(a) )$ for any $\kX \in \Der(\algA)$ and $a \in \algA$, is an automorphism of the Lie algebra $\Der(\algA)$ and $\PsiDer(f \kX) = \Psi(f) \PsiDer(\kX)$ for any $f \in \calZ(\algA)$ (so $\PsiDer$ is not necessary an automorphism for the structure of $\calZ(\algA)$-module). For inner derivations, one has $\PsiDer(\ad_a) = \ad_{\Psi(a)}$.
	\medskip
	\par 	
	The maps $\Psi : \OmegaDer^p(\algA) \to \OmegaDer^p(\algA)$ defined by 
	\begin{align*}
		\Psi(\omega) (\kX_{1}, \dots , \kX_{p}) 
		\defeq \Psi\left( \omega( \PsiDer^{-1}(\kX_{1}), \dots, \PsiDer^{-1}(\kX_{p}) ) \right)
	\end{align*}
	for any $\omega \in \OmegaDer^p(\algA)$ and $\kX_i \in \Der(\algA)$, defines an automorphism of the graded differential algebra $(\OmegaDer^\grast(\algA), \dd)$.
\end{proposition}

For $p=0$, $\Psi$ defined on $\OmegaDer^0(\algA) = \algA$ is exactly the original automorphism $\Psi$ of $\algA$, so that the notation is justified.

\begin{proof}
	For any $f \in \calZ(\algA)$ and $a \in \algA$, one has $\Psi(f) a = \Psi( f \Psi^{-1}(a) ) = \Psi( \Psi^{-1}(a) f ) = a \Psi(f)$ so that $\Psi(f) \in \calZ(\algA)$.
	\medskip
	\par 	
	With obvious notations, one has 
	\begin{align*}
		\PsiDer(\kX) \cdotaction (a b) 
		&= \Psi\left( \kX \cdotaction \Psi^{-1}(a b) \right) 
		= \Psi\left( \kX \cdotaction (\Psi^{-1}(a) \Psi^{-1}(b)) \right) 
		= \Psi\left( \kX \cdotaction \Psi^{-1}(a) \right) b + a \Psi\left( \kX \cdotaction \Psi^{-1}(b)\right)\\ 
		&= (\PsiDer(\kX) \cdotaction a) b + a (\PsiDer(\kX) \cdotaction b)
	\end{align*}
	so that $\PsiDer(\kX)$ is a derivation. 
	\medskip
	\par 	
	In the same way, one has $\PsiDer(f \kX) \cdotaction a = \Psi( f (\kX \cdotaction \Psi^{-1}(a)) ) = \Psi(f) \Psi( \kX \cdotaction \Psi^{-1}(a) ) = \Psi(f) \PsiDer(\kX) \cdotaction a$. 
	\medskip
	\par 	
	For $\kX, \kY \in \Der(\algA)$, one has 
	\begin{align*}
		\PsiDer([\kX, \kY]) \cdotaction a 
		&= \Psi( \kX \cdotaction (\kY \cdotaction \Psi^{-1}(a)) ) - \Psi( \kY \cdotaction (\kX \cdotaction \Psi^{-1}(a)) ) = \Psi( \kX \cdotaction \Psi^{-1} ( \PsiDer(\kY) \cdotaction a) ) - \Psi( \kY \cdotaction \Psi^{-1} ( \PsiDer(\kX) \cdotaction a) )
		\\
		&= \PsiDer(\kX) \cdotaction ( \PsiDer(\kY) \cdotaction a ) - \PsiDer(\kY) \cdotaction ( \PsiDer(\kX) \cdotaction a )= [\PsiDer(\kX), \PsiDer(\kY)] \cdotaction a
	\end{align*}
	so that $\PsiDer([\kX, \kY]) = [\PsiDer(\kX), \PsiDer(\kY)]$. The inverse $\PsiDer^{-1}$ is defined by $\PsiDer^{-1}(\kX) \cdotaction a \defeq \Psi^{-1}( \kX \cdotaction \Psi(a) )$ as it can be easily checked. For inner derivations, one has $\PsiDer(\ad_a) \cdotaction b = \Psi( [a, \Psi^{-1}(b)]) = [\Psi(a), b] = \ad_{\Psi(a)} \cdotaction b$.
	\medskip
	\par 	
	For any $\omega \in \OmegaDer^p(\algA)$, it is easy to check that $\Psi(\omega)$ is a $\calZ(\algA)$-multilinear antisymmetric map from $\Der(\algA)^p$ to $\algA$. For any $\omega \in \OmegaDer^p(\algA)$ and any $\eta \in \OmegaDer^q(\algA)$, the relation $\Psi(\omega) \wedge \Psi(\eta) = \Psi( \omega \wedge \eta )$ is a direct consequence of the definition of $\Psi$ on forms. The proof of $\Psi(\dd a) = \dd \Psi(a)$ is a straightforward computation: $\Psi(\dd a)(\kX) = \Psi( \dd a (\PsiDer^{-1}(\kX))) = \Psi( \PsiDer^{-1}(\kX) \cdotaction a) = \kX \cdotaction \Psi(a)$ on the one hand and $(\dd \Psi(a))(\kX) = \kX \cdotaction \Psi(a)$ on the other hand. To prove $\dd \Psi(\omega) = \Psi(\dd \omega)$ for $\omega \in \OmegaDer^p(\algA)$, one has to use a similar computation and the fact that $\PsiDer^{-1}$ is a morphism of Lie algebras.
\end{proof}

\begin{example}[Transport of derivations by inner automorphisms]
	\label{ex transport derivations inner automorphisms}
	Let $u \in \algA$ be an invertible element (one can take $u$ to be unitary when $\algA$ has an involution). The map $\Psi(a) \defeq u a u^{-1}$ defines an automorphism of $\algA$ and a simple computation shows that $\PsiDer(\kX) = \kX + \ad_{u (\kX \cdotaction u^{-1})}$ for any $\kX \in \Der(\algA)$. In particular, if $\kX = \ad_a$ is an inner derivation, then $\PsiDer(\ad_a) = \ad_{u a u^{-1}} = \ad_{\Psi(a)}$ as expected. Notice also that $\Psi(f) = f$ for any $f \in \calZ(\algA)$ so that $\PsiDer : \Der(\algA) \to \Der(\algA)$ is an automorphism of $\calZ(\algA)$-module in that case.
\end{example}

\section{Dirac-based Differential Structures}
\label{DiDS}
The notion of Spectral triple, elaborated by A. Connes, mainly motivated by the Atiyah-Singer index theorem, is a set of data that permits to encode  algebraically and analytically all geometric features. In this subsection, we recall some main facts about the construction of differential structures using spectral triples. We refer to \cite{ConnMarc08b, Suij15a, Mass12a} for further details.
\medskip
\par 
The spectral triple way to construct differential structure consists in adding an operator $D$ to the couple $(\algA, \calH)$, this is a differential operator called Dirac operator in analogy with the Dirac operator introduced by P. Dirac to make the Schrödinger equation consistent with relativity. It can also be viewed as the square root of the Laplacian, the inverse line element $D=ds^{-1}$, and contain the data of the metric.
\medskip
\par
\begin{definition}[Spectral triple]
	A Spectral triple $(\algA, \calH,D)$ is the data of an involutive unital algebra
	$\algA$ represented by bounded operators on a Hilbert space $\calH$, and of a self-adjoint operator $D$ acting on $\calH$ such that the resolvent $(i + D^2)^{-1}$ is compact and that for any $a\in\algA$, $[D, a]$
	is a bounded operator.
\end{definition}

Let $(\algA, \hs, D)$ be a spectral triple and denote by $\pi : \algA \to \calB(\hs)$ the representation on the Hilbert space $\hs$. This makes $\hs$ a left $\algA$-module. 
\medskip
\par 
An even spectral triple $(\algA, \hs, D, \gamma)$ is equipped with a $\bbZ_2$-grading linear map $\gamma$ on $\hs$ such that $\gamma^\dagger = \gamma$, $\gamma^2 = 1$, $\gamma D + D \gamma = 0$ ($D$ is odd), $\gamma \pi(a) = \pi(a)\gamma$ for any $a \in \algA$ ($\algA$ is even). The grading $\gamma$ induces a decomposition $\hs = \hs^{+} \oplus \hs^{-}$ according to the eigenvalues $\pm 1$ of $\gamma$. Spectral triples without such a grading are referred to as odd spectral triples.
\medskip
\par 
A real spectral triple $(\algA, \hs, D, J)$ is equipped with a map $J :\hs \to \hs$ which is an anti-unitary operator: $\langle J \psi_1, J \psi_2 \rangle = \langle \psi_2, \psi_1 \rangle$ for any $\psi_1, \psi_2 \in \hs$ such that $\forall a, b \in \algA$:
\begin{flalign*}
	&[a, J b^* J^{-1}] = 0&  &\text{commutant property}\qquad\qquad\qquad&\\
	&[[D, a], J b^\ast J^{-1}] = 0&   &\text{first-order condition.}&
\end{flalign*}
The map $\hs \times \algA \to \hs$ defined by $(\xi, a) \mapsto J a^\ast J^{-1} \xi$ defines a right module structure on $\hs$ so that $\hs$ is a $\algA$-bimodule. We denote by $a^\circ$ the element in the opposite algebra $\algA^\circ$ which corresponds to $a \in \algA$ ($\algA \simeq \algA^\circ$ as vector spaces by the formal map $\algA \ni a \mapsto a^\circ \in \algA^\circ$ and the new product in $\algA^\circ$ is $a^\circ b^\circ \defeq (b a)^\circ$). Then, using $a^\circ \mapsto J a^\ast J^{-1}$ as a left representation of $\algA^\circ$, $\hs$ becomes a left $\algA \otimes \algA^\circ$-module. We will frequently write $a^\circ \psi =  J a^\ast J^{-1} \psi = \psi a$ for any $a \in \algA$ and $\psi \in \hs$. We define $\algA^{e} \defeq \algA \otimes \algA^\circ$. An even real spectral triple is an uplet $(\algA, \hs, D, J, \gamma)$ with $\gamma$ as before. Notice then that $\gamma a^\circ = a^\circ \gamma$ for any $a \in \algA$, and so $\gamma$ commutes with the left representation of $\algA^{e}$ on $\hs$. 
\medskip
\par 
In the odd and even cases, the $KO$-dimensions $n \mod 8$ are given in Table~\ref{table KO dimensions}, where the numbers $\epsilon, \epsilon', \epsilon'' = \pm 1$ are defined by the requirements:
\begin{align*}
	J^2 = \epsilon\qquad\qquad J D = \epsilon ' D J\qquad\qquad J \gamma = \epsilon'' \gamma J
\end{align*}
The last requirement holding only in the even case. 

\begin{table}[h]
	\label{TableKO}
	\centering
	\begin{tabular}{rrrrrrrrr}
		\toprule
		n & 0 & 1 & 2 & 3 & 4 & 5 & 6 & 7 \\
		\midrule
		$\epsilon$ & 1 & 1 & -1 & -1 & -1 & -1 & 1 & 1 \\
		$\epsilon'$ & 1 & -1 & 1 & 1 & 1 & -1 & 1 & 1 \\
		$\epsilon''$ & 1 & & -1 & & 1 & & -1 & \\
		\bottomrule
	\end{tabular}
	\caption{$KO$-dimensions of real spectral triples.}
	\label{table KO dimensions}
\end{table}
\medskip
\begin{remark}
	When $J^2 = -1$ and $\hs$ is finite dimensional, the dimension of the spectral triple is even (see \cite[Lemma~3.8]{Suij15a} for instance).
\end{remark}
\medskip
\par
Given a finite spectral triple $(\calA, \calH, D)$, the $\algA$-bimodule of Connes's differential 1-forms is given by
\begin{align}
	\label{EqForm}
	\Omega_{D}^{1}(\algA):=\left\{\sum_{k} a_{k}\left[D, b_{k}\right]: a_{k}, b_{k} \in \algA\right\}.
\end{align}
Thus, we can define the differential as the map $d:\calA\to \Omega^1(\calA)$ defined by $d(.)=\left[D,.\right]$ and $d^2\neq 0$.
\begin{remark}
	This definition connects with the usual differential presented in section \ref{UsualSD}, if we take the usual Dirac operator $D_M=i\gamma^\mu\partial_\mu$, $f\in \calC^\infty(\Man)$ and $\psi$ in the corresponding Hilbert space, then:
	\begin{align*}
		[D_M,f]\psi=D_M(f. \psi)-f. (D_M\psi)=i\gamma^\mu\partial_\mu(f).\psi
	\end{align*}
	We can see that even if the Dirac operator acts on the Hilbert space, it is equivalent to an action on the algebra (through the action on the Hilbert space acted on by the algebra), and therefore equivalent to an Heisenberg picture of the differential structure.
\end{remark}

Therefore, as mentioned in the end of section \ref{UDS}, we can construct the exterior algebra of differential forms associated to the Dirac operator using the corresponding representation $\piD\, :\, \Omega^n_U(\algA)\,\to\, \calB(\calH)$ of the universal differential algebra:
\begin{align*}
	&\piD(a\ddU b_1\ddU b_2\dots \ddU b_n)=a[D, b_1][D, b_2]\dots [D, b_n]\qquad\qquad a, b_1, b_2, \dots b_n\,\in\, \algA&
\end{align*}

\medskip
\par
The Dirac operators gives a geometric structure to $(\calA, \calH, D)$, allowing to reconstruct lengths between states. Indeed, taking two states $\psi_1$ and $\psi_2$ with $\psi_{(1,2)}:\,\algA\,\to\, \bbC$, we can define the distance between these two states:
\begin{align*}
	d(\psi_1,\psi_2)=sup\{|\psi_1(a)-\psi_2(a)|; \, a\,\in\, \algA,\,\,\, ||[D,a]||\leq 1\}
\end{align*}
If we consider the commutative algebra $\algA=\calC^\infty(\Man)$, and the usual Dirac operator $D_\Man=i\gamma^\mu\partial_\mu$, a pure state leads to the notion of points trough Gelfand Naimark theorem $\psi_{1,2}\,\to\, x_{1,2}\in \Man$, then we recover the usual distance corresponding to the metric $g$ on $\Man$:
\begin{align*}
	d_g(x_1,x_2)=\sup\{|f(x_1)-f(x_2)|; \, f\,\in\, \calC^\infty(\Man),\,\,\, ||[D_\Man,f]||\leq 1\}.
\end{align*}
The link between $\hat{x}(f)$ and $f(x)$ being given by the Gelfand transform eq.(\refeq{GelfTf}). The proof of this can be found in \cite{Suij15a}[page$\sim$66].The link between the Dirac operator and the metric is given by the gamma matrices trough $\{\gamma^\mu,\gamma^\nu\}=2g^{\mu\nu}$. This illustrates how the slope of functions and the distance are linked trough the Dirac operator.

\chapter{Differential Structure on Matrix Algebras}
\label{ChapDiffMatrixAlg}
In this chapter, we will make the differential structures presented in chapter \ref{DifCalc} more “concrete" by showing how they find realization on finite algebras in matrix representations, the first case being $M_n(\bbC)$ and the second $\bigoplus_{i=1}^nM_{n_i}(\bbC)$. This will be done only for derivation and spectral triples based differential calculus since it has already been done in a general way for universal differential calculus in section \ref{UDS}. This is a very important chapter since as mentioned in chapter \ref{FiniteAlg}, finite algebras are the algebras of interest here, and that constructed NC gauge theories will be based on such differential calculus defined on these finite algebras. The case of AF-algebras is left to chapters \ref{sec DBA} and \ref{sec STA} since it requires the notion of $\phi$-compatibility which will be presented in chapter \ref{sec AFA} in order to relate differential structures between two steps in the inductive sequence.

\section{\texorpdfstring{Differential Structure on $M_n(\bbC)$ and $\bigoplus_{i=1}^nM_{n_i}(\bbC)$ using Derivations}{Differential structure on Mn(C) and Oplus Mni(C) using derivations}}
\label{ResMatrDer}

Since the understanding the situation $\algA_i = M_{n_i}(\bbC)$ is our main objective for $AF$-algebras, we give here a series of notations and results for later use when this specific situation will be considered in the case of NC gauge theories in the derivation framework. This is for instance the case in Sect.~\ref{DBNCGFT} and chapter \ref{sec DBA}. We refer to \cite{DuboKernMado90b, Mass95a, DuboMass98a, Mass08b, Mass12a} for more details.

\subsection{\texorpdfstring{Derivations, Differential Structure, Metric, Integration and Hodge $\hstar$-operator on $M_n(\bbC)$}{Derivations, differential structure, metric, integration and Hodge star-operator on Mn(C)}}
\label{sec metric hodge}

The center of the algebra $M_n \defeq M_n(\bbC)$ is $\calZ(M_n) = \bbC \bbbone_n$ where $\bbbone_n$ is the unit matrix in $M_n$. Let $\ksl_n$ be the Lie algebra (for the commutator) of traceless matrices in $M_n$. Then the map $\ksl_n \ni a \mapsto \ad_a \in \Int(M_n)$ realizes an isomorphism $\ksl_n \simeq \Der(M_n) = \Int(M_n)$ (with $\Out(M_n)=0$).
\medskip
\par 
Let $\{ E_\alpha \}_{\alpha \in I}$ be a basis of $\ksl_n$, where $I$ is a totally ordered set with $\card(I) = n^2 -1 = \dim \ksl_n$. Choosing an abstract totally ordered set $I$ to label this basis will be convenient when the inductive sequence defining the $AF$-algebra will be considered since then the $\alpha$'s will be constructed as cumulative multi-indices. The ordering will be used to order basis forms (for instance to define volume forms). Let us introduce the unique multiplet $(\alpha^0_1, \dots, \alpha^0_{n^2-1}) \in I^{n^2-1}$ such that $\alpha^0_1 < \cdots < \alpha^0_{n^2-1}$.  We will use the notation $C(n)_{\alpha\beta}^\gamma = C_{\alpha\beta}^\gamma$ for the structure constants of the Lie algebra $\ksl_n$ in the basis $\{ E_\alpha \}_{\alpha \in I}$: $[E_\alpha, E_\beta] = C_{\alpha\beta}^\gamma E_\gamma$.
\medskip
\par 
The basis $\{ E_\alpha \}_{\alpha \in I}$ induces a basis $\{ \partial_\alpha \defeq \ad_{E_\alpha} \}_{\alpha \in I}$ of $\Der(M_n) = \Int(M_n)$.  Let $\{ \theta^\alpha \}_{\alpha \in I}$ be its dual basis in $\ksl_n^*$. The derivation $\partial_\alpha$ is real if and only if $E_\alpha$ is anti-Hermitean and one has $[\partial_\alpha, \partial_\beta] = C_{\alpha\beta}^\gamma \partial_\gamma$.
\medskip
\par 
The space of NC forms on $M_n$ has a simple structure:
\begin{align*}
	\OmegaDer^\grast(M_n)
	&= M_n \otimes \exter^\grast \ksl_n^*
\end{align*}
and the differential is the Chevalley-Eilenberg differential for the differential graded algebra associated to the Lie algebra $\ksl_n$ with values in $M_n$ using the adjoint representation. Identifying $\theta^\gamma$ with $\bbbone_n \otimes \theta^\gamma \in \OmegaDer^1(M_n) = M_n \otimes \exter^1 \ksl_n^*$, one has $\dd \theta^\gamma = - \tfrac{1}{2} C_{\alpha\beta}^\gamma \theta^\alpha \wedge \theta^\beta$.
\medskip
\par 
Let us consider the canonical metric $g : \Der(M_n) \times \Der(M_n) \to \calZ(M_n) \simeq \bbC$ defined by $g(\ad_a, \ad_b) \defeq \tr(a b)$ for $a, b \in \ksl_n$. This is not the metric defined in \cite{DuboKernMado90b} where a factor $\tfrac{1}{n}$ was put in front of the trace (to get the \emph{normalized} trace). The reason for this convention is linked to our constructions on AF-algebras and will be explained below (see \eqref{eq gB and gA} and comments after).  Once the basis $\{ \partial_\alpha \}_{\alpha \in I}$ is given, one introduces the components $g_{\alpha \beta} \defeq g(\partial_\alpha, \partial_\beta) = \tr(E_\alpha E_\beta)$ of $g$.
\medskip
\par 
Let $\abs{g}$ be the determinant of the matrix $g$. We define the (NC) integral $\int_{M_n}$ on $\OmegaDer^\grast(M_n)$ by the following rule. For any $\omega \in \OmegaDer^p(M_n)$ with $p<n^2-1$, $\int_{M_n} \omega = 0$. Any $\omega \in \OmegaDer^{n^2-1}(M_n)$ can be written as $\omega = a \sqrt{\abs{g}} \theta^{\alpha^0_1} \wedge \cdots \wedge \theta^{\alpha^0_{n^2-1}}$ for a unique $a \in M_n$ which is independent of the chosen basis $\{ E_\alpha \}_{\alpha \in I}$ and we define 
\begin{align}
	\label{NewIntegralDer}
	\int_{M_n} \omega 
	&= \int_{M_n} a \sqrt{\abs{g}} \theta^{\alpha^0_1} \wedge \cdots \wedge \theta^{\alpha^0_{n^2-1}} 
	\defeq \tr(a)
\end{align}
Once again, this is not the convention used in \cite{DuboKernMado90b} where a factor $\tfrac{1}{n}$ was put in front of the RHS. In our convention, $\omega_{\vol} \defeq \sqrt{\abs{g}} \theta^{\alpha^0_1} \wedge \cdots \wedge \theta^{\alpha^0_{n^2-1}}$ is the volume form whose integral is normalized to $n$.
\medskip
\par 
The metric permits to define the Hodge $\hstar$-operator 
\begin{align*}
	\hstar : \OmegaDer^p(M_n) \to \OmegaDer^{n^2-1-p}(M_n)
\end{align*}
defined by
\begin{align}
	\label{eq def hodge star}
	\hstar (\theta^{\alpha_1} \wedge \cdots \wedge \theta^{\alpha_p})
	&\defeq
	\tfrac{1}{(n^2-1-p) !} \sqrt{\abs{g}} g^{\alpha_1\beta_1} \cdots g^{\alpha_p\beta_p} \epsilon_{\beta_1, \dots, \beta_{n^2-1}} \theta^{\beta_{p+1}} \wedge \cdots \wedge \theta^{\beta_{n^2-1}}
\end{align}
where $\epsilon_{\beta_1, \dots, \beta_{n^2-1}}$ is the completely antisymmetric tensor such that $\epsilon_{\alpha^0_1, \dots, \alpha^0_{n^2-1}} = 1$.

\begin{example}[Transport by inner automorphisms]
	\label{ex transport matrix by inner automorphisms}
	Let us consider the situation described in Example~\ref{ex transport derivations inner automorphisms} in the context of the matrix algebra. Let $\omega = \frac{1}{p!} \omega_{\alpha_1, \dots, \alpha_p} \theta^{\alpha_1} \wedge \cdots \wedge \theta^{\alpha_p}$ be a $p$-form. To compute $\omega^u \defeq \Psi(\omega)$, let us introduce the matrix $U = (U_\alpha^{\beta})$ defined by $u^{-1} E_\alpha u = U_\alpha^{\beta} E_{\beta}$, so that $\PsiDer^{-1}(\partial_\alpha) = \ad_{u^{-1} E_\alpha u} = U_\alpha^{\beta} \partial_{\beta}$. Since $SL_n$ \GN{$GL_n$ à la base}\GN{attention indices $\alpha$ pour mult plong AF et aussi pr base ici}is unimodular, one has $\det(U) = 1$. Notice also that $u^{-1} [E_{\alpha_1}, E_{\alpha_2}] u = [u^{-1} E_{\alpha_1} u, u^{-1} E_{\alpha_2} u] = U_{\alpha_1}^{\beta_1} U_{\alpha_2}^{\beta_2} [E_{\beta_1}, E_{\beta_2}] = U_{\alpha_1}^{\beta_1} U_{\alpha_2}^{\beta_2} C_{\beta_1 \beta_2}^{\beta_3} E_{\beta_3}$ on the one hand and $u^{-1} [E_{\alpha_1}, E_{\alpha_2}] u = C_{\alpha_1 \alpha_2}^{\alpha_3} u^{-1} E_{\alpha_3} u = C_{\alpha_1 \alpha_2}^{\alpha_3} U_{\alpha_3}^{\beta_3} E_{\beta_3}$ on the other hand, so that $U_{\alpha_1}^{\beta_1} U_{\alpha_2}^{\beta_2} C_{\beta_1 \beta_2}^{\beta_3} = C_{\alpha_1 \alpha_2}^{\alpha_3} U_{\alpha_3}^{\beta_3}$. By definition, $(\theta^\alpha)^u(\partial_{\alpha'}) = u \theta^\alpha(\PsiDer^{-1}(\partial_{\alpha'})) u^{-1} = u U_{\alpha'}^{\beta'} \delta_{\beta'}^{\alpha} u^{-1} = U_{\alpha'}^{\alpha} = U_{\beta}^{\alpha} \theta^{\beta}(\partial_{\alpha'})$ so that $(\theta^\alpha)^u = U_{\beta}^{\alpha} \theta^{\beta}$. In the same way, $\omega^u_{\alpha_1, \dots, \alpha_p} = \omega^u(\partial_{\alpha_1}, \dots, \partial_{\alpha_p}) = u \omega(\PsiDer^{-1}(\partial_{\alpha_1}), \dots, \PsiDer^{-1}(\partial_{\alpha_p})) u^{-1} = U_{\alpha_1}^{\beta_1} \cdots U_{\alpha_p}^{\beta_p} u \omega_{\beta_1, \dots, \beta_p} u^{-1}$, so that $\omega^u = \frac{1}{p!} U_{\alpha_1}^{\beta_1} \cdots U_{\alpha_p}^{\beta_p} u \omega_{\beta_1, \dots, \beta_p} u^{-1} \theta^{\alpha_1} \wedge \cdots \wedge \theta^{\alpha_p} = \frac{1}{p!} u \omega_{\alpha_1, \dots, \alpha_p} u^{-1} (\theta^{\alpha_1})^u \wedge \cdots \wedge (\theta^{\alpha_p})^u$.\\ Now, the metric $g(\ad_a, \ad_b) = \tr(a b)$, for $a, b \in \ksl_n$, is invariant by the transport associated to the inner automorphism $\Psi(a) = u a u^{-1}$, and so one has $g_{\alpha \beta} = U_{\alpha}^{\alpha'} U_{\beta}^{\beta'} g_{\alpha' \beta'}$ and $g^{\alpha \beta} = U_{\alpha'}^{\alpha} U_{\beta'}^{\beta} g^{\alpha' \beta'}$ for the inverse metric. In particular, all the conditions and properties concerning orthonormality associated to $g$ are transported by $\Psi$. For $\omega = a \sqrt{\abs{g}} \theta^{\alpha^0_1} \wedge \cdots \wedge \theta^{\alpha^0_{n^2-1}}$, one has $\omega^u = u a u^{-1} \sqrt{\abs{g}} (\theta^{\alpha^0_1})^u \wedge \cdots \wedge (\theta^{\alpha^0_{n^2-1}})^u = u a u^{-1} \sqrt{\abs{g}} U_{\beta_1}^{\alpha^0_1} \cdots U_{\beta_{n^2-1}}^{\alpha^0_{n^2-1}} \theta^{\beta_1} \wedge \cdots \wedge \theta^{\beta_{n^2-1}} = u a u^{-1} \sqrt{\abs{g}} \theta^{\alpha^0_1} \wedge \cdots \wedge \theta^{\alpha^0_{n^2-1}}$ where we have used $\theta^{\beta_1} \wedge \cdots \wedge \theta^{\beta_{n^2-1}} = \epsilon^{\beta_1, \dots, \beta_{n^2-1}} \theta^{\alpha^0_1} \wedge \cdots \wedge \theta^{\alpha^0_{n^2-1}}$ and $\epsilon^{\beta_1, \dots, \beta_{n^2-1}} U_{\beta_1}^{\alpha^0_1} \cdots U_{\beta_{n^2-1}}^{\alpha^0_{n^2-1}} = \det(U) = 1$. This implies that $\int_{M_n} \omega^u = \tr(u a u^{-1}) = \tr(a) = \int_{M_n} \omega$. Since the metric $g$ is invariant, the Hodge $\hstar$-operator is also invariant according to \eqref{eq def hodge star}, and since the inverse metric is also invariant under the action of $U$, a straightforward computation of $\hstar ( (\theta^{\alpha_1})^u \wedge \cdots \wedge (\theta^{\alpha_p})^u)$ shows that the relation \eqref{eq def hodge star} is also valid when one replaces all the $\theta^\alpha$ by $(\theta^\alpha)^u$ on both sides. Combining all these results and the explicit relation \eqref{eq hodge star omega omega'}, one can show that for any $p$-forms $\omega$ and $\omega'$, one has $\int_{M_n} (\omega \wedge \hstar \omega')^u = \int_{M_n} \omega^u \wedge \hstar \omega'^u = \int_{M_n} \omega \wedge \hstar \omega'$.
\end{example}

\subsection{\texorpdfstring{Derivations, Differential Structure, Metric, Integration and Hodge $\hstar$-operator on $\bigoplus_{i=1}^nM_{n_i}(\bbC)$}{Derivations, differential structure, metric, integration and Hodge star-operator on OplusMni(C)}}
\label{ResDirMatrDer}
In this subsection we consider the derivation-based differential calculus on algebras decomposed as 
\begin{align*}
	\algA = \algA_1 \toplus \cdots \toplus \algA_r = \toplus_{i=1}^{r} \algA_i
\end{align*}
The results are given for these general algebras in order to see how direct sum of matrix algebras are particular. We define respectively
\begin{align*}
	&\pi^i : \algA \to \algA_i,
	\text{ and }
	\iota_i : \algA_i \to \algA
\end{align*}
as the natural projection on the $i$-th term and the natural inclusion of the $i$-th term\footnote{The notation $\pi$ is taken to be the same as the notation for representation, but always with indices}.
\medskip
\par 
Some results are presented using this full generality but others will require $\algA_i = M_n$.
\medskip
\par 
Let's start with center and derivations.

\begin{lemma}[Center of $\algA$]
	The center of $\algA$ is $\calZ(\algA) = \toplus_{i=1}^{r} \calZ(\algA_i)$.
\end{lemma}

\begin{proof}
	Every $a = \toplus_{i=1}^{r} a_i \in \calZ(\algA)$ must commute with any $b = 0 \toplus \cdots 0 \toplus b_j \toplus 0 \toplus 0$ for any $j =1, \dots, r$ and any $b_j \in \algA_j$. This implies that $a_j \in \calZ(\algA_j)$ for any $j$. The result follows since $\toplus_{i=1}^{r} \calZ(\algA_i) \subset \calZ(\algA)$.
\end{proof}

Let us introduce the convenient notation for the elements 
\begin{align*}
	\hbbbone_i \defeq \iota_i(\bbbone) = 0 \toplus \cdots \toplus 0 \toplus \overbrace{\bbbone}^{\text {i}}  \toplus 0 \toplus \cdots \toplus 0 \in \calZ(\algA) \subset \algA. 
\end{align*}
Notice that we use the fact that the $\algA_i$'s are unital.

\begin{proposition}[Decomposition of derivations]
	\label{prop decomposition derivations}
	One has
	\begin{align}
		\label{eq Der(A)}
		\Der(\algA) = \toplus_{i=1}^{r} \Der(\algA_i),
	\end{align}
	\textsl{\textit{i.e.}} for any $a = \toplus_{i=1}^{r} a_i \in \algA$ and $\kX = \toplus_{i=1}^{r} \kX_i \in \Der(\algA)$, one has $\kX (a) = \toplus_{i=1}^{r} \kX_i (a_i)$.
	\medskip
	\par 
	This decomposition holds true as Lie algebras and modules over $\calZ(\algA)$ on the left and over $\toplus_{i=1}^{r} \calZ(\algA_i)$ on the right.
	\medskip
	\par 
	If   $\,\Der(\algA_i) = \Int(\algA_i)$ for any $i=1, \dots, r$, then 
	\begin{align}
		\label{eq Der(A) = Int(A)}
		\Der(\algA) = \Int(\algA) = \toplus_{i=1}^{r} \Int(\algA_i)
	\end{align}
\end{proposition}

\begin{proof}
	The vector space decomposition \eqref{eq Der(A)} can be established using the maps $\kX_i^j \defeq \pi^j \circ \kX \circ \iota_i : \algA_i \to \algA_j$. For any $a = \toplus_{i=1}^{r} a_i$ and $b = \toplus_{i=1}^{r} b_i$, one has
	\begin{align*}
		\kX(a) 
		&= \toplus_{j=1}^{r} \left( \tsum_{i=1}^{r} \kX_i^j (a_i) \right)
	\end{align*}
	and the Leibniz rule $\kX(a b) = \kX(a) b + a \kX(b)$ can then be written as
	\begin{align*}
		\toplus_{j=1}^{r} \left( \tsum_{i=1}^{r} \kX_i^j (a_i b_i) \right)
		&= \toplus_{j=1}^{r} \left( \tsum_{i=1}^{r} \kX_i^j (a_i) \right)  b_j
		+ \toplus_{j=1}^{r} a_j \left( \tsum_{i=1}^{r} \kX_i^j (b_i) \right).
	\end{align*}
	For a fixed $k$, take $a_i = b_i = 0$ for $i \neq k$. Then this relation reduces to
	\begin{align*}
		\toplus_{j=1}^{r} \kX_k^j (a_k b_k)
		&= 0 \toplus \cdots \toplus \kX_k^k(a_k) b_k \toplus 0 \toplus \cdots \toplus 0
		+
		0 \toplus \cdots \toplus a_k \kX_k^k(b_k) \toplus 0 \toplus \cdots \toplus 0.
	\end{align*}
	The $k$-th term shows that $\kX_k^k \in \Der(\algA_k)$, which implies in particular that $\kX_k^k(\bbbone) = 0$. Then, with $b_k = \bbbone$, one gets $\kX_k^j = 0$ for $j \neq k$. This shows that $\kX(a) = \toplus_{i=1}^{r} \kX_i (a_i)$ with $\kX_i \defeq \kX_i^i \in \Der(\algA_i)$ and we write
	\begin{align*}
		\kX = \toplus_{i=1}^{r} \kX_i 
	\end{align*}
	
	For any $f = \toplus_{i=1}^{r} f_i \in \calZ(\algA) = \toplus_{i=1}^{r} \calZ(\algA_i)$, one has
	\begin{align*}
		(f \kX) (a)
		&= f (\toplus_{i=1}^{r} \kX_i (a_i))
		= \toplus_{i=1}^{r} (f_i \kX_i) (a_i)
	\end{align*}
	so that \eqref{eq Der(A)} holds true as modules over $\calZ(\algA)$ on the left and over $\toplus_{i=1}^{r} \calZ(\algA_i)$ on the right.
	\medskip
	\par 
	For $\kY = \toplus_{i=1}^{r} \kY_i \in \Der(\algA)$, one has
	\begin{align*}
		[\kX, \kY] (a)
		&= \kX \circ \kY(a) - \kY \circ \kX(a)
		= \kX \left( \toplus_{i=1}^{r} \kY_i(a_i) \right) - \kY \left( \toplus_{i=1}^{r} \kX_i(a_i) \right)
		\\
		&= \toplus_{i=1}^{r} \kX_i \circ \kY_i(a_i) - \toplus_{i=1}^{r} \kY_i \circ \kX_i(a_i)
		\\
		&= \toplus_{i=1}^{r} [\kX_i, \kY_i](a_i)
	\end{align*}
	so that 
	\begin{align*}
		[\kX, \kY] = \toplus_{i=1}^{r} [\kX_i, \kY_i]
	\end{align*}
	This shows that \eqref{eq Der(A)} holds true as Lie algebras.
	\medskip
	\par 
	The proof of \eqref{eq Der(A) = Int(A)} is then a direct consequence: assuming $\Der(\algA_i) = \Int(\algA_i)$, one has $\Int(\algA) \subset \Der(\algA) = \toplus_{i=1}^{r} \Int(\algA_i) \subset \Int(\algA)$ where the last inclusion follows from $\ad_{a_1} \toplus \cdots \toplus \ad_{a_r} = \ad_{a_1 \toplus \cdots \toplus a_r}$.
\end{proof}

We define $\iotaDer_i : \Der(\algA_i) \to \Der(\algA)$ as the inclusion $\kX_i \mapsto 0 \toplus \cdots \toplus 0 \toplus \kX_i \toplus 0 \cdots \toplus 0$ in the $i$-th term. This is a morphism of Lie algebras and for any $a_i \in \algA_i$ and $f_i \in \calZ(\algA_i)$, one has $\iota_i(\kX_i(a_i)) = \iotaDer_i( \kX_i)(\iota_i(a_i))$ and $\iotaDer_i( f_i \kX_i) = \iota_i(f_i) \iotaDer_i( \kX_i)$. Notice also that for any $\kX = \toplus_{i=1}^{r} \kX_i$, one has $\hbbbone_i  \kX = \iotaDer_i( \kX_i) = \hbbbone_i \iotaDer_i( \kX_i)$.

\bigskip
Using results given in Sect.~\ref{sec metric hodge}, we obtain the direct Corollary of Prop.~\ref{prop decomposition derivations}:
\begin{corollary}
	For $\algA = M_{n_1} \toplus \cdots \toplus M_{n_r}$, one has $\Der(\algA) = \Int(\algA) \simeq \ksl_{n_1} \toplus \cdots \toplus \ksl_{n_r}$.
\end{corollary}

Let's see now how it works for the Derivation-based differential calculus. A useful result concerning the structure of the derivation-based differential calculus associated to $\algA$ is the following.
\begin{proposition}[Decomposition of forms]
	\label{prop decomposition forms}
	For any $p \geq 0$, one has
	\begin{align*}
		\OmegaDer^p(\algA) &= \toplus_{i=1}^{r} \OmegaDer^p(\algA_i),
	\end{align*}
	that is, any $\omega \in \OmegaDer^p(\algA)$ decomposes as $\omega = \toplus_{i=1}^{r} \omega_i$ with $\omega_i \in \OmegaDer^p(\algA_i)$ and 
	\begin{align*}
		\omega(\kX_1, \dots, \kX_p)
		&= \toplus_{i=1}^{r} \omega_i( \kX_{1, i}, \dots, \kX_{p, i} )
		\quad\text{for any $\kX_k = \toplus_{i=1}^{r} \kX_{k, i} \in \Der(\algA)$.}
	\end{align*}
	
	This decomposition is compatible with the $\calZ(\algA)$-linearity on the left and the $\toplus_{i=1}^{r} \calZ(\algA_i)$-linearity on the right, and it is compatible with  the products in $\OmegaDer^\grast(\algA)$ and $\OmegaDer^\grast(\algA_i)$.
	\medskip
	\par 
	The differential $\dd$ on $\OmegaDer^\grast(\algA)$ decomposes along the differentials $\dd_i$ on $\OmegaDer^\grast(\algA_i)$ as
	\begin{align*}
		\dd \omega 
		&= \toplus_{i=1}^{r} \dd_i \omega_i
	\end{align*}
\end{proposition}

We will extend the projection map $\pi^i : \OmegaDer^\grast(\algA) \to \OmegaDer^\grast(\algA_i)$ with the same notation.

\begin{proof}
	Let $\omega \in \OmegaDer^p(\algA)$ and define, for any $i_1, \dots, i_p, j = 1, \dots, r$,
	\begin{align*}
		\omega^j_{i_1, \dots, i_p} : \Der(\algA_{i_1}) \times \cdots \times \Der(\algA_{i_p}) \to \algA_j
	\end{align*}
	by
	\begin{align*}
		\omega^j_{i_1, \dots, i_p}( \kX_{1, i_1}, \dots, \kX_{p, i_p} ) = \pi^j \circ \omega ( \iotaDer_{i_1}(\kX_{1, i_1}), \dots, \iotaDer_{i_p}(\kX_{p, i_p}) )
	\end{align*}
	for any $\kX_{k, i_k} \in \Der(\algA_{i_k})$ ($k=1, \dots, p$). Then one has
	\begin{align*}
		\omega(\kX_1, \dots, \kX_p)
		&= \toplus_{j=1}^{r} \left( \tsum_{i_1, \dots, i_p = 1}^{r} \omega^j_{i_1, \dots, i_p}( \kX_{1, i_1}, \dots, \kX_{p, i_p} ) \right)
	\end{align*}
	with $\kX_k = \toplus_{i_k=1}^{r} \kX_{k, i_k}$. Let $f_k = \toplus_{i_k=1}^{r} f_{k, i_k} \in \calZ(\algA)$, then applying $\omega$ on the $p$ derivations $f_k \kX_k$, the $\calZ(\algA)$-linearity of $\omega$ gives
	\begin{align}
		\toplus_{j=1}^{r} & \left( \tsum_{i_1, \dots, i_p = 1}^{r} \omega^j_{i_1, \dots, i_p}( f_{1,i_1} \kX_{1, i_1}, \dots, f_{p,i_p} \kX_{p, i_p} ) \right)
		\nonumber
		\\
		\label{eq Z(A) linearity omega general}
		&=
		\toplus_{j=1}^{r} f_{1,j} \cdots f_{p,j} \left( \tsum_{i_1, \dots, i_p = 1}^{r} \omega^j_{i_1, \dots, i_p}( \kX_{1, i_1}, \dots, \kX_{p, i_p} ) \right)
	\end{align}
	The arbitrariness on the $f_k$'s permits to simplify this relation in successive steps. Let us fix $j_1$ and take $f_{1} = \hbbbone_{j_1}$.
	Then on the LHS, the term at $j$ in $\toplus_{j=1}^{r}$ reduces to
	\begin{align*}
		\tsum_{i_2, \dots, i_p = 1}^{r} \omega^j_{j_1, i_2, \dots, i_p}( \kX_{1, j_1}, f_{2,i_2} \kX_{p, i_p}, \dots, f_{p,i_p} \kX_{p, i_p} ). 
	\end{align*}
	On the RHS, the only non zero term along $\toplus_{j=1}^{r}$ occurs at $j = j_1$ and gives 
	\begin{align*}
		f_{2,j_1} \cdots f_{p,j_1} \left( \tsum_{i_1, \dots, i_p = 1}^{r} \omega^{j_1}_{i_1, \dots, i_p}( \kX_{1, i_1}, \dots, \kX_{p, i_p} ) \right).
	\end{align*}
	By arbitrariness on the $\kX_k$'s, this implies that for $j \neq j_1$ one has $\omega^j_{i_1, \dots, i_p} = 0$ and the remaining non trivial relation becomes (substituting $j_1$ to $j$)
	\begin{align*}
		\tsum_{i_2, \dots, i_p = 1}^{r} & \omega^j_{j, i_2, \dots, i_p}( \kX_{1, j}, f_{2,i_2} \kX_{2, i_2}, \dots, f_{p,i_p} \kX_{p, i_p} )
		=
		f_{2,j} \cdots f_{p,j} \left( \tsum_{i_2, \dots, i_p = 1}^{r} \omega^j_{j, i_2, \dots, i_p}( \kX_{1, j}, \kX_{2, i_2}, \dots, \kX_{p, i_p} ) \right).
	\end{align*}
	Let us now fix $j_2$ and take $f_{2} = \hbbbone_{j_2}$. Then the relation first gives
	\begin{align*}
		\tsum_{i_3, \dots, i_p = 1}^{r} \omega^j_{j, j_2, i_3, \dots, i_p} & ( \kX_{1, j}, \kX_{2, j_2}, f_{3,i_3} \kX_{3, i_3}, \dots, f_{p,i_p} \kX_{p, i_p} ) = 0 \qquad \text{ for $j \neq j_2$},
	\end{align*}
	which implies $\omega^j_{j, i_2, i_3, \dots, i_p} = 0$ for $i_2 \neq j$. Then, with this relation, the non vanishing term (at $j=j_2$) simplifies to (substituting $j_2$ to $j$)
	\begin{align*}
		\tsum_{i_3, \dots, i_p = 1}^{r} & \omega^{j}_{j, j, i_3, \dots, i_p}( \kX_{1, j}, \kX_{2, j}, f_{3,i_3} \kX_{3, i_3}, \dots, f_{p,i_p} \kX_{p, i_p} )=
		f_{3,j} \cdots f_{p,j} \left( \tsum_{i_3, \dots, i_p = 1}^{r} \omega^{j}_{j, j, i_3, \dots, i_p}( \kX_{1, j}, \kX_{2, j}, \kX_{3, i_3},\dots, \kX_{p, i_p} ) \right).
	\end{align*}
	We can repeat this argument for $i_k$ up to $k=p$ and conclude that the only non zero maps $\omega^j_{i_1, \dots, i_p}$ are $\omega^j_{j, \dots, j}$ for $j=1, \dots, r$. Defining $\omega_i \defeq \omega^i_{i, \dots, i}$, one then gets
	\begin{align*}
		\omega(\kX_1, \dots, \kX_p)
		&= \toplus_{i=1}^{r} \omega_i( \kX_{1, i}, \dots, \kX_{p, i} )
	\end{align*}
	and \eqref{eq Z(A) linearity omega general} reduces to
	\begin{align*}
		\toplus_{i=1}^{r}  \omega_i( f_{1,i} \kX_{1, i}, \dots, f_{p,i} \kX_{p, i} )
		&= \toplus_{i=1}^{r} f_{1,i} \cdots f_{p,i} \omega_i( \kX_{1, i}, \dots, \kX_{p, i} ).
	\end{align*}
	This shows that for any $i$, $\omega_i$ is $\calZ(\algA_i)$-linear. Finally, the antisymmetry of $\omega$ implies antisymmetry of the $\omega_i$'s. This proves that $\omega_i \in \OmegaDer^p(\algA_i)$.
	\medskip
	\par 
	To prove the compatibility of this decomposition with the products, consider $\omega = \toplus_{i=1}^{r}  \omega_i \in \OmegaDer^p(\algA)$ and $\eta = \toplus_{i=1}^{r}  \eta_i \in \OmegaDer^q(\algA)$, and $p+q$ derivations $\kX_{k} = \toplus_{i_k=1}^{r} \kX_{k, i_k} \in \Der(\algA)$. Then by definition
	\begin{align*}
		(\omega &\wedge \eta) (\kX_{1}, \dots, \kX_{p+q})
		\\
		&= \tfrac{1}{p! q!} \sum_{\sigma \in \kS_{p+q}} (-1)^{\abs{\sigma}} \omega (\kX_{\sigma(1)}, \dots, \kX_{\sigma(p)}) \eta (\kX_{\sigma(p+1)}, \dots, \kX_{\sigma(p+q)})
		\\
		&= \begin{multlined}[t]
			\tfrac{1}{p! q!} \sum_{\sigma \in \kS_{p+q}} (-1)^{\abs{\sigma}} 
			\left( \toplus_{i=1}^{r} \omega_i (\kX_{\sigma(1), i}, \dots, \kX_{\sigma(p), i}) \right)
			\left( \toplus_{j=1}^{r} \eta_j (\kX_{\sigma(p+1), j}, \dots, \kX_{\sigma(p+q), j}) \right)
		\end{multlined}
		\\
		&= \tfrac{1}{p! q!} \sum_{\sigma \in \kS_{p+q}} (-1)^{\abs{\sigma}} 
		\left( \toplus_{i=1}^{r} 
		\omega_i (\kX_{\sigma(1), i}, \dots, \kX_{\sigma(p), i})
		\eta_i (\kX_{\sigma(p+1), i}, \dots, \kX_{\sigma(p+q), i}) 
		\right)
		\\
		&= \toplus_{i=1}^{r} \left( 
		\tfrac{1}{p! q!} \sum_{\sigma \in \kS_{p+q}} (-1)^{\abs{\sigma}} 
		\omega_i (\kX_{\sigma(1), i}, \dots, \kX_{\sigma(p), i})
		\eta_i (\kX_{\sigma(p+1), i}, \dots, \kX_{\sigma(p+q), i}) 
		\right)
		\\
		&= \toplus_{i=1}^{r}
		(\omega_i \wedge \eta_i) (\kX_{1, i}, \dots, \kX_{p+q, i})
	\end{align*}
	so that $\omega \wedge \eta =  \toplus_{i=1}^{r} \omega_i \wedge \eta_i$.
	\medskip
	\par 
	Using similar notations, one has
	\begin{align*}
		(\dd \omega) (\kX_{1}, \dots, \kX_{p+1})
		&=  \sum_{k=1}^{p+1} (-1)^{k+1} \kX_{k} \cdotaction \omega (\kX_{1}, \dots \omi{k} \dots, \kX_{p+1})
		+ \sum_{1 \leq k < \ell \leq p+1} (-1)^{k+\ell} \omega ([\kX_{k}, \kX_{\ell}], \dots \omi{k} \dots \omi{\ell} \dots, \kX_{p+1})
		\\
		&= \begin{multlined}[t]
			\sum_{k=1}^{p+1} (-1)^{k+1} \toplus_{i=1}^{r} \kX_{k, i} \cdotaction \omega_i (\kX_{1, i}, \dots \omi{k} \dots, \kX_{p+1, i})
			\\
			+ \sum_{1 \leq k < \ell \leq p+1} (-1)^{k+\ell} \toplus_{i=1}^{r} \omega_i ([\kX_{k, i}, \kX_{\ell, i}], \dots \omi{k} \dots \omi{\ell} \dots, \kX_{p+1, i})
		\end{multlined}
		\\
		&= \toplus_{i=1}^{r} \Big( \begin{multlined}[t]
			\sum_{k=1}^{p+1} (-1)^{k+1} \kX_{k, i} \cdotaction \omega_i (\kX_{1, i}, \dots \omi{k} \dots, \kX_{p+1, i})
			\\
			+ \sum_{1 \leq k < \ell \leq p+1} (-1)^{k+\ell} \omega_i ([\kX_{k, i}, \kX_{\ell, i}], \dots \omi{k} \dots \omi{\ell} \dots, \kX_{p+1, i})
			\Big)
		\end{multlined}
		\\
		&= \toplus_{i=1}^{r} (\dd_i \omega_i) (\kX_{1, i}, \dots, \kX_{p+1, i})
	\end{align*}
	so that $\dd \omega = \toplus_{i=1}^{r} \dd_i \omega_i$.
\end{proof}
\medskip
\par
For the metric, the integration and the Hodge $\hstar$-operator, we will consider only the situation $\algA_i = M_{n_i}$. This permits to limit the study of metrics and Hodge $\hstar$-operators to the structures defined in Sect.~\ref{sec metric hodge} (see also \cite{DuboKernMado90b, Mass95a, DuboMass98a, Mass08b, Mass12a}).
\medskip
\par 
For every $\algA_i = M_{n_i}$, one introduces a basis $\{ E^{i}_{\alpha} \}_{\alpha \in I_{i}}$ of $\ksl_{n_i}$ where $I_{i}$ is a totally ordered set of cardinal $n_i^2-1$. Let $\{ \partial^{i}_{\alpha} \defeq \ad_{E^{i}_{\alpha}} \}_{\alpha \in I_{i}}$ be the induced basis of $\Der(M_{n_i})$. The dual basis is denoted by $\{ \theta_{i}^{\alpha} \}_{\alpha \in I_{i}}$.
\medskip
\par 
We consider the metric $g$ on $\algA = \toplus_{i=1}^{r} M_{n_i}$ defined by $g( \partial^{i}_{\alpha}, \partial^{i'}_{\alpha'} ) = 0$ for $i \neq i'$ and $g^{i}_{\alpha \alpha'} \defeq g( \partial^{i}_{\alpha}, \partial^{i}_{\alpha'}) = \tr ( E^{i}_{\alpha} E^{i}_{\alpha'} )$ as in Sect.~\ref{sec metric hodge}. Then, by construction, $\Der(\algA_i)$ is orthogonal to $\Der(\algA_{i'})$ when $i \neq i'$. 
\medskip
\par 
A natural way to define a (NC) integral of forms on $\algA$ is to decompose it along the $\algA_i$ as
\begin{align*}
	\int_{\algA} \omega
	& \defeq \tsum_{i=1}^{r} \int_{i} \omega_i
\end{align*}
for any $\omega = \toplus_{i=1}^{r} \omega_i \in \OmegaDer^\grast(\algA)$. Here $\int_{i} = \int_{\algA_i}$ is defined as in Sect.~\ref{sec metric hodge} using the volume form $\omega_{\vol, i} \defeq \sqrt{\abs{g^i}} \theta_i^{\alpha^0_1} \wedge \cdots \wedge \theta_i^{\alpha^0_{n_i^2-1}}$ where $(\alpha^0_1, \dots, \alpha^0_{n^2-1}) \in I_i^{n_i^2-1}$ is such that $\alpha^0_1 < \cdots < \alpha^0_{n_i^2-1}$.
\medskip
\par 
In order to compute $\int_{\algA} \omega$, one has to find the unique element $a = \toplus_{i=1}^{r} a_i \in \algA$ such that $\toplus_{i=1}^{r} a_i \omega_{\vol, i}$ captures the highest degrees in every $\OmegaDer^\grast(\algA_i)$, and then one has $\int_{\algA} \omega = \tsum_{i=1}^{r} \tr(a_i)$. In particular, with $\omega_{\vol}
\defeq \toplus_{i=1}^{r} \omega_{\vol, i}$, one has $\int_{\algA} \omega_{\vol} = \tsum_{i=1}^{r} n_i$.
\medskip
\par 
The metric $g$ on $\algA$ gives rise to a well defined Hodge $\hstar$-operator, which, according to the proof of Lemma~\ref{lemma *omega decomposed}, can be written using the Hodge $\hstar$-operators on each $\algA_i$ for the metric $g^i$. For any $\omega = \toplus_{i=1}^{r} \omega_i, \omega' = \toplus_{i=1}^{r} \omega'_i \in \OmegaDer^\grast(\algA)$, one has 
\begin{align*}
	\omega \wedge \hstar \omega'
	&= \tsum_{i=1}^{r} \omega_i \wedge \hstar_i \omega'_i
\end{align*}
where $\hstar_i$ is defined on $\OmegaDer^\grast(M_{n_i})$ as in Sect.~\ref{sec metric hodge}, and we define the NC scalar product of forms on $\algA = \toplus_{i=1}^{r} M_{n_i}$ by
\begin{align}
	\label{eq def scalar product forms}
	(\omega, \omega') \defeq
	\int_\algA \omega \wedge \hstar \omega'
	= \tsum_{i=1}^{r} \int_i \omega_i \wedge \hstar_i \omega'_i
\end{align}
This expression will be used to define action functional out of a connection $1$-form.

\subsection{\texorpdfstring{A technical result on Hodge $\hstar$-operators}{A technical result on Hodge *-operators}}
\label{sec technical result on Hodge * operators}

Let $V$ be a finite dimensional vector space, with $\dim V = n$ and let $V^*$ its dual. Denote by $\{ e_k \}_{k}$ and $\{ \theta^k \}_{k}$ a basis of $V$ and its dual basis in $V^*$. Let $g$ be a metric on $V$ and let $g_{k\ell} \defeq g(e_k, e_\ell)$. Denote by $(g^{k\ell})$ its inverse matrix and by $\abs{g}$ the determinant of $g$. 
\medskip
\par 
Let $\algA$ be an associative unital algebra equipped with a linear form $\tau : \algA \to \bbC$. We define forms on $V$ with values in $\algA$ as elements in $\algA \otimes \exter^\grast V^*$. There is then a natural multiplication: for any $\omega \in \algA \otimes \exter^p V^*$ and $\eta \in \algA \otimes \exter^q V^*$, $\omega \wedge \eta \in \algA \otimes \exter^{p+q} V^*$ is defined by
\begin{align*}
	(\omega \wedge \eta)(e_1, \dots, e_{p+q})\defeq
	\frac{1}{p!q!} \sum_{\sigma\in \kS_{p+q}} (-1)^{\abs{\sigma}} \omega(e_{\sigma(1)}, \dots, e_{\sigma(p)}) \eta(e_{\sigma(p+1)}, \dots, e_{\sigma(p+q)})
\end{align*}
for any $e_I, \dots, e_{p+q} \in V$ where $\kS_{n}$ is the group of permutations of $n$ elements.
\medskip
\par 
Given an orientation $\theta^1 \wedge \cdots \wedge \theta^n$ of the basis $\{ \theta^k \}_{k}$, the metric $g$ and the linear form $\tau$ define an “integration” $\int_V : \algA \otimes \exter^\grast V^* \to \bbC$ which is non zero only on $\algA \otimes \exter^n V^*$ where it is defined, for any $\omega$ uniquely written as $\omega = \sqrt{\abs{g}} a \otimes \theta^1 \wedge \cdots \wedge \theta^n$, by $\int_V \omega \defeq \tau(a)$. This definition does not depend on the basis $\{ \theta^k \}_{k}$ (only on its orientation up to a sign). The $n$-form $\omega_{\vol} \defeq \sqrt{\abs{g}} \bbbone \otimes \theta^1 \wedge \cdots \wedge \theta^n$ is called the volume form.
\medskip
\par 
The metric $g$ defines also a Hodge $\hstar$-operator on $\algA \otimes \exter^\grast V^*$ defined on $\omega = \tfrac{1}{p!} \omega_{\ell_1, \dots, \ell_p} \otimes \theta^{\ell_1} \wedge \cdots \wedge \theta^{\ell_p}$ by the usual formula
\begin{align}
	\hstar ( \tfrac{1}{p!} \omega_{\ell_1, \dots, \ell_p} \otimes \theta^{\ell_1} \wedge \cdots \wedge \theta^{\ell_p} )
	\nonumber
	= \tfrac{1}{(n-p)!} \tfrac{1}{p!} \sqrt{\abs{g}} \omega_{\ell'_1, \dots, \ell'_p} g^{\ell_1 \ell'_1} \cdots g^{\ell_p \ell'_p} \epsilon_{\ell_1, \dots, \ell_n} \otimes \theta^{\ell_{p+1}} \wedge \cdots \wedge \theta^{\ell_n}
	\label{eq hodge on V}
\end{align}
where $\epsilon_{\ell_1, \dots, \ell_n}$ is the completely antisymmetric tensor such that $\epsilon_{1, \dots, n} = 1$.  For any $\omega, \omega' \in \algA \otimes \exter^p V^*$, a standard computation gives (see for instance~\cite[Sect.~2.4]{Bert96a})
\begin{align}
	\label{eq hodge star omega omega'}
	\omega \wedge \hstar \omega'
	&= \tfrac{1}{p!} \sqrt{\abs{g}} \omega_{\ell_1, \dots, \ell_p} \omega'^{\ell_1, \dots, \ell_p} \otimes \theta^{1} \wedge \cdots \wedge \theta^{n}
\end{align}
with $\omega'^{\ell_1, \dots, \ell_p} \defeq g^{\ell_1 \ell'_1} \cdots g^{\ell_p \ell'_p} \omega'_{\ell'_1, \dots, \ell'_p}$.
\medskip
\par 
\begin{lemma}
	\label{lemma *omega decomposed}
	Suppose that $V = \toplus_{i=1}^{r} V_i$ is an orthogonal decomposition for $g$. Denote by $g_i$ the restriction of $g$ to $V_i$, denote by $\hstar_i$ the corresponding Hodge star operator on $\algA \otimes \exter^\grast V_i^*$, and denote by $\int_{V_i}$ the corresponding integration with volume form $\omega_{\vol, i}$ such that $\omega_{\vol} =  \omega_{\vol, 1} \wedge \cdots \wedge \omega_{\vol, r}$. 
	
	Let $\omega_i, \omega'_i \in \algA \otimes \exter^{p_i} V_i^*$ and $\omega = \tsum_{i=1}^{r} \omega_i, \omega' = \tsum_{i=1}^{r} \omega'_i \in \algA \otimes \exter^\grast V^*$. Then
	\begin{align*}
		\int_{V} \omega \wedge \hstar \omega'
		&=
		\tsum_{i=1}^{r} \int_{V_i} \omega_i \wedge \hstar_i \omega'_i
	\end{align*}
\end{lemma}

\begin{proof}
	Let us introduce some notations. Let $n_i = \dim V_i$ (so that $n = \tsum_{i=1}^{r} n_i$); let $\{ e^i_k \}_{k=1, \dots, n_i}$ be a basis of $V_i$; for $\ell = 1, \dots, n$ written as $\ell = n_1 + \cdots + n_{i-1} + k$ with $k=1, \dots, n_i$, let $e_\ell \defeq e^i_k$ be the elements of a basis of $V$; let $\{ \theta_i ^k \}_{k=1, \dots, n_i}$ and $\{ \theta^\ell \}_{\ell=1, \dots, n}$ be the corresponding dual basis. Let $I_i$ be the set of indices $\ell = n_1 + \cdots + n_{i-1} + k$ with $k=1, \dots, n_i$, so that $e_\ell \in V_i$ for $\ell \in I_i$, and let $I_i^c$ be its complement in $\{1, \dots, n\}$.
	\medskip
	\par 
	The matrix $(g_{k\ell})$ is block diagonal, and so is its inverse $(g^{k\ell})$ with blocks $(g_i^{k\ell})$ and $\sqrt{\abs{g}} = \prod_{i=1}^{r} \sqrt{\abs{g_i}}$. The orientation of the basis $\{ \theta^\ell \}_{\ell=1, \dots, n}$ is chosen such that $\omega_{\vol} =  \omega_{\vol, 1} \wedge \cdots \wedge \omega_{\vol, r}$ with $\omega_{\vol, i} \defeq \sqrt{\abs{g_i}} \bbbone \otimes \theta_i^1 \wedge \cdots \wedge \theta_i^{n_i}$.
	\medskip
	\par 
	By linearity and the fact that $\omega_i \wedge \hstar \omega'_j = 0$ for $i \neq j$, one has $\omega \wedge \hstar \omega' = \tsum_{i=1}^{r} \omega_i \wedge \hstar \omega'_i$. For fixed $i$, to compute $\hstar \omega'_i$ we use \eqref{eq hodge on V}: in the RHS, only components of $(g^{k\ell})$ belonging to the block $(g_i^{k\ell})$ can appear, so that $\epsilon_{\ell_1, \dots, \ell_n}$ must contain all the indices $\ell \in I_i^c$. This implies that the RHS contains all the volume forms $\omega_{\vol, j} =\sqrt{\abs{g_j}} \bbbone \otimes \theta_j ^1 \wedge \cdots \wedge \theta_j ^{n_j}$ for $j \neq i$.
	\medskip
	\par 
	Let us consider a fixed $n$-uplet $(\ell_1, \dots, \ell_n)$ in the sum in the RHS of \eqref{eq hodge on V}. Our strategy is to collect all the $\ell_r \in I_i$ in front of the $n$-uplet. Since for $r=1, \dots, p_i$ one has $\ell_r \in I_i$, we need only to use a permutation of the remaining indices to collect the other $n_i - p_i$ indices which belong to $I_i$. As a permutation, we use the unique $(n_i - p_i, n - n_i)$-shuffle\footnote{A $(r,s)$-shuffle is a permutation $\sigma \in \kS_{r+s}$ such that $\sigma(1) < \sigma(2) < \cdots < \sigma(r)$ and $\sigma(r+1) < \sigma(r+2) < \cdots < \sigma(r+s)$. It is well-known that there are $\tfrac{(r+s)!}{r! s!}$ $(r,s)$-shuffles in $\kS_{r+s}$.} that maps $(\ell_{p_i + 1}, \dots, \ell_n)$ into $(\ell'_{p_i + 1}, \dots, \ell'_{n_i}, \ell''_{1}, \dots, \ell''_{n - n_i})$ such that $\ell'_r \in I_i$ for  $r = p_i + 1, \dots, n_i$ and $\ell''_{r'} \in I_i^c$ for $r' = n_i + 1, \dots, n$. This shuffle transforms the term (no summation) $\epsilon_{\ell_1, \dots, \ell_n} \theta^{\ell_{p_i+1}} \wedge \cdots \wedge \theta^{\ell_n}$ into $\epsilon_{\ell_1, \dots, \ell_{p_i}, \ell'_{p_i + 1}, \dots, \ell'_{n_i}, \ell''_{1}, \dots, \ell''_{n - n_i}} \theta^{\ell'_{p_i + 1}} \wedge \cdots \wedge \theta^{\ell'_{n_i}} \wedge \theta^{\ell''_{1}} \wedge \cdots \wedge \theta^{\ell''_{n - n_i}}$ (since the shuffle acts on both the $\epsilon$ indices and the $\theta^\ell$, there is no sign).
	\medskip
	\par 
	The summation on all the $n - p_i$-uplets $(\ell_{p_i + 1}, \dots, \ell_n)$ can now be performed in 3 steps: first using the $(n_i - p_i, n - n_i)$-shuffle which separates the indices belonging to $I_i$ and $I_i^c$ (there are $\tfrac{(n-p)!}{(n_1-p)! n_2!}$ such shuffles to use); then using a permutation on the $n - n_i$ indices $\ell''_{r'} \in I_i^c$ for $r' = n_i + 1, \dots, n$ to order them in increasing order (there are $(n - n_i)!$ such permutations) so that we can make appear the $\omega_{\vol, j}$ for $j \neq i$; finally managing the summation on the indices $\ell'_r \in I_i$ for  $r = p_i + 1, \dots, n_i$. This gives the series of equalities (with the previous notations and convention for the indices):
	\begin{align*}
		\tfrac{1}{(n - p_i)!} 
		\sqrt{\abs{g}}
		\epsilon_{\ell_1, \dots, \ell_n} 
		\theta^{\ell_{p_i + 1}} \wedge \cdots \wedge \theta^{\ell_n}
		&= \begin{multlined}[t]
			\tfrac{(n - p_i)!}{(n - p_i)! (n_i - p_i)! (n - n_i)!} 
			\sqrt{\abs{g}}
			\epsilon_{\ell_1, \dots, \ell_{p_i}, \ell'_{p_i + 1}, \dots, \ell'_{n_i}, \ell''_{1}, \dots, \ell''_{n - n_i}} 
			\\
			\theta^{\ell'_{p_i + 1}} \wedge \cdots \wedge \theta^{\ell'_{n_i}} 
			\wedge 
			\theta^{\ell''_{1}} \wedge \cdots \wedge \theta^{\ell''_{n - n_i}}
		\end{multlined}
		\\
		&= \begin{multlined}[t]
			\tfrac{(n - n_i)!}{(n_i - p_i)! (n - n_i)!}
			\sqrt{\abs{g}}
			\epsilon_{\ell_1, \dots, \ell_{p_i}, \ell_{p_i + 1}, \dots, \ell_{n_i}, 1, \dots, n_1 + \cdots + n_{i-1}, n_1 + \cdots + n_{i} + 1, \dots, n}
			\\
			\theta^{\ell_{p_i + 1}} \wedge \cdots \wedge \theta^{\ell_{n_i}}
			\wedge 
			(\wedge_{\ell' \in I_i^c} \theta^{\ell'})
		\end{multlined}
		\\
		&= \begin{multlined}[t]
			\tfrac{1}{(n_i - p_i)!} 
			\sqrt{\abs{g_i}}
			\epsilon_{\ell_1, \dots, \ell_{p_i}, \ell_{p_i + 1}, \dots, \ell_{n_i}, 1, \dots, n_1 + \cdots + n_{i-1}, n_1 + \cdots + n_{i} + 1, \dots, n}
			\\
			\theta^{\ell_{p_i + 1}} \wedge \cdots \wedge \theta^{\ell_{n_i}}
			\wedge 
			(\wedge_{j=1, \dots, r ; j\neq i} \omega_{\vol, j})
		\end{multlined}
		\\
		&= \begin{multlined}[t]
			(-1)^{p_i(n_1 + \cdots + n_{i-1})}
			\tfrac{1}{(n_i - p_i)!} 
			\sqrt{\abs{g_i}}
			\epsilon_{1, \dots, n_1 + \cdots + n_{i-1}, \ell_1, \dots, \ell_{n_i}, n_1 + \cdots + n_{i} + 1, \dots, n}
			\\
			\omega_{\vol, 1} \wedge \cdots \wedge \omega_{\vol, i-1}
			\wedge
			\theta^{\ell_{p_i + 1}} \wedge \cdots \wedge \theta^{\ell_{n_i}}
			\wedge 
			\omega_{\vol, i+1} \wedge \cdots \wedge \omega_{\vol, r}
		\end{multlined}
	\end{align*}
	To deal with the summation on the indices $\ell_r \in I_i$, we shift their values by $-(n_1 + \cdots + n_{i-1})$ to get indices $k_r = 1, \dots, n_i$. Then we can replace the $\epsilon$ tensor by the tensor $\epsilon_{k_1, \dots, k_{n_i}}$ and at the same time replacing the $\theta^{\ell_r}$'s by the $\theta_i^{k_r}$'s. Using the factor $\tfrac{1}{(n_i - p_i)!} \sqrt{\abs{g_i}}$ in front, this makes appears $\hstar_i$:
	\begin{align*}
		\hstar \omega'_i
		&=
		(-1)^{p_i(n_1 + \cdots + n_{i-1})}
		\omega_{\vol, 1} \wedge \cdots \wedge \omega_{\vol, i-1}
		\wedge
		(\hstar_i \omega'_i)
		\wedge 
		\omega_{\vol, i+1} \wedge \cdots \wedge \omega_{\vol, r}
	\end{align*}
	so that, with $\omega_i \wedge \hstar_i \omega'_i = a_i \omega_{\vol,i}$, where $a_i = \tfrac{1}{p_i !} \omega_{i, k_1, \dots, k_{p_i}} {\omega'_i}^{k_1, \dots, k_{p_i}}$,
	\begin{align*}
		\omega_i \wedge \hstar \omega'_i
		&=
		\omega_{\vol, 1} \wedge \cdots \wedge \omega_{\vol, i-1}
		\wedge
		(\omega_i \wedge \hstar_i \omega'_i)
		\wedge 
		\omega_{\vol, i+1} \wedge \cdots \wedge \omega_{\vol, r}=
		a_i
		\omega_{\vol, 1} \wedge \cdots \wedge \omega_{\vol, r}
		=
		a_i \omega_{\vol}
	\end{align*}
	Notice that moving $\omega_i$ inside the product of the volume forms $\omega_{\vol, j}$ and then putting $a_i$ in front of this product is possible since volume forms have commutative values in $\algA$. Since $\int_{V_i} \omega_i \wedge \hstar_i \omega'_i = \tau(a_i)$, one gets $\int_{V} \omega \wedge \hstar \omega' = \tsum_{i=1}^{r} \int_{V} \omega_i \wedge \hstar \omega'_i = \tsum_{i=1}^{r} \tau(a_i) = \tsum_{i=1}^{r} \int_{V_i} \omega_i \wedge \hstar_i \omega'_i$.
\end{proof}

\section{\texorpdfstring{Differential Structure on $M_n(\bbC)$ and $\bigoplus_{i=1}^nM_{n_i}(\bbC)$ using Finite Spectral Triples}{Differential structure on Mn(C) and Oplus Mni(C) using Finite Spectral Triples}}

\label{ResMatrDirac}

In this section, we recall all the important facts about finite real spectral triples that will be needed later on. In particular, their classification by Krajewski diagrams \cite{Kraj98e} which diagrammatically encode the essential algebraic data used to characterize spectral triples (see also \cite{Suij15a}, in which a sketch of this classification is given). The result are given only for $\bigoplus_{i=1}^nM_{n_i}(\bbC)$ since in this framework the case of $M_{n}(\bbC)$ can be seen as a simple restriction of the above more general situation. We will set up the normal form in which spectral triples express, according to a basis of irreducible representations of the Hilbert space, and differential structure being constructed from the Dirac operator that will be derived according to this basis, using the formula eq.(\refeq{EqForm}). The role played by metric, integration, and Hodge $\hstar$-operator in the derivation framework, will be respectively played by the Dirac operator, the integration and the spectral action as we will see in subsection \ref{SpectralAction} and chapter \ref{NCSMPP}.  In the following, we will not need to consider the analytic axioms since we consider only finite dimensional algebras and representations.
\medskip
\par
In some proofs we will use the following  well-known technical result, which comes from the existence of cyclic vectors in $\bbC^n$ for the matrix multiplication:
\begin{lemma}
	\label{lemma technical module endo reduction}
	For any $n \geq 1$ and any vector space $V$, a linear map $\Psi : \bbC^n \otimes V \to \bbC^n \otimes V$ such that $\Psi(a \xi \otimes v) = a \Psi(\xi \otimes v)$ for any $a \in M_n(\bbC)$, $\xi \in \bbC^n$ and $v \in V$, reduces to a linear map $\varphi : V \to V$ such that $\Psi(\xi \otimes v) = \xi \otimes \varphi(v)$.
\end{lemma}

\begin{proof}
	Let $\{ \xi^\alpha \}_{1 \leq \alpha \leq n}$ be the canonical basis of $\bbC^{n}$. For any $\xi \in \bbC^n$, let $a \in M_n(\bbC)$ be the matrix with first column $\xi$ and all other columns zero. Then one has $\xi = a \xi^1$ and $a \xi^\alpha = 0$ for any $\alpha > 1$. So, $\Psi(\xi \otimes v) = a \Psi(\xi^1 \otimes v) = a \sum_{\alpha=1}^{n} \xi^\alpha \otimes \varphi_{\alpha}(v)$ for a family of linear maps $\varphi_\alpha : V \to V$. With the chosen $a$, the sum reduces to its first term only, and so $\Psi(\xi \otimes v) =  a \xi^1 \otimes \varphi_1(v) = \xi \otimes \varphi_1(v)$. So $\varphi \defeq \varphi_1$ is the desired linear map.
\end{proof}

\subsection{Finite Spectral Triples}
\label{subsec finite spectral triples}

A spectral triple $(\algA, \hs, D)$ is said to be finite if $\algA$ is a finite dimensional involutive $\bbC$-algebra, $\hs$ is a finite dimensional Hilbert space on which $\algA$ is represented, and $D = D^\dagger$ is a self-adjoint operator on $\hs$. The faithful representation $\pi$ of $\algA$ on $\hs$ will be omitted when no confusion is possible. In the following, we will write $\algA_i = M_{n_i} = M_{n_i}(\bbC)$ since no other matrix algebras will be considered. Let $\inj^i : \algA_i \to \algA$  be the canonical inclusion and $\proj_i : \algA \to \algA_i$ be the canonical projection.
\medskip
\par
Consider the set $\Lambda \defeq \{ \bn_1, \dots, \bn_r \}$ of irreducible representations (irreps) of $\algA$, where $\bn_i$ is a short notation that designates at the same time the integer $n_i$ defining the irrep (on the space $\bbC^{n_i}$) and the integer $i$ (the same that appears in the presentation $\toplus_{i=1}^{r} M_{n_i}$ of $\algA$). $\Lambda$ is completely defined by $\algA$ and, reciprocally, $\algA = \toplus_{i=1}^{r} M_{n_i}$ can be recovered from $\Lambda$.  Notice that the same dimension $n = n_{i} = n_{i'} = \ldots$ can appear several times, but labeled by different integers $i, i', \dots$. We will also use the ordering on the labels $i$'s. Denote by $\hs_{\bn_i} \defeq \bbC^{n_i}$ the irreducible representations (irreps) of the $\algA_i$'s, and so of $\algA$. 
\medskip
\par
The Hilbert space $\hs$ can be decomposed into orthogonal components $\hhs_{\bn_i} \defeq \inj^i(\algA_i) \hs$, so that $\hs = \toplus_{i=1}^{r} \hhs_{\bn_i}$. Define $\injhs^i : \hhs_{\bn_i} \to \hs$ and $\projhs_i : \hs \to \hhs_{\bn_i}$ the natural inclusions and (orthogonal) projections. Then there are integers $\mu_i$, the multiplicities of the irreps, such that $\hhs_{\bn_i} \simeq \hs_{\bn_i} \otimes \bbC^{\mu_i} = \bbC^{n_i} \otimes \bbC^{\mu_i}$. So, up to unitary equivalence, the Hilbert space $\hs$ can be decomposed as $\hs \simeq \toplus_{i=1}^{r} \bbC^{n_i} \otimes \bbC^{\mu_i}$ and we now suppose that a unitary map has been chosen such that $\hhs_{\bn_i} = \bbC^{n_i} \otimes \bbC^{\mu_i}$.\footnote{For sake of completeness, let us mention that the scalar product of this decomposition is the usual one:  $\langle \psi, \psi' \rangle_{\hs} = \tsum_{i=1}^{r} \langle \xi_i, \xi'_i \rangle_{\bbC^{n_i}} \langle \vm_i, \vm'_i \rangle_{\bbC^{\mu_i}}$ for any $\psi = \toplus_{i=1}^{r} \xi_i \otimes \vm_i$ (and the same for $\psi'$) where $\xi_i \in \bbC^{n_i}$ and $\vm_i \in \bbC^{\mu_i}$.} If one requires a faithful representation of $\algA$, then $\mu_i \geq 1$ for all $i$.
\medskip
\par
In the even case, one has:
\begin{lemma}
	\label{lemma gamma non real}
	$\gamma$ decomposes along a family of linear maps $\ell_{i} : \bbC^{\mu_{i}} \to  \bbC^{\mu_{i}}$ such that $\gamma(\xi_i \otimes \vm_i) = \xi_i \otimes \ell_i(\vm_i)$ for any $\xi_i \otimes \vm_i \in \bbC^{n_{i}} \otimes \bbC^{\mu_{i}}$. This family satisfies $\ell_{i}^\dagger = \ell_{i}$ and $\ell_{i}^2 = 1$.
\end{lemma}

\begin{proof}
	First, consider $a = \toplus_{i=1}^{r} a_i$ and $\psi = \toplus_{i=1}^{r} \xi_i \otimes \vm_i$ with only one non-zero component at fixed $i$ for $\psi$ and $a$ with $a_i = \bbbone_{n_i}$ (unit matrix in $M_{n_i}$): then one has $\gamma( \xi_i \otimes \vm_i) \in \bbC^{n_i} \otimes \bbC^{\mu_i}$. Since $\gamma^2 = 1$, this implies that $\gamma$ decomposes along the isomorphisms $\gamma_i : \bbC^{n_i} \otimes \bbC^{\mu_i} \to \bbC^{n_i} \otimes \bbC^{\mu_i}$ such that $\gamma_i^\dagger = \gamma_i$, $\gamma_i^2 = 1$ and $\gamma_i a_i  = a_i \gamma_i$ for any $a_i \in M_{n_{i}}$.
	\medskip
	\par
	Then, by Lemma~\ref{lemma technical module endo reduction}, since $\gamma$ commutes with $\pi$, there is a family of linear maps $\ell_{i} : \bbC^{\mu_{i}} \to  \bbC^{\mu_{i}}$ such that $\gamma(\xi_i \otimes \vm_i) = \xi_i \otimes \ell_i(\vm_i)$ for any $\xi_i \otimes \vm_i \in \bbC^{n_{i}} \otimes \bbC^{\mu_{i}}$. The relations $\ell_{i}^\dagger = \ell_{i}$ and $\ell_{i}^2 = 1$ are direct consequences of those on $\gamma_i$.
\end{proof}

In the odd situation, let us consider any orthonormal basis $\{ \vm_{i}^{p} \}_{1 \leq p \leq \mu_{i}}$ of $\bbC^{\mu_{i}}$. In the even case, we require this basis to be eigenvectors of $\ell_{i}$ with eigenvalues $s_{i}^{p} = \pm 1$. Then, for any $1 \leq i \leq r$, let $\Gamma^{(0)}_{\bn_i} \defeq \{ (i, p) \mid 1 \leq p \leq \mu_{i} \}$, and for any $v = (i,p) \in \Gamma^{(0)}_{\bn_i}$, define $\lambda : \Gamma^{(0)}_{\bn_i} \to \Lambda$ as $\lambda(v) \defeq \bn_i$. Notice that $\mu_{i} = \# \Gamma^{(0)}_{\bn_i}$. For any $v \in \Gamma^{(0)}_{\bn_i}$, we then define
\begin{align*}
	\hs_{v} \defeq \Span \{ \xi_i \otimes \vm_{i}^{p} \mid \xi_i \in \bbC^{n_i} \} \simeq \hs_{\bn_i}
\end{align*}
Then, in the even case, $\gamma$ restricts to the multiplication by $s_{i}^{p}$ on $\hs_{v}$ with $v = (i,p)$. We define $s (v) = s_{i}^{p}$ for any $v$.
\medskip
\par
The map $\lambda$ is extended in an obvious way on the set
\begin{align*}
	\Gamma^{(0)} \defeq \cup_{i=1}^{r} \Gamma^{(0)}_{\bn_i}
\end{align*}
and there is an orthogonal decomposition of $\hs$ into irreps
\begin{align*}
	\hs = \toplus_{v \in \Gamma^{(0)}} \hs_{v}
\end{align*}
Let $e = (v_1, v_2) \in \Gamma^{(0)} \times \Gamma^{(0)}$, then the Dirac operator decomposes along maps $D_{e} : \hs_{v_1} \to \hs_{v_2}$. With $\Be \defeq (v_2, v_1)$, $D^\dagger = D$ is equivalent to $D_{\Be} = D_{e}^\dagger$. In the even case, $\gamma D = - D \gamma$ implies that $s(v_2) D_e = - s(v_1) D_e$, so that $D_{e}$ is non-zero only when $s(v_2) = - s(v_1)$.
\medskip
\par
The previous decomposition of the spectral triple $(\algA, \hs, D)$ or $(\algA, \hs, D, \gamma)$ can be summarized using a decorated graph $\Gamma$, a so-called Krajewski Diagram, together with $\Lambda$:
\begin{enumerate}
	\item The set of vertices $\Gamma^{(0)}$ of the graph is equipped with a map $\lambda : \Gamma^{(0)} \to \Lambda$. By a slight abuse of notation, the map $\lambda$ will sometimes be used in the compact notation $\bbC^{\lambda(v)} = \bbC^{n_i}$. We will also use the map $i(v) \defeq i$ for $\lambda(v) = \bn_i$.
	
	\item For any vertex $v \in \Gamma^{(0)}$, define $\hs_v \defeq \hs_{\lambda(v)} = \bbC^{\lambda(v)}$. The element $\lambda(v) \in \Lambda$ is a decoration of the vertex $v$.
	
	\item For any $\bn_i \in \Lambda$, define $\Gamma^{(0)}_{\bn_i} \defeq \{ v \in \Gamma^{(0)} \mid \lambda(v) = \bn_i \} = \lambda^{-1}(\bn_i)$ and $\mu_{i} \defeq \# \Gamma^{(0)}_{\bn_i}$.
	
	\item In the even case, a second decoration is the assignment of a grading map $s(v) = \pm 1$.
	
	\item For every $e = (v_1, v_2) \in \Gamma^{(0)} \times \Gamma^{(0)}$, let $\Be \defeq (v_2, v_1)$.
	
	\item The space $\Gamma^{(1)} \subset \Gamma^{(0)} \times \Gamma^{(0)}$ of edges of the graph are couples $e = (v_1, v_2)$ such that:
	\begin{enumerate}
		\item  there is a non-zero linear map $D_e : \hs_{v_1} \to \hs_{v_2}$ such that $D_{\Be} = D_e^\dagger : \hs_{v_2} \to \hs_{v_1}$. 
		\item $s(v_2) = - s(v_1)$ in the even case;
	\end{enumerate}
	Then $D_e$ defines a decoration of $e$.
\end{enumerate}
\medskip
\par
Given such a Krajewski Diagram, one can construct a spectral triple up to unitary equivalence in the following way. As already mentioned, $\Lambda$ determines the algebra $\algA \simeq \toplus_{i=1}^{r} M_{n_i}$. A vertex $v \in \Gamma^{(0)}$ designates a copy of an irrep $\hs_{v} = \hs_{\lambda(v)} = \bbC^{\lambda(v)}$, and $\mu(v) \defeq \# \Gamma^{(0)}_{\lambda(v)}$ is the multiplicity of this irrep in $\hs$. So, the Hilbert space decomposes as $\hs \defeq \toplus_{v \in \Gamma^{(0)}} \hs_{v} = \toplus_{i=1}^{r} \bbC^{n_i} \otimes \bbC^{\mu_i}$.
\medskip
\par
Then the representation $\pi$ of $\algA$ on $\hs$ decomposes as $\pi(a) \psi = \toplus_{v \in \Gamma^{(0)}} a_{i(v)} \psi_v = \toplus_{i=1}^{r} (a_i \xi_i) \otimes \vm_i$, for any $a = \toplus_{i=1}^{r} a_i \in \algA$ and any $\psi = \toplus_{v \in \Gamma^{(0)}} \psi_v = \toplus_{i=1}^{r} \xi_i \otimes \vm_i$,  with $\xi_i \otimes \vm_i \in \bbC^{n_{i}} \otimes \bbC^{\mu_{i}}$, where $a_{i(v)} \psi_v$ (resp. $a_i \xi_i \otimes \vm_i$) is the multiplication of the matrix $a_{i(v)}$ on the vector $\psi_v \in \bbC^{\lambda(v)}$ (resp. $a_i$ on $\xi_i \in \bbC^{n_i}$).
\medskip
\par
In the even case, $\gamma$ is determined as the multiplication by the decoration $s(v) = \pm 1$ on $\hs_{v}$. The Dirac operator $D$ is reconstructed by the decorations $D_{e}$ of the edges $e \in \Gamma^{(1)}$.
\medskip
\par
It is useful to write an explicit reconstruction of $D$ on the decomposition $\hs = \toplus_{i=1}^{r} \bbC^{n_i} \otimes \bbC^{\mu_i}$. First, introduce an orthonormal basis for each $\bbC^{\mu_i}$, for instance its canonical basis. Since $v \in \Gamma^{(0)}$ designates a specific copy of an irrep $\bn_i = \lambda(v)$, one can label all the basis vectors in the union of all the $\bbC^{\mu_i}$'s as $\{ \vm_v \}_{v \in \Gamma^{(0)}}$: for any $v \in \Gamma^{(0)}_{\bn_i}$, $\vm_v$ is an element of an orthonormal basis of $\bbC^{\mu_i}$. Then we can use the identification $\hs_v = \Span \{ \xi \otimes \vm_v \mid \xi \in \bbC^{\lambda(v)} \}$. For any $e = (v_1, v_2) \in \Gamma^{(1)}$ with $\lambda(v_1) = \bn_i$ and $\lambda(v_2) = \bn_j$, define $\hD_{e} : \bbC^{n_{i}} \otimes \bbC^{\mu_{i}} \to \bbC^{n_{j}} \otimes \bbC^{\mu_{j}}$, for any $\xi \in \bbC^{n_{i}}$, as
\begin{align*}
	\hD_{e} (\xi \otimes \vm_{v})
	&=
	\begin{cases}
		0 & \text{if $v \neq v_1$} \\
		(D_e \xi) \otimes \vm_{v_2} & \text{if $v = v_1$}
	\end{cases}
\end{align*}
Then define 
\begin{align*}
	D_{j}^{i} \defeq 
	\sum_{\substack{e = (v_1, v_2) \in \Gamma^{(1)} \\ \lambda(v_1) = \bn_{i},  \lambda(v_2) = \bn_{j}}} \hD_{e}
	: \bbC^{n_{i}} \otimes \bbC^{\mu_{i}} \to \bbC^{n_{j}} \otimes \bbC^{\mu_{j}}
\end{align*}
Notice that the summation can be written on $e \in \Gamma^{(1)} \cap (\Gamma^{(0)}_{\bn_{i}} \times \Gamma^{(0)}_{\bn_{j}})$. All the operators $D_{j}^{i}$ can be collected in an operator $D : \hs \to \hs$ which is automatically self-adjoint.

\subsection{Finite Real Spectral Triples}

Let us now consider (odd) finite (resp. even) real spectral triples $(\algA, \hs, D, J)$ (resp. $(\algA, \hs, D, J, \gamma)$). The Hilbert space $\hs$ is then a bimodule over $\algA = \toplus_{i=1}^{r} M_{n_i}$, or equivalently a left $\algA^{e}$-module, with $\algA^{e} = \toplus_{{i}, {j} = 1}^{r} M_{n_{i}} \otimes M_{n_{j}}^\circ$. This implies that $\hs$ decomposes into orthogonal components $\hhs_{\bn_i \bn_j} \defeq \inj^{i}(\algA_{i}) \inj^{j}(\algA_{j})^\circ \hs$, so that $\hs = \toplus_{{i}, {j}=1}^{r} \hhs_{\bn_i \bn_j}$.
\medskip
\par
Denote by ${\bbC^m}^\top$ (${}^\top$ for transpose) the $m$-dimensional $\bbC$-vector space of row vectors, which is a natural right $M_m$-module, and denote by $\bbC^{m \circ}$ its corresponding left $M_m^\circ$-module ($\bbC^m \simeq \bbC^{m \circ}$ as column vectors by the formal map $\bbC^m \ni \xi \mapsto \xi^\circ \in \bbC^{m \circ}$ and, for any $a \in M_m$ and $\xi \in \bbC^m$, $a^\circ \xi^\circ \defeq (\xi^\top a)^\top$). \label{footnote circ action}
\medskip
\par
Let us recall the following result:
\begin{lemma}
	For any integers $n, m \geq 1$, the irreducible left $M_n \otimes M_m^\circ$-representations are isomorphic to $\bbC^n \otimes \bbC^{m \circ}$.
\end{lemma}

\begin{proof}
	The proof relies on the identification of $M_n \otimes M_m^\circ = \End(\bbC^n) \otimes \End(\bbC^{m \circ})$ with $\End(\bbC^n \otimes \bbC^{m \circ})$ by the natural map $\End(\bbC^n) \otimes \End(\bbC^{m \circ}) \ni \psi_n \otimes \psi_m^\circ \mapsto [ \xi_n \otimes \xi_m^\circ \mapsto \psi_n(\xi_n) \otimes \psi_m^\circ(\xi_m^\circ)]$, which implies that $M_n \otimes M_m^\circ$ is isomorphic to a matrix algebra over the vector space $\bbC^n \otimes \bbC^{m \circ}$. Then one applies the usual result on irreducible left representations of matrix algebras.
\end{proof}

Let $\mu_{i j}$ be the multiplicity of the irrep $\hs_{\bn_i \bn_j} \defeq \bbC^{n_{i}} \otimes \bbC^{n_{j} \circ}$ of $M_{n_{i}} \otimes M_{n_{j}}^\circ$ and so of $\algA^{e}$, in $\hs$. Then one has $\hhs_{\bn_i \bn_j} \simeq \hs_{\bn_i \bn_j} \otimes \bbC^{\mu_{i j}}  \simeq \bbC^{n_{i}} \otimes \bbC^{\mu_{i j}} \otimes \bbC^{n_{j} \circ}$, so that $\hs \simeq \toplus_{{i}, {j} = 1}^{r} \bbC^{n_{i}} \otimes \bbC^{\mu_{i j}} \otimes \bbC^{n_{j} \circ}$. In the following, we suppose that a unitary map has been chosen such that $\hhs_{\bn_i \bn_j} = \bbC^{n_{i}} \otimes \bbC^{\mu_{i j}} \otimes \bbC^{n_{j} \circ}$. \footnote{The factor $\bbC^{\mu_{i j}}$ has been positioned in the middle to put forward the bimodule structure. In the proof of Prop.~\ref{prop diagonalization spm for KO dims} it will be convenient to change this convention.}
\medskip
\par
Denote by $J_0$ the anti-unitary operator on $\bbC^n \otimes \bbC^\mu \otimes {\bbC^m}^\circ$ defined by $\xi \otimes \vm \otimes \eta^\circ \mapsto \Bxi \otimes \Bvm \otimes \Beta^\circ$ where $\Bxi$ is the entrywise complex conjugated vector (the same for $\Bvm$ and $\Beta^\circ$). Then $J_0$ extends naturally to $\hs$ as an anti-unitary operator which preserves each summand $\hhs_{\bn_i \bn_j}$ and one has $J_0^{-1} = J_0$. Notice that $J_0$ depends on the canonical basis for the vector spaces $\bbC^n$, $\bbC^\mu$ and $\bbC^m$ (But any fixed orthonormal basis could have been used). We will use the natural notation $J_0(\psi) = \Bpsi$ for any $\psi \in \hs$. Define $K \defeq J J_0$, so that $J = K J_0$. For any $a = \toplus_{i=1}^{r} a_i \in \algA$, define $a^\top = \toplus_{i=1}^{r} a_i^\top$ where $a_i^\top = J_0 a_i^\ast J_0$ is the transpose of $a_i \in M_{n_i}$. For any operator $A$ on $\hs$, define $\BA \defeq J_0 A J_0$ (if $A$ is written as a matrix, $\BA$ is the entrywise complex conjugate matrix, whence the notation).

\begin{proposition}
	\label{prop J K L}
	$K$ is a unitary operator on $\hs$ such that $K \BK = \BK K = \epsilon$ and $a K(\psi) b = K( b^\top \psi a^\top)$ for any $a,b \in \algA$ and $\psi \in \hs$.
	\medskip
	\par
	For any $1 \leq {i}, {j} \leq r$, $K(\hhs_{\bn_i \bn_j}) = \hhs_{\bn_j \bn_i}$, so that $\hhs_{\bn_i \bn_j}$ and $\hhs_{\bn_j \bn_i}$ have the same dimension, \textit{i.e.} they correspond to the same multiplicity $\mu_{i j} = \mu_{{j} {i}}$.
	\medskip
	\par
	There is a linear map $L_{i j} : \bbC^{\mu_{i j}} \to \bbC^{\mu_{{j} {i}}}$ satisfying $L_{i j}^\dagger = L_{i j}^{-1}$ and $L_{{j} {i}} \BL_{i j} = \BL_{{j} {i}} L_{i j} = \epsilon$, such that, for any $\xi_{i} \otimes \vm_{i j} \otimes \eta_{j}^\circ \in \hhs_{\bn_i \bn_j}$, $K(\xi_{i} \otimes \vm_{i j} \otimes \eta_{j}^\circ) = \eta_{j} \otimes L_{i j}(\vm_{i j}) \otimes \xi_{i}^\circ$.
\end{proposition}

\begin{proof}
	Since $K$ is the composition of two anti-unitary operators on $\hs$, it is unitary. One has $\BK = J_0 (J J_0) J_0 = J_0 J$ so that $K \BK = J J_0 J_0 J =  J^2 = \epsilon$ and the same for $\BK K = \epsilon$.
	\medskip
	\par
	For any $a \in \algA$ and $\psi \in \hs$, one has $\psi a = J a^\ast J^{-1} \psi = K J_0 a^\ast J_0 K^{-1} \psi$. From this we get two relations. 
	For the first one, replace $\psi$ by $K \psi$ to get $(K \psi) a = K J_0 a^\ast J_0 \psi$, which can be written as $(K \psi) a = K ( a^\top \psi)$. For the second one, act on both sides with $K^{-1} = \epsilon\, J_0 K J_0$ (consequence of $J^2 = \epsilon$) to get $\epsilon\, J_0 K J_0 (\psi a) = \epsilon\, J_0 a^\ast J_0 J_0 K J_0 \psi$, that we can simplify as $K J_0 (\psi a) = a^\ast K J_0 \psi$. Replacing $a^\ast$ by $a$ and $\Bpsi$ by $\psi$, one gets $K( \psi a^\top) = a ( K \psi)$.	These two relations can be combined as $a K(\psi) b = K( b^\top \psi a^\top)$ for any $a,b \in \algA$ and $\psi \in \hs$. The map $K^{-1}$ satisfies the same relation.
	\medskip
	\par
	Consider an element $\psi$ which has only a non-zero component $\xi_{i} \otimes \vm_{i j} \otimes \eta_{j}^\circ \in \hhs_{\bn_i \bn_j}$ and consider a unique non-zero component $\bbbone_{n_{i}}$ for $a$ and a unique non-zero component $\bbbone_{n_{j}}$ for $b$. Then $K(\xi_{i} \otimes \vm_{i j} \otimes \eta_{j}^\circ) = b^\top K(\xi_{i} \otimes \vm_{i j} \otimes \eta_{j}^\circ) a^\top \in \hhs_{\bn_j \bn_i}$, so that $K(\hhs_{\bn_i \bn_j}) \subset \hhs_{\bn_j \bn_i}$. Using the same line of reasoning with $K^{-1}$, one gets $K^{-1}(\hhs_{\bn_i \bn_j}) \subset \hhs_{\bn_j \bn_i}$, and so $K(\hhs_{\bn_i \bn_j}) = \hhs_{\bn_j \bn_i}$. 
	\medskip
	\par
	One can use a slight adaptation of Lemma~\ref{lemma technical module endo reduction} to show that there is a family of linear maps $L_{i j} : \bbC^{\mu_{i j}} \to \bbC^{\mu_{{j} {i}}}$ such that $K$ decomposes as $K(\xi_{i} \otimes \vm_{i j} \otimes \eta_{j}^\circ) = \eta_{j} \otimes L_{{i} {j}}(\vm_{i j}) \otimes \xi_{i}^\circ$ for any $\xi_{i} \in \bbC^{n_{i}}$, $\vm_{i j} \in \bbC^{\mu_{i j}}$, and $\eta_{j}^\circ \in \bbC^{n_{j} \circ}$. It is easy to show that $K^\dagger$ decomposes along the family of linear maps $L_{i j}^\dagger : \bbC^{\mu_{{j} {i}}} \to \bbC^{\mu_{i j}}$ and  $L_{i j}^\dagger = L_{i j}^{-1}$ since $K$ is unitary. The relations $L_{{j} {i}} \BL_{i j} = \BL_{{j} {i}} L_{i j} = \epsilon$ are consequence of $K \BK = \BK K = \epsilon$ since $\BK$ decomposes along the $\BL_{i j}$'s.
\end{proof}

\begin{corollary}
	For any $\xi_{i} \otimes \vm_{i j} \otimes \eta_{j}^\circ \in \hs_{\bn_i \bn_j}$, one has $J(\xi_{i} \otimes \vm_{i j} \otimes \eta_{j}^\circ) = \Beta_{j} \otimes L_{i j}(\Bvm_{i j}) \otimes \Bxi_{i}^\circ$.
\end{corollary}

\begin{lemma}
	Let $\vm \in \bbC^{\mu_{ii}}$ be an eigenvector of $L_{ii}$ with eigenvalue $\lambda \in U(1)$. Then $\Bvm$ is also an eigenvector with eigenvalue $\epsilon\, \lambda$.
\end{lemma}

\begin{proof}
	The complex conjugate of the relation $L_{ii} \vm = \lambda \vm$ gives $\BL_{ii} \Bvm = \Blambda \Bvm$, which is equivalent to $\epsilon\, L_{ii}^{-1} \Bvm = \lambda^{-1} \Bvm$, so that $L_{ii} \Bvm = \epsilon\, \lambda \Bvm$.
\end{proof}

\begin{proposition}
	\label{prop gamma to ell}
	In the even case, there is a family of linear maps $\ell_{i j} : \bbC^{\mu_{i j}} \to \bbC^{\mu_{i j}}$ such that $\gamma(\xi_{i} \otimes \vm_{i j} \otimes \eta_{j}^\circ) = \xi_{i} \otimes \ell_{i j}(\vm_{i j}) \otimes \eta_{j}^\circ$ for any $\xi_{i} \otimes \vm_{i j} \otimes \eta_{j}^\circ \in \hhs_{\bn_i \bn_j}$. This family satisfies $\ell_{i j}^\dagger = \ell_{i j}$ and $\ell_{i j}^2 = 1$.
\end{proposition}

\begin{proof}
	For any $\psi \in \hs$ and any $a, b \in \algA$, one has $\gamma(a \psi b) = a \gamma(\psi) b$. Then one can use a slight adaptation of Lemma~\ref{lemma technical module endo reduction} to deduce the family of linear maps $\ell_{i j}$. The relations satisfied by this family of linear maps are consequence of the relations satisfied by $\gamma$.
\end{proof}

\begin{lemma}[Technical results on the families $L_{i j}$ and $\ell_{i j}$]
	\label{prop ell and L}
	The families of linear maps $L_{i j}$ and $\ell_{i j}$ satisfy the following properties for any $i, j$. 
	\begin{enumerate}
		\item For any $p = 1, 2$, let $\vm_{i j}^{p} \in \bbC^{\mu_{i j}}$ and define $\vm_{j i}^{p} \defeq L_{i j}(\Bvm_{i j}^{p}) \in \bbC^{\mu_{j i}}$. Then $\langle \vm_{j i}^{1}, \vm_{j i}^{2} \rangle_{\bbC^{\mu_{j i}}} = \langle \vm_{i j}^{2}, \vm_{i j}^{1} \rangle_{\bbC^{\mu_{i j}}}$. In particular, $\vm_{i j}^{p}$ and $\vm_{j i}^{p}$ have the same norm and $\vm_{i j}^{1}$ and $\vm_{i j}^{2}$ are orthogonal if and only if $\vm_{j i}^{1}$ and $\vm_{j i}^{2}$ are orthogonal. 
		
		\item $L_{i j} \circ \Bell_{i j} = \epsilon'' \ell_{j i} \circ L_{i j}$. 
		
		\item Let $\vm_{i j} \in \bbC^{\mu_{i j}}$ be an eigenvector of $\ell_{i j}$ with eigenvalue $s_{i j} = \pm 1$. Then $\vm_{j i} \defeq L_{i j}(\Bvm_{i j}) \in \bbC^{\mu_{j i}}$ is an eigenvector of $\ell_{j i}$ with eigenvalue $s_{j i} =  \epsilon'' s_{i j}$.
	\end{enumerate}
\end{lemma}

\begin{proof} These relations are straightforward computations using previously proved properties on the families of linear maps $L_{i j}$ and $\ell_{i j}$.
	
	1. One has $\langle \vm_{j i}^{1}, \vm_{j i}^{2} \rangle = \langle L_{i j}(\Bvm_{i j}^{1}), L_{i j}(\Bvm_{i j}^{2}) \rangle = \langle \Bvm_{i j}^{1}, \Bvm_{i j}^{2} \rangle = \overline{\langle \vm_{i j}^{1}, \vm_{i j}^{2} \rangle} = \langle \vm_{i j}^{2}, \vm_{i j}^{1} \rangle$.
	
	2. The relation $J \gamma(\xi_i \otimes \vm_{i j} \otimes \eta_j^\circ) = \epsilon'' \gamma J(\xi_i \otimes \vm_{i j} \otimes \eta_j^\circ)$ can be written has $\Beta_{j} \otimes L_{i j}(\overline{\ell_{i j} (\vm_{i j})}) \otimes \Bxi_{i}^\circ = \epsilon'' \Beta_{j} \otimes \ell_{j i} (L_{i j}(\Bvm_{i j})) \otimes \Bxi_{i}^\circ$ for any $\xi_i$, $\vm_{i j}$, and $\eta_j^\circ$.  This implies the relation.
	
	3. Using $\ell_{i j}(\vm_{i j}) = s_{i j} \vm_{i j}$, one gets $\ell_{j i}(\vm_{j i}) = \ell_{j i}(L_{i j}(\Bvm_{i j})) = \epsilon'' L_{i j}(\Bell_{i j}(\Bvm_{i j})) = \epsilon'' L_{i j}(\overline{\ell_{i j}(\vm_{i j})}) = \epsilon'' s_{i j} L_{i j}(\Bvm_{i j}) = \epsilon'' s_{i j} \vm_{j i}$.
\end{proof}

Let us now describe, in the two following propositions, the key constructions which lead to the classification of finite real spectral triples. The content of these two propositions will be useful in Sect.~\ref{sec AF algebras} (mainly in Prop.~\ref{prop diagonalization spm for KO dims}).

\begin{proposition}
	\label{prop basis odd case}
	Consider the odd case situation. 
	
	For $1 \leq i \neq j  \leq r$, there is an orthonormal basis $\{ \vm_{i j}^{p} \}_{1 \leq p \leq \mu_{i j}}$ of $\bbC^{\mu_{i j}}$ such that $\vm_{j i}^{p} = L_{i j}(\Bvm_{i j}^{p})$ and $\vm_{i j}^{p} = \epsilon\, L_{j i}(\Bvm_{j i}^{p})$ for any $i<j$ and any $1 \leq p \leq \mu_{j i} = \mu_{i j}$. 
	
	For $i = j$ and $\epsilon = 1$ ($KO$-dimensions $1$ and $7$), there is an orthonormal basis $\{ \vm_{ii}^{p} \}_{1 \leq p \leq \mu_{ii}}$ of $\bbC^{\mu_{ii}}$ such that $\vm_{ii}^{p} = L_{ii}(\Bvm_{ii}^{p})$.
	
	For $i = j$ and $\epsilon = -1$ ($KO$-dimensions $3$ and $5$), $\mu_{ii}$ is even and there is an orthonormal basis $\{ \vm_{ii}^{p} \}_{1 \leq p \leq \mu_{ii}}$ of $\bbC^{\mu_{ii}}$ such that $\vm_{ii}^{2a} = L_{ii}(\Bvm_{ii}^{2a-1})$ and $\vm_{ii}^{2a-1} = \epsilon\, L_{ii}(\Bvm_{ii}^{2a})$ for any $a = 1, \dots, \mu_{ii}/2$.
\end{proposition}

\begin{proof}
	For $1 \leq i < j  \leq r$, consider any orthonormal basis $\{ \vm_{i j}^{p} \}_{1 \leq p \leq \mu_{i j}}$ of $\bbC^{\mu_{i j}}$ and define $\vm_{j i}^{p} \defeq L_{i j}(\Bvm_{i j}^{p}) \in \bbC^{\mu_{j i}}$ for all $1 \leq p \leq \mu_{j i} = \mu_{i j}$. These vectors form an orthonormal basis satisfying the relations.
	\medskip
	\par
	For $i = j$, the proof is an adaptation of the proof in \cite{Wign60n} or \cite[Lemma~3.8]{Suij15a} to the endomorphism $L_{ii}$ of $\bbC^{\mu_{ii}}$. Let us simplify the notations by replacing $\bbC^{\mu_{ii}}$ by $\bbC^{\mu}$ and $L_{ii}$ by $L$. Recall that $\BL L = L \BL = \epsilon$.
	\medskip
	\par
	Suppose $\epsilon = 1$. Consider any vector $v \in \bbC^{\mu}$ of norm $1$ and define $\vm^{1} =  c (v + L(\Bv))$ if $L(\Bv) \neq - v$ and $\vm^{1} = i v$ if $L(\Bv) = -v$, where $c \in \bbR$ is chosen so that $\vm^{1}$ is of norm $1$. Then $L(\Bvm^{1}) = c ( L(\Bv) + L ( \overline{L(\Bv)})) = c ( L(\Bv) + v) = \vm^{1}$ in the first situation, while $L(\Bvm^{1}) = -i L(\Bv) = i v = \vm^{1}$ in the second situation. Consider now a second vector $v' \in \bbC^{\mu}$ of norm $1$ that is orthogonal to $\vm^{1}$. Then, by Point~1 in Lemma~\ref{prop ell and L}, $L(\Bv')$ is also orthogonal to $L(\Bvm^{1}) = \vm^{1}$. This implies that $\vm^{2}$, defined from $v'$ as $\vm^{1}$ was defined from $v$, is orthogonal to $\vm^{1}$. Iterating this construction, one gets an orthonormal basis of $\bbC^{\mu}$ with the required property.
	\medskip
	\par
	Suppose $\epsilon = -1$. Let $\vm^{1} \in \bbC^{\mu}$ be any vector of norm $1$, and define $\vm^{2} \defeq L(\Bvm^{1})$, which, by Lemma~\ref{prop ell and L}, has norm $1$. Then $\langle \vm^{2}, \vm^{1} \rangle = \langle L(\Bvm^{1}), \vm^{1} \rangle = - \langle L(\Bvm^{1}), L ( \BL(\vm^{1})) \rangle = - \langle \Bvm^{1}, \BL(\vm^{1}) \rangle = - \langle \Bvm^{1}, \Bvm^{2} \rangle = - \overline{\langle \vm^{1}, \vm^{2} \rangle} = - \langle \vm^{2}, \vm^{1} \rangle$ so that $\langle \vm^{2}, \vm^{1} \rangle = 0$. Consider now a vector $\vm^{3}$ of norm $1$ which is orthogonal to $\vm^{1}$ and $\vm^{2}$, and let $\vm^{4} \defeq L(\Bvm^{3})$. From the previous computation, we know that the norm $1$ vector $\vm^{4}$ is orthogonal to $\vm^{3}$. One has $\langle \vm^{4}, \vm^{1} \rangle = - \langle L(\Bvm^{3}), L(\BL(\vm^{1})) \rangle = - \langle \Bvm^{3}, \BL(\vm^{1}) \rangle = - \langle \Bvm^{3}, \Bvm^{2} \rangle = - \langle \vm^{2}, \vm^{3} \rangle = 0$. So $\vm^{4}$ is orthogonal to $\vm^{1}$, and then of $\vm^{2} = L(\Bvm^{1})$. Iterating this construction, one gets a basis of $\bbC^{\mu}$ with the required property, and it shows also that $\mu$ is even.
\end{proof}

\begin{proposition}
	\label{prop basis even case}
	Consider the even case situation. 
	\medskip
	\par
	For $1 \leq i \neq j \leq r$, there is an orthonormal basis $\{ \vm_{i j}^{p} \}_{1 \leq p \leq \mu_{i j}}$ of $\bbC^{\mu_{i j}}$ of eigenvectors of $\ell_{i j}$ with eigenvalues $s_{i j}^{p} = \pm 1$ such that $\vm_{j i}^{p} = L_{i j}(\Bvm_{i j}^{p})$ and $\vm_{i j}^{p} = \epsilon\, L_{j i}(\Bvm_{j i}^{p})$ for any $i<j$, and $s_{j i}^{p} = \epsilon'' s_{i j}^{p}$.
	\medskip
	\par
	For $i = j$, $\epsilon = 1$, and $\epsilon'' = 1$ ($KO$-dimension $0$), there is an orthonormal basis $\{ \vm_{ii}^{p} \}_{1 \leq p \leq \mu_{ii}}$ of $\bbC^{\mu_{ii}}$ of eigenvectors of $\ell_{ii}$ with eigenvalues $s_{i}^{p} = \pm 1$ such that $\vm_{ii}^{p} = L_{ii}(\Bvm_{ii}^{p})$.
	\medskip
	\par
	For $i = j$ and $\epsilon = -1$ ($KO$-dimensions $2$ and $4$), or $\epsilon = 1$ and $\epsilon'' = -1$ ($KO$-dimension $6$), $\mu_{ii}$ is even and there is an orthonormal basis $\{ \vm_{ii}^{p} \}_{1 \leq p \leq \mu_{ii}}$ of $\bbC^{\mu_{ii}}$ of eigenvectors of $\ell_{ii}$ with eigenvalues $s_{i}^{p} = \pm 1$ such that $\vm_{ii}^{2a} = L_{ii}(\Bvm_{ii}^{2a-1})$, $\vm_{ii}^{2a-1} = \epsilon\, L_{ii}(\Bvm_{ii}^{2a})$, and $s_{i}^{2a} = \epsilon'' s_{i}^{2a-1}$ for any $a = 1, \dots, \mu_{ii}/2$.
	In $KO$-dimensions $2$ and $6$, one can choose the basis such that $s_{i}^{2a} = +1$ and $s_{i}^{2a-1} = -1$.\end{proposition}

\begin{proof}
	We will use results from Lemma~\ref{prop ell and L}. For $1 \leq i < j \leq r$, the orthonormal basis $\{ \vm_{i j}^{p} \}_{1 \leq p \leq \mu_{i j}}$ of $\bbC^{\mu_{i j}}$ constructed above can be chosen such that it is a basis of eigenvectors of $\ell_{i j}$, so that $\ell_{i j}(\vm_{i j}^{p}) = s_{i j}^{p} \vm_{i j}^{p}$ with $s_{i j}^{p} = \pm 1$. Then the vectors $\vm_{j i}^{p} \defeq L_{i j}(\Bvm_{i j}^{p})  \in \bbC^{\mu_{j i}}$ form a basis of eigenvectors of $\ell_{j i}$ for the eigenvalues $s_{j i}^{p} = \epsilon'' s_{i j}^{p}$.
	\medskip
	\par
	Let us consider the cases for $i = j$. As before, we will omit the index $i$ to simplify the notations. 
	\medskip
	\par
	For $\epsilon = -1$ ($KO$-dimensions $2$ and $4$), the basis $\{ \vm^{p} \}_{1 \leq p \leq \mu}$ of $\bbC^{\mu}$ in Prop.~\ref{prop basis odd case} can be constructed step by step such that, for any $a = 1, \dots, \mu/2$,  $\ell(\vm^{2a-1}) = s^{2a-1} \vm^{2a-1}$ with $s^{2a-1} = \pm 1$ so that $\vm^{2a} = L(\Bvm^{2a-1})$ satisfies $\ell(\vm^{2a}) = s^{2a} \vm^{2a}$ with $s^{2a} = \epsilon'' s^{2a-1}$. 
	\medskip
	\par
	For $\epsilon = 1$, let us consider two situations. Firstly, suppose $\epsilon'' = 1$ ($KO$-dimension $0$). In the construction of the basis for the corresponding situation in Prop.~\ref{prop basis odd case}, we can choose, step by step, the vectors $v^{p}$ in such a way that $\ell(v^{p}) = s^{p} v^{p}$ for $s^{p} = \pm 1$ and  $v^{p}$ is orthogonal to the $\vm^{p'}$ already defined for $p' < p$ (It is easy to check that $\ell$ restricts to an endomorphism on the orthogonal complement of the $\vm^{p'}$ for $p' < p$). Then, since $\epsilon'' = 1$,  $L(\Bv^{p})$ is also an eigenvectors of $\ell$ with the same eigenvalue, the associated vector $\vm^{p} = c (v^{p} + L(\Bv^{p}))$ or $\vm^{p} = i v^{p}$ is an eigenvector of $\ell$ with eigenvalue $s^{p}$ (and, as in the proof of Prop.~\ref{prop basis odd case}, they are two by two orthogonal). So the orthonormal basis $\{ \vm^{p} \}_{1 \leq p \leq \mu}$ is a basis of eigenvectors of $\ell$.
	\medskip
	\par
	Secondly, suppose $\epsilon'' = -1$ ($KO$-dimension $6$). Here, we cannot use the same argument as before since $L(\Bv)$ would be an eigenvector of $\ell$ with eigenvalue $-s$. Let $\vm^{1} \in \bbC^{\mu}$ be any eigenvector of $\ell$ of norm $1$ with eigenvalue $s^{1}$, and define $\vm^{2} \defeq L(\Bvm^{1})$, which is an eigenvector of $\ell$ with eigenvalue $s^{2} = -s^{1}$. Then $\langle \vm^{2}, \vm^{1} \rangle = \langle \ell(\vm^{2}), \ell(\vm^{1}) \rangle = - \langle \vm^{2}, \vm^{1} \rangle$ so that $\langle \vm^{2}, \vm^{1} \rangle = 0$ and therefore $\vm^{1}$ and $\vm^{2}$ are orthogonal. Consider now a vector $\vm^{3}$ of norm $1$ which is orthogonal to $\vm^{1}$ and $\vm^{2}$ and which is an eigenvector of $\ell$ with eigenvalue $s^{3}$. Then $\vm^{4} \defeq L(\Bvm^{3})$ is an eigenvector of $\ell$ with eigenvalue $s^{4} = -s^{3}$, which, by the previous computation, is orthogonal to $\vm^{3}$. It is also orthogonal to $\vm^{1}$: $\langle \vm^{4}, \vm^{1} \rangle = \langle L(\Bvm^{3}), \vm^{1} \rangle = \langle \BL(L(\Bvm^{3})), \BL(\vm^{1}) \rangle = \langle \Bvm^{3}, \Bvm^{2} \rangle = \overline{\langle \vm^{3}, \vm^{2} \rangle} = 0$ and so is also orthogonal to  $\vm^{2} = L(\Bvm^{1})$. Iterating this construction, one first shows that $\mu$ is even and one gets an orthonormal basis of $\bbC^{\mu}$ with eigenvectors of $\ell$ such that $\vm^{2a} = L(\Bvm^{2a-1})$ for any $a = 1, \dots, \mu/2$.
	\medskip
	\par
	In $KO$-dimensions $2$ and $6$, one has $\epsilon'' = -1$. At each step of the construction of the couples of basis vectors $\vm^{2a-1}$ and $\vm^{2a}$, one is such that $s^{p} = +1$ and the other $s^{p} = -1$, so that, up to a permutation $(\vm^{2a-1}, \vm^{2a}) \rightsquigarrow (\vm^{2a}, \epsilon \vm^{2a-1})$, one can force $s^{2a} = +1$ and $s^{2a-1} = -1$.
\end{proof}

We are now in position to use these results to decompose in a suitable way the Hilbert space $\hs$ into irreps. We already know that $\hs = \toplus_{{i}, {j}=1}^{r} \hhs_{\bn_i \bn_j}$ and that $\hhs_{\bn_i \bn_j} = \bbC^{n_{i}} \otimes \bbC^{\mu_{i j}} \otimes \bbC^{n_{j} \circ}$. 
\medskip
\par
Consider first the odd case. For any $1 \leq i, j \leq r$, define the set $\Gamma^{(0)}_{\bn_i \bn_j} \defeq \{ (i, p, j) \mid 1 \leq p \leq \mu_{i j} \}$, and for any $v = (i, p, j) \in \Gamma^{(0)}_{\bn_i \bn_j}$, define $\lambda, \rho : \Gamma^{(0)}_{\bn_i \bn_j} \to \Lambda$ as $\lambda(v) \defeq \bn_i$ and $\rho(v) \defeq \bn_j$. Notice that $\mu_{i j} = \# \Gamma^{(0)}_{\bn_i \bn_j}$. Define $\Jim : \Gamma^{(0)}_{\bn_i \bn_j} \to \Gamma^{(0)}_{\bn_j \bn_i}$ as $\Jim(v) \defeq (j, p, i)$ for any $v = (i, p, j)$. Using the orthonormal basis $\{ \vm_{i j}^{p} \}_{1 \leq p \leq \mu_{i j}}$ of $\bbC^{\mu_{i j}}$ given in Prop.~\ref{prop basis odd case}, let us define 
\begin{align}
	\label{hsv from vmijp basis}
	\hs_v \defeq \Span \{ \xi_{i} \otimes \vm_{i j}^{p} \otimes \eta_{j}^\circ \mid \xi_{i} \in \bbC^{n_{i}} \text{ and }  \eta_{j}^\circ \in \bbC^{n_{j} \circ} \} \simeq \hs_{\bn_i \bn_j}
\end{align}
Then, for $i<j$, $J(\xi_{i} \otimes \vm_{i j}^{p} \otimes \eta_{j}^\circ) = \Beta_{j} \otimes L_{i j}(\Bvm_{i j}^{p}) \otimes \Bxi_{i}^\circ = \Beta_{j} \otimes \vm_{j i}^{p} \otimes \Bxi_{i}^\circ \in \hs_{\Jim(v)}$ for any $\xi_{i} \in \bbC^{n_i}$, $\eta_{j}^\circ \in \bbC^{n_j \circ}$, and $v \in \Gamma^{(0)}_{\bn_i \bn_j}$, while $J(\xi_{j} \otimes \vm_{j i}^{p} \otimes \eta_{i}^\circ) = \Beta_{i} \otimes L_{j i}(\Bvm_{j i}^{p}) \otimes \Bxi_{j}^\circ = \epsilon\, \Beta_{i} \otimes \vm_{i j}^{p} \otimes \Bxi_{j}^\circ \in \hs_{\Jim(v)}$ for any $\xi_{j} \in \bbC^{n_j}$, $\eta_{i}^\circ \in \bbC^{n_i \circ}$ and $v \in \Gamma^{(0)}_{\bn_j \bn_i}$. 
In the same way, for $i = j$ and $\epsilon = 1$, $J(\xi_{i} \otimes \vm_{i}^{p} \otimes \eta_{i}^\circ) = \Beta_{i} \otimes \vm_{i}^{p} \otimes \Bxi_{i}^\circ \in \hs_{\Jim(v)}$ where $\Jim(v) = v \in \Gamma^{(0)}_{\bn_i \bn_i}$, and for $\epsilon = -1$, with $v = (i, 2a -1, i)$ for any $a = 1, \dots, \mu_{ii}/2$, $J(\xi_{i} \otimes \vm_{i}^{2a-1} \otimes \eta_{i}^\circ) = \Beta_{i} \otimes \vm_{i}^{2a} \otimes \Bxi_{i}^\circ \in \hs_{\Jim(v)}$ where $\Jim(v)  = (i, 2a, i) \in \Gamma^{(0)}_{\bn_i \bn_i}$. Since $J^2 = \epsilon$, the maps $\Jim : \Gamma^{(0)}_{\bn_i \bn_j} \to \Gamma^{(0)}_{\bn_j \bn_i}$ induce an involution on 
\begin{align*}
	\Gamma^{(0)} \defeq \cup_{i,j=1}^{r} \Gamma^{(0)}_{\bn_i \bn_j}
\end{align*}
with the property $\lambda \circ \Jim = \rho$ (and so $\rho \circ \Jim = \lambda$), where $\lambda, \rho : \Gamma^{(0)} \to \Lambda$ are defined in an obvious way. This involution encodes some properties of the family of maps $L_{i j}$, and so of the map $J : \hs_v \to \hs_{\Jim(v)}$ for any $v \in \Gamma^{(0)}$. We have also an orthogonal decomposition of $\hs$ along irreps:
\begin{align*}
	\hs = \toplus_{v \in \Gamma^{(0)}} \hs_v
\end{align*}
To fully reconstruct $J$, one needs to keep track of the parity of $p$ for $i=j$ and $KO$-dimensions $2$, $3$, $4$, or $5$ ($\epsilon = -1$): this is done below.
\medskip
\par
Knowing the $KO$-dimension (in particular the value of $\epsilon$) and $\Jim$, one can reconstruct the map $J$, up to a unitary on $\hs$ defining the explicit identifications $\hs_v = \hs_{\bn_i \bn_j}$ for $(\bn_i, \bn_j) = (\lambda(v), \rho(v))$, since then, for any $\xi_{i} \otimes \eta_{j}^\circ \in \hs_v$, one has $J(\xi_{i} \otimes \eta_{j}^\circ) = \Beta_{j} \otimes \Bxi_{i}^\circ \in \hs_{\Jim(v)}$ for $i < j$ and $J(\xi_{i} \otimes \eta_{j}^\circ) = \epsilon\, \Beta_{j} \otimes \Bxi_{i}^\circ \in \hs_{\Jim(v)}$ for $i > j$. For $i=j$ and $KO$-dimensions $1$ and $7$, one has $J(\xi_{i} \otimes \eta_{i}^\circ) = \Beta_{i} \otimes \Bxi_{i}^\circ$.
\medskip
\par
In the even case, one can use in the same way the orthonormal basis given in Prop.~\ref{prop basis even case}. Then one obtains the same results concerning the map $J$, in particular the existence of $\Jim$ with the same properties and the reconstruction formulas for $J$ for $i \neq j$ and $i=j$ in $KO$-dimension $0$. But, in addition, by Prop.~\ref{prop gamma to ell}, the basis in Prop.~\ref{prop basis even case} are composed of eigenvectors of $\gamma$, and by construction, $\gamma$ is then just the multiplication by $\pm 1$ on every $\hs_v$. As before, we define a grading decoration of $v$ as $s(v)  = \pm 1$, which is the eigenvalue of the associated eigenvector. Notice then that $s \circ \Jim = \epsilon'' s$ as can be checked in Prop.~\ref{prop basis even case}. The grading decoration $s$ fully determines $\gamma$.
\medskip
\par
In $KO$-dimensions $2$, $3$, $4$, or $5$, we have to take into account the parity of $p$ when $i=j$. We define a parity decoration of $v$ by $\chi(v) = 0$ if $p$ is odd and $\chi(v) = 1$ if $p$ is even. Then for $i=j$ and $KO$-dimensions $2$, $3$, $4$, $5$, or $6$, one has $J(\xi_{i} \otimes \eta_{i}^\circ) = \epsilon^{\chi(v)} \Beta_{i} \otimes \Bxi_{i}^\circ$ (since $\epsilon = 1$ in $KO$-dimension $6$, this relation holds also in that case). In all these $KO$-dimensions, $\mu_{ii}$ is even and since $\chi(\Jim(v)) = 1- \chi(v)$, half of the $v$ in $ \Gamma^{(0)}_{\bn_i \bn_i}$ are decorated by $0$ (resp. $1$).
\medskip
\par
The Dirac operator decomposes along the orthogonal subspaces $\hs_v$ as $D_e : \hs_{v_1} \to \hs_{v_2}$ where we define $e \defeq (v_1, v_2) \in  \Gamma^{(0)} \times  \Gamma^{(0)}$. With $\Be \defeq (v_2, v_1)$, $D^\dagger = D$ is equivalent to $D_{\Be} = D_e^\dagger$. Moreover, let $(\bn_{i_k}, \bn_{j_k}) = (\lambda(v_k), \rho(v_k))$ for $k=1,2$, then the first-order condition, written with $a = \toplus_{i=1}^{r} \lambda_i \bbbone_{n_{i}}$ and $b = \toplus_{i=1}^{r} \mu_i \bbbone_{n_i}$, implies $(\lambda_{i_1} - \lambda_{i_2}) (\mu_{j_1} - \mu_{j_2}) D_e = 0$, so that $D_e = 0$ when $i_1 \neq i_2$ and $j_1 \neq j_2$, or, equivalently, $D_e$ can be non-zero only when $i_1 = i_2$ or $j_1 = j_2$. With $b$ as before and any $a$, one gets $[a_{i_1}, D_e] = 0$ in the situation $i_1 = i_2$ and $j_1 \neq j_2$, and $[a_{j_1}^\circ, D_e] = 0$ in the situation $i_1 \neq i_2$ and $j_1 = j_2$. Using the same arguments as in the proof of Prop.~\ref{prop J K L}, one gets that $D_e$ reduces to a linear map $D_{R,e} : \bbC^{n_{j_1}\circ} \to \bbC^{n_{j_2}\circ}$ when $i_1 = i_2$ and $j_1 \neq j_2$ as $D_e = \bbbone_{n_{i_1}} \otimes D_{R,e}$ and to a linear map $D_{L,e} : \bbC^{n_{i_1}} \to \bbC^{n_{i_2}}$ when $i_1 \neq i_2$ and $j_1 = j_2$ as $D_e = D_{L,e} \otimes \bbbone_{n_{j_1}}$. When $i_1 = i_2$ and $j_1 = j_2$, nothing general can be said about $D_e$. Finally, define $\Jim(e) \defeq (\Jim(v_1), \Jim(v_2))$, then the relation $J D = \epsilon' D J$ implies that $D_{e}$  and $D_{\Jim(e)}$ are related by $J$ and $\epsilon'$ (an explicit expression is given below). In particular, they are both zero or non-zero at the same time. As before, in the even case, the relation $\gamma D = - D \gamma$ implies that $D_{e}$ is non-zero only when $s(v_2) = - s(v_1)$.
\medskip
\par
Let us abstract the construction using a decorated graph $\Gamma$, together with $\Lambda$ and the $KO$-dimension $d$. 
\begin{enumerate}
	\item The set of vertex $\Gamma^{(0)}$ of the graph is equipped with two maps $\lambda, \rho : \Gamma^{(0)} \to \Lambda \times \Lambda$, that we write as a single map $\pi_{\lambda\rho} \defeq \lambda \times \rho$, and define $i(v) \defeq i$ and $j(v) \defeq j$ for $\pi_{\lambda\rho}(v) = (\bn_i, \bn_j)$.
	
	\item There is an involution $\Jim : \Gamma^{(0)} \to \Gamma^{(0)}$ such that $\lambda \circ \Jim = \rho$ and such that $\Jim(v) = v$ when $\lambda(v) = \rho(v)$ in $KO$-dimensions $0$, $1$, and $7$. 
	
	\item For any vertex $v \in \Gamma^{(0)}$ with $\pi_{\lambda\rho}(v) = (\bn_i, \bn_j)$, define $\hs_v \defeq \hs_{\lambda(v) \rho(v)} = \bbC^{\lambda(v)} \otimes \bbC^{\rho(v) \circ} = \bbC^{n_{i}} \otimes \bbC^{n_{j} \circ}$. The element $(\bn_i, \bn_j) \in \Lambda \times \Lambda$ is a decoration of the vertex $v$. 
	
	\item Define the set $\Gamma^{(0)}_{\bn_i \bn_j} \defeq \{ v \in \Gamma^{(0)} \mid \pi_{\lambda\rho}(v) = (\bn_i, \bn_j) \} = \pi_{\lambda\rho}^{-1}(\bn_i, \bn_j)$ and $\mu_{ij} \defeq \# \Gamma^{(0)}_{\bn_i \bn_j}$.
	
	\item  Define the map $\hJim_{v} : \hs_v \to \hs_{\Jim(v)}$ as $\hJim_{v}(\xi^{(v)} \otimes \eta^{(v)\circ}) = \eta^{(v)} \otimes \xi^{(v)\circ}$ for any $\xi^{(v)} \in \bbC^{\lambda(v)}$ and $\eta^{(v)\circ} \in \bbC^{\rho(v)\circ}$. Notice that $\hJim_{\Jim(v)} \circ \hJim_{v} = \Id_{\hs_v}$.
	
	\item If the $KO$-dimension is even, a second decoration of each vertex is the assignment of a grading map $s(v) = \pm 1$ such that $s \circ \Jim = \epsilon'' s$.
	
	\item If the $KO$-dimension is $2$, $3$, $4$, $5$, or $6$, then $\mu_{ii}$ is even and another decoration of each vertex $v \in\Gamma^{(0)}_{\bn_i \bn_i}$  is the parity $\chi(v) = 0, 1$ such that $\chi(\Jim(v)) = 1 - \chi(v)$, so that half of the vertices in $\Gamma^{(0)}_{\bn_i \bn_i}$ are decorated by the value $0$ or $1$.
	
	\item For any $v \in \Gamma^{(0)}$, define
	\begin{align}
		\label{eq epsilon(v, d)}
		\epsilon(v, d) \defeq 
		\begin{cases}
			1 & \text{for $i(v) < j(v)$},\\
			\epsilon & \text{for $i(v) > j(v)$},\\
			1 & \text{for $i(v) = j(v)$ and $d = 0, 1, 7$},\\
			\epsilon^{\chi(v)} & \text{for $i(v) = j(v)$ and $d = 2, 3, 4, 5, 6$.}
		\end{cases}
	\end{align}
	One can check that $\epsilon(v, d) \epsilon(\Jim(v), d) = \epsilon$ for any $v \in \Gamma^{(0)}$.
	
	\item For every $e = (v_1, v_2) \in \Gamma^{(0)} \times \Gamma^{(0)}$, let $\Be \defeq (v_2, v_1)$ and $\Jim(e) \defeq (\Jim(v_1), \Jim(v_2))$.
	
	\item The space $\Gamma^{(1)} \subset \Gamma^{(0)} \times \Gamma^{(0)}$ of edges of the graph are couples $e = (v_1, v_2)$ such that:
	\begin{enumerate}
		\item $\lambda(v_1) = \lambda(v_2)$ or $\rho(v_1) = \rho(v_2)$ (or both);
		\item $s(v_2) = - s(v_1)$ in the even case;
		\item  there is a non-zero linear map $D_e : \hs_{v_1} \to \hs_{v_2}$ such that:
		\begin{enumerate}
			\item $D_{\Be} = D_e^\dagger : \hs_{v_2} \to \hs_{v_1}$;
			
			\item $D_{\Jim(e)} = \epsilon' \epsilon(v_1, d) \epsilon(v_2, d)\, \hJim_{v_2} J_0 D_e J_0 \hJim_{\Jim(v_1)} : \hs_{\Jim(v_1)} \to \hs_{\Jim(v_2)}$;
			\label{item DJim(e)}
			
			\item For $\lambda(v_1) = \lambda(v_2)$ and $\rho(v_1) \neq \rho(v_2)$, $D_e = \bbbone_{n_{i_1}} \otimes D_{R,e}$ with $D_{R,e} : \bbC^{n_{j_1}\circ} \to \bbC^{n_{j_2}\circ}$; \label{item De = 1 DRe}
			
			\item For $\lambda(v_1) \neq \lambda(v_2)$ and $\rho(v_1) = \rho(v_2)$,  $D_e = D_{L,e} \otimes \bbbone_{n_{j_1}}$ with  $D_{L,e} : \bbC^{n_{i_1}} \to \bbC^{n_{i_2}}$. \label{item De = DLe 1}
		\end{enumerate}
		Then $D_e$ defines a decoration of $e$.
		
	\end{enumerate}
\end{enumerate}

\medskip
For any $\xi_{i} \otimes \eta_{j}^\circ \in \hs_{v_1}$, it is convenient to write $D_e (\xi_{i_1} \otimes \eta_{j_1}^\circ) = D_{L,e}^{(1)} \xi_{i_1} \otimes D_{R,e}^{(2)} \eta_{j_1}^\circ$ as a sumless Sweedler-like notation\footnote{This notation is usual for computations on coalgebras.} where there is an implicit summation over finite families of operators $D_{L,e}^{(1)} : \bbC^{n_{i_1}} \to \bbC^{n_{i_2}}$ and $D_{R,e}^{(2)} : \bbC^{n_{j_1}\circ} \to \bbC^{n_{j_2}\circ}$. In the previous points~\ref{item De = 1 DRe} and \ref{item De = DLe 1}, this decomposition is explicitly given (summation reduced to a single term).
\medskip
\par
\begin{figure}[h]
	{\centering \includegraphics[]{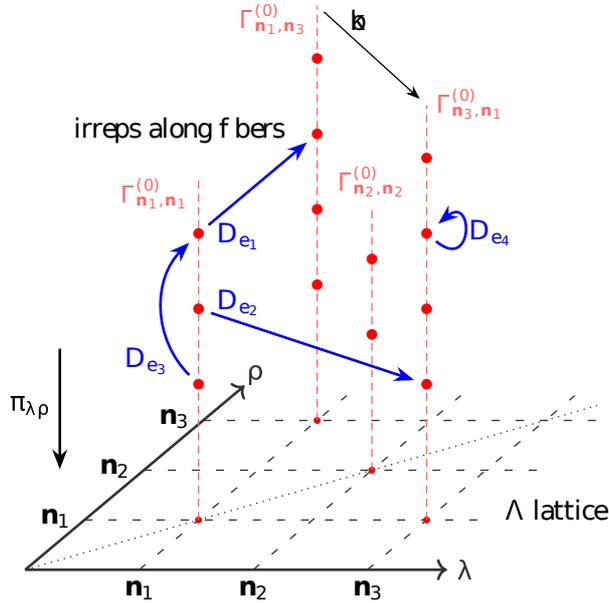}\par}
	\caption{A Krajewski diagram for the algebra $M_{n_1} \oplus M_{n_2} \oplus M_{n_3}$ in the $3$-dimensional representation explained in the text. The component $D_{e_1}$ (resp. $D_{e_2}$) of the Dirac operator joins two irreps with same $\lambda$ (resp. same $\rho$); $D_{e_3}$ and $D_{e_4}$ join irreps with the same $\lambda$ and $\rho$, but $D_{e_4}$ can be non-zero only in the odd case. The maps $\hJim$ realize the axial symmetry defined by the dotted line in the $\Lambda$ lattice.}
	\label{fig krajewskiDiagram}
\end{figure}

One can see $\Gamma^{(0)}$ as a set of points on top of the points $\Lambda \times \Lambda$, where the (down) projection is $\pi_{\lambda\rho}$. Each point in $\Gamma^{(0)}_{\bn_i \bn_j} = \pi_{\lambda\rho}^{-1}(\bn_i, \bn_j)$ is a copy of the irrep $\hs_{\bn_i \bn_j}$. So, instead of considering $(\bn_i, \bn_j)$ as a decoration of $\hs_v$, one could treat $v$ as an element of the “fiber” $\Gamma^{(0)}_{\bn_i \bn_j}$ on top of $(\bn_i, \bn_j)$. The edges in $\Gamma^{(1)}$, once projected in $\Lambda \times \Lambda$, connect points horizontally, vertically  or self-connect a (projected) point. A convenient representation of $\Gamma$ is then a $3$-dimensional set of points decorated by some values (as seen above) and linked by decorated lines, see Fig.~\ref{fig krajewskiDiagram}.
\medskip
\par

These data determine a real (odd or even) spectral triple. As explained in Sect.~\ref{subsec finite spectral triples}, the algebra $\algA$ is determined by $\Lambda$. A vertex $v \in \Gamma^{(0)}$ defines the irrep $\hs_v$ with multiplicity $\mu(v) \defeq \# \Gamma^{(0)}_{\pi_{\lambda\rho}(v)}$, so the Hilbert space is $\hs \defeq \toplus_{v \in \Gamma^{(0)}} \hs_v = \toplus_{i,j=1}^{r} \hhs_{\bn_i \bn_j}$ with $\hhs_{\bn_i \bn_j} = \bbC^{n_i} \otimes \bbC^{\mu_{i j}} \otimes \bbC^{n_j \circ}$.  Any operator $A$ on $\hs$ decomposes into linear maps $A_{v_2}^{v_1} : \hs_{v_1} \to \hs_{v_2}$.
\medskip
\par
As before, we can describe the representation $\pi$ along these two decompositions. For any $a = \toplus_{i=1}^{r} a_i \in \algA$, any $v=(i,p,j)$, and any $\psi = \toplus_{v \in \Gamma^{(0)}} \psi_v = \toplus_{i,j=1}^{r} \xi_{i} \otimes \vm_{i j}^{p} \otimes \eta_{j}^\circ$ with $\psi_v \in \hs_v = \bbC^{\lambda(v)} \otimes \bbC^{\rho(v) \circ}$ and $\xi_{i} \otimes \vm_{i j}^{p} \otimes \eta_{j}^\circ \in \bbC^{n_i} \otimes \bbC^{\mu_{i j}} \otimes \bbC^{n_j \circ}$, one has $\pi(a) \psi = \toplus_{v \in \Gamma^{(0)}} a_{i(v)} \psi_v = \toplus_{i, j=1}^{r} (a_i \xi_{i}) \otimes \vm_{i j}^{p} \otimes \eta_{j}^\circ$ where $a_{i(v)} \psi_v$ is the multiplication of the matrix $a_{i(v)}$ on the left factor of $\bbC^{\lambda(v)} \otimes \bbC^{\rho(v) \circ}$ and $a_i \xi_{i}$ is the usual matrix multiplication on $\bbC^{n_i}$. In other words, the decomposition of the operator $\pi(a)$ along the $\hs_v$'s is
\begin{align}
	\label{eq pi(a) decomposition along Hv}
	\pi(a)_{v_2}^{v_1} = a_{i(v_1)} \delta_{v_2}^{v_1} : \hs_{v_1} \to \hs_{v_2}
\end{align} 
(where $\delta_{v_2}^{v_1}$ is the Kronecker symbol). In the real case, for any $a = \toplus_{i=1}^{r} a_i \in \algA$, any $b = \toplus_{j=1}^{r} b_j \in \algA$, and any $\psi_v \in \hs_v$, one has $a b^\circ \psi_v = a_{i(v)} b_{j(v)}^\circ \psi_v \in \hs_v$ ($\pi$ omitted) where $b_{j(v)}^\circ$ acts on the right factor of $\bbC^{\lambda(v)} \otimes \bbC^{\rho(v) \circ}$ (see footnote~\ref{footnote circ action}). A similar relation holds on $\bbC^{n_i} \otimes \bbC^{\mu_{i j}} \otimes \bbC^{n_j \circ}$.
\medskip
\par
In the even case, $\gamma$ is determined as the multiplication by the decoration $s(v) = \pm 1$ on $\hs_{v}$. The real operator $J$ is reconstructed by the family of maps
\begin{align*}
	J_v \defeq \epsilon(v, d)\, J_0 \hJim_{v} = \epsilon(v, d)\, \hJim_{v} J_0 : \hs_v \to \hs_{\Jim(v)}
\end{align*}
or, equivalently, with $i = i(v)$ and $j = j(v)$, and any $\xi_{i} \otimes \eta_{j}^\circ \in \hs_v$,
\begin{align*}
	J(\xi_{i} \otimes \eta_{j}^\circ) = \epsilon(v, d)\, \Beta_{j} \otimes \Bxi_{i}^\circ \in \hs_{\Jim(v)}.
\end{align*}
In other words, $J_{v_2}^{v_1} = \epsilon(v_1, d)\delta_{\Jim(v_1)}^{v_1} \, J_0 \hJim_{v_1}$.
\medskip
\par
The Dirac operator is reconstructed by the decorations $D_{e}$ of the edges $e \in \Gamma^{(1)}$. All the properties of a (even) real spectral triple are encoded in the relations required above. One has $J_{\Jim(v)} J_{v} = \epsilon(\Jim(v), d) \epsilon(v, d) \hJim_{\Jim(v)} \hJim_{v} J_0^2 = \epsilon$. In the same way, $D_{\Jim(e)} J_{v_1} = \epsilon' \epsilon(v_1, d) \epsilon(v_2, d) \, \hJim_{v_2} J_0 D_{e} J_0  \hJim_{\Jim(v_1)} ( \epsilon(v_1, d)\, \hJim_{v_1} J_0) = \epsilon' \epsilon(v_2, d) \, \hJim_{v_2} J_0 D_{e}$ while $J_{v_2} D_e = (\epsilon(v_2, d)\, J_0 \hJim_{v_2}) D_e$, so that $J D = \epsilon' D J$. Finally, $J_v \gamma_v = s(v) \epsilon(v, d) J_0 \hJim_{v}$ while $\gamma_{\Jim(v)} J_{v} = s(\Jim(v)) \epsilon(v, d) J_0 \hJim_{v}$ so that $J \gamma = \epsilon'' \gamma J$.
\medskip
\par
Let us describe the Dirac operator for the decomposition $\hs = \toplus_{i,j=1}^{r} \hhs_{\bn_i \bn_j}$. Introduce an orthonormal basis for each $\bbC^{\mu_{ij}}$ and label all these basis vectors in the union of all the $\bbC^{\mu_{ij}}$'s as $\{ \vm_v \}_{v \in \Gamma^{(0)}}$: for any $v \in \Gamma^{(0)}_{\bn_i \bn_j}$, $\vm_v$ is an element of an orthonormal basis of $\bbC^{\mu_{ij}}$. We use the identification $\hs_v = \Span \{ \xi \otimes \vm_v \otimes \eta^\circ \mid \xi \in \bbC^{\lambda(v)}, \eta^\circ \in \bbC^{\mu(v) \circ} \}$. For any $e = (v_1, v_2) \in \Gamma^{(1)}$ with $v_1 \in \Gamma^{(0)}_{\bn_{i_1} \bn_{j_1}}$ and $v_2 \in \Gamma^{(0)}_{\bn_{i_2} \bn_{j_2}}$,  define $\hD_{e} : \hhs_{\bn_{i_1} \bn_{j_1}} \to \hhs_{\bn_{i_2} \bn_{j_2}}$, for any $\xi \in \bbC^{n_{i_1}}$ and $\eta^\circ \in \bbC^{n_{j_1}\circ}$, as
\begin{align*}
	\hD_{e} (\xi \otimes \vm_{v} \otimes \eta^\circ)
	&=
	\begin{cases}
		0 & \text{if $v \neq v_1$} \\
		( D_{L,e}^{(1)} \xi) \otimes \vm_{v_2} \otimes (D_{R,e}^{(2)} \eta^\circ) & \text{if $v = v_1$}
	\end{cases}
\end{align*}
Then define 
\begin{align*}
	D_{i_2 j_2}^{i_1 j_1} \defeq 
	\sum_{\substack{e = (v_1, v_2) \in \Gamma^{(1)} \\ v_1 \in \Gamma^{(0)}_{\bn_{i_1} \bn_{j_1}}, v_2 \in \Gamma^{(0)}_{\bn_{i_2} \bn_{j_2}} }} \hD_{e}
	: \hhs_{\bn_{i_1} \bn_{j_1}} \to \hhs_{\bn_{i_2} \bn_{j_2}}
\end{align*}
The operators $D_{i_2 j_2}^{i_1 j_1}$ can be collected in an operator $D : \hs \to \hs$ which is self-adjoint.
\medskip
\par
One can write a specific version of \eqref{eq piD sum algebras} for the decomposition $\hs = \toplus_{v \in \Gamma^{(0)}} \hs_v$ in terms of the operators $D_e$ for $e \in \Gamma^{(1)}$. For any $\bomega \in \bOmega^n_U(\algA) \subset \kT^n \algA$ which decomposes along a sum of typical terms $\toplus_{i_1, \dots, i_{n-1} =1}^{r} \big( a^0_{i} \otimes a^1_{i_1} \otimes \cdots \otimes a^{n-1}_{i_{n-1}} \otimes a^n_{j} \big)_{i, j = 1}^{r} \in \kT^n \algA$ and any $v_0, v_n \in \Gamma^{(0)}$, one has
\begin{multline}
	\label{eq piD AF algebras}
	\piD(\bomega)_{v_0}^{v_n} = 
	\!\!\!\!\!\!\!\sum_{ \substack{ \text{all terms at the}\\ \text{$(i(v_0), i(v_n))$ entry in $\bomega$} } } \!\!\!\!\!\!\!
	\tsum_{v_1, \dots, v_{n-1} \in \Gamma^{(0)}}
	a^{0}_{i(v_0)} D_{(v_1, v_0)} a^{1}_{i(v_1)}  D_{(v_2, v_1)} \cdots
	\\
	\cdots a^{n-1}_{i(v_{n-1})} D_{(v_n, v_{n-1})} a^{n}_{i(v_n)} : 
	\hs_{v_n} \to \hs_{v_0}
\end{multline}
In this formula, one supposes $D_{(v_{i+1}, v_i)} = 0$ when $(v_{i+1}, v_i) \not\in \Gamma^{(1)}$.
\medskip
\par
A complete basis for any $i,j = 1, \dots, r$, let $\{ \vm_{i j}^{p} \}_{1 \leq p \leq \mu_{i j}}$ be an orthonormal basis of $\bbC^{\mu_{i j}}$ (for instance as in Prop.~\ref{prop basis odd case} or \ref{prop basis even case}), to which we associate the irreps $\hs_v$ defined as in \eqref{hsv from vmijp basis} for any $v = (i, p, j) \in \Gammasub[\bn_i \bn_j]^{(0)}$. One can then fix an orthonormal basis $\{ e_{ij, \alpha} = \xi_{i, \alpha}^{(1)} \otimes \eta_{j, \alpha}^{\circ (2)} \}_{1 \leq \alpha \leq n_i n_j}$ (sumless Sweedler-like notation) of $\bbC^{n_i} \otimes \bbC^{n_j \circ}$. Let
\begin{align*}
	\TGammasub[\bn_i \bn_j]^{(0)}
	&\defeq \Gammasub[\bn_i \bn_j]^{(0)} \times \{ 1, \dots, n_i n_j \}
	& \text {and}&&
	\TGamma^{(0)} &\defeq \cup_{i,j= 1}^{r} \TGammasub[\bn_i \bn_j]^{(0)}
\end{align*}
Then for any $\Tv = (v, \alpha) \in \TGammasub[\bn_i \bn_j]^{(0)}$, let $e_{\Tv} \defeq \xi_{i, \alpha}^{(1)} \otimes \vm_{i j}^{p} \otimes \eta_{j, \alpha}^{\circ (2)} \in \hs_v$. The family $\{ e_{\Tv} \}_{\Tv \in \TGamma^{(0)}}$ defines an orthonormal basis of $\hs$. We define $v : \TGammasub^{(0)} \to \Gamma^{(0)}$ as $v(\Tv) = v$ for $\Tv = (v, \alpha)$.

\part{Noncommutativity, Spacetime, and Noncommutative Geometry}
\label{part2ThinkingNCG}

\chapter{General Outline}
\label{GeneralOutline}
Now that some conceptual and mathematical grounds have been established to understand the origin of NCG and how it works on the mathematical side, let us try to see which meaning can be given to this noncommutativity, as well as for the possible NCG that could accompany it. Since this part is not essential to understanding the technical developments which follow and present assertions of a different nature than the other three parts, the reader can directly move on to part \ref{TowardNCGFT}.
\medskip
\par 
Welcome to those readers who have been willing to keep reading, before going any further, I would like to warn you that the ideas developed here are personal ideas, not developed in the literature for most of them, so there may not be a consensus about them. However, I have tried to make my statements as clear as possible, demonstrate them as much as possible, and support my statements with quotations to connect them to the literature and other author's thoughts, whenever possible.
\medskip
\par
\GN{eventuel intro}I believe that a full-fledged mathematical physicist is both someone who can understand and develop the mathematical structures that define our representations of reality, and make the theory clear at that level. But I think that she/he must also be someone who understands how the mathematical objects in the theories relate to observables, concrete physical objects, and how the intuitive/ontological picture we have of the world and the one we get from its actual mathematical representation go on together. These are not passive philosophical claims, because these clarifications are essential to guide the physicist's vision and work in a meaningful way. Many great scientists like Einstein, Bohr having done huge and fundamental contributions to physics are examples of researchers who elaborate theories using something outside the scope of pure mathematical thinking, with through experiments, philosophical principles about nature, intuitive feelings... That is why this part is for me a very important and needed part that can be considered as an attempt to complete the mathematical presentation made in the first part.
\medskip
\par
Attempting to give an ontology or an intuitive explanation of what NCG is, seems not to be an easy task. A partial attempt in this direction was done in \cite{huggett2018philosopher}. Intuitive objects such as the notion of a point no longer make sense. But can we think of anything outside our usual spatial representations? Can we think of a structure without assuming a set of points that supports our intuition? The NCG forces us to try to understand the origin of the notion of point which appears so precious to our minds. 
\medskip
\par
There is a certain line of interpretation of NCG, especially developed by A. Connes. These are mainly based on wave considerations, with the wave number $k$ eigenvalue of the Laplacian formed by the Dirac operator, with distance thought as linked with wave nature for example. This branch of interpretation happens to be based on a formalism that reveals part of reality, and which is formally equivalent to matrix mechanics as Shrödinger showed. In what follows, I will try to highlight the contributions that this dual approach of matrix mechanics can provide to the interpretation of NCG. 
\medskip
\par
NCG historically comes from the formalism used to describe QM, operator algebras acting on Hilbert spaces through the Gelfand-Naimark theorem. The NC of observables in QM is seen as a kind of validation of the potential relevance of the use of NCGs in physics. Usual NCGs are thought to correspond to the notion of a “space” underlying an algebra of NC observables considered as continuous functions on this “space”. Then, if there is an interpretation for the NC of observables in QM, this can lead to an interpretation for the corresponding NCG. There are two ways of looking at the QM formalism, more precisely concerning the $\psi$ state. These two are the epistemic view and the ontic view. The first one postulates that $\psi$ is not relative to an objective physical entity, but to a potential knowledge, the second one says that $\psi$ corresponds to a real physical state of nature. If QM is an epistemic theory, then there are no objective geometrical consequences to be determined from its formalism, because it does not refer to anything substantial. But if the theory is ontic, then (via Gelfand Naimark theorem and the equivalence of categories between algebra and geometry highlighted in the first part) there must be geometric consequences. This is why I will often talk about QM and the ways of interpreting it, to make sense of what I will say in section \ref{NCObsEf} and chapter \ref{ObjectiveNCGFromQM}. We will see that the interpretation we have of the state (done in section \ref{Realism}) permits us to deduce the interpretation of the NC in QM (done in \ref{NCObsEf}), and finally to deduce an interpretation of the deduced NCG (in chapter \ref{ObjectiveNCGFromQM}). This will be supported by the physical arguments given in chapter \ref{GeomAndSpacetime} inducing that only observables can lead to the space-time's geometric representation and that these observables must therefore be taken to be more fundamental than space-time for the experimental physicist and thinker.
\medskip
\par
First of all, in chapter \ref{NCShadow} we will have a general discussion about what we mean by NC, NCG, and why NCG remains largely unknown outside. I will then define some important notions on which the following statements will be based. To introduce some useful terminology, we will first define what is a physical theory in chapter \ref{PhysTh}, and what are the elements of reality. This will be necessary for the sections \ref{Realism} and \ref{GeomAndST} where more advanced considerations will come into play in the discussion. Then, I will give arguments to support these different points:
\begin{enumerate}[label=\upshape(\Roman*),ref=(\Roman*)]
	\item\label{CstarUnif} The formalism of $C^*$-algebras acting on Hilbert spaces encompasses both Classical and QM, thus offering a framework to think about their unification (section \ref{KvN}).
	\item\label{QMont} There are strong indications that the QM formalism refers directly to ontic states, and should therefore be considered seriously as reflecting something objective \textit{i.e.} this is not just an algorithm that leads to good predictions (section \ref{Realism}).
	\item\label{KnowChang} That point \ref{QMont} induces that NC is linked to non-passivity of the measurement process, and therefore to contextual state change (section \ref{NCObsEf}).
	\item\label{ObsFirstST} Observables are first to the notion of space-time. The space-time must not be thought of otherwise than as a deduction of the observables that we collect (section \ref{GeomAndST}).
	\item\label{STNotObjective} Giving objective existence to space-time structures seems to be misleading (section \ref{PointOnPoint}).
	\item\label{AlgebraizationOfGeom} Geometry can be entirely algebraized, this procedure culminating with the Gelfand-Naimark theorem and the birth of NCG. Leading to the fact that the structures of the algebra of observables and space-time contain just as much information, and are equivalent (sections \ref{BackToObs} and \ref{AlgeGeom}). 
	\item\label{QMToGeom} Because the formalism and observables of QM reflect something objective according to point \ref{QMont}, observables are prior to space-time structures trough points \ref{ObsFirstST}, \ref{STNotObjective} and \ref{AlgebraizationOfGeom}. We then have to understand what are the geometric consequences behind this formalism. The first one concerns the origin of the notion of point, the second one of connectedness, and the last one is about a length scale at which  the notion of point no longer stands, opening the door to NCG (chapter \ref{ObjectiveNCGFromQM}). 
	\item\label{QMToNCG} Considering such a $\psi$-ontic interpretation, using \ref{KnowChang} for the consequence on the interpretation of the NC of observables, this implies that a natural interpretation of the NCG can be that of a measured geometry, this measurement process being non-passive, changing the underlying notion of “point” coming from the pure state undergoing change (section \ref{PhiloNCG}).
\end{enumerate} 
This can be viewed as three blocks, the first one containing points \ref{CstarUnif}, \ref{QMont}, and \ref{KnowChang} is about the ontic meaning of the formalism of QM and the interpretation of the NC. The second includes points \ref{ObsFirstST}, \ref{STNotObjective} and \ref{AlgebraizationOfGeom} defends the fact that observables are prior to space-time representation, and that they contain all the information we need. The last one with points \ref{QMToGeom} and \ref{QMToNCG} proposes to set up the consequences of the assumption of the first two blocks, when their respective implications are put together.

\chapter{Why NCG remains in the Shadow for Physicists?}
\label{NCShadow}
It is surprising to note how unknown NCG is to researchers in theoretical physics, and to what extent the minimum basis of understanding of the subject remains out of reach for many of them. Indeed, it is not easy to enter NCG, the language used is very limited to the mathematical one, and few efforts are made to go deeper in the understanding. The specification of the supposed kind of NC with which NCG deals with, is not clarified anywhere to my knowledge (which I will try to do in section \ref{WhichNC} and afterward). Even worse, the terms chosen to talk about NCG are often extensions of the terms of usual geometry, which have no more sense in this new context, as we will see in section \ref{PBWorld}. What is a NCG? How to represent it? And why is it a relevant extension of the geometric framework? Few simple answers are given to these elementary questions. It is not enough to say that two observables do not commute and that mathematically this is written $[a,b]\neq 0$ to have a complete intuition of what NCG is. One of the main problems in giving an intuition of the NCG lies in the notion of point, which loses all meaning in NCG. Yet, this notion is the support of our intuition of “tangible” things, of our representations of reality. So how do we go without it?

\section{Are Sets and their Elements Primary Objects?}
\label{AreSetsPrimary} 
\epigraph{I know that the great Hilbert said “We will not be driven out of the paradise Cantor has created for us” and I reply “I see no reason for walking in!”}{\textit{R. Hamming}}

Set theory is currently regarded as a foundation of mathematics. Most fields of mathematics are based on the notion of a set of objects whose relations and operations are studied. The elements can be called points, vectors, functions… This way of proceeding, by identifying objects, and looking at the notions they have between them, is embedded in the innate functioning of our minds and provides us with a grid for reading the world, through different systems of representations. When it comes to representations aimed at describing the “objective” phenomena of the physical world, we formulate physical theories whose symbolic objects are related to physical objects, phenomena and events. To describe our space-time, the representation is that of a space-time where the set is called space, and the elements are called points, then points events. The same is true for many physical theories, where the notions of set and element manifest themselves in different representations. But what is the proof that this reading grid offers a complete picture of reality?  
\medskip
\par
A hint of the insufficiency of this general framework of thinking is given by the non-commutativity of observables in QM. For instance, if we consider an operational way to define the elements of a set, according to the determination of their properties (the experimental way being the only one accessible to us), and that these properties no longer commute during this determination, it will no longer be possible to separate this set into sub-sets corresponding to these properties. Let's take a simple example to illustrate this. If we consider that our set is a box and its “elements” are particles that can have two shapes (square and spherical) corresponding to the observable $S$ for shape, and two colors (red and blue) for the observable $C$, then if $[C,S]\neq 0$, it will not be possible to partition this set into for example the set of red and spherical particles, the other set blue and square... This is not possible because these particles cannot have the status of set elements for the properties defined by $S$ and $C$ jointly, and thus, the procedure for creating sets by specifying the properties of their elements becomes meaningless, as we no longer obtain sets. \GN{eclaircir}
\medskip
\par
Thus, if the observables do not commute in QM, then we cannot classify “entities”  as elements of a set for these attributes taken together, as we only have access to observables, and we are not allowed to say more than what observables tell us. Thus as shown in section \ref{PhaseVSConfigSpace}, when partitioning the physical reality according to the two properties deduced from these observables, the notion of phase space will no longer be appropriate. This is intimately linked to the notions of realism/anti-realism, and therefore to the notion of the element of reality which will be presented in section \ref{Realism}. We will see, with the arguments of section \ref{GeomAndST} and chapter \ref{ObjectiveNCGFromQM} that for the portion of these observables that leads to geometrical representations of the world, the usual geometrical framework of space with points is no longer consistent. In this line of thought, another reason resulting from the formalism of QM, in disagreement with the notion of element and thus of point, will be presented in section \ref{EntangConnect}.

\section{Which NC are we talking about?}
\label{WhichNC}
What do we mean by non-commutativity? There is the answer of the mathematician, and the one of the physicist, which are in some way connected. 
\medskip 
\par 
For the mathematician, two notions of noncommutativities are encountered:
\begin{itemize}
	\item The one of the groups, we will say non-abelian for groups. Most often, the concept of group is accompanied by the concept of transformations of “spaces”. 
	\item The one of algebras, we will say non-commutative for algebras. There is not necessarily the idea of transformation.
\end{itemize}
As mentioned before, only NC of algebras is connected to NCG and more generally to NC mathematics. The one of group is usually thought as linked to transformation, and if we consider a non-abelian group whose algebra of function is commutative, then it will belong to commutative mathematics. This is important to note that no mention of NC coordinates was given as input to define what is NCG, we only talk about NC of observables.
\medskip
\par 
In the same way, the physicist encounters many non-commutativities in his everyday life. Some of them are thought to be normal (classical), and others to be not. But what distinguishes normal non-commutativities from those that could be attributed to an NCG for example? I haven't found any clarification on this important subject. I will try to give one in what follow. 
\medskip
\par 
Indeed, a question that naturally arises is: “of which non-commutativity are we talking about?” There are various kinds of non-commutativities, in the present state of affairs, when they correspond to physical theories, these can be classified into two categories:
\begin{enumerate}
	\item Process-like NCs in space and time (Type 1 NCs): NCs of rotations, irreversible processes, operations on solids, an example given by Connes being the operations of closing a bottle and filling it. These have extensions in space-time.
	\item NCs which are not attributed to processes in space and time (Type 2 NCs): they are perceived by the physicist as belonging to processes without any spatial and temporal extension. They are thought (by many of them) as intrinsic. These are the NC of observables in QM, of all quantized observables, or of non-abelian gauge theories. These do not have extensions in space-time.
\end{enumerate}
If this first category corresponds to classical intuition, it is to the second that the special character of the NCG is clearly attributed, because these observables can well be, within the framework of these models, considered as functions with values on “points” (in quantum mechanical models), thus without extension in space-time. We can see that the categories for the physicist are linked to the ones of the mathematician since group action is thought of as a kind of transformation in physics, and that algebras are connected to observables in physics, usually not thought of as transformation. We will see in sections \ref{Realism} and \ref{NCObsEf} that according to Bohr, the NC in QM does not represent a process in space and time. And that he is probably at the origin of this intuition concerning the NC of observables in QM, as not being seen as processes in space-time.
\medskip
\par 
We will see in section \ref{NCObsEf}, that if Type 1 NCs are fully compatible with the realistic thesis (ontic interpretation) of the state $\psi$, on the other hand, type 2 NCs induce an anti-realistic thesis regarding the nature of the symbolic state, and thus (according to the PBR theorem that will be presented) of physical reality by rejecting the idea that there is an element of physical reality, showing how NC and realism are intimately connected. We will discuss to which extent the anti-realist hypothesis is actually tenable in section \ref{Realism}. In section \ref{NCObsEf}, we will see that type 2 NCs are in fact type 1 NCs if one adopts an ontic interpretation of the state in QM.

\section{The Problem of the Words used in NCG}
\label{PBWorld}
\epigraph{The limits of my language mean the limits of my world.}{\textit{L. Wittgenstein}}
The use of words that make good sense is important when trying to understand new pieces of knowledge. Physics is not only about equations and mathematical structures, but about common language words and interpretations, which guide our intuition and the elaboration of mathematical structures that are relevant to physics. Words carry a heavy load of cultural connotations, and conceptual associations, both conscious and unconscious. If we consider any conceptual framework (it could be a physical theory), it is associated with a set of words, subjective representations, and epistemological concepts that provide us with a 'satisfactory' understanding of that framework. When considering an attempt to extend this framework, care must be taken with this set of words associated with the old framework, when using them to understand the new one, as some of these concepts may lose their meaning in the process. As has often happened in the history of physics, new words were invented to accompany new perceptions whose relevance became obvious.
\medskip
\par
However, some physicists who have left an undeniable mark on the history of physics, such as Bohr, have helped to mark a philosophical break with this tendency to create new words to understand physics, particularly in quantum physics. Indeed, as J-M. Lévy-Leblond says in \cite{levy2000mots} Bohr considered the forms of expressions of classical physics to be unsurpassable, and considered the desire to find forms of expression specific to the quantum domain unattainable and condemnable. For him, it was an impossibility in principle. I find this position extremely difficult to justify, as it appears both surprising and dogmatic. It is difficult to understand why, unlike many great physicists in history who have encountered counter-intuitive phenomena, Bohr could allow himself to create and spread the idea that these counter-intuitive characteristics were in fact fundamental principles (such as the principle of complementarity), expressible in the usual language of classical physics, as counter-intuitive as they appear to us. Despite his influence, physicists of the following generations have nevertheless created new words, specific to quantum physics, such as coherence, entanglement, beable... But the influence of this epistemological break continues to strongly influence physics, and in particular it's current teaching, where there is no longer any question of having an understanding other than mathematical, the rest of the modes of understanding knowledge being quite widely considered as subjective and illusory.
\medskip 
\par 

The creation of new words is a constitutive process of scientific knowledge, which must accompany the emergence of new notions. This is why, starting from the usual geometrical framework, and all the words and concepts that surround it, such as the notion of point, distance, coordinate, and differential structure... and going towards the extension of this framework which is NCG, it will be necessary to pay particular attention to the words chosen. I have noticed that some words commonly used in NCG do not really make sense in this framework, and thus disturb its comprehension. Here are some of them:
\begin{itemize}
	\item The notion of point, which remains present in the notion of finite $n$-point space, does not seem adequate. These terms designate the finite Hilbert space on which the algebra is represented, but as we have seen with the Gelfand-Naimark theorem, these elements only acquire the status of points for commutative algebras. The fact that the elements in Hilbert spaces are intuitively considered as points comes from the status they have in their use in QM, as classical outcomes, where they only acquire point status at the end of the measurement process. One suggestion is to call these elements pre-points, reserving point status for them when the algebra living on them is or becomes commutative.
	\item In mathematics, a space is a set with remarkable additional structures, allowing us to define objects analogous to those of usual geometry. The elements can be called points, vectors, functions... The notion of space does not make sense in NCG, since we can no longer think of this reality as based on the precious notions of elements of a set, which are points in the geometry. The Hilbert space does not have the status of a set (of points) from the point of view of the algebra, it only has one for the classical observer who perceives well-defined outcomes, there is thus an error in terminology.
	\item In NCG, coordinates are replaced by generators of the algebra. As they do not commute they cannot be simultaneously diagonalized and the space disappears. But people continue to name these generators coordinates. Take a look at the meaning of the word coordinate, this comes from the Latin “co” for “together” with “ordinare” for arranging or put in order. Therefore, the meaning of coordinate can be seen as “put in relative order objects together” potentially with numbers. In NCG, neither objects/points nor order relations are given in a non-contextual way because of the NC of the observables providing such numbers used for this ordering. Therefore, this word has no meaning in NCG. 
	
\end{itemize}
The first two points are deductions from the arguments presented in section \ref{AreSetsPrimary} concerning the limitations of the concepts of set and set's elements (space and points respectively).
\medskip
\par 
There are other obstacles to the conceptual understanding of NCG, and this at the basis of its study, as mentioned in section \ref{WhichNC}, about which NC is concerned. In addition, regarding the status of NCG in physics, most physics researchers seem to consider, unconsciously at least, that NCG is a theory, just like loop quantum gravity or string theory. However, NGC is not a theory, it is a framework of mathematical thoughts, in the same way as Riemannian manifolds were. General relativity has not been called a theory of Riemannian manifolds, it is a theory whose mathematical basis is based on this mathematical formalism. So it seems natural to me that researchers who want to set up a theory of quantum gravitation should try to acquire some knowledge of NCG as a framework, in order to try to understand how to quantify space-time in a relevant way. NCG should not be considered as a theory, and therefore a rival theory, but rather as representing a mathematical framework for formulating such theories. Because theories that want to go beyond general relativity must belong to a formalism that is broader than that of Riemannian manifolds while having links with it since the theory of general relativity must be found in the limiting cases of large scales and small energy densities.
\medskip
\par
The problem is probably even deeper than that, connected to the very structure of our language. In everyday life, we identify objects, which can be thought of as elements of a set, and we attribute specific words to them. But if the notion of element/object is no longer relevant, is it still possible to put words on things? If, as with many words, they refer in their definitions to properties of the thing referred to, then to the extent that these properties become contextual, the usual words lose their meaning. I don't think there is anything in the structure of our language to capture in essence the nature of an object whose properties are revealed contextually. But as always, in science, we can create new words, and new meanings, to improve and extend our representations of reality.
\medskip
\par 
In section \ref{PhiloNCG} I will bring a personal answer resulting from the arguments collected in the different sections which will follow, as to how one can represent the NCG, and how one can come to get rid of the notions of points, space, and coordinates, by changing one's perspective, this change of perspective being among other things allowed by the theorem of Gelfand-Naimark via the relation \ref{GelfTf}.
I would finish this section by saying that in my opinion, the words' problem extends well beyond NCG, and contaminates many fields of the current research in theoretical physics.

\chapter{NC and Physical Theories}
\label{PhysTh}
In QM, three phenomena can be considered as really new. The principle of superposition, entanglement, and the contextuality due to the NC of observables. All of these are due to the replacement (in QM) of functions by operators. In this chapter, we will try to understand how NC can be understandable in the context of physical theories (about nature), which change occurs at the level of the formalism, and what can this tells us about nature, in the context of QM.

\section{General Structure of Physical Theories}

\epigraph{Experience without theory is blind, but theory without experience is mere intellectual play.}{\textit{I. Kant}}

A physical theory consists of two things: a mathematical structure $\StrMath$ and a mapping of the elements of this structure to the operational measurements process of the corresponding elements in nature. Given a physical state $ER$ (an element of reality), our representation in the mathematical structure $\StrMath$ will be called the symbolic state $\pi_\StrMath(ER)$. The measurement of an $ER$ is the determination of the magnitude of something through the comparison of this unknown quantity associated with the $ER$ with a reference quantity of equivalent nature (carried by the element of reality of the object being used as a measuring device), known as the measurement unit. The informational content we get from the measurement of an $ER$ is called an observable $O(ER)$. The physical theory must give the link between $\pi_\StrMath(ER)$ and the measurement process of the physical state $ER$ such represented:
\begin{align*}
	\pi_\StrMath(ER)\overset{}{\longrightarrow} \text{Measurement of } ER\longrightarrow O(ER)
\end{align*}
If there is no such measurement process, the fact that $\pi_\StrMath(ER)$ is representing something real may be questionable. We will call $Mes(ER)$ the measurement process of the $ER$, and $\pi_\StrMath(Mes(ER))$ it's representation in the mathematical structure $\StrMath$. $O(ER)$ is the only thing that we see (as experimenters), $ER$ and $Mes(ER)$ being out of reach. As the measurement process corresponds to an interaction, it can change the element of reality:
\begin{align*}
	ER\overset{Measurement}{\longrightarrow}ER'
\end{align*}
If the mode of observation of this $ER$ is non-passive, that is to say, that it disturbs the $ER$, then, we will have to distinguish two notions of physical state, the one which would correspond to the physical state before measurement and the one after the process, which we will call $F(ER)=ER'$. $F(ER)$ can be seen as what we see, and $ER$ as ``what is'' (before measurement). A physical theory can then be considered as describing what we observe, not what is, but we often do this through shortcuts, because in our classical world, these two categories are confused. We can complete this picture by specifying how and by which apparatus/observer the measurement is done since observational content is both a function of the $ER$ and of the measurement apparatus. We will encode all these data into one symbol, “$a$” for apparatus (and then for observable in the framework of QM as we already see in chapter \ref{DifCalc}). The previous notations become $Mes_a(ER)$, $O_a(ER)$ and $F_a(ER)$.
\medskip
\par
The act of (classical) measurement should be seen as an interaction between physical objects and is therefore the restriction to classical observers of a more complete picture of the interaction pattern underlying the measurements process. This interaction extracts some information about the physical state, this information taking the form of $\pi_\StrMath(ER)$. For some theories, $\pi_\StrMath(ER)$  does not describe directly an $ER$ but the data we have on it, for example statistical data. It's also possible that $\pi_\StrMath(ER)$ can be an incomplete representation of an $ER$, that some hidden variable exists. It is also possible for some theories to reject the idea of an existing $ER$. 
\medskip
\par 
In the special case of QM, $\pi_\StrMath(ER)$ will be replaced by $\psi$ or $\rho=|\psi\rangle\langle \psi |$, and $O_a(ER)$ by $\lambda_i$ with a probability weight $\tr(P_{\lambda_i}\rho)$ (or by $O_a^{av}(\psi)=\langle\psi|a|\psi\rangle=\tr(a\rho)$ in average), for discrete spectrum as defined in \eqref{spectralDec}. The actual QM theory does not predict deterministically the outcome. There will be two manners to interpret its mathematical structure $\StrMath$. Indeed, ontological models of QM can be categorized in two ways: the $\psi$-ontic way and $\psi$-epistemic way. $\psi$-ontic models consider the symbolic state $\psi$ ($\pi_\StrMath(ER)$) as referring directly to objective physical reality, whereas the $\psi$-epistemic ones consider it as simple knowledge of the observer on it. As we will see in section \ref{Realism}, this second kind of interpretation can be split into two categories, the one that admits the existence of $ER$, and the one that rejects it. For models admitting the existence of a physical state ($ER$), we will call $\psi$ its symbolic state's representation.
\medskip
\par
Let's $a$ and $b$ design two different measurement processes, potentially linked to different observables (or to the same). Therefore, it is possible to encounter non-commutative relations like:
\begin{align*}
	[\,O_a\,,\,O_b\,](ER)\,\neq\, 0.
\end{align*}
This means that  $[\,O_a\,,\,O_b\,](ER)=O_a(F_b(ER))-O_b(F_a(ER))\neq 0$. As said, the physicist doesn't know $ER$ (then $F(ER)$) nor the way in which a classical observation $O(ER)$ is given from this $ER$. Therefore an interpretation of non-commutativity may concern either the link between $ER$ and $F(ER)$, or the way in which the measurement process can transmit the information $O(ER)$, in classical information format. We will see in section \ref{NCObsEf}, that in the framework of QM, an ontic interpretation of the symbolic state implies that $ER\neq F_{a\,\text{or}\, b}(ER)$, putting the interpretation at the level of the link between $ER$ and $F(ER)$, whereas for epistemic interpretation (defended by Bohr), it is at the level of the transmission in a classical form that the interpretation must be centered.
\medskip
\par
The basic mathematical structures $\StrMath$ corresponding to a huge part of the physical theories are in the field of algebra. Algebra is the study of symbols and the rules for manipulating and combining these symbols in formulas, the basic example being the product. These symbols will correspond directly or indirectly to elements of reality, and the rules will correspond to how these elements of reality interact or combine into bigger systems. Therefore it is interesting to play with different symbols (representations), or different products, like the universal/tensorial, star, matrix, or inner products for example.

\section{Classical Phase Space vs. Quantum Configuration Space}
\label{PhaseVSConfigSpace}

In general, a physical theory consists of four components: a convex state space $\EspPhas$, a real vector space of observables $\mathcal{U}_\mathbb{R}$, a function connecting these two spaces $\langle\, , \rangle\ : \EspPhas\times\mathcal{U}_\mathbb{R}\to \mathbb{R}$ allowing to obtain the expected value of the given observable for the state in question, and finally laws specifying the dynamics of the theory, such as a Lagrangian, the Schrödinger or Dirac equation \cite{landsman1998lecture}. 
\medskip
\par
In classical mechanics, the phase space $\EspPhas$ is a space of points interpreted as the pure states of the system. The mixed states are identified by measurement probabilities on $\EspPhas$. The observables of the theory are real functions on the phase space $\EspPhas$, most often continuous and bounded. Thus the space $\mathcal{U}_\mathbb{R}$ of observables can be taken to be equal to $C^\infty(\EspPhas, \mathbb{R})$, $C_0(\EspPhas, \mathbb{R})$ or $C_b(\EspPhas, \mathbb{R})$.
There is a function $\langle\, ,\rangle : \EspPhas\times \mathcal{U}_\mathbb{R}\to \mathbb{R}$ between the state space $\EspPhas$ of measurement probabilities $\mu$ sur $\EspPhas$,and the space $\mathcal{U}_\mathbb{R}$ of observables $f $:
\begin{align*}
	\langle\mu,f\rangle=\mu(f)=\int_S d_\mu(\sigma)f(\sigma)\equiv \frac{\sum A_ne^{-\beta E_n}}{Z}
\end{align*}
with $f(\sigma)$ the value of the observable $f$ in the state $\sigma$, and $Z$ the partition function. This function gives the expected value for the observable $f$. In general, we have $\langle\mu,f\rangle^2\neq \langle\mu,f^2\rangle$. 
For a pure state $\sigma$, the measure is the Dirac one and we have $\delta_\sigma (f)=f(\sigma)$ and $\langle\mu,f\rangle^2= \langle\mu,f^2\rangle$.
\medskip
\par
In QM, because of the commuting property of the canonically associated variable of the state space, this one cannot be taken to be a state's phase space anymore, it became a configuration space $\EspConf$. The state space will then consists of the set of all density matrices $\hat{\rho}$, constructed on a given Hilbert space $\mathcal{H}$:
\begin{align*}
	\EspConf=\{\hat{\rho}=|\psi\rangle\langle\psi|\quad ,\quad \psi\in\mathcal{H},\quad \| \psi\|=1 \}
\end{align*} 
these are the pure states which are in correspondence with the unitary $\psi$ states. Similarly, the observables are the bounded self-adjoint operators on $\calH$. The function linking the states to the observables is given by:
\begin{align*}
	\langle\hat{\rho}, A\rangle=Tr(\hat{\rho} A)
\end{align*}
equal to $(\psi, A\psi)$ for pure states. Thus we have respectively for the classical case, the statistical physics case and the quantum case:
\begin{align*}
	\langle\mu,f\rangle=\int_S d_\mu(\sigma)f(\sigma)\quad\leftrightarrow\quad \langle Z,A\rangle=\frac{\sum A_ne^{-\beta E_n}}{Z}\quad\leftrightarrow\quad \langle\hat{\rho}, A\rangle=Tr(\hat{\rho} A)
\end{align*} 
\medskip
\par
The main difference between quantum and classical systems is that for the quantum, the property $\langle\hat{\rho}, A\rangle^2\neq \langle\hat{\rho}, A^2\rangle$, will generally be true, just as in the classical case, but it will be true even for pure states, with equality if $\psi$ is an eigenvector of $A$.
\medskip
\par
In general, a quantum system is obtained from a classical system by the quantization process described by the linear application $Q : \mathcal{U}_\mathbb{R}^0\to \mathcal{L}(\mathcal{H})$ with $\mathcal{U}_\mathbb{R}^0$ belonging to $C^\infty(\EspPhas,\mathcal{R})$, and $\mathcal{L}(\mathcal{H})$ the space of self-adjoint operators on $\mathcal{H}$. The quantization procedure therefore consists in establishing the relationship:
$$\text{Classical observable $f$}\leftrightarrow \text{Quantum observable $Q(f)$}.$$

It is interesting to observe that the set of classical observables is supposed to constitute $\calC^0(\Man)$, where the procedure for expressing this algebra as being based on $\Man$ can be provided by the Gelfand Naimark theorem. Thus, the quantization of an observable belonging to $\calC^0(\Man)$ will no longer be thought of as being based on a manifold $\Man$ but as based on a Hilbert space, which makes the processes of QM impossible to visualize as occurring in the usual space-time. The observables thus quantized cannot be thought of as living on a manifold. Thus, given a locally compact manifold $\Man$ of Hausdorff kind, a “quantization” of $\Man$ consists in giving a Hilbert space $\calH$ and a positive map $Q : \calC^0(\Man)\to \mathcal{B}(\calH)$. When $\Man$ is compact, we want $Q(1_\Man)=\bbbone$. More detail on the quantization procedure can be found in \cite{landsman2006between}.

\section{KvN formalism and the Classical to Quantum Unification}
\label{KvN} 
Many physicists believe that the classical world must be thought of as an emergence of the quantum world, in certain limit cases. For instance, with phenomena involving large numbers of particles, whose action goes far beyond $\hbar$ are in which decoherence occurs. Therefore, it  seems interesting to find a formalism that deals with both classical mechanics (\textit{i.e.} classical observables) and QM (\textit{i.e.} quantum observables). It is important to notice that in the language of $C^*$-algebra, these two algebras of observables correspond to:
\begin{itemize}
	\item Algebra of continuous functions $\calC^\infty(\Man)$ over a particular topological space (Classical observables)
	\item Algebras of bounded operators $\calB(\calH)$ on a Hilbert space.
\end{itemize}
It turns out that such formalism was known very quickly after the development of QM. Indeed, the formalism of $C^*$-algebras and Hilbert spaces seems to be the most general framework to describe physics. A little-known fact is that a reformulation of classical physics has been made in this framework so that it encompasses classical and quantum in the same formalism. Koopman-von Neumann (KvN) mechanics, first presented by Koopman and von Neumann in 1931, provides a description of classical mechanics in terms of Hilbert spaces. It opens up an interesting path for studying quantum-to-classical correspondence. An elementary point that emerges is that only two distinctions between the classical and the quantum descriptions are needed to go from classical to the actual formulation of the quantum formalism. This transition needs to add non-commutativity between conjugate variables and a collapse to render observational content from the formalism. A good introduction to the KvN formalism can be found in \cite{bondar2012operational}
\medskip
\par 
On the algebraic level, such a reconciliation of quantum and classical concepts would result in particular in a formalism that contains the explanation of the $\calB(\calH)\,\to\, C^\infty(\Man)$ transition. The problem is that we have a Hilbert space on the one hand and a manifold on the other hand. Some physicists justify the fact that we cannot represent ourselves quantum mechanical phenomena because they don't live in a manifold $\Man$ (space-time) but on a Hilbert space $\calH$. However, as we have seen with the Gelfand-Naimark theorem, there is a way to find the points of $\Man$ from the pure states of its algebra, when it becomes commutative. This last fact seems to offer a tool for the path towards the unification of classical and quantum, and to recover $\Man$ from $\calH$ and the algebra. More detail on the quantum to classical correspondence can be found in \cite{landsman2006between}.

\section{What is Measurement?}
\label{Measurement}
What is our contact with elements of reality? The basic answer is through measurement. We use to think of measurement and observation as direct perception of what is, but measurement is an interaction, and this must have consequences. Indeed empirical knowledge and knowledge in general are not knowledge of reality in itself, but of our interaction with this reality. A measurement process can be seen as an interaction between the object we want to measure and an object of reference, on which, the corresponding perturbations are interpreted as caused by this interaction. It is therefore logical to perceive the measurement scheme as an interaction and then a reading of the information, and not a direct reading of reality as it actually is. However, since usual classical measurements correspond to interactions that disturb the object being measured in a negligible way, the intuition that emerges is that we are observing reality as it is, which is a shortcut to phenomenal thinking, since the reality is that we are observing a disturbance on our sensors and that this disturbance is the signature of what the object is, after this interaction.
\medskip
\par 
This is intuitive insofar as any measurement corresponds to a physical interaction where the disturbance of the physical quantities of a reference object (photon, measuring device, etc.) testifies to the presence or of information on the measured object. This perturbation is linked to conserved physical quantities, such as energy, momentum, spin, or other quantum numbers. Thus, any change in the properties of the standard object will necessarily be accompanied by an equivalent change in the physical state being measured. Or the minimal element of reality for a measurement apparatus is the photon of light, which has at least a certain impulse... Thus, the interaction that is intended to imprint on the mediator a trace testifying to the presence of the measured object is always accompanied by the negative reflection of this interaction on the measured particle, thus an alteration of the corresponding quantities, this is a fact that cannot be contested and which is fully inscribed in the usual classical thinkings of the world. This argument has a simple consequence, it is that any measurement acquiring non-trivial information disturbs the physical state, a proof of this last assertion can be found in \cite{busch2009no}, where the author shows that any measurement process that acquires nontrivial information on a physical object disturbs this object. 
This is quite classical reasoning. Then there is no instrument that leaves unchanged all states of the system unless the associated observable is trivial. 
\medskip
\par 
The fact that measurement in QM cannot be so straightforwardly accompanied by the idea that it is an observation of the world as it is is at the origin of the creation of the concept of beable introduced by J. Bell to replace the notion of observable \cite{bell2001john}.

\newpage
\section{Realism, Anti-Realism and the PBR Theorem}
\label{Realism}
\epigraph{Naive realism leads to physics, and physics, if true, shows naive realism to be false. Therefore naive realism, if true, is false; therefore it is false.}{\textit{B. Russell}}

Many scientists have the conception that QM's formalism does not describe reality (only give us experimental results), they have accepted to be mystified, and resign themselves to say that $\psi$ and $a$ are formal tools that give the measurement's result, but without physical meaning, as they do not directly represent real objects. For them, the only view is that of a Hilbert space, and physical objects do not live elsewhere than in this mathematical representation, or perhaps that there are no real physical objects behind this description...
\medskip
\par 
In this section, we will discuss the interpretations of QM's formalism. It will be a question of conceptually partitioning the set of possible ontological options, according to whether realism is admitted or not, whether it is at the level of the physical state or of what is observed, and finally whether the theory is of the $\psi$-ontic or $\psi$-epistemic type. The motivations and concepts behind these postures will be presented throughout. Then arguments will be presented to isolate one of the possibilities of this classification as agreeing with the results of QM's experiments. This section is of special significance, as the conclusion reached by these arguments will be a very important element of the argument presented in section \ref{NCObsEf}, and for the conclusions drawn in chapter \ref{ObjectiveNCGFromQM} on the interpretation of the NCG. 

\medskip
\par  
As mentioned before, two views can be adopted regarding the ontological nature of the objects described by QM, the Realistic (R) and the Non-Realistic (NR) postures. To define the relevance of the notion of realism, let us start by mentioning the original criteria that led to the conjecture of the existence of an element of reality.
Two criteria of reality can be defined:
\begin{enumerate}
	\item The Repeatability Criterion (RC): It can be defined as follows (EPR paper \cite{einstein1935can}): if without disturbing the system in any way, we can predict with certainty (with a probability of 1) the value of a physical quantity, then there is an element of physical reality $ER$ corresponding to this physical quantity.
	\item The Kickability Criterion (KC): If a reference system is disturbed in a way that is not understandable as its free behavior, then we will say that something physically real has interacted with it, in other worlds, it possesses a causal power. 
\end{enumerate}
Put like that, it is obvious that these realism criteria refer to $O(ER)$, because it is a question about what we observe. The NR view denies the existence of objective observed observable $O(ER)$ before the measurement, if RC is not checked, the R position is thus hardly tenable. The KC is a much less powerful criterion than the RC, because its negation is not clearly definable, as the notion of free behavior is difficult to specify, and sometimes a matter of convention. The non-verification of the KC is therefore not a direct refutation of the R position.
\begin{remark}
	By element of reality (or equivalently physical state), we only refer to a piece of hypothetical information “carried” by the physical entity in question.
\end{remark}
\begin{remark}
	\label{remarkProveNonDisturb}
	It is interesting to stress that (RC) assumes that the measurement process must be non-disturbing. Thus, disproving the RC will require proving that no disturbance has been caused to the system during the measurement. 
\end{remark}
\begin{remark}
	The (KC) is the physical principle at the heart of the very notion of measurement, as discussed in section \ref{Measurement}.
\end{remark} 
\medskip
\par 
Although the notion of realism was initially defined with observational results, it is important to distinguish between the R/NR thesis which concerns the physical state ($\psi$ in ontic models) itself (ER), and those that concern the measurement result ($a\psi$ in ontic models for a given observable $a$), thus $O(ER)$. We are therefore left with four categories of realism posture, the one about the physical state, which will be called SR and SNR thesis (for State Realist or State Non-Realist), and the ones concerning observations will be called OR and ONR thesis. SNR trivially induces ONR, but the converse is not true as we will see with the NC observables. The state and observation views on realism are sometimes confused by different authors, this is the origin of many misconceptions. One could summarize these two positions in this way:
\begin{enumerate}
	\item Reality exists independently of our observation, we can consider it conceptually, and the act of observation is a more or less precise transmission of this reality, through an interaction which is the measurement process.
	\item Reality is what is observed.
\end{enumerate}
\begin{remark}
	As mentioned in section \ref{Measurement}, if a theory only knows disruptive measurement process, therefore, we have to make the distinction between what is, and what we observe. This is not the case in classical mechanics. 
\end{remark} 
\medskip
\par 
The $OR$ and $ONR$ categories are not really classifying, as they are secondary in the causal chain of the measurement process, and can be considered as (classical) deductions of the physical state. Therefore we will only consider the $SR$ and $SNR$ categories in the classification.
It is also possible to divide the set of theories according to the link between the possible physical state, and the symbolic state representing it. The theories concerning QM are thus distributed according to 3 categories:
\begin{itemize}
	\item $\psi$-ontic theory admitting SR ($\psi$-OSR): $\psi$ is a state that corresponds in a bi-uniquely way to this ontic physical state.
	\item $\psi$-epistemic theory admitting SR ($\psi$-ESR): Realism is admitted, a real physical state exists, but the $\psi$ state only represents information about this state.
	\item $\psi$-epistemic theory admitting only SNR ($\psi$-ESNR): Realism is refuted, there is no ontic physical state, $\psi$ is a kind of information, about potentialities but there is no defined physical state.
\end{itemize}
This classification is summarised in the following diagram: 
\begin{figure}[h]
	\begin{center}
		\begin{tikzpicture}[sibling distance=13em,
			node/.style = {shape=rectangle,
				rounded corners, 
				draw, align=center,
				top color=white, bottom color=blue!20},
			]]
			
			\node {Observations} 
			child {node {RC $\&$ KC}
				child { node[name=or] {OR} }
				child { node[name=onr]  {ONR}
					child { node[name=f] {SR}
						child { node[name=osr]  {$\psi$-OSR} }
						child { node {$\psi$-ESR} } edge from parent 
						node[left] { }}
					child { node {SNR} 	child { node {$\psi$-ESNR} } } }  };
			
			\draw (or) edge node[sloped, anchor=center, above, text width=0.2cm] {C} (f);
			\draw (onr) edge node[sloped, anchor=center, above, text width=0.4cm] {NC} (f);

			\node[left of=or, node distance=3.6cm] {Observable $O$ $\longrightarrow$};
			\node[left of=f, node distance=5.6cm] {Physical state $ER$ $\longrightarrow$};
			\node[left of=osr, node distance=3.2cm] {Link with $\pi_\StrMath(ER)$ $\longrightarrow$};
		\end{tikzpicture}
	\end{center} 
\end{figure}

This diagram is presented as starting on the top from the observations, \textit{i.e.} from what an experimenter can be sure of, and going deeper into the speculations we can make about nature within the theory, first about the realism of what is observed, then about the possible underlying physical state, and finally about the different types of symbolic $\psi$ states describing it, thus offering a classification of possible theories in QM.  
\medskip
\par 
Bohr was perhaps the most influential physicist involved in the development, and especially the interpretation of QM. He was probably the most aware of the subtleties of QM, and the most invested in trying to develop a consistent picture of it. So we will try to understand how his views allow us to locate its position in this diagram, which is of great importance, since he is, broadly speaking, followed by most of the physicists past and present, with the Copenhagen interpretation. We will proceed in the logical order of the diagram, from top to bottom. 
\medskip
\par
The first question concerns his position in terms of realism regarding the observational results $OR$ and $ONR$. One can unambiguously say that he was in the $ONR$ category, his principle of complementarity being the incarnation of it, as this quotation from Bohr in \cite{bohr1950notions} shows:
\medskip
\par 
\textit{A sentence like “we cannot know both the momentum and the position of an atomic object” raises at once questions as to the physical reality of two such attributes of the object...}
\medskip
\par
It is interesting to note that his statements are strictly at the level of $ONR$, without leaving any clue as to its positioning $SR$/$SNR$, as he uses the term “attributes of the object”. What Bohr disputes is that the reality criterion can be applied jointly to position and momentum. This dependence on the context of measurement is called contextuality. One cannot speak of certain predictions concerning both position and momentum, these are revealed not as intrinsic attributes of the quantum object, but as relative to the apparatus context. Therefore, he positions himself at the level of $ONR$. The main argument which made him adopt such a position is the NC of observables and the contextuality which is deduced from it, which he interpreted with the principle of complementarity which he supported (As we shall see in section \ref{NCObsEf}) because of his attachment to wave-particle duality, which he considered as a fundamental duality \cite{mehra1987niels}.
\medskip
\par
Concerning his positioning regarding the realism of the physical state ($SR$/$SNR$), a lot of confusion has been made about his claims, sometimes by attributing him ideas that he never had, but also by authors who do not distinguish between the $ONR$ thesis and the $SNR$ thesis, or between the claims concerning the link between physical and symbolic states, \textit{i.e.} in a more general way between reality and our representation of reality. Indeed, it is commonly attributed to the defenders of the Copenhagen interpretation, and thus to Bohr, to deny the existence of an objective reality concerning the quantum world, thus displaying a $\psi$-ESNR posture, which was a novelty in history. This can be seen, for example, in this quote from Heisenberg: 
\medskip
\par 
\textit{“atoms or elementary particles themselves are not real; they form a world of potentialities or possibilities rather than one of things or facts”.}
\medskip
\par 
Similarly, in \cite{petersen1963philosophy} A. Petersen refers to the following “shocking” quotation of Bohr, which was very often attributed to him afterward:
\medskip
\par
\textit{'There is no quantum world. There is only an abstract quantum physical description.'}
\medskip
\par 
This would put the defenders of Copenhagen, as well as Bohr, in the category of $SNR$, denying the existence of an objective physical state. However, as N.D. Mermin says in \cite{mermin2004s}, Bohr has never published such an assertion, in any of his articles where he affirms his position on quantum reality, nor in any of his other writings. Several authors (Folse in \cite{folse1986niels}, and Faye in \cite{faye2012niels}) argue that Bohr had a realistic view of the physical state (SR). They say that Bohr always implicitly supported the $SR$ thesis, as this quote from Bohr in \cite{bohr1929atomic} proves:
\medskip
\par 
\textit{[...] the extraordinary development in the methods of experimental physics has made known to us a large number of phenomena which in a direct way inform us of the motions of	atoms and of their number. We are aware even of phenomena which with certainty may be assumed to arise from the action of a single atom. However, at the same time as every doubt regarding the reality of atoms has been removed and as we have gained a detailed knowledge of the inner structure of atoms, we have been reminded in an instructive manner of the natural limitation of our forms of perception.}
\medskip
\par
He never went back on these statements. Similarly, in \cite{folse1986niels} Folse argues that the reality of atoms has been established by experiments that directly cause phenomena that inform us of the existence and behavior of atoms. This is the $KC$ criterion: the proof that there exist real quantum objects is an experimental fact, the simple fact being that we measure them, or that we deduce causal consequences from their properties, as Bohr underlines, this is a proof of the objective existence of these quantum objects. For Bohr the existence of atomic systems is “the given”, which theory must describe in a way allowing prediction of those phenomena caused by these entities. Bohr is then, in fact, clearly convinced about the objective reality of the physical state. Many authors say that Bohr's words are confused or even contradictory concerning his ontic position, but his words on this subject are always accurate. They are in terms of “objective communication” and not in terms of “truth”. The confusion comes, in my opinion, from the readers, whose innate unconscious reflex is to associate with what one observes, the impression that it is pure reality, without assuming that it is only a piece of information and that this observation is moreover an interaction. Other authors like Hacking and Cartwright have supported this view about the reality of quantum physical states, which they call Entity realism. 
\medskip
\par 
Let us come back to the $SNR$ posture, which is justified by its defenders by saying that $SNR$ represents only information about potentialities and that this is a complete description of reality \textit{i.e.} that there is no hidden $ER$ below this. A very simple argument can provide a serious counter-indication to the $SNR$ posture thus described.
\begin{proposition}
	\label{NoGoMaisonSNR}
	The SNR view is not in accordance with the results of quantum mechanical experiments.
\end{proposition}
\begin{proof}
	The $SNR$ posture rejects the idea that there is an element of reality corresponding to the quantum world and therefore carrying the information that determines the measurement result. However, the repeatability of an identical measurement made at short time intervals forces one to consider that information is conserved. So the only way left by this position is, as they do, to say that this information is on the side of the observer, only our abstract description, and the information thus symbolized exists, not the physical state. This is the position defended by Heisenberg in particular. Now let's take any quantum experiment, with a $\psi$ state, and two observers $O_1$ and $O_2$, these two observers are sensibly the same, they are two identical machines. We then choose to measure the state with $O_1$, we collect the information by looking at the result of the measurement, then we remove $O_1$, and we do the same experiment with $O_2$, immediately after, and we collect the information again. If there is no $ER$, and $\psi$ only represents the information we have, on a world of potentialities, that is to say that there is no information stored in a $ER$, then knowing that $O_2$ does not have the information that $O_1$ has collected, it should have every chance to find another result, since this information cannot be stored in a $ER$. However, the experimental results show repeatability of the collected information, which is in disagreement with the $SNR$ posture. 
\end{proof}
We will give more arguments against the $SNR$ view later.
\medskip
\par 
Now, concerning his position on the link between the physical state and the symbolic state, Bohr was clearly epistemic, as were most of the advocates of the Copenhagen interpretation. The remarks of the preceding paragraph thus place him in the $\psi$-ESR position. The epistemic interpretation of the symbolic state is supported by various physicists, for several reasons:
\begin{itemize}
	\item Because of the multidimensionality (The fact that we are left with nothing but a description of a state living in a multidimensional Hilbert space.) of the $\psi$-state in Hilbert space, which seems not to offer a realistic vision of the phenomenon as existing in space-time.
	\item To solve the collapse problem, becoming an actualization of the observer's knowledge in the Copenhagen interpretation. For Bohr, the $\psi$ function is a predictive tool and does not reflect any physical reality. It is therefore not necessary to say that the subject would provoke any physical process during the measurement. It merely defines a new form of objectivity, by including the mention of the experimental context, while refraining from including the human subject. This new objectivity is thus fully linked to the thesis of contextuality.  In this view (Copenhagen) the 'collapse' of the wave function is not a physical process, and it just reflects an update of our information about the system.
	\item Because of the EPR paper which emphasizes that the formalism of QM is in contradiction with the thesis of local realism. 
\end{itemize}

His belonging to the epistemic category is underlined by an important aspect of his thought, concerning the formalism of which he had a purely symbolic vision \cite{bohr1950notions}:
\medskip
\par
\textit{The entire formalism is to be considered as a tool for deriving predictions, of definite or statistical character, as regards information obtainable under experimental conditions described in classical terms and specified by means of parameters entering into the algebraic or differential equations of which the matrices or the wave-functions, respectively, are solutions. These symbols themselves, as is indicated already by the use of imaginary numbers, are not susceptible to pictorial interpretation; and even derived real functions like densities and currents are only to be regarded as expressing the probabilities for the occurrence of individual events observable under well-defined experimental conditions.}
\medskip
\par 
Therefore as Faye says in \cite{faye2012niels}:
\medskip
\par 
\textit{Bohr also considered the state vector to have no ontological status but merely to be a heuristic device for the calculation of the probability of a specific outcome of the measurement.}
\medskip
\par 
Bohr is definitely on the $\psi$-ESR side. For him and many others, the QM formalism has no direct correspondence with the reality of real objects, it is a symbolic representation that is not faithful to reality, allowing to set up a kind of algorithm enabling to find the results of experiments. Most of the physicists seem to have this posture, and do not consider $\psi$ as real, but rather as a kind of information, which is supported by the fact that the result provided is probabilistic. This point is of importance for the chapter \ref{ObjectiveNCGFromQM}, because depending on whether one attributes a concrete character to the symbolic state or not, speaking of geometrical consequences deduced from the QM formalism will have a meaning, or will not have any. 
\medskip
\par
It seems, therefore, that the only positions which are relatively tenable are the $ONR$, the $\psi$-$OSR$, and the $\psi$-$ESR$, the first being non-disjoint from the other two.  Although it is not intuitively obvious, the  $\psi$-$OSR$ and $\psi$-$ESR$'s postures are not only different at the philosophical level. Indeed the quantum wavefunction $\psi$ is said to be $\psi$-ontic if any physical/ontic state in the theory is consistent with only one pure quantum state and $\psi$-epistemic if there exist physical/ontic states that are consistent with more than one pure quantum state. This point will be important to formally distinguish these two ontological issues and will be used in the proof of the PBR theorem which I will present now. This will be of major significance in order to decide between the postures, \textit{i.e.} between the $\psi$-OSR and $\psi$-ESR's  interpretations, and then in section \ref{NCObsEf} to interpret the NC of observable in QM. To finish, in chapter \ref{ObjectiveNCGFromQM} it will permit to interpret the associated geometry we can infer of such an observable, according to the arguments that I will present in chapter \ref{GeomAndSpacetime} about the objective path (trough measurement) we have to test physical theories about space-time, in the mathematical framework $\StrMath$ of geometry.
\medskip
\par
A theorem of capital importance allows us to determine between the  $\psi$-OSR and $\psi$-ESR positions. Indeed, in 2012, a powerful no-go theorem against $\psi$-epistemic interpretations $\psi$-ESR was found \cite{pusey2012reality}. This is called the PBR theorem (for Pusey, Barrett and Rudolph). This theorem is considered as being of similar importance to the Bell, and  Bell–Kochen–Specker ones. As these last theorems, PBR theorem is a statement about hidden variable theories. It constrains hidden variable theories in a very strong way, almost killing the relevance of their potential existence.  
\medskip
\par
What the PBR theorem shows is that assuming the existence of a physical state ($ER$), the idea that the associated symbolic state $\pi_\StrMath(ER)=\psi$ does not represent reality in a one-to-one way (then this physical state), but our (probabilistic) knowledge on it is no longer tenable. Indeed, if there is such a physical state, the pure symbolic state that we associate with it corresponds to a definite statement about this reality, and not to a probabilistic one. $\psi$ is therefore no less “real” than the analogous objects used in classical theories to designate states of the system.
\medskip
\par 
More precisely the PBR theorem states that if we assume that:
\begin{enumerate}
	\item A quantum system always does have some objective physical state.
	\item For any repeatable experimental procedure, there exists a well-defined probability distribution over the set of possible final outcome states, and a quantum system can be measured projectively on any basis.
	\item  Statistical independence of experiments, and tendency to tend to the probability distribution just mentioned when the experiment is repeated many times. 
	\item Even if two systems are brought together for a joint measurement on an entangled basis, they may until then be treated as separate systems, with their own space of states, and prepared so that the respective probability distributions are independent.
	\item Quantum predictions are correct.
\end{enumerate}
Then, for a QM description done in a given Hilbert space, taking two preparation procedures that place the system into two different (symbolic) pure states, they can't represent the same physical state.
If we have two quantum symbolic states $\pi_\StrMath(ER_1)$ and $\pi_\StrMath(ER_2)$ with a physical state as an unknown, then:
\begin{align}
	\label{ConseqPBR}
	\pi_\StrMath(ER_1)\neq \pi_\StrMath(ER_2)\qquad \to \qquad ER_1\neq ER_2
\end{align}
Two different symbolic pure quantum states are directly linked to different physical states, whatever the reality in which these physical states take sense may be. This is a really strong statement, it basically says that the symbolic quantum states $\psi$ we use in QM are as real as our usual classical symbolic states, since they refer both to physical states in a faithful and complete manner. The symbolic quantum state is then ontic, and the only survivor of this battle and the happy winner is $\psi$-$OSR$.
\medskip
\par 
\begin{remark}
	\label{RQDualitéF}
	In the $\psi$-OSR theories, the wave-particle duality cannot be a fundamental duality (which is at the basis of Bohr's philosophy). Indeed, if the aspects of wave and corpuscular behavior can be associated with two different $\psi$ states (one spread out in the space of positions and the other not), knowing that these states are in one-to-one correspondence with two physical states, then these two physical states will be different too. There is not, as Bohr represents it to himself, a quantum reality, whose properties can be manifested as undulatory or corpuscular according to our mode of observation, but two distinct quantum elements of reality. Thus, it is an effective duality, not a fundamental one, and the principle of complementary that accompanies it is no longer tenable.
\end{remark} 
\medskip
\par
This theorem is of crucial importance for the interpretation of the quantum mechanical formalism, but it says nothing about $\psi$-ESNR type interpretations since it is based on the assumption of the existence of a physical state. I have presented above a strong argument against this posture in the proposition \ref{NoGoMaisonSNR}. This point being important, I wish to give here other more qualitative arguments against the SNR thesis, some of these arguments are also arguments against the $\psi$-ESR posture: 
\begin{enumerate}
	\item \textbf{Against $\psi$-ESNR:} If QM is the basis of the classical as usually thought\footnote{In the sense that classical theories must be recovered as approximations of quantum theory, in certain regimes, i.e. large number of particles, large action, particle dense environment...}, then the quantum state must have the same status of realism as our classical states. I don't see how it can be otherwise, we are constituted of particles with quantum behavior! And it happens that a collapsed particle can find coherences again \cite{bouchard2015observation}. So the fact that its status of realism is going to pass from realism to nonrealism and then realism by regaining coherences seems to me doubtful. Another argument comes from the KvN formalism, indeed the implicit message is that the unification of the classical and the quantum can eventually be done in a single formalism. So if quantum and classical states were of different natures, as the $\psi$-ESNR interpretation suggests, then this unification would not be thinkable. I find this relatively inconceivable. In connection with the previous argument, the wave function after the collapse is of the same nature as the one before the collapse in the formalism, the only difference (in the case of a position measurement for example) is that it is more localized. So there is no information about potentialities on one side and a concrete actualized result on the other, it appears strange to say that the ontic status of $\psi$ can have changed during the measurement process, going from $\psi$-epistemic to $\psi$-ontic. I add to this the fact that what we call real objects, are made of these same particles, localized, and interacting, so how could it be that the ontic status of $\psi$ is different from that of our machines, of all classical objects? I don't see how. This continuity of vision argument is for me a strong reason to believe that $\psi$ designates an entity no less real than any object whose reality is considered without doubt. 
	\item \textbf{Against $\psi$-ESNR:} The multidimensionality of the Hilbert space in which the wave function lives is not an argument against realism, it is only that it is no longer thinkable as a phenomenon in a usual space-time, as mentioned in \ref{KvN}.
	\item \textbf{Against $\psi$-ESNR:} In connection with the previous point, the EPR paper as well as Bell's inequalities formulated later cannot be taken as proof that realism must be abandoned, it is only local realism that is faulted, there are many other notions of realism, such as structural realism, or ontic structural realism which remain possible. 
	\item  \textbf{Against $\psi$-ESNR and $\psi$-ESR:} As information does not correspond to something real, then if $\psi$ only describes information, then it cannot act on the environment. Thus, the second criterion of realism (KC) is an argument for me that decoherence is proof that the wave function is objective because it interacts with the particles in the environment wherever it has coherences. If $\psi$ was only information, it would not interact with the environment.
	\medskip
	\par 
	I have a simple question to ask if we consider that a criterion of existence for an entity is its ability to interact with its environment, that $\psi$ does not describe anything real, only information, and that information that is not material cannot interact with the environment. How is it then that a theory such as decoherence, verified experimentally, is based on the fact that before the collapse, $\psi$ interacts with all the particles of the environment contained in $D(\psi)$? Isn't this a clue that $\psi$ is in itself a real entity?   
	
	\item \textbf{Against $\psi$-ESR:} 	The epistemic interpretation of $\psi$ was promoted to solve the problem of the collapse postulate, taking it as an actualization of the observer's knowledge. This was, together with the wave-particle duality, a strong reason that led them to think that $\psi$ refers only to information are potentialities, and that we should abandon the assumption of realism. But today, the collapse postulate is more commonly considered to be a problem of the actual QM's measurement theory.
	
	\item \textbf{Against $\psi$-ESR:} Wave-particle duality was taken to be a proof of the fact that $\psi$ does not refer directly to a physical state, but to information we have on it. But as mentioned in remark \ref{RQDualitéF}, which was a deduction of the PBR theorem, wave-particle duality can no longer be taken as a fundamental duality. Therefore, wave-particle cannot be taken to validate $\psi$-ESR interpretations. 
	
\end{enumerate}

\medskip
\par 
The PBR theorem, as well as these arguments, provide a counter indication to the $\psi$-ESR and $\psi$-ESNR theories, thus putting ahead, as far as this classification is complete, the $\psi$-OSR theories. The $\psi$-state thus designates a physical state, \underline{the collapse must correspond to a non passive physical process} that remains to be discovered. Therefore, surprising properties such as noncommutativity and entanglement must be taken seriously as to what they say about our world, at its deepest level because they concern the nature of the fundamental constituents of all things in our universe.

\newpage
\section{Knowing is Changing: NC as an Observer Effect?}
\label{NCObsEf}
\epigraph{The Very Perception Is Action.}{\textit{J. Krishnamurti}}

As mentioned in section \ref{KvN}, one of the main discrepancies between QM and classical mechanics comes from the emergence of observables that do not commute. This fact is the mathematical origin of the Uncertainty Principle (UP) which was introduced by Heisenberg in 1927. How we make intuitive sense of this noncommutativity depends on the interpretation of the wave function we use. There are two almost equivalent ways to consider the UP, via the wave approach, or via the matrix approach. These are comparable representations, but they highlight different aspects of reality. 
\medskip
\par
In wave formalism, it is easy to see to which extent a spread according to the momentum modes leads to a possible interferential pattern, which allows the particle to be localized, but, if the wave packet consists of only one $p$ mode, then, just like a planar wave, it is spread all over space, and the particle is delocalized. There is therefore a relationship between the spread in position and the one in momentum. It is a wave property, $\psi$ cannot be localized in both momentum and the position basis at the same time. The bases of the Fourier modes in position and momentum representations are therefore connected.
\medskip
\par
Let's see to which extent the matrix formalism can complete this picture. In QM, measurement is taken to collapse the state of the system to one of the eigenstates. When the state is not an eigenstate of the observable, this collapse is not a projection, but something like a transformation then a projection. For an operator $a$, we will call $a_c$ the composition of $a$ with the collapse operator (plus normalization, which will not be taken into consideration here, because it does not change the argumentation). As in \eqref{spectralDec}, if we take two (normal) operators $a$ and $b$ with discrete spectrum, then we obtain their following spectral decomposition:
\begin{align*}
	a=\sum_{i\in S_a}\lambda_i P^a_{\lambda_i}\qquad\qquad\text{and}\qquad\qquad b=\sum_{j\in S_b}\lambda_j P^b_{\lambda_j}.
\end{align*}
This represents the operation that corresponds to the associated measurement process in QM. Let us see what NC of $a$ and $b$ means in first time, and what it implies when we add the collapse postulate \textit{i.e.} what is the consequence for $[a_c,b_c]$.
\medskip
\par
If $\psi$ is in an eigenstate of $a$ then it is not affected by the measurement of $a$ : $a\psi=\lambda\psi$ with $\lambda$ the eigenvalue, result of the observation. Otherwise, $\psi$ undergoes a transformation: $a\psi=\psi'$. 

\medskip
\par
The NC of the observables $a$ and $b$ on $\psi$ induces that $\psi$ cannot be the eigenstate of both simultaneously. Thus, at least one of these two observables is a \underline{non-passive measure of the symbolic state $\psi$}. 
\medskip
\par
To illustrate this, let us take $a=\hat {x}$ and $b=\hat {p}$, and assume that $\psi$ is eigenvector of the operator position: $\hat {x}|\psi \rangle =x_{0}|\psi \rangle$, and that it is not the case for the momentum $\hat {p}|\psi \rangle=|\psi' \rangle$, thus: 
\begin{align*}
	[\hat {x},\hat {p}]|\psi \rangle =(\hat {x}\hat {p}-\hat {p}\hat {x})|\psi \rangle =(\hat {x}-x_{0}\hat {I})\hat {p}|\psi \rangle =i\hbar |\psi \rangle
\end{align*}

so that $(\hat {x}-x_{0}\hat {I})\hat {p}|\psi \rangle=(\hat {x}-x_{0}\hat {I})|\psi' \rangle\neq 0$ which explicitly means that $|\psi' \rangle$ is not an eigenvector of $\hat{x}$ with eigenvalue $x_0$ ($\hat {x}|\psi' \rangle\neq x_0|\psi' \rangle$). Then it allows us to get a result different from $x_0$ as if the position had changed. The conclusion is the same if we take $\psi$ to be a linear combination of eigenvectors of $\hat{x}$, and that at least one of these elements is not an eigenvector of $\hat{p}$.
\medskip
\par
If we add the collapse in this picture: $a\to a_c$ and $b\to b_c$ (to obtain the true measurement process's induced transformation), then the effect on $\psi$ is to collapse it onto one of the eigenspaces associated to $P^a_{\lambda_i}$ for $a$ or $P^b_{\lambda_i}$ for $b$. A basic result of spectral theory is that if you have $[a,b]\neq 0$, then their exists at least one couple $(i,j)\in (S_a,S_b)$ such that $[P^a_{\lambda_i},P^b_{\lambda_j}] \neq 0$. This induces $[a_c, b_c]\neq 0$ and the previous conclusion works also here. 
\medskip
\par
Then NC of observables (operators in general) is always connected to the non-passive action of at least one of the two measurement processes on the symbolic state $\psi$\footnote{Note that these conclusions only apply to the symbolic state, then not directly to the physical one.}. An example for the first case (without collapse) can be the NC of gauge potential in non-abelian gauge field theories (if we replace $(a,b)$ by $(A_\mu,A_\nu)$), and the second one (with collapse) corresponds to QM's measurement process. For this last case, the non-passiveness is not only induced by the NC since the collapse represents a change of the symbolic state too. We can say that NC adds a contextual property to this collapse-induced change. We cannot directly conclude that NC is linked to the non-passiveness of the measurement process because the previous conclusions are at the level of the symbolic state $\psi$. Therefore, according to our ontologic posture, we are left with two ways of interpreting this NC:
\begin{itemize}
	\item The one coming from $\psi$-ESR interpretation schemes: There is change at the level of the symbolic state $\psi$, but as it is not directly linked to the physical state, we are not forced to said that the latter undergoes an observer effect. It is even possible to suppose that there is no change at all in this physical state, by postulating that this NC is induced by the fact that the classical properties in question have no real meaning for the quantum object, which can thus be, without ambiguity, both a wave and a particle, whereas this has no meaning in the classical world. This avoids having to consider discontinuous changes in the physical state in question. It's a statement about reality itself. Particles don't have precisely defined momentum and position to measure. This is a NC interpretation of type 2. There is no Observer effect.
	
	\item The less represented one coming $\psi$-OSR interpretation schemes: the revelation of the classical properties has nothing mysterious, it is in bijection with the real physical state. But here it is not a question of a single physical state, since this one is in bijection with the symbolic state... there are thus two physical states, one transiting to the other one during the measurement, and thus non passivity of the act of measurement! This is caused by the fact that collapse must be considered to be a physical process which transforms the physical state in this kind of interpretation, non-commutativity rendering this process contextual. It's a statement about what we can know about reality. The act of measuring position changes the momentum, and vice-versa. This is an interpretation for type 1 NC. There is an Observer effect.
\end{itemize}
\medskip
\par
Before going further, let's try to understand how this NC has been historically interpreted, mainly by Heisenberg, then by Bohr.
\medskip
\par
Heisenberg's first explanation of his UP was based on the idea that any measurement process must change the physical state, and that this leads to significant perturbation when considering the measurement of quantum states. Indeed, Heisenberg originally illustrated the intrinsic impossibility of violating the UP by using the observer effect of an imaginary microscope as an elementary measuring device. The argument goes as follows. Let us take the example of the electron that we want to localize in space, the minimal measurement scheme to localize it implies shining at least one photon on it. If the electron was a classical object, like a bird in the sky, the photon will affect its state in a so tiny way that we neglect this fact in all days life, which leads to the feeling of seeing “what is”. However, this is not true for quantum objects. Most of the time, the energy of a single photon is ridiculously small (in comparison to the one of the electron), and affects it in a negligible way, but it's not true for very energetic photons. 
\medskip
\par
Let's explain this measurement scheme more concretely, the microscope experiment consists in observing how a light ray is deviated by the Compton effect when it scatters with an electron, the measurement of the deviation angle permit to determine the localization of the electron with a certain accuracy. But this ray of light has a certain wavelength $\lambda$, below which it is not possible to measure the position of the particle, and this wavelength is linked with its energy by the relation $E=hc/\lambda$. Thus, improving the accuracy of the measurement of such a position requires to increase the energy of the photon used. But such a photon is also associated with momentum $p=h c/\lambda$, and a portion of that momentum will be transmitted to the particle during the interaction process at play, by conservation of this quantum number. Therefore, the more you measure the position with accuracy, the more you change the momentum. The fact that the momentum of the particle is changed makes the notion of phase space meaningless, this being at the origin of the introduction of the configuration space mentioned in section \ref{PhaseVSConfigSpace}.
\medskip
\par
This is a very important argument, it is obviously a purely experimental fact of QM and it as not to be deduced by any outer logical argument, I will come back on this later in section \ref{ComptonScale}. More details on the microscope experiment can be founded in Heisenberg book \cite{heisenberg1949physical}, and \cite{tipler2012modern}, and \cite{greenstein2006quantum}. This phenomenon was understood by him as a manifestation of the UP. This argument of the non-passivity of measurement seems quite reasonable, especially since, as said before, it has been demonstrated in \cite{busch2009no}, that a measurement that acquires information on a given system, necessarily disturbs the measured state. But these argument work at the level of the symbolic state, therefore we cannot conclude directly like Heisenberg does that it is connected with an objective observer effect.
\medskip
\par
Indeed, it was argued by Bohr, that the Heisenberg UP has nothing to do with an observer effect. According to Bohr, he did not make clear the distinction between a position measurement merely disturbing the momentum value of the particle and the more radical idea that momentum is meaningless or undefinable in a context where the position was measured instead. As mentioned in \cite{mehra1987niels}:
\medskip 
\par 
\textit{Bohr felt that Heisenberg had not treated the thought-experiments with the X-ray microscope and the investigations on the Compton effect and resonance fluorescence light quite properly; his fundamental objection was something quite different, As Heisenberg recalled: “The main point was that Bohr wanted to take this dualism between waves and corpuscles as the central point of the problem.”}
\medskip
\par 
It is important to stress the fact that Bohr has no fundamental criticism to do to Heisenberg's paper, but its attachment to wave-particle duality and its epistemic interpretation of $\psi$, which are linked in the large picture in which he tries to make sense of QM lead him to see things differently than Heisenberg for this experiment and it's linked to non-commuting observables. For Bohr, the wave-corpuscle duality was a fundamental duality, which he made a consequence of the Complementary Principle (CP), which he proposed as a replacement for Heisenberg's UP. 
\medskip
\par
In his view of CP, Bohr does not refer to discontinuous changes of the corresponding quantities during the measurement process, on the contrary, he emphasizes the possibility of defining these quantities. He also rejects the common view (that Heisenberg adopts for example) about the fact that measurement creates definite results: 
\medskip
\par
\textit{The unaccustomed features of the situation with which we are confronted in quantum theory necessitate the greatest caution as regard all questions of terminology. Speaking, as it is often done of disturbing a phenomenon by observation, or even of creating physical attributes to objects by measuring processes is liable to be confusing, since all such sentences imply a departure from conventions of basic language which even though it can be practical for the sake of brevity, can never be unambiguous. (Bohr 1939: 24)}
\medskip
\par
Bohr always stressed that uncertainty relations are first and foremost an expression of complementarity \cite{mehra1987niels}. This may seem odd since complementarity is a dichotomic relation between two types of description whereas the uncertainty relations allow for intermediate situations between two extreme potentialities (wave and corpuscular behaviors), and therefore causal chain, contrary to the CP which defends the idea that causal view is a classical restriction that doesn't make sense in QM:
\medskip 
\par
\textit{A causal description of the process cannot be attained; we have to content ourselves with complementary descriptions. “The viewpoint of complementarity may be regarded”, according to Bohr, “as a rational generalization of the very ideal of causality”}
\medskip
\par
For Bohr, there is no time lapse during which a physical object behaving as a wave would become a particle or the opposite. He means that these two visions of things are only valid at the classical level and that the particle does not carry these properties in itself. Rather, the symbolic $\psi$ state being only there to give us a sort of classical information useful to understand this inexpressible reality in a raw way as possessing in an unambiguous way these (classical) properties called complementary. This is why, in this same line of thought, he defended that causal thinking was also limited to our classical understanding of phenomena and that it reached its limit in a very obvious way in the experiments involving these complementary observables.
\medskip
\par
In a more general way, Bohr defended the idea that the usual language, impregnated with the classical vision, was not able to make quantum phenomena understandable, and that this was probably a fundamental limitation that could not be overcome for our intuitive understanding of QM within our classical language. Indeed, in \cite{bohr1950notions}, in order to defend the idea that the NC, in connection with the principle of complementarity, is not a matter of processes in space and time, and that therefore the classical causal analysis is no longer valid, Bohr does not hesitate to say that the ambiguity of such an interpretation lies in the fact that the classical language, which is intrinsically causal, cannot describe the reality that lies behind such a NC:
\medskip 
\par
\textit{These so-called indeterminacy relations	explicitly bear out the limitation of causal analysis, but it is important to recognize that no unambiguous interpretation of such relations can be given in words suited to describe a situation in which physical attributes are objectified in a classical way.}
\medskip
\par
To summarise Bohr's view on the interpretation of the NC of observables, the complementary between such phenomena implies that there is only one object which is making different phenomenal appearances in different experimental arrangements, which lead to different interactions scheme between the observer apparatus, and the physical state. There is no causal chain, this physical state is not changed by the measurement process, it is only the measurement of certain properties simultaneously that cannot make sense, and different experimental arrangements will provide complementary evidence about this unique object. This phenomenon cannot be intuitively represented because the classical words, which he considers unsurpassable, do not allow us to intuitively capture its nature.
\medskip
\par
For me, this is not a sufficient explanation, he did not explicitly show that it was not a matter of processes in space and time, whereas it was the result of Heisenberg's initial demonstration, of which he did not refute the mathematical content. If he had sought to make this demonstration, he would have realized that the conservation of quantum numbers during the process of measurement induces that the perception of an information that testifies to the presence or properties of an object, during the process of measurement, necessarily induces the perturbation of the corresponding properties that the measured object would carry, this argument having been more fully discussed in section \ref{Measurement}. On the other hand, he relies on the idea that classical language is unsurpassable and provides a limitation to this understanding, which seems questionable from the arguments presented in section \ref{PBWorld}. Furthermore, his interpretation of NC, which he accompanies with his principle of complementarity, is not satisfactory, as the use of the concept that conventional language and associated classical notions cannot deal with the quantum is not a convincing argument. It is like a black box that allows us to say that we cannot hope to find a satisfactory explanation for quantum phenomena, even by creating new words. This makes it acceptable to abandon the idea of obtaining an intuitive understanding of reality. The concern to obtain an intuitive and conceptual coherence is no longer put forward. In my opinion, this idea has strongly polluted the scientific thoughts developed later. Moreover, as JM. Lévy-Leblond points out \cite{levy2000mots}, the principle of complementarity was not considered a relevant and convincing idea by researchers such as Einstein and Schrödinger, and even among his close collaborators such as Heisenberg and Pauli. Yet it was the basis of Bohr's philosophy, and of the commonly supported Copenhagen interpretation of which he was the principal architect. 
\medskip
\par 
As we have seen in section \ref{Realism}, there is a much stronger argument to defeat his view. Indeed, Bohr's statements and interpretations are based on his epistemic interpretation of the symbolic state, without which, if he were to admit an ontic view of formalism, he could not say that there is no transformation of the state as mentioned above, and worse than that as mentioned in remark \ref{RQDualitéF}, the wave-corpuscle duality can then no longer be a fundamental duality, which is at the heart of the interpretation he defends. These two aspects are then connected to two distinct physical states, and there are therefore intermediate stages of the physical state to consider.
\medskip
\par
But, as we see in section \ref{Realism}, only $\psi$-OSR interpretations can be retained to be in accordance with experimental results. So if the argumentation presented in section \ref{Realism} is correct, the arguments of Bohr cannot be well founded. Then, NC of QM is of the first kind, linked to process in space and time, caused by the non-passivity of the act of measurement, then of the same kind as “classical” ones. However, they differ in that we are unable, as classical observers, to see beyond these NCs, to see the intermediate stages of the process. In QM, processes such as collapse, decoherence, the measurement process, quantum jumps, etc. were initially considered to be instantaneous and discontinuous. These phenomena of discontinuous jumps from one symbolic state to another were called quantum jumps, and it was the very nature of this discontinuity in evolution that made Bohr (among others) prefer to opt for an epistemic interpretation, as he could not consider that the physical state could undergo such a discontinuous evolution, which seems reasonable. Schrödinger, who thought that the wave function is a faithful representation of reality, directly opposed this idea of quantum jump \cite{schrodinger1952there} by also rejecting the principle of complementarity. Nevertheless, at the time, nothing could have foreshadowed a possible non-continuity of the evolution of the symbolic state, and therefore Bohr's interpretation could make sense in the light of the knowledge of the time, even if he could have concluded that the QM formalism could possibly be either incomplete or that certain developments were missing to draw all the consequences.  Today, it is experimentally demonstrated that decoherence, quantum jumps \cite{minev2019catch} and therefore the measurement process is not instantaneous or discontinuous, but continuous and over very short periods of time. There is as yet no theory explaining collapse in a continuous manner, but history seems to show that the apparent discontinuities in the evolution of $\psi$ are gradually disappearing from the scientific landscape, suggesting the possibility of a strictly continuous evolution for the state (physical and symbolic).
\medskip
\par
To summarise the information provided by wave and matrix's representations of reality, the wave aspect tells us that the distribution of the state according to the impulse and position bases are related. In particular for their spread. And the matrix interpretation clearly shows us how the state will undergo an evolution for its distribution according to one of these two bases, during the act of measuring one of the two observables. Thus, we can now see how any non-commutativity is linked to the non-passivity of the act of measuring one of the two observables over the symbolic state $\psi$, and then on the physical one. This can also be seen as a simple consequence of what was said in section \ref{Measurement} about the fact that any measurement that obtains nontrivial data is disruptive for the state.
\medskip
\par
\begin{remark}
	According to remark \ref{remarkProveNonDisturb}, the negation of observable realism induces proof of the fact that there is no state change during the process. Bohr skipped this step.
\end{remark}
\begin{remark}
	According to PBR theorem if there is an ER, it will be directly associated with a symbolic state in a one-to-one correspondence, then supposing that the NC of observable is of type 2 lead to the fact that both the symbolic and physical state has not been changed since it's not linked to process in space-time, but as we just see, for $\psi$-OSR interpretations, NC is linked to state change. Therefore, as mentioned in section \ref{WhichNC}, NC of type 2 and state realism SR are not compatible. But NC will always be linked to observable nonrealism (ONR).
\end{remark}

We thus see how the NCs of QM and gauge theory formalism arise from the same fact, the non-passivity of the interaction that is the measurement in QM, or the non-passivity of the interaction with a boson vector in gauge theory. Therefore, by the several arguments I have presented so far, I would adopt the ontic point of view and its implications on the equivalence between noncommutativity and nonpassivity of the act of measurement in what follows.

\chapter{Geometry and the Physical Theories of Space-Time} 
\label{GeomAndSpacetime}
\epigraph{Geometry is not true, it is advantageous.}{\textit{H. Poincaré}}

History shows us that the extension of the geometrical framework used in physical theory has several times allowed us to better understand natural processes. The greatest advances in this direction were, in my opinion, the passage from Euclidean geometry to Riemannian geometry. Each of these steps was motivated by physical arguments. The first geometry thought to describe space-time was a geometry of absolute space-time, as a background theater for events. Then with Galileo, space lost its absolute character by becoming that of Galilean reference frames. Then came Einstein who made geometry dependent on the observer, the notions of distances and durations relative to the latter, and the notion of simultaneity also, thus breaking the idea of an absolute time. Later, geometry became actively determined by the fields of matter.
\medskip
\par
I believe that the time of questioning the geometrical notions and their link with the concept of space-time is not over. Therefore, we can ask the following question: is there any reason to think that we need a new extended geometric picture to understand physical processes and space-time? Several arguments can be given for an affirmative answer. The purpose of the following considerations is to explore how the algebraic framework in which NCG was developed, and the concepts surrounding it may be relevant when we try to go even further in understanding what space-time is. In what follows, we will question the epistemological status of geometric descriptions of space-time, starting by asking how it is measured, and then considering how geometric concepts describing space-time, for example, have been progressively algebraized throughout history, this process culminating with the Gelfand-Naimark theorem and the creation of the NCG's framework.

\newpage
\section{Geometry and Space-Time}
\label{GeomAndST}
\epigraph{Time and space are modes by which we think and not conditions in which we live.}{\textit{A. Einstein, 1944}}

Geometry is a representation of the mind where the processes of reality are made intelligible, it aims to represent space-time. But what exactly is geometry, and what is the relation between its objects $\StrMath_{st}$ and those of physics? 
\begin{align*}
	\text{geometric object }\StrMath_{st}\overset{  ?  }{\longleftrightarrow}\text{Spacetime structure }ER_{st}:\StrMath_{st}\sim \pi_{\StrMath_{st}}(ER_{st}).
\end{align*} 
Is $\StrMath_{st}$ connected to a space-time element of reality $ER_{st}$? There is no proof of this, on the contrary, the supposition of the existence of an $ER_{st}$ leads to problems as we will see in section \ref{PointOnPoint}.
\medskip
\par
This question is more complicated than it seems at first sight. For example, a geometric object can be a distance, a duration, a point, or a coordinate. It must correspond to some structure of space-time. This means that we have to find an operational experimental way to deduce what corresponds to the given geometric object. This is the only objective access we can have to this structure and giving meaning to these concepts outside of this scope is probably misleading. For example, if we want to make sense of the concept of location for a material object, then we have to find an experimental operational way by which what we refer by this concept can be measured. If there is no such path, this concept becomes meaningless.  
\begin{remark}
	Note that these considerations on operational ways of measurement of space-time concepts often concern the use of photons and material objects. We will try to do the same in section \ref{ComptonScale}.  
\end{remark}
\medskip
\par
An important observation can then be made, the spatiotemporal (physical) structures $ER_{st}$ are not measurable. As Poincaré said:
\medskip
\par 
\textit{``Experiments only show us the relationships of bodies to each other; none of them deals, or can deal, with the relationships of bodies to space, or with the mutual relationships of the various parts of space.''} 
\medskip
\par   
We cannot directly observe space and its properties. We can only observe material bodies or light (and other bosons) phenomena. It is only through the observation of these material bodies and light phenomena that we can obtain information about the properties of space. Then these corresponding $ER$ are those of material objects. Therefore, the conclusions we draw about the properties of space-time concern the use of concepts and experiments about its material bodies and light experiment. 
\medskip
\par  
As I said, these experiments do not give us direct access to these notions, but to a set of observables $\{O_i\}$ with relations $\{R_{ij}\}$ between them, that allow us to deduce these notions. We must therefore conclude that observables (of material objects) are (on the operational level) primary to the notion of space-time.
$$\textbf{Spacetime is not an observable.}$$
It is important to stress that these deductions do not represent space-time in itself, but space-time as experienced by the processes of nature. This distinction becomes important when these processes start to affect the structure of space-time, or, more exactly, of the entities whose measurement makes it possible to deduce it. The experienced space-time becomes different from the ``pre-existing'' one. Thus, what we observe is space-time as it is experienced by material objects. The “measurement of space-time structure” can then be seen like this:
\begin{align*}
	\pi_\StrMath(ER)\overset{Measurement}{\longrightarrow}O(ER) \overset{Deduction}{\longrightarrow}\text{idealized “space-time structure” }\pi_{\StrMath_{st}}(ER).
\end{align*}
Notice that we use $ER$ a material element of reality instead of $ER_{st}$ a space-time element of reality since the latter are not measurable, only deducible from measurement on material objects, and then potentially are man-made creation. A space-time concept can need more than one element of reality to be deduced.  
\medskip
\par
In physics, we have to let the phenomena speak by themselves and take care not to allow our conscious representations to replace them, and eventually mislead us. What about the representation of space-time? We implicitly believe that space-time exists, but we never measure it directly. It is therefore a man-made concept whose fundamental nature can be questioned. We can thus ask ourselves what led us to the sensation/feeling of space-time. Ernst Mach, in \cite{mach1890analysis} thinks that space is constructed from the association, the ordering of our sensations (i.e collection of measurements made by the body). In the same philosophy an intuition about this was given by Poincaré: 
\medskip
\par 
\textit{The notion of space cannot be an integral part of any of our sensations taken in isolation. It is only when we observe the order in which these sensations follow one another that this notion can be born.}
\medskip
\par 
In other words, the collection data of observation change (of material objects) leads to the intuitive feeling of space-time. This set can be written $\Oset=\{\delta O_{a_i}(ER^j)\}^{i\in I_O}_{j\in I_{ER}}$, with $I_O$ the index set of all observer/measurement contexts and $I_{ER}$ the one of physical states/elements of reality. It's now important to distinguish between two kinds of observation change:
\begin{itemize}
	\item The ones which lead to the elaboration of our geometric intuition of space-time, we will call the set of all these observable changes $\Oset_{st}$.
	\item The ones which do not correspond to this usual geometric description, we will call the set of all these observable changes $\Oset_{nst}$ for non-space-time likable.
\end{itemize}
With the total set $\Oset=\Oset_{st}+\Oset_{nst}$. I will come back on this in section \ref{BackToObs}.
\medskip
\par 
The question is then to know how this collection of observables and observable change leads to the intuitive, then geometric representation we have of our world, based on concepts like distance, time, point, and coordinates. A first observation, which will be useful later is that (in many cases, maybe all) the collection of these observables together with the composition law form an algebra, and that, as mentioned in chapter \ref{DifCalc}, the observable variations are encoded in something called differential structure, which as we have seen in chapter \ref{DifCalc} enabled to complete the geometry to algebra unification. Therefore the intuition of Poincaré seems to be proved on the mathematical side since it is implemented in the differential structure, which was the key ingredient to recover something like geometry in the NCG framework.
\medskip
\par 
An element of response to explain how an impression of spatial representation, and then of geometry, can emerge from such a collection of observations is again given by Poincaré, (where he meets Lie and Helmholtz for the basic intuition). Indeed, for Poincaré, \underline{the intuition of transformation groups is primary to that of geometry}. For him, the study of geometry is only the study of a group (of 
transformations). To say that Euclidean geometry is the most suitable means that among the possible groups, the one constituted by the transformations of Euclidean geometry corresponds the most to our experience and in particular to the motion of natural solid bodies. As he said:
\medskip
\par 
\textit{Like Lie, I believe that the more or less unconscious notion of the continuous group is the only logical basis of our geometry. Like Helmholtz, I believe that the observation of the movements of solid bodies is its psychological origin.}
\medskip
\par 
Taking the group of transformations as the fundamental origin of geometric intuition, he adventured to imagine other worlds, where a different experience, with the observation of different groups could lead to the elaboration of other geometries, such as non-Euclidean geometry.
\medskip
\par 
Other researchers agree with him on this intuition.
According to Klein, geometry is a space, or a set of points, with a structure, and the bijective applications of the space onto itself that preserve the structure form a group which can be called the group of automorphisms. For Riemann, an automorphism of a Riemannian space is a distance-preserving application of the space to itself. We will show in chapter \ref{NCGFTGener} how the automorphisms of the algebra play a fundamental role in NCG, in order to find on the one hand the usual coordinate changes of general relativity, and on the other hand the local gauge transformations.  
\medskip
\par 
We thus see how the notion of observable and symmetry group on it is seen as the fundamental “reason” behind the intuition of space and its geometrization. As mentioned in section \ref{UsualSD}, these groups of symmetries on the observables form the automorphism group of the algebra of observables, which is connected (for infinitesimal automorphisms) to the notion of derivation, and hence of differential calculus. The argument that groups of symmetries on observables are more fundamental than the notion of geometry, and that we should deduce our notion of space accordingly, will be a crucial point in the argument of the sections \ref{BackToObs} and \ref{AlgeGeom}, where it will be a question of defending the relevance of geometrizing Gauge's theories, and of discarding the notions of geometry and space-time, in order to work on these concepts on the side of the algebra of observables.
\medskip
\par 
However, these statements assume that we are talking about measured space-time, not space-time in itself.  It is not clear then whether our theories speak of space-time in itself, or of the one we measure. If the way we measure it does not affect its structure, it is not essential, for all practical purposes, to differentiate these cases. But if this is not the case, we must know it, because the two cases will give a different description of what space-time is. I have found no indication in the literature on special and general relativities regarding this point of interpretation. But if we go back to the method used by Einstein and Poincaré to discover special relativity \cite{hladik2004comment}, we notice that the concepts of duration and distance are studied according to the operational mode of measurement by the photons of light, whose speed is independent of any reference frame. This led them to a deeper understanding of the concepts of duration, distance, and simultaneity. This is for me a strong indication that relativistic theories are descriptions of measured space, another indication being that, if this theory is a physical theory, then it must predict what we observe, then the measured space-time. In this line of thought, we will test the notion of point in section \ref{ComptonScale}, using a similar measurement scheme, with photon and material bodies.
\medskip
\par 
More details and the origin of the citations used in this section can be found in \cite{michel2004reflexion}, \cite{nabonnand2010theorie}, and \cite{bachtold2014geometrie}.

\section{The Point on the Point: is Space-Time Substantial?}
\label{PointOnPoint}
\epigraph{Formerly, people thought that if matter disappeared from the universe, space and time would remain. Relativity declares that space and time would disappear with matter.}{\textit{A. Einstein}}
Whether space-time corresponds to something having tangible existence/substance (manifold substantialism view of general relativity) or not is a very difficult question. The theory of ether was supposed to describe “what fills” space-time as a kind of fluid in which material objects undergo displacement. But according to Einstein in \cite{einstein2007ether}, the relativities show us that if there is such an ether, it must not possess any mechanical properties like velocity for example. This seems to be far away from what we can think of as something being substantial (i.e. corresponding to an element of reality). 
\medskip
\par 
Let's suppose that there are such elements of physical reality corresponding to elementary space-time structures $ER_{st}$. Let's suppose that this elementary structure corresponds to what we call space-time points, we will then call it $ER_x$. Let's try to test whether such an element of reality can truly be consistent with observations. 
\medskip
\par 
As mentioned in section \ref{Realism}, the origin of the concept of element of reality comes from the fact that the RC (repeatability criterion) is respected, \textit{i.e.} one or more observers will make predictions in agreement with each others when they do repeated measurements on the same object at not-so-distant time intervals. This therefore equally applies to space-time structures. Take the notion of point and the corresponding element of reality $ER_x$. If there is such a physical notion, \underline{independent of any material object}, then, a series of idealized measurements \footnote{Because as said before, space-time is not measurable since only observation of material objects is possible, and that there are no material objects to measure in this idealized measurement scheme} of this notion of point by different observers should all coincide with the following result \textit{i.e.} that $ER_x$ describes a substantive point in space-time with a probability of 1. But we know that the notion of a point depends on the chosen reference frame, so different observers will see either a point or trajectories in space-time. Indeed, any test of the notion of a spatial point without the notion of a material object of reference is doomed to failure, because of the physical equivalence of all Galilean reference frames. The question of the difficulty of conceiving physically an empty space, without material objects of reference had already been raised by Descartes, a similar argument using a box containing space as the elementary tool can be found in \cite{einstein1990theorie}.
\medskip
\par
Similarly as mentioned in section \ref{GeomAndST}, spatial structures are not measurable, only the material objects from which we deduce spatial concepts are. Thus, from the point of view of the physicist's thinking, concepts of space-time are derived from those of observation of material objects and thus are not necessarily supposed to represent a real physical entity in their own right, as distinct from what matter is. There are many other arguments against the idea that space-time is an autonomous physical entity.
\medskip
\par 
A strong argument against giving physical meaning to points ($x$) in general relativity was given by Einstein with the hole argument. The general covariance requirement on the Einstein field equations implies that if we consider a hole in space-time \textit{i.e.} a region where the stress-energy tensor $T$ vanishes, a change in the coordinate system $x\,\to\, x'$ taken to be only inside the hole will induce different metrics $\tilde{g}$, for only one $T$. But the new metric is also a solution in the first coordinate system $x$. Therefore, according to the “points” $x$ of $\Man$, there can be as many distinct trajectories as they are different metrics, but for only one matter field configuration given by $T$. This is problematic if we consider these points as being true objective (and then distinct) locations. Thus, either we restrict ourselves in the set of possible coordinate changes and lose the general covariance, or we lose the deterministic property of the field equations concerning the trajectories which are its solution. These two solutions are not acceptable. This can also be seen as a consequence of diffeomorphism invariance, see \cite{stachel2014hole, giovanelli2021nothing} for technical details.
\medskip
\par
Some time later, Einstein finally realized that this argument is based on an implicit assumption we have on space-time, and therefore on its symbolic representation as being constituted of points $x$ on which physical properties like the value of a gravitational field (or equivalently of the $T$, then of any field associated to $x$) are defined in a meaningful way. By relaxing this requirement, he makes general covariance and determinism compatible. But how do we provide meaning to “points” $x$ if they are not therefore intuitively related to “locations” of physical quantities? How do these coordinates relate to objective locations?
\medskip
\par
The answer was given by Einstein as “the point coincidence argument”. This means that true objective locations are not given by coordinates $x$, but by crossings of world lines which are solutions of the Einstein fields equations. Indeed, the only observable structures of space-time are point coincidence events, interaction of a part of the observer apparatus, and the corresponding structure. As he said in \cite{einstein1923grundlage}:
\medskip
\par
\textit{Our space-time verifications invariably amount to a determination of space-time coincidences. If, for example, events consisted merely in the motion of material points, then ultimately nothing would be observable but the meetings of two or more of these points. Moreover, the results of our measurements are nothing but verifications of such meetings of the material points of our measuring instruments with other material points,
	coincidences between the hands of a clock and points on the clock-dial, and observed point-events happening at the same place at the same time. The introduction of a system of reference serves no other purpose than to facilitate
	the description of the totality of such coincidences.}
\medskip
\par
He considers this argument as one of its deepest insights. This was a strong counter indication to the manifold substantialism view since its points are not referring by their intrinsic structure to locations. Then, the coordinate systems have no intrinsic meaning. The physical significance of the notion of space-time point is then dependent on the concept of matter as moving inside them. The notion of point must therefore be operationally based on such coincidence, this will be used in section \ref{ComptonScale}. 
\medskip
\par
Therefore, is it still possible (according to these arguments) to think of empty space, with locations, as mentioned above? Going back to the hole argument, if we say that space-time does exist where there is no matter, we realize that the only way to specify an adequate notion of point is to invoke the geodesic crossings, and thus from a physical perspective, to reintroduce at least one particle into the hole which must interact with our test particle. Then, if we implement operationally the point coincidence argument, this means that to implement an objective notion of point, the hole is no longer a hole since it must contain a particle with which coincidence permits to set up this notion of point.
\medskip
\par 
So we can see that the idea of thinking of space-time physics as based on a notion of a manifold $\Man$ with points leads to paradoxes and strange postulates about what these points are. One way to get rid of this problem could be to find a formalism where $\Man$ does not play a primary role. The formalism of general relativity consists roughly in supposing the existence of a smooth manifold $\Man$ endowed with a metric, on which we consider the set of tensorial operations that we can do on it, and thus construct equations whose structural properties encode the relevant structures of $\Man$. An interesting observation is that $\Man$ only comes into play in formalism as a tool to introduce the ring of smooth functions on $\Man$ \cite{geroch1972einstein}. Everything that follows in the developments is purely algebraic. It is this remark that led to the development of Einstein's algebras, which are exclusively based on this ring. By taking this object as first, without any reference to a manifold, one can reformulate general relativity, without being troubled by $\Man$ and the paradoxes that come with it when it is taken as input.  
\medskip
\par 
As mentioned by N. Huggett in \cite{huggett2018philosopher}, Earman suggested that taking the algebra as more fundamental than $\Man$ and its points would avoid the hole problem about the meaningfulness of coordinates by virtue of locations. This was seen as a replacement of manifold substantivalism by algebraic substantivalism. In this view, points are seen as deductions of matter fields, and of the elements of the corresponding algebra.
The idea that algebra should be taken as more fundamental than space-time and these points was defended by J. Earman and in Geroch's work in an attempt to bring arguments around the substantivalist-antisubstantivalist debate revived with general relativity and its conceptual problems.   

\medskip
\par 
I cannot conclude from these arguments that only the antisubstantivalist position makes sense, but insofar as this would be true, and that physical elements of reality are needed to explain the origin of space-time, then only the elements of reality offered by material objects can be used to derive such a space-time structure and it's representation. As we see, this assertion is supported by the point coincidence argument. In chapter \ref{ObjectiveNCGFromQM}, I will present how some essential structures at the heart of space-time's concept such as the notion of point and continuity between two points can be offered by pure matter concepts. In section \ref{ComptonScale}, I will use these arguments to show the limits at which such an operationally implemented notion of point failed, opening the door to NCG... A nice review about the Relationalism versus Substantivalism debate can be found in \cite{field1984can}.

\section{Back to what is Observable, Consequence for Gauge Theories}
\label{BackToObs}

In section \ref{GeomAndST}, arguments have been put for emphasizing the primary status of observables over our usual representation of space-time as an entity in itself. Let us then try to reproduce the path of creation of the intuition of space proposed by Poincaré, without presuming anything. A first observation is that we will acquire, by a set of measurements, a whole state of observables, and data on their evolution. Our mind then begins to make representations in which this shapeless mass of observables, as well as their evolutions, can make sense of these data, according to the symmetries it perceives in the structure and evolution of these observables. Let's start with the notion of point. Usually, points are taken as input in all physical theories, \textit{i.e.} we suppose point $x\in\Man$ as pre-existing entities on which events can occur. But we have to admit that these notions of point are necessarily reconstructions from the observables we collect, and concecuntly, these observables have to be considered as more fundamental than these points. As mentioned in section \ref{GNMath}, the Gelfand Naimark theorem provides such a link between observable and points of an underlying “hypothetical manifold”. This theorem has a deep meaning, it implements a part of Poincaré's idea presented in section \ref{GeomAndST}, \textit{i.e.} space-time geometrical notions have to be deduced from the set of observations we can collect. Indeed, if we take again the equality \refeq{GelfTf}, $\hat{f}\in\algA$ being such an observable, $x$ a character on $\algA$, and $f$ the inverse Gelfand Transform of $\hat{f}$, then we have: 
\begin{align*}
	\hat{f}(x)=x(f)
\end{align*}
This shows that we can see these two views as equivalent:
\begin{itemize}
	\item space-time as fundamental $\to\,\hat{f}(x)$: Take the points $x\in\Man$ as fundamental, and then interpret measurement results as evaluations of observable on these points.
	\item Algebra of observables as fundamental $\to\, x(f)$: Take observables $\hat{f}\in \algA$ as fundamental, and points as special functions evaluating these observables.
\end{itemize}
We saw in section \ref{PointOnPoint} that the notion of point being a source of problems, the approach starting from observables (Einstein Algebra) could be considered as more fundamental. Moreover, as defended in section \ref{GeomAndST}, the structures of space-time are not measurable, only the observables on the matter are. And finally, as mentioned in sections \ref{PhaseVSConfigSpace} and \ref{KvN}, we saw that the first approach which consists in the evaluation of observables pointwisely is not defined (because of the NC), whereas the algebra of observables is. This is why I think the second approach is more fundamental. Then, as mentioned in \cite{coquereaux1998espaces}, the measurement process can be taken as a replacement for the first given consideration of the geometric background space. 
\medskip
\par
Let us thus resume our initial apprehension of what space-time is, starting from the observables and their variations as mentioned in section \ref{GeomAndST}. As said above, a given proportion ($\Oset_{st}$) of these observables (and their evolutions) leads to our usual representation of space-time, while the other ($\Oset_{nst}$) can rather be perceived as moving in non-spatialized dimensions, in purely formal structures, remaining attached to points of space-time. These non-spatialized evolutions can be for example the evolution of color at a given point, of a fermionic field according to its different charge components, of a spin, or simply of any properties of matter without spatial extension, and manifesting a temporal evolution. In the first case, we speak of reconstructing notions of trajectories in space-time and in the second case of evolution in a non-spatial sense. But what determines these trajectories and evolutions?
\medskip
\par 
The most advanced theoretical view is to consider that trajectories are the results of the path for which the phase interferences of the wave function of a particle are the more constructive. It is then the evolution of the phase of the particle that becomes the primary theoretical object of these notions of trajectory and evolution. The phase is the evolution parameter taken as an argument of the function whose variation is considered.
\medskip
\par 
In geometry, the evolution of this phase along paths is encoded by the notion of parallel transport, implemented by connections. We therefore observe trajectories, which are derived from the evolution of the phase, which is parameterized by the connections that encode the parallel transport. More generally, given the collection of phase observations, we are free to interpret the infinitesimal \underline{irregular} changes of the phase in two ways:
\begin{itemize}
	\item Either as real phase changes (those induced by gauge potentials), which we will say are induced by forces.
	\item Either as false phase changes (those induced by gravitation), thought to be simply induced in appearance by the non-regular structure of the background space.
\end{itemize}
\medskip
\par 
Apart from technical conveniences, I have not found any conceptual or theoretical reasons for geometrizing one force over another. What is objective, in short, are the observational changes in the material objects we collect. A certain portion of these changes ($\Oset_{st}$) can make sense in a geometric scheme, in this case Riemannian (for gravitation), and we are free to do the same for the others ($\Oset_{nst}$) if we find such a geometric representation. A little-known fact is that Einstein himself did not defend the idea that General Relativity was a geometrization of gravitation, he rejected this hypothesis, which made no sense to him, more details on his position can be found in \cite{lehmkuhl2014einstein}. This point will be further defended in the next section \ref{AlgeGeom}. 
\medskip
\par 
Another argument that makes connections as non-concrete as coordinates in terms of the idea that they represent something physical is provided in \cite{healey2007gauging}, where it is shown that gauge potential cannot be thought of as related to a physical (element of) reality, only curvature can be.
\medskip
\par  
As we will see in part \ref{TowardNCGFT}, gauge theories offer a theoretical framework where the whole set of forces, and thus phase evolutions, is encoded. Many researchers such as Einstein, Kaluza and Klein, and others have sought to describe the interactions of the SMPP in a unifying scheme with gravitation, but this has not been successful. As mentioned in chapter \ref{DifCalc}, differential structures can be understood as the study of variations in the general meaning, this variation can be thought of as being along some degrees of freedom, abstract or concrete, like space-time degrees of freedom (coordinates), another observable degree of freedom, or abstract numbers. Here the variations of interest being the ones of the phase, we will see that the NC extension of these differential structures and then of the geometric framework offers a way to make gravitation and the other interactions on the same footing, in a single unified formalism, making the interactions of the SMPP appear like pseudo forces induced by the underlying NCG.

\newpage
\section{The Way of the Algebraization of Geometrical Concepts}
\label{AlgeGeom}
\epigraph{Algebra is nothing more than geometry, in words; geometry is nothing more than algebra, in pictures.}{\textit{S. Germain}}

In the previous sections, I have highlighted the links between algebraic and geometric structures, presenting arguments for the fact that observables were primary to the geometric intuition of space-time. However, it is not yet clear that all geometric properties can be encoded algebraically. For example, we have seen that the notion of point, having long resisted algebraization, finally yielded this tendency thanks to the Gelfand Naimark theorem. In this section, I propose to show the historical path of this algebraization of geometrical notions, whose culminating point can be considered as the creation of the NCG framework, previously presented in part \ref{PartCalToNCG}.
\medskip
\par 
Together with arithmetic, geometry is one of the oldest mathematical fields. It comes from ancient Greeks, with the first meaning being “land measurement”. The first purpose of geometry was to understand spatial properties such as distance, relative positions, and size... It was initially done with compasses and rulers. Its objects (points, straight lines,...) are familiar to us, certainly, because they are connected to \underline{“a priori” given representations of our mind.} We can see that these objects are the basic axiomatic objects with which Euclid initially played to give axiomatic foundations to geometry. But I think that we have to admit that these notions rely on very human-like representations probably inherited through the process of natural selection. For example, the Euclid axioms are based on primary objects such as points, lines, distances, and relative locations... which only make sense for our inherited representations of the world, and which are very difficult to explain in detail. The notion of a point, for example, which appears simple, is a profound concept, very difficult to define. This is why I think that geometry is a more subjective mathematical science than others. 
\medskip
\par 
History shows that geometry, as a mathematical discipline, was gradually absorbed by algebra and that each of the primary objects qualified as subjective has been incorporated and made more meaningful in the algebraic frameworks. This can be seen as a 4-step process (to simplify). 
\medskip
\par 
The first step can be seen as the invention of coordinate systems by Descartes, implementing distance and relative locations. Descartes is the creator of analytic geometry; perceiving this discipline as an “algebraic presentation of the ancient's geometry”. This means that he reduced geometric problems to calculations of length and translated geometric questions into algebraic equations. Before Descartes, it was understood that algebra and geometry were completely separate branches of mathematics with no connection between them. His work is the first to propose the idea of uniting algebra and geometry in a single discipline. The key to this unification was the notion of coordinates. 
\medskip
\par 
In mathematics, coordinates are numbers used to locate a point in relation to a coordinate system. In physics, these numbers are the result of observables, such as the measurement on a graduated ruler, the time taken for light to travel between two material objects, etc. In mathematics, a system of coordinates allows each point in an N-dimensional space to correspond to one (and only one) N-tuple of scalars. In many cases, the scalars considered are real numbers, but it is possible to use complex numbers or elements of any commutative field. 
\medskip
\par 
It is important to stress that this definition is based (in input) on the existence of the notion of point and commutative field for the associated coordinates. These coordinates are our starting point for understanding the different generalizations of geometry, as were the Euclidean coordinates in an orthonormal frame of reference, and then the curvilinear coordinates for manifolds.
\medskip
\par 
Thus, analytic geometry is an approach to geometry in which objects are described by equations or inequations using a coordinate system. Then in the case of Euclidean geometry, a choice of Cartesian coordinates allows the sophisticated axiomatic deductions of geometry to be transported into the field of algebraic calculations.
\medskip
\par 
A second step can be seen as containing all attempts to find deeper characterizations of topology and geometry of a manifold $\Man$ (and more general geometries later) through algebraic objects like those based on differential operators. Spectral geometry can be seen as one of the most interesting and prolific research areas in this direction, providing a lot of characteristic information about topology and geometry from differential operators defined on $\Man$. To be more precise, Spectral geometry is the study of the relationships between geometric structures on $\Man$ and the spectra of differential operators like the Laplacian. Two directions were followed as attempts to answer to these two different questions:
\begin{itemize}
	\item The inverse problems direction: Which geometric feature can be recovered from the knowledge of the eigenvalues of the Laplacian? 
	\item The direct problems direction: Knowing the geometry of a Riemannian manifold $\Man$, what is the behavior of the eigenvalues of the Laplacian?
\end{itemize} 
The first question is more commonly expressed as “Can one hear the shape of a drum?”. The answer was the negative, these differential operators don't fully characterize the underlying geometric structure; two different objects with nonequivalent geometries were found to have the same spectrum.  
\medskip
\par
It is interesting to underline that such differential operators are based on the differential structures, which as mentioned in chapter \ref{DifCalc} where essential structures to complete the equivalence between algebra and geometry. Similarly, such operators appear almost systematically in all of the most important equations in physics. Indeed the spectrum of the Laplacian, or more generally of the d'Alembertian, is connected to important physical properties of sounds, light, heat, and atomic process. It is present in:
\begin{itemize}
	\item The Heat equation: $\partial_tu=\alpha\Delta u$ whose eigenvalues give the time decay rates of eigenfunctions in time.
	\item The wave equation: $\square u =0$ governs light wave propagation for example, harmonic frequencies being given by the eigenvalues of the Laplacian.
	\item The Schrodinger equation: $i\hbar\partial_t\psi =-(\hbar^2/2m)\Delta \psi + V\psi$.
\end{itemize}
It is interesting to note that in the case of the Heat equation, the computation of the heat kernel (function of the spectrum) and then something called the heat trace (spectral invariant of interest), lead to an asymptotic expansion of this trace where the first three moments were connected to important geometric data:
\begin{align*}
	a_0=\operatorname{vol}(\Man), \quad a_1=\frac{1}{6} \int_\Man s, \quad a_2=\frac{1}{360} \int_\Man\left(5 s^2-2| R_{\mu\nu}|^2-10|R_{\mu\nu\rho}^\lambda|^2\right)
\end{align*}
$s$ being the scalar curvature, with Ricci and curvature tensors given by $R_{\mu\nu}$ and  $R_{\mu\nu\rho}^\lambda$.
\medskip
\par 
This is very similar to the way in which a spectral invariant is constructed in the spectral triple framework as we will see in subsection \ref{SpectralAction}, which will in the same way connect to important geometrical data as we will see. More details about spectral geometry and this asymptotic expansion can be founded in \cite{cruz2003spectrum} and \cite{urakawa2017spectral}.
\medskip
\par 
A third step encompasses the work that tries to find topological information from algebra, \textit{i.e.} by deducing from algebra the notion of set, and therefore of points, and then of topology. The first realization was to find a duality between complete Boole algebras and sets. After this around 1930, Marshall Stone set up a link between mutually commuting projections on the Hilbert space and Boole algebras, making a link between these projections and elements of a set, but this link is not complete since it is not implying that this Boole algebra is complete. Later Stone duality shows the equivalence between compact completely discontinuous spaces and Boolean algebras. The remaining question is what kind of algebra can correspond to non-discontinuous compact spaces, such as $\bbR$ or $\Man$? As mentioned in section \ref{GNMath}, the answer was given by the Gelfand duality, which provides a complete link between topological data, and the algebra of observable. This was a crucial step since it absorbs the concept of point which was previously taken as an input.
\medskip
\par 
The final step can be considered as the elaboration of the NCG framework, giving a differential structure and then the equivalent of a geometrical structure to the algebra of observables as shown in chapter \ref{DifCalc}. NCG only exists in the algebraic setting, mainly because inputs of the geometric framework like points are relegated to the secondary level in the restrained situation of commutative algebras. The algebra of observables establishes itself as a background (by replacing $\Man$), providing a framework that encompasses the equivalent of all that was done in the usual geometric one, and goes far beyond.

\chapter{\texorpdfstring{From Quantum Theory to NCG: Geometric Consequences of the $\psi$-ontic Interpretation}{From Quantum theory to NCG: geometric consequences of the psi-ontic interpretation}}
\label{ObjectiveNCGFromQM}

\epigraph{When forced to summarize the general theory of relativity in one sentence: Time and space and gravitation have no separate existence from matter.}{\textit{A. Einstein}}
The Gelfand Naimark theorem establishes a one-to-one correspondence between the algebra of continuous “functions” and the topology of an underlying space. As we already said, our intuition of space comes from the observations we gather about our environment. The question is therefore whether we should seriously consider the information possessed by the quantum states and find the correct intuitive geometrical representation coming with it. 
\medskip
\par
If we look at QM with $\psi$-epistemic eyes,  $\psi$ does not refer to an objective reality, but to a tool for predicting the results of experiments. Then, there are no geometrical consequences to be deduced from it. But with $\psi$-ontic eyes, as the state corresponds to something objective, then, we have to consider it seriously, and try to understand what is the corresponding geometrical view.
\medskip
\par
In section \ref{Realism}, I provided arguments in favour of the $\psi$-ontic interpretation. Then, in sections \ref{GeomAndST} and \ref{BackToObs}, I gave arguments defending the fact that the observables should be considered by the physicists as primary to the notion of space-time, which would be an “a posteriori” deduction. These arguments provide a strong indication of the fact that one must take the formalism of QM as seriously as that of classical theories regarding its representational fidelity to elements of reality, and consider the induced geometrical consequences. This will be partially done in what follows, by relying on the links between topological and algebraic realities presented in section \ref{GNMath}.
\medskip
\par
In this chapter, we will see how the elementary constituent of the notion of space, \textit{i.e.} the notion of point, can be deduced from the QM formalism in section \ref{PurToPoint}, and then, in section \ref{EntangConnect} I will outline a way to understand what can fundamentally connect these points and thus be the potential origin of the notion of continuity for spaces. Then, in section \ref{ComptonScale}, I will show how, when and at what scale this notion of point loses meaning, in connection with the NC. We will see in particular that this happens for scales larger than the Planck scale $\lambda_p$. As I will discuss in section \ref{NCSMPPGeomForces} these arguments could eventually provide a physical justification for the NCSMPP... Then in section \ref{NaturInner} I will show how, when restricted to unitary operators (without collapse), the notion of inner derivation presented in section \ref{DeDS}, acquires a natural meaning. To finish, I will make the synthesis of all what has been said in this part, and of the vision of the NCG that I deduce in section \ref{PhiloNCG}.

\newpage
\section{Gelfand Naimark Theorem, Pure States and Points}
\label{PurToPoint}
In general relativity, $x$ represents point-events, \textit{i.e.} potential supports of an interaction between particles, or the simple presence of one of them. The implicit view is that there is a space-time $\Man$, a point $x\in\Man$, and that a particle can eventually be found at this point. In this view, the point is primary to the presence of a particle. In section \ref{PointOnPoint}, I have presented some arguments that go against this view, \textit{i.e.} defining a notion of point, independent of the presence of a material object in it, while remaining objective (using the realism criteria set up in section \ref{Realism}) seems to be unsuccessful. A possible conclusion of such a reasoning is that the event particle is upstream of this notion of point, which would be in line with the arguments presented in sections \ref{GeomAndST} and \ref{AlgeGeom} concerning the primacy of the algebra of observables (concerning material objects) over space-time notions. In QM, it is the $\psi$ states that represent the famous point-event particles of which the current theory of gravitation speaks. As we have seen, the Gelfand-Naimark theorem tells us that it is the pure states that provide the notion of point, from the algebra. Hence, given these arguments, it seems likely that these pure states are those of material particles, and that the notion of point to which an observer has access thereby arises like this. In what follows, it is assumed that the pure states at the origin of the notion of points considered correspond to states of material particles.
\medskip
\par 
Gelfand Naimark theorem said that if there is a pure state $\pi_\StrMath(ER)$ (corresponding in a bijective way through PBR theorem to a physical state $ER$ as seen in \ref{Realism}), then in the case of commutative algebras, we can construct a point $x$ from this $ER$. As coordinates are deducible from these observables, then, only commuting observables will lead to a good notion of coordinate (associating numbers to points in a non-contextual way):   
\begin{align*}
	\text{Point}\equiv \text{pure state}+\text{commuting observables}\to \text{Notion of coordinates}
\end{align*}
As this depends on the projection property of the state, it will not be possible to associate such point events with mixed states. 
\begin{remark}
	\label{RQInnerGeomFiniteRepSpace}
	Therefore, we can see that pure states provide the notion of space-time points. But these pure states also contain degrees of freedom that are not spatializable (Dirac fermion and anti-fermion component of the wave-function, left/right component, spin, and other quantum numbers being in superposed states into the total pure state). Then if the category equivalence behind the Gelfand-Naimark theorem is to be taken seriously, these internal degrees of freedom must therefore know their geometric counterpart. As we will see, this is partially done in the SMPP, where these internal degrees of freedom are taken to be the element of a finite space $F$, which will be named the fermionic representation space. Then any pure state can be associated with a notion of point, with inner degrees of freedom given by these inner structures, that can be geometrized as we will see in remark \ref{rqInner} and chapter \ref{NCGFTGener}.
\end{remark}
\begin{remark}
	Making such a link between a pure (symbolic) state and objective points is meaningless in $\psi$-ESR theories ($\psi$-epistemic theories admitting the existence of a physical state) because one physical state can correspond to two distinct symbolic states, then to two distinct points \footnote{Without any mention to coordinate change or move in space-time to explain the two different symbolic formulations of the same point given to be attached to the physical state}. This cannot be linked to any objective physical notion of point, since a particle is defined to be able to interact with others at a given location, not at two at the same time.
\end{remark}
An interesting terminology issue will be to no longer state that a particle is at a point $x$, but that the particle is the point. This connects with the point coincidence argument presented in section \ref{PointOnPoint}, where the impossibility of defining a relevant notion of point without the pre-existence of a particle with which to interact in this “place” was considered. This finally appears to be logical since the only observed points are material objects...
\medskip
\par  
Taking $\psi$ to be a pure state, we call $\Dst(\psi)$ the space-time domain where the presence density $\rho=\psi\psi^*$ of $\psi$ is non-zero. Then the only notion of point event we can associate to $\psi$ is one $\Dst(\psi)$. This extends the notion of point: usual points are recovered when decoherence and collapse occur during classical measurement processes, making $\Dst(\psi)$ very “localized”.
\medskip
\par 
Then we can see that we have points, coming from the physical states of material objects, which can offer candidates for the points events described in general relativity. But how do these points relate between themselves, and how connectedness between such points can be implemented, using concepts belonging only to the realm of material objects?

\newpage
\section{Entanglement and Connectedness}
\label{EntangConnect}

Entanglement is usually conceived (at the formal level) as “non-separability” of the tensor product in the Hilbert space. Two states $\psi_1$ and $\psi_2$ are said to be entangled if the joined state $\psi_{1,2}$ is not the tensor product state $\psi_1\otimes\psi_2$. Now, what does the $\psi$-ontic interpretation tell us? It shows that this description is not only a formal one, but that it represents directly a kind of physical reality. Let's see what this means for our algebraic to topology correspondence.
\medskip
\par
Entanglement has a direct geometric correspondence among the equivalences listed in table \ref{figGN}. This is given in a transparent way by the equivalence provided between connectedness and projectionless. If we consider the previous join pure quantum state $\psi_{1,2}$, its reduced states (in $\psi_{1}$ and/or $\psi_{2}$'s basis, when taking the partial trace) are mixed so that each of them cannot be associated with a point event as mentioned in section \ref{PurToPoint}, only the total state can. Therefore, the point is the union of the two particles, and $\psi_1$ and $\psi_2$'s elements of reality cannot be seen as disconnected pieces (in the topological meaning). Then two entangled particles form a point, even if they are far away in our usual space-time representation of the world.  
\medskip
\par 
Then, entanglement between $ER_1$ and $ER_2$ leads to connectedness of the associated space-time elements $\tilde{x}_1$ and $\tilde{x}_2$ which are not points, only the joined physical state $ER_{1,2}$ being one. 
\medskip
\par 
Many physicists believe that any mixed state $\psi$ is the projected sub-state of a larger pure state $\psi_{tot}$, where it shares entanglements with other sub-states of this $\psi_{tot}$. There is no consensus on the reality of such an assumption. However, it is certain that a pure state which interacts with its environment becomes mixed, and I do not know of any other process which makes it possible to go from a pure to a mixed state. In the previous section \ref{PurToPoint}, the fact that mixed states cannot be linked to the notion of point was mentioned. Then if we claim that any mixed state $\psi$ can be seen as a sub-state of a bigger pure state $\psi_{tot}$, then there is a possibility to reconstruct such a notion of point (then location), through this purification procedure. 
\medskip
\par 
The fact that entanglement may be at the origin of the continuous structure of space-time is at the very heart of the Einstein–Rosen $=$ Einstein–Podolsky–Rosen conjecture. This conjecture proposes an ambiguous connection between two concepts, one existing in the field of observables (then algebra), the other, in that of spatial concepts (then manifold, with metric). Nothing explains the link between these two concepts in the claims of its advocates. Fortunately, the previous assertions give an unambiguous connection between these two concepts and their representations. I think that the concepts of space-time in general relativity, and of observable in QM must be unified in the purely algebraic framework offered by NCG. 
\medskip
\par 
Points of the manifold correspond to point-events in physics, two entangled particles thus form a single event. I take such an entanglement as a potential candidate for the implementation of the concept of point coincidence mentioned in section \ref{PointOnPoint}. This means that concerning a material object, an associated objective notion of point can only be built using its link with another material object, this link being interaction or entanglement (which are often linked since entanglement is created by interactions). 

\newpage
\section{Position Measurement, Compton Scale and NCG}
\label{ComptonScale}
\epigraph{If one wants to be clear about what is meant by “position of an object,” for example of an
	electron..., then one has to specify definite experiments by which the “position of an electron”
	can be measured; otherwise this term has no meaning at all.}{\textit{W. Heisenberg, 1927}}

Space-time is represented as a manifold made of points-event. The mind likes it very much because the points are easy to think about, they are concrete objects and support our intuition of things. But does this view resist to experience?
\medskip
\par 
It is commonly admitted that at the Planck scale $\lambda_p$, the notion of localization via the coordinates' measurement of a particle loses its meaning. The argument used, developed in \cite{doplicher1995quantum}, deduces from general relativity and the wave nature of particles (linked to the UP) that the measurement of the position of a particle beyond a certain precision requires the intervention of a high-energy mediator which generates a black hole, for dimensions of the order of $\lambda_p$. This paper seems to be a reference taken by Connes in \cite{connes1997noncommutative} to physically motivate NCG, and represents the general line of thought regarding the conjecture of a limit scale at which the notion of space-time supported by a Riemannian manifold loses credibility. It is interesting to note in this approach that it is the measured space-time that is no longer a Riemannian manifold, which seems to be in the same line of thought as the arguments presented in section \ref{GeomAndST}. Moreover, these statements clearly seem to imply that the act of measurement modifies geometry, but this important conclusion is not made explicit.
\medskip
\par 
In this section, we will test the spatio-temporal notion of a point event. To test the notion of a point, we need to implement an experiment that gives us access to this notion. A first observation is that if we do not measure absolute position, but the position of a material object, then the notion of a point in space-time has no objective meaning without that of the measurement of a material object with respect to a given object taken as reference. Imagine that we want to test the notion of position for the most elementary material object we have, a particle. The only way to do this, without absorbing or totally disturbing the physical state of this particle, is to send a light beam on it, which will be scattered by the Compton effect (for charged particles). This deviation will provide an indication of the presence of the particle in the corresponding space-time area. This is the essence of the microscope experiment argument mentioned in section \ref{NCObsEf} and an implementation of the point coincidence argument presented in section \ref{PointOnPoint}.
\medskip
\par
As we see, the photon possesses a wavelength $\lambda$, which must be smaller than the desired scale of localization to permit to do this measurement. But the more you want to know precisely the localization of the electron, the more you need a photon with a small wavelength (then big energy), and the more it affects the state of the electron, changing its measured momentum. If we take the particle designed by $\psi$ to have a mass $m_\psi$, then its mass-energy is given by $E_\psi=m_\psi c^2$. When the energy of the photon used to measure the position starts to be of this order, then this measurement scheme starts to change the momentum of the particle. The wavelength of the photon is connected to its momentum through $p_\gamma=h/\lambda$. If we take $\lambda$ to be the scale at which we want to measure the location of $\psi$, then we see that it starts to affect in a non-negligible way the momentum $p_\psi$ when $p_\gamma$ starts to be close to the mass-energy $E_\psi$. Equalizing these two parts give the Compton wavelength:
\begin{align*}
	E_\psi=p_\gamma c\qquad\to\qquad \lambda_\psi^c=\frac{h}{m_\psi c}
\end{align*}
In other words, the Compton wavelength is nothing but the wavelength of a photon whose energy is the same as the rest-mass energy of the particle. Then if we want to test the notion of location for a fermionic particle $\psi$, it starts to change its momentum when trying to measure this location under $\lambda_\psi^c$ scale precision. As mentioned in section \ref{NCObsEf}, this is the origin of the NC of position and momentum observables. But as proved by the Gelfand Naimark theorem, and defended in chapter \ref{GeomAndSpacetime}, observables are not only linked to the space-time picture but first to it. Then, this NC is linked to a NCG which starts to occur at $\lambda_\psi^c$ scales.  
\medskip
\par 
The same appears if we measure momentum, and that $\psi$ is not in momentum eigenstate basis, $\psi$ undergoes change, and the position measurement before and after the momentum measurement reveals different positions. Since momentum and position are observables that we can obtain in space-time, then, it is not only a NC in phase space but also in something which can be seen as an extension (in an unknown meaning) of $\calC^\infty(\Man)$. Therefore, if as Heisenberg we take physical theories to give the prediction of what we observe (“measurement=meaning principle”) so that we restrict our notion of reality to what is accessible by experiments, then, a correct description of the structure of space-time at Compton scales cannot be made in a geometrical framework where the fundamental elements are points.  
\begin{remark}
	It is interesting to note that when Einstein and Poincaré studied the notions of duration, distance, and simultaneity, they proceeded similarly: by testing these notions with experiments where light rays are used to “measure” these notions. It is in this regard that the previous comments on the test of the notion of a point can be seen in this line of thought. 
\end{remark} 
More details on the minimal length and the link between Compton scale and uncertainties can be found in \cite{MinLenghtMoni} and \cite{ReducedComptWavPremov}.
\medskip
\par
The Compton wavelength of a particle is also the scale at which quantum field theory starts to be needed to obtain accurate predictions. The fact that this coincides with the second quantization (the one of bosonic and fermionic fields) can be understood thanks to the above arguments. Such processes, using gauge bosons for example, become so energetic that particle creation and annihilation start to be possible. 
\medskip
\par
Thus, the two pure NC encountered in theoretical physics (that of QM, and that of quantized Gauge connections) can be associated with the non-passivity of the act of interaction represented by the multiplication of the mathematical representations $\pi_\StrMath(ER_1)$ and $\pi_\StrMath(ER_2)$ of the interacting entities $ER_1$ and $ER_2$. The interaction process is represented mathematically by the multiplication operation.
\medskip
\par
Therefore, the notion of point, and then of Riemannian manifold, loses meaning well before reaching the Planck scales $\lambda_p$. It loses effect at different scales, depending on the mass of the fermionic field taken into consideration for the test of this notion of point. Thus, to each fermionic field $\psi_f$ will be associated a mass $m_f$ and a length scale $\lambda_c^f$, the associated Compton wavelength, considered as the scale at which the notion of point and Riemannian manifold ceases to be relevant. These are therefore scales, which according to the arguments developed in the section \ref{NCObsEf}, are accompanied by a NC at the level of the observables, and thus by a NCG at the level of the deduced geometry. As we will see in chapter \ref{NCSMPP}, these are curiously the characteristic scales of the NC spaces created to geometrize the set of forces in the NCSMPP.

\section{Almost Naturalness of Inner Derivations}
\label{NaturInner}

In section \ref{NCObsEf} we saw that the NC of the observables could be associated with the non-passivity of the act of measurement. The claim of this section concerns QM theory without collapse (therefore not concerning physics, unfortunately). If before the measurement of the observable $b$, we have  $O_a(\psi)=\langle\psi|a|\psi\rangle$, after the measurement of $b$, the state is transformed: $|\psi\rangle\to b|\psi\rangle$ and thus $O_a(\psi)$ becomes: 
\begin{align*}
	O_a(b\psi)=\langle\psi|b^*ab|\psi\rangle
\end{align*}
Thus the variation of the observable can be written:
\begin{align*}
	\delta_b O_a(\psi)=\langle\psi|b^*ab|\psi\rangle-\langle\psi|a|\psi\rangle
\end{align*}
The fate of the observation coming from $a$ when the observable $b$ is realized is therefore given by the inner automorphism $\ad_{b^*}(a)=b^*ab$. If $b$ is unitary, then we have: 
\begin{align*}
	\ad_{b^*}(a)=b^*ab=\ad_{b^{-1}}(a)=b+b^{-1}ab-b=b+b^{-1}[a,b]=b+b^{-1}d_b(a)
\end{align*}
With the derivation $d_b(a)=[a,b]$. This expression is analogous to the one of usual variation of functions along space coordinates: $f(x+dx)=f(x)+\partial_xf(x)dx$.
\medskip
\par 
An example of a unitary operator is the translation operator (temporal or spatial), it is interesting to note that this is connected to the momentum, as well as to the degrees of freedom of space-time. As we can see in the Ehrenfest equation, these inner derivations are the derivations of QM:
\begin{align*}
	i\hbar\partial_t\rho =[H,\rho].
\end{align*}
The fact that many observables are not unitary operators (not only thanks to collapse) makes it impossible to build a differential structure from these observables. The obstacle to unitarity in the measurement process goes further in QM, with the collapse coming with the process. However, the collapse is considered as a problem of measurement and therefore does not have to be considered an unalterable fact. It is moreover possible that our observables are approximations of a larger operator, which may be unitary. In any case, the obstacle to unitarity for the evolution of $\psi$ is accompanied by an artificial renormalization of the wave function because its norm is not preserved, which from the point of view of elegance and conceptual simplicity can be considered as a defect of the current formalism. I therefore do not despair of the possibility that inner derivations may acquire a physical meaning.
\medskip
\par 
An interesting feature of inner derivations is that they implement an important operational physical reality of the study of variations. Indeed, in practice, the study of variations is always done by collecting the observables of an object whose variations are being studied. But it is not possible to study variations in an abstract way, it is always done by comparing these observables with those of a reference object, \textit{i.e.} the observables of a clock, a ruler, or any other physical object presenting a range of possible observations relevant to the study of this evolution. The study of variations is thus fundamentally that of the evolution of the links of relative observables between objects, which is indeed the case for the inner derivations where it is a question of the evolution of the observable linked to $a$ when the observation $b$ is made.

\newpage
\section{Geometry through Algebraic Eyes:}
\label{PhiloNCG}

\epigraph{It is common to find the anti-realist arguing that the success of quantum theory has dealt the death blow to realism, while simultaneously some realists herald the quantum revolution as demanding a reconstruction of the conceptual framework of physics to attain a fuller understanding of nature’s structure.}{\textit{H. Folse}}

In this section I will synthesize what has been said in this part, deducing some consequences in the same time, and building (in an incomplete manner) the emerging general picture. This will be done in a different order now that some of the ideas have been presented, so as to highlight the logical path to the NCG and to offer a better understanding of the proposed algebraic substantialism philosophy behind it.
\medskip
\par 
By NCG, I mean the algebraic reformulation of geometry, which can go beyond the scope of the latter through NC algebras. Therefore I speak also of Riemannian geometry seen from the algebraic eyes.
\medskip
\par 
We do not observe space-time, but observables of material objects, from which we derive concepts of space-time. Space-time is therefore a concept deduced from observables of material objects. We can then suppose that there is a space-time which is an entity in itself, where material objects arise in a subordinate way: this is the geometrical substantialism view of space-time. But we have seen that general relativity and various arguments undermine this view. 
\medskip
\par 
Is it then possible to completely determine space-time from observables? In physics, observables can be considered as elements of an algebra, while space-time is best described in terms of geometry. History shows that concepts of geometry may have been absorbed into those of algebra, culminating in the absorption of the notions of point and tangent space into the framework of NCG. It then seems plausible that our theories of space-time can be entirely deduced from the observables of material objects.
\medskip
\par 
We can therefore see that the geometric and algebraic representations of space-time are equivalent. It has been shown that the algebraic framework probably offers a more appropriate and fundamental framework. Because it is both less limited and closer to physics (observables), potentially allowing to overcome some internal inconsistencies in general relativity. This is why algebraic substantialism will be taken as a replacement for the geometric one.
\medskip
\par 
Let us therefore admit that observables are primary to the notion of space-time, and then look at what this might imply. It turns out that the observables accessible to us can be divided into two categories, classical observables, which are perfectly related to the idea of space-time, and quantum observables, which as we have seen are problematic at this level. In the algebraic framework, we have seen that pure states defined on an algebra provide the notion of a point and that observables that do not commute do not provide such notions of point. If the QM formalism designates objective physical states, then the mathematical equivalence between observable and geometry can be used to extract an objective geometrical meaning from the quantum formalism. If not, then no relevant geometrical consequences can be extracted. We have presented arguments to defend that this is the case: QM's formalism refers to objective physical entities. Then, it remains to understand what “geometric” consequences can be deduced from this formalism.
\medskip
\par 
Geometry is a deduction from observables, we are not allowed to say more.  But for a certain class of phenomena, the measurement corresponding to the objects $\StrMath$ (of the material physical theories) which permits to determine the represented spatial elements $\StrMath_{st}$ can no longer be considered as passive
\begin{align*}
	\pi_\StrMath(ER_1)\overset{Measurement}{\longrightarrow}\pi_\StrMath(ER_2)\overset{Deduction}{\longrightarrow}\text{space-time structure }\StrMath_{st}
\end{align*}
But if the physical theory of space-time is a theory that is intended to account for what is observed, then $\StrMath_{st}$ must be modified accordingly. Because our model of reality must fit with what is observed, and that, according to the PBR theorem, if we take two distinct symbolic states, corresponding to two physical states, then
\begin{align*}
	\pi_\StrMath(ER_1)\neq \pi_\StrMath(ER_2)\qquad \to \qquad ER_1\neq ER_2 
\end{align*}
But as we saw with Gelfand Naimark theorem in section \ref{PurToPoint}, this will correspond to two different points $x_1$ and $x_2$ given by $ER_1$ and $ER_2$ respectively. However, we have seen in section \ref{NCObsEf}, that the NCG is interpreted as a perturbation of the symbolic and thus the physical state, which means that at least one of the two observables causes a change of the physical state. Therefore, if we suppose that this change is from $ER_1\,\to\, ER_2$, then, this corresponds to a change $x_1\,\to\, x_2$ of the associated notion of point. The NCG can thus be interpreted as geometry in motion, this motion being induced by the act of measurement, which is, as mentioned in section \ref{Measurement}, always an interaction. The NGC is thus a perceived geometry, where the very act of observation disturbs the geometry, \textit{i.e.} the notion of point. In the algebraic substantialist view, we can see the measurement process can be taken as a replacement for the process of considering the geometric background space. If points can be seen as secondary to the process of measurement, then, I have no problems with NCG since nothing forces this process to lead to this notion. 
\medskip
\par 
It is interesting to note that according to the arguments developed in section \ref{Measurement} and in \cite{busch2009no}, the non-passivity of the act of measurement is universal (concern any measurement processes, classical and quantum). Thus, if noncommutativities are associated with such a nonpassivity of the act of measurement as defended in section \ref{NCObsEf}, then the noncommutativities of some observables are universal but asymptotically null for macroscopic objects (giving classical theories with commuting observables as an “asymptotic” theory). The interesting point that emerges is that insofar as the classical observables are only asymptotically commutative and that they allow us to deduce our usual representation of space-time, then the latter is only asymptotically Riemannian, without being fundamentally so. This invites us to reconsider the NCG as a potentially universal framework (in line with the idea presented in the introduction that NC is more fundamental), of which the eventual Riemannian space-time is only an approximation when the observables tend asymptotically to commute. 
\medskip
\par 
Other interesting topological consequences such as the link between entanglement and connectedness have been seen, as well as the extension of the notion of point. This suggests that where the QM formalism seems to offer great novelties (superposed states, NC, entanglement), notable geometrical consequences are to be found.
\medskip
\par
In section \ref{AreSetsPrimary}, we talk about the limitation of the thinking of sets as primary objects, this being linked to NC of characteristic properties used to build the set thanks to conditions given by these properties. It is interesting to note that the second big novelty of the QM world, \textit{i.e.} entanglement, is also connected to the end of the set's thinking as being fundamentals. Indeed, two entangled particles cannot be seen as linked to distinct elements of a set, neither at the level of their states nor at the level of the characteristic properties of these states, collected during the measurement process. This is for me a strong indication that nature shows us the limitation of set's thinking as being universal.
\medskip
\par 
For those who believe that QM is the ground of any theories about nature (because this talks about the mechanics of the elementary components of everything), this signifies that set's thinking is no longer fundamental, but asymptotically true. This may also be the case for mathematics.
\medskip
\par 
An order is a conceptual representation that we use to make sense of what we observe in the world. It is a tool, to find the “good order” in which observed things seem to make sense. It is an “a priori” given reading grid of the world used to interpret the events we observe. An order can be seen as better than another if it requires less information to understand and predict what happens in a meaningful way. Space-time is the main order, its causal structure gives the order in which we make sense of processes. This order was first described to be Euclidean with absolute space-time, then Euclidean without absolute space, then after some steps Riemannian. Each time, these upgrades  help to obtain more consistent picture of the process which occurs in nature.
\medskip
\par
In the above, I have argued that the formalism of QM should be taken seriously in terms of the geometrical consequences that can be deduced from it. This way of thinking opens up a surprising perspective. QM presently does not admit any satisfactory ontological scheme, it defies our intuition in various ways, in particular via noncommutativity and entanglement whose geometrical consequences have been discussed. Even worse, it seems to be in inadequacy with certain spatial and temporal notions such as locality, simultaneity... It is then in conflict with the actual notion of order. Can these difficulties of interpretation and visualization of quantum processes come from a limitation of our space-time representation to fully capture the nature of quantum events? Does QM invite us by its very formalism and its geometric consequences to revisit the notion of 'order' that we use to make sense of observed processes (space-time)? If it is the case, then the NCG framework can potentially offer a picture (deduced from quantum formalism) in which quantum processes will express intuitively. This potential algebraic framework can be used as a tool to understand the quantum to classical unification, and unify geometry (space) to observable (matter), closing the path undertaken by Einstein. This would be achieved in a theory in which all space-time concepts are absorbed into matter ones, providing a theory in which all elements are observable.

\part{Toward Noncommutative Gauge Field Theories}
\label{TowardNCGFT}
\chapter{The SMPP in the Framework of Gauge Field Theories (GFT)}
\label{SMPPGFT}
In this chapter, the foundations of gauge theories and the current SMPP will be presented. This will introduce the different notions that will be generalized later in chapters \ref{NCGFTGener} and \ref{NCSMPP}, such as connection, curvature, action, and the formalism of gauge theory. A mention will be made on the possible links between the current theory of gravitation, and the treatment of fundamental interactions in the SMPP. Then I will finish by presenting the generally followed procedure to find theories beyond the SMPP.

\section{Principle of Gauge Theory}
\label{PrincipGT}
Maxwell's formulation of electromagnetism can be considered as the birth of gauge theory. The elementary observation was that electric and magnetic fields can be seen as secondary to a vector potential and that adding suitable vector fields to this vector potential doesn't change the physics i.e electric and magnetic fields which affect the observed motion of charged particles. This can be considered as a symmetry of the theory. Gauge transformations can then
 be seen as field transformations that left unchanged some important quantities like the Lagrangian, which correspond to an important physical quantity. Later, symmetry arguments and gauge principle were at the earth of many of the biggest achievements of theoretical physics, such as particle physics, in QM, and in relativity theories. 
\medskip
\par
Let's see now how the gauge principle is implemented in particle physics, leading to the covariant derivative and its curvature.
\medskip
\par
Let $\lig$ be a finite-dimensional compact Lie group parameterized by $\Man$. The field $\psi(x)$ is a multiplet representation of  $\lig$, it represents the fermion field. Taking $g(x)\in\lig$, it's representation $\pi(g(x))$ must be a unitary, to preserve inner product on the fermion field, and therefore physical probabilities. In what follows, we will only speak about the representation, then we will use the notation $g(x)$ for $\pi(g(x))$.
\medskip
\par
A Lagrangian $\lag(\psi,\partial \psi)$ can be defined for this field. Physical Lagrangians are composed of elements like $\psi\overline{\psi}$ and $\psi\partial_\mu\overline{\psi}$. The requirement is that this Lagrangian must be invariant under the action of $\lig$:
\begin{align*}
	\lag(\psi,\partial \psi)=\lag(g\psi,\partial (g\psi))\qquad \qquad\forall g\in\lig
\end{align*} 
But as $g(x)$ is a function on $\Man$: 
\begin{align*}
	\partial (g(x)\psi(x))=(\partial  g(x))\psi(x)+g(x)(\partial \psi(x))\neq g(x)\partial \psi(x).
\end{align*}
Then as the Lagrangian is a polynomial function of $\psi$ and $\partial \psi$ it will not necessarily be invariant under the action of $\lig$. 
\medskip
\par 
This is solved by introducing a connection field $A=A_\mu^a(x)T_a$ which is an element of the Lie-algebra based on $\Man$, with $T_a$ its generators. It transforms in an inhomogeneous way: 
\begin{align}
	\label{eq GTF}
	A\to \tilde{A}=gAg^{-1}+g\partial g^{-1}
\end{align} 
this represents bosons which are responsible of the different interactions. The derivative on $\Man$ is then generalized to a covariant derivative in a fiber bundle:
\begin{align} 
	\label{CovarDer}
	D=\partial +(ie/\hbar)A 
\end{align}
where $e$ is the coupling constant of the theory (electric charge in the case of electromagnetism). This implements the minimal coupling between fermions and bosons fields. Then we have that $D (g\psi)=g D\psi$ and therefore: 
\begin{align*} 
	\lag(\psi, D \psi)=\lag(g\psi,g D\psi)\qquad \qquad\forall g\in\lig
\end{align*}
Let's go back to coordinate representation to better understand the link with physics. Let's take $g(x)=\exp(-(ie/\hbar)\Lambda(x))$ for the unitary transformation, with $\Lambda(x)=\Lambda(x)^aT_a$ the generator of the phase transformation. Then we obtain the new fermion field $\tilde{\psi}(x)=g(x)\psi(x)=\psi(x) -(ie/\hbar)\Lambda(x)\psi(x)$
and \ref{eq GTF} becomes
$\tilde{A}_\mu(x)= A_\mu(x)+\partial_\mu \Lambda(x)$. 
\medskip
\par
Let us now explore the equivalent transformations, experienced when $\psi(x)$ and $A_\mu(x)$ undergo infinitesimal displacements in space-time. The gauge potential aims to define a notion of parallel transport, \textit{i.e.} to specify between two neighboring points in $\Man$ what will be considered as an equivalent phase, in view of its transformation which is not attributed to an intrinsic change of the phase, but to an extrinsic one induced by the curvature of the folding space. Thus, the phases at $x^\mu$ and $x^\mu+dx^\mu$ are considered to be parallel transports of each other if they differ by an amount $(e/\hbar)A_\mu(x)dx^\mu$. Thus we get that
\begin{align*}
	\hat{\psi}(x+dx)=\exp(i(e/\hbar)A_\mu(x)dx^\mu)\psi(x)
\end{align*}
which is considered as an objective transformation of the state since it undergoes spatial displacement. After some calculations, we obtain that the total phase accumulation around an infinitesimal path represented by the square of sides $(dx^\mu, dx^\nu)$ is given by:
\begin{align*}
	\hat{\psi}(x)= (1+(ie/\hbar)F_{\mu\nu}dx^\mu dx^\nu)\psi(x)
\end{align*}
With the term:
\begin{align*}
	F_{\mu\nu}=(\partial_\mu A_\nu-\partial_\nu A_\mu)-i(e/\hbar)[A_\mu,A_\nu]
\end{align*}
that is interpreted as the curvature of the gauge potential, directly connected to the phase accumulation. In physics phase is not an observable, only the difference of phases is, and thus corresponds to a physical invariant. We can see that $F_{\mu\nu}$ is connected to this infinitesimal difference of phases between the two paths ($dx^\mu$ then $dx^\nu$) and ($dx^\nu$ then $dx^\mu$). As $F_{\mu\nu}$ corresponds to something physical (not depending on the way we parameterize it with coordinates), a natural requirement will be to find gauge transformations $g(x)$ preserving the scalar associated to the curvature: $\Tr(F_{\mu\nu}(x))=\Tr(g(x)F_{\mu\nu}(x)g^{-1}(x))$. Another way to calculate the curvature is obtained from the covariant derivative $D_\mu=\partial_\mu + (ie/\hbar)A_\mu$ introduced earlier:
\begin{align*}
	(ie/\hbar)F_{\mu\nu}=[D_\mu,D_\nu]-D_{[\partial_\mu,\partial_\nu]}=[D_\mu,D_\nu]
\end{align*}
because $[\partial_\mu,\partial_\nu]=0$ with holonomic basis in Riemannian geometry.
\medskip
\par 
Thus, the connection is interpreted as the phase difference accumulated during an infinitesimal displacement, this quantity cannot be associated with an observable \textit{i.e.} a physical quantity, a then fundamental invariant is obtained by calculating the phase accumulation on an infinitesimal loop. 
\medskip
\par
The symmetry principle provides a strong thinking line for developing physical theories. The way of proceeding can be seen as follows. Let's take a more general setting than fermion fields and Lagrangian based on these fields. Let $\calV$ be a vector space (degrees of freedom of a theory), and $K(\calV)$ be the space of polynomial functions on $\calV$ (potential functional of these degrees of freedom representing something physically objective). Consider now $\lig$ a group acting on $\calV$, $g\in \lig$, and $f\in K(\calV)$. The group action is $(g.f)(x):=f(g^{-1}(x))$. Two ways to construct models have been used:
\begin{enumerate}
	\item Given $\lig$ we name $K(\calV)^\lig$ the space of polynomials on $\calV$ which are left invariant by $\lig$ : $\forall g\in \lig$ we have $g.f=f$.
	\item Given $k_1(\calV)\in K(\calV)$ we name $\lig_{k_1(\calV)}$ the set of group elements which let $k_1(\calV)$ unchanged.
\end{enumerate}
The first one takes the group as fundamental and tries to find the correct invariant polynomial which is left invariant by this group, the second takes the polynomial as prior, trying to find the correct symmetry group, to catch other structures for example. These two ways of doing have greatly contributed to the advancement of physics, but the first one was the most efficient way to build the SMPP, by constraining its structure using symmetry arguments. More generally, the procedure followed by physicists to build and improve the SMPP can be seen as follows:
\begin{enumerate}
	\item Choose the structure group $G$ with $n$ generators and make it be locally parameterized by $\Man$: $\lig=Map(\Man, G)$, this provides a space of local symmetries.
	\item Choose the (locally defined) matter fields (Dirac fermions, scalars, etc.) as representations of $\lig$. It is an implementation of the local symmetries just mentioned.
	\item Set up a notion of derivation, leading to the differential structure upon which the Lagrangian $\lag$ and then motion equations can be expressed.
	\item Construct a covariant derivative extending the previous derivation by choosing the gauge bosons in a representation of $\lig$. It is an implementation of the minimal coupling between gauge and matter fields.
	\item Write the more general Lagrangian which is invariant under $\lig$ action (gauge principle) and satisfies other requirements such as being renormalizable.
	\item If there is a Higgs like mechanism, compute the minimum of the Higgs potential and deduce the corresponding masses.
	\item Use QFT tools based on this Lagrangian to obtain predictions and compare them to experimental results.
\end{enumerate}

Thus, any framework implementing in a clear and coherent way the gauge principle will be welcome. The general structure of this framework must make in relation 3 main structures. A geometric structure given by the space-time manifold $\Man$, an algebraic structure $\algA$ which uses to be a finite-dimensional algebra, and a global structure $G(\Man,\algA)$ that joins the two in one piece. The geometric structure can be obtained through a projection from the global structure, and the algebraic one from inclusion into the global. 
\medskip
\par 
As we will see in the next section \ref{GFTInFiberBundle}, the formalism of fiber bundle and differential geometry proved to be particularly suitable to implement this gauge principle. However, one of its limitations, related to the Higgs mechanism, will be presented in this same section. Then in Chapter \ref{NCGFTGener}, I will show how the NC extension of gauge theories allows to overcome this limitation, and thus make the structure from which the SMPP Lagrangian is extracted in a more mathematically consistent way. Other formalisms implement the gauge principle and address this particular limitation, such as the transitive Lie algebroids \cite{jordan2014gauge} for example.

\section{Gauge Theory in Fiber Bundle's Framework and Yang–Mills Theory}
\label{GFTInFiberBundle}
A fiber bundle $E$ is a topological construction that is locally expressible as a product space between elements of a base space $\Man$ and a fiber $\Fin$. A continuous surjective projection map $\pi\, :\, E\,\to\, \Man$ permits to express this local structure. When it is globally expressible as $E=\Man\times \Fin$, it is said to be trivial. The formalism of fiber bundles has proved to be particularly suitable to provide a mathematical structure for gauge theories, gradually establishing itself as the appropriate formalism from the 1975s. It is in this context in particular that the gauge potential acquired the status of connection in the fiber bundle, providing, in some way, a geometric interpretation of the gauge theories. Actually, all the forces of the SMPP are expressed as induced by gauge fields which are expressed in such a formalism. 
\medskip
\par 
\begin{definition}[Section in a fiber bundle]
	Taking the fiber bundle $\pi : E\,\to\, \Man$, a section on this fiber is a continuous function $f:\Man\,\to\, E$ such that $\pi(f(x))=x$ for all $x\in\Man$.
\end{definition}
$E$ is the total space, $\Man$ is the base space, and $\Gamma(E)$ is the space of global sections of $E$. These can be vector, tensor, or spinor fields, and correspond to (fermionic) matter fields. An important property to consider in order to find the NC generalization of $\Gamma(E)$ is that it is a module on the algebra $\calC^\infty(\Man)$.
\medskip
\par 
Let $G_E$ be a Lie group acting on $E$, taking an element $e\in E$ and $g\in G_E$, the right action of $G_E$ on $E$ is given by $R_g(e)=e.g$. The gauge group is the group $G$ of “vertical” automorphisms of $E$, i.e the ones which respect fibers: $\pi(e.g)=\pi(e)$. The fiber above $x$ will be then denoted by $\Fin_x=\pi^{-1}(x)$, it represents all the values the fermionic vector field defined over $x$ can take. 
\medskip
\par 
Let $\lieAlg$ be the Lie algebra of $G$. A connection on $E$ is a one form on $E$ with values in $\lieAlg$: $\omega\,\in\, \Omega^1(E)\otimes\lieAlg$ such that $\forall \xi\in \lieAlg$\GN{pb notation xi} and $\forall g\in G$:
\begin{align*}
	Ad_g(R_g^*(\omega))=\omega  \qquad\qquad\text{and}\qquad\qquad \omega(\xi^E)=\xi   
\end{align*}
with $Ad_g(e)=geg^{-1}$ the adjoint representation, and $\xi^E$ the fundamental vector field on $E$, linked to $R_{e^{t\xi}}$, $t$ being a real parameter. The associated curvature $\Omega\,\in\,\Omega^2(E)\otimes\lieAlg$ is given by $\Omega=\dd\omega +\frac{1}{2}[\omega,\omega]$ with the graded bracket at the level of $\lieAlg$. It is then possible to locally trivialize $E$ on small open sets of $x\in\Man$\GN{jord p 8 si nescessaire} in order to obtain the usual gauge connection expressed here in spacetime coordinates $A_\mu\,\in\, \Omega^1(x)\otimes\lieAlg$ and it's curvature $F_{\mu\nu}\,\in\, \Omega^2(x)\otimes\lieAlg$ which transform in the same way as in section \ref{PrincipGT}, $g\in \lig$ being parameterized by $\Man$.
\medskip
\par 
The SMPP consists in the understanding of two things, what are matter fields, and through which mediators they interact. In the fiber bundle framework, these two fields will correspond to well identified structures in the fiber bundle formalism:
\begin{itemize}
	\item Matter fields will correspond to sections in the fiber bundle
	\item Interaction fields will correspond to connection in the principal bundle 
\end{itemize} \GN{lien mathematical structure $\calS$}
Because of their statistics, matter fields will correspond to fermions and interaction fields to bosons. 
\medskip
\par
In the field of developments in particle physics, Yang-Mills theories can be considered as one of the most important breakthroughs coming from the application of the Gauge principle, they are now at the heart of the structure of the SMPP. Indeed, exploring how non-abelian Lie groups like $SU(n)$ are implemented in the gauge potential can lead to building gauge theories that permit a better understanding of strong and weak forces, later allowing the electro-weak unification. Now, all forces of the SMPP are expressed as Yang-Mills theories.
\medskip
\par 
In the case of gauge symmetry under the action of a non-Abelian (locally parameterized) group $\lig$, the situation is slightly more complicated, as said, it is described by a Yang-Mills theory. If we consider $\lig=SU(n)$, it acts on the fermionic field $\psi(x)\in\bbC^n$, we note $\{T^a\}_{a=1\dots n-1}$ the infinitesimal generators, then we have: 
\begin{align}
	\label{notationGauge}
	\psi(x)=\left(\begin{array}{c}
		\psi_{1}(x) \\
		\vdots \\
		\psi_{n}(x)
	\end{array}\right) ; \qquad\qquad A_{\mu}(x)=A_{\mu}^{a}(x) T^{a} ;\qquad\qquad  g(x)=e^{i\alpha^a(x)T^a}.
\end{align}
The curvature being given by $F_{\mu\nu}=F_{\mu\nu}^aT^a$ with $F_{\mu\nu}^a=\partial_\mu A_\nu^a-\partial_\nu A_\mu^a-(ie/\hbar)f^{abc}A_\mu^bA_\nu^c$ with $f^{abc}$ the structure constants of $\lieAlg$.
\medskip
\par  
To build an invariant from $F$, we have to understand how the inner product work on indices forms. Taking the couple $(\Man, \tilde{g})$ of a manifold with it's metric, then $\tilde{g}$ induce the bilinear form $\langle\,\,|\,\,\rangle\, : \Omega^k(\Man)\otimes \Omega^k(\Man)\,\to\, \Omega^0(\Man)$. For compact manifold, an integral against the volume form $vol_{\tilde{g}}$\GN{verif exist pas macro ,annex} can be defined, this is called the Hodge inner product $( \,\, | \,\, )\, : \, \Omega^k(\Man)\otimes \Omega^k(\Man)\,\to\, \bbR$, it is defined by:
\begin{align*}
	( F_1 | F_2 )\ \,\defeq \, \int_\Man \langle F_1|F_2\rangle vol_{\tilde{g}}
\end{align*}
This allows to define the Hodge star operator $\hstar\, :\, \Omega^k(\Man)\,\to\, \Omega^{n-k}(\Man)$, \textit{i.e.} this is the unique linear function such that:
\begin{align*}
	F_1\wedge \hstar F_2 = \langle F_1|F_2\rangle  vol_{\tilde{g}}
\end{align*}
Then the bosonic action associated with $A$ can be given by the Yang-Mills action:
\begin{align*}
	\act_b=\int_\Man F\wedge \hstar F
\end{align*}
This number must be left invariant by two kinds of process, gauge transformations, change of coordinate (according to general relativity framework), and trough any transformation corresponding to the equarions of motion, these last ones being given by computing the solutions of:
\begin{align*}
	\delta\act_b = 2\int_\Man \delta F\wedge \hstar F = 0.
\end{align*}
for one shell solutions. The fermionic action is given by:
\begin{align*}
	\act_f=-\int_\Man i\overline{\psi}(\gamma^\mu D_\mu +imc/\hbar)\psi d^4x
\end{align*}
\medskip
\par 
This provides a nice link between physical objects and mathematical structures. But as mentioned in \cite{jordan2014gauge}, the Higgs field is an object with ambiguous status in terms of the mathematical structure representing it. Indeed, it is both considered as a Boson, and as a section in a vector bundle, which typically represents Fermions fields. Moreover, its potential is not deduced from the formalism but added by hand. This appears to be slightly inconsistent from the mathematical structures' point of view. As we will see in section \ref{ACManifold}, and then in sections \ref{DBNCGFT} and \ref{STNCGFT}, the reformulation of the SMPP in the NC framework will make the Higgs field more consistent as a mathematical structure, with a natural potential coming from the formalism.

\section{The Standard Model of Particle Physics }
\label{SMPPLag}
The purpose of this section is to highlight the structure of the SMPP, making the connection between it's Lagrangian, the meanings and interpretations of its terms, it's symmetries according to the gauge principle, and in which way equations of motion can be obtained from it.
\medskip
\par 
A Lagrangian $\lag$ is a quantity that takes a particular value at each point in space and time. The Lagrangian is one of the most fundamental objects in theoretical physics. It is a functional which can depend for example on fields (in field theory) or on the metric of space-time (in general relativity), and which has the property of having its integral on the space-time (\textit{i.e.} the action) minimized for a physical state and its evolution. This condition is expressed by the Euler-Lagrange equation, the so-called equation of motion. 
The Lagrangian also allows us to define and study the crucial notion of symmetry. We call global symmetry under the action of a group $\calG$ a transformation that leaves the Lagrangian invariant and depends on a single parameter $g\in \calG$ for the whole space-time. Conversely, a local symmetry depends on a parameter $g\in \calG$ for each
point in space-time or, more precisely, on a smooth function $g(x)$.
\medskip
\par 
The Lagrangian of the SMPP is the following:
\begin{align}
	\label{LagSMPP}
	\lag=-\frac{1}{4} F_{\mu \nu} F^{\mu \nu}+i \overline{\psi} \slashed{D} \psi+\overline{\psi}_{i} y_{i j} \Phi \psi_{j}+h . c .+\left|D^{\mu} \Phi\right|^{2}+V(\Phi)
\end{align}  
with $\overline{\psi}=\psi^\dagger\gamma_0$, $\slashed{D}=i\gamma^\mu D_\mu$, $y_{i j}$ the Yukawa coupling matrix, $\Phi$ the Higgs field and $V(\Phi)=\lambda\Phi^2+\mu\Phi^4$ its potential.
\medskip
\par 
The quantum version of the Standard Model Lagrangian consists of several quantum fields, each associated with a Standard Model particle. There are actually 3 types of fields classified by their spins:
\begin{itemize}
	\item those for Fermions (quarks and leptons) designated by $\psi$ with spin $1/2$.
	\item those for the mediator bosons (photon, gluons, $W^{\pm}$ and $Z^0$ bosons), designated by $A_\mu$, with spin $1$.
	\item The Higgs bosonic field is designated by $\Phi $ with spin $0$. 
\end{itemize}
These fields are functions of the space-time points. They are not simple functions with numerical values, but operators acting on field's state, adding or removing particles. Consequently the Lagrangian and its associated action become operators too, this is called quantization and consists in promoting these fields, which are for the moment only functions, to the status of operators on a Hilbert space, the Fock space, which then verifies certain imposed commutation relations. There is no canonical method to quantize fields, this leads to many difficulties such as the need of renormalization. An important physical point that the Lagrangian makes it possible to determine is how the various particles propagate between points in space-time. The corresponding terms are those defining the creation of a particle at a given point and its annihilation at another one, the ones with two fields.
\medskip
\par 
Einstein's relation ($E=mc^2$) allows the conversion between energy and mass, and thus the possibility of changing the number and nature of particles during a process. Unlike in classical and pure QM, the nature of the system can then change, and visit many intermediate states. The possible transitions and their probabilities are constrained by the structure of the Lagrangian. All independent groupements of terms of $\lag$ describe the interaction between particles (potentials and interactional terms) or propagators (kinetics terms): 
\begin{itemize}
	\item  \textbf{Kinetic terms:} They correspond to the free propagation of the particle along space-time degrees of freedom, it implies spacetime derivatives and mass terms. For Fermions $i \overline{\psi} \slashed{\partial} \psi$, for Bosons these are terms like $(\partial^{\mu}A_\nu)^2$ and for the Higgs Boson $\left|\partial^{\mu} \Phi\right|^{2}$
	\item \textbf{Potential and international terms:} For Fermions  $i \overline{\psi} \slashed{D} \psi-i \overline{\psi} \slashed{\partial} \psi$ and $\overline{\psi}_{i} y_{i j} \Phi \psi_{j}$, for Bosons $-\frac{1}{4} F_{\mu \nu} F^{\mu \nu}-(\partial^{\mu}A_\nu)^2$ and $\left|D^{\mu} \Phi\right|^{2}-\left|\partial^{\mu} \Phi\right|^{2}$ and $V(\Phi)$
	
\end{itemize}
$\lag$ correspond then to the difference between potential and Kinetics terms.

\begin{remark}
	$\overline{\psi}_{i} y_{i j} \Phi \psi_{j}$ containing the Yukawa coupling matrix $y_{i j}$ which specifies the strength of the interaction (and then it's probability to occur) between fermions and Higgs field. This term gives both the fermion's masses and the probability strength of the flavor-changing process. We will see in section \ref{NCSMPPGeomForces}, mainly in remarks \ref{Rq propagGNC} and \ref{LComptPropagNCG}, that in the context of the NCSMPP, this term acquires a kind of kinetic interpretation in the same way as $i \overline{\psi} \slashed{D} \psi$.
\end{remark}
The SMPP is written in the framework of gauge theory, we will denote by $\text{GFT}_\text{SMPP}$ the actual gauge field theory of the SMPP.

\section{Gravitation and Gauge Theories}
\label{GravAndGT}
Actually, we know 4 fundamental forces, the ones of the SMPP, and Gravitation. Many physicists try to unify Gravitation to the three other forces, in order to understand how to quantize gravitation and obtain a more unified picture. Some basic hints can make us think that these two blocks share some similarities:
\begin{itemize}
	\item The position coordinate is not observable, only the difference of coordinate between two positions is. The same hold for phases and then connections, only relative difference in phases are observable. Making gauge potential and coordinate on the same footing for the fact that they do not directly represent an observable, but rather a relationship between observables.
	\item Like gauge transformations, general coordinate transformations have no consequences on physics, they represent symmetries of the theory.
	\item The curvature of spacetime curves the trajectories of test particles. Interactions symbolized by gauge potentials acting on fermionic fields of test particles curve their trajectories, in a similar way as gravitation when we compute the curvature of the gauge potential (see the gravitational Aharonov-Bohm effect). 
\end{itemize}
But the problem is that gravitation is a geometric theory contrary to other forces and that they are expressed in different mathematical fields.
\medskip
\par 
Historically, gravitation was elaborated by Einstein in the framework of pseudo-Riemannian geometry and not in the one of gauge theory. The basic principle of this framework is to define an affine connection $\nabla\, :\, \Gamma(\TanBun)\times \Gamma(\TanBun)\,\to\, \Gamma(\TanBun)$ by the constraints:
\begin{align*}
	\nabla_{f\partial_\mu}\partial_\nu=f\nabla_{\partial_\mu}\partial_\nu\qquad\text{and}\qquad\nabla_{\partial_\mu}f\partial_\nu=\partial_\mu f\partial_\nu+f\nabla_{\partial_\mu}\partial_\nu   
\end{align*} 
$\forall f\in \calC^\infty(\Man)$ and $\partial_\mu,\partial_\nu\in \Gamma(\TanBun)$. These locally connect tangent spaces.
\medskip
\par 
In order to make this connection compatible with the metric structure $\tilde{g}$, we define the metric connection by
\begin{align}
	\label{ConnCompMetricRG}
	\partial_\lambda\tilde{g}(\partial_\mu,\partial_\nu)=\tilde{g}(\nabla_\lambda\partial_\mu,\partial_\nu)+\tilde{g}(\partial_\mu,\nabla_\lambda\partial_\nu)
\end{align} 
which can be seen as a sub-case of equation \eqref{ConnCompMetric} that will be defined later.
\medskip
\par 
Adding the torsion-free condition ($\nabla_{\partial_\mu}\partial_\nu-\nabla_{\partial_\nu}\partial_\mu=[\partial_\mu,\partial_\nu]$) we obtain the Levi-Civita connection, which is at the heart of general relativity. 
\medskip
\par 
Using these Levi-Civita connection, we can construct a covariant derivative (in local coordinate) on the tangent bundle $\nabla_{\partial_\mu}\, :\, \Gamma(\TanBun)\,\to\, \Gamma(\TanBun)$, with curvature being given by $[\nabla_{\partial_\mu},\nabla_{\partial_\nu}]-\nabla_{[\partial_\mu, \partial_\nu]}$. Using the Christoffel symbols defined by $\nabla_{\partial_\mu}\partial_\nu=\Gamma^\lambda_{\mu\nu}\partial_\lambda$, the curvature can be computed, giving the curvature tensor $R_{\mu\nu\rho}^\lambda$, then the Ricci tensor $R_{\mu\nu}=R_{\mu\nu\lambda}^\lambda$, and then the scalar curvature $\tilde{g}^{\mu\nu}R_{\mu\nu}$. This last scalar can be taken to define the action of the Gravitation theory: 
\begin{align*}
	\act[\tilde{g}]\propto\int_\Man R{\sqrt {|\tilde{g}|}}\,\mathrm {d} ^{4}x  
\end{align*} 
But the fact that the primary field taken to be the metric $\tilde{g}$ is not a 1-form makes the Lagrangian given by $R$ only invariant under the group $\Diff(\Man)$, taken to be the gauge group of the theory. Therefore, $R$ is only a scalar for the geometric structure defined on $\Man$, but not for general Gauge transformation, then different to gauge theory in the general meaning.  
\medskip
\par
In the same way, we can see that forces in gauge theories differ from the gravitation one since this last has been given a geometric interpretation, promoting it to the status of pseudo-force \textit{i.e.} fictitious force induced by the effect of an irregular background geometry. This is not the case with Gauge's theories, for which forces are considered as pure forces, without any underlying geometrical representation explaining them.   
\medskip
\par 
But subsequent developments reformulated this theory in the field of Cartan geometry making this theory be of gauge type. More details on the construction can be found in \cite{jordan2014gauge,bennett2021pedagogical}. We will see in chapter \ref{NCSMPP} that the NCSMPP offers such a way to reformulate both Gravitation and SMPP forces into the same framework, at the same time being gauge theory and pseudo forces induced by pure geometry.

\section{Open Doors to go Beyond the SMPP}
\label{OpenDoorBeyondSMPP}

In physical theories, unification can be seen as how two theories, previously understood as independent become sub-manifestations of a larger theory, according to some parameters like energy change, formalism change inside the same framework, framework change... It is perceived as an enhancement of the theoretical understanding.
\medskip
\par
Taking a physical theory and its mathematical structure $\StrMath$, not all theoreticians will agree on the fact that it is complete and consistent or not, and in this last case, they can differ on which way this enhancement must be done. The enhancement of a physical theory can be thinking according to several criteria and arguments. I propose here 3 kinds of enhancement arguments:
\begin{enumerate}
	\item The first concern is the diminution of the number of constant inputs. Like for example the 18 inputs of the SMPP.
	\item The second one is the diminution of the number of independent formal inputs. The equations of the theory are deduced from minimal principle, the number of “put by hand” elements is minimal. A good framework can help to obtain this result.
	\item Consistency with experiment, with other theories on their crossings area, and on the conceptual and epistemological level.
\end{enumerate}
In what will follow (see section \ref{NCSMPPLag}), I will highlight according to which one of these arguments NCG improves our understanding of the SMPP. 
\medskip
\par
Concerning the SMPP, here is a non-complete list of what can be seen as missing-points with potential enhancement needed in the actual description:
\begin{itemize}
	\item Find a mechanism that determines the origin of fermion masses, and thus explains the three-generation existence. (12 inputs over 18 are concerned)
	\item About the Cabibbo Kobayashi-Maskawa (CKM) matrix $M_{CKM}$:
	
	\begin{enumerate}
		\item The eigenstates of the free Hamiltonian are also the ones of the gravitational, strong and electromagnetic interactions, but they are not eigenstates of weak one. There are no explanations for this fact \cite{hubsch2015advanced}.
		\item The non-unity of the measured CKM matrix. According to the theoretical framework in which this matrix is constructed, it must be unitary.
		\item There is no theory predicting the value of the angles parameterizing the CKM matrix. 
	\end{enumerate}
	\item Neutrino mixing and oscillations
	\item Potential unification with gravitation, find a better formalism implementing naturally the Higgs field and its potential as mentioned, explain and/or unify the coupling constants...
\end{itemize}
Some of these points are questionable on whether they are truly problems in the actual description. 
\medskip
\par 
In this regard, Grand Unified Theories (GUT) can be considered as the most significant attempt for solving such problems. The search for unification in field theory is a very old and fruitful field of investigation in theoretical physics. Some nice achievements are the unification of electric and magnetic fields in the electromagnetic one, and later of this last with weak force in the electroweak unification. GUT are extensions of the SMPP with larger symmetry groups than the existing ones. They must contain the SMPP group as a subgroup and possess complex representations (fermionic fields) in accordance with the ones of the SMPP... Actually, GUTs theorists try to exploit the successful arguments of gauge symmetries and group theory to unify electroweak and strong interactions into a single one. A hint for such hope of unification is that the flows of the gauge couplings “constants” of the SMPP at a high energy scale seem to converge.
\medskip
\par 
Therefore, the question is what are the ways in which the SMPP fields can be embedded into an unified field? This unified field (and its groups) must be large enough to encompass the current groups of the SMPP, but not too, so that there will not be too much new physics, likely not to fit with the experiment. It must correspond to the effective theory at higher energies, and render the actual SMPP within a certain limit, with some physical process (like spontaneous symmetry breaking for example). By introducing new fields with their symmetries, GUTs can improve many features of the SMPP. For example, they can provide a gauge coupling unification, give an explanation for the lightness of neutrino masses, introduce contributions that can solve some flavor anomalies, potentially explain charge quantization and predict values for weak mixing angles.
\medskip
\par 
The first and simplest GUT being explored was the one of the Georgi–Glashow model corresponding to the gauge group $SU(5)$. Many other groups like $SU(5)\times U(1), SO(10),E_6\dots$ have been explored, without succeeding. A nice review of GUTs attempts can be found in \cite{GUTCroon}.
\medskip
\par 
Let us try to explain why this kind of attempt seems to offer a set of natural paths for unification, by trying to think in a general way about unification. The first assumption to make is that this can be done in the current formalism of gauge theories, or in any more or less equivalent formalisms based on field theories, such as the NCSMPP presented in chapter \ref{NCSMPP}.
The basic ingredients of the SMPP are gauge potentials, not fermionic fields, which are secondary representations, so the structure of the standard model is primarily driven by the nature of these bosonic fields, this is the second assumption. A procedure to create such a model is to choose the right gauge field according to these invariance groups and symmetry arguments, and then to choose among the possible set of representations (fermionic fields, chiral structure...), which one is the most adequate. Assumption three is that the same procedure should be used in attempts at unification, but taking into account the basic ingredients of the old theory, \textit{i.e.} those fermionic and bosonic degrees of freedom (dofs), which should be linked according to certain criteria. It is now a question of understanding the (conceptual) passage from the gauge theory of the SMPP, \textit{i.e.} $\text{GFT}_\text{SMPP}$, to the grand unified theory:
\begin{align*}
	\text{GFT}_\text{SMPP}\, \rightarrow\, \text{GFT}_\text{GUT}
\end{align*}
in the first time, because the $\text{GFT}_\text{GUT}$ must be built from the one of the SMPP. Then we have to understand the (physical) mechanism $\text{GFT}_\text{SMPP}\, \leftarrow\, \text{GFT}_\text{GUT}$ which renders $\text{GFT}_\text{SMPP}$ from $\text{GFT}_\text{GUT}$ trough a suitable physical process, like spontaneous symmetry breaking mechanism SSBM. This can be seen into a 3 step process: first find the good gauge field algebra, then a suitable (fermionic field) representation and check that the corresponding mechanism restitutes the actual SMPP. There are many ways to imagine GUTs in this way.
In this thesis, the construction of gauge fields based on inductive sequences of algebras (which correspond to these gauge potentials taken as input) will offer a framework to elaborate such GUT and understand the $\text{GFT}_\text{SMPP}\, \rightarrow\, \text{GFT}_\text{GUT}$ transition. This transition will correspond to a step in the inductive sequence. This will be presented in section \ref{sec AFA}, and done in part \ref{partNCGFTAF}. The implementation of assumption three (condition between the two theories, at least on its elementary degrees of freedom \textit{i.e.} gauge and fermionic fields) will be given by the $\phi$-compatibility condition defined along the sequence and presented in section \ref{sec AFA}. 

\chapter{Noncommutative Gauge Field Theories (NCGFT)}
\label{NCGFTGener}
In this chapter, I propose to show how the concepts presented in the previous chapter can be generalized in NCG. It will be done in a general way for the notions of connection, curvature, fiber bundle, and symmetries, but using derivation in section \ref{DBNCGFT} then spectral triples in section \ref{STNCGFT} based GFT. Then the actual NCSMPP will be presented in chapter \ref{NCSMPP}, with some discussions. From now on, we will always suppose that $\algA$ is unital, with unit $\bbbone$.

\section{Fiber Bundle and Serre-Swan Theorem}
\label{SerSwann}
Another important topological structure possess a well identified algebraic counterpart, that of vector bundles. Indeed, the Serre-Swan theorem states that any vector bundle on a manifold $\Man$ defines a $\calC^\infty(\Man)$-projective module of finite type on the algebra $\calA=\calC^\infty(\Man)$ by considering the set of smooth sections of this fiber; and conversely, if $\Man$ is connected, any $\calC^\infty(\Man)$-projective module of finite type comes from a vector bundle on $\Man$. As for the Gelfand-Naimark theorem this duality is more than a coincidence. It results from the category equivalence between that of vector bundles on $\Man$ and that of $\calC^\infty(\Man)$-projective modules of finite type on $\calA=\calC^\infty(\Man)$. Let first define what is a projective module. 
\begin{definition}
	\label{DefProjMod}
	Let $\calA$ be an algebra, a module $\calM$ over $\calA$ is projective iff we have $\calM\oplus\calN =\calA^n$, for some module $\calN$ (over $\calA$), and an integer $n\geq 1$.
\end{definition}
We build a free module $\calA^n$ over $\calA$.
\begin{lemma}
Taking a projection $p\in M_n(\calA)$ such that $p^2=p=p^*$, then we can define the projective module $\calM_p=\calA^np$.
\end{lemma}

\begin{proof}
	If $p$ is a projection, then $1-p$ is also a projection, and they are mutually orthogonal. Therefore $\calM_{1-p}=\calA^n(1-p)$ is such that we have $\calA^n=\calM_p\oplus\calM_{1-p}$ so that according to definition \ref{DefProjMod}, this two modules are projective.
\end{proof}
This module is said to be finitely generated when $n$ is finite. 
\medskip
\par
Let illustrate this by using our favorite commutative algebra $\calA=\calC^\infty(\Man)$, $p$ is therefore a function of $x\in M$. Thus $\calM_{p(x)}=\calC^\infty(\Man)^np(x)$ define the module associated to $x$. 
\medskip
\par
Then any module $\modM$ on $\calA$ will be considered as the non-commutative equivalent of an associated vector bundle, through the associated section. The set of sections of a trivial vector bundle whose fiber is of dimension $n$ is isomorphic to the module $(\calC^\infty(\Man))^n$. When the fiber is not trivial, we place ourselves in a larger space, which means adding dimensions to trivialize it, the initial fiber is then obtained using a projection $p$: $(\calC^\infty(\Man))^np$. In the commutative case, the space of matter fields can therefore be represented by a right module on $\calA$, in the NC case, care should be taken to distinguish between the left and right modules. In NCG, bi-module will frequently be more adapted than module.
\medskip
\par 
As we shall see, this new extending framework for the usual gauge theories is entirely algebraic, and thus leaves us with more freedom. This motivates the introduction of $C^*$-algebras as non-commutative generalizations of locally compact topological spaces, and modules as non-commutative generalizations of vector bundles.
\medskip
\par 
It is interesting to notice that when $\calA$ is unital, there is a correspondence between projective modules of finite type and idempotents of $M_n(\calA)$. We can define (in an equivalent way to what has been done in section \ref{Ktheory}) a notion of algebraic equivalence between idempotents which is equivalent to the notion of isomorphism between projective modules. As we have seen in section \ref{Ktheory}, these idempotents allow us to classify algebras via K-theory, and thus in our context to classify the associated vector bundles. This is why the Serre-Swan theorem allowed to relate topological and algebraic K-theory (see \cite{swan2006algebraic} for more details).
\medskip
\par
Then if we take $\calA=\calC^\infty(\Man)$, we have $M_n(\calA)=\calC^\infty(M, M_n(\bbC))= \calC^\infty(\Man)\otimes M_n(\bbC)$, we recognize the AC-manifold. Thus the Serre-Swan theorem tells us that it is possible to construct a vector bundle using this AC-manifold, with base space $\Man$. We will see in section \ref{ACManifold} that the AC-manifold is the basic structure of NCGFT, and that it has led to the current non-commutative standard model of particle physics in chapter \ref{NCSMPP}. The fact that K-theory allows to classify $AF$-algebras (which are generalizations of $M_n(\bbC)$) provides a strong mathematical motivation to develop NCGFTs on AC-manifolds where the finite part algebra is taken to be an $AF$ algebra. Indeed, K-theory and Serre-Swan theorem are intimately related through their use of the idempotents of $M_n(\calA)$. Then K-theory can be used to classify the corresponding gauge theories and “even more” the ones based on AF-algebras offering a strong mathematical reason to do NCGFTs based on AF-algebras.

\section{Generalization of Vector Fields, Connections and Curvature}
\label{NCConnectionModCurv}

As said before, the Serre-Swan theorem asserts that finitely generated projective modules are linked to vector bundles.
Therefore we have to replace usual vector fields $\calV$ constituting sections by modules $\modM$ over an algebra. A very nice fact, proved by Cuntz and Quillen in \cite{cuntz1995algebra} is that every projective module admits the existence of connection and conversely, any module having a connection is projective. Defining a connection requires the pre-existence of a differential structure. In the usual framework, this can be the covariant derivative or the de Rham differential. As mentioned in chapter \ref{DifCalc}, these can be replaced by derivation and spectral triples-based differential structures.
\medskip
\par 
In the case of the universal differential calculus $(\Omega^\grast_U(\algA), \ddU)$, a NC connection can be defined as a $1$-form $\omega = \sum_i a^{0}_i \ddU a^{1}_i \in \Omega^1_U(\algA)$, but it is a too general definition. Now if we take an algebra $\algA$ with an involution $^*$ and the associated differential calculus $(\Omega^\grast(\calA),\dd)$ with $\omega\in\Omega^n$ and $\eta\in\Omega^m$, we will suppose that the graded algebra $\Omega^\grast$ possess an involution such that $(\omega\eta)^*=(-1)^{nm} \eta^*\omega^*$, and that $\dd$ is real $(\dd\omega)^*=\dd(\omega^*)$.

\begin{definition}
	let $\modM$ be a left $\algA$-module, and $\Omega^1(\algA)$ be a first-order differential calculus over $\calA$, with differential $d$. A connection is defined by a map $\nabla\, :\, \modM\,\to\, \Omega^1(\algA)\otimes_\calA\modM$ such that, $\forall a\in\calA$ and $e\in\modM$:
	\begin{align*}
		\nabla(ae)=a\nabla(e)+da\otimes_\calA e.
	\end{align*}
\end{definition}
Then the associated curvature is given by $R=\nabla^2$ with $R\, :\, \modM\,\to\, \Omega^2(\algA)\otimes_\calA\modM$.
\medskip
\par 
A Hermitian structure on $\modM$ is a $\bbR$-linear map $h : \modM \otimes \modM \to \algA$ such that $h(a_1 e_1, a_2 e_2) = a_1 h(e_1, e_2) a_2^*$ for any $a_1, a_2 \in \algA$ and $e_1, e_2 \in \modM$. A connection $\nabla$ is Hermitian if for any real $\kX$ of $\algA$ and any $e_1, e_2 \in \modM$, one has
\begin{align}
	\label{ConnCompMetric}
	\dd h(e_1, e_2)
	&= h(\nabla (e_1), e_2) + h(e_1, \nabla(e_2))
\end{align}
This can be viewed as a generalization of the metric condition given in equation \eqref{ConnCompMetricRG}.
\medskip
\par
It turns out that the module concept is not sufficient to elaborate nice structures, in particular concerning hermiticity problems. The bimodule structure (a module that is both left and right, making left and right multiplications compatible) is therefore preferable, especially in the context of spectral triples.

\section{Automorphisms of Finite Noncommutative Spaces}
\label{AutomGen}
\epigraph{If you cannot change your condition, change your perception.}{\textit{D. Mridha}}

In the framework of NCGFT, automorphisms will provide gauge and coordinate change. As mentioned in chapter \ref{DifCalc}, they are two ways to consider such changes. The first can be seen as change at the level of the Hilbert space (\textit{i.e.} it's generalization as module here), and the other as being at the level of the algebra.
\medskip
\par 
To be general, let $\algA$ be an $^*$ algebra with unit, and $\modM$ a left $\algA$-module. Let's start with the automorphisms which are implemented at the level of $\modM$.
\medskip
\par 
\begin{definition}[Automorphism of a module $\modM$]
	An automorphism of a module $\modM$ is a $\algA$-linear map $\Pi\, :\, \modM\,\to\,\modM$ which is invertible and preserves the module structure:
	\begin{align*}
		\Pi(ae)=a\Pi(e)\qquad\qquad \forall e\in\modM\qquad\text{and}\qquad a\in\algA.
	\end{align*}
\end{definition}
\medskip
\par 
Let's suppose that $\modM$ is equipped with a Hermitian structure $h$. The gauge group $\calG$ of $\modM$ is the group of automorphisms of $\modM$ as a left module that preserve the Hermitian structure such that for any $a \in \algA$, $e, e' \in \modM$, $g \in \calG$ satisfies 
\begin{align}
	\label{GaugeModGroup}
	g(a e) &= a g(e),
	\quad
	h(g(e),g(e')) = h(e, e').
\end{align}
The action of $g$ on a connection is defined by the compositions $\nabla \mapsto \nabla^{g} \defeq g \circ \nabla\circ g^{-1}$. It is easy to check that $\nabla^{g}$ is a connection and that $\nabla^{g_2 \circ g_1} = (\nabla^{g_1})^{g_2}$ for any $g_1, g_2 \in \calG$. 
\medskip
\par
Now, for the automorphisms implemented on the algebra:

\begin{definition}[$*$-automorphism of a $*$-algebra $\calA$]
	An $*$-automorphism of a $*$-algebra $\calA$ is a linear map $\Lambda\, :\, \calA\,\to\,\calA$ which is invertible and preserve the $*$-algebra structure:
	\begin{align*}
		\Lambda(ab)=\Lambda(a)\Lambda(b)\qquad\text{and}\qquad \Lambda(a^*)=\Lambda(a)^*\qquad\qquad\qquad \forall a\in\algA .
	\end{align*}
	Given $\calA$ a $*$-algebra, we denote by $\Aut(\calA)$ its automorphism group.
\end{definition}
$\Aut(\calA)$ will be related to the gauge group associated to $\algA$. 
We call $\calU(\calA)$ the group of unitary elements in $\calA$:
\begin{align*}
	\calU(\calA)=\left\{u \in \calA: u u^{*}=u^{*} u=\bbbone\right\}
\end{align*}
\begin{definition}[Inner automorphism of $\calA$]
	An automorphism $\Lambda$ is called inner if it is of the form $\Lambda(a)=\ad_u(a)=uau^*$ with $u\in\calU(\calA)$ and $a\in\calA$. The group of inner automorphisms is denoted
	by $\Inn(\calA)$. 
\end{definition}
In a similar way as for derivations with equation \eqref{InQuotCenter}, the map $\calU(\calA)\,\to\, \Inn(\calA)$ defined by $u\,\to\, \ad_u$ is surjective with kernel the elements $u\in \calU(\calA)$ such that $\ad_u(a)=a\,\, \forall a$, which correspond to $\calU(\calZ(\calA))$. Then we have:
\begin{align*}
	\Inn(\calA)\simeq \calU(\calA)/\calU(\calZ(\calA))
\end{align*}

\begin{definition}[Outer automorphism of $\calA$]
	The group of outer automorphisms of $\calA$ is defined by the quotient
	\begin{align*}
		\Out(\calA):=\Aut(\calA)/\Inn(\calA)
	\end{align*} 
	
\end{definition}

\medskip
\par
For $\Man$ a smooth compact manifold, and $\calA=C^\infty(\Man)$, we have $\Aut(\calA)=\Out(\calA)\simeq\Diff(\Man)$, the group of diffeomorphisms of $\Man$. Taking a diffeomorphism $\Xi\, : \, \Man\,\to\, \Man$\GN{verif Xi}, the link between diffeomorphism and (outer) automorphism is given by the pullback over function space:
\begin{align*}
	\Xi^*(f)(x)=f(\Xi(x))\qquad\qquad \forall f\in\algA\qquad\text{and}\qquad \forall x\in\Man
\end{align*}
Conversely for $\calA=M_n(\bbC)\simeq \calB(\calH)$ for a given Hilbert space $\calH$ of dimension $n$, all automorphisms will be inner.\GN{ref proof ? walter} 
\medskip  
\par
Taking an algebra $\algA$, we have the short exact sequence of groups:
\begin{align*}
	1\, \longrightarrow \, \Inn(\calA)\,\longrightarrow\, \Aut(\calA)\,\longrightarrow\, \Out(\calA)\,\longrightarrow\, 1
\end{align*}
from which \ref{ExactSeqDer} can be seen as an infinitesimal version.
\medskip
\par 

In sections \ref{DBNCGFT} and \ref{STNCGFT}, we will see that the first kind of automorphisms will be implemented for NCGFT based on derivations and that the second kind will be used for its spectral triples counterpart.

\section{General Principles of NCGFT and the AC-Manifold}
\label{ACManifold}

Let's try to understand how the principle of gauge theory can be implemented using this extended framework. The general procedure (which includes usual gauge theories) will be to:\\ 

\begin{figure}[h]
	\begin{center}
		\smartdiagramset{back arrow disabled=true, uniform color list=white!80!black for 6 items,text width=3cm, module minimum width=3.6cm, module minimum height=2.4cm, module x sep=4.5cm, font=\fontsize{8pt}{13pt}\selectfont
		}
		
		\smartdiagram[flow diagram:horizontal]{Find a differential structure linked to degrees of freedom associated with some “space”,
			Build connection to implement parallel transport along these degrees of freedom, Find an automorphism invariant scalar built from this connection}
	\end{center}
\end{figure}
Now, suppose that we are interested in finding a space for which the diffeomorphism group is a compact Lie group. But they are no such a manifold. As mentioned in the previous section \ref{AutomGen}, taking a manifold $\Man$, its diffeomorphism group is equivalent to $\Aut(\calC^\infty(\Man))$. Thus, by moving into the algebraic equivalent, a very important observation can be made, automorphisms of algebras can be extended to provide compact Lie groups, these being the automorphisms of matrix algebras. The question then becomes, what is the appropriate structure of “space” \textit{i.e.} algebra that could both contain $\Diff(\Man)$ and the gauge Lie group $\lig$ (potentially the one of the SMPP) in its automorphisms group?   
\medskip
\par 
An answer was given by the almost commutative manifold, alternatively designed by the “space” $\hMan=\Man\times\Fin$ or its algebra $\halgA \defeq \calC^\infty(\Man) \otimes \af$, $\Fin$ being a finite space. The automorphisms group $\Aut(\halgA)$ is then equivalent to $\Diff(\Man)\ltimes\lig$, with $\lig=Map(\Man, G)$, $G$ being the standard model group $U(1) \times S U(2) \times S U(3)$ for example. We recognize the different structures at the heart of the implementation of the gauge principle mentioned in section \ref{PrincipGT} \textit{i.e.} the geometric structure $\Man$, the algebraic structure $\af$ and the global one $\halgA \defeq \calC^\infty(\Man) \otimes \af$.
\medskip
\par 
Subsequently, the AC-manifold's algebra was taken to be the fundamental structure that allowed to obtain the NCSMPP. The first use of the AC-Manifold was made in 1990, in \cite{DuboKernMado90a} and \cite{DuboKernMado90b} by M. Dubois-Violette, R. Kerner and J. Madore with the algebra $\calC^\infty(\calM)\otimes M_n(\bbC)$.  This model shows for the first time that gauge theories arise naturally from a NC structure, with the advantage to consider the gauge group as the set of unitary elements of an $^*$-algebra. In addition, it allowed us to understand that Higgs and gauge fields could be gathered in the same non-commutative object, with a more natural interpretation for the Higgs field, the potential of the Higgs being naturally given in the formalism, addressing the problem raised in section \ref{GFTInFiberBundle}. Indeed, the Higgs field can be seen as a connection along the internal NC degrees of freedom, its potential is equivalent to both the “Einstein-Hilbert action” and the usual bosonic potential. Higgs scalar then acquires a geometrical interpretation, it gives both a connection in the finite part of the AC-manifold and the fermionic mass matrix. This model was the basis of later developments, and in particular of the NCSMPP mainly developed by A. Connes about which I will speak in chapter \ref{NCSMPP}.
\begin{remark}
	\label{rqInner}
	Notice that we chose the terminology $\Fin$ both for the finite space and the fiber. This is because the finite space is a kind of replacement for this fiber in the context of NCG. The finite space is constituted of $n$ elements, each one being able to take several values. These elements correspond to the different distinguishable characteristics (like charge) the fermionic field can possess \textit{i.e.} the $i$ of the $\psi_{i}(x)$ defined in section \refeq{notationGauge} which represents the value taken by the field for the element $i$ at spacetime location $x$. Fixing its values for each $i$ and $x\in\Man$ will define a section. And if we visit all its potential values for every $i=1\dots n$ at a given $x$, then we recover the fiber $F_x$ defined in section \ref{GFTInFiberBundle}. The finite space and the fiber are the same spaces but seen differently by the formalism. We will call $F$ the fermionic representation space. 
\end{remark}
Therefore, doing NCGFT will consist in working with the algebraic counterpart of the usual gauge theory structures defined in section \ref{GFTInFiberBundle}, the main equivalences being given in table \ref{figEquivTopoAlgGFT}.
\begin{figure}[h]
	\begin{center}
		{\setlength{\doublerulesep}{0pt}
			\begin{tabular}{|c|c|}
				\hline
				\textbf{Usual gauge field theories:} & \textbf{NCGFT:}\\ (differential geometry) & (algebra)\\
				\hline
				symmetry group $G$ (Lie group) & associative algebra $\calA$\\ 
				\hline
				algebra $C^\infty(\Man)$ & center of the algebra\\ 
				\hline
				vector field/sections on the fiber bundle & finitely generated projective module $\modM$ on $\calA$\\ 
				\hline
				gauge group: vertical automorphisms $G$ & gauge group: $\Aut(\modM)$\\ 
				\hline
				differential structure: de Rham & differential structure: universal,\\ 
				& derivations, spectral triples\\
				\hline
				Covariant derivative & NC connection 1-form associated to the\\ & differential structure and based on $\modM$ \\
				\hline
				
			\end{tabular} 
		}
	\end{center}
	\caption{Equivalences between topological and algebraic properties for gauge field theories.}
	\label{figEquivTopoAlgGFT}
\end{figure}\\
The general procedure will then be to find a differential calculus based on $\af$ or $\modM$, then find a notion of connection, and then compute the action functional. Having done this, the road will be taken to obtain more concrete gauge theories, using the AC-manifold's algebra $\halgA \defeq \calC^\infty(\Man) \otimes \af$, and then construct the associated NCGFT which will be called $\text{NCGFT}_{\af}$, the finite algebra $\af$ being the only variable input. Then we will obtain the associated action functional $\act_{\af}$ (which is not only a function of $\af$ as we will see), doing this in the derivation framework in section \ref{DBNCGFT} and in the spectral triples one in section \ref{STNCGFT}. In chapter \ref{NCSMPP}, using spectral triples-based NCGFT and taking an appropriate algebra $\af$, we will see that the action of the SMPP can be extracted, this NCGFT thus being named the NCSMPP.

\section{Derivation-based NCGFT}
\label{DBNCGFT}
As we will see, the derivation's framework offers a way to do NCGFT, implementing the Higgs boson in its generalized connection, and  naturally implementing a SSBM-like mechanism. Initial developments can be found in \cite{DuboKernMado90a, DuboKernMado90b}. 
\medskip
\par
Let $\algA$ be a unital associative algebra equipped with an involution $a \mapsto a^*$ and let $\modM$ be a left $\algA$-module. A (NC) connection $\nabla$ on $\modM$ is a family of linear maps $\nabla_\kX : \modM \to \modM$ defined for any $\kX \in \Der(\algA)$ such that
\begin{enumerate}
	\item $\nabla_{f \kX} = f \nabla_{\kX}$ and $\nabla_{\kX + \kY} = \nabla_{\kX} + \nabla_{\kY}$ for any $f \in \calZ(\algA)$ and $\kX, \kY \in \Der(\algA)$.
	
	\item $\nabla_{\kX} (a e) = (\kX \cdotaction a) e + a \nabla_{\kX} e$ for any $a \in \algA$, $e \in \modM$ and $\kX \in \Der(\algA)$.
\end{enumerate}
The connection must be Hermitian: $\kX \cdotaction h(e_1, e_2)= h(\nabla_\kX e_1, e_2) + h(e_1, \nabla_\kX e_2)$ for any $a_1, a_2 \in \algA$ and $e_1, e_2 \in \modM$.
\medskip
\par 
The curvature of $\nabla$ is the family of maps $R(\kX, \kY) : \modM \to \modM$ defined for any $e \in \modM$ and $\kX, \kY \in \Der(\algA)$ by
\begin{align*}
	R(\kX, \kY) e \defeq ( \nabla_{\kX} \nabla_{\kY} - \nabla_{\kY} \nabla_{\kX} - \nabla_{[\kX, \kY]}) e
\end{align*}
It can be easily shown that $R(\kX, \kY) (a e) = a R(\kX, \kY) e$ for any $a \in \algA$ so that $R(\kX, \kY) \in \Hom_{\algA}(\modM, \modM)$ (space of homomorphisms of left modules). 

\medskip
\par 
A special case of interest is the left module $\modM = \algA$ for the multiplication in $\algA$ equipped with the canonical Hermitian structure $h(a,b) \defeq a b^*$ for any $a,b \in \modM = \algA$. Then, since $\algA$ is unital, the connection $\nabla$ is completely given by its values on the unit $\bbbone \in \algA$: for any $e \in \modM = \algA$ and any $\kX \in \Der(\algA)$,
\begin{align*}
	\nabla_{\kX} e 
	= \nabla_{\kX} (e \bbbone)
	= (\kX \cdotaction e) + e (\nabla_{\kX} \bbbone)
	= (\kX \cdotaction e) + e \omega(\kX)
\end{align*}
where we define $\omega(\kX) \defeq \nabla_{\kX} \bbbone$. Then one has $\omega \in \OmegaDer^1(\algA)$ and $\omega$ is called the connection $1$-form of $\nabla$. The compatibility of $\nabla$ with $h$ implies that for any real derivation $\kX$, one has $\omega(\kX) + \omega(\kX)^* = 0$ since $0 = \kX \cdotaction \bbbone = \kX \cdotaction h(\bbbone, \bbbone) = \omega(\kX) + \omega(\kX)^*$.
\medskip
\par 
The curvature can be computed in terms of $\omega$ as $R(\kX, \kY) e = e \Omega(\kX, \kY)$ where
\begin{align*}
	\Omega(\kX, \kY) \defeq (\dd \omega)(\kX, \kY) - [\omega(\kX), \omega(\kY)]
\end{align*}
using the fact that $\modM = \algA$ is also a right $\algA$-module. The $2$-form $\Omega \in \OmegaDer^2(\algA)$ is the curvature $2$-form of $\nabla$.
\medskip
\par 
The gauge group $\calG$ is the space $\calU(\algA)$ of unitary elements in $\algA$ which act on the right on $\modM$ and respect conditions defined in \ref{GaugeModGroup}. Indeed, any $g \in \calG$ is defined by its value $u \defeq g(\bbbone) \in \algA$ so that $g(e) = g(e \bbbone) = e g(\bbbone) = e u$. Since $\calG$ is a group, the element $u$ is invertible in $\algA$ and the unitary condition comes from the compatibility with the Hermitian structure: $\bbbone = h(\bbbone, \bbbone) = h(\bbbone u, \bbbone u) = u u^\ast$. A computation shows that the connection $1$-form associated to $\nabla^g_{\kX}=g \circ \nabla_{\kX} \circ g^{-1}$ is given by:
\begin{align*}
	\omega^u \defeq u^{-1} \omega u - u^{-1} (\dd u)
\end{align*}
and its curvature $2$-form is $\Omega^u \defeq u^{-1} \Omega u$. A very simple NC connection is given by $\nabla_{\kX}^0 \defeq a\,\to\, \kX a$.

\begin{proposition}[Transport of connections by automorphisms]
	\label{prop transport of connections by automorphisms}
	Let us consider the hypothesis of Prop.~\ref{prop transport forms}. 
	\medskip
	\par
	Let $\PsiMod : \modM \to \modM$ be an invertible linear map such that $\PsiMod(a e) = \Psi(a) \PsiMod(e)$ for any $a \in \algA$ and $e \in \modM$.
	\medskip
	\par
	Let $\nabla$ be a connection on $\modM$ compatible with a Hermitian structure $h$ on $\modM$. Then, for any $\kX \in \Der(\algA)$ and $e \in \modM$, the maps $\nabla^{\PsiMod}_\kX e \defeq \PsiMod \left( \nabla_{\PsiDer^{-1}(\kX)} \PsiMod^{-1}(e) \right)$ define a connection on $\modM$ which is compatible with the Hermitian structure $h^{\PsiMod}$ defined by $h^{\PsiMod}(e_1, e_2) \defeq \Psi \left( h( \PsiMod^{-1}(e_1), \PsiMod^{-1}(e_2)) \right)$. Its curvature $R^{\PsiMod}$ satisfies $R^{\PsiMod}(\kX, \kY) e = \PsiMod \left( R(\PsiDer^{-1}(\kX), \PsiDer^{-1}(\kY) ) \PsiMod^{-1}(e) \right)$ where $R$ is the curvature of $\nabla$.
	\medskip
	\par
	Let $g \in \calG$ be a gauge transformation on $\modM$. Then $g^{\PsiMod} (e) \defeq \PsiMod \circ g \circ \PsiMod^{-1}(e)$ belongs to $\calG$. If $g$ is compatible with $h$ then $g^{\PsiMod}$ is compatible with $h^{\PsiMod}$. One has $(\nabla^g)^{\PsiMod} = (\nabla^{\PsiMod})^{g^{\PsiMod}}$.
	\medskip
	\par
	For $\modM = \algA$, let $\PsiMod = \Psi$. Let $\omega$ (resp. $\omega^\Psi$) be the connection $1$-form of $\nabla$ (resp. of $\nabla^{\Psi}$). Then one has $\omega^\Psi = \Psi(\omega)$. Let $u = g(\bbbone)$ and $u^\Psi = g^\Psi(\bbbone)$, then $u^\Psi = \Psi(u)$.
\end{proposition}

\begin{proof}
	For any $a \in \algA$, $f \in \calZ(\algA)$, $\kX \in \Der(\algA)$ and $e \in \modM$, one has 
	\begin{align*}
		\nabla^{\PsiMod}_\kX (a e)
		&= \PsiMod \left( \nabla_{\PsiDer^{-1}(\kX)} \PsiMod^{-1}(a e) \right)
		= \PsiMod \left( \nabla_{\PsiDer^{-1}(\kX)} \Psi^{-1}(a) \PsiMod^{-1}(e) \right)
		\\
		& = \begin{multlined}[t]
			\PsiMod \left( (\PsiDer^{-1}(\kX) \cdotaction \Psi^{-1}(a) ) \PsiMod^{-1}(e) \right) + \PsiMod \left( \Psi^{-1}(a) \nabla_{\PsiDer^{-1}(\kX)} \PsiMod^{-1}(e) \right)
		\end{multlined}
		\\
		&= \PsiMod \left( \Psi^{-1}(\kX \cdotaction a) \PsiMod^{-1}(e) \right)  + a \PsiMod \left( \nabla_{\PsiDer^{-1}(\kX)} \PsiMod^{-1}(e) \right)= (\kX \cdotaction a) e + a \nabla^{\PsiMod}_\kX e
	\end{align*}
	and
	\begin{align*}
		\nabla^{\PsiMod}_{f \kX} e
		&= \PsiMod \left( \nabla_{\PsiDer^{-1}(f \kX)} \PsiMod^{-1}(e) \right)
		= \PsiMod \left( \nabla_{\Psi^{-1}(f )\PsiDer^{-1}(\kX)} \PsiMod^{-1}(e) \right)= \PsiMod \left( \Psi^{-1}(f ) \nabla_{\PsiDer^{-1}(\kX)} \PsiMod^{-1}(e) \right)
		\\
		&= f \PsiMod \left( \nabla_{\PsiDer^{-1}(\kX)} \PsiMod^{-1}(e) \right)= f \nabla^{\PsiMod}_{\kX} e
	\end{align*}
	so that $\nabla^{\PsiMod}$ is a connection. The compatibility with $h^{\PsiMod}$ is proved by
	\begin{align*}
		\kX \cdotaction h^{\PsiMod}(e_1, e_2)
		&= \kX \cdotaction \Psi \left( h( \PsiMod^{-1}(e_1), \PsiMod^{-1}(e_2)) \right)
		= \Psi \left( \PsiDer^{-1}(\kX) \cdotaction h( \PsiMod^{-1}(e_1), \PsiMod^{-1}(e_2)) \right)
		\\
		&= \begin{multlined}[t]
			\Psi \left( h( \nabla_{\PsiDer^{-1}(\kX)} \PsiMod^{-1}(e_1), \PsiMod^{-1}(e_2)) \right)
			+ \Psi \left( h( \PsiMod^{-1}(e_1), \nabla_{\PsiDer^{-1}(\kX)} \PsiMod^{-1}(e_2)) \right)
		\end{multlined}
		\\
		&= \begin{multlined}[t]
			\Psi \left( h( \PsiMod^{-1} (\nabla^{\PsiMod}_{\kX} e_1), \PsiMod^{-1}(e_2)) \right)
			+ \Psi \left( h( \PsiMod^{-1}(e_1), \PsiMod^{-1} (\nabla^{\PsiMod}_{\kX} e_2)) \right)
		\end{multlined}
		\\
		&= h^{\PsiMod}( \nabla^{\PsiMod}_{\kX} e_1, e_2)
		+ h^{\PsiMod}(e_1, \nabla^{\PsiMod}_{\kX} e_2) .
	\end{align*}
	The relation for the curvature $R^{\PsiMod}$ is a straightforward computation:
	\begin{align*}
		R^{\PsiMod}(\kX, \kY) e
		&= \left(  \nabla^{\PsiMod}_{\kX}  \nabla^{\PsiMod}_{\kY} -  \nabla^{\PsiMod}_{\kY}  \nabla^{\PsiMod}_{\kX} -  \nabla^{\PsiMod}_{[\kX, \kY]} \right) e
		\\
		&= \begin{multlined}[t]
			\PsiMod \circ \nabla_{\PsiDer^{-1}(\kX)} \circ \PsiMod^{-1} \circ \PsiMod \circ \nabla_{\PsiDer^{-1}(\kY)} \circ \PsiMod^{-1}(e)\\
			- \PsiMod \circ \nabla_{\PsiDer^{-1}(\kY)} \circ \PsiMod^{-1} \circ \PsiMod \circ \nabla_{\PsiDer^{-1}(\kX)} \circ \PsiMod^{-1}(e)
			- \PsiMod \circ \nabla_{\PsiDer^{-1}([\kX, \kY])} \circ \PsiMod^{-1}(e)\qquad\qquad\qquad\qquad\qquad\qquad\qquad\qquad\qquad\qquad\qquad\qquad\qquad\qquad\qquad\qquad
		\end{multlined}
		\\
		&= \begin{multlined}[t]
			\PsiMod \circ \nabla_{\PsiDer^{-1}(\kX)} \circ \nabla_{\PsiDer^{-1}(\kY )} \circ \PsiMod^{-1}(e)
			- \PsiMod \circ \nabla_{\PsiDer^{-1}(\kY)} \circ \nabla_{\PsiDer^{-1}(\kX)} \circ \PsiMod^{-1}(e)
			\\
			- \PsiMod \circ \nabla_{[\PsiDer^{-1}(\kX), \PsiDer^{-1}(\kY)]} \circ \PsiMod^{-1}(e)\qquad\qquad\qquad\qquad\qquad\qquad\qquad\qquad\qquad\qquad\qquad\qquad
		\end{multlined}
		\\
		&= \PsiMod \left( R(\PsiDer^{-1}(\kX), \PsiDer^{-1}(\kY)) \PsiMod^{-1}(e) \right) 
	\end{align*} 
	
	The map $g^{\PsiMod}$ is obviously invertible with inverse $(g^{\PsiMod})^{-1} = \PsiMod \circ g^{-1} \circ \PsiMod^{-1}$. It is a morphism of modules: $g^{\PsiMod}(a e) = \PsiMod \circ g \circ \PsiMod^{-1} (a e) = \PsiMod \circ g \left( \Psi^{-1}(a) \PsiMod^{-1}(e) \right) = \PsiMod \left( \Psi^{-1}(a) g \circ \PsiMod^{-1}(e) \right) = a g^{\PsiMod}(e)$. One has
	\begin{align*}
		h^{\PsiMod}( g^{\PsiMod}(e_1), g^{\PsiMod}(e_2) )
		&= \Psi \left( h( \PsiMod^{-1} \circ \PsiMod \circ g \circ \PsiMod^{-1}(e_1), \PsiMod^{-1} \circ \PsiMod \circ g \circ \PsiMod^{-1}(e_2) \right)
		\\
		&= \Psi \left( h( g \circ \PsiMod^{-1}(e_1), g \circ \PsiMod^{-1}(e_2) \right)= \Psi \left( h( \PsiMod^{-1}(e_1), \PsiMod^{-1}(e_2) \right)
		= h^{\PsiMod}(e_1, e_2)
	\end{align*}
	and
	\begin{align*}
		(\nabla^g)^{\PsiMod}_\kX
		&= \PsiMod \circ \nabla^g_{\PsiDer^{-1}(\kX)} \circ \PsiMod^{-1}
		= \PsiMod \circ g \circ \nabla_{\PsiDer^{-1}(\kX)} \circ g^{-1} \circ \PsiMod^{-1}
		\\
		&= (\PsiMod \circ g \circ \PsiMod^{-1}) \!\circ\! (\PsiMod \circ \nabla_{\PsiDer^{-1}(\kX)} \circ \PsiMod^{-1}) \!\circ\! (\PsiMod \circ g \circ \PsiMod^{-1})^{-1}
		\\
		&= g^{\PsiMod} \circ \nabla^{\PsiMod}_\kX \circ (g^{\PsiMod})^{-1}
		= (\nabla^{\PsiMod})^{g^{\PsiMod}}_\kX .
	\end{align*}
	Finally, one has 
	\begin{align*}
		\omega^\Psi(\kX) 
		&= \nabla^\Psi_\kX \bbbone 
		= \Psi \circ \nabla_{\PsiDer^{-1}(\kX)} \Psi^{-1} \bbbone 
		= \Psi \circ \nabla_{\PsiDer^{-1}(\kX)} \bbbone = \Psi \left( \omega(\PsiDer^{-1}(\kX)) \right) 
		= \Psi(\omega) (\kX)
	\end{align*}
	and $u^\Psi = g^\Psi(\bbbone) = \Psi \circ g \circ \Psi^{-1}(\bbbone) = \Psi \circ g(\bbbone) = \Psi(u)$.
\end{proof}

\subsection{Action in the Derivation-based Framework}
To define a gauge field theory on $\modM$, one considers the “fields” defining a connection $\nabla$ on $\modM$ and a Lagrangian $\lag(\nabla)$ for these fields. This Lagrangian is usually constructed for the left module $\modM = \algA$ using a Hodge star operator $\hstar$ on the space of forms on $\algA$: \begin{align*}
	\lag(\nabla) \defeq - \Omega \wedge \hstar \Omega.
\end{align*}
Then, using a trace $\int_{\algA}$ (which sends forms to scalars) we can define an action:
\begin{align*}
	\act[\nabla] \defeq \int_{\algA} \lag(\nabla) =  - \int_{\algA} \Omega \wedge \hstar \Omega
\end{align*}
the sign is necessary for positivity. The matter Lagrangian can be defined in a similar way. One first considers $\nabla$ as a map $\nabla : \modM \to \OmegaDer^1(\algA) \otimes_{\algA} \modM$. Using a natural involution on $\OmegaDer^\grast(\algA)$ which extends the involution on $\algA$ (see \cite{FranLazzMass14a} for instance), one can extend $h$ to $(\OmegaDer^p(\algA) \otimes_{\algA} \modM )  \otimes (\OmegaDer^q(\algA) \otimes_{\algA} \modM ) \to \OmegaDer^{p+q}(\algA)$ by $h(\omega_p \otimes e, \omega_q \otimes e') \defeq \omega_p  h(e, e') \wedge \omega_q^*$. Then $\int_{\algA} h( \nabla e, \hstar \nabla e)$ defines a Klein-Gordon type action for matter fields $e \in \modM$.
\medskip
\par 
Since we restrict our analysis to matrix algebras, we refer to Sect.~\ref{sec metric hodge} for the construction of an explicit Hodge star operator and a trace.

\medskip
Let us describe the degrees of freedom in the gauge sector of a NCGFT defined on $M_n$ and on $C^\infty(\Man) \otimes M_n$, see \cite{DuboKernMado90a, DuboKernMado90b, DuboMass98a, Mass12a, FranLazzMass14a} for some details.  We use some notations from Sect.~\ref{sec metric hodge}.

\subsection{\texorpdfstring{Example of $\algA = M_n$}{Example of A = Mn}}

\label{example Mn}
Let us consider $\algA = M_n$. Then $\Der(\algA) = \Der(M_n) \simeq \ksl_n$, and, for $k=1, \ldots, n^2-1$, let $\{ E_k \}$ be a basis of anti-Hermitean traceless matrices in $\ksl_n$ so that $\{ \partial_k \defeq \ad_{E_k} \}$ is a basis of real derivations of $M_n$.\footnote{We depart here from the conventions in many papers where the $E_k$ are chosen to be Hermitean and $\partial_k$ are defined as $\ad_{i E_k}$.} Let us consider the left module $\modM = \algA$ with the canonical Hermitian structure $h(a,b) \defeq a b^*$. There is a canonical connection $\mrnabla$ on $\modM$ defined by $\mrnabla_{\partial_k} a \defeq E_k a$ for any $k=1, \ldots, n^2-1$ and $a \in \algA = \modM$ with connection $1$-form  $\mromega(\partial_k) = E_k$. This canonical connection satisfies two important properties: firstly, its curvature is zero; secondly, it is gauge invariant (see also \cite{CagnMassWall11a} for another occurrence of such a canonical connection).
 It is then convenient to compare any connection $1$-form $\omega$ on $\modM$ to this canonical connection, by writing $\omega = \omega_k \theta^k = \mromega - B_k \theta^k = (E_k - B_k) \theta^k$. Then the curvature $2$-form $\Omega = \tfrac{1}{2} \Omega_{k\ell} \theta^k \wedge \theta^\ell$ has components $\Omega_{k\ell} \defeq \Omega(\partial_k, \partial_\ell) = - ([B_k, B_\ell] - C_{k\ell}^m B_m)$. This curvature vanishes iff $E_k \mapsto B_k$ is a representation of the Lie algebra $\ksl_n$ (for instance $B_k = 0$ or $B_k = E_k$). The connection $\omega$ is compatible with $h$ iff $\omega_k + \omega^*_k = 0$ for any $k$. Since the $E_k$'s are anti-Hermitean, this compatibility condition is equivalent to $B_k + B_k^* = 0$ for any $k$ and then $\Omega_{k\ell}^* = - \Omega_{k\ell}$. We can then decompose $B_k = B_k^\ell E_\ell + i B_k^0 \bbbone_n$ with real functions $B_k^\ell$, $\ell = 0, \ldots, n^2-1$, so that the number of degrees of freedom (number of real functions) in $\omega$ is $n^2 (n^2-1)$. The action of a gauge transformation $g \in U(n)$ induces the transformation $B_k \mapsto g^{-1} B_k g$ (the inhomogeneous part of the gauge transformation is absorbed by $\mromega$).
\medskip
\par
Notice that this approach is only interesting for $n \geq 2$ since for $n=1$, $M_1 = \bbC$ is commutative and so there is no derivation and so no degree of freedom $B_k$'s.
\medskip
\par
Suppose that the basis $\{ \partial_k \}$ is orthonormal for the metric $g$ defined as in Sect.~\ref{sec metric hodge}. Since $\Omega \wedge \hstar \Omega = \tfrac{1}{2} \Omega_{k\ell} \Omega^{k\ell} \omega_{\vol}$ for $\Omega^{k\ell} = g^{kk'} g^{\ell \ell'} \Omega_{k'\ell'}$, the action is then $- \tfrac{1}{2} \sum_{k,\ell} \tr (\Omega_{k\ell})^2 = - \tfrac{1}{2} \sum_{k,\ell} \tr ([B_k, B_\ell] - C_{k\ell}^m B_m)^2$. Notice that $- \tr (\Omega_{k\ell})^2 = \tr (\Omega_{k\ell} \Omega_{k\ell}^*) \geq 0$.
\medskip
\par
From Prop.~\ref{prop transport of connections by automorphisms} and Examples~\ref{ex transport derivations inner automorphisms} and \ref{ex transport matrix by inner automorphisms}, an inner automorphism defined by a unitary element $u$ in $M_n$ produces a transport of all the structures defining the NCGFT on $M_n$. One has $\omega^u = U_k^\ell u \omega_\ell u^{-1} \theta^k$, and since $\mromega^u = U_k^\ell u E_\ell u^{-1} \theta^k = \mromega$, this implies that $B_k$ is mapped to $B^u_k = U_k^\ell u B_\ell u^{-1}$. One then has $[B^u_k, B^u_\ell] - C_{k\ell}^m B^u_m = U_k^{k'} U_\ell^{\ell'} u^{-1} \left( [B_{k'}, B_{\ell'}] - C_{k'\ell'}^{m' }B_{m'} \right) u$ and the Lagrangian in the $B^u_k$ is the same as the Lagrangian in the $B_k$. We conclude that such an action of inner automorphisms is not relevant from a physical point of view.

\subsection{\texorpdfstring{Example of $\halgA = C^\infty(\Man) \otimes M_n$}{Example of A = Mn otimes C(M)}}

\label{example C(\Man) otimes Mn}
Let us consider the algebra $\halgA = C^\infty(\Man) \otimes M_n$ for a manifold $\Man$. The space of derivations is $\Der(\halgA) = [\Gamma(\Man) \otimes \bbbone_n] \oplus [C^\infty(\Man) \otimes \ksl_n]$ where $\Gamma(\Man) = \Der(C^\infty(\Man))$ is the space of vector fields on $\Man$, and $\calZ(\halgA)=C^\infty(\Man)$. For $\mu = 1, \ldots, \dim M$, let $\{ \partial_\mu \}$ (usual partial derivatives in a coordinate system given by a chart of $\Man$) be a basis of real derivations on the geometric part, and let $\{ \dd x^\mu \}$ be the dual basis of $1$-forms. We can define a metric on this almost commutative manifold:
\begin{align*}
	\hat{g}(\partial_\mu+\partial_\alpha,\partial_\nu+ \partial_\beta)=\tilde{g}(\partial_\mu,\partial_\nu)+g(\partial_\alpha, \partial_\beta) =\tilde{g}_{\mu\nu}+g_{k \ell}
\end{align*}
where we recognize $\tilde{g}_{\mu\nu}$ the usual metric for manifold, and $g_{k \ell}$ the metric for the inner degrees of freedom defined in section \ref{ResMatrDer}.
\GN{verif metric, pas de terme renormalisant en $\Lambda$ entre deux ?}
\medskip
\par 	
Let us consider as before the left module $\modM = \halgA$ with the canonical Hermitian structure $h(a,b) \defeq a b^*$. Then a connection $1$-form $\omega$ can be written as $\omega = \omega_\mu \dd x^\mu + \omega_k \theta^k = A_\mu \dd x^\mu + (E_k - B_k) \theta^k$ with $A_\mu, B_k \in \halgA$ and this connection is compatible with $h$ when $A_\mu + A_\mu^* = 0$ and $B_k + B_k^* = 0$ (since the $E_k$'s are anti-Hermitean). As before, we can decompose $A_\mu = A_\mu^\ell E_\ell + i A_\mu^0 \bbbone_n$ and $B_k = B_k^\ell E_\ell + i B_k^0 \bbbone_n$ so that the number of degrees of freedom in $\omega$ is $n^2 (\dim M + n^2-1)$. A gauge transformation given by $g \in C^\infty(\Man) \otimes U(n)$ induces the transformations $A_\mu \mapsto g^{-1} A_\mu g - g^{-1} \dddR g$ and $B_k \mapsto g^{-1} B_k g$ where $\dddR$ is the ordinary de~Rham differential on $\Man$ (to simplify, we used the notation $\dd x^\mu$ instead of $\dddR x^\mu$). So $A_\mu \dd x^\mu$ can be identified with an ordinary $U(n)$-connection.
\medskip
\par
The curvature of $\omega$ can be decomposed into three parts: $\Omega = \tfrac{1}{2} \Omega_{\mu\nu} \dd x^\mu \wedge \dd x^\nu + \Omega_{\mu k} \dd x^\mu \wedge \theta^k + \tfrac{1}{2}  \Omega_{k\ell} \theta^k \wedge \theta^\ell$ with
\begin{align*}
	\Omega_{\mu\nu}
	&= \partial_\mu A_\nu - \partial_\nu A_\mu - [A_\mu, A_\nu],
	\\
	\Omega_{\mu k}
	&= - ( \partial_\mu B_k - [A_\mu, B_k] ),
	\\
	\Omega_{k\ell}
	&= - ( [B_k, B_\ell] - C_{k\ell}^m B_m ).
\end{align*}
The term $\Omega_{\mu\nu}$ is the usual field strength of $A_\mu$, $\Omega_{\mu k}$ is (up to a sign) the covariant derivative of $B_k$ along the connection $A_\mu$ and $\Omega_{k\ell}$ is the expression obtained for the algebra $\algA = M_n$. Using natural notions of metric and Hodge $\hstar$-operator in this context, a natural Lagrangian is the sum of 3 (positive) terms $- \tfrac{1}{2} \tr(\Omega_{\mu\nu} \Omega^{\mu\nu}) - \tr(\Omega_{\mu k} \Omega^{\mu k}) - \tfrac{1}{2} \tr(\Omega_{k\ell} \Omega^{k\ell})$. Finding a minimal configuration for such a Lagrangian is equivalent to minimizing independently these 3 terms. The last one vanishes if and only if $E_k \mapsto B_k$ is a representation of $\ksl_n$. One possibility is the take $B_k = 0$ for all $k$ (referred to as the “null-configuration” in the following), which cancels also the second term. Then one reduces the theory to massless gauge fields $A_\mu$. Another more stimulating configuration is to consider $B_k = E_k$ (referred to as the “basis-configuration” in the following), and then the second term reduces to $- \tr ([A_\mu, E_k] [A^\mu, E^k])$, which, after developing, produces mass terms for the $A_\mu$ fields, see Lemma~\ref{lem mass rep-config}. This is similar to the SSBM implemented in the SMPP to give masses to some gauge fields.
\medskip
\par
Notice that for an ordinary Yang-Mills theory in the framework of fiber bundles and connections, with structure group $U(n)$, we have only the fields $A_\mu^\ell$, $\ell = 0, \ldots, n^2-1$ since there is no “algebraic part” which produces the $B_k$'s. With the structure group $SU(n)$, there is no field $A_\mu^0$ (the matrices $E_\ell$, $\ell = 1, \ldots, n^2-1$, generate the real Lie algebra $\ksu(n)$).
\medskip
\par
Contrary to Example~\ref{example Mn}, this case is also interesting for $n=1$. In that case, the degrees of freedom are only in the spatial direction (the $A_\mu$'s) and they can be used to construct an ordinary $U(1)$ gauge field theory.
\medskip
\par
Once again, one can ask about the action of an inner automorphism defined by a unitary element in $\halgA$. The action of such an automorphism on a “spatial” derivation $X \in \Gamma(\Man)$ is given by $\PsiDer(X) = X + \ad_{u(X\cdotaction u^{-1})}$ (see Example~\ref{ex transport derivations inner automorphisms}). If $u$ is a unitary in $M_n$ (not depending on $\Man$), then one gets $\PsiDer(X) = X$. This implies that the spatial directions (the $\partial_\mu$'s) are only affected by $u$ through the action of $\Psi$, $\omega^u_\mu = u \omega_\mu u^{-1}$, while the “inner” directions (the $\partial_k$'s) change according to the rules given in Example~\ref{ex transport matrix by inner automorphisms}. This implies that the Lagrangian in the new fields is the same as the one in the original fields and so such an action of inner automorphism is not relevant from a physical point of view. When $u$ is depending on $\Man$, the second term in $\PsiDer(X)$ does not vanish and it produces mixing between spatial directions and inner directions: some degrees of freedom in the $B_k$'s are sent in the spatial part $\omega^u_\mu$. This situation will not be considered in the following.

\begin{lemma}
	\label{lem mass rep-config}
	Let us consider a NCGFT as previously given. In the basis-configuration for the $B_k$'s, the masses induced on the fields $A_\mu^\ell$ for $\ell = 1, \dots, n^2-1$, in the decomposition $A_\mu = A_\mu^\ell E_\ell + i A_\mu^0 \bbbone_n$, are all the same and equal to $m_{\text{basis-config}} = \sqrt{2n}$, while the field $A_\mu^0$ is mass-less.
\end{lemma}

\begin{proof}
	Using the metric $g$ defined as $g(E_k, E_\ell) = \tr(E_k E_\ell)$ (see Sect.~\ref{sec metric hodge}), the masses for the fields $A_\mu^\ell$, $\ell = 1, \dots, n^2-1$, are given by the term 
	\begin{align*}
		M^2_{\ell_1 \ell_2} A_\mu^{\ell_1} A^{\mu, \ell_2}
		&= - g^{k_1 k_2} \tr ([A_\mu, E_{k_1}] [A^\mu, E_{k_2}]) 
		= - A_\mu^{\ell_1} A^{\mu, \ell_2} g^{k_1 k_2} \tr( [E_{\ell_1}, E_{k_1}] [E_{\ell_2}, E_{k_2}] )
		\\
		&= - A_\mu^{\ell_1} A^{\mu, \ell_2} g^{k_1 k_2} C_{\ell_1 k_1}^{m_1} C_{\ell_2 k_2}^{m_2} \tr( E_{m_1} E_{m_2})
		= - A_\mu^{\ell_1} A^{\mu, \ell_2} g^{k_1 k_2} g_{m_1 m_2} C_{\ell_1 k_1}^{m_1} C_{\ell_2 k_2}^{m_2}
	\end{align*}
	where $A_\mu = A_\mu^\ell E_\ell + i A_\mu^0 \bbbone_n$. Since the field $A_\mu^0$ disappears, its mass is $0$.
	\medskip
	\par
	For any $X, Y \in \ksu(n)$, the Killing form $K(X, Y) = \tr( \ad_{X} \circ \ad_{Y} )$ satisfies $K(X,Y) = 2n \tr(XY)$ so that, on the one hand, $K_{k\ell} \defeq K(E_k, E_\ell) = 2 n \, g_{k\ell}$ and on the other hand, $K_{k\ell} = C_{k m}^n C_{\ell n}^m$. Let us define $C_{k \ell m} \defeq g_{m n} C_{k \ell}^n$, so that $C_{k \ell}^n = g^{m n} C_{k \ell m}$ and $C_{k \ell m}$ is completely antisymmetric in $(k, \ell, m)$. This leads to $g^{k_1 k_2} g_{m_1 m_2} C_{\ell_1 k_1}^{m_1} C_{\ell_2 k_2}^{m_2} = g^{k_1 k_2} C_{\ell_1 k_1 m} C_{\ell_2 k_2}^{m} = - g^{k_1 k_2} C_{\ell_1 m k_1} C_{\ell_2 k_2}^{m} = - C_{\ell_1 m}^k C_{\ell_2 k}^{m}$, so that $M^2_{\ell_1 \ell_2} = C_{\ell_1 m}^k C_{\ell_2 k}^{m} = K_{\ell_1 \ell_2} = 2 n \, g_{\ell_1 \ell_2}$. This proves that the diagonalization of $(M^2_{\ell_1 \ell_2})$ gives a unique eigenvalue $2 n$ so that there is a unique mass $m_{\text{basis-config}} = \sqrt{2n}$.
\end{proof}
\medskip
\par 
\begin{remark}
	The way to do NC gauge theories in the derivation framework can be seen as a geometrization of the generators $T_a$ of the gauge potential introduced in chapter \ref{SMPPGFT}. Making this object being at the same level of the $\partial_\mu$, in the covariant derivative \eqref{CovarDer} which becomes even more similar to a simple derivation, in a NCG. This looks like a nice and direct implementation of NCG in the gauge field theory framework.
\end{remark}

This model lacks a fermionic action. More details can be found in \cite{Mass08b, jordan2014gauge}.

\section{Spectral Triple-based NCGFT}
\label{STNCGFT} 
For models using spectral triples, we take the module to be a Hilbert space $\modM=\calH$. As we will see, here the connection is obtained from the inner fluctuations of the Dirac operator. As the elaboration of spectral triples is intimately linked to the elaboration of a NCSMPP, we let the presentation of the history of the elaboration of spectral triples based NCGFT to chapter \ref{NCSMPP}. 

\medskip
\par 
Two spectral triples $(\algA, \hs, D, J, \gamma)$ and $(\algA^{\prime}, \hs^{\prime}, D^{\prime} ,J^{\prime}, \gamma^{\prime})$ are unitary equivalent when $\algA' = \algA$, and there exists a unitary operator $U : \hs \rightarrow \hs'$ and an algebra isomorphism $\Lambda : \algA \rightarrow \algA'$ such that $\pi' \circ \Lambda = U \pi U^{-1}$, $D' = U D U^{-1}$, $J' = U J U^{-1}$, and $\gamma' = U \gamma U^{-1}$.
\medskip
\par 
A symmetry of a spectral triple is a unitary equivalence between two spectral triples such that $\hs' = \hs$, $\algA' = \algA$, and $\pi' = \pi$, so that $U : \hs \to \hs$ and $\Lambda \in \Aut(\algA)$, \textit{\textit{i.e.}} a symmetry acts only on $D$, $J$ and $\gamma$. In the following, we will only consider automorphisms $\Lambda$ which are $\algA$-inner, that is, there is a unitary $u \in \calU(\algA)$ such that $\Lambda_u(a) = u a u^\ast$. This unitary in $\algA$ defines the unitary $U = \pi(u) J \pi(u) J^{-1} : \hs \rightarrow \hs$, which can be interpreted as the conjugation with $\pi(u)$ for the bimodule structure. A straightforward computation shows that such $U$ acts as an automorphism on $\algA$, leaves $J$ and $\gamma$ invariant:
\begin{align*}
	&UaU^{-1}=\pi(u)\pi(u)^\circ a (\pi(u)^\circ)^*\pi(u)^*=\pi(u)a\pi(u)^*\pi(u)^\circ(\pi(u)^\circ)^*=\pi(u)a\pi(u)^*=\Lambda_{\pi(u)}(a)  & \\
	&UJU^{-1}=\pi(u)J\pi(u)(\pi(u)^*(\pi(u)^*)^\circ)=\pi(u)JJ\pi(u)^*J^{-1}=\epsilon J^{-1}=J& \\
	& U\gamma U^{-1}=\epsilon^{\prime\prime 2}\gamma UU^{-1}=\gamma
\end{align*}
and modifies the Dirac operator $D$ as:
\begin{equation}
	\label{fluctuDi}
	D^u =U D U^*= D + \pi(u) [D, \pi(u)^\ast] + \epsilon' J\left( \pi(u) [D, \pi(u)^\ast] \right) J^{-1}.
\end{equation}
The usual way to look at this relation is to interpret the commutator with $D$ as a kind of differential: this expression tells us that $D$ is modified by the addition of two inhomogeneous terms of the form ``$\pi(u) \dd \pi(u)^{-1}$''.  
\medskip
\par 
By definition, gauge transformations are inner symmetries of a spectral triple. In order to compensate for the inhomogeneous terms, we can use the same trick as in ordinary gauge field theory: add to the first order differential operator $D$ a gauge potential. To do that, we need a convenient notion of NC connections.
\medskip
\par 
We now suppose that there is an orthogonal decomposition of the Hilbert space $\hs = \toplus_{i=1}^{r} \hs_i$ such that the representation decomposes along $\pi = \toplus_{i=1}^{r} \pi_i$ where $\pi_i$ is a representation of $\algA_i$ on $\hs_i$: for any $\psi = \toplus_{i=1}^{r} \psi_i \in \hs$ and $a = \toplus_{i=1}^{r} a_i \in\algA$, $\pi(a) \psi = \toplus_{i=1}^{r} \pi_i(a_i) \psi_i$. Then the Dirac operator $D$ decomposes as a $r \times r$ matrix of operators $D_{j}^{i} : \hs_i \to \hs_j$.
We propose to write the representation $\piD$ as follows. Let us consider the universal differential calculus $(\Omega^\grast_U(\algA), \ddU)$ and consider any $\bomega \in \bOmega^n_U(\algA) \subset \kT^n \algA$ which decomposes along a sum of typical terms $\toplus_{i_1, \dots, i_{n-1} =1}^{r} \big( a^0_{i} \otimes a^1_{i_1} \otimes \cdots \otimes a^{n-1}_{i_{n-1}} \otimes a^n_{j} \big)_{i, j = 1}^{r} \in \kT^n \algA$. Then $\piD(\bomega)$ is the $r \times r$ matrix of operators 
\begin{align}
	\label{eq piD sum algebras}
	\piD(\bomega)_i^j = 
	\!\!\!\!\!\sum_{ \substack{ \text{all terms at the}\\ \text{$(i,j)$ entry in $\bomega$} } } \!\!\!\!\!
	\tsum_{i_1, \dots, i_{n-1} = 1}^{r} 
	a^{0}_{i} D_{i}^{i_1} a^{1}_{i_1}  D_{i_1}^{i_2} \cdots D_{i_{n-2}}^{i_{n-1}} a^{n-1}_{i_{n-1}} D_{i_{n-1}}^{j} a^{n}_{j} : 
	\hs_j \to \hs_i
\end{align}
Notice that, since $\bomega \in \bOmega^n_U(\algA)$, these sums define bounded operators because only commutators $[D, a]$ could appear in $\piD(\bomega)$ (this is not necessarily the case for a generic element in $\kT^n \algA$). 

\medskip
\par 
As mentioned in section \ref{NCConnectionModCurv}, a NC connection is defined as a $1$-form $\omega = \sum_i a^{0}_i \ddU a^{1}_i \in \Omega^1_U(\algA)$ (finite sum). Elements in the vector spaces $\Omega^n_U(\algA)$ can be represented as bounded operators on $\hs$ by
\begin{equation*}
	\piD \big(\tsum_i a^{0}_i \ddU a^{1}_i \cdots \ddU a^{n}_i \big) \defeq \tsum_i \pi(a^{0}_i) [ D, \pi(a^{1}_i)] \cdots [ D, \pi(a^{n}_i)].
\end{equation*}
Notice that the map $\piD$ is not a representation of the graded differential algebra $\Omega^\grast_U(\algA)$. In particular, $\ddU$ is not represented by the commutator $[D, -]$ as a differential. The representation $\piD$ can also be used to represent $n$-forms on the right module structure of $\hs$ by $\tsum_i a^{0}_i \ddU a^{1}_i \cdots \ddU a^{n}_i \mapsto J \piD\left(\tsum_i a^{0}_i \ddU a^{1}_i \cdots \ddU a^{n}_i \right) J^{-1}$. The map $\piD$ may have a non trivial kernel, this is why we will prefer to use $\omega \in \Omega^1_U(\algA)$ instead of $\piD(\omega)$ in some forthcoming constructions. We notice that the term $\pi(u) [D, \pi(u)^\ast]$ in equation \eqref{fluctuDi} can be interpreted as a particular connection.
\medskip
\par 
Given $D$ and $\omega \in \Omega^1_U(\algA)$, one defines the operator $D_\omega \defeq D + \piD(\omega) + \epsilon' J \piD(\omega) J^{-1}$ called the fluctuated Dirac operator. Where $\piD(\omega)$ is an element of the space of Connes differential 1-forms $\Omega_{D}^{1}(\algA)$. By a gauge transformation $u \in \calU(\algA)$, $D_\omega$ is transformed into
\begin{align*}
	(D_\omega)^u = D + \pi(u) \piD(\omega) \pi(u)^\ast + \pi(u) [D, \pi(u)^\ast]  + \epsilon' J \pi(u) \piD(\omega) \pi(u)^\ast J^{-1} + \epsilon' J \pi(u) [D, \pi(u)^\ast] J^{-1}.
\end{align*}
This relation can be written as $D_{\omega^u}$, where $\omega^u \in \Omega^1_U(\algA)$ is a gauge transformation of $\omega$ defined as $\omega^u \defeq u \omega u^\ast + u \ddU u^\ast$.
\medskip
\par 
This is how a gauge field (in the general meaning of the term) is extracted from the Dirac when it is undergoing inner fluctuation. Therefore, considering $\halgA=\calC^\infty(\Man)\otimes \af$ we see that $\Out(\calC^\infty(\Man))$ the group of Outer automorphisms of $\halgA$ is enhanced by the inner automorphisms of $\af$. As the metric comes out from the Dirac, these new automorphisms give us deformation of the Dirac and then of metric for the spectral geometry of $(\halgA, \calH_{\halgA}, D_{\halgA})$, these deformations are interpreted as the gauge fields of Yang-Mills theory.  
\medskip
\par 
Thus, any real spectral triple  $(\algA, \hs, D, J, \gamma)$ possesses a gauge group deduced from the unitaries of $\algA$, linking this triple to other triples wherein the algebra is unitary equivalent to $\algA$ and where the Hilbert space is left unchanged. The gauge group of the triple can be defined as follows:  

\begin{definition}[Gauge group of a spectral triple]
	We define by $\calG(\calA, \calH; J)$ the gauge groupe of the given spectral triple by: 
	\begin{align*}
		\calG(\calA, \calH; J)=\{U=uJuJ^{-1}|u\in\calU(\calA)\} 
	\end{align*}
	
\end{definition}
It is connected to the gauge group acting on $\omega$ by the unitary $u:\omega \mapsto u \omega u^{*}+u\left[D, u^{*}\right]$. The element $U=uJuJ^{-1}$ corresponds to the composition of the left and right actions
of $u$ on $\calH$: $U\, :\, \xi\,\to\, u.\xi. (u^*)^\circ$ which is called the adjoint action
of $u$.
\medskip
\par  
Taking a spectral triple $(\algA, \hs, D, J, \gamma)$, it is possible to construct another spectral triple with a given commutative subalgebra of $\algA$. Let's define:
\begin{align*}
	\algA_J\defeq \{\, a\, \in\, \algA\, |\, aJ=Ja^*\}
\end{align*}  
\begin{proposition}
	Then $(\algA_J, \hs, D, J, \gamma)$ is also a spectral triple, we have that is $a^*\in\algA_J$ for any $a\in \algA_J$, that $\algA_J\in\calZ(\algA)$, that for any $\omega\in \Omega_{D}^{1}(\algA)$: $[a,\omega]=0$, and $\calG(\calA, \calH; J)=\calU(\algA)/\calU(\algA_J)$.
\end{proposition} 
\begin{proof}
	If we take $a\in\algA_J$ and $b\in\algA$ then:
	\begin{align*}
		J^{-1}a^*J=(J^{-1}aJ)^*=(a^*)^*=a\qquad\qquad\text{and then}\qquad a^*J=Ja
	\end{align*} 
	Using the fact that $a=J^{-1}a^*J$ we have $JaJ^{-1}=a^*$ and then $a\epsilon^2=a=Ja^*J^{-1}$, so that:
	\begin{align*}
		[a,b]=[Ja^*J^{-1},b]=0
	\end{align*}
	thanks to commutant property. The fact that $\algA_J$ commutes with $\Omega_{D}^{1}(\algA)$ is a direct consequence of the first order condition. For the last point, if we consider the surjective map $\calU(\algA)\,\to\, \calG(\calA, \calH; J)$ given by $u\,\to\, uJuJ^{-1}$, then this map is a group morphism with kernel given by:
	\begin{align*}
		\{u\in\calU(\algA)\,|\, uJuJ^{-1}=\bbbone\,\leftrightarrow\, uJ=Ju^*\}=\calU(\algA_J)
	\end{align*}
	which proves the last point.
\end{proof}
Because of the Gelfand-Naimark theorem, $\algA_F$ can be seen as a subalgebra of $\calC(X)$, $X$ being a background space.

\subsection{Dirac Operator on a Spin Manifold}
\label{DirSpinMfld}
\epigraph{No one fully understands spinors. Their algebra is formally understood but their general significance is mysterious. In some sense they describe the 'square root' of geometry and, just as understanding the square root of -1 took centuries, the same might be true of spinors.}{\textit{M. Atiyah}}
The motivation of this subsection is to introduce the structures behind GR from the NCG framework point of view and introduce some notations that will be important in subsection \ref{SpectralAction} and then chapter \ref{NCSMPPLag}. 
\medskip
\par
Let's consider the tangent bundle $\TanBun$ over $\Man$, as said the Levi-Civita connection $\nabla$ is the only connection on $\TanBun$ which is both compatible with the inner product defined by the metric, and torsion-free. It can be written in local coordinate as $\nabla_{\partial_\mu}(\partial_\nu)=\Gamma_{\mu\nu}^\lambda\partial_\lambda$, with $\Gamma_{\mu\nu}^\lambda$ the Christoffel's symbols.
\medskip
\par
Let $E$ be a vector bundle over $\Man$, and $\nabla^E$ the associated connection, the associated Laplacian is given by:
\begin{align}
	\label{Laplacian}
	\Delta^E=-g^{\mu\nu}(\nabla_\mu^E\nabla_\nu^E-\Gamma_{\mu\nu}^\lambda\nabla_\lambda^E).
\end{align} 
Now let $\Man$ be a Riemannian spin$^c$ manifold, $\SpinBun$ be the spinor bundle define on $\Man$, and $\{e_i(x)\}$ be a local orthonormal ($g(e_i,e_j)=\delta_{ij}$) basis of $\Gamma(\TanBun)|_x$. In this basis, the Christoffel symbols became $\nabla_{e_i}=\Gamma_{\mu i}^jdx_\mu\otimes e_j$. The spin connection $\nabla^\SpinBun$ is defined to be a lift of the Levi-Civita connection to the spinor bundle:
\begin{align*}
	\nabla_\mu^\SpinBun=\partial_\mu-\frac{1}{4}\Gamma_{\mu i}^j\gamma^i\gamma_j
\end{align*}
It acts on elements $\psi(x)$ of the Hilbert space. Then, the Dirac operator $\DM\, :\, \Gamma^\infty(\SpinBun)\,\to\,\Gamma^\infty(\SpinBun) $ on $\Man$ is given by the composition of the spin connection with Clifford multiplication $c$ by $\DM=-i\gamma^\mu\nabla_\mu^\SpinBun$. 
\begin{remark}
	Taking $f\in\calC^\infty(\Man)$, the Dirac operator $\DM$ is linked to the usual differential:
	\begin{align*}
		[\DM, f]\psi=-ic(\nabla^\SpinBun(f\psi)-f\nabla^\SpinBun\psi)=-ic(df)\psi
	\end{align*}
\end{remark}
As wanted, this Dirac operator connects to the Laplacian, taking $s=R_{\mu\nu}g^{\mu\nu}$ the scalar curvature, and the Laplacian of the spin bundle $\Delta^\SpinBun$ defined according to the equation \eqref{Laplacian}, we have that: 
\begin{align}
	\label{LaplacianSpinc}
	\DM^2=\Delta^\SpinBun+\frac{1}{4}s.
\end{align}
The scalar product on $\Man$ is defined as follows, taking the fiber of $\SpinBun$ above $x\in\Man$, we define $\langle \, , \, \rangle_x$ the scalar product parameterized by $\Man$, and which varies continuously along $\Man$. Then we have the following scalar product on $\Gamma(\SpinBun)$:
\begin{align}
	\label{ScalProd}
	(\psi_1 , \psi_2)=\int_\Man \langle \psi_1(x), \psi_2(x)\rangle_x\sqrt{\det g}dx
\end{align}
The completion of $\Gamma(\SpinBun)$ for the scalar product $(\, , \, )$ is called the space of square integrable spinors, and will be denoted $L^2(\SpinBun)$. As mentioned in \cite{besnard2021noncommutative, lawson2016spin}, the scalar product defined in \eqref{ScalProd} is not defined for pseudo Riemannian manifolds. More details can be found in \cite{Suij15a}.

\subsection{Action in the Spectral Triples Framework, case of the AC-Manifold}
\label{SpectralAction}
\epigraph{Math and music are intimately related. Not necessarily on a conscious level, but sure.}{\textit{S. Sondheim}}

Given a spectral triple $(\calA, \calH, D)$ it is possible to extract spectral invariants from this structure. These quantities do not vary under the action of unitaries, and thus of the gauge transformations. We will retain two invariants here, for their capacity to give precisely the action of the standard model coupled to gravity, as we will see in chapter \ref{NCSMPP} when an adequate spectral triple is chosen.
\medskip
\par 
The first invariant, originally introduced by A. Chamseddine and A. Connes in \cite{chamseddine1997spectral} is called the bosonic spectral action. Given $D_\omega=D+\omega+\epsilon^\prime J\omega J^{-1}$  the (bosonic) spectral action is defined by:
\begin{align}
	\label{BosAction}
	\act_b[\omega]
	&\defeq \Tr f(D_\omega / \Lambda)
\end{align}
where $f:\bbR\,\to\, \bbR$ is a positive and even function which decay at $\pm\infty$ and provided a trace class operator, $\Lambda\in\bbR$ being a cutoff parameter. This is the only natural spectral invariant that respects additivity. More conditions on $f$ can be founded in \cite{eckstein2018spectral} and \cite{Suij15a}. 
\medskip
\par 
The second one will be called the fermionic spectral action. In the even case, when the spectral triple is equipped with $\gamma$ and $J$ operators, for any $\Tpsi \in \Ths^+$, where $\Ths^+$ corresponds to Grassmann vectors associated to vectors $\psi \in \hs^+ = \ker (\gamma - 1)$ (even elements in $\hs$), the fermionic spectral action is defined by:
\begin{align}
	\label{FerAction}
	\act_f[\omega, \Tpsi] 
	&\defeq \frac{1}{2}\langle J \Tpsi, D_\omega \Tpsi \rangle_{\Ths}
\end{align}
\begin{proposition}
	The fermionic action is invariant under gauge transformation.
\end{proposition}
\begin{proof}
	Taking $U=uJuJ^{-1}$, we have $\act_f[\omega^u,U\Tpsi] = \frac{1}{2}\langle J U\Tpsi, U D_\omega U^* U\Tpsi \rangle_{\Ths}=\frac{1}{2}\langle J U\Tpsi, U D_\omega  \Tpsi \rangle_{\Ths}=\frac{1}{2}\langle UJ \Tpsi, U D_\omega  \Tpsi \rangle_{\Ths}=\act_f[\omega, \Tpsi]$.
\end{proof}
\medskip
\par
For the bosonic action a “good” function $f$ can be taken to be $f:D_\omega\,\to\, \exp(-tD_\omega^2)$. 
\begin{remark}
	It is interesting to note that this invariant involves a function of the Laplacian constructed from the Dirac. This way of producing characteristic invariants of an underlying geometry is not new. As mentioned in section \ref{AlgeGeom}, the study of the eigenvalues of differential operators such as the Laplacian defined on a manifold (to see what geometric data they contain) was at the heart of investigations in the field of Spectral geometry. The spectral action can therefore be seen as a way to sum the eigenvalues of the Laplacian, these eigenvalues reflecting objective physical quantities, such as energy, that are conserved during unitary transformations. As mentioned in section \ref{AlgeGeom}, this operator is both connected to the geometry of the underlying space and to the stable planar waves allowed in that geometry. One can therefore interpret this spectral action as a means of summing up the possible stables modes of the fields in the geometry in question, these modes being linked to energies. It is thus quite credible that this action corresponds to a physical invariant. An advantage of the spectral triple approach is that it allows us to completely characterize the underlying geometry in the commutative case at least, unlike the original work done in spectral geometry on Riemannian manifolds where the same spectrum of the Laplacian could correspond to two different manifolds. What is very curious is that, as we will see in chapter \ref{NCSMPP}, this action provides the Lagrangian of the SMPP whereas the latter is made from the curvatures of the connections, which have a completely different interpretation as seen in section \ref{PrincipGT}.    
\end{remark}

To compute $\act_b$, an asymptotic expansion of this action can be computed from the existence of its heat kernel expansion, taking the limit $t\,\to\, 0$:
\begin{align*}
	\Tr(\exp(-tH))\sim\sum_{i}t^{i}c_i
\end{align*}
All the formulas used here work also if we take $D$ instead of $D_\omega$, but the induced spectral action is the one that leads to the SMPP. More analytic details can be found in \cite{Suij15a} and \cite{eckstein2018spectral}.\GN{plus de résultats généraux ?}
\medskip
\par 
It is interesting now to see what form this action takes in the case of an almost commutative manifold $\halgA \defeq \calC^\infty(\Man) \otimes \af$. The corresponding spectral triple can be seen as the tensor product of $(\calC^\infty(\Man), L^2(\SpinBun), \DM, \JM, \gammaM)$ with $(\af, \hsAF, \DAF, \JAF, \gammaAF)$ defined by:
\begin{align*}
	&(\halgA \defeq \calC^\infty(\Man) \otimes \af ,\, \hshA \defeq L^2(\SpinBun) \otimes \hsAF,\, \DhA \defeq \DM \otimes \bbbone + \gammaM \otimes \DAF,&\\ 
	&\JhA \defeq \JM \otimes \JAF,\, \gammahA \defeq \gammaM \otimes \gammaAF).
\end{align*}

It is important to emphasize that given two real spectral triples, their product will not always be a spectral triple. For example, if we take the two triplets $(\calC^\infty(\Man),  L^2(\SpinBun),\DM , \JM, \gammaM )$ and $(\af,\hsAF,\DAF, \JAF, \gammaAF)$
with corresponding $(\epsilon_M,\epsilon^{\prime}_M,\epsilon^{\prime\prime}_M)$ and $(\epsilon_{F},\epsilon^{\prime}_{F},\epsilon^{\prime\prime}_{F})$, then a direct calculation shows that the signs $(\epsilon,\epsilon^{\prime},\epsilon^{\prime\prime})$ of the almost commutative manifold are defined if this relations are satisfied:
\begin{align*}
	\epsilon^{\prime}=\epsilon^{\prime}_M=\epsilon^{\prime\prime}_M\epsilon^{\prime}_{F}\qquad \& \qquad \epsilon=\epsilon_{M}\epsilon_{F}\qquad\&\qquad\epsilon^{\prime\prime}=\epsilon^{\prime\prime}_M\epsilon^{\prime\prime}_{F}
\end{align*}
and if $(\epsilon,\epsilon^{\prime},\epsilon^{\prime\prime})$ exist in the table of KO-dimensions. Any AC-manifold must respect these relations to be associated with a definite spectral triple. We will then suppose that this is the case.
\medskip
\par
Now, taking any $\omega\in \Omega^1_U(\halgA)$, the fluctuated Dirac operator takes the form: 
\begin{align*}
	\DhA[,\omega] = \DhA + \omega + \epsilonhA' \JhA \omega \JhA^{-1}
\end{align*}
$\omega$ can be represented as $\piDhA(\omega)=a[\DhA,b]$, with $a$ and $b$ in $\halgA$, we will take the notation $\omega$ for $\piDhA(\omega)$ now. Computing this connection gives:
\begin{align}
	\label{fluctuDir}
	a[\DhA,b]=\overbrace{a[\DM \otimes \bbbone,b]}^{\text { first term }}+\overbrace{a[\gammaM \otimes \DAF,b]}^{\text { second term }}
\end{align} 
The first term give:
\begin{align}
	\label{fluctuDirFirst}
	a[-i\gamma^\mu\partial_\mu\otimes\bbbone,b]=[-i\gamma^\mu\partial_\mu\otimes\bbbone,b]a=-i\gamma^\mu\otimes a\partial_\mu b\defeq\gamma^\mu \otimes \tilde{A_\mu}
\end{align}
Thanks to first-order condition and the fact that $[\gamma, \pi(a)]=0$. Then the second term: 
\begin{align*}
	a[\gammaM \otimes \DAF,b]=\gammaM\otimes a[\DAF,b]\defeq \gammaM\otimes \tilde{\Phi}  
\end{align*}
Then $\omega=\gamma^\mu \otimes \tilde{A}_\mu + \gammaM\otimes \tilde{\Phi} $ with the Hermitian operators $\tilde{A}_\mu$ and $\tilde{\Phi}$ on $C^\infty(M) \otimes \hsA$. Let's compute the fluctuated Dirac operator $\piDhA(\omega)$, it split into three terms. The first is $ \DM \otimes \bbbone$, then we have: 
\begin{align*}
	&\gamma^\mu \otimes \tilde{A}_\mu + (\epsilon_\Man'\otimes\epsilon_\Fin')(\JM \otimes \JAF)  \gamma^\mu \otimes \tilde{A}_\mu (\JM^{-1} \otimes \JAF^{-1})&\\
	& =\gamma^\mu \otimes (\tilde{A}_\mu+\epsilon_\Man'\epsilon_\Man''\epsilon_\Fin'\JAF \tilde{A}_\mu\JAF^{-1})&\\
	& \defeq \gamma^\mu\otimes A_\mu&
\end{align*}
the last term being:
\begin{align*}
	&\gammaM \otimes \DAF+\gammaM\otimes \varphi + (\epsilon_\Man'\otimes\epsilon_\Fin')(\JM \otimes \JAF)\gammaM\otimes \tilde{\Phi} (\JM^{-1} \otimes \JAF^{-1})&\\
	&= \gammaM \otimes (\DAF+\tilde{\Phi}+\epsilon_\Man'\epsilon_\Man''\epsilon_\Fin'\JAF\tilde{\Phi}\JAF^{-1})&\\
	&\defeq \gammaM \otimes \Phi.  &
\end{align*} 
We obtain $\DhA[, \omega] = \DM \otimes 1 + \gamma^\mu \otimes A_\mu + \gammaM \otimes \Phi$. As we see, the action of an element $U\in \calG(\halgA , \hshA; \JhA )$ is equivalent to a transformation at the level of the connection given by $\omega^u=u\omega u^*+u[\DhA,u^*]$. This gives:
\begin{align*}
	\omega^u=u(\gamma^\mu \otimes A_\mu + \gammaM \otimes \Phi)u^*+u[\DM \otimes \bbbone + \gammaM \otimes \DAF,u^*]
\end{align*}
Using $[u,\gamma]=0$ and $[\nabla_\mu^\SpinBun,u^*]=\partial_\mu u^*$, we recover the usual gauge transformation:
\begin{align*}
	A_\mu\,\to\, uA_\mu u^* -i u\partial_\mu u^*\qquad\qquad\qquad \Phi\,\to\, u\Phi u^*+u[\DAF,u^*]
\end{align*}
Now let $\nabla_\mu^\SpinBun$ be the spin connection on $\SpinBun$ introduced in the previous section \ref{DirSpinMfld}, and consider the vector bundle $E = \SpinBun \otimes (M \times \hsA)$ such that $L^2(E) = \hshA$, and let $\nabla_\mu^E \defeq \nabla_\mu^\SpinBun \otimes 1 + 1 \otimes (\partial_\mu + i A_\mu)$ and $\Omega^E_{\mu\nu}=[\nabla_\mu^E,\nabla_\nu^E]$ be the natural twisted connection on $E$ defined by the spectral triple, and it's curvature. Then we have that $\DhA[, \omega] = -i \gamma^\mu \nabla_\mu^E + \gammaM \otimes \Phi$.
\medskip
\par
Finally, let $D_\mu \defeq \partial_\mu + i \ad(A_\mu)=\ad(\nabla_\mu^E)$ and its curvature $F_{\mu\nu} \defeq \partial_\mu A_\nu - \partial_\nu A_\mu + i[A_\mu, A_\nu]$ or equivalently $[D_\mu, D_\nu]=i \ad(F_{\mu\nu})$. $F_{\mu\nu}$ and $\Omega^E_{\mu\nu}$ are linked by the relation:
\begin{align*}
	\Omega^E_{\mu\nu}=\Omega^\SpinBun_{\mu\nu}\otimes\bbbone+i\bbbone\otimes F_{\mu\nu}.
\end{align*}
As we will see, $\Phi$ will correspond to the Higgs field. More complete computational details can be found in \cite{Suij15a}.
\medskip
\par
Let's compute $\DhA[, \omega]^2=(\DM \otimes \bbbone + \gamma^\mu \otimes A_\mu + \gammaM \otimes \Phi)^2$ now. After some computation (details in \cite{Suij15a}[p.146]) we find that
\begin{align*}
	\DhA[, \omega]^2&=\Delta^E+\frac{1}{4}s\otimes \bbbone-i\frac{1}{2}\gamma^\mu\gamma^\nu\otimes F_{\mu\nu}+i\gammaM\gamma^\mu\otimes D_\mu\Phi+\bbbone\otimes \Phi^2 &\\
	&\defeq \Delta^E-F&
\end{align*}
which is called the generalized Laplacian, $\Delta^E$ being given by \ref{Laplacian}.
\medskip
\par 
Inserting this in the heat kernel expansion, we have:
\begin{align*}
	\Tr(\exp(-t\DhA[, \omega]^2))\sim\sum_{k\geq 0}t^{\frac{k-n_\Man}{2}}a_k(\DhA[, \omega]^2)
\end{align*}
where $n_\Man$ corresponds to the dimension of $\Man$, and 
\begin{align*}
	a_k(\DhA[, \omega]^2)\defeq \int_\Man a_k(x,\DhA[, \omega]^2)\sqrt{g}d^4x
\end{align*}
with the Seeley-DeWitt coefficients $a_k(x,\DhA[, \omega]^2)$. The $a_k$ with odd $k$ being zero, and $f_i=\int_0^\infty f(p)p^{i-1}dp$ the moments of $f$. Doing the computation, we find: 
\begin{align*}
	&a_0(x, \DhA[, \omega]^2)=(4\pi)^{\frac{-n_\Man}{2}}\Tr(\bbbone)&\\
	&a_2(x, \DhA[, \omega]^2)=(4\pi)^{\frac{-n_\Man}{2}}\Tr((s/6)+F)&\\
	&a_4(x, \DhA[, \omega]^2)=((4\pi)^{\frac{-n_\Man}{2}}/360)\Tr(-12\Delta s+5s^2-2R_{\mu\nu}R^{\mu\nu}+2R_{\mu\nu\lambda\rho}R^{\mu\nu\lambda\rho}+60sF&\\
	&\qquad\qquad\qquad\qquad\qquad\qquad\qquad\quad +180 F^2-60\Delta F+30\Omega^E_{\mu\nu}(\Omega^E)^{\mu\nu})&\\
	&a_6(x, \DhA[, \omega]^2)=\propto \text{43 terms of order 3: } F^3, FR^2,(\Omega^E_{\mu\nu})^3\dots&\\
	& \dots & 
\end{align*}
More details about this asymptotic expansion can be found in \cite{gilkey2011spectral}. This permits us to compute the heat kernel expansion of the bosonic spectral action, taking $n_\Man=4$ we have: 
\begin{align}
	\label{eqSpectralMom}
	\Tr f(\DhA[, \omega] / \Lambda)\sim f(0)a_4(\DhA[, \omega]^2)+2f_2\Lambda^2a_2(\DhA[, \omega]^2)+2f_4\Lambda^4a_0(\DhA[, \omega]^2)+\calO(\Lambda^{-1})
\end{align}
After some computations, we can see that if we define $n$ to be the dimension of $\hsAF$, and define: 
\begin{align}
	\label{LagSpectral}
	\lag(g_{\mu\nu}, A_\mu, \Phi)\defeq n\lag_\Man(g_{\mu\nu})+\lag_b(A_\mu)+\lag_\Phi(g_{\mu\nu}, A_\mu, \Phi)
\end{align}
with:
\begin{align*}
	&\lag_\Man(g_{\mu\nu})=\frac{f_4\Lambda^4}{2\pi^2}-\frac{f_2\Lambda^2}{24\pi^2}s+\frac{f(0)}{480\pi^2}(s^2-3R_{\mu\nu}R^{\mu\nu})&\\
	&\lag_A(A_\mu)=\frac{f(0)}{24\pi^2}\Tr(F_{\mu\nu}F^{\mu\nu})&\\
	&\lag_\Phi(g_{\mu\nu}, A_\mu, \Phi)=-\frac{2f_2\Lambda^2}{4\pi^2}\Tr(\Phi^2)+\frac{f(0)}{8\pi^2}\Tr(\Phi^4)+\frac{f(0)}{24\pi^2}\Delta(\Tr(\Phi^2))+\frac{f(0)}{48\pi^2}s\Tr(\Phi^2)+\frac{f(0)}{8\pi^2}\Tr((D_\mu\Phi)(D^\mu\Phi))&
\end{align*} 
then:
\begin{align*}
	\Tr f(\DhA[, \omega] / \Lambda)\sim \int_\Man\lag(g_{\mu\nu}, A_\mu, \Phi)\sqrt{g}d^4x +\calO(\Lambda^{-1})
\end{align*}
\begin{remark}
	Because of the Laplacian, the term $\frac{f(0)}{24\pi^2}\Delta(\Tr(\Phi^2))$ will make no contribution in this boundary term integral. 
\end{remark}

\begin{remark}
	We can see that in the case of the AC-manifold, $\DhA[, \omega]^2$ contains the Laplacian $\Delta^E$ over the Riemannian manifold defined in subsection \ref{DirSpinMfld}. We can therefore see that it extends in some way the usual Laplacian, and then the physics behind it, in the sense of the discussion made in section \ref{AlgeGeom}. 
\end{remark}
More computational details can be found in \cite{Suij15a}.
\medskip
\par 
Let's see what happens for the fermionic action. If we consider an element $\psi\in \hshA$, then it takes the form $\psi=\chi \otimes \xi$ with $\chi\in L^2(\SpinBun)$ and $\xi\in\hsAF $. Then we have $\hshA^+=L^2(\SpinBun)^+\otimes \hsAF^+\oplus L^2(\SpinBun)^-\otimes \hsAF^-= \ker (\gammahA - 1)$, so that  any element $\psi \in \hshA^+$ will take the form:
\begin{align}
	\label{VectAC}
	\psi=\tilde{\chi}_L \otimes \xi +\tilde{\chi}_R \otimes \eta 
\end{align}
with $\tilde{\chi}_L\in L^2(\SpinBun)^+$ for left handed spinors, $\tilde{\chi}_R\in L^2(\SpinBun)^-$ for right handed spinors, $\xi\in \hsAF^+$, $\eta\in \hsAF^-$. Taking $\{e_v\}_{v\in\Gamma^{(0)}_+}$ the orthonormal basis of $\hsAF^+$ and $\{e_v\}_{v\in\Gamma^{(0)}_-}$ the one of $\hsAF^-$, equation \eqref{VectAC} can be written:
\begin{align*}
	\psi=\bigoplus_{(v,\, v')\in(\Gamma^{(0)}_+,\,\Gamma^{(0)}_-)}\chi_L^{(v)} \otimes e_v+\chi_R^{(v)}  \otimes e_{v'} 
\end{align*}
We restrict ourselves to the case $\epsilon''=-1$ such that $\JAF\,:\hsAF^\pm\to\hsAF^\mp$, this being the only interesting case in order to obtain the good fermionic action for the NCSMPP. In this case, we can consider that $\JAF$ connects the basis of $\hsAF^-$ and $\hsAF^+$ \textit{i.e.} $\{\JAF e_v\defeq e_{\Jim(v)}\}_{v\in\Gamma^{(0)}_-}$ is an orthonormal basis of $\hsAF^+$. Then we have $\psi=\bigoplus_{v\in\Gamma^{(0)}_-}\chi_L^{(v)}  \otimes e_v+\chi_R^{(v)}  \otimes e_{\Jim(v)} $. Because $\gammahA\DhA=-\DhA\gammahA$ the operator $\DAF$ and then $\Phi$ will act as $\DAF\, :\hsAF^\pm\to\hsAF^\mp$. Therefore, taking $e = (v_1, v_2)$, we can define $\Phi_e\, : \calH_{v_1}\,\to\, \calH_{v_2}$ as the restriction of $\Phi$ on these corresponding subspaces such that $\Phi_{e}e_{v_1}=m_ee_{v_2}$ with $m_e$ a scalar. Furthermore, we define $\Jim(e)$ by  $\Jim(e)=(\Jim(v_1),\Jim(v_2))$.
\medskip
\par 
Taking the corresponding Grassmann vector $\Tpsi$, the fermionic spectral action becomes:
\begin{align}
	\label{FermAct}
	\act_f[\omega, \Tpsi] = \frac{1}{2}\langle \JhA \Tpsi, \DhA[, \omega] \Tpsi \rangle_{\Ths}
\end{align} 
Using the fact that $A_\mu$ take the form $\begin{psmallmatrix} Y_\mu & 0 \\ 0 & -Y_\mu \end{psmallmatrix}$ in the $\{(e_v,e_{\Jim(v)})\}_{v\in\Gamma^{(0)}_-}$ basis, such that $Y_\mu e_v=Y_\mu^{(v)}e_v$ and $Y_\mu e_{\Jim(v)}=-Y_\mu^{(v)}e_{\Jim(v)}$ with $Y_\mu^{(v)}$ a scalar and $v\in\Gamma^{(0)}_-$, 
\eqref{FermAct} then contains the 4 terms:
\begin{align*}
	&\JhA \tilde{\psi}=\bigoplus_{v\in\Gamma^{(0)}_-}\JM\tilde{\chi}_L^{(v)}  \otimes e_{\Jim(v)}+\epsilon_F\JM\tilde{\chi}_R^{(v)}  \otimes e_v&\\
	&(\DM \otimes \bbbone)\tilde{\psi}=\bigoplus_{v\in\Gamma^{(0)}_-}\DM\tilde{\chi}_L^{(v)}  \otimes e_v+ \DM\tilde{\chi}_R^{(v)}  \otimes e_{\Jim(v)}  &\\
	&(\gamma^\mu \otimes A_\mu)\tilde{\psi}=\bigoplus_{v\in\Gamma^{(0)}_-}\gamma^\mu\tilde{\chi}_L^{(v)} \otimes Y_\mu e_v-\gamma^\mu\tilde{\chi}_R^{(v)}  \otimes Y_\mu e_{\Jim(v)}  &\\
	&(\gammaM \otimes \Phi)\tilde{\psi}=\bigoplus_{v'\in\Gamma^{(0)}_+}\sum_{\substack{v\in\Gamma^{(0)}_- \\ e=(v,v') \\ e'=(\Jim(v),v')}}\gammaM\tilde{\chi}_L^{(v)}\otimes\Phi_e e_v+\gammaM\tilde{\chi}_R^{(v)}\otimes\Phi_{e'}e_{\Jim(v)} &
\end{align*}
Taking the corresponding scalar products (the one defined in \eqref{ScalProd} for the spinor field, and the usual one associated with the inner product for the Hilbert space of the finite part), after some computations, we find that:
\begin{align*}
	\act_f[\omega, \Tpsi]=\sum_{v\in\Gamma^{(0)}_-}\Big(&(\JM\tilde{\chi}_L^{(v)},\DM\tilde{\chi}_R^{(v)})-(\JM\tilde{\chi}_L^{(v)},\gamma^\mu Y_\mu^{(v)}\tilde{\chi}_R^{(v)})&\\
	& +\sum_{\substack{v'\in\Gamma^{(0)}_- \\ e=(v',\Jim(v) ) }}\big(m_e(\JM\tilde{\chi}_L^{(v)},\gammaM\tilde{\chi}_L^{(v')})+\epsilon_Fm_{\Jim(e)}(\JM\tilde{\chi}_R^{(v)}, \gammaM\tilde{\chi}_R^{(v')}) \big)\Big)&
\end{align*}
Or equivalently $\act_f[\omega, \Tpsi]=\sum_{v\in\Gamma^{(0)}_-}\act_f^v[\omega, \Tpsi] $, with $(\,\, ,\,\, )$ the spinor space's scalar product defined in \eqref{ScalProd}.

\chapter{The NCSMPP (in the Framework of Spectral Triples)}
\label{NCSMPP}

In the context of NCG's applications in physics, the elaboration of a NCGFT yielding the Lagrangian of the SMPP can be considered as one of the most important achievements. This NCGFT is written using spectral triple techniques. This realization is the result of a long work in which 7 major steps can be highlighted:

\begin{enumerate}
	\item In 1990, M. Dubois-Violette, R.  Kerner and J. Madore in \cite{DuboKernMado90a} \cite{DuboKernMado90b} showed that building NCGFTs on the AC-manifold model allowed to obtain a situation analogous to the usual gauge theories, with a natural interpretation of the Higgs field, and the potential naturally coming from the generalized curvature. This work was done via the framework of derivations of the algebra for the study of the purely NC part.
	
	\item Just after that, in 1990, A. Connes and J. Lott, inspired by this work, set up an equivalent thinking framework, but with the spectral triple technique, to build a non-commutative Standard Model \cite{ConnLott90a}. The so-called Connes-Lott model did not incorporate gravitation.
	
	\item In 1996, an important step was reached by A. Chamseddine and A. Connes. In \cite{chamseddine1997spectral} they propose an action principle (presented in section \ref{SpectralAction}), associated with the spectral triple of a NC space, which allows to incorporate gravitation into the previous model, but it suffers from the problem of fermion doubling and imposes null-mass for neutrinos. 
	
	\item In 1996, F. Lizzi, G. Mangano, G. Miele and G. Sparano provide a way to overcome the fermion doubling problem \cite{lizzi1997fermion}.
	
	\item In 2006, J. Barrett and A. Connes, in \cite{barrett2007lorentzian} and \cite{connes2006noncommutative} solved these two problems in one move, by allowing the metric dimension of a space to be independent of its KO-dimension, and making the choice of KO-dimension 6 for the finite NC part. It additionally gives a mass relation, which must be imposed at the unification scale, and therefore constrains potential models beyond the SMPP.
	
	\item In 2007, thanks to A. Chamseddine, A. Connes, and M. Marcolli, this works culminate in the full derivation of the (minimal) Standard Model, minimally coupled with gravity's Lagrangian, directly computed from the spectral action on a given AC-manifold \cite{ChamConnMarc07a}, predicting a mass of 170 GeV for the Higgs boson, and giving a postdiction of the Top quark mass. I will succinctly describe this model in chapter \ref{NCSMPP}.
	
	\item In 2012 after the discovery of the Higgs boson, with a mass of 125 GeV, A. Chamseddine and A. Connes revisited their model to find an adequate Higgs mass, proposing not to neglect a certain scalar field appearing in the calculations of the NCSMPP \cite{chamseddine2012resilience}.  
	
	\item The achievement of the NCSMPP, corresponding to actual phenomenology opens the path to go beyond the SMPP in this framework. An important open door in this way was given in 2013 by A. Chamseddine, A. Connes, and W. van Suijlekom \cite{chamseddine2013beyond}.   
\end{enumerate}
Other very important contributions to this programme can be mentioned, mainly due to D. Kastler, T. Schücker, B. Iochum, J. M. Gracia-Bondia, L. Carminati, C. Stephan, see for instance this nonexhaustive list \cite{carminati1996connes, iochum1997universal, iochum1995riemannian, gracia1998standard, iochum2004classification}.\\

In this chapter, I will show the elementary structure behind this NCSMPP, making some general remarks and highlighting how it allows a geometrization of all forces. Then I will mention how this non-commutative reformulation offers an adequate framework to think and constrain the creation of GUTs, preparing the last part of this thesis where I will explore a way to do it, using this framework and $AF$-algebras.

\section{Obtaining the Full Lagrangian of the SMPP}
\label{NCSMPPLag}
As mentioned in section \ref{STNCGFT}, the development of the NCGFT in the framework of spectral triple allowed to set up a NC structure from which the SMPP Lagrangian can be extracted in a natural way. The goal of this section is to present the general structure behind the NCGFT which leads to the actual NCSMPP. This section aims to present the general idea of the NC structure behind this reformulation. Then a lot of technical details will not figure here in order to focus on the essential structures. A more complete presentation of the NCSMPP can be found in \cite{chamseddine2007gravity,Suij15a,jureit2007noncommutative}.
\medskip
\par

The NCSMPP is based on the following AC-manifold:

\begin{flalign}
	\label{ACSMPP}
	ST_{\calA_{SM}} =\, & (\halgA \defeq \calC^\infty(\Man) \otimes \algA_{SM} , \hshA \defeq L^2(\SpinBun) \otimes \calH_{SM}, \DhA \defeq \DM \otimes \bbbone + \gammaM  \otimes D_{SM},&\nonumber\\ 
	& \JhA \defeq \JM \otimes J_{SM}, \gammahA \defeq \gammaM \otimes \gamma_{SM})
\end{flalign} 

\par
We note that $ST_{\calA_{SM}}$ is here only parameterized by the finite part, because it will be the only variable part in what will follow, \textit{i.e.} the manifold part will always be the same as here. $\JM$ and $\gammaM$ define the charge conjugation and chirality operator used in particle physics. To obtain a definite spectral triple for the AC-manifold, with good spectral action, KO dimension 6 will be the good choice for the finite part (the space of fermionic representations).

\medskip
\par 
The algebra of the finite part is given by $\algA_{SM}\simeq \bbC\oplus \bbH\oplus M_3(\bbC)$, an element $a\in \algA_{SM}$ is given by the elements $(a_1, a_2, a_3)\in (\bbC, \bbH, M_3(\bbC))$. The Hilbert space is given by $\calH_{SM}=(\calH_l\oplus \calH_{\overline{l}}\oplus \calH_q\oplus \calH_{\overline{q}})^{\oplus 3}$ with $\calH_l$ for leptons, $\calH_{\overline{l}}$ for anti-leptons, $\calH_q$ for quarks, $\calH_{\overline{q}}$ for anti-quarks, and $\oplus 3$ for the 3 generations. The ways elements of $\algA_{SM}$ act on the different parts of $\calH_{SM}$ can be found in \cite{Suij15a}[p.188]. Imposing the unimodularity condition, the Gauge group of the finite part is then given by the quotient on the finite cyclic subgroup $\mu_6$:
\begin{align*}
	\calG( \algA_{SM} , \calH_{SM}; J_{SM})=U(1)\times SU(2)\times SU(3)/\mu_6.
\end{align*}
To build the connection, we use the same procedure as done in section \ref{DirSpinMfld} concerning equation \eqref{fluctuDir}. Taking $a,b\in\algA_{SM}$, the associated one form $a[\DhA,b]$ then split into two terms, one coming from $D_{SM}$ and the other from $\DM$. Then, the equation \eqref{fluctuDirFirst} which corresponds to this second term now splits into 3 parts $\gamma^\mu \otimes \tilde{A}_{\mu,j}$ with $\tilde{A}_{\mu,j}=-ia_j\partial_\mu b_j$ and $j\in\{1,2,3\}$. Computing the fluctuated Dirac operator, we recover $\DhA[, \omega] = \DM \otimes 1 + \gamma^\mu \otimes (A_{\mu,j}) + \gammaM \otimes \Phi$ where $A_{\mu,1}$ (resp $A_{\mu,2}$ and $A_{\mu,3}$) correspond to the $U(1)$ (resp $SU(2)$ and $SU(3)$) gauge bosons. Having obtain $\DhA[, \omega]$, we can therefore compute the bosonic spectral action $\act_b[\omega]$
\medskip
\par
For the fermionic action, the novelty is that the Hilbert spaces $\hs_{SM}^\pm$ now split into two pieces $\hs_{SM, R/L}^\pm$, by pairing with the spinor field's chiralities. Then, we have the basis $\hs_{SM, R}^-$ given by $\{e_v^R\}_{v\in\Gamma^{(0)}_{-,R}}$ and $\{J_{SM} e_v^R\defeq e^R_{\Jim(v)}\}_{v\in\Gamma^{(0)}_{-,R}}$ for a basis of $\hs_{SM, R}^+$ (Because we choose KO dimension 6 so that $\epsilon_F''=-1$, then $J_{SM}$ change the chirality). The same being done with $\{e_v^L\}_{v\in\Gamma^{(0)}_{+,L}}$ for  $\hs_{SM,L}^\pm$. 
\medskip
\par
It is possible to connect  $\hs_{SM, R}^\pm$ and  $\hs_{SM,L}^\mp$ basis using an operator $h\, :\,\Gamma^{(0)}_{\pm,R/L} \to \Gamma^{(0)}_{\mp,L/R}$ such that $h^2=\bbbone$, $h\circ\Jim\circ h\circ \Jim=\bbbone $ and that the basis of $\hs_{SM, L/R}^\pm$ are also given by $\{e_{h(v)}^{L/R}\}_{v\in\Gamma^{(0)}_{\mp,R/L}}$. Another splitting can be given by the type of fermion field, with leptons ($\nu$ for neutrinos type and $e$ for the other leptons), and by quarks ($u$ for quarks up type and $d$ for quarks down). Because there are only 3 generations, each of the basis of the four Hilbert spaces $\hs_{SM, R/L}^\pm$ will possess 3 elements. Furthermore, $h$ and $\Jim$ operators cannot connect different types of fermions. $\Gamma^{(0)}_{-}$ will then be the index set of fermions (resp. $\Gamma^{(0)}_{+}$ for anti fermions), and $\Gamma^{(0)}_{R/L}$ the index set of Right/Left handed fermions. Then any $\psi$ in $\hshA^+$ will take the general form:
\begin{align*}
	\psi=\bigoplus_{v\in\Gamma^{(0)}_{-,R}}&\chi_{e,R}^{(v)} \otimes e_v^R+\chi_{e,L}^{(v)}  \otimes e_{h(v)}^L +\chi_{\bar{e},R}^{(v)} \otimes e_{\Jim(v)}^R+\chi_{\bar{e},L}^{(v)}  \otimes e_{\Jim(h(v))}^L&\\
	&+\chi_{\nu,R}^{(v)} \otimes \nu_v^R+\chi_{\nu,L}^{(v)}  \otimes \nu_{h(v)}^L +\chi_{\bar{\nu},R}^{(v)} \otimes \nu_{\Jim(v)}^R+\chi_{\bar{\nu},L}^{(v)}  \otimes \nu_{\Jim(h(v))}^L &\\
	&+\chi_{u,R}^{(v)} \otimes u_v^R+\chi_{u,L}^{(v)}  \otimes u_{h(v)}^L +\chi_{\bar{u},R}^{(v)} \otimes u_{\Jim(v)}^R+\chi_{\bar{u},L}^{(v)}  \otimes u_{\Jim(h(v))}^L &\\
	&+\chi_{d,R}^{(v)} \otimes d_v^R+\chi_{d,L}^{(v)}  \otimes d_{h(v)}^L +\chi_{\bar{d},R}^{(v)} \otimes d_{\Jim(v)}^R+\chi_{\bar{d},L}^{(v)}  \otimes d_{\Jim(h(v))}^L &
\end{align*}
All Dirac spinors are independent. The fermionic spectral action $\act_f[\omega, \Tpsi] $ can then be computed.
\medskip
\par 
A complete computation of these two actions leads to the SMPP's Lagrangian as presented in \eqref{LagSMPP}, coupled to gravitation.  

\begin{remark}
	\label{RQ dofBPhi}
	In the NCSMPP, $A_\mu$ and $\Phi$ are considered as independent degrees of freedom, taken as input in calculations. But on the formal level of the construction, we can see that they are deductions of $a[\DAF,b]=\tilde{\Phi}$ and $-i a\partial_\mu b= \tilde{A}_\mu$. Then, they are formally linked by their dependence on $a$ and $b$. As far as I know, the fact that $A_\mu$ and $\Phi$ are linked by the algebra's degrees of freedom was not mentioned, and I haven't found any authors mentioning the reason for such an omission. 
	\medskip
	\par 
	Insofar as there may be no reason for such an omission, one advantage of such an observation would be to decrease the number of independent degrees of freedom in the current theory and more strongly constrain the structure of the SMPP.
\end{remark}

\begin{remark}
	We can see that $n_\Man$, the number of dimensions of space-time is linked to the structure of the Lagrangian in this approach. In equation \eqref{eqSpectralMom}, we see that the therms $\{a_k\}_k$ which give non-zero contribution in the spectral action at the origin of the Lagrangian of the SMPP are the ones such that $k\leq n_\Man$. If the space-time was of dimension 2 or 6, the Lagrangian should be very different. This is a strong constraint. 
\end{remark}

\begin{remark}
	The terms in $\Tr(\Phi^2)$ and $\Tr(\Phi^4)$ in the Lagrangian can be seen as weight associated with loops in Krajewski's diagram (in the irreducible's representation space of the fermionic field) because of the trace properties. This can be put in analogy with the curvature $F_{\mu\nu}F^{\mu\nu}$ where $F_{\mu\nu}$ was given in section \ref{PrincipGT} by computing the total phase accumulation of a fermion field along a loop in space-time. Therefore, we can see that in the two cases, the curvature of the connections is associated with loops in their corresponding geometries, commutative and not commutative. As mentioned in section \ref{PurToPoint}, if the pure state of a given particle gives the notion of point, the irreducible representations which correspond to Left/Right, fermion/anti-fermion, and charge parts of this pure state can be seen as internal degrees of freedom living within this point.
\end{remark}
I think that according to the three criteria of enhancement of a theory listed in section \ref{OpenDoorBeyondSMPP}, the NCSMPP can be considered as potentially better than the actual SMPP in the usual framework, and must therefore replace it. Let's see how it satisfies these different points.
\medskip
\par
For the first point, the number of independent inputs in the NCSMPP is inferior to the one of the SMPP. News relations between the SMPP's parameters are obtained, like the GUT relation of the couplings constants, are other relations giving a prediction for the Higgs mass and a post-diction for the Top quark. This fact comes from the more algebraically constrained structure of the NC framework. 
\medskip
\par 
For the diminution of the number of independent formal inputs, the fact that the Higgs is now a well-defined object in the mathematical structure and that its potential is deduced from the formalism is a great advance. Another enhancement concerns the fact that the right hypercharges are naturally implementes by the NC framework. In fact, all the structures of the Lagrangian are deduced from minimal principles: extending the notion of space, studying the general automorphisms of its algebra, and computing an invariant thanks to the obtained fluctuated Dirac operator. Furthermore, it offers a link between the whole structure of the Lagrangian and the dimension of space-time. 
\medskip
\par
To complete the praise for this NC reformulation, with the third kind of enhancement (Consistency with other theories on their crossings area, and on the conceptual and epistemological level), the NCSMPP is not less than a unification of all forces in one framework, where similar structures can be found by extending the “geometric” picture, such as generalization of the line element, of the notion of propagation along these line element, and then of motion with it's associated kinetic energy as we will see in section \ref{NCSMPPGeomForces}. 
\medskip
\par
Nevertheless, this can't be considered a complete reformulation, since as mentioned in subsection \ref{DirSpinMfld}, the spacetime manifold is taken to be Riemannian whereas it is Minkowskian in relativity theory. The compact Riemannian assumption is essential to obtain the asymptotic development of the bosonic spectral action which cannot therefore be Lorentz invariant.\GN{citation essais} 
\medskip
\par
Now that the NCSMPP has been set up, what can we do with this structure? Are there some advantages to study the SMPP in this framework? One of the main prospects of the NCSMPP concerns the creation of GUTs, because of the strong constraints offered by the formalism. This will be more fully discussed in section \ref{NCSMPPGeomForces} and in the whole part \ref{partNCGFTAF} where the purpose will be to set up a general formalism to create GUTs in the framework of NCGFT. Another perspective of interest concerns the potentiality to offer ways to quantize gravitation since it is now presented in a formalism equivalent to that of the other forces, for which we know how to do.
\medskip
\par
It is not excluded that the SMPP can be found in the derivation framework too, but few theoretical efforts have been made in this direction. More details on the NCSMPP can be found in \cite{ChamConnMarc07a,jureit2007noncommutative,Suij15a}.

\section{Beyond the SMPP in the NCSMPP's Framework}
\label{NCSMPPGeomForces} 
\epigraph{Where there is matter, there is geometry.}{\textit{J. Kepler}}

The NCSMPP in the framework of spectral triples can be seen as a “classical” (no quantization mentioned) unification of general relativity and the SMPP, all in one pure geometric framework. In this section we will discuss the meaning of this geometrization, and how this NCSMPP offers a nice and constrained way to make models beyond the SMPP. This section can be viewed as the NC (partial) reply to sections \ref{GravAndGT} and \ref{OpenDoorBeyondSMPP}.
\medskip
\par 
The NCSMPP is a powerful way to geometrize the interaction of elementary particles, in the same way as Gravitation was geometrized with general relativity. Let's first try to highlight how the NCSMPP offers a unifying framework for all forces, then a unification at the formal level. 
\medskip
\par 
Let's consider the Dirac equation:
\begin{align*}
	(-i(\hbar/c)\gamma^\mu\partial_\mu+m)\psi=0
\end{align*}
The idea of the NCSMPP is to “geometrize” this mass term such that the Dirac equation appears as a simple kinetic term linked to a move in something bigger than spacetime. This is done by adding a new term into the Dirac operator $\DhA \defeq \DM \otimes \bbbone + \gammaM  \otimes D_{SM}$ which represents the correspondings new degrees of freedom.
\medskip
\par
Take a look at another expression of the Dirac equation using the reduced Compton wavelength $\lambdabar_c=\hbar/mc$:
\begin{align*}
	(-i\gamma^\mu\partial_\mu+1/\lambdabar_c)\psi=0
\end{align*}
We recognize  $\DM$ and the scalar coming from $D_{SM}$. We can see that the mass term is associated with an inverse length scale term, proportional to the Compton wavelength defined in section \ref{ComptonScale}. In the NCSMPP, this mass term is taken to be given by the Dirac operator $D_{SM}$ of a pure NCG. These Compton scales will then correspond to the natural scales of this NCG. 
\medskip
\par 
\begin{remark}
	In section \ref{ComptonScale}, I give physical arguments about the fact that the usual notion of point fails at such scales, which can then induce a NCG. We have therefore this nice coincidence, on one side, we have the NCSMPP built from mathematical arguments stating that the natural scale associated to a NCG is the Compton scale. And on the other side, we have the physical arguments developed in part \ref{part2ThinkingNCG} which induce the same thing. I believe that it is an open door to find a physical justification for such a geometrization of all forces. 
\end{remark}
\begin{remark}
	\label{LComptPropagNCG}
	It is interesting to note that the Dirac operator $\DM$ is related to the momentum of the particle, and thus to the de Broglie characteristic scale $\lambdabar_b=\hbar/mv$, whereas the mass is related to the Compton characteristic scale $\lambdabar_c=\hbar/mc$. These two scales are equal when $v=c$. If $\lambdabar_b$ is usually associated with a wavelength which is observable in a space-time propagation, we cannot say the same for $\lambdabar_c$. However, the framework of NCG makes it possible to perceive this scale as associated with propagation, but along the NCG's degrees of freedom.    
\end{remark}
\begin{remark}
	\label{Rq propagGNC}
	$\overline{\psi}_{i} y_{i j} \phi \psi_{j}$ and  $i \overline{\psi} \slashed{D} \psi$ are very close in form, an idea of the NCSMPP is to make the thirst term be a kinetic one by creating new pure NC degrees of freedom (in the same way as $\partial_\mu$ are those of space-time). This is done by constructing the corresponding ``covariant derivative'' $D_{SM}$ which will give $y_{i j} \phi$ and then the fermions and bosons masses. In this way, we can say that $\overline{\psi}_{i} y_{i j} \phi \psi_{j}$ describes the kinetic term associated to the propagation (corresponding to $y_{i j} \phi$) of the fermions in these pure NC degrees of freedom \textit{i.e.} in the representation space associated to the finite space $\Fin_{SM}$ (the finite space associated to $\algA_{SM}$). This implement the idea presented in remark \ref{RQInnerGeomFiniteRepSpace} about the continuation of the Gelfand-Naimark equivalence for these inner degrees of freedom.  
	\medskip
	\par
	This offers in the same way a framework to geometrize the set of observable change $\Oset_{nst}$ mentioned in section \ref{GeomAndST} with the link made with gauge theory done in section \ref{BackToObs}. The NCSMPP offers then a geometrical picture for all observable variations $\Oset$, taking $\Oset_{st}$ to be encoded by the gravitational part of the SMPP.
\end{remark}
The parts of the Lagrangian which are linked to the Dirac equation can then be associated with kinetics terms associated with space-time and non-space-time propagation:
\begin{itemize}
	\item $i \overline{\psi} \slashed{\partial} \psi$ corresponds to the part linked to displacements in space-time degrees of freedom, linked to the variation observable set $\Oset_{st}$
	\item $\overline{\psi}_{i} y_{i j} \phi \psi_{j}$ corresponds to the part linked to displacements in “non space-time” degrees of freedom, linked to the variation observable set $\Oset_{nst}$. This corresponds to an elementary displacements in the inner representation space of the fermion fields, from the element indexed by $i$ to the one indexed by $j$.
\end{itemize}

$\lambdabar_c$ can then be considered as the scale at which the notion of point fails according to section \ref{ComptonScale}, such that no-space-time propagation is possible below, since the notion of point fails at such scales. New degrees of freedom are then visited at these scales, these are those of a NCG. 
\medskip
\par 
Energy is a quantity deducible from the observation of the state's rate of change that occurs during physical processes (then of $\Oset$). For free particles, it can therefore be seen as the addition of the kinetic energies associated to move along space-time (“C" for commutative) and NC degrees of freedoms (dofs): 
\begin{align*}
	E^2=\, \overbrace{p^2c^2}^{\text{C dofs}}\, +\, \overbrace{m^2c^4}^{\text{NC dofs}}.
\end{align*}
We have the momentum corresponding to the observation of the state change according to space degrees of freedom, and the mass-energy which will then corresponds to the state change according to inner degrees of freedom \textit{i.e.} in the fermionic representation space $\Fin_{SM}$.
\medskip
\par 
Inertia can then be seen as induced by the proportion of the total kinetic energy (linked to the spacetime and non-spacetime degrees of freedom), which leaks into the inner (non-spatialized) degrees of freedom and then can be seen as a kind of non-total transmission of the potential energy encoded in gauge fields (which leads to the acceleration) into kinetic energy for the spacetime degrees of freedom.
\medskip
\par 
This appears to be in line with the program in search for unified field theory (UFT), in which Kaluza–Klein's theory and Einstein's works around the unification of gravitation and electromagnetism into one geometric picture are well-known attempts. The NCSMPP made it possible to reformulate the SMPP in a way that unifies it with gravitation so that the four fundamental forces are expressed as pure gravitation on a space $\hMan=\Man\times\Fin$. $\Diff(\Man)$ now fits into a generalization of the diffeomorphisms $\Diff(\hMan)=\Diff(\Man)\ltimes\lig$ on this extended space, defined trough the automorphisms of the algebra $\halgA \defeq \calC^\infty(\Man) \otimes \af$ with $\lig=Map(\Man, G)$, and $G$ being the standard model symmetry group for example. In general relativity (commutative geometry), gravity emerged as a pseudo-force associated with general coordinate transformations which are equivalent to diffeomorphisms. In this extended almost-commutative geometry, the other forces become pseudo-forces too, induced by the automorphisms of the algebra, the diffeomorphisms of $\Man$ being equivalent to the outer ones.
\medskip
\par 
The usual line element $ds$ being connected to inverse of the Dirac operator $\DM$ (which contains the gravitational forces data) is now extended to $d\tilde{s}$ trough the new Dirac operator $\DhA \defeq \DM \otimes \bbbone + \gammaM  \otimes D_{SM}$. Therefore, this new line element contains all the forces in its very structure, where gauge bosons appear as inner fluctuations of the associated metric. 
\medskip
\par 
If this framework offers a nice classical unification for gravitation and the other forces, it seems to be very adapted to the elaboration of GUT-like unification too. This is probably because of its algebraic and very constrained structure, making links between the fermionic content and the GUT relation for example. The fact that the SMPP is now interpreted geometrically makes GUTs extensions (at higher energies) similar to a change in the geometry, just like for the geometry in general relativity which is a function of the stress-energy tensor and can therefore change with it. In what follows, a general framework to do such GUTs will be constructed, using AF-algebras as base algebra, and suitable compatibility conditions to implement the link between the theory and its GUT. This will be done using derivation and then spectral triples-based gauge theories. More details about the constraints offered by this NCSMPP approach can be found in \cite{jureit2008classification,iochum2004classification,van2013going,chamseddine2008standard}.

\part{\texorpdfstring{Noncommutative Gauge Field Theories based on $AF$-Algebras}{Noncommutative gauge field theories based on AF-algebras}}
\label{partNCGFTAF}
\chapter{\texorpdfstring{GUTs, $AF$-Embedding and $\phi$-Compatibility Condition}{GUTs, AF-Embedding and phi-Compatibility Condition}}
\label{sec AFA}
The main purpose of this thesis is to construct NCGFTs based on $AF$ $C^*$-algebras. As mentioned in section \ref{AFalg}, because of the composition property of $\phi$ \textit{i.e.} $\phi_{m,p} \circ \phi_{n,m} = \phi_{n,p}$, one needs only to describe the homomorphisms $\phi_{n,n+1} : \algA_n \to \algA_{n+1}$. Then, the heart of the problem is to understand what happens during a single step in the inductive sequence, and so to identify how NCGFTs based on the algebra at a given step ($\calA=\calA_n$) transform into the next one ($\calB=\calA_{n+1}$):
\begin{align*}
	\text{NCGFT}_\calA\leadsto \text{NCGFT}_\calB.
\end{align*}

The $\text{NCGFT}_\calA$ does not only depend on the algebra $\algA$. According to the case (derivations or spectral triples), NCGFTs will also depend on the chosen representation, on the chosen automorphisms, on the selected connection, and the choices concerning the additional structural operators ($D$, $J$, $\gamma$ ...). It is thus necessary to connect all these essential ingredients along the inductive sequence to relate the NCGFTs at different steps. It is the $\phi$-compatibility condition (presented in section \ref{PhiComp}) that implements this constraint between structures. This part represents the main realization of the work done during this thesis, for which the articles \cite{MassNieu21q} and \cite{masson2022lifting} are the achievements, using the method of derivations for the first one, and of spectral triples for the second.    

\section{\texorpdfstring{GUTs as a Motivations to do NCGFT's based on $AF$-Algebras}{Grand Unified Field theories as a motivations to do NCGFT's based on AF algebras}}
\label{GUTandAF}
In section \ref{OpenDoorBeyondSMPP}, we highlighted the fact that the creation of GUT-type models should take the algebra describing the gauge field as input, and give an explanation of the transformation $\text{GFT}_\text{SMPP}\, \rightarrow\, \text{GFT}_\text{GUT}$. In this part, we will try to show how the construction of NCGFTs based on algebras of type AF offers an interesting way to understand this move, using suitable conditions to link NCGFTs along the inductive sequence.
\medskip
\par 
There is no general way to understand how to get NCGFT$_\calB$ from NCGFT$_\calA$. But this seems reasonable to expect that the structure of NCGFT$_\calA$ would be recovered from NCGFT$_\calB$, under certain conditions since we expect to obtain a GUT-like process. Thus, an insight can be that the NCGFT$_\calB$ must contain all degrees of freedom of NCGFT$_\calA$, and that essential structures must be related.
\medskip
\par 
We will then observe the consequences on the algebra of the AC-manifold whose finite part is taken to be this $AF$ $C^*$-algebra, $\halgA_{AF} \defeq C^\infty(\Man) \otimes \algA_{AF}$, in order to identify how NCGFTs based on the algebra at a given step $\halgA \defeq C^\infty(\Man) \otimes \algA$ transform into the next $\halgB \defeq C^\infty(\Man) \otimes \algB$. The Lagrangian obtained at a given step can be interpreted as a unified NCGFT of the previous one. For instance, if we take $\halgA \defeq C^\infty(\Man) \otimes \algA$ to be the one which gives the NCSMPP ($\halgA \defeq C^\infty(\Man) \otimes \algA_{SM}$ with the corresponding structure presented in chapter \ref{NCSMPP}), then the purpose is to see how according to suitable conditions $\halgB \defeq C^\infty(\Man) \otimes \algB$ can represent a GUT like extension of the NCSMPP. The inherited structures will be identified in order to see how the old degrees of freedom (those from the previous step) and the new ones interact, and how we recover parts of the “old” Lagrangian in the new.
\medskip
\par
We propose to explore how the general embedding given by $\phi$ of $\calA$ into $\calB$ can provide such a link between the two NCGFT, and how $AF$'s embedding special characteristics can help to obtain ``more concrete'' class of models. A key ingredient to do this sequence of NCGFT will be the $\phi$-compatibility condition. It will permit to relate of the structures of NCGFT$_\calA$ and NCGFT$_\calB$. As we will see, one of the main properties of AF embedding is that they do not allow the embedding of two different block algebras in the direct sum of $\calA$ at the same position within a block of $\calB$ (see equation \eqref{eq product phiijell}). The novelty with the spectral triples approach is that this will not be the case for some irreps in the Hilbert space which can be “mixed” during the embedding (see equation \eqref{eq scalar production phiH u}). 
\medskip
\par
In our approach to NCGFT based on a “sequence” of finite dimensional NCGFT on the $\algA_n$'s, we will not suppose that an approximation at a level $n_0$  gives us all the information about the “limiting” NCGFT in the $AF$ algebra. In other words, some new inputs (in addition to the $A_{n,n+1}$'s) could be “added” at every step. This implies (obviously) that many non-equivalent NCGFTs could be constructed on top of a unique $AF$ algebra. This relies on the fact that there may be physical motivations to construct one sequence rather than another and that the chosen embedding at every step could participate in the phenomenology. This is similar to, but also a departure from, GUT where some information is encoded in the SSBM reducing the large group to the group of the SMPP: in our research program, we can look at our embeddings as being in duality with the SSBM, in a way that will be illustrated in Sects.~\ref{sect direct limit NCGFT} and \ref{sec spectral actions AF AC manifold}.

\section{\texorpdfstring{Structure of the $AF$ Embedding}{Structure of the AF Embedding}}

We will study the lift of an inclusion $\phi : \algA \to \algB$ regarding some of the structures defined on $\algA$ and $\algB$. Here we consider the special case of sums of matrix algebras, $\algA = \toplus_{i=1}^{r} M_{n_i}$ and $\algB = \toplus_{k=1}^{s} M_{m_k}$. For reasons that will be explained in Sect.~\ref{sect direct limit NCGFT} for derivations and Sect.~\ref{sec one step in the sequence ST} for spectral triples, $\phi$ is not necessarily unital. We define the corresponding projection and injection maps $\pi^\algA_i$, $\pi^\algB_k$, $\iota_\algA^i$ and $\iota_\algB^k$. We also suppose that there are (orthogonal) decompositions $\hsA = \toplus_{i=1}^{r} \hsiA[,i]$ and $\hsiB = \toplus_{k=1}^{s} \hsB[,k]$ such that the $\hsiA[,i]$ (resp. $\hsiB[,k]$) are Hilbert spaces on which $\algA_i$ (resp. $\algB_k$) are represented. In other words, the (left) module structures are compatible with the direct sums of algebras and Hilbert spaces: for any $a = \toplus_{i=1}^{r} a_i \in \algA$ and $\psi = \toplus_{i=1}^{r} \psi_i \in \hsA$, one has $a \psi = \toplus_{i=1}^{r} a_i \psi_i$ (and similarly for $\algB$). 
\medskip
\par

\medskip
\par
An operator $A$ on $\hsA$ can be decomposed along the operators $A_{j}^{i} \defeq \projHA_{j} \circ A \circ \injHA^{i} : \hsiA[,i] \to \hsiA[,j]$. The same holds for operators on $\hsB$. For computational purposes, we recall that one has
\begin{align*}
	A \psi &= \toplus_{j=1}^{r} \big( \tsum_{i=1}^r  A_{j}^{i}(\psi_{i}) \big)
	= \tsum_{i, j=1}^{r} \injHA^{j} \circ A_{j}^{i}(\psi_{i}).
\end{align*}
In the same way, a (general, not necessarily the one of the $AF$-algebra) morphism of algebras $\phi : \algA \to \algB$ decomposes along the maps $\phi_k^i \defeq \projB_k \circ \phi \circ \injA^i : \algA_i \to \algB_k$ and a morphism of Hilbert spaces $\phiH : \hsA \to \hsB$ decomposes along the $\phiH[,k]^i \defeq \projHB_k \circ \phiH \circ \injHA^i : \hsiA[,i] \to \hsiB[,k]$. One has
\begin{align*}
	\phi(a) &= \toplus_{k=1}^{s} \big( \tsum_{i=1}^r  \phi_k^i(a_i) \big),
	\quad \text{and} \quad
	\phiH(\psi) = \toplus_{k=1}^{s} \big( \tsum_{i=1}^r  \phiH[,k]^i (\psi_i) \big).
\end{align*}
Notice also that $\phi(a a') = \phi(a) \phi(a')$ implies
\begin{align}
	\label{eq phik product as sum ij}
	\tsum_{i=1}^{r} \phi_k^i(a_i a'_i) = \tsum_{i,j=1}^{r} \phi_k^i(a_i) \phi_k^j(a'_j) \quad \text{for any $k=1, \dots, s$}
\end{align}
\medskip
\par 
Let's now consider the inclusion given by the $AF$-algebra. The inclusion $\phi$ is taken in its simplest form, and we normalize it such that, for any $a = \toplus_{i=1}^{r} a_i$,
\begin{align}
	\label{eq phi-k(a)}
	\phi_k (a) \defeq \projB_k\circ \phi (a)
	&= 
	\begin{pmatrix}
		a_1 \otimes \bbbone_{\alpha_{k1}} & 0 & \cdots & 0 & 0\\
		0 & a_2 \otimes \bbbone_{\alpha_{k2}} & \cdots & 0 & 0\\
		\vdots & \vdots & \ddots & \vdots & \vdots\\
		0 & 0 & \cdots & a_r \otimes \bbbone_{\alpha_{kr}} & 0 \\
		0 & 0 & \cdots & 0 & \bbbzero_{n_{0,k}}
	\end{pmatrix}
\end{align}
where the integer $\alpha_{ki} \geq 0$ is the multiplicity of the inclusion of $M_{n_i}$ into $M_{m_k}$,  $\bbbzero_{n_0}$ is the $n_0 \times n_0$ zero matrix such that $n_0 \geq 0$ satisfies $m_k = n_0 + \tsum_{i=1}^{r} \alpha_{ki} n_i$, and 
\begin{align*}
	a_i \otimes \bbbone_{\alpha_{ki}}
	&= \left.
	\begin{pmatrix}
		a_i & 0 & 0 & 0 \\
		0 & a_i & 0 & 0 \\
		\vdots & \vdots & \ddots & \vdots \\
		0 & 0 & \cdots & a_i \\
	\end{pmatrix}
	\right\} \text{$\alpha_{ki}$ times.}
\end{align*}
We define the maps $\phi_{k}^{i} \defeq \phi_{k} \circ \injA^i : M_{n_i} \to M_{m_k}$, which take the explicit form
\begin{align}
	\label{eq phi-k-i(ai)}
	\phi_{k}^{i}(a_i)
	&= \begin{pmatrix}
		0  & \cdots & 0 & 0 & 0 & \cdots & 0 \\
		\vdots  & \ddots & \vdots & \vdots  & \vdots & \cdots & \vdots \\
		0  & \cdots & 0 & 0 & 0 & \cdots & 0 \\
		0  & \cdots & 0 & a_i \otimes \bbbone_{\alpha_{ki}} & 0 & \cdots & 0 \\
		0  & \cdots & 0 & 0 & 0 & \cdots & 0 \\
		\vdots  & \vdots & \vdots & \vdots  & \vdots & \ddots & \vdots \\
		0  & \cdots & 0 & 0 & 0 & \cdots & 0
	\end{pmatrix}
\end{align}
The maps $\phi_{k}^{i}$ satisfy a stronger relation than \eqref{eq phik product as sum ij}: for any $i,j = 1, \dots, r$ and $k=1, \dots, s$, 
\begin{align}
	\label{AFSpecialCharact}
	\phi_k^i(a_i) \phi_k^j(a'_j)
	&= \begin{cases}
		0 & \text{if $i \neq j$}\\
		\phi_k^i(a_i a'_i) & \text{if $i = j$}.
	\end{cases}
\end{align}
This is the main specific characteristic of $AF$-algebra's embedding structures.
\medskip
\par
If $\algA = \algA_{n}$ and $\algB = \algA_{n+1}$ for a $AF$-algebra $\varinjlim \algA_n$, then the multiplicities $\alpha_{ki}$ define the Bratteli diagram of this $AF$-algebra and vice versa. The integers $n_{0,k}$ are defined by complementarity at each step.
\medskip
\par
When $\alpha_{ki} > 0$, for $1 \leq \alpha \leq \alpha_{ki}$ we define the maps $\phi_{k, \alpha}^{i} : M_{n_i} \to M_{m_k}$ which insert $a_i$ at the $\alpha$-th entry on the diagonal of $\bbbone_{\alpha_{ki}}$ in the previous expression, so that $a_i$ appears only once on the RHS. The maps $\phi_{k}$, $\phi_{k}^{i}$, and $\phi_{k, \alpha}^{i}$ are morphisms of algebras and one has
\begin{align}
	\phi &= \toplus_{k=1}^{s} \phi_k : \toplus_{i=1}^{r} M_{n_i} \to \toplus_{k=1}^{s} M_{m_k},
	\nonumber
	\\
	\label{eq decompositions phi}
	\phi_k &= \tsum_{i=1}^{r} \phi_{k}^{i} \circ \projA_i : \toplus_{i=1}^{r} M_{n_i} \to M_{m_k},
	\\
	\phi_{k}^{i} &= \tsum_{\alpha=1}^{\alpha_{ki}} \phi_{k, \alpha}^{i} : M_{n_i} \to M_{m_k}.
	\nonumber
\end{align}
Notice then that $\phi_k(\bbboneA) = \tsum_{i=1}^{r} \tsum_{\alpha=1}^{\alpha_{ki}} \phi_{k, \alpha}^{i}(\bbbone_{\algA_i})$ fills the diagonal of $M_{m_k}$ with $\tsum_{i=1}^{r} \alpha_{ki} n_i$ copies of $1$ except for the last $n_{0,k}$ entries. When $n_{0,k} = 0$, one gets $\phi_k(\bbboneA) = \bbbone_{\algB_k}$, otherwise, let 
\begin{align}
	\label{eq def pn0k}
	p_{n_{0,k}} \defeq \bbbone_{m_k} - \phi_k(\bbboneA) \in M_{m_k}
	\quad \text{ and } \quad
	p_{n_0} \defeq \toplus_{k=1}^{s} p_{n_{0,k}} \in \algB.
\end{align}
The $p_{n_{0,k}}$'s are diagonal matrices with zero entries except for the last $n_{0,k}$ diagonal entries (bottom right) which are equal to $1$. 

\begin{lemma}
	\label{lem product phiijell}
	For any $a,b \in \algA$, any  $i_1, i_2 \in \{1, \dots, r\}$, any $\alpha_1 \in \{1, \dots, \alpha_{k i_1} \}$ and any $\alpha_2 \in \{1, \dots, \alpha_{k i_2} \}$,
	\begin{align}
		\label{eq product phiijell}
		\phi^{i_1}_{k, \alpha_1}(a)\phi^{i_2}_{k, \alpha_2}(b) = \delta_{i_1, i_2} \delta_{\alpha_1, \alpha_2} \phi^{i_1}_{k, \alpha_1}(a b).
	\end{align}
\end{lemma}

\begin{proof}
	This is a direct consequence of the definition of $\phi^{i}_{k, \alpha}$ and the multiplications of block diagonal matrices.
\end{proof}

\section{\texorpdfstring{$\phi$-Compatibility in the Derivation Framework}{phi-compatibility in the derivation framework}}
\label{PhiComp}

\begin{definition}
	An one-to-one map $\phiMod : \modM \to \modN$ between a left $\algA$-module $\modM$ and a left $\algB$-module $\modN$ is $\phi$-compatible if $\phiMod(a e) = \phi(a) \phiMod(e)$ for any $a \in \algA$ and $e \in \modM$.
\end{definition}
\begin{remark}
	\label{RQ physMean}
	A “physical meaning” for the $\phi$-compatibility condition is to conserve the action of embedded algebras elements on embedded Module's spaces, meaning in some ways that bosonic fields (elements in the algebra) continue to act in the same way on their associated fermionic fields (elements in the module).
\end{remark}
In the following, since we want to construct a direct limit of modules accompanying a direct limit of algebras with one-to-one maps, we always suppose that $\phiMod$ is also one-to-one. As before we define the corresponding projection and injection maps $\pi^\modM_i$, $\pi^\modN_k$, $\iota_\modM^i$ and $\iota_\modN^k$.  
\medskip
\par
From \cite[Cor.~III.1.2]{Davi96a}, we know that all the left modules on $\algA$ and $\algB$ are of the form $\modM = \bbC^{n_1} \otimes \bbC^{\alpha_1} \toplus \cdots \toplus \bbC^{n_r} \otimes \bbC^{\alpha_r}$ and $\modN = \bbC^{m_1} \otimes \bbC^{\beta_1} \toplus \cdots \toplus \bbC^{m_s} \otimes \bbC^{\beta_s}$ for some integers $\alpha_i$ and $\beta_k$. Two situations are easily handled to construct an one-to-one $\phi$-compatible map $\phiMod : \modM \to \modN$.
\begin{enumerate}
	\item The case $\alpha_i = \beta_k = 1$ for any $i$ and $k$. In that situation, $\phiMod$ can be constructed in a natural way using the multiplicities $\alpha_{ki}$ of $\phi$. Denote by $\boldone_n \in \bbC^n$ (resp. $\boldzero_{n} \in \bbC^{n}$) the vector with all the entries equal to $1$ (resp. $0$). Then one can define $\phiMod$ by
	\begin{align*}
		\pi^\modN_k \circ \phiMod (e)
		&=
		\begin{pmatrix}
			e_1 \otimes \boldone_{\alpha_{k1}} \\
			e_2 \otimes \boldone_{\alpha_{k2}}\\
			\vdots \\
			e_r \otimes \boldone_{\alpha_{kr}}\\
			\boldzero_{n_0}
		\end{pmatrix}
		\in \modN_k = \bbC^{m_k}
	\end{align*}
	
	\item The case $\alpha_i = n_i$ and $\beta_k = m_k$ for any $i$ and $k$. This situation corresponds to $\modM = \algA$ and $\modN = \algB$. The canonical map $\phiMod$ is taken to be $\phi$ itself.
\end{enumerate}
By construction, these two maps are $\phi$-compatible. In the more general situation, we have to inject $\alpha_{ki}$ times (as rows) $\bbC^{n_i} \otimes \bbC^{\alpha_i}$ into $\bbC^{m_k} \otimes \bbC^{\beta_k}$ ($\bbC^{n_i} \otimes \bbC^{\alpha_i}$ is injected only  when $\alpha_{ki}>0$). A necessary condition is that $\beta_k$ is large enough to accept the largest $\alpha_i$. This necessary condition leaves open the possibility of constructing many modules and many maps $\phiMod$ which are $\phi$-compatible.
\medskip
\par
Similarly to $\phi$, we decompose $\phiMod$ as $\phiMod[,k]^i \defeq \pi^\modN_k \circ \phiMod \circ \iota_\modM^i : \modM_i \to \modN_k$ and for any $1 \leq \alpha \leq \alpha_{ki}$, $\phiMod[,k,\alpha]^{i} : \modM_i \to \modN_k$ which insert $e_i \in \modM_i$ at the $\alpha$-th row.

\section{\texorpdfstring{$\phi$-Compatibility in the Spectral Triples Framework}{phi-compatibility in the spectral triple framework}}
\label{PhiCompST}

The first structure to consider are  the Hilbert spaces $\hsA$ and $\hsB$, that we can consider as left modules on $\algA$ and $\algB$ via their corresponding representations that are not explicitly written in the following.  

\begin{definition}
	\label{def phi phiH}
	A morphism of Hilbert spaces $\phiH : \hsA \to \hsB$ is $\phi$-compatible if $\phiH(a \psi) = \phi(a) \phiH(\psi)$ for any $a \in \algA$ and $\psi \in \hsA$ (the representations $\piA$ and $\piB$ are omitted in this relation).
\end{definition}
For bimodules, the associated $\phi$-compatibility relation will be set up in subsection \ref{sec AF algebras}. Given the morphism $\phiH : \hsA \to \hsB$, one can decompose $\hsB$ as $\hsB = \phiH(\hsA) \oplus \phiH(\hsA)^{\perp}$ in a unique $\phiH$-dependent way, where $\phiH(\hsA) = \Ran(\phiH)$ is the range of $\phiH$. This implies that any operator $B$ on $\hsB$ can be decomposed as $B = \smallpmatrix{ B_{\phi}^{\phi} & B_{\phi}^{\perp} \\ B_{\perp}^{\phi} & B_{\perp}^{\perp} }$ with obvious notations, for instance $B_{\phi}^{\perp} : \phiH(\hsA)^{\perp} \to \phiH(\hsA)$. In this orthogonal decomposition, one has $B^\dagger = \smallpmatrix{ B_{\phi}^{\phi \dagger} & B_{\perp}^{\phi \dagger} \\ B_{\phi}^{\perp \dagger} & B_{\perp}^{\perp \dagger} }$. 

\begin{definition}[$\phi$-compatibility of operators]
	\label{def phi compatibility operators}
	Given two operators $A$ on $\hsA$ and $B$ on $\hsB$, we say that they are $\phi$-compatible if $\phiH( A \psi) = B_{\phi}^{\phi} \phiH( \psi)$ for any $\psi \in \hsA$ (equality in $\phiH(\hsA)$).
\end{definition}

This definition makes sense since both sides belong to $\phiH(\hsA)$. Notice that, by an abuse of notation, we use the terminology “$\phi$-compatibility” but this notion depends on the couple of maps $(\phi, \phiH)$.
\medskip
\par
One can define a stronger $\phi$-compatibility between $A$ and $B$:
\begin{definition}[Strong $\phi$-compatibility of operators]
	\label{def strong phi compatibility operators}
	Given two operators $A$ on $\hsA$ and $B$ on $\hsB$, we say that they are strong $\phi$-compatible if $\phiH( A \psi) = B \phiH( \psi)$ for any $\psi \in \hsA$ (equality in $\hsB$).
\end{definition}

Notice that these two $\phi$-compatibility conditions imply that $\Ker \phiH \subset \Ker \phiH \circ A$, since, if $\psi \in \Ker \phiH$, then $0 = B_{\phi}^{\phi} \phiH( \psi) = \phiH( A \psi)$ in the first case, and similarly in the second case. A sufficient condition for this to hold for every $A$ is to require $\phiH$ to be one-to-one.

\begin{remark}
	\label{rmk piA(a) st phi comp piB(phi(a))}
	Definition~\ref{def phi phiH} implies that $\piA(a)$ and $\piB(\phi(a))$ are strong $\phi$-compatible for any $a \in \algA$.
\end{remark}

The following Proposition gives other consequences of the two definitions, where diagonality refers to the previously defined $2\times 2$ matrix decomposition.

\begin{proposition}\phantom{A}
	\label{prop strong and not strong phi compatibility}
	\begin{enumerate}
		\item $\phi$-compatibility and strong $\phi$-compatibility are stable under sums of operators.\label{item phi comp and st phi comp sums}
		
		\item Compositions of strong $\phi$-compatible operators are strong $\phi$-compatible (this is not necessarily true for $\phi$-compatible operators). \label{item st phi comp composition}
		
		\item If $A$ on $\hsA$ and $B$ on $\hsB$ are strong $\phi$-compatible then $B_{\perp}^{\phi} = 0$. \label{item st phi comp Bperpphi = 0}
		
		\item Strong $\phi$-compatibility implies $\phi$-compatibility. \label{item st phi comp implies phi comp}
		
		\item If $B_{\perp}^{\phi} = 0$, the $\phi$-compatibility implies the strong $\phi$-compatibility. \label{item Bperpphi = 0 and phi comp implies st phi comp}
		
		\item When $B$ is self-adjoint, strong $\phi$-compatibility implies that $B$ is diagonal.\label{item B self-adjoint str phi comp}
		
		\item If $A$ on $\hsA$ and $B$ on $\hsB$ are strong $\phi$-compatible and $A$ and $B$ are unitaries, then $A^\dagger$ and $B^\dagger$  are strong $\phi$-compatible and $B$ is diagonal.\label{item A B unitaries str phi comp}
		
		\item For any $a \in \algA$, the operator $\piB \circ \phi(a)$ on $\hsB$ reduces to a diagonal matrix $\piB \circ \phi(a) = \smallpmatrix{ \piB \circ \phi(a)_{\phi}^{\phi} & 0 \\ 0 & \piB \circ \phi(a)_{\perp}^{\perp} }$.\label{item piBphi(a) diagonal}
	\end{enumerate}
\end{proposition}

\begin{proof}
	Point~\ref{item phi comp and st phi comp sums} is obvious by linearity of the compatibility conditions and the matrix decompositions. For point~\ref{item st phi comp composition}, let $A_1, A_2$ be two operators on $\hsA$ and $B_1, B_2$ two operators on $\hsB$ which are strong $\phi$-compatible with $A_1$ and $A_2$ respectively. Then for any $\psi \in \hsA$, one has $\phiH(A_1 A_2 \psi) = B_1 \phiH(A_2 \psi) = B_1 B_2 \phiH(\psi)$ so that $A_1 A_2$ is strong $\phi$-compatible with $B_1 B_2$. For $\phi$-compatibility, this line of reasoning is not possible in general.
	\medskip
	\par
	One can identify $\phiH( \psi)$ with $\smallpmatrix{\phiH( \psi) \\ 0} \in \phiH(\hsA) \oplus \phiH(\hsA)^{\perp} = \hsB$ (resp. $\phiH( A \psi)$ with $\smallpmatrix{\phiH( A \psi) \\ 0}$), so that $B_{\phi}^{\phi} \phiH( \psi)$ identifies with $\smallpmatrix{ B_{\phi}^{\phi} \phiH( \psi) \\ 0 }$ while $B \phiH( \psi)$ identifies with $\smallpmatrix{ B_{\phi}^{\phi} \phiH( \psi) \\ B_{\perp}^{\phi} \phiH( \psi) }$. The $\phi$-compatibility condition implies that the map $B_{\phi}^{\phi} : \phiH(\hsA) \to \phiH(\hsA)$ is completely determined by $A$ and $\phiH$, while the strong $\phi$-compatibility condition implies firstly that $\phiH( A \psi) = B_{\phi}^{\phi} \phiH( \psi)$, and secondly that $B_{\perp}^{\phi} : \phiH(\hsA) \to \phiH(\hsA)^{\perp}$ is the zero map, which is point~\ref{item st phi comp Bperpphi = 0}. So, using these results, one gets that the strong $\phi$-compatibility implies the $\phi$-compatibility condition (which only constrains the  $B_{\phi}^{\phi}$ component of $B$), which is point~\ref{item st phi comp implies phi comp}. For point~\ref{item Bperpphi = 0 and phi comp implies st phi comp}, from $B_{\perp}^{\phi} = 0$ and $\phiH( A \psi) = B_{\phi}^{\phi} \phiH( \psi)$, one gets $B \phiH( \psi) = \smallpmatrix{ B_{\phi}^{\phi} \phiH( \psi) \\ B_{\perp}^{\phi} \phiH( \psi) } = \smallpmatrix{ B_{\phi}^{\phi} \phiH( \psi) \\ 0 } = \smallpmatrix{\phiH( A \psi) \\ 0} = \phiH( A \psi)$, which is the strong $\phi$-compatibility condition.
	\medskip
	\par
	Point~\ref{item B self-adjoint str phi comp}: if $B$ is self-adjoint, the condition $B = B^\dagger$ implies $B_{\phi}^{\perp \dagger} = B_{\perp}^{\phi} = 0$, so that $B$ is diagonal.
	\medskip
	\par
	Point~\ref{item A B unitaries str phi comp}: if $A$ and $B$ are unitaries, then $\phiH( \psi) = \phiH( A^\dagger A \psi)$ on the one hand and $\phiH( \psi) = B^\dagger B \phiH( \psi)$ on the other hand, so that $\phiH( A^\dagger A \psi) = B^\dagger B \phiH( \psi) = B^\dagger \phiH( A \psi)$. Since $A$ is invertible, any $\psi' \in \hsA$ can be written as $\psi' = A \psi$, so that $\phiH( A^\dagger \psi) = B^\dagger \phiH( \psi)$ for any $\psi$, which proves that $A^\dagger$ and $B^\dagger$  are strong $\phi$-compatible. The strong $\phi$-compatibilities implies $B_{\perp}^{\phi} = 0$ and $(B^\dagger)_{\perp}^{\phi} = B_{\phi}^{\perp \dagger} = 0$, and so $B$ is diagonal.
	\medskip
	\par
	Point~\ref{item piBphi(a) diagonal}: let us use the notation $\piB \circ \phi(a) = \smallpmatrix{ \piB \circ \phi(a)_{\phi}^{\phi} & \piB \circ \phi(a)_{\phi}^{\perp} \\ \piB \circ \phi(a)_{\perp}^{\phi} & \piB \circ \phi(a)_{\perp}^{\perp} }$ for any $a \in \algA$. From Definition~\ref{def phi phiH}, $\piB \circ \phi(a)$ is strong $\phi$-compatible with $\piA(a)$, so that $\piB \circ \phi(a)_{\perp}^{\phi} = 0$. Since $\piB \circ \phi(a^\ast) = \piB \circ \phi(a)^\dagger$, this implies that $\piB \circ \phi(a^\ast)_{\phi}^{\perp} = 0$ for any $a$, so that $\piB \circ \phi(a)$ reduces to a diagonal matrix.
\end{proof}

One can associate to $B = \smallpmatrix{ B_{\phi}^{\phi} & B_{\phi}^{\perp} \\ B_{\perp}^{\phi} & B_{\perp}^{\perp} }$ the operator $\widehat{B}_{\phi}^{\phi} = \smallpmatrix{ B_{\phi}^{\phi} & 0 \\ 0 & 0 }$. Then the $\phi$-compatibility between $A$ and $B$ is equivalent to the strong $\phi$-compatibility between $A$ and $\widehat{B}_{\phi}^{\phi}$.

\begin{lemma}
	The $\phi$-compatibility of $\phiH$ is equivalent to $\phiH[,k]^i (a_i \psi_i) = \phi_k^i(a_i) \phiH[,k]^i (\psi_i)$ for any $1 \leq i \leq r$, $1 \leq k \leq s$, $a_i \in \algA_i$ and $\psi_i \in \hsiA[,i]$.
\end{lemma}

\begin{proof}
	One has $\phiH(a \psi) = \toplus_{k=1}^{s} \big( \tsum_{i=1}^r  \phi_k^i(a_i \psi_i) \big)$ and $\phi(a) \phiH(\psi) = \toplus_{k=1}^{s} \big( \tsum_{i=1}^r  \phi_k^i(a_i) \phiH[,k]^i (\psi_i) \big)$ so that $\phiH(a \psi) = \phi(a) \phiH(\psi)$ is equivalent to $\tsum_{i=1}^r  \phi_k^i(a_i \psi_i) = \tsum_{i=1}^r  \phi_k^i(a_i) \phiH[,k]^i (\psi_i)$ for any $k$. Taking $a_i$ and $\psi_i$ non-zero only for one value of $i$, this implies that $\phi_k^i(a_i \psi_i) = \phi_k^i(a_i) \phiH[,k]^i (\psi_i)$ for any $i$. Reciprocally, if this last equally is satisfied for any $i$, it implies the previous one by linearity.
\end{proof}

\begin{lemma}
	Two operators $A$ on $\hsA$ and $B$ on $\hsB$ are strong $\phi$-compatible if and only if $\sum_{j=1}^{r} \phiH[,k]^{j} \circ A_{j}^{i} (\psi_i) = \tsum_{\ell=1}^{s} B_{k}^{\ell} \circ \phiH[,\ell]^{i} (\psi_i)$ for any $1 \leq i \leq r$, $1 \leq k \leq s$, and $\psi_i \in \hsiA[,i]$.
	\medskip
	\par
	Two operators $A$ on $\hsA$ and $B$ on $\hsB$ are $\phi$-compatible if and only if $\sum_{j=1}^{r} \phiH[,k]^{j} \circ A_{j}^{i} (\psi_i) = \tsum_{\ell=1}^{s} B_{\phi, k}^{\phi, \ell} \circ \phiH[,\ell]^{i} (\psi_i)$ for any $1 \leq i \leq r$, $1 \leq k \leq s$, and $\psi_i \in \hsiA[,i]$.
\end{lemma}

\begin{proof}
	On the one hand, one has $\phiH(A \psi) = \toplus_{k=1}^{s} \big( \sum_{i,j=1}^{r} \phiH[,k]^{j} \circ A_{j}^{i} (\psi_i) \big)$ and on the other hand $B \phiH(\psi) = \toplus_{k=1}^{s} \big( \tsum_{\ell=1}^{s} \tsum_{i=1}^{r} B_{k}^{\ell} \circ \phiH[,\ell]^{i} (\psi_i) \big)$. So, the relation $\phiH(A \psi) = B \phiH(\psi)$ is equivalent to $\sum_{i,j=1}^{r} \phiH[,k]^{j} \circ A_{j}^{i} (\psi_i) = \tsum_{\ell=1}^{s} \tsum_{i=1}^{r} B_{k}^{\ell} \circ \phiH[,\ell]^{i} (\psi_i)$ for any $k$. Taking $\psi_i$ non-zero only for one value of $i$, this implies $\sum_{j=1}^{r} \phiH[,k]^{j} \circ A_{j}^{i} (\psi_i) = \tsum_{\ell=1}^{s} B_{k}^{\ell} \circ \phiH[,\ell]^{i} (\psi_i)$ for any $i$ and $k$. By linearity, this relation implies the previous one.
	\medskip
	\par
	Concerning the $\phi$-compatibility, one can replace $B$ by $\widehat{B}_{\phi}^{\phi}$ in the previous result. Since $\widehat{B}_{\phi}^{\phi}$ acts only on $\phiH(\hsA)$, one can replace $\widehat{B}_{\phi, k}^{\phi, \ell}$ by the operators $B_{\phi, k}^{\phi, \ell} : \projHB_{\ell} \circ \phiH(\hsA) \to \projHB_{k} \circ \phiH(\hsA)$ in the final relation.
\end{proof}
\medskip
\par
We can extend the maps $\phi_k^i$ as $\phi_{k_0, \dots, k_n}^{i_0, \dots, i_n} : \algA^\otimes_{i_0, \dots, i_{n}} \to \algB^\otimes_{k_0, \dots, k_{n}}$ by $\phi_{k_0, \dots, k_n}^{i_0, \dots, i_n}(a^0_{i_0} \otimes \cdots \otimes a^n_{i_n}) \defeq \phi_{k_0}^{i_0}(a^0_{i_0}) \otimes  \cdots \otimes \phi_{k_n}^{i_n}(a^n_{i_n})$ for any $i_0, \dots, i_n$ and $k_0, \dots, k_n$, and then we define maps $\hphi : \kT^n \algA \to \kT^n \algB$, for any $n \geq 1$, by $\toplus_{i_1, \dots, i_{n-1} = 1}^{r} \big( a^0_{i} \otimes a^1_{i_1} \otimes \cdots \otimes a^{n-1}_{i_{n-1}} \otimes a^n_{j} \big)_{i,j=1}^{r} \mapsto \toplus_{k_1, \dots, k_{n-1} = 1}^{s} \big( \tsum_{i_1, \dots, i_{n-1}=1}^{r} \phi_{k, k_1, \dots, k_{n-1}, \ell}^{i, i_1, \dots, i_{n-1}, j}( a^0_{i} \otimes a^1_{i_1} \otimes \cdots \otimes a^{n-1}_{i_{n-1}} \otimes a^n_{j}) \big)_{k,\ell=1}^{s}$, and, for $n=0$, the diagonal matrix with entries $a_i$ at $(i,i)$ is sent to the diagonal matrix with entries $\tsum_{i=1}^{r} \phi_{k}^{i}(a_i)$ at $(k,k)$. Using \eqref{eq product kT n np} and  \eqref{eq phik product as sum ij}, one can check that $\hphi : \kT^\grast \algA \to \kT^\grast \algB$ is a morphism of graded algebras and that $\hphi(\bOmega^1_U(\algA)) \subset \bOmega^1_U(\algB)$, so that $\hphi : \bOmega^\grast_U(\algA) \to \bOmega^\grast_U(\algB)$ is a morphism of graded algebras. These properties are consequences of the general situation that will be described in Sect.~\ref{sec general situations}.
\medskip
\par
From a purely mathematical point of view, the strong $\phi$-compatibility condition appears more natural, mainly because of properties given in points~\ref{item st phi comp composition}, \ref{item B self-adjoint str phi comp} and \ref{item A B unitaries str phi comp} and as we will see later with proposition \ref{prop KO dim strong phi compatibility} because it preserves spectral triples KO dimension. This condition implements equality of actions at the level of all $\hsB$, making the relation dependent on the new degrees of freedom. But it is too restrictive for physics as it imposes a diagonal operator $\calD_\calB$ and thus no couplings between new and old degrees of freedom along the sequence, which must lead to very limited models for physics.
\medskip
\par 
On the contrary, the $\phi$-compatibility condition is only an equality at the level of $\phiH(\hsA)$, then being less restrictive. It appears to be more natural since it is only based on inherited degrees of freedom and then appears closer to the relation of definition \ref{def phi phiH} which was first in this construction. Consequently, this condition appears to be more natural and in line with the $\phi$-compatibility condition on Hilbert spaces. Moreover, from a physical point of view, the $\phi$-compatibility condition is natural in the sense that it focuses on the conservation of the action of NCGFT's operators at a given step on the embedding of the Hilbert space on which they previously act. As mentioned in remark \ref{RQ physMean}, this condition embodies the physical idea of the transport of the action of a physical operator $A\in \calA$ on $\calH_\calA$, when this one is embedded in $\calH_\calB$. Moreover, it allows the Dirac $\calD_\calB$ to have off-diagonal terms. This permits couplings between the degrees of freedom of the injected and added fermionic fields, which seem to be a natural requirement for GUT models. This is why we have chosen to focus our study on this $\phi$-compatibility condition of operators to obtain NCGFTs of interest, as will be done later in section \ref{sec spectral actions AF AC manifold}.

\chapter{\texorpdfstring{Derivation-based Approach to NCGFT on $AF$-Algebras}{Derivation-based approach to NCGFT on AF algebras}}

\label{sec DBA}
In this chapter, we propose to set up NCGFTs based on $AF$-Algebras using the derivation based approach. This will be done for one step in this two cases, $\phi : \algA = \toplus_{i=1}^{r} M_{n_i} \to \algB = \toplus_{k=1}^{s} M_{m_k}$ and $\widehat{\phi} : \halgA \defeq C^\infty(M) \otimes \algA \to \halgB \defeq C^\infty(M) \otimes \algB$, then computing the link between the action at successive steps in subsection \ref{sec phi compatibility of NCGFT}. In section \ref{sec numerical exploration of the SSBM}, some physical applications concerning mass spectra generated by Spontaneous Symmetry Breaking Mechanisms (SSBM) are proposed using numerical computations for specific situations. This chapter is an account of the results given in \cite{MassNieu21q}.

\section{Modules and Connections for Direct Sum Algebras}
\label{ModAndConnectionDirectSum}

We consider left modules on $\algA$ of the form $\modM = \toplus_{i=1}^{r} \modM_i$ where $\modM_i$ is a left module on $\algA_i$. This requirement is sufficient for the particular situation $\algA = M_{n_1} \toplus \cdots \toplus M_{n_r}$ since, according to \cite[Cor.~III.1.2]{Davi96a}, the modules of this algebra are of the form $\bbC^{n_1} \otimes \bbC^{\alpha_1} \toplus \cdots \toplus \bbC^{n_r} \otimes \bbC^{\alpha_r} = M_{n_1 \times \alpha_1} \toplus \cdots \toplus M_{n_r \times \alpha_r}$ for some integers $\alpha_i$, where $M_{n_i \times \alpha_i}$ is the vector space of $n_i \times \alpha_i$ matrices over $\bbC$.

\medskip
Define $\piMod_i : \modM \to \modM_i$ as the projection on the $i$-th term and $\iotaMod^i : \modM_i \to \modM$ as the natural inclusion. Then, $\piMod_i \circ \iotaMod^i = \Id_{\modM_i}$ and for any $a =\toplus_{i=1}^{r} a_i$ and $e = \toplus_{i=1}^{r} e_i$, one has $\piMod_i (a e) = \pi^i(a) \piMod_i (e)$ and $\piMod_i (\iota_i(a_i) e) = a_i \piMod_i (e)$.

\begin{proposition}[Decomposition of connections]
	\label{prop decomposition connections}
	A connection $\nabla$ on the left $\algA$ module $\modM$ defines a unique family of connections $\nabla^i$ on the left $\algA_i$ modules $\modM_i$ such that for any $e = \toplus_{i=1}^{r} e_i$ and any $\kX = \toplus_{i=1}^{r} \kX_i$, one has
	\begin{align*}
		\nabla_\kX e
		&= \toplus_{i=1}^{r} \nabla^i_{\kX_i} e_i.
	\end{align*}
	Denote by $R_i$ the curvature associated to $\nabla^i$, then, for any $\kX = \toplus_{i=1}^{r} \kX_i$, any $\kY = \toplus_{i=1}^{r} \kY_i$, and any $e = \toplus_{i=1}^{r} e_i$, one has
	\begin{align*}
		R(\kX, \kY) e 
		&=  \toplus_{i=1}^{r} R_i(\kX_i, \kY_i) e_i
	\end{align*}
\end{proposition}

\begin{proof}
	Since $\kX = \toplus_{i=1}^{r} \kX_i = \tsum_{i=1}^{r} \iotaDer_i( \kX_i)$, one has $\nabla_{\kX} e = \tsum_{i=1}^{r} \nabla_{\iotaDer_i( \kX_i)} e$ for any $e \in \modM$. This implies that $\nabla_{\kX}$ is completely given by the $r$ maps $\nabla_{\iotaDer_i(\kX_i)} : \modM \to \modM$.
	\medskip
	\par 
	So, for a fixed $i$ and for any $\kX_i \in \Der(\algA_i)$, let us study the map $\nabla_{\iotaDer_i(\kX_i)} : \modM \to \modM$. Since $\iotaDer_i( \kX_i) = \hbbbone_i \iotaDer_i( \kX_i)$, one has, for any $e \in \modM$, $\nabla_{\iotaDer_i(\kX_i)} e = \nabla_{\hbbbone_i  \iotaDer_i(\kX_i)} e = \hbbbone_i  \nabla_{\iotaDer_i(\kX_i)} e$ so that $\piMod_j \circ \nabla_{\iotaDer_i(\kX_i)} = 0$ for $j \neq i$. In other words, $\nabla_{\iotaDer_i(\kX_i)} e$ takes its values in $\modM_i$.
	\medskip
	\par 
	For a fixed $j$, take now $e = \iotaMod^j(e_j)$ for some $e_j \in \modM_j$. Since $\hbbbone_j \iotaMod^j(e_j) = \iotaMod^j(e_j)$ one has $\nabla_{\iotaDer_i(\kX_i)} \iotaMod^j(e_j) = \nabla_{\iotaDer_i(\kX_i)} (\hbbbone_j \iotaMod^j(e_j)) = (\iotaDer_i(\kX_i) \cdotaction \hbbbone_j) \iotaMod^j(e_j) + \hbbbone_j \nabla_{\iotaDer_i(\kX_i)} e_j = \hbbbone_j \nabla_{\iotaDer_i(\kX_i)} e_j$ since $\iotaDer_i(\kX_i) \cdotaction \hbbbone_j = 0$ whatever $i$ and $j$. If $j \neq i$, then $\nabla_{\iotaDer_i(\kX_i)} \iotaMod^j(e_j) = \hbbbone_j \nabla_{\iotaDer_i(\kX_i)} e_j = 0$ since $\nabla_{\iotaDer_i(\kX_i)} e_j$ has only components in $\modM_i$. This implies that  $\nabla_{\iotaDer_i(\kX_i)}$ is only non zero on components in $\modM_i$, and so defines a map
	\begin{equation*}
		\nabla^i_{\kX_i} \defeq \piMod_i \circ \nabla_{\iotaDer_i(\kX_i)} \circ \iotaMod^i :  \modM_i \to \modM_i.
	\end{equation*}
	Then, by construction, one has, for $e = \toplus_{i=1}^{r} e_i$ and $\kX = \toplus_{i=1}^{r} \kX_i$, $\nabla_\kX e = \toplus_{i=1}^{r} \nabla^i_{\kX_i} e_i$.
	\medskip
	\par 
	Now, let $f_i \in \calZ(\algA_i)$, then 
	\begin{align*}
		\nabla^i_{f_i \kX_i} e_i 
		&= \piMod_i \left( \nabla_{\iotaDer_i(f_i \kX_i)} \circ \iotaMod^i(e_i) \right) 
		= \piMod_i \left( \nabla_{\iota_i(f_i) \iotaDer_i(\kX_i)} \circ \iotaMod^i(e_i)\right) 
		\\
		&= \piMod_i \left( \iota_i(f_i) \nabla_{\iotaDer_i(\kX_i)} \circ \iotaMod^i(e_i) \right) 
		= f_i \piMod_i \circ \nabla_{\iotaDer_i(\kX_i)} \circ \iotaMod^i(e_i)
		\\
		&= f_i \nabla^i_{\kX_i} e_i 
	\end{align*}
	Let $a_i \in \algA_i$, then
	\begin{align*}
		\nabla^i_{\kX_i} a_i e_i 
		&= \piMod_i \left( \nabla_{\iotaDer_i(\kX_i)} \circ \iotaMod^i(a_i e_i) \right)
		= \piMod_i \left( \nabla_{\iotaDer_i(\kX_i)} \circ \iota_i(a_i) \iotaMod^i(e_i) \right)
		\\
		&= \piMod_i \left( (\iotaDer_i(\kX_i) \cdotaction \iota_i(a_i)) \iotaMod^i(e_i) \right)
		+ \piMod_i \left( \iota_i(a_i) \nabla_{\iotaDer_i(\kX_i)} \circ \iotaMod^i(e_i) \right)
		\\
		&= (\kX_i \cdotaction a_i) \piMod_i \circ \iotaMod^i(e_i)
		+ a_i \piMod_i \circ \nabla_{\iotaDer_i(\kX_i)} \circ \iotaMod^i(e_i)
		\\
		&= (\kX_i \cdotaction a_i) e_i + a_i \nabla^i_{\kX_i} e_i
	\end{align*}
	These two relations show that $\nabla^i$ defines a connection on the left $\algA_i$ module $\modM_i$.
	\medskip
	\par 
	Concerning the curvature, one has $\nabla_\kX \nabla_\kY e = \nabla_\kX ( \toplus_{i=1}^{r}  \nabla^i_{\kY_i} e_i) = \toplus_{i=1}^{r} \nabla^i_{\kX_i} \nabla^i_{\kY_i} e_i$ and $\nabla_{[\kX, \kY]} e = \toplus_{i=1}^{r} \nabla^i_{[\kX_i, \kY_i]} e_i$ so that $R(\kX, \kY) e = \toplus_{i=1}^{r} ( [\nabla^i_{\kX_i}, \nabla^i_{\kY_i}] - \nabla^i_{[\kX_i, \kY_i]} ) e_i = \toplus_{i=1}^{r} R_i(\kX_i, \kY_i) e_i$.
\end{proof}

Let us now consider the special case $\modM = \algA$ with the natural left module structure. In that situation, we can characterize $\nabla$ by its connection $1$-form $\omega \in \OmegaDer^1(\algA)$ defined by $\omega(\kX) \defeq \nabla_{\kX} \bbbone$ and its curvature takes the form of the multiplication on the right by the curvature $2$-form $\Omega \in \OmegaDer^2(\algA)$ defined by $\Omega(\kX, \kY) \defeq (\dd \omega)(\kX, \kY) - [\omega(\kX), \omega(\kY)]$.

\begin{proposition}
	In the previous situation, the decomposition of the connection $\nabla_\kX = \toplus_{i=1}^{r} \nabla^i_{\kX_i}$ in Prop~\ref{prop decomposition connections} is related to the decomposition of the connection $1$-form $\omega =  \toplus_{i=1}^{r} \omega_i$ in Prop.~\ref{prop decomposition forms} where $\omega_i \in \OmegaDer^1(\algA_i)$ is the connection $1$-form associated to the connection $\nabla^i$.
	
	In the same way, the connection $2$-form $\Omega$ of $\nabla$ decomposes along the connection $2$-forms $\Omega_i$ of $\nabla^i$: $\Omega = \toplus_{i=1}^{r} \Omega_i$.
\end{proposition}

\begin{proof}
	One has $\bbbone = \toplus_{i=1}^{r} \bbbone_i$ where $\bbbone_i$ is the unit in $\algA_i$. With $\kX = \toplus_{i=1}^{r} \kX_i$, one then has $\omega(\kX) = \nabla_{\kX} \bbbone = \toplus_{i=1}^{r} \nabla^i_{\kX_i} \bbbone_i = \toplus_{i=1}^{r} \omega_i(\kX_i)$.
	\medskip
	\par 
	The curvature $2$-forms are defined in terms of differentials and Lie brackets (commutators in the respective algebras) from the connection $1$-forms. We have shown that these operations respect the decomposition of forms. This proves the relation of the curvature $2$-form of $\nabla$.
\end{proof}

\section{Lifting one Step of the Defining Inductive Sequence}
\label{sec one step in the sequence der}

As expected, $\phi$ does not relate the centers of $\algA = \toplus_{i=1}^{r} M_{n_i}$ and $\algB = \toplus_{k=1}^{s} M_{m_k}$. This implies in particular that we can't expect to find or to construct a “general” map to inject $\Der(\algA)$ into $\Der(\algB)$ as modules over the centers, or, with less ambition, to inject a sub module and sub Lie algebra of $\Der(\algA)$ into a sub module and sub Lie algebra of $\Der(\algB)$. This strategy may indeed require very specific situations. 
\medskip
\par
Since it is convenient to consider all the derivations of $\algA$ and $\algB$, our approach is to keep track of the derivations in $\Der(\algB)$ which “come from” (to be defined below) derivations in $\Der(\algA)$. These derivations will propagate along the sequence of the direct limit, while new derivations will be introduced at each step of the limit. 
\medskip
\par
For any $i$, let us chose an \emph{orthogonal basis} $\{ \partial^{i}_{\!\!\algA, \alpha} \defeq \ad_{E^{i}_{\!\!\algA, \alpha}} \}_{\alpha \in I_{i}}$ of $\Der(\algA_i) = \Int(M_{n_i})$ where $E^{i}_{\!\!\algA, \alpha} \in \ksl_{n_i}$ and $I_{i}$ is a totally ordered set of cardinal $n_i^2-1$. For any $k$, we can introduce a basis of $\Der(\algB_k) = \Int(M_{m_k})$ in two steps. Let us define the set
\begin{align*}
	J^\phi_{k} \defeq 
	\left\{
	(i, \alpha, \kappa) \, / \, i \in \{ 1, \dots, r\}, \, \alpha \in \{1, \dots, \alpha_{ki}\}, \, \kappa \in I_{i}  
	\right\}
\end{align*}
and for any $\beta = (i, \alpha, \kappa) \in J^\phi_{k}$, define
\begin{align*}
	E^{k}_{\algB, \beta} \defeq \phi^{i}_{k, \alpha}(E^{i}_{\!\!\algA, \kappa}) \in \ksl_{m_k}
	\text{ and } \partial^{k}_{\algB, \beta} \defeq \ad_{E^{k}_{\algB, \beta}} \in \Der(\algB_k).
\end{align*}
The set $J^\phi_{k}$ is totally ordered for $\beta = (i,\alpha,\kappa) < \beta' = (i',\alpha',\kappa')$ iff $i<i'$ or [$i=i'$ and $\alpha < \alpha'$] or [$i=i'$ and $\alpha=\alpha'$ and $\kappa < \kappa' \in I_{i}$]. 
\medskip
\par
Denote by $g_{\algA}$ and $g_{\algB}$ the metrics on $\algA$ and $\algB$, defined as in Sect.~\ref{sec metric hodge}. We know that $g_{\algB}( \ad_{E^{k}_{\algB, \beta}}, \ad_{E^{k'}_{\algB, \beta'}} ) = 0$ for $k \neq k'$. So, let us consider a fixed value $k$. With the previous notations, one then has $g_{\algB}( \ad_{E^{k}_{\algB, (i,\alpha,\kappa)}}, \ad_{E^{k}_{\algB, (i',\alpha',\kappa')}} ) = 0$ when $i \neq i'$ or [$i = i'$ and $\alpha \neq \alpha'$] since then the product of matrices $E^{k}_{\algB, (i,\alpha,\kappa)} E^{k}_{\algB, (i',\alpha',\kappa')}$ is zero. Then one has 
\begin{align}
	g^{k}_{\algB, \beta\beta'}
	& \defeq
	g_{\algB}( \ad_{E^{k}_{\algB, (i,\alpha,\kappa)}}, \ad_{E^{k}_{\algB, (i',\alpha',\kappa')}} ) 
	= \tr( E^{k}_{\algB, (i,\alpha,\kappa)} E^{k}_{\algB, (i',\alpha',\kappa')} ) 
	\nonumber= \delta_{i i'} \delta_{\alpha \alpha'} \tr( E^{i}_{\!\!\algA, \kappa} E^{i}_{\!\!\algA, \kappa'} ) 
	\nonumber
	\\
	&= \delta_{i i'} \delta_{\alpha \alpha'} g_{\algA}( \ad_{E^{i}_{\!\!\algA, \kappa}}, \ad_{E^{i}_{\!\!\algA, \kappa'}} )= \delta_{i i'} \delta_{\alpha \alpha'} g^{i}_{\algA, \kappa\kappa'}
	\label{eq gB and gA}
\end{align}
In the following, we will use the fact that the metric $( g^{k}_{\algB, \beta\beta'} )_{\beta, \beta' \in J^\phi_{k}}$ (and also its inverse $( g_{\algB, k}^{\beta\beta'} )_{\beta, \beta' \in J^\phi_{k}}$) is diagonal by blocks along the divisions induced by the choice of a couple $(i, \alpha)$ for which $\beta = (i, \alpha, \kappa)$. Notice also that if the $\partial^{i}_{\!\!\algA, \kappa}$'s are orthogonal (resp. orthonormal) for $g_{\algA}$, so are the $\partial^{k}_{\algB, \beta}$'s for the metric $g_{\algB}$ for any $\beta \in J^\phi_{k}$. This is the reason we chose to remove the “$\tfrac{1}{n}$” factor in front of the definition of the metrics (see Sect.~\ref{sec metric hodge}).
\medskip
\par
We can now complete the family $\{ \partial^{k}_{\algB, \beta} \}_{\beta \in J^\phi_{k}}$ into a full basis of $\Der(\algB_k)$ with the same notation, $\beta \in J_{k} = J^\phi_{k}  \cup J^c_{k}$ where $J^c_{k}$ is a complementary set to get $\card(J_{k}) = m_k^2 -1$, in such a way that 
\begin{align}
	\label{eq block orthogonality basis}
	g_{\algB} ( \partial^{k}_{\algB, \beta}, \partial^{k}_{\algB, \beta'} ) = 0 
	\text{ for any $\beta \in J^\phi_{k}$ and $\beta' \in J^c_{k}$.}
\end{align}
In other words, the metric $g_{\algB}$ is block diagonal and decomposes $\Der(\algB_k)$ into two orthogonal summands. We choose any total order on $J_{k}$ which extends the one on $J^\phi_{k}$. One knows that such a procedure is always possible, but for practical applications (for instance to construct gauge field theories, as in Sect.~\ref{sec numerical exploration of the SSBM}), such a concrete basis could be useful, in particular since the basis that we construct is adapted to the map $\phi : \algA \to \algB$. Lets then see how to construct such a basis adapted to $\phi$.
\medskip
\par
Recall that the maps $\phi^{i}_{k, \alpha}$ send the matrix algebras $M_{n_i}$, for $i=1,\dots, r$ and $\alpha = 1, \dots, \alpha_{ki}$, on the diagonal of $M_{m_k}$, with a possible remaining block $\bbbzero_{n_0}$ on this diagonal. In order to manage this last block in the same way as the others, let us add the value $i=0$ to refer to this block, with $\alpha_{k0}=1$. In the following, we will use the notation $\phi^{i}_{k, \alpha}(M_{n_i})$ with $i=0$ (and $\alpha = 1$) to refer to this block.
\medskip
\par
Consider any matrix $E \in M_{m_k}$ which have only non zero entries outside the blocks $\phi^{i}_{k, \alpha}(M_{n_i})$ for $i=0, \ldots, r$. Then a straightforward computation using block matrices shows that $\tr( E^{k}_{\algB, \beta} E ) = \tr( E E^{k}_{\algB, \beta} ) = 0$ for any $\beta \in J^\phi_{k}$ (the product in the trace is off diagonal). In the same way, for any $E$ in the block $\phi^{0}_{k, 1}(M_{n_0})$, one has $E^{k}_{\algB, \beta} E = E E^{k}_{\algB, \beta} = 0$.  This implies that $\ad_E$ for any $E$ outside of the blocks $\phi^{i}_{k, \alpha}(M_{n_i})$ for $i=1, \ldots, r$ is orthogonal (for the metric induced by the trace) to $\ad_{E^{k}_{\algB, \beta}}$ for any $\beta \in J^\phi_{k}$. 
\medskip
\par
To complete the free family $\{ \ad_{E^{k}_{\algB, \beta}} \}_{\beta \in J^\phi_{k}}$ in $\Der(\algB_k)$, it is then sufficient to describe the elements $E \in \ksl_{m_k} \simeq \Der(\algB_k)$ “outside” of the blocks $\phi^{i}_{k, \alpha}(M_{n_i})$ for $i=1, \ldots, r$. We will do that using the block decomposition induced by the $\phi^{i}_{k, \alpha}(M_{n_i})$.
\medskip
\par
We can introduce a first family of matrices $E^{k}_{\algB, \beta}$ for new indices $\beta$ (in a set $J^c_{k}$) as (traceless) matrices with non zero entries in $\phi^{0}_{k, 1}(M_{n_0})$. There are $n_0^2 -1$ such elements.
\medskip
\par
Then, let us notice that for fixed $i=0, \dots,r$, the $\phi^{i}_{k, \alpha}(M_{n_i})$ for $\alpha =1, \dots, \alpha_{ki}$ are in diagonal blocks $\alpha_{ki} n_i \times \alpha_{ki} n_i$. We call these blocks the \emph{enveloping blocks} of $M_{n_i}$ inside $M_{m_k}$. We introduce a second family of matrices $E^{k}_{\algB, \beta}$ as matrices with non zero entries outside the enveloping blocks of the $M_{n_i}$'s. For every $i=0, \dots, r$, the row containing the enveloping block of $M_{n_i}$ contains $\tsum_{i,i', i\neq i'} (\alpha_{ki} n_i)(\alpha_{ki'} n_{i'})$ entries (non zero outside the enveloping block).
\medskip
\par
The next level of blocks embedding we consider is the one inside the enveloping block of $M_{n_i}$, for every $i$. In such a block, $\phi^{i}_{k, \alpha}$ maps $M_{n_i}$ into the diagonal. For every $i = 0, \dots, r$, we can then introduce a family of matrices  $E^{k}_{\algB, \beta}$ with non zero entries inside the enveloping blocks of $M_{n_i}$  but outside the blocks $\phi^{i}_{k, \alpha}(M_{n_i})$ (for $\alpha = 1, \dots, \alpha_{ki}$). For a fixed $i$, there are $\alpha_{ki}(\alpha_{ki}-1)$ blocks of size $n_i \times n_i$, so that one can construct $\tsum_{i} \alpha_{ki}(\alpha_{ki}-1) (n_i)^2$ such matrices in the third family (as expected, for $i=0$ there is no contribution since $\alpha_{k0} = 1$).
\medskip
\par
Inside the enveloping block of $M_{n_i}$, we can also construct matrices with non zero entries in the blocks $\phi^{i}_{k, \alpha}(M_{n_i})$. Indeed, for fixed $i$ and $\alpha$, let us consider the matrix $E_i^\alpha$ with $\bbbone_{n_i}$ in the block $\phi^{i}_{k, \alpha}(M_{n_i})$. For $\beta = (i,\alpha, \kappa)$ one has $E^{k}_{\algB, \beta} E_i^\alpha = E^{k}_{\algB, \beta}$ and for $\beta' = (i,\alpha', \kappa)$ with $\alpha \neq \alpha'$ one has $E^{k}_{\algB, \beta'} E_i^\alpha = 0$. Notice that $E_i^\alpha \notin \ksl_{m_k}$, but, for $\alpha = 1, \dots, \alpha_{ki}-1$, the matrices $E_{i}^{\alpha} - E_{i}^{\alpha+1}$ belong to $\ksl_{m_k}$ and, by the previous remark, are orthogonal to the $E^{k}_{\algB, \beta}$ for $\beta \in J^\phi_{k}$. These matrices constitute the fouth family: there are $\tsum_{i} (\alpha_{ki}-1)$ such matrices (once again, there is not contribution for $i=0$) 
\medskip
\par
The fifth and last family of matrices are constructed as $n_{i+1} E_i^1 - n_i E_{i+1}^1 \in \ksl_{m_k}$ for $i=0, \dots, r-1$. These $r$ matrices have entries in different enveloping blocks.
\medskip
\par
Let us collect the number of matrices $E^{k}_{\algB, \beta}$ for $\beta \in J^c_{k}$ that we have constructed:
\begin{align*}
	\card(J^c_{k})
	&= \begin{multlined}[t]
		n_0^2 - 1 
		+ \tsum_{i \geq 0} \tsum_{i' \geq 0 ; i\neq i'} (\alpha_{ki} n_i)(\alpha_{ki'} n_{i'}) 
		+ \tsum_{i \geq 0} \alpha_{ki}(\alpha_{ki}-1) (n_i)^2
		+ \tsum_{i \geq 0} (\alpha_{ki}-1)
		+ r
	\end{multlined}
	\\
	&= \begin{multlined}[t]
		n_0^2
		+ 2 \tsum_{i \geq 1} (\alpha_{ki} n_i) n_0 
		+ \tsum_{i,i' \geq 1} (\alpha_{ki} n_i)(\alpha_{ki'} n_{i'})
		- \tsum_{i \geq 1} \alpha_{ki} [ (n_i)^2 - 1 ]
		- 1
	\end{multlined}
	\\
	&= (m_k)^2 - 1 - \card(J^\phi_{k})
\end{align*}
where we have used $m_k = n_0 + \tsum_{i \geq 1} \alpha_{ki} n_i$ and $\card(J^\phi_{k}) = \tsum_{i \geq 1} \alpha_{ki} [ (n_i)^2 - 1 ]$. This shows that $\card(J_{k}) = (m_k)^2 - 1 = \dim \ksl_{m_k}$ for $J_{k} \defeq J^\phi_{k} \cup J^c_{k}$ and that the free family $\{ E^{k}_{\algB, \beta} \}_{\beta \in J_{k}}$ is a basis for $\ksl_{m_k}$. This construction and this computation can be adapted to the case $n_0 = 0$.
\medskip
\par
Notice that the derivations $\partial^{k}_{\algB, \beta} = \ad_{E^{k}_{\algB, \beta}}$ for $\beta \in J^c_{k}$ are not necessarily orthogonal. One can apply the Gram-Schmidt process to transform this basis into an orthonormal one. 

\medskip
\par
Let us return to the main construction. Notice that \eqref{eq block orthogonality basis} implies that the inverse of the matrix $( g^{k}_{\algB, \beta\beta'} )_{\beta,\beta' \in J_{k}}$, denoted by $( g_{\algB, k}^{\beta\beta'} )_{\beta,\beta' \in J_{k}}$, is also block diagonal  and is such that $( g_{\algB, k}^{\beta\beta'} )_{\beta,\beta' \in J^\phi_{k}}$ is the inverse of $( g^{k}_{\algB, \beta\beta'} )_{\beta,\beta' \in J^\phi_{k}}$ with $g_{\algB, k}^{(i,\alpha,\kappa)(i',\alpha',\kappa')} = \delta^{i i'} \delta^{\alpha \alpha'} g_{\algA, i}^{\kappa \kappa'}$.
\medskip
\par
The derivations $\partial^{k}_{\algB, \beta}$ for $\beta \in J^\phi_{k}$ are the one “inherited” from the derivations on $\algA$. We will use the convenient notation $\partial^{k}_{\algB, \beta} = \phi^{i}_{k, \alpha}(\partial^{i}_{\!\!\algA, \kappa})$ for $\beta = (i,\alpha,\kappa)$. A key ingredient to introduce the $\phi$-compatibility condition on forms is the following.

\begin{lemma}
	\label{lem inherited derivations relations}
	For any $1 \leq k \leq s$, $1 \leq i, i' \leq r$, $1 \leq \alpha \leq \alpha_{ki}$, $1 \leq \alpha' \leq \alpha_{ki'}$, $\kappa \in I_{i}$, $\kappa' \in I_{i'}$, $a_{i'} \in \algA_{i'}$, one has
	\begin{align*}
		\partial^{k}_{\algB, (i,\alpha,\kappa)} \cdotaction \phi^{i'}_{k, \alpha'}(a_{i'})
		&=
		\phi^{i}_{k, \alpha}(\partial^{i}_{\!\!\algA, \kappa}) \cdotaction \phi^{i'}_{k, \alpha'}(a_{i'})
		= \delta_{i, i'} \delta_{\alpha, \alpha'} \phi^{i}_{k, \alpha}(\partial^{i}_{\!\!\algA, \kappa} \cdotaction a_{i'})
	\end{align*}
	and
	\begin{align*}
		[\partial^{k}_{\algB, (i,\alpha,\kappa)}, \partial^{k}_{\algB, (i',\alpha',\kappa')}  ]
		&= [ \phi^{i}_{k, \alpha}(\partial^{i}_{\!\!\algA, \kappa}), \phi^{i'}_{k, \alpha'}(\partial^{i'}_{\!\!\algA, \kappa'}) ]
		= \delta_{i, i'} \delta_{\alpha, \alpha'} \phi^{i}_{k, \alpha}( [\partial^{i}_{\!\!\algA, \kappa}, \partial^{i}_{\!\!\algA, \kappa'}])
	\end{align*}
\end{lemma}

\begin{proof}
	$\phi^{i}_{k, \alpha}(\partial^{i}_{\!\!\algA, \kappa})$ is an inner derivation for the matrix $E^{k}_{\algB, (i,\alpha,\kappa)}$ in which the only non zero part $E^{i}_{\!\!\algA, \kappa}$ is located on the diagonal of $M_{m_k}$ at a position depending on $i$ and $\alpha$, see above. In the same way, the non zero part of $\phi^{i'}_{k, \alpha'}(a_{i'})$ is $a_{i'}$ on the diagonal of $M_{m_k}$. When $i \neq i'$ or $\alpha \neq \alpha'$, the commutator of these two matrices is zero. When $i = i'$ and $\alpha = \alpha'$, the commutator is $[E^{i}_{\!\!\algA, \kappa}, a_i] = \partial^{i}_{\!\!\algA, \kappa} \cdotaction a_i$ on the diagonal of $M_{m_k}$ at the position designated by $i$ and $\alpha$. This matrix is obviously $\phi^{i}_{k, \alpha}(\partial^{i}_{\!\!\algA, \kappa} \cdotaction a_i)$. This proves the first relation.
	
	The proof of the second relation relies on the same kind of argument since the $\partial^{i}_{\!\!\algA, \kappa}$ are inner derivations.
\end{proof}

\begin{definition}[$\phi$-compatible forms]
	A form $\omega = \toplus_{i=1}^{r} \omega_i \in \OmegaDer^\grast(\algA)$ is $\phi$-compatible with a form $\eta = \toplus_{k=1}^{s} \eta_k \in \OmegaDer^\grast(\algB)$ if and only if for any $1 \leq i \leq r$, $1 \leq k \leq s$, $1 \leq \alpha \leq \alpha_{ki}$, $\omega_i$ and $\eta_k$ have the same degree $p$ and for any $\partial^{i}_{\!\!\algA, \kappa_1}, \dots, \partial^{i}_{\!\!\algA, \kappa_p} \in \Der(\algA_i)$ ($\kappa_k \in I_{i}$) , one has
	\begin{align}
		\label{eq forms compatibility}
		\phi^{i}_{k, \alpha}\left( \omega_i( \partial^{i}_{\!\!\algA, \kappa_1}, \dots, \partial^{i}_{\!\!\algA, \kappa_p}) \right)
		&=
		\eta_k \left( \phi^{i}_{k, \alpha}(\partial^{i}_{\!\!\algA, \kappa_1}), \dots,  \phi^{i}_{k, \alpha}(\partial^{i}_{\!\!\algA, \kappa_p}) \right)
	\end{align}
\end{definition}

Notice that the LHS of \eqref{eq forms compatibility} has only non zero values in block matrices on the diagonal of $M_{m_k}$ at a position which depends on $i$ and $\alpha$ (see above). This implies that the RHS has the same structure.

\begin{proposition}
	\label{prop compatible forms product differential}
	Let $\omega = \toplus_{i=1}^{r} \omega_i, \omega' = \toplus_{i=1}^{r} \omega'_i \in \OmegaDer^\grast(\algA)$ and $\eta = \toplus_{k=1}^{s} \eta_k, \eta' = \toplus_{k=1}^{s} \eta'_k \in \OmegaDer^\grast(\algB)$ such that $\omega$ is $\phi$-compatible with $\eta$ and $\omega'$ is $\phi$-compatible with $\eta'$. Then $\omega \wedge \omega'$ is $\phi$-compatible with $\eta \wedge \eta'$ and $\dd \omega$ is $\phi$-compatible with $\dd \eta$.
\end{proposition}

\begin{proof}
	Since the product of forms decompose along the indices $i$ and $k$, we fix these indices in the proof and we suppose that the degres of $\omega_i$ and $\omega'_i$ are $p$ and $p'$ respectively. Inserting the RHS of \eqref{eq forms compatibility} into \eqref{eq def form product} one has
	\begin{align*}
		(\eta_k \wedge \eta'_k) & \left( \phi^{i}_{k, \alpha}(\partial^{i}_{\!\!\algA, \kappa_1}), \dots,  \phi^{i}_{k, \alpha}(\partial^{i}_{\!\!\algA, \kappa_{p+p'}}) \right)
		\\
		&= \begin{multlined}[t]
			\frac{1}{p!p'!} \!\! \sum_{\sigma\in \kS_{p+p'}} (-1)^{\abs{\sigma}} 
			\eta_k \left( \! \phi^{i}_{k, \alpha}(\partial^{i}_{\!\!\algA, \kappa_{\sigma(1)}}), \dots,  \phi^{i}_{k, \alpha}(\partial^{i}_{\!\!\algA, \kappa_{\sigma(p)}}) \! \right)
			\eta'_k \left( \! \phi^{i}_{k, \alpha}(\partial^{i}_{\!\!\algA, \kappa_{\sigma(p+1)}}), \dots,  \phi^{i}_{k, \alpha}(\partial^{i}_{\!\!\algA, \kappa_{\sigma(p+p')}}) \! \right)
		\end{multlined}
		\\
		&= \begin{multlined}[t]
			\frac{1}{p!p'!} \sum_{\sigma\in \kS_{p+p'}} (-1)^{\abs{\sigma}} 
			\phi^{i}_{k, \alpha} \Big(
			\omega_i ( \partial^{i}_{\!\!\algA, \kappa_{\sigma(1)}}, \dots, \partial^{i}_{\!\!\algA, \kappa_{\sigma(p)}} )
			\omega'_i ( \partial^{i}_{\!\!\algA, \kappa_{\sigma(p+1)}}, \dots, \partial^{i}_{\!\!\algA, \kappa_{\sigma(p+p')}} )
			\Big)
		\end{multlined}
		\\
		&= \phi^{i}_{k, \alpha} \left( (\omega_i \wedge \omega'_i) ( \partial^{i}_{\!\!\algA, \kappa_1}, \dots, \partial^{i}_{\!\!\algA, \kappa_{p+p'}} ) \right)
	\end{align*}
	In the same way, inserting the RHS of \eqref{eq forms compatibility} into \eqref{eq def differential} one has
	\begin{align*}
		(\dd_k \eta_k) &\left( \phi^{i}_{k, \alpha}(\partial^{i}_{\!\!\algA, \kappa_1}), \dots,  \phi^{i}_{k, \alpha}(\partial^{i}_{\!\!\algA, \kappa_{p+1}}) \right)
		\\
		&= 
		\begin{multlined}[t]
			\sum_{k=1}^{p+1} (-1)^{k+1}
			\phi^{i}_{k, \alpha}(\partial^{i}_{\!\!\algA, \kappa_{k}}) \cdotaction \eta_k \left(
			\phi^{i}_{k, \alpha}(\partial^{i}_{\!\!\algA, \kappa_1}), \dots \omi{k} \dots, \phi^{i}_{k, \alpha}(\partial^{i}_{\!\!\algA, \kappa_{p+1}})
			\right)
			\\
			+ \sum_{1 \leq k < k' \leq p+1} (-1)^{k+k'}
			\eta_k \Big(
			\left[ \phi^{i}_{k, \alpha}(\partial^{i}_{\!\!\algA, \kappa_{k}}), \phi^{i}_{k, \alpha}(\partial^{i}_{\!\!\algA, \kappa_{k'}}) \right], 
			\dots \omi{i} \dots \omi{k} \dots,
			\phi^{i}_{k, \alpha}(\partial^{i}_{\!\!\algA, \kappa_{p+1}})
			\Big)
		\end{multlined}
		\\
		&= \begin{multlined}[t]
			\sum_{k=1}^{p+1} (-1)^{k+1}
			\phi^{i}_{k, \alpha} \left(
			\partial^{i}_{\!\!\algA, \kappa_{k}} \cdotaction \omega_i( \partial^{i}_{\!\!\algA, \kappa_1}, \dots \omi{k} \dots, \partial^{i}_{\!\!\algA, \kappa_{p+1}} )
			\right)\\
			+ \sum_{1 \leq k < k' \leq p+1} (-1)^{k+k'}
			\phi^{i}_{k, \alpha} \left(
			\omega_i( [\partial^{i}_{\!\!\algA, \kappa_{k}}, \partial^{i}_{\!\!\algA, \kappa_{k'}}], \dots \omi{k} \dots, \partial^{i}_{\!\!\algA, \kappa_{p+1}} )
			\right)
		\end{multlined}
		\\
		&= \phi^{i}_{k, \alpha} \left(
		(\dd_i \omega_i) ( \partial^{i}_{\!\!\algA, \kappa_1}, \dots, \partial^{i}_{\!\!\algA, \kappa_{p+1}} )
		\right)
	\end{align*}
\end{proof}

Let $\{ \theta^\kappa_{\!\!\algA, i} \}_{\kappa \in I_i}$ be the dual basis of $\{ \partial^{i}_{\!\!\algA, \kappa} \}_{\kappa \in I_{i}}$. Then one has $\theta^\kappa_{\!\!\algA, i} ( \partial^{i'}_{\!\!\algA, \kappa'} ) = \delta_{i}^{i'} \delta_{\kappa'}^{\kappa'}$. In the same way, denote by $\{ \theta^\beta_{\algB, k} \}_{\beta \in J_{k}}$ the dual basis of $\{ \partial^{k}_{\algB, \beta} \}_{\beta \in J_{k}}$.

\begin{remark}[$\phi$-compatibility and components of forms]
	\label{ex one forms compatibility}
	Let us first illustrate $\phi$-compatibility for $1$-forms. One has $\omega = \toplus_{i=1}^{r} \omega^{i}_{\kappa} \otimes \theta^\kappa_{\!\!\algA, i}$ for $\omega^{i}_{\kappa} \in \algA_i$ and $\eta = \toplus_{k=1}^{s} \eta^{k}_{\beta} \otimes \theta^\beta_{\algB, k}$ for $\eta^{k}_{\beta} \in \algB_k$. Then \eqref{eq forms compatibility} reduces to $\phi^{i}_{k, \alpha} ( \omega^{i}_{\kappa} ) = \eta^{k}_{(i,\alpha, \kappa)}$. This means that the components $\eta^{k}_{\beta}$ of $\eta$ in the “inherited directions” $\phi^{i}_{k, \alpha}(\partial^{i}_{\!\!\algA, \kappa})$'s are inherited from $\omega$.
	
	In the same way, for $p$-forms, the LHS of \eqref{eq forms compatibility} is non zero only for components along the $\beta$'s of the form $(i,\alpha, \kappa)$ \emph{with the same couple} $(i, \alpha)$, and these components are given by the RHS. So, all the components $\eta^{k}_{\beta_1 \dots \beta_p}$ in the “inherited directions” $\beta_1, \dots, \beta_p$ for $\beta_k = (i, \alpha, \kappa_k)$ (same $i$ and $\alpha$) are constrained by the $\phi$-compatibility condition.
\end{remark}
\medskip
\par
Let $\modM = \toplus_{i=1}^{r} \modM_i$ and $\modN = \toplus_{k=1}^{s} \modN_k$ and let $\nabla^{\modM}$ and $\nabla^{\modN}$ be two connections on the $\algA$-module $\modM$ and a $\algB$-module $\modN$ with an one-to-one $\phi$-compatible map $\phiMod : \modM \to \modN$. We will used the maps $\phiMod[,k,\alpha]^{i} : \modM_i \to \modN_k$. These connections define connections $\nabla^{\modM,i}$ on $\modM_i$ and $\nabla^{\modN,k}$ on $\modN_k$.

\begin{definition}
	The two connections $\nabla^{\modM}$ and $\nabla^{\modN}$ are said to be $\phi$-compatible if and only if, for any $1 \leq i \leq r$, $1 \leq k \leq s$, $1 \leq \alpha \leq \alpha_{ki}$, $\kappa \in I_{i}$, one has
	\begin{align}
		\label{eq def compatibility connections}
		\phiMod[,k,\alpha]^{i} \left(
		\nabla^{\modM,i}_{\partial^{i}_{\!\!\algA, \kappa}} e_i
		\right)
		&= \nabla^{\modN,k}_{\phi^{i}_{k,\alpha}(\partial^{i}_{\!\!\algA, \kappa})} \phiMod[,k,\alpha]^{i} (e_i).
	\end{align}
\end{definition}

When $\modM = \algA$ and $\modN = \algB$, one can introduce the connection $1$-forms $\omega_\modM$ and $\omega_\modN$ for $\nabla^{\modM}$ and $\nabla^{\modN}$. Then one has
\begin{lemma}
	If $\omega_\modM$ and $\omega_\modN$ are $\phi$-compatible, then $\nabla^{\modM}$ and $\nabla^{\modN}$ are $\phi$-compatible. 
\end{lemma}

\begin{proof}
	Here we have $\phiMod = \phi$. Using $\algA_i \ni e_i = e_i \bbbone_{\algA_i}$ and the definitions of the connection $1$-forms, one has $\nabla^{\modM,i}_{\partial^{i}_{\!\!\algA, \kappa}} e_i = (\partial^{i}_{\!\!\algA, \kappa} \cdotaction e_i) \bbbone_{\algA_i} + e_i \omega_{\modM,i}( \partial^{i}_{\!\!\algA, \kappa} )$, so that
	\begin{align*}
		\phi^{i}_{k,\alpha} \left(
		\nabla^{\modM,i}_{\partial^{i}_{\!\!\algA, \kappa}} e_i
		\right)
		&= \phi^{i}_{k,\alpha} \left( (\partial^{i}_{\!\!\algA, \kappa} \cdotaction e_i) \bbbone_{\algA_i} \right)
		+ \phi^{i}_{k,\alpha} \left( e_i \omega_{\modM,i}( \partial^{i}_{\!\!\algA, \kappa} ) \right)
		\\
		&= \left( \phi^{i}_{k,\alpha}(\partial^{i}_{\!\!\algA, \kappa}) \cdotaction \phi^{i}_{k,\alpha}(e_i) \right) \phi^{i}_{k,\alpha}( \bbbone_{\algA_i} )
		+ \phi^{i}_{k,\alpha}(e_i) 
		\omega_{\modN,k} \left( \phi^{i}_{k,\alpha}(\partial^{i}_{\!\!\algA, \kappa}) \right)
		\\
		&= \left( \phi^{i}_{k,\alpha}(\partial^{i}_{\!\!\algA, \kappa}) \cdotaction \phi^{i}_{k,\alpha}(e_i) \right) \phi^{i}_{k,\alpha}( \bbbone_{\algA_i} )
		+ \phi^{i}_{k,\alpha}(e_i) 
		\nabla^{\modN,k}_{\phi^{i}_{k,\alpha}(\partial^{i}_{\!\!\algA, \kappa})} \bbbone_{\algB_k}
		\\
		&= \phi^{i}_{k,\alpha}(\partial^{i}_{\!\!\algA, \kappa}) \cdotaction \phi^{i}_{k,\alpha}(e_i)
		+ \phi^{i}_{k,\alpha}(e_i) \nabla^{\modN,k}_{\phi^{i}_{k,\alpha}(\partial^{i}_{\!\!\algA, \kappa})} \bbbone_{\algB_k}= \nabla^{\modN,k}_{\phi^{i}_{k,\alpha}(\partial^{i}_{\!\!\algA, \kappa})} \phi^{i}_{k,\alpha}(e_i)
	\end{align*}
	where we have used $\phi^{i}_{k,\alpha}(e_i) = \phi^{i}_{k,\alpha}(e_i) \phi^{i}_{k,\alpha}( \bbbone_{\algA_i} )$ and $\phi^{i}_{k,\alpha}(\partial^{i}_{\!\!\algA, \kappa}) \cdotaction \phi^{i}_{k,\alpha}( \bbbone_{\algA_i} ) = \phi^{i}_{k,\alpha}( \partial^{i}_{\!\!\algA, \kappa} \cdotaction \bbbone_{\algA_i} ) = 0$ so that, using the Leibniz rule, $\left( \phi^{i}_{k,\alpha}(\partial^{i}_{\!\!\algA, \kappa}) \cdotaction \phi^{i}_{k,\alpha}(e_i) \right) \phi^{i}_{k,\alpha}( \bbbone_{\algA_i} ) = \phi^{i}_{k,\alpha}(\partial^{i}_{\!\!\algA, \kappa}) \cdotaction \phi^{i}_{k,\alpha}(e_i)$.
\end{proof}

\begin{remark}
	Let us stress that the reverse of this Lemma is not true: $\phi$-compatibility between connections is weaker than $\phi$-compatibility between their connection $1$-forms. Indeed, let us assume that the two connections are $\phi$-compatible, that is that the first and last expressions are equal in the computation in the above proof. Then one can extract an equality at the second line between the connection $1$-forms: Take $e_i = \bbbone_{\algA_i}$ and since then the first terms in both sides are zero, one gets
	\begin{align}
		\label{eq weak form compatibility}
		\phi^{i}_{k,\alpha} \left( \omega_{\modM,i}( \partial^{i}_{\!\!\algA, \kappa} ) \right)
		&= \phi^{i}_{k,\alpha}(\bbbone_{\algA_i}) 
		\omega_{\modN,k} \left( \phi^{i}_{k,\alpha}(\partial^{i}_{\!\!\algA, \kappa}) \right)
	\end{align}
	This a weaker relation than \eqref{eq forms compatibility}. In Remark~\ref{ex one forms compatibility}, we noticed that the $\phi$-compatibility \eqref{eq forms compatibility} between forms implies that all the values $\omega_{\modN,k} \left( \phi^{i}_{k,\alpha}(\partial^{i}_{\!\!\algA, \kappa}) \right)$ come exactly from the values $\omega_{\modM,i}( \partial^{i}_{\!\!\algA, \kappa} )$. What \eqref{eq weak form compatibility} says is that $\phi^{i}_{k,\alpha}(\bbbone_{\algA_i})$, as a projector, selects only a part of the matrix $\omega_{\modN,k} \left( \phi^{i}_{k,\alpha}(\partial^{i}_{\!\!\algA, \kappa}) \right)$ to be compared with the matrix $\omega_{\modM,i}( \partial^{i}_{\!\!\algA, \kappa} )$. So some parts of the matrix $\omega_{\modN,k} \left( \phi^{i}_{k,\alpha}(\partial^{i}_{\!\!\algA, \kappa}) \right)$ may not be inherited. 
	
	For practical reasons, we prefer to deal with the stronger $\phi$-compatibility condition, since it permits to “trace” (to “follow”) the degrees of freedom of the $\omega_{\modM,i}$'s “inside” the $\omega_{\modN,k}$'s. The weaker $\phi$-compatibility condition mix up these degrees of freedom into the matrices $\omega_{\modN,k} \left( \phi^{i}_{k,\alpha}(\partial^{i}_{\!\!\algA, \kappa}) \right)$.
\end{remark}
\medskip
\par
Let us consider the following specific situation for $\phi : \algA \to \algB$, where $\algA = \toplus_{i=1}^{r} M_{n_i}$ is embedded into $\algB = M_m$ with $m \geq \tsum_{i=1}^{r} n_i$ in such a way that each $M_{n_i}$ appears once and only once on the diagonal of $M_m$. We consider on $\algA$ and $\algB$ the integral defined in Sect.~\ref{sec metric hodge}.
\begin{proposition}
	\label{prop eta start eta omega star omega}
	Let $\omega = \toplus_{i=1}^{r} \omega_i \in \OmegaDer^\grast(\algA)$ and $\eta \in \OmegaDer^\grast(\algB)$ be such that $\eta$ is $\phi$-compatible with $\omega$ and $\eta$ vanishes on every derivation $\partial_{\algB, \beta}$ with $\beta \in J^c$ (here we omit the $k=1$ index). Then
	\begin{align*}
		\int_{\algB} \eta \wedge \hstar_\algB \eta
		&=
		\int_{\algA} \omega \wedge \hstar_\algA \omega
	\end{align*}
\end{proposition}

\begin{proof}
	Since $k=1$ and $\alpha =1$ are the only possible values, to simplify the notations we write $\phi^i$ for $\phi^i_{k, \alpha}$. We will refer to $\phi^i(M_{n_i}) \subset M_m$ as the $i$-th block on the diagonal of $M_m$. The metric $g_\algB$ induces an orthogonal decomposition of the derivations of $\algB$ which is compatible with these blocks.  Two derivations $\ad_E$ and $\ad_{E'}$ such that $E \in \phi^i(M_{n_i})$ and $E'$ has no non vanishing entries in $\phi^i(M_{n_i})$ are orthogonal. So, $\Der(\algB) = \toplus_{i=1}^{r} \Der( \phi^i(M_{n_i}) ) \oplus \Der(\algB)^{c}$ where the decomposition is orthogonal for $g_\algB$ and $\Der(\algB)^{c}$ are derivations “outside” of the $\phi^i(M_{n_i})$'s.
	\medskip
	\par
	For any $p$ one has $\eta( \phi^i(\partial^{i}_{\!\!\algA, \kappa_1}), \dots, \phi^i(\partial^{i}_{\!\!\algA, \kappa_{p}}) ) = \phi^i \left( \omega_i( \partial^{i}_{\!\!\algA, \kappa_1}, \dots, \partial^{i}_{\!\!\algA, \kappa_{p}} ) \right)$.\footnote{Since the $\omega_i$ and $\eta$ are not supposed to be homogeneous in $\OmegaDer^\grast(\algA_i)$ and $\OmegaDer^\grast(\algB)$, one has to evaluate them against any number of derivations.} These are the only possible values for $\eta$. In particular $\eta$ vanishes on any derivation in $\Der(\algB)^{c}$. Notice that $\phi^i \left( \omega_i( \partial^{i}_{\!\!\algA, \kappa_1}, \dots, \partial^{i}_{\!\!\algA, \kappa_{p}} ) \right)$ is $\omega_i( \partial^{i}_{\!\!\algA, \kappa_1}, \dots, \partial^{i}_{\!\!\algA, \kappa_{p}} ) \in M_{n_i}$ at the $i$-th block on the diagonal of $M_m$. We can then write $\eta = \tsum_{i=1}^{r} \eta_i$ where $\eta_i \in \phi^i(M_{n_i}) \otimes \exter^\grast \Der( \phi^i(M_{n_i}) )^* \subset M_m \otimes \exter^\grast \Der( \phi^i(M_{n_i}) )^*$. 
	\medskip
	\par
	Let us define the metric $g_{\phi^i(M_{n_i})}$ as the restriction of $g_\algB$ to $\Der( \phi^i(M_{n_i}) )$. From this metric we construct its Hodge $\hstar$-operator $\hstar_{\phi^i(M_{n_i})}$ and its NC integral $\int_{\phi^i(M_{n_i})}$ along the derivations of the matrix block $\phi^i(M_{n_i})$. Then we can apply Lemma~\ref{lemma *omega decomposed}, where the linear form $\tau$ is the ordinary trace on $M_m$, to get
	\begin{align*}
		\int_{\algB} \eta \wedge \hstar_\algB \eta
		&= \tsum_{i=1}^{r} \int_{\phi^i(M_{n_i})} \eta_i \wedge \hstar_{\phi^i(M_{n_i})} \eta_i
	\end{align*}
	One has $\Der( \phi^i(M_{n_i}) ) \simeq \Der(M_{n_i})$ and, using \eqref{eq gB and gA}, $g_{\phi^i(M_{n_i})}$ becomes $g_{\algA, i}$ in this identification, so that  $\hstar_{\phi^i(M_{n_i})}$ corresponds to $\hstar_{M_{n_i}}$. In this identification, the form $\eta_i$ is then $\omega_i$, so that $\int_{\phi^i(M_{n_i})} \eta_i \wedge \hstar_{\phi^i(M_{n_i})} \eta_i = \int_{M_{n_i}} \omega_i \wedge \hstar_{M_{n_i}} \omega_i$. This shows that $\int_{\algB} \eta \wedge \hstar_\algB \eta = \tsum_{i=1}^{r} \int_{M_{n_i}} \omega_i \wedge \hstar_{M_{n_i}} \omega_i = \int_{\algA} \omega \wedge \hstar_\algA \omega$ by \eqref{eq def scalar product forms}.
\end{proof}

A slight adjustment of the proof of Prop.~\ref{prop eta start eta omega star omega} gives (with the same notations):
\begin{corollary}
	Suppose that $\phi : \algA \to \algB$ includes $\kappa_i$ times $M_{n_i}$ on the diagonal of $M_m$. Then, with the same assumptions, one has
	\begin{align*}
		\int_{\algB} \eta \wedge \hstar_\algB \eta
		&=
		\tsum_{i=1}^{r} \kappa_{i} \int_{i} \omega_i \wedge \hstar_i \omega_i
	\end{align*}
\end{corollary}

\section{Direct Limit of NC Gauge Field Theories}
\label{sect direct limit NCGFT}

\subsection{\texorpdfstring{$\phi$-Compatibility of NC Gauge Field Theories}{phi-compatibility of NC gauge field theories}}
\label{sec phi compatibility of NCGFT}

As mentioned in Sect.~\ref{AFalg}, we will consider non unital $*$-homomorphisms $\phi : \algA = \toplus_{i=1}^{r} M_{n_i} \to \algB = \toplus_{k=1}^{s} M_{m_k}$. The reason for this choice is that we would like to cover physical situations where the gauge group are enlarged at each step of the defining inductive sequence $\{ (\algA_n, \phi_{n,m}) \, / \,  0 \leq n < m \}$. For instance, one may ask for the inclusion of $U(2)$ into $U(3)$, which can be performed in our framework by considering a natural inclusion $\phi : M_2 \to M_3$. This inclusion cannot be unital. A unital morphism would require for instance to consider the inclusion $\phi : M_1 \oplus M_2 \to M_3$, which may not correspond to a phenomenological requirement.
\medskip
\par
We first consider a NCGFT on the algebra $\algA = \toplus_{i=1}^{r} M_{n_i}$. Let us use the notations of Sect.~\ref{sec one step in the sequence der}: for any $i$, let $\{ \partial^{i}_{\!\!\algA, \kappa} \defeq \ad_{E^{i}_{\!\!\algA, \kappa}} \}_{\kappa \in I_{i}}$ be an orthogonal basis  of $\Der(\algA_i) = \Int(M_{n_i})$ where $E^{i}_{\!\!\algA, \kappa} \in \ksl_{n_i}$ and $I_{i}$ is a totally ordered set of cardinal $n_i^2-1$, and let $\{ \theta^\kappa_{\!\!\algA, i} \}_{\kappa \in I_i}$ be the dual basis of $\{ \partial^{i}_{\!\!\algA, \kappa} \}_{\kappa \in I_{i}}$.
\medskip
\par
Since we are interested in the manipulation of connections as $1$-forms, we restrict our analysis to the left module $\modM = \algA$. From Example~\ref{example Mn} and the results in Sect.~\ref{ModAndConnectionDirectSum}, with obvious notations, a connection $1$-form can be written as $\omega =  \toplus_{i=1}^{r} \omega_i$ and its curvature $2$-form as $\Omega = \toplus_{i=1}^{r} \Omega_i$, with $\omega_i = \omega^{i}_{\kappa} \theta^\kappa_{\!\!\algA, i} = \mromega_{i} - B^{i}_{\!\!\algA, \kappa} \theta^\kappa_{\!\!\algA, i} = (E^{i}_{\!\!\algA, \kappa} - B^{i}_{\!\!\algA, \kappa}) \theta^\kappa_{\!\!\algA, i}$ and $\Omega_i = \tfrac{1}{2} \Omega^{i}_{\kappa_1\kappa_2} \theta^{\kappa_1}_{\!\!\algA, i} \wedge \theta^{\kappa_2}_{\!\!\algA, i}$ with $\Omega^{i}_{\kappa_1\kappa_2} = - ([B^{i}_{\!\!\algA, \kappa_1}, B^{i}_{\!\!\algA, \kappa_2}] - C(n_i)_{\kappa_1\kappa_2}^{\kappa_3} B^{i}_{\!\!\algA, \kappa_3})$ where $C(n_i)_{\kappa_1\kappa_2}^{\kappa_3}$ are the structure constants for the basis $\{ E^{i}_{\!\!\algA, \kappa} \}$ of $\ksl_{n_i}$.
\medskip
\par
The natural action for this NCGFT is then 
\begin{align*}
	\act &=
	- \tsum_{i=1}^{r} \int_i \Omega_i \wedge \hstar_i \Omega_i
	= - \tsum_{i=1}^{r} \tfrac{1}{2} \tr ( \Omega^{i}_{\kappa_1\kappa_2} \Omega^{i, \kappa_1\kappa_2})= - \tsum_{i=1}^{r} \tfrac{1}{2} \tsum_{\kappa_1\kappa_2 \in I_i} \tr ( \Omega^{i}_{\kappa_1\kappa_2} )^2
	\\
	&= - \tsum_{i=1}^{r} \tfrac{1}{2} \tsum_{\kappa_1\kappa_2 \in I_i} \tr ( [B^{i}_{\!\!\algA, \kappa_1}, B^{i}_{\!\!\algA, \kappa_2}] - C(n_i)_{\kappa_1\kappa_2}^{\kappa_3} B^{i}_{\!\!\algA, \kappa_3} )^2
\end{align*}
where in the last line we have used the fact that the metric is diagonal.
\medskip
\par
As in Subsect~\ref{example C(M) otimes Mn}, one can also consider a NCGFT on the algebra $\halgA \defeq C^\infty(M) \otimes \algA = \toplus_{i=1}^{r} C^\infty(M) \otimes M_{n_i}$ for a manifold $M$. Then $\omega_i = A^{i}_{\!\!\algA,\mu} \dd x^\mu + (E^{i}_{\!\!\algA, \kappa} - B^{i}_{\!\!\algA, \kappa}) \theta^\kappa_{\!\!\algA, i}$ and $\Omega_i = \tfrac{1}{2} \Omega^{i}_{\mu_1\mu_2} \dd x^{\mu_1} \wedge \dd x^{\mu_2} + \Omega^{i}_{\mu \kappa} \dd x^\mu \wedge \theta^\kappa_{\!\!\algA, i} + \tfrac{1}{2}  \Omega^{i}_{\kappa_1\kappa_2} \theta^{\kappa_1}_{\!\!\algA, i} \wedge \theta^{\kappa_2}_{\!\!\algA, i}$ with
\begin{align*}
	\Omega^{i}_{\mu_1\mu_2}
	&= \partial_{\mu_1} A^{i}_{\!\!\algA, \mu_2} - \partial_{\mu_2} A^{i}_{\!\!\algA, \mu_1} - [A^{i}_{\!\!\algA, \mu_1}, A^{i}_{\!\!\algA, \mu_2}],
	\\
	\Omega^{i}_{\mu \kappa}
	&= - ( \partial_\mu B^{i}_{\!\!\algA, \kappa} - [A^{i}_{\!\!\algA, \mu}, B^{i}_{\!\!\algA, \kappa}] ),
	\\
	\Omega^{i}_{\kappa_1\kappa_2}
	&= - ( [B^{i}_{\!\!\algA, \kappa_1}, B^{i}_{\!\!\algA, \kappa_2}] - C(n_i)_{\kappa_1\kappa_2}^{\kappa_3} B^{i}_{\!\!\algA, \kappa_3} ).
\end{align*}
In that case, the natural action is
\begin{align*}
	\act &= \begin{multlined}[t]
		- \tsum_{i=1}^r \int_M \Big(
		\tfrac{1}{2} \tr( \Omega^{i}_{\mu_1\mu_2} \Omega^{i, \mu_1\mu_2} ) 
		+ \tr( \Omega^{i}_{\mu \kappa} \Omega^{i, \mu \kappa} ) 
		+ \tfrac{1}{2} \tr( \Omega^{i}_{\kappa_1\kappa_2} \Omega^{i, \kappa_1\kappa_2} ) 
		\Big) \sqrt{\abs{g_M}} \dd x
	\end{multlined}
\end{align*}
where $g_M$ is a metric on $M$.
\medskip
\par
This action shares the same main features as mentioned in Subsect~\ref{example C(M) otimes Mn}: it makes appear a SSBM thanks to the presence of the scalar fields $B^{i}_{\!\!\algA, \kappa} = B^{i, \kappa'}_{\!\!\algA, \kappa} E^{i}_{\!\!\algA, \kappa'} + i B^{i, 0}_{\!\!\algA, \kappa} \bbbone_{n_i}$ which can be non zero for a minimal configuration of the Higgs potential $- \tfrac{1}{2} \tsum_{i=1}^r \tr( \Omega^{i}_{\kappa_1\kappa_2} \Omega^{i, \kappa_1\kappa_2} )$. Then the couplings in $- \tsum_{i=1}^r \tr( \Omega^{i}_{\mu \kappa} \Omega^{i, \mu \kappa} )$ induce mass terms for the (gauge bosons) fields $A^{i}_{\!\!\algA,\mu} = A^{i, \kappa}_{\!\!\algA,\mu} E^{i}_{\!\!\algA, \kappa} + i A^{i, 0}_{\!\!\algA,\mu} \bbbone_{n_i}$. We will concentrate of this feature in the following.
\medskip
\par
In order to simplify the analysis of the relation between NCGFT defined at each step of an inductive sequence of finite dimensional algebras $\{ (\algA_n, \phi_{n,m}) \, / \,  0 \leq n < m \}$, we will consider a unique inclusion $\phi : \algA \to \algB$ with $\algA = \toplus_{i=1}^{r} M_{n_i}$ and $\algB = \toplus_{k=1}^{s} M_{m_k}$ as in Sect.~\ref{sec one step in the sequence der}.
\medskip
\par
Let $\eta = \toplus_{k=1}^{s} \eta_k$ be a connection $1$-form on $\algB$ for the module $\modN = \algB$. Denote by $\Theta = \toplus_{k=1}^{s} \Theta_k$ its curvature $2$-form. We use the notation $\eta_k = \eta^{k}_{\beta} \theta^\beta_{\!\algB, k} = (E^{k}_{\!\algB, \beta} - B^{k}_{\!\algB, \beta}) \theta^\beta_{\!\algB, k}$ and $\Theta_k = \tfrac{1}{2} \Theta^{k}_{\beta_1\beta_2} \theta^{\beta_1}_{\!\algB, k} \wedge \theta^{\beta_2}_{\!\algB, k}$ with $\Theta^{k}_{\beta_1\beta_2} = - ([B^{k}_{\!\algB, \beta_1}, B^{k}_{\!\algB, \beta_2}] - C(m_k)_{\beta_1\beta_2}^{\beta_3} B^{k}_{\!\algB, \beta_3})$ where $C(m_k)_{\beta_1\beta_2}^{\beta_3}$ are the structure constants for the basis $\{ E^{k}_{\!\algB, \beta} \}$ of $\ksl_{m_k}$.
\medskip
\par
We suppose that $\eta$ is $\phi$-compatible with $\omega$. This implies that, for all $\beta = (i, \alpha, \kappa) \in  J^\phi_{k}$, $\eta^{k}_{\beta} = \phi^{i}_{k, \alpha}(\omega^{i}_{\kappa})$ and that $B^{k}_{\!\algB, \beta} = \phi^{i}_{k, \alpha}(B^{i}_{\!\!\algA, \kappa})$.

\begin{lemma}
	\label{lem curvature inherited indices}
	For any $\beta_1 = (i_1, \alpha_1, \kappa_1), \beta_2 = (i_2, \alpha_2, \kappa_2) \in J^\phi_{k}$, $\Theta^{k}_{\beta_1\beta_2} = 0$ for $i_1 \neq i_2$ or $\alpha_1 \neq \alpha_2$ and $\Theta^{k}_{\beta_1\beta_2} =\phi^{i}_{k, \alpha}( \Omega^{i}_{\kappa_1\kappa_2})$ for $i = i_1 = i_2$ and $\alpha = \alpha_1 = \alpha_2$.
\end{lemma}

\begin{proof}
	One has to evaluate 
	\begin{align*}
		\Theta_{k} & (\phi^{i_1}_{k, \alpha_1}(\partial^{i_1}_{\!\!\algA, \kappa_1}), \phi^{i_2}_{k, \alpha_2}(\partial^{i_2}_{\!\!\algA, \kappa_2})) 
		\\
		&= \begin{multlined}[t]
			\phi^{i_1}_{k, \alpha_1}(\partial^{i_1}_{\!\!\algA, \kappa_1}) \cdotaction \eta_{k}(\phi^{i_2}_{k, \alpha_2}(\partial^{i_2}_{\!\!\algA, \kappa_2}))
			- \phi^{i_2}_{k, \alpha_2}(\partial^{i_2}_{\!\!\algA, \kappa_2}) \cdotaction \eta_{k}(\phi^{i_1}_{k, \alpha_1}(\partial^{i_1}_{\!\!\algA, \kappa_1}))
			\\
			- \eta_{k}( [\phi^{i_1}_{k, \alpha_1}(\partial^{i_1}_{\!\!\algA, \kappa_1}), \phi^{i_2}_{k, \alpha_2}(\partial^{i_2}_{\!\!\algA, \kappa_2})] )
			- [ \eta_{k}( \phi^{i_1}_{k, \alpha_1}(\partial^{i_1}_{\!\!\algA, \kappa_1}) ), \eta_{k}( \phi^{i_2}_{k, \alpha_2}(\partial^{i_2}_{\!\!\algA, \kappa_2}) ) ]
		\end{multlined}
		\\
		&= \begin{multlined}[t]
			\phi^{i_1}_{k, \alpha_1}(\partial^{i_1}_{\!\!\algA, \kappa_1}) \cdotaction \phi^{i_2}_{k, \alpha_2}(\omega_{i_2}(\partial^{i_2}_{\!\!\algA, \kappa_2}))
			- \phi^{i_2}_{k, \alpha_2}(\partial^{i_2}_{\!\!\algA, \kappa_2}) \cdotaction \phi^{i_1}_{k, \alpha_1}(\omega_{i_1}(\partial^{i_1}_{\!\!\algA, \kappa_1}))
			\\
			- \eta_{k}( [\phi^{i_1}_{k, \alpha_1}(\partial^{i_1}_{\!\!\algA, \kappa_1}), \phi^{i_2}_{k, \alpha_2}(\partial^{i_2}_{\!\!\algA, \kappa_2})] )
			- [ \phi^{i_1}_{k, \alpha_1}(\omega_{i_1}( \partial^{i_1}_{\!\!\algA, \kappa_1}) ), \phi^{i_2}_{k, \alpha_2}(\omega_{i_2}( \partial^{i_2}_{\!\!\algA, \kappa_2}) ) ]
		\end{multlined}
	\end{align*}
	For $i_1 \neq i_2$ or $\alpha_1 \neq \alpha_2$, from Lemma~\ref{lem inherited derivations relations}, all the terms in the first line vanish while the last commutator is zero since the two matrices involved do not occupy the same position on the diagonal of $M_{m_k}$. For $i_1 = i_2 = i$ and $\alpha_1 = \alpha_2 = \alpha$, the expression reduces to $\phi^{i}_{k, \alpha} ( \Omega_{i}( \partial^{i}_{\!\!\algA, \kappa_1}, \partial^{i}_{\!\!\algA, \kappa_2}) )$ (which is also a consequence of Prop.~\ref{prop compatible forms product differential}).
	
	Notice that this result is also a direct consequence of the expression of $\Theta^{k}_{\beta_1\beta_2}$ in terms of the $B^{k}_{\!\algB, \beta}$'s.
\end{proof}

The curvature components $\Theta^{k}_{\beta_1\beta_2}$ for $\beta_1, \beta_2 \in J_{k}$ can be separated according to the 3 possibilities: (1) $(\beta_1, \beta_2)$ in $J^\phi_{k} \times J^\phi_{k}$; (2) $(\beta_1, \beta_2)$ or $(\beta_2, \beta_1)$ in $J^\phi_{k} \times J^c_{k}$; (3) $(\beta_1, \beta_2)$ in $J^c_{k} \times J^c_{k}$. From \eqref{eq block orthogonality basis}, the metric $g_\algB$ (and its inverse) is block diagonal for these subsets of indices. The natural action on $\algB$ can then be decomposed as
\begin{align*}
	\act &=
	- \tsum_{k=1}^{s} \tfrac{1}{2} \tsum_{\beta_1\beta_2 \in J_{k}} \tr ( \Theta^{k}_{\beta_1\beta_2} \Theta^{k, \beta_1\beta_2} )
	\\
	&= \begin{multlined}[t]
		- \tsum_{k=1}^{s} \Big(
		\tfrac{1}{2} \tsum_{\beta_1\beta_2 \in J^\phi_{k}} \tr ( \Theta^{k}_{\beta_1\beta_2} \Theta^{k, \beta_1\beta_2} )
		+ \tsum_{\beta_1 \in J^\phi_{k}, \beta_2 \in J^c_{k}} \tr ( \Theta^{k}_{\beta_1\beta_2} \Theta^{k, \beta_1\beta_2} )
		+ \tfrac{1}{2} \tsum_{\beta_1\beta_2 \in J^c_{k}} \tr ( \Theta^{k}_{\beta_1\beta_2} \Theta^{k, \beta_1\beta_2} )
		\Big)
	\end{multlined}
\end{align*}
For fixed $k$, let us consider the first summation on $J^\phi_{k}$. By Lemma~\ref{lem curvature inherited indices}, the two indices $\beta_1 = (i_1, \alpha_1, \kappa_1), \beta_2 = (i_2, \alpha_2, \kappa_2) \in J^\phi_{k}$ must satisfy $i_1 = i_2$ and $\alpha_1 = \alpha_2$ to get a non zero contribution, so that 
\begin{align*}
	\tfrac{1}{2} \tsum_{\beta_1\beta_2 \in J^\phi_{k}} 
	\tr ( \Theta^{k}_{\beta_1\beta_2} \Theta^{k, \beta_1\beta_2} )& = 
	\tfrac{1}{2} \tsum_{i=1}^{r} \tsum_{\alpha=1}^{\alpha_{ki}} \tsum_{\kappa_1, \kappa_2 \in I_i} 
	\tr ( \Theta^{k}_{(i, \alpha, \kappa_1)(i, \alpha, \kappa_2)} \Theta^{k, (i, \alpha, \kappa_1)(i, \alpha, \kappa_2)} ) 
	\\
	& =
	\tfrac{1}{2} \tsum_{i=1}^{r} \tsum_{\alpha=1}^{\alpha_{ki}} \tsum_{\kappa_1, \kappa_2 \in I_i} 
	\tr ( \Omega^{i}_{\kappa_1\kappa_2} \Omega^{i, \kappa_1\kappa_2} )=
	\tfrac{1}{2} \tsum_{i=1}^{r} \alpha_{ki} \tsum_{\kappa_1, \kappa_2 \in I_i} 
	\tr ( \Omega^{i}_{\kappa_1\kappa_2} \Omega^{i, \kappa_1\kappa_2} )
\end{align*}
where we have used \eqref{eq gB and gA} (which holds true also for the inverse metrics) to make an equivalence between raising the indices $\beta$ and raising the indices $\kappa$.
\medskip
\par
This relation tells us that the \emph{action on $\algB$ contains copies of terms from the action on $\algA$}. As expected, these terms involve the degrees of freedom which are inherited on $\algB$ from those on $\algA$. They involve also the multiplicities of the inclusions of $M_{n_i}$ into $M_{m_k}$. This implies that the relative weights of these terms are not the same on $\algB$ as they are on $\algA$.
\medskip
\par
Let us now consider the algebra $\halgB \defeq C^\infty(M) \otimes \algB = \toplus_{k=1}^{s} C^\infty(M) \otimes M_{m_k}$. The connection $1$-form is parametrized as $\eta_k = A^{k}_{\algB,\mu} \dd x^\mu + (E^{k}_{\!\algB, \beta} - B^{k}_{\!\algB, \beta}) \theta^\beta_{\!\algB, k}$. We extend the $\phi$-compatibility between $\eta$ and $\omega$ on the geometrical part by the condition that $A^{k}_{\algB,\mu} = \tsum_{i=1}^{r} \phi^{i}_{k}(A^{i}_{\!\!\algA,\mu}) + A^{k, c}_{\algB,\mu}$ where $A^{k, c}_{\algB,\mu} \in M_{m_k}$ has zero entries in the image of $\phi_{k}$ (which is concentred as blocks on the diagonal). In other words, all the degrees of freedom in $A^{i}_{\!\!\algA,\mu}$ are copied into $A^{k}_{\algB,\mu}$ according to the map $\phi$.
\medskip
\par
From this decomposition, the components of the curvature $2$-forms $\Theta_k = \tfrac{1}{2} \Theta^{k}_{\mu_1\mu_2} \dd x^{\mu_1} \wedge \dd x^{\mu_2} + \Theta^{k}_{\mu \beta} \dd x^\mu \wedge \theta^\beta_{\!\algB, k} + \tfrac{1}{2}  \Theta^{k}_{\beta_1\beta_2} \theta^{\beta_1}_{\!\algB, k} \wedge \theta^{\beta_2}_{\!\algB, i}$ can be separated into inherited components from the curvature $2$-forms $\Omega_{i}$ on $\halgA$, interactions terms between inherited components of the $\omega_i$ $1$-forms with new components of the $\eta_k$ $1$-forms, and completely new terms from new components of the $\eta_k$. Explicitly, one has
\begin{align*}
	\Theta^{k}_{\mu_1\mu_2}
	&= \partial_{\mu_1} A^{k}_{\algB, \mu_2} - \partial_{\mu_2} A^{k}_{\algB, \mu_1} - [A^{k}_{\algB, \mu_1}, A^{k}_{\algB, \mu_2}]
	\\
	&= \begin{multlined}[t]
		\tsum_{i=1}^{r} \phi^{i}_{k}\left(
		\partial_{\mu_1} A^{i}_{\!\!\algA,\mu_2} - \partial_{\mu_2} A^{i}_{\!\!\algA,\mu_1} - [ A^{i}_{\!\!\algA,\mu_1}, A^{i}_{\!\!\algA,\mu_2}]
		\right)
		\\
		- \tsum_{i=1}^{r} \left(
		[ \phi^{i}_{k}( A^{i}_{\!\!\algA,\mu_1} ), A^{k, c}_{\algB,\mu_2} ]
		+ [ A^{k, c}_{\algB,\mu_1}, \phi^{i}_{k}( A^{i}_{\!\!\algA,\mu_2} ) ]
		\right)
		+ \partial_{\mu_1} A^{k, c}_{\algB,\mu_2} - \partial_{\mu_2} A^{k, c}_{\algB,\mu_1} - [ A^{k, c}_{\algB,\mu_1}, A^{k, c}_{\algB,\mu_2} ]
	\end{multlined}
\end{align*}
In the same way, for any $\beta = (i, \alpha, \kappa) \in J^\phi_{k}$, one has
\begin{align*}
	\Theta^{k}_{\mu \beta}
	&= - ( \partial_\mu B^{k}_{\!\algB, \beta} - [A^{k}_{\algB, \mu}, B^{k}_{\!\algB, \beta}] )
	= - \phi^{i}_{k, \alpha}\left( \partial_\mu B^{i}_{\!\!\algA, \kappa} - [ A^{i}_{\!\!\algA,\mu} , B^{i}_{\!\!\algA, \kappa}]  \right)
	+ [A^{k, c}_{\algB,\mu}, \phi^{i}_{k, \alpha}(B^{i}_{\!\!\algA, \kappa})]
\end{align*}
where the last term is off diagonal. For $\beta \in J^c_{k}$, $\Theta^{k}_{\mu \beta}$ depends on the fields $A^{k, c}_{\algB,\mu}$ which couple with $B^{k}_{\!\algB, \beta}$. Finally, $\Theta^{k}_{\beta_1\beta_2}$ has been explored before.
\medskip
\par
Denote by $\algA^\times$ and $\algB^\times$ the groups of invertible elements in $\algA$ and $\algB$ and let us define $\tphi : \algA^\times \to \algB^\times$, for any $a = \toplus_{i=1}^{r} a_i \in \algA^\times$, as
\begin{align*}
	\tphi^k (a) \defeq \pi_\algB^k \circ \tphi (a)
	&= 
	\begin{pmatrix}
		a_1 \otimes \bbbone_{\alpha_{k1}} & 0 & \cdots & 0 & 0\\
		0 & a_2 \otimes \bbbone_{\alpha_{k2}} & \cdots & 0 & 0\\
		\vdots & \vdots & \ddots & \vdots & \vdots\\
		0 & 0 & \cdots & a_r \otimes \bbbone_{\alpha_{kr}} & 0 \\
		0 & 0 & \cdots & 0 & \bbbone_{n_0}
	\end{pmatrix}
\end{align*}
It is easy to check that $\tphi$ is a morphism of groups: for $a \in \algA^\times$, one has $\tphi(a) \in \algB^\times$ and $\tphi(a)^{-1} = \tphi(a^{-1})$. One has also $\tphi(a)^* = \tphi(a^*)$ so that if $u \in \calU(\algA)$ is a unitary element in $\algA$, so is $\tphi(u)$ in $\algB$.

\begin{lemma}
	\label{lem hphi product phi}
	For any $a \in \algA$ and any $u \in \algA^\times$, one has $\tphi(u)\phi(a) = \phi(ua)$ and $\phi(a)\tphi(u) = \phi(au)$. For any $k \in \{1, \dots, s\}$, any $i \in \{1, \dots, r\}$, and any $\alpha \in \{1, \dots, \alpha_{ki} \}$, one has $\tphi(u)\phi^{i}_{k, \alpha}(a) = \phi^{i}_{k, \alpha}(ua)$ and $\phi^{i}_{k, \alpha}(a)\tphi(u) = \phi^{i}_{k, \alpha}(au)$.
\end{lemma}

\begin{proof}
	From \eqref{eq decompositions phi}, it is sufficient to prove the last relations involving $\phi^{i}_{k, \alpha}$. Since $\tphi(u)$ differs only from $\phi(u)$ at the last $n_0 \times n_0$ block entry on the diagonal where $\phi^{i}_{k, \alpha}(a)$ is zero, one has $\tphi(u)\phi^{i}_{k, \alpha}(a) = \phi(u)\phi^{i}_{k, \alpha}(a) = \phi^{i}_{k, \alpha}(ua)$ by \eqref{eq decompositions phi} and \eqref{eq product phiijell}. The same proof applies for the right multiplication by $u$.
\end{proof}

\begin{proposition}
	\label{prop gauge transformations phi A and B}
	Let $\omega$ be a connection $1$-form on $\algA$ and let $\eta$ be a $\phi$-compatible connection $1$-form on $\algB$. Let $u \in \calU(\algA)$ and $v \defeq \tphi(u) \in \calU(\algB)$. Then $\omega^u$ and $\eta^v$ are $\phi$-compatible.
\end{proposition}

\begin{proof}
	Recall that $\omega^u = u^{-1} \omega u - u^{-1} (\dd u)$ and $\eta^v = v^{-1} \eta v - v^{-1} (\dd v)$. Notice that $\pi_\algA^i(\omega^u) = u_i^{-1} \omega_i u_i - u_i^{-1} (\dd u_i) \in \OmegaDer^1(\algA_i)$ and similarly $\pi_\algB^k(\eta^v) = v_k^{-1} \eta_k v_k - v_k^{-1} (\dd v_k) \in \OmegaDer^1(\algB_k)$. For any $k$ and any $\beta = (i, \alpha, \kappa) \in J^\phi_{k}$, one has
	\begin{align*}
		\pi_\algB^k(\eta^v) ( \phi^{i}_{k, \alpha}(\partial^{i}_{\!\!\algA, \kappa}) )
		&=
		v_k^{-1} \eta_k( \phi^{i}_{k, \alpha}(\partial^{i}_{\!\!\algA, \kappa}) ) v_k - v_k^{-1} [\phi^{i}_{k, \alpha}(E^{i}_{\!\!\algA, \kappa}), v_k]
		\\
		&= v_k^{-1} \phi^{i}_{k, \alpha} \left( \omega_i(\partial^{i}_{\!\!\algA, \kappa}) \right) v_k  - v_k^{-1} \phi^{i}_{k, \alpha}([E^{i}_{\!\!\algA, \kappa}, u_i])
		\\
		& =
		\phi^{i}_{k, \alpha} \left( u_i^{-1} \omega_i(\partial^{i}_{\!\!\algA, \kappa}) u_i - u_i^{-1} (\dd u_i)(\partial^{i}_{\!\!\algA, \kappa}) \right)
		= \phi^{i}_{k, \alpha} \left( \pi_\algA^i(\omega^u)(\partial^{i}_{\!\!\algA, \kappa}) \right)
	\end{align*}
	where we have used Lemma~\ref{lem hphi product phi}.
\end{proof}

We have a similar result for connections on $\halgA$ and $\halgB$:
\begin{proposition}
	\label{prop gauge transformations phi hA and hB}
	Let $\omega$ be a connection $1$-form on $\halgA$ and let $\eta$ be a $\phi$-compatible (in the extended version) connection $1$-form on $\halgB$. Let $u \in \calU(\halgA)$ and $v \defeq \tphi(u) \in \calU(\halgB)$. Then $\omega^u$ and $\eta^v$ are $\phi$-compatible (in the extended version).
\end{proposition}

\begin{proof}
	Concerning the algebraic parts of $\omega_i$ and $\eta_k$, the proof is the same as for Prop.~\ref{prop gauge transformations phi A and B}. It remains to show that the $A^{i, u_i}_{\!\!\algA,\mu} \defeq u_i^{-1} A^{i}_{\!\!\algA,\mu} u_i - u_i^{-1} \partial_\mu u_i$ are copied into $A^{k, v_k}_{\algB,\mu} \defeq v_k^{-1} A^{k}_{\algB,\mu} v_k - v_k^{-1} \partial_\mu v_k$ according to the map $\phi$. Using the fact that $v_k$ is block diagonal, and that these blocks are $\phi^{i}_{k, \alpha}(u_i)$ or $\bbbone_{n_0}$, the diagonal part of the first term $v_k^{-1} A^{k}_{\algB,\mu} v_k$ is exactly $\tsum_{i=1}^{r} \phi^{i}_{k}(u_i^{-1} A^{i}_{\!\!\algA,\mu}  u_i)$. The second term $v_k^{-1} \partial_\mu v_k$ contains only blocks on the diagonal: the zero block from the block $\bbbone_{n_0}$ and blocks $\phi^{i}_{k, \alpha}(u_i)^{-1} \partial_\mu \phi^{i}_{k, \alpha}(u_i) = \phi^{i}_{k, \alpha}(u_i^{-1} \partial_\mu u_i)$ otherwise. This proves that the blocks on the diagonal of $A^{k, v_k}_{\algB,\mu}$ are copies of the $A^{i, u_i}_{\!\!\algA,\mu}$ according to the map $\phi$. Obviously, the off diagonal part of $A^{k, v_k}_{\algB,\mu}$ mixes the degrees of freedom from the $A^{i}_{\!\!\algA,\mu}$'s and $u_i$'s.
\end{proof}

Prop.~\ref{prop gauge transformations phi A and B} and \ref{prop gauge transformations phi hA and hB} show that $\phi$-compatibility of connections is compatible with gauge transformations.
\medskip
\par
We have now at hand all the technical ingredients to discuss NCGFT on the $AF$ $C^*$-algebra defined by a sequence $\{ (\algA_n, \phi_{n,m}) \, / \,  0 \leq n < m \}$. This NCGFT uses the derivation-based differential calculus constructed on the dense “smooth” subalgebra $\algA_\infty \defeq \cup_{n\geq 0} \algA_n$ as the inductive limit of the differential calculi $(\OmegaDer^\grast(\algA_n), \dd)$, and a natural module is the algebra $\algA_\infty$ itself. With obvious notations, the same holds for $\halgA_\infty$. All these constructions are canonical. A connection is constructed as a limit of connections on each $\algA_n$ (with module the algebra itself). If we insist the connection $1$-form on $\algA_{n+1}$ to be $\phi_{n,n+1}$-compatible with the connection $1$-form on $\algA_{n}$, then some degrees of freedom in this connection are inherited from those of the connections on $\algA_{n}$, and new degrees of freedom are added. This limiting procedure is compatible with a good notion of gauge transformations (see Prop.~\ref{prop gauge transformations phi A and B} and Prop.~\ref{prop gauge transformations phi hA and hB}). 
\medskip
\par
Concerning the dynamics, the terms in the action functional on $\algA_{n}$ can be found (with possible different weights) as terms in the action functional on $\algA_{n+1}$. If a solution for the gauge field degrees of freedom has been found on $\algA_{n}$, then these degrees of freedom appear as \emph{fixed fields} in the action on $\algA_{n+1}$, and so as constrains when one solves the field equations on $\algA_{n+1}$ for the new fields (non inherited degrees of freedom). The same applies to an inductive sequence $\{ (\halgA_n, \phi_{n,m}) \, / \,  0 \leq n < m \}$. The Lagrangian on $\algA_\infty$ (or $\halgA_\infty$) should be constructed as a limiting procedure by adding new terms at each step in order to take into account the new degrees of freedom. But then this Lagrangian could contain an infinite number of terms. From a physical point of view, we do not expect to reach that point: only some finite dimensional “approximations” (at some levels $n$) can be considered and tested in experiments. \emph{In other word, the purpose of our construction is not to define a “target” NCGFT (which could be quite singular) but to construct a direct sequence of finite dimensional NCFGT.} We expect all the empirical data to be encoded into this sequence which \emph{formally defines} a NCFGT on $\algA_\infty$ (or $\halgA_\infty$).
\medskip
\par
As already mentioned at the end of Sect.~\ref{AFalg}, the $*$-homomorphisms $\phi_{n,n+1} : \algA_n \to \algA_{n+1}$ are only characterized up to unitary equivalence in $\algA_{n+1}$. We have shown that the action of such an unitary equivalence, which takes the form of an inner automorphism on $\algA_{n+1}$, is a transport of structures that does not change the physics. This is why it is convenient to work with the standard form used in this thesis for these $*$-homomorphisms.

\section{Numerical Exploration of the SSBM}
\label{sec numerical exploration of the SSBM}

We would like now to concentrate on the SSBM in our framework. Using previous notations, when the Higgs potential $- \tfrac{1}{2} \tsum_{i=1}^r \tr( \Omega^{i}_{\kappa_1\kappa_2} \Omega^{i, \kappa_1\kappa_2} )$ for $\halgA$ is minimized, the degrees of freedom in the $B^{i}_{\!\!\algA, \kappa} = B^{i, \kappa'}_{\!\!\algA, \kappa} E^{i}_{\!\!\algA, \kappa'} + i B^{i, 0}_{\!\!\algA, \kappa} \bbbone_{n_i}$ are fixed (possibly with a choice in many possible configurations) and the $\phi$-compatibility transports these values into the $B^{k}_{\!\algB, \beta} = B^{k, \beta'}_{\!\algB, \beta} E^{k}_{\!\algB, \beta'} + i B^{k, 0}_{\!\algB, \beta} \bbbone_{m_i}$'s. Then, using these fixed values, minimizing the Higgs potential for $\halgB$ only concerns a subset of all the $B^{k, \beta'}_{\!\algB, \beta}$'s. The configuration they define is not necessarily the minimum of the Higgs potential for $\halgB$ if it were computed along all the $B^{k, \beta'}_{\!\algB, \beta}$'s. 
\medskip
\par
The configuration of the fields $B^{i, \kappa'}_{\!\!\algA, \kappa}$'s on $\halgA$ (resp. the fields $B^{k, \beta'}_{\!\algB, \beta}$'s on $\halgB$) induces a mass spectrum for the gauge fields $A^{i, \kappa}_{\!\!\algA,\mu}$'s (resp. the gauge fields $A^{k, \beta}_{\algB, \mu}$'s). In order to illustrate the way the masses of the $A^{k, \beta}_{\algB, \mu}$'s are related to the masses of the $A^{i, \kappa}_{\!\!\algA,\mu}$'s by the constraints induced by $\phi$, we have produced numerical computations of the mass spectra for simple situations $\phi : \algA \to \algB$. These computations have been performed using the software \textit{Mathematica}. The following situations have been considered:
\begin{enumerate}
	\item $\algA = M_2$ and $\algB = M_3$. This is the minimal non trivial situation one can consider. It illustrates many features of the other situations concerning the masses of the fields $A^{k, \beta}_{\algB, \mu}$'s.
	
	\item $\algA = M_2 \oplus M_2$ and $\algB = M_4$. This situation is used to illustrate how two different configurations for the fields $B^{i, \kappa'}_{\!\!\algA, \kappa}$'s (one for each $M_2$) can conflict to produce a rich typology for the masses of the fields $A^{k, \beta}_{\algB, \mu}$'s.
	
	\item $\algA = M_2 \oplus M_2$ and $\algB = M_5$. This situation is used to show, by comparison with the preceding one, how the target algebra influences the mass spectrum.
	
	\item $\algA = M_2 \oplus M_3$ and $\algB = M_5$. This situation is used to show, by comparison with the preceding one, how the source algebra influences the mass spectrum.
\end{enumerate}

Due to the large number of parameters involved in the mathematical expressions, the numerical computations cannot explore the full space of configurations for the fields $B^{i, \kappa'}_{\!\!\algA, \kappa}$'s. This is why we have chosen to work with a very simplified situation: for every $i$, the fields $B^{i, \kappa'}_{\!\!\algA, \kappa}$'s are parametrized by a single real parameter $\lambda_i$ which interpolates, on the interval $[0,1]$, between the null-configuration and the basis-configuration (see Sect.~\ref{example C(M) otimes Mn}) as $B^{i}_{\!\!\algA, \kappa} = \lambda_i E^{i}_{\!\!\algA, \kappa}$. For each value of the $\lambda_i$'s, the minimum of the Higgs potential on $\halgB$ along the fields $B^{k, \beta'}_{\!\algB, \beta}$'s which are not inherited (via $\phi$) is computed. Then this minimum configuration is inserted into the couplings with the fields $A^{k, \beta}_{\algB, \mu}$'s to compute the mass spectrum. A comparison with the mass spectrum of the fields $A^{i, \kappa}_{\!\!\algA,\mu}$'s is easily done, since this spectrum is fully degenerate: all these fields have the same mass $\lambda_i = \sqrt{2 n_i}$ according to Lemma~\ref{lem mass rep-config}.
\medskip
\par
Let us first make some general remarks on the expected results. When $\lambda_i = 0$ for all $i=1, \dots, r$ (in our examples $r=2$ at most), the configuration on $\halgA$ is the null-configuration. So, we expect the constraints due to the values of the fields $B^{i, \kappa'}_{\!\!\algA, \kappa} = 0$ when computing the minimum of the Higgs potential on $\halgB$ to produce the null-configuration for the fields $B^{k, \beta'}_{\!\algB, \beta}$'s. In the same way, when $\lambda_i = 1$ for all $i=1, \dots, r$, the configuration on $\halgA$ is the basis-configuration, and since $\phi$ preserves Lie brackets, the minimum of the Higgs potential on $\halgB$ is expected to be the basis-configuration for the fields $B^{k, \beta'}_{\!\algB, \beta}$'s. All the numerical computations presented below are consistent with these expected results.
\medskip
\par
When $\lambda_i$ is neither $0$ nor $1$, the configuration for the fields $B^{i, \kappa'}_{\!\!\algA, \kappa}$'s is not a minimum of the Higgs potential on $\halgA$ (with value $0$ in that case). Nevertheless, we consider these configuration as “possible” since $\algA$ is not necessarily the first algebra in the sequence $\{ (\algA_n, \phi_{n,m}) \, / \,  0 \leq n < m \}$. Indeed, as the results will show, and as already mentioned, the minimum of the Higgs potential on $\halgB$ is not the minimum along all the $B^{k, \beta'}_{\!\algB, \beta}$'s, and its value can be non zero. So, we are not reduced to considering only zero minima on $\halgA$ and it is legitimate to explore other configurations for the fields $B^{i, \kappa'}_{\!\!\algA, \kappa}$'s on $\halgA$.
\medskip
\par
\begin{figure}[h]
	\centering
	\subfloat[Minimum values for the Higgs potential with details in insert.]{%
		\includegraphics[width=0.48\linewidth]{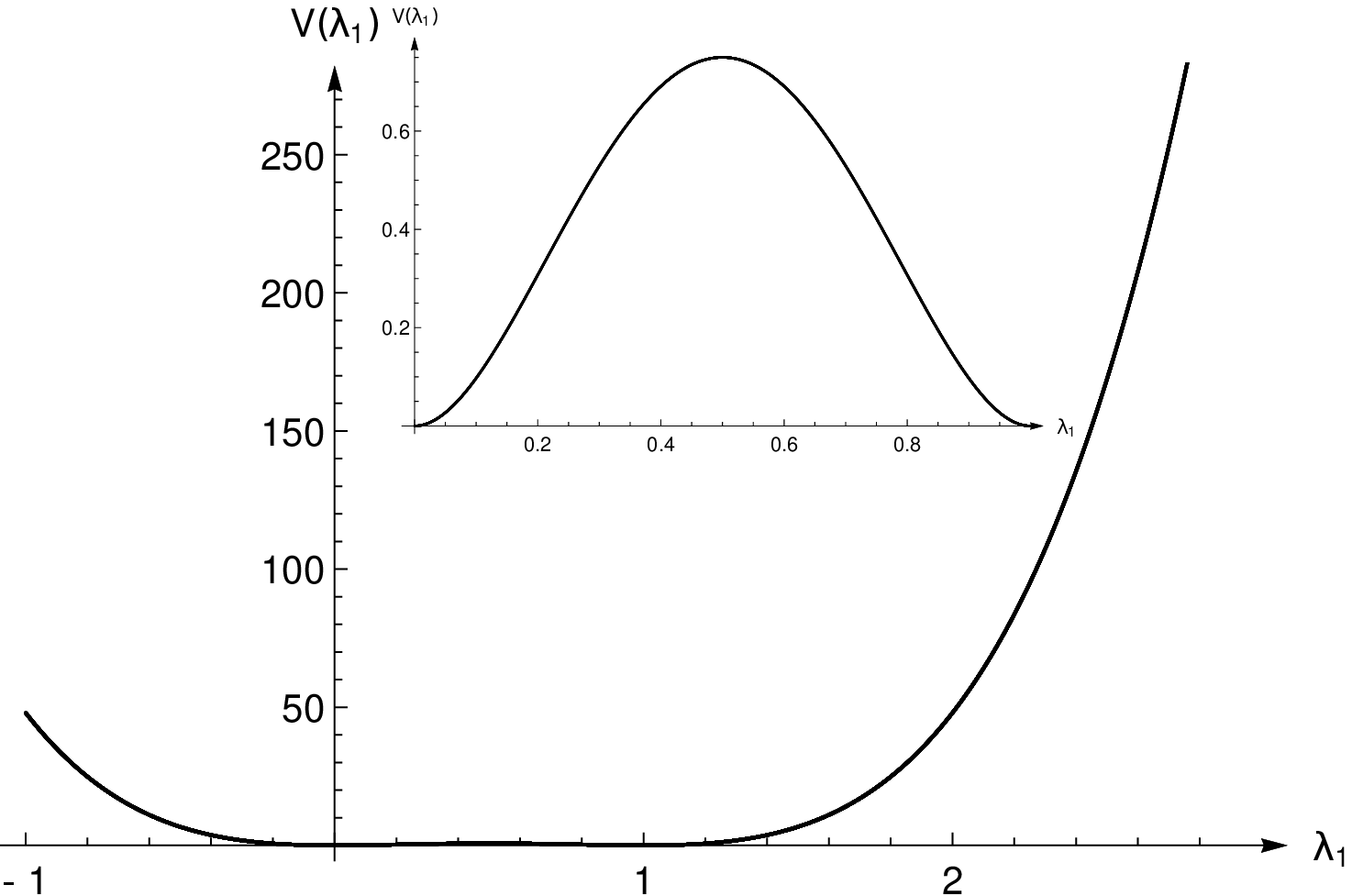}
		\label{fig-5Vd}
	}
	\subfloat[Masses for the gauge fields.]{%
		\includegraphics[width=0.48\linewidth]{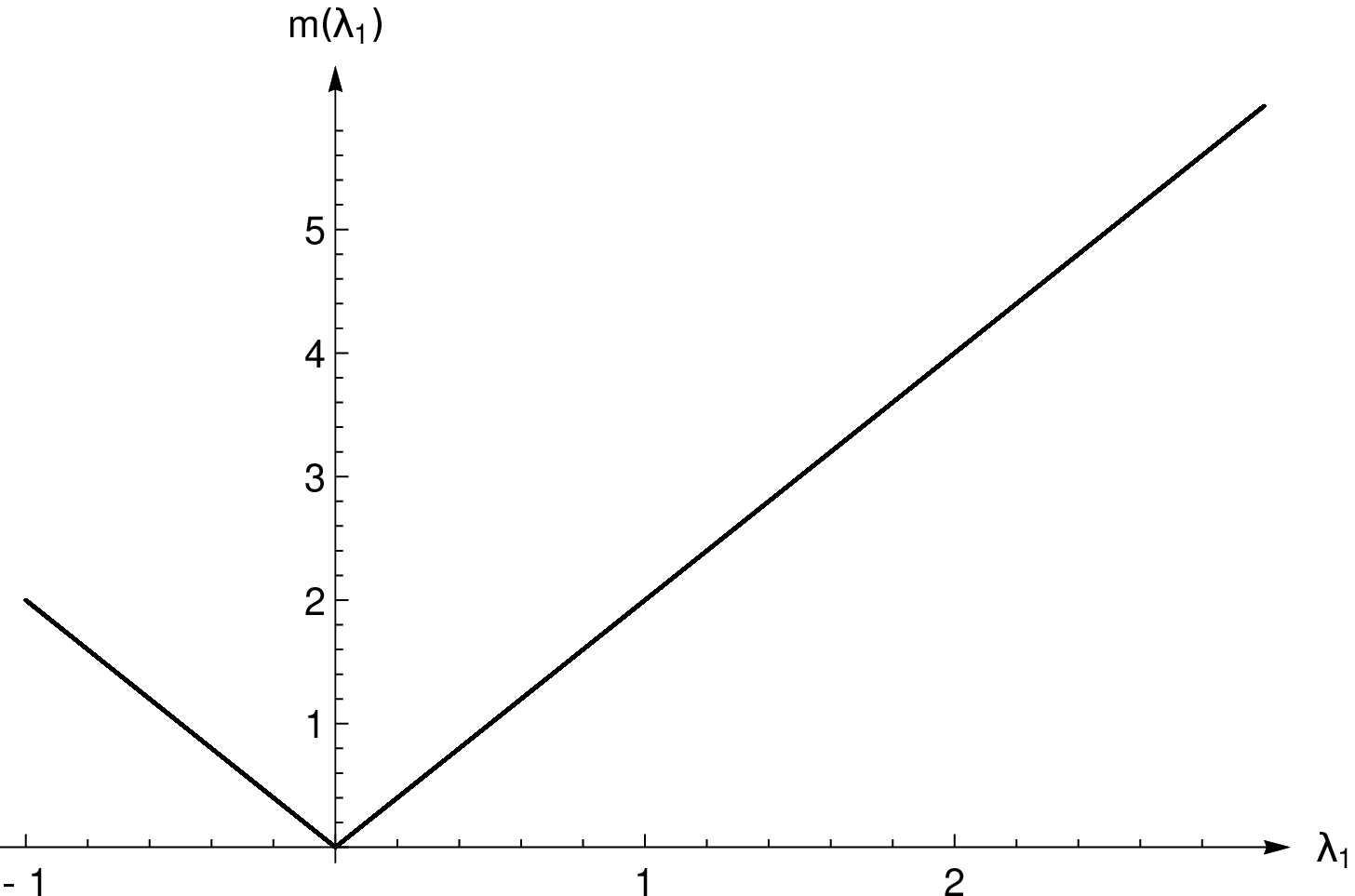}
		\label{fig-5Md}
	}
	\caption{$\algA = M_2$: plots for $\lambda_1 \in [-1, 3]$.}
	\label{fig-5Vd-5Md}
\end{figure}

Before describing the four cases, let us consider the situation in Fig.~\ref{fig-5Vd-5Md} which concerns the algebra $M_2$ only. The plot in Fig.~\ref{fig-5Vd} is the Higgs potential for the fields $B^{1}_{\!\!\algA, \kappa} = \lambda_1 E^{1}_{\!\!\algA, \kappa}$ depending only on $\lambda_1$. It is a quadratic polynomials in $\lambda_1$ and it looks very much like the Higgs potential of the SMPP in this approximation (reduction to a $1$-parameter dependency). The plot in Fig.~\ref{fig-5Md} is the mass spectrum for the $A^{1, \kappa}_{\!\!\algA,\mu}$ fields. As proved in Lemma~\ref{lem mass rep-config}, it is fully degenerated and depends linearly on $\lambda_1$ with slope $\sqrt{2 n_1}$ where $n_1 = 2$ in the present case. Similar plots can be obtained for any value $n_1$. These plots can be compared to the ones obtained in the four cases numerically explored.
\medskip
\par

All the numerical computations have been performed using orthonormal basis. We have noticed that the mass matrices, which have been computed in terms of an orthonormal basis $E^{1}_{\algB, \beta}$ (here we have only $k=1$) constructed as in subsection \ref{example C(\Man) otimes Mn}, are almost diagonal, up to terms of order $10^{-6}$ (these small values could be considered as numerical artifacts). This motivates the introduction of the following nomenclature for the labels appearing in the plots of the mass spectra for the gauge bosons. These labels refer to directions defined for specific subsets of matrices $E^{1}_{\!\algB, \beta} \in M_{m_1} = \algB$. The labels $a^i$ will refer to the inherited directions in $\phi(M_{n_i}) \subset \algB$. Let $\widetilde{\phi(\algA)}$ be the smallest square matrix block in $M_{m_1}$ that contains all the $\phi(M_{n_i})$. The label $b$ (resp. $d$) will refer to non diagonal (resp. diagonal) directions in $\widetilde{\phi(\algA)}$ that are not labeled by the $a^i$. When $\widetilde{\phi(\algA)} \neq \algB$, the labels $c^i$ will refer to non diagonal directions in $\algB \backslash \widetilde{\phi(\algA)}$ that commute with all the $\phi(M_{n_{i'}})$ for $i' \neq i$: this means that these directions are matrices in $M_{m_1}$ with non zero entries only in $\algB \backslash \widetilde{\phi(\algA)}$ at the same rows and columns occupied by $\phi(M_{n_i})$. In the same situation, the label $e$ will refer to diagonal directions with non zero entries in $\algB \backslash \widetilde{\phi(\algA)}$ and which commute with $\widetilde{\phi(\algA)}$.  
\medskip
\par
\begin{figure}[h]
	\centering
	\subfloat[Minimum values for the Higgs potential with details in insert.]{%
		\includegraphics[width=0.48\linewidth]{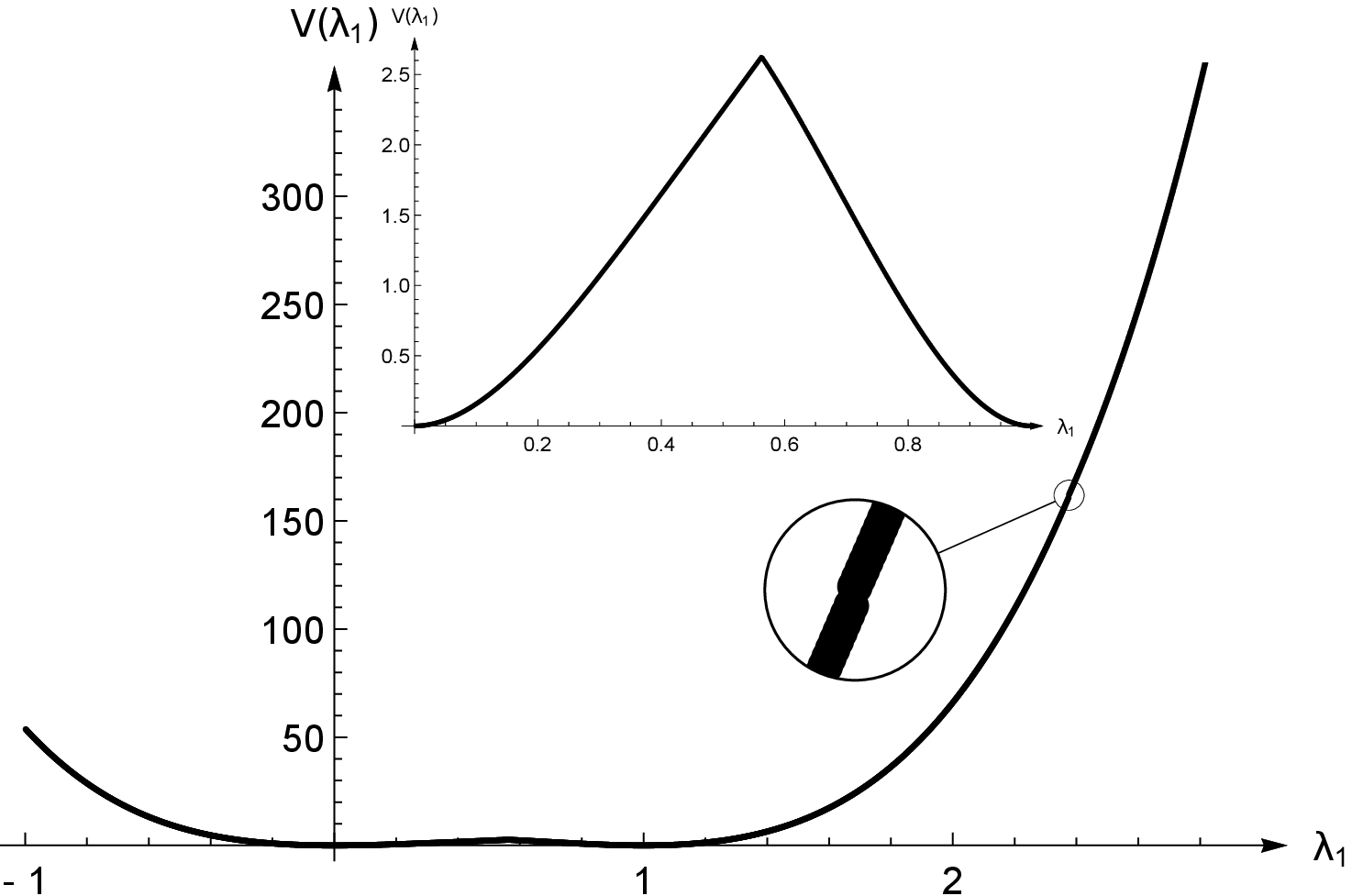}
		\label{fig-1Vd}
	}
	\subfloat[Masses for the gauge fields.]{%
		\includegraphics[width=0.48\linewidth]{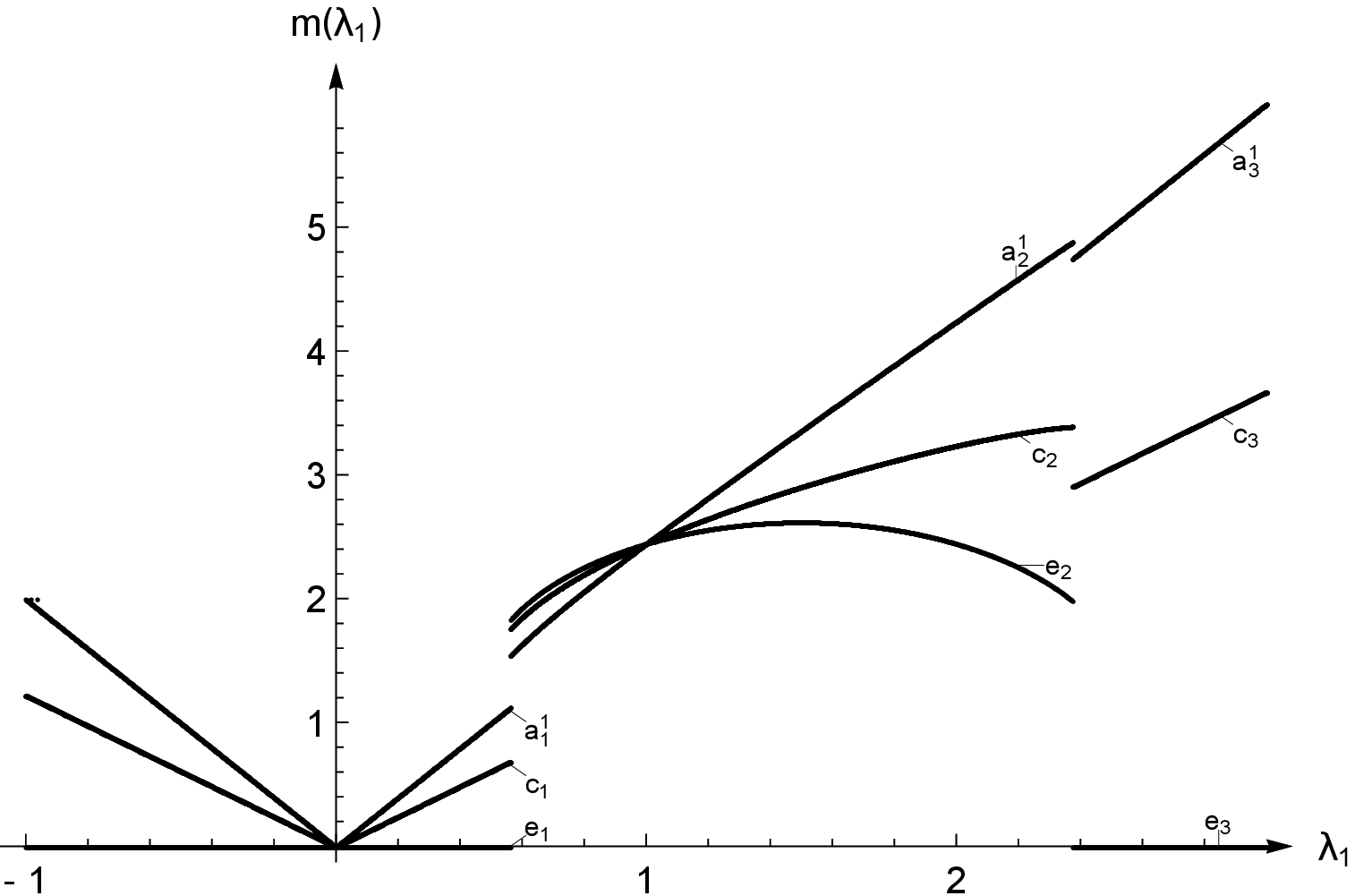}
		\label{fig-1Md}
	}
	\caption{$\algA = M_2 \to \algB = M_3$: plots for $\lambda_1 \in [-1, 3]$.}
	\label{fig-1Vd-1Md}
\end{figure}

For the first case $\algA = M_2$ and $\algB = M_3$, see Fig.~\ref{fig-1Vd-1Md}, there is only one real parameter $\lambda_1$. On the plots, this parameter is restricted to $\lambda_1 \in [-1,3]$.\footnote{A numerical exploration on the interval $[-100, 100]$ shows that the lines presented on Fig.~\ref{fig-1Md} are extended linearly.} On Fig.~\ref{fig-1Vd}, one sees that minimum values for the configurations of the $B^{1}_{\!\algB, \beta}$ fields in the Higgs potential, taking into account the fixed values of the inherited fields $B^{1}_{\!\!\algA, \kappa} = \lambda_1 E^{1}_{\!\!\algA, \kappa}$, are zero only at $\lambda_1 = 0, 1$. These values show a maximum near $\lambda_1 \simeq 0.563$ and grow rapidly outside of $\lambda_1 \in [0,1]$. 

This plot shows a similar global conformation as the one in Fig.~\ref{fig-5Vd}. 
On Fig.~\ref{fig-1Md}, the induced masses for the gauge bosons $A^{1}_{\algB, \mu}$ are presented. This mass spectrum is richer than the one in Fig.~\ref{fig-5Md}. It is not continuous, and one of its discontinuities coincides with the maximum of the values of the Higgs potential minima near $\lambda_1 \simeq 0.563$. The second discontinuity, near $\lambda_1 \simeq 2.376$, corresponds to a discontinuity in the Higgs potential minima plot that is visible at larger scale, as shown in the zoom effect circle. This mass spectrum is organized as follows: the $a^1$-lines have degeneracy $3$, the $c^1$-lines have degeneracy $4$, and the $e$-lines have degeneracy $1$, which amounts to the $8$ fields $A^{1, \beta}_{\algB, \mu}$ on $\halgB$. Notice that the $a^1_1$ and $a^1_3$ (resp. $c^1_1$ and $c^1_3$) straight lines are part of the same straight line (as shown by the dotted lines) with slope $2 = \sqrt{2 n_1}$ (resp. $\sqrt{3/2}$). Up to these small off-diagonal  values in the mass matrix, the $3$ fields $A^{1, \beta}_{\algB, \mu}$ belonging to the $a^1$-lines are the inherited $3$ fields $A^{1, \kappa}_{\!\!\algA,\mu}$ on $M_2$. The slope of the $a^1_1$ and $a^1_3$ lines shows that the inherited fields retain their masses when they are induced by $\phi : \algA \to \algB$. The $a^1_2$-line reveals that there is a slight breaking of this invariance for a specific range in $\lambda_1$. The $e$-lines correspond to the diagonal direction $\diag(1,1,-2)/\sqrt{6} \in M_3$.
\medskip
\par
For the next three cases, there are two parameters $\lambda_i$, $i=1,2$ and the plots explore the square  $(\lambda_1, \lambda_2) \in [0,1]^2$. Concerning the minimum values for the Higgs potential, all the points in the square can be displayed. But concerning the mass spectra, all the points in the square would give a cloud of points impossible to interpret. This is why we have chosen to display what happens along 7 specific lines in the square $[0,1]^2$, which are given in Fig.~\ref{fig-lambda-square}:
\begin{figure}[h]
	\centering
	\includegraphics[width=0.25\linewidth]{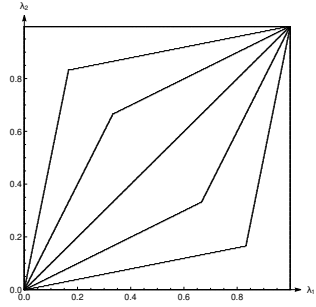}
	\caption{The 7 lines for $(\lambda_1, \lambda_1) \in [0,1]^2$ along which computations of masses have been performed.}
	\label{fig-lambda-square}
\end{figure}

Since the resulting plots in $3d$ may be quite difficult to read nevertheless, we have displayed two specific directions: the first one is the diagonal in the plane $(\lambda_1, \lambda_2)$, for $\lambda_1 = \lambda_2 \in [-1, 3]$; the second one is the anti-diagonal $\lambda_1 + \lambda_2 = 0.5$ in the square $[0,1]^2$. The diagonal plots can be directly compared to the first case, and they display a comparable rich structure: a restricted number of degenerated masses and some discontinuities. The anti-diagonal plots can be used to better understand how the inherited and new degrees of freedom behave in relation to each other (as encoded in the nomenclature for the labels). The choice of the parameter $0.5$ for the anti-diagonal line $\lambda_1 + \lambda_2 = 0.5$ is justified by the fact that we then explore in the 3 cases a region without discontinuity.
\medskip
\par
\begin{figure}[h]
	\centering
	\subfloat[Minimum values for the Higgs potential.]{%
		\includegraphics[width=0.48\linewidth]{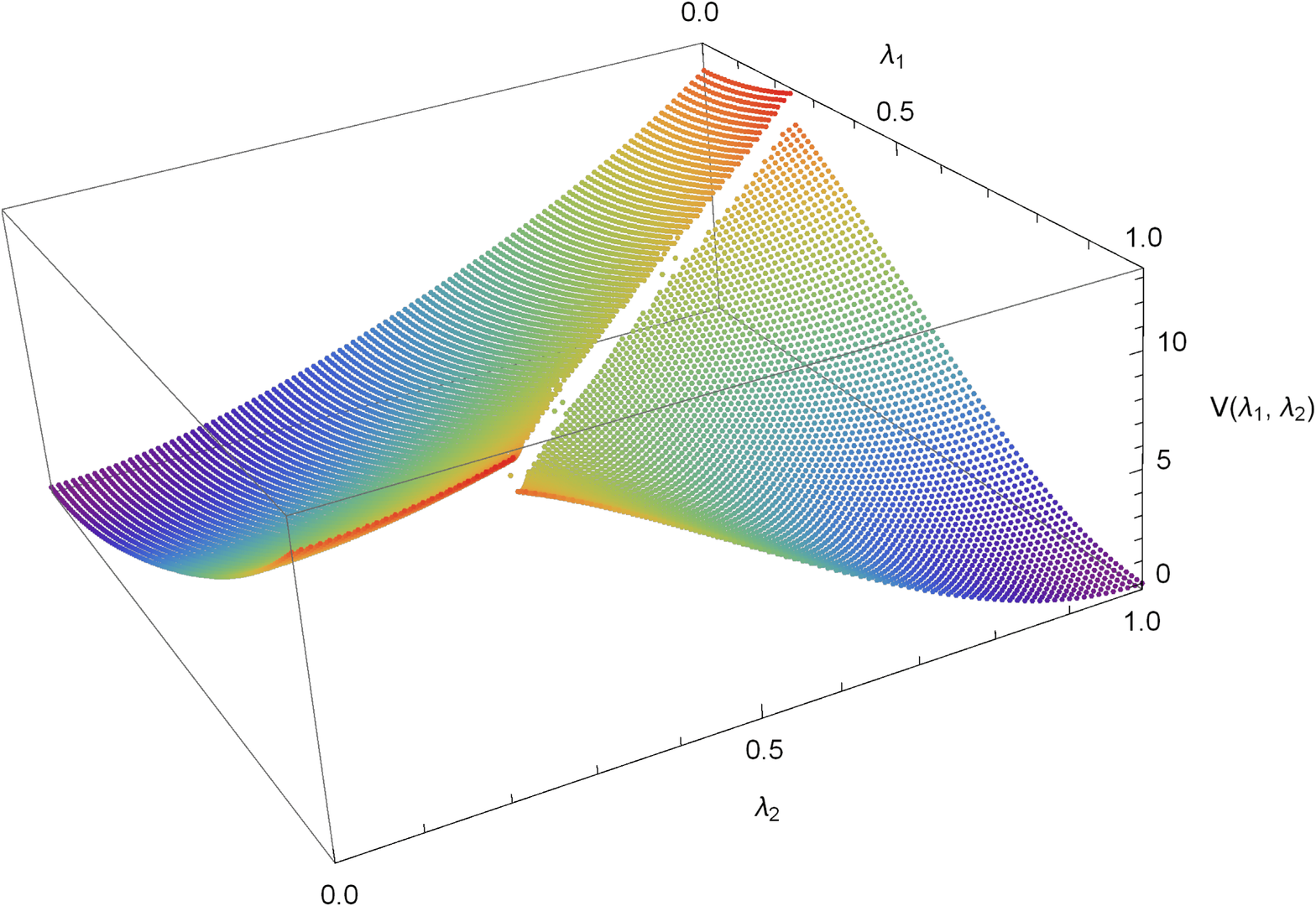}
		\label{fig-2Vn.pdf}
	}
	\subfloat[Masses for the gauge fields.]{%
		\includegraphics[width=0.48\linewidth]{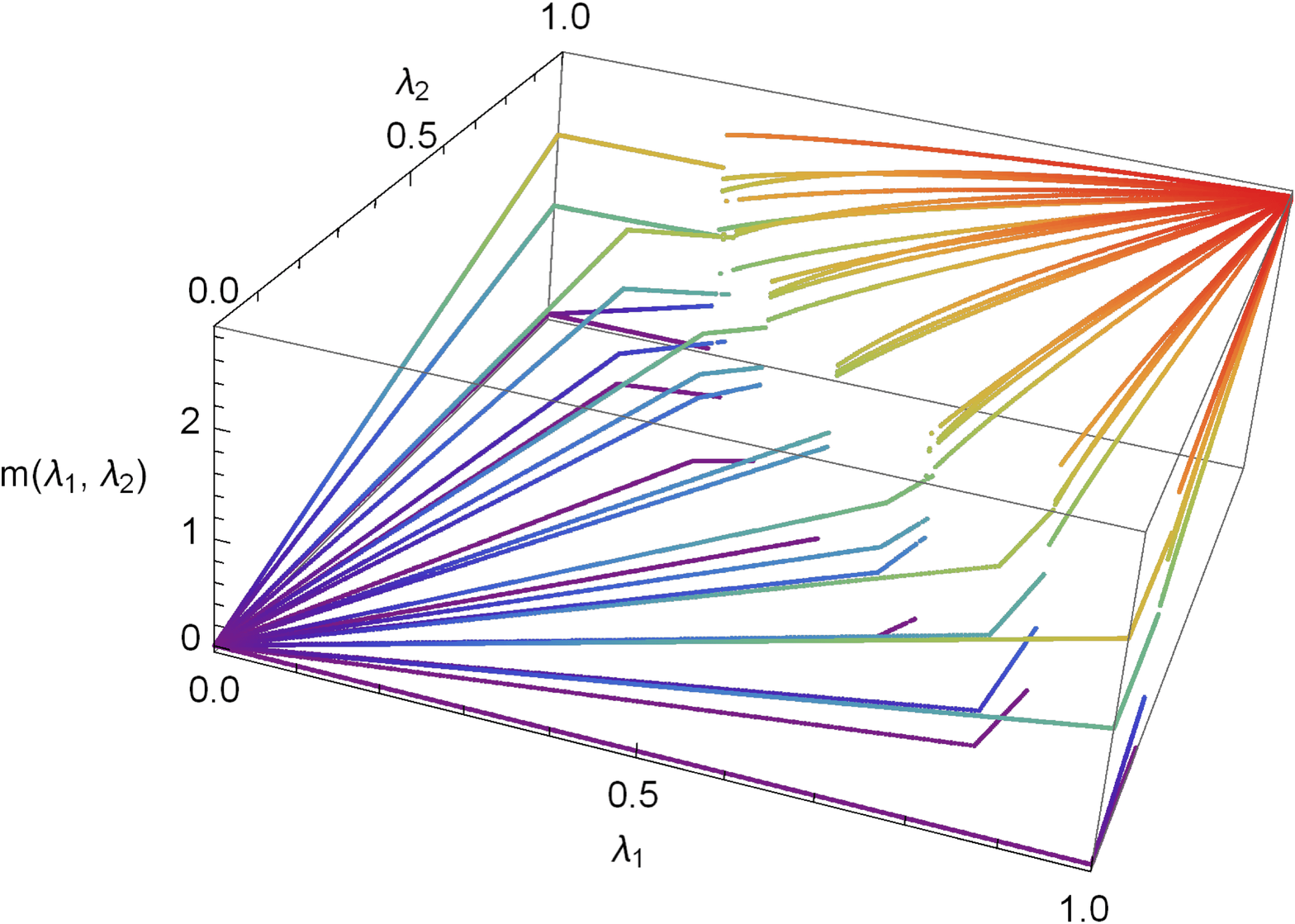}
		\label{fig-2Mlc.pdf}
	}
	\caption{$\algA = M_2 \oplus M_2 \to \algB = M_4$: plots for the square $[0,1]^2$ in the plane $(\lambda_1, \lambda_2)$.}
	\label{fig-2Vn-2Mlc}
\end{figure}
\begin{figure}[h]
	\centering
	\subfloat[Minimum values for the Higgs potential.]{%
		\includegraphics[width=0.48\linewidth]{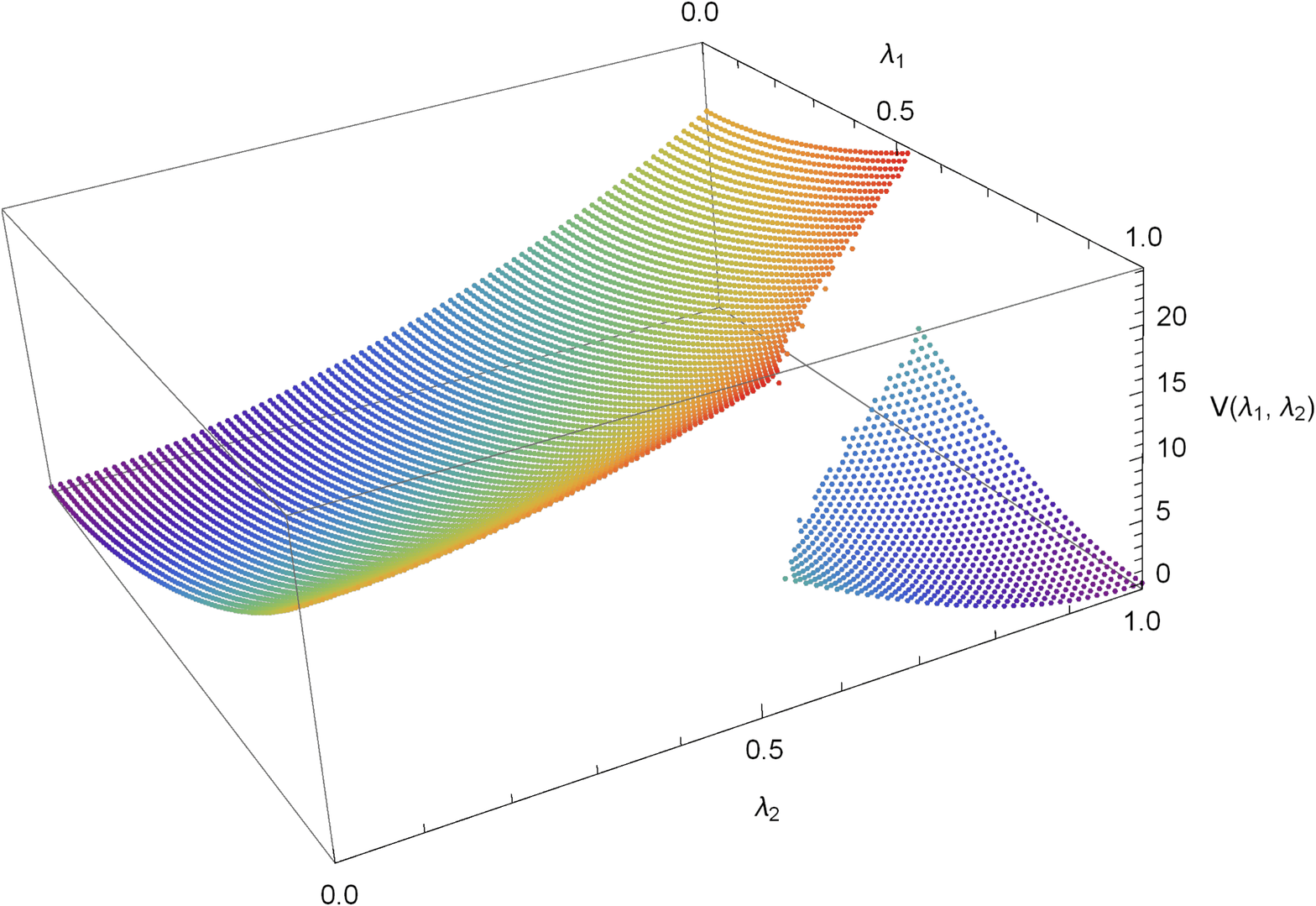}
		\label{fig-3Vn.pdf}
	}
	\subfloat[Masses for the gauge fields.]{%
		\includegraphics[width=0.48\linewidth]{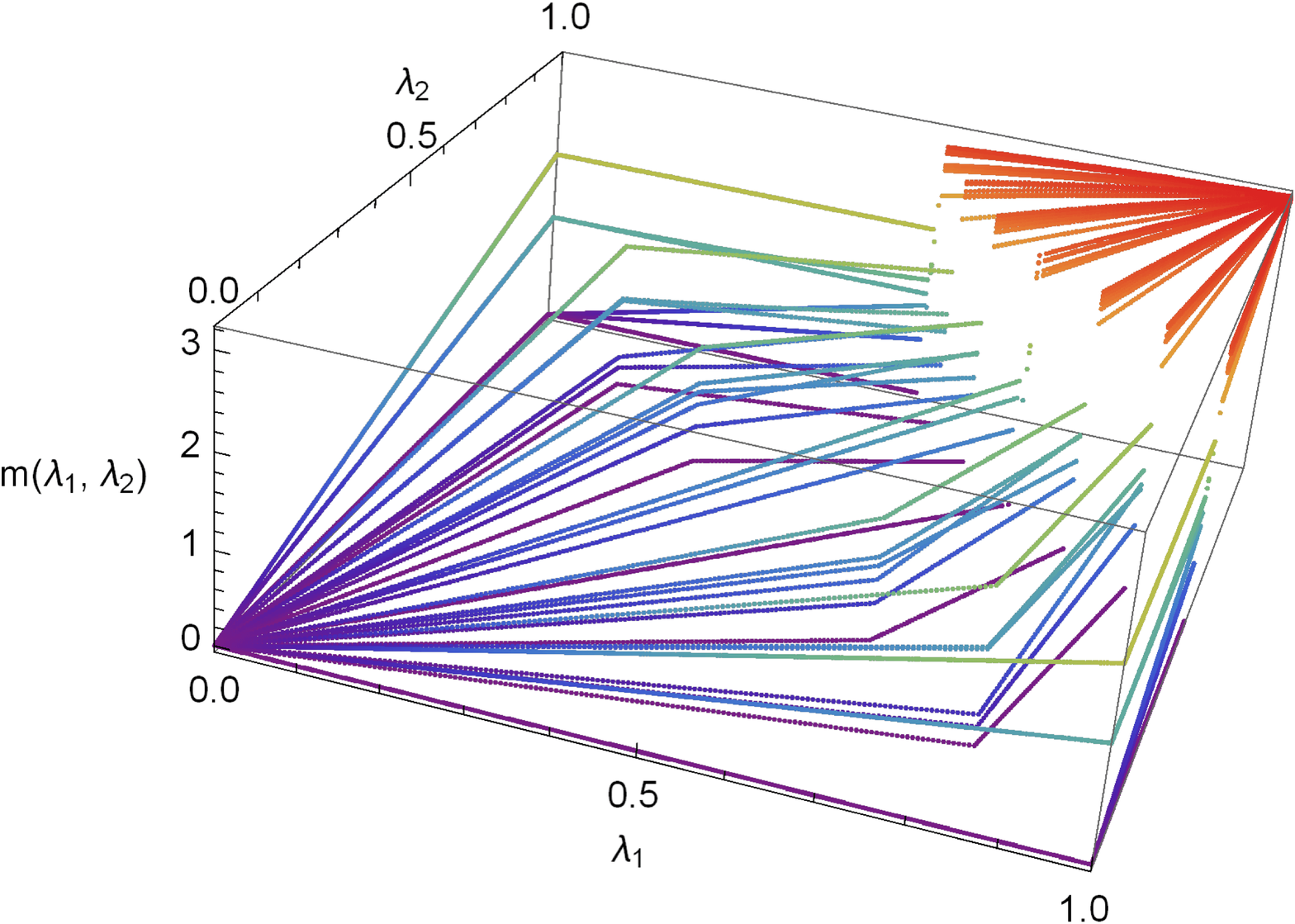}
		\label{fig-3Mlc.pdf}
	}
	\caption{$\algA = M_2 \oplus M_2 \to \algB = M_5$: plots for the square $[0,1]^2$ in the plane $(\lambda_1, \lambda_2)$.}
	\label{fig-3Vn-3Mlc}
\end{figure}
\begin{figure}[h]
	\centering
	\subfloat[Minimum values for the Higgs potential.]{%
		\includegraphics[width=0.48\linewidth]{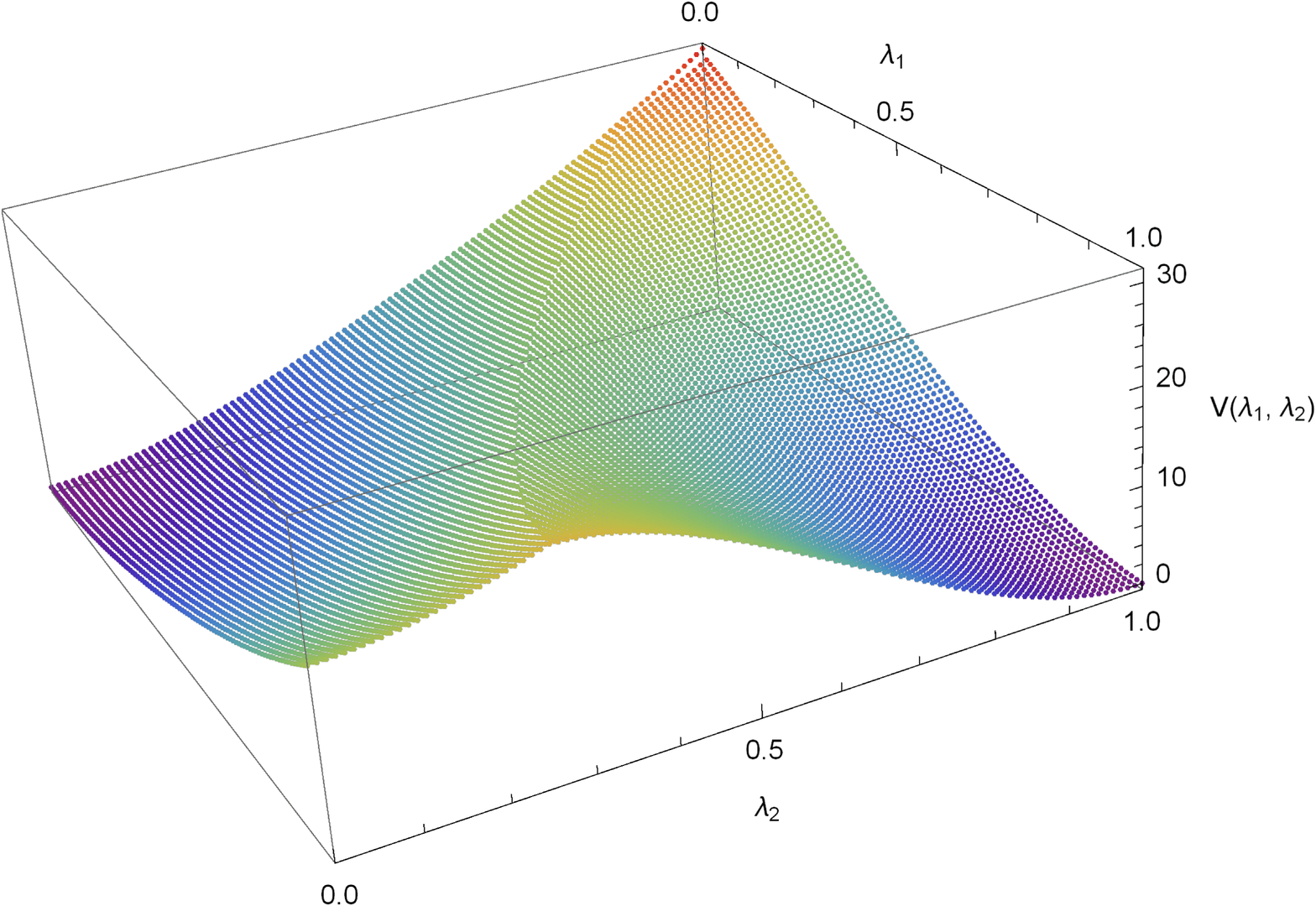}
		\label{fig-4Vn}
	}
	\subfloat[Masses for the gauge fields.]{%
		\includegraphics[width=0.48\linewidth]{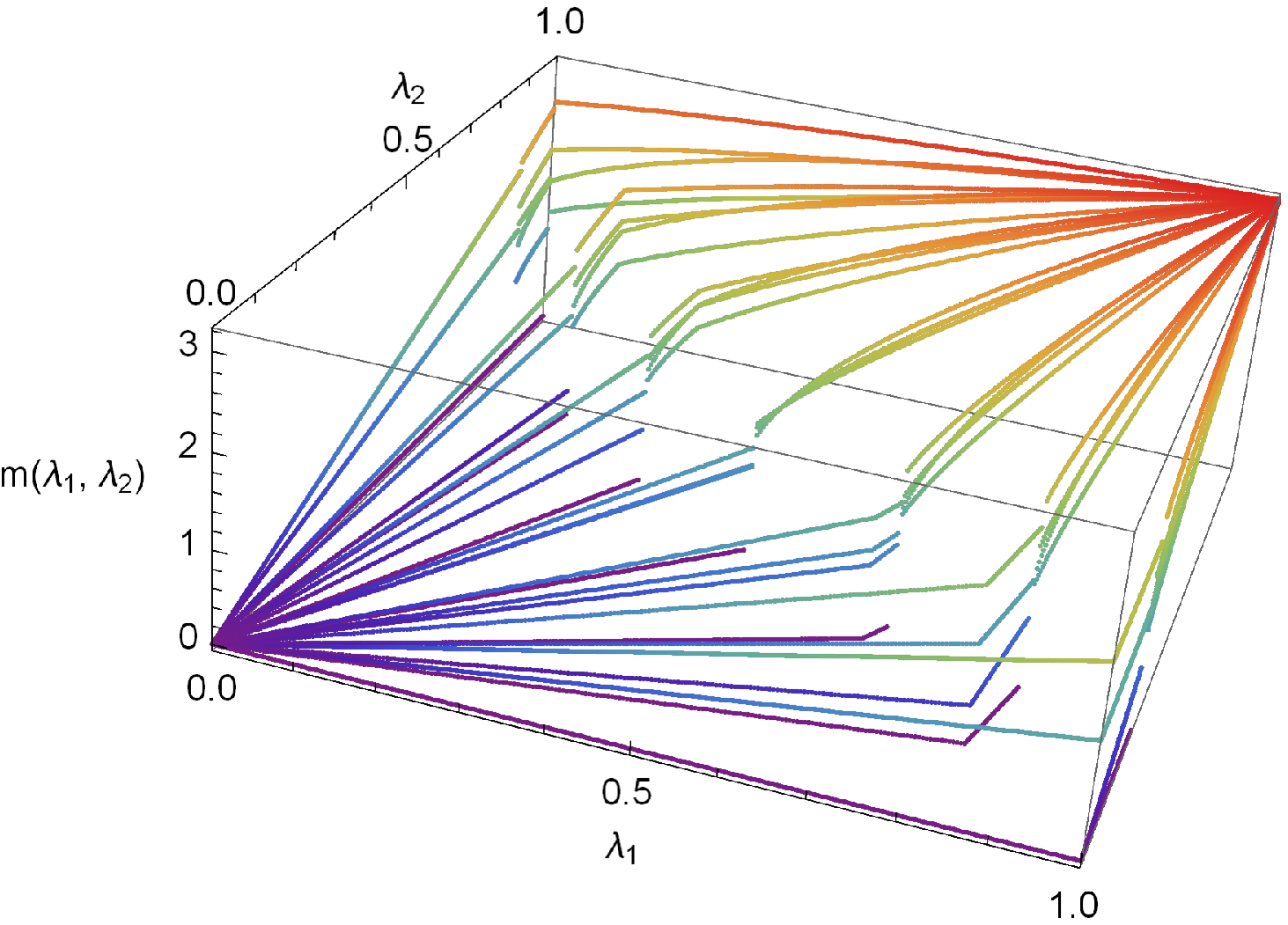}
		\label{fig-4Mlc}
	}
	\caption{$\algA = M_2 \oplus M_3 \to \algB = M_5$: plots for the square $[0,1]^2$ in the plane $(\lambda_1, \lambda_2)$.}
	\label{fig-4Vn-4Mlc}
\end{figure}
\begin{figure}[h]
	\centering
	\subfloat[Minimum values for the Higgs potential with details in insert.]{%
		\includegraphics[width=0.48\linewidth]{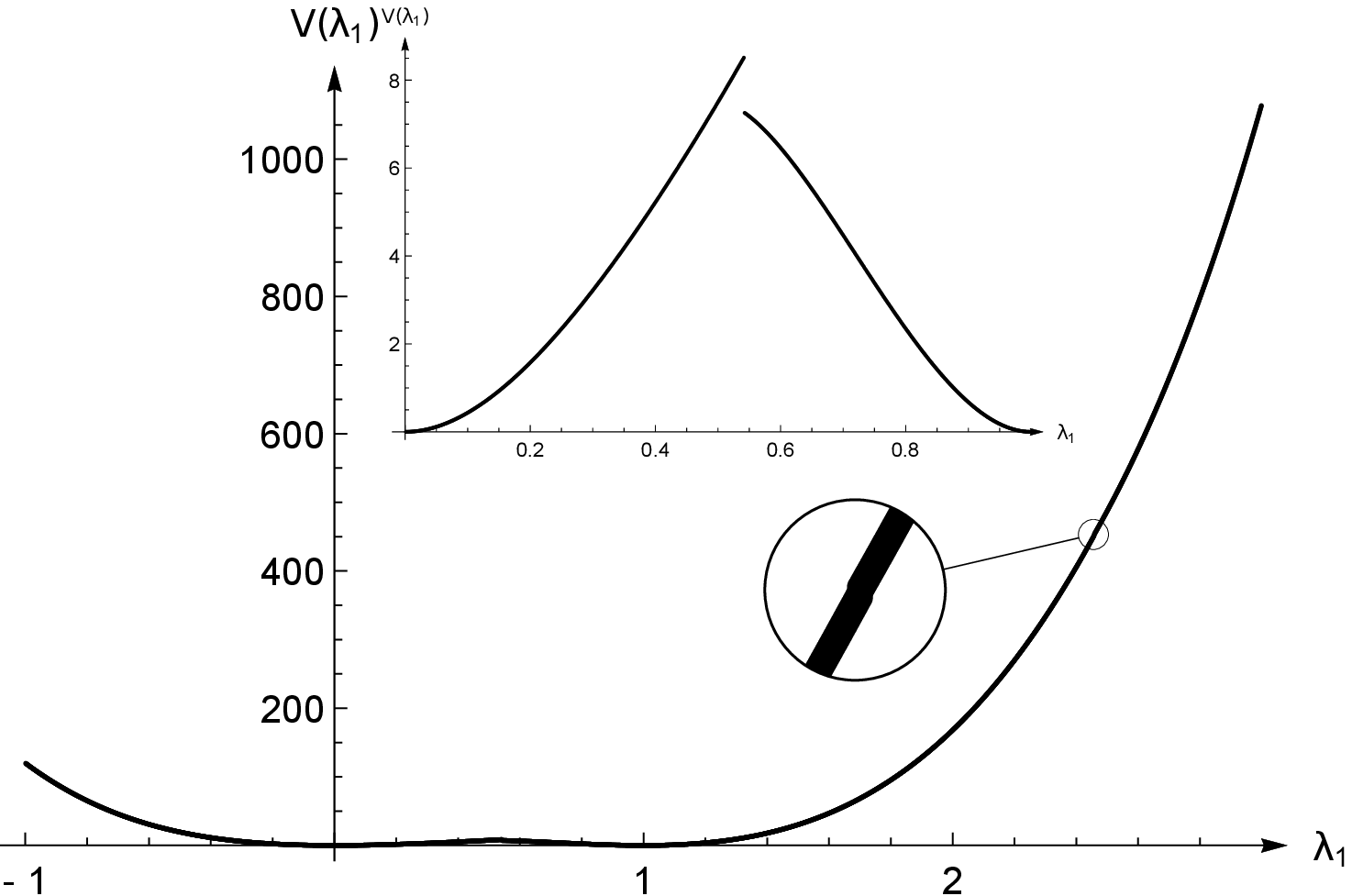}
		\label{fig-2Vd}
	}
	\subfloat[Masses for the gauge fields.]{%
		\includegraphics[width=0.48\linewidth]{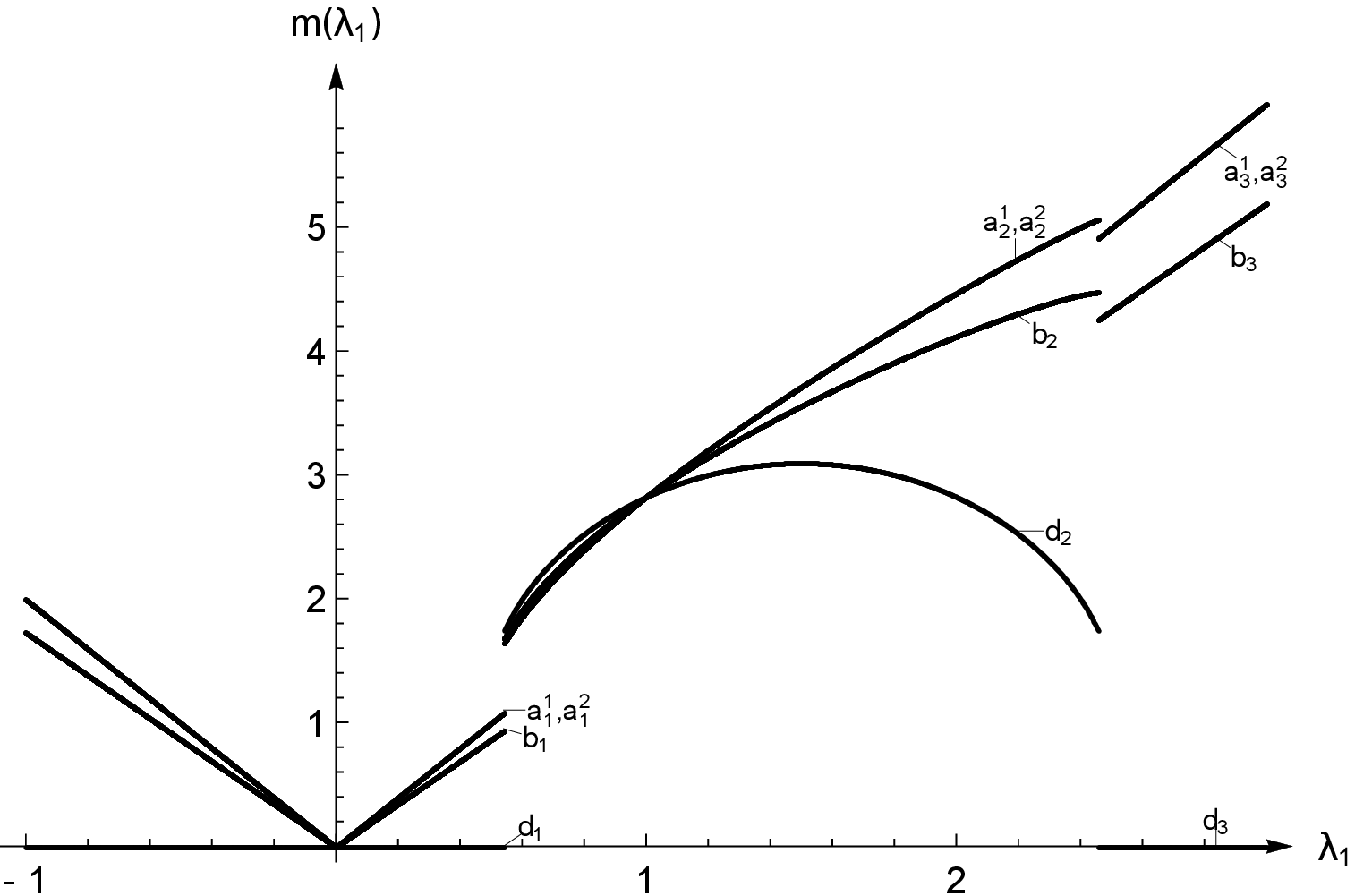}
		\label{fig-2Md}
	}
	\caption{$\algA = M_2 \oplus M_2 \to \algB = M_4$: plots on the diagonal $(\lambda_1, \lambda_1)$ for $\lambda_1 \in [-1, 3]$.}
	\label{fig-2Vd-2Md}
\end{figure}
\begin{figure}[h]
	\centering
	\subfloat[Minimum values for the Higgs potential with details in insert.]{%
		\includegraphics[width=0.48\linewidth]{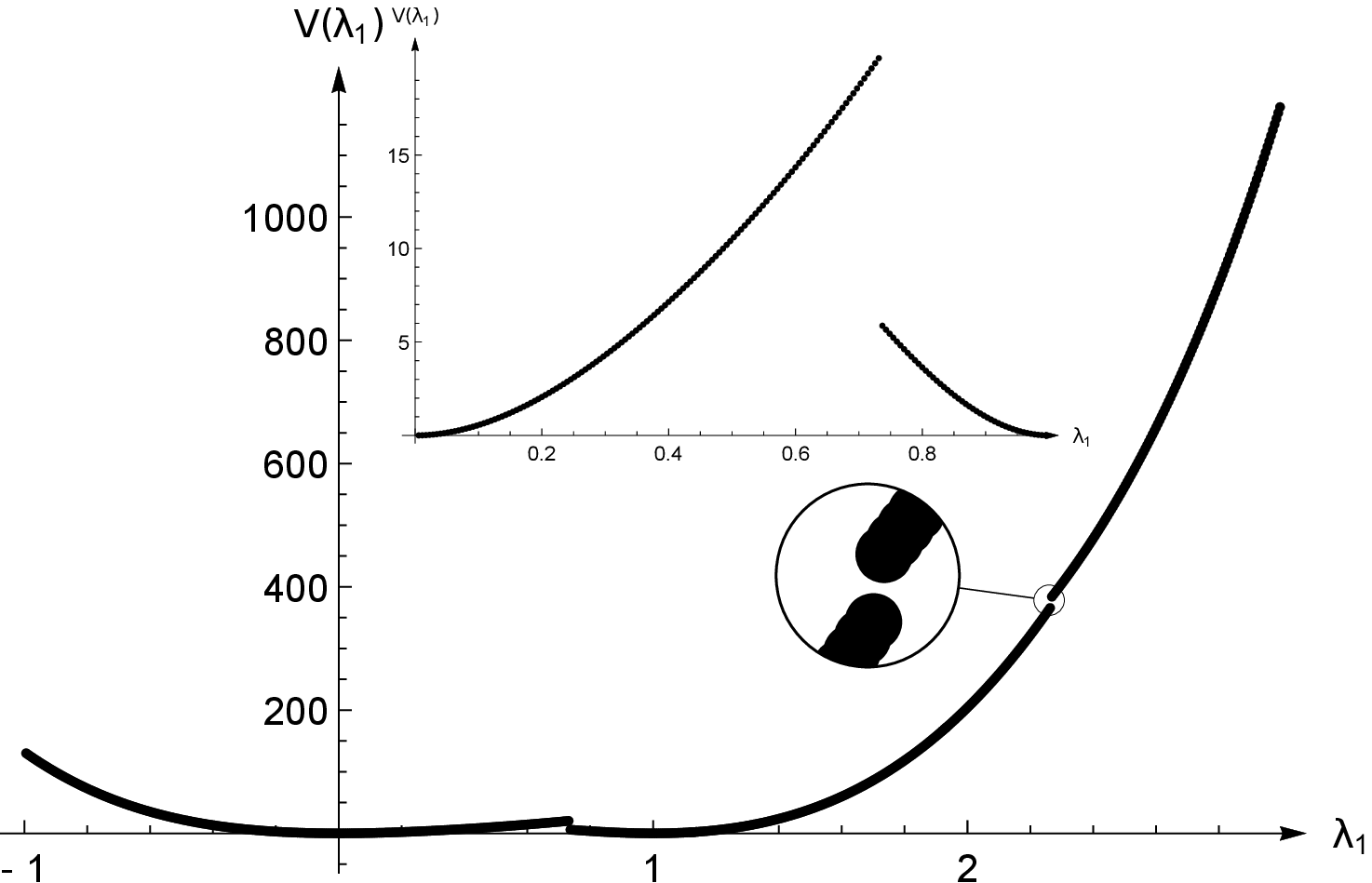}
		\label{fig-3Vd}
	}
	\subfloat[Masses for the gauge fields.]{%
		\includegraphics[width=0.48\linewidth]{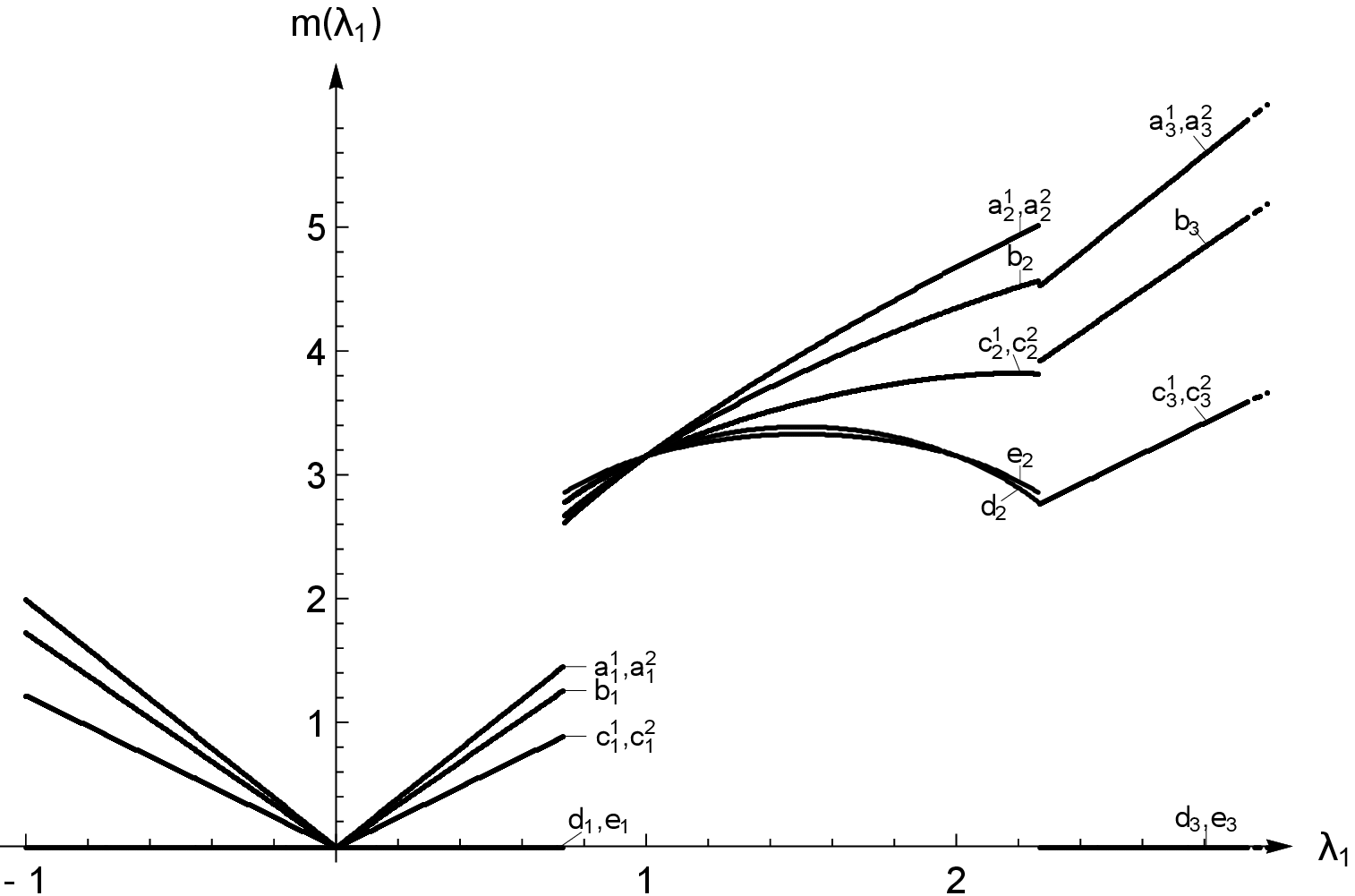}
		\label{fig-3Md}
	}
	\caption{$\algA = M_2 \oplus M_2 \to \algB = M_5$: plots on the diagonal $(\lambda_1, \lambda_1)$ for $\lambda_1 \in [-1, 3]$.}
	\label{fig-3Vd-3Md}
\end{figure}
\begin{figure}[h]
	\centering
	\subfloat[Minimum values for the Higgs potential with details in insert.]{%
		\includegraphics[width=0.48\linewidth]{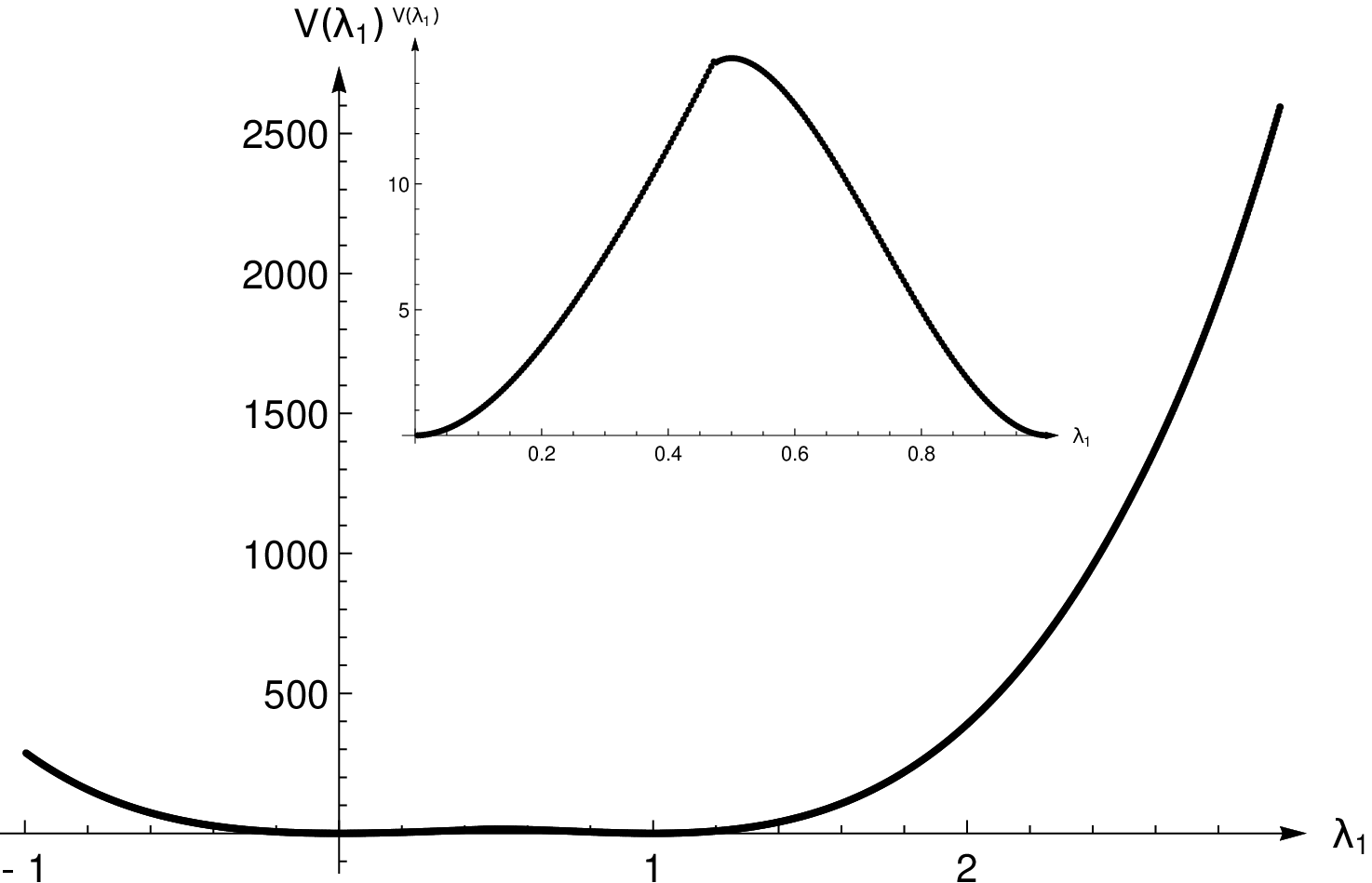}
		\label{fig-4Vd}
	}
	\subfloat[Masses for the gauge fields.]{%
		\includegraphics[width=0.48\linewidth]{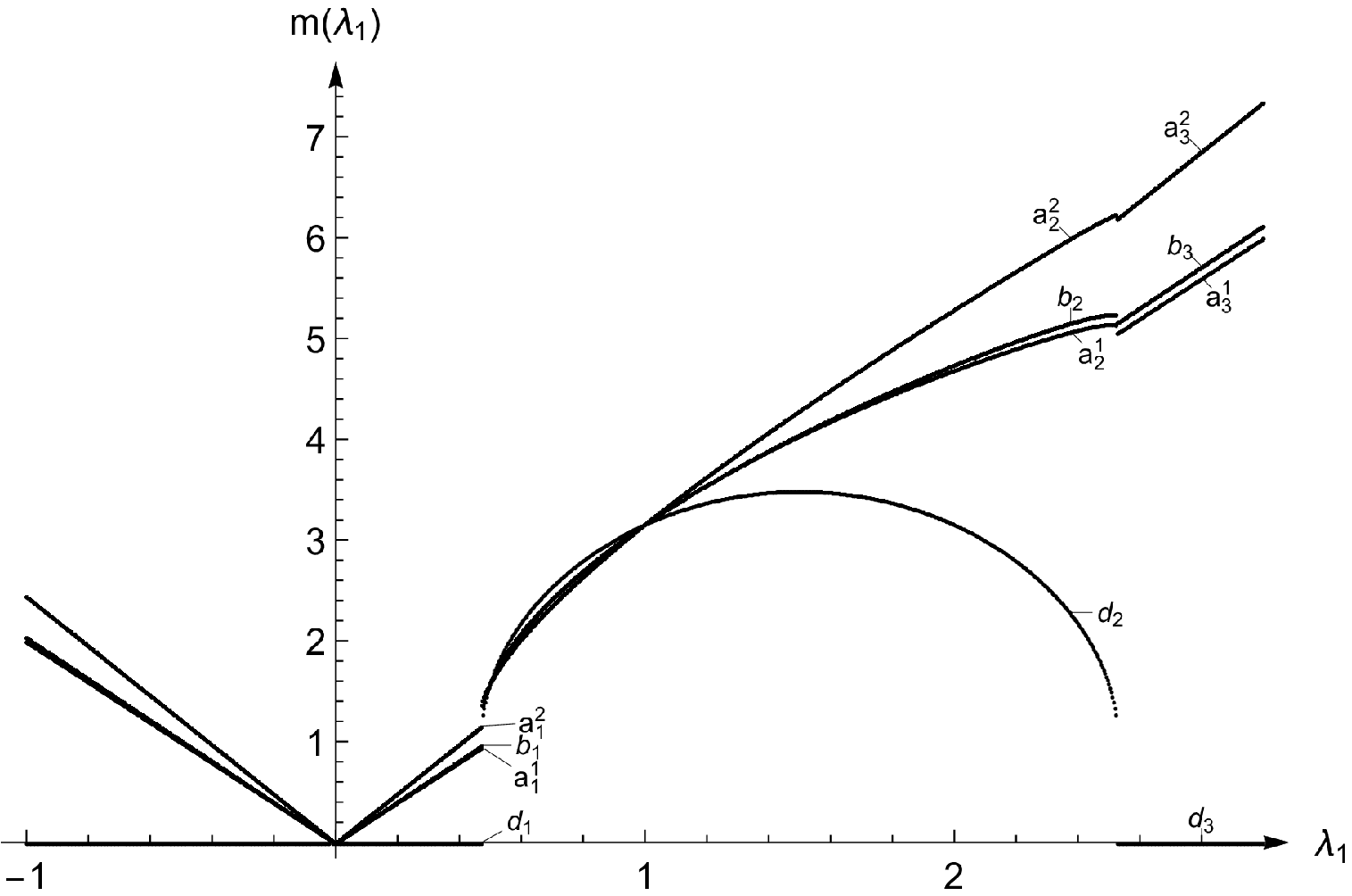}
		\label{fig-4Md}
	}
	\caption{$\algA = M_2 \oplus M_3 \to \algB = M_5$: plots on the diagonal $(\lambda_1, \lambda_1)$ for $\lambda_1 \in [-1, 3]$.}
	\label{fig-4Vd-4Md}
\end{figure}
\begin{figure}[h]
	
	\subfloat[$\algA = M_2 \oplus M_2 \to \algB = M_4$.]{%
		\includegraphics[width=0.32\linewidth]{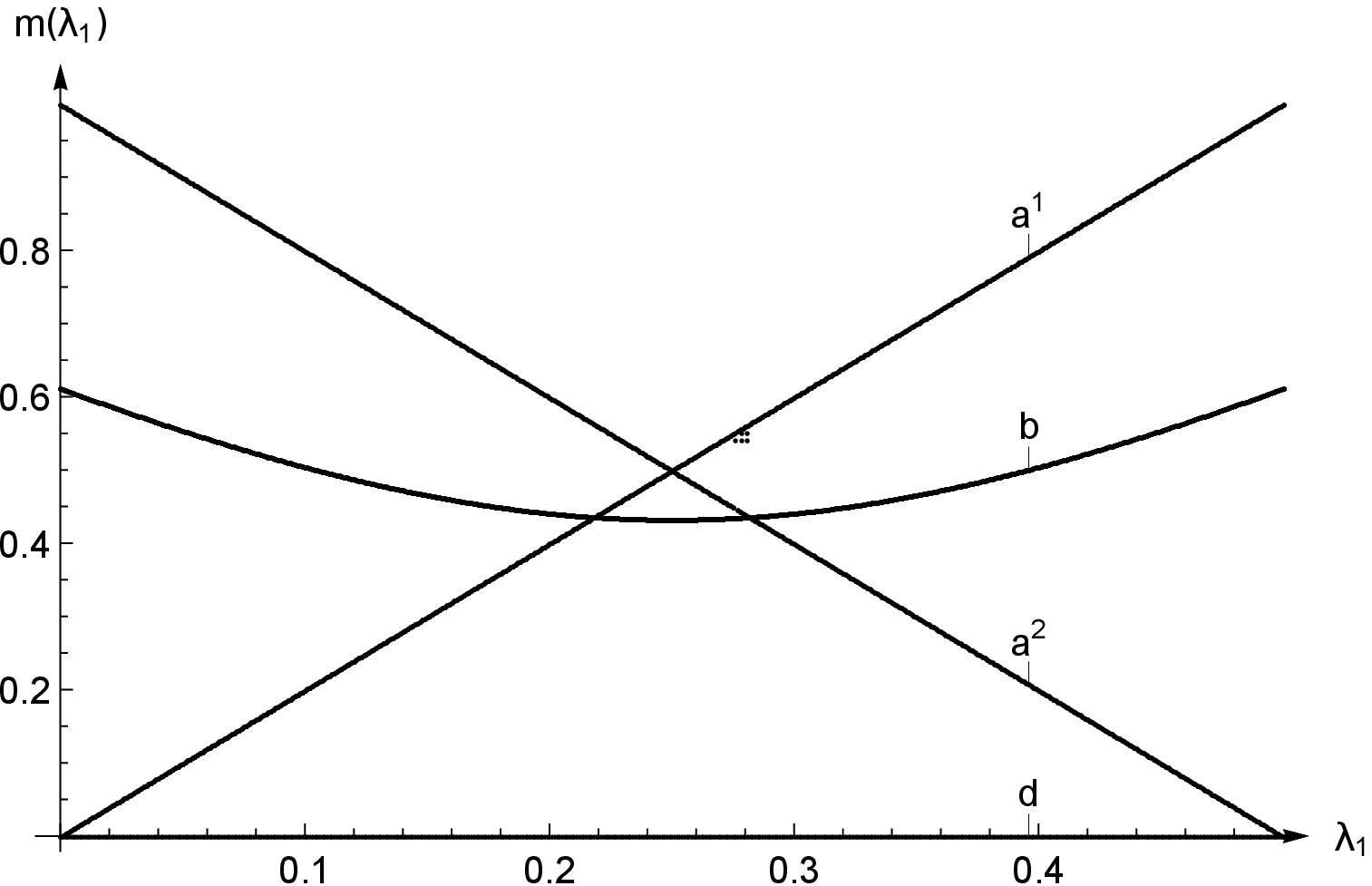}
		\label{fig-2Mad_0.5}
	}
	\subfloat[$\algA = M_2 \oplus M_2 \to \algB = M_5$.]{%
		\includegraphics[width=0.32\linewidth]{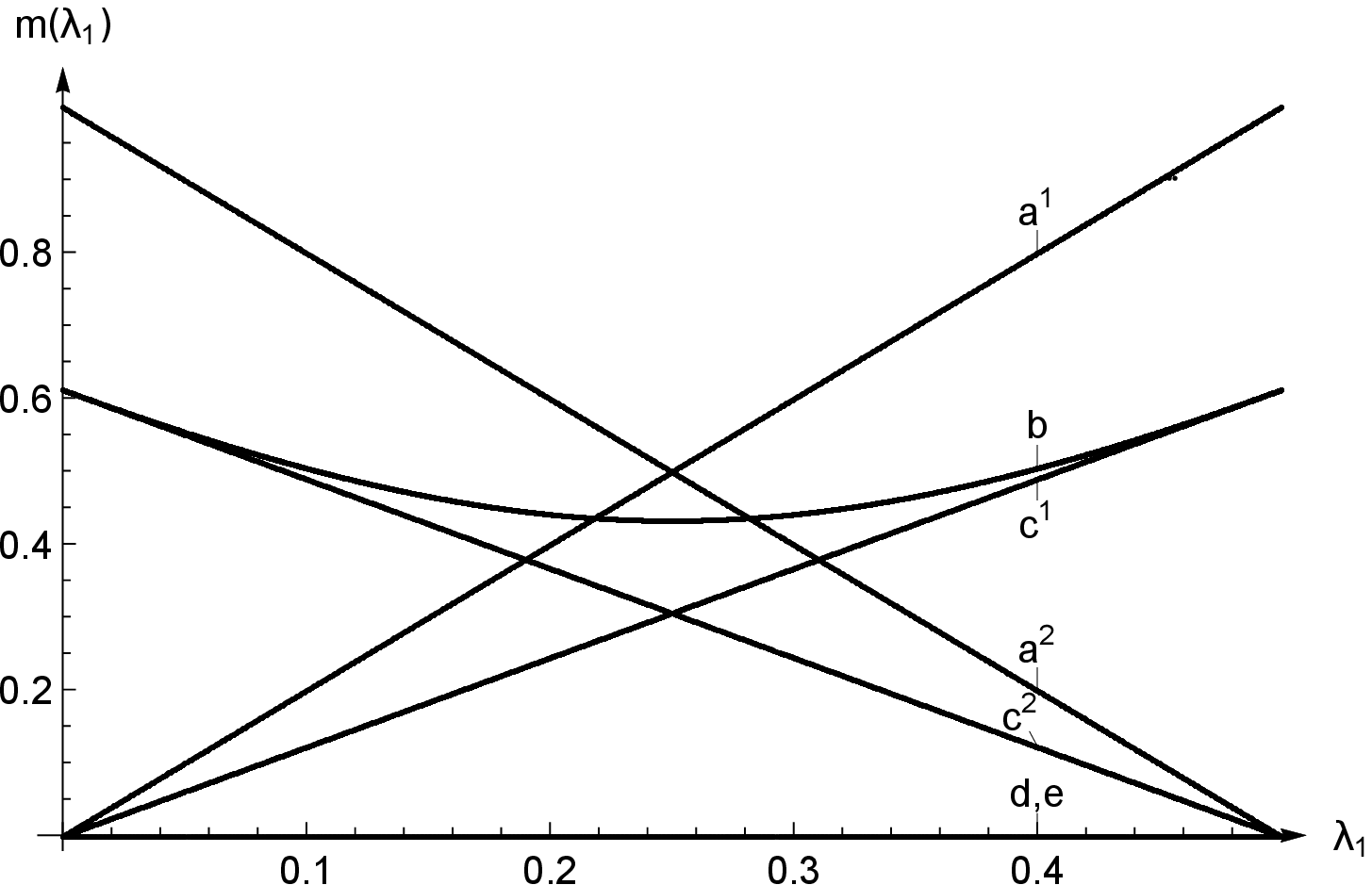}
		\label{fig-3Mad_0.5}
	}
	\subfloat[$\algA = M_2 \oplus M_3 \to \algB = M_5$.]{%
		\includegraphics[width=0.32\linewidth]{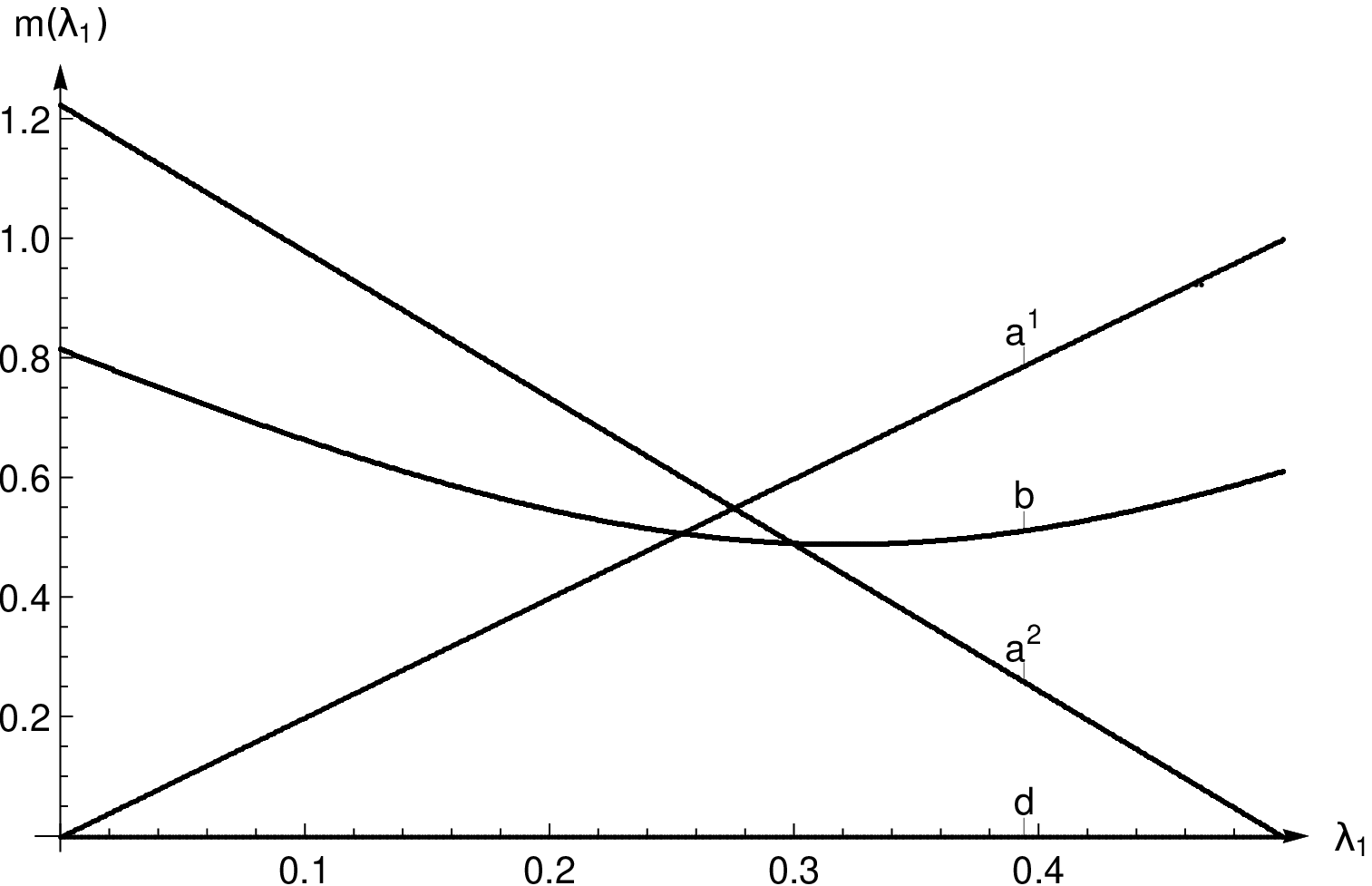}
		\label{fig-4Mad_0.5}
	}
	\caption{Masses for the gauge fields along the line $\lambda_1 + \lambda_2 = 0.5$.}
	\label{fig-2Mad-3Mad-4Mad}
\end{figure}

All results are given in the figures \ref{fig-2Vn}, \ref{fig-2Mlc}, \ref{fig-2Vn-2Mlc}, \ref{fig-3Vn}, \ref{fig-3Mlc}, \ref{fig-3Vn-3Mlc}, \ref{fig-4Vn}, \ref{fig-4Mlc}, \ref{fig-4Vn-4Mlc}, \ref{fig-2Vd}, \ref{fig-2Md}, \ref{fig-2Vd-2Md}, \ref{fig-3Vd}, \ref{fig-3Md}, \ref{fig-3Vd-3Md}, \ref{fig-4Vd}, \ref{fig-4Md}, \ref{fig-4Vd-4Md}, \ref{fig-2Mad_0.5}, \ref{fig-3Mad_0.5}, \ref{fig-4Mad_0.5}, \ref{fig-2Mad-3Mad-4Mad}.

Let us consider the second case $\algA = M_2 \oplus M_2$ and $\algB = M_4$. The minimum values for the Higgs potential in Fig.~\ref{fig-2Vn} show a line of discontinuity that have a counterpart in the mass spectrum in Fig.~\ref{fig-2Mlc}. Exploring $\lambda_1 = \lambda_2 \in [-1, 3]$ in Fig.~\ref{fig-2Vd-2Md} shows that there are other discontinuities (at least one in the range considered) as in Fig.~\ref{fig-1Vd-1Md}. In the mass spectrum in Fig.~\ref{fig-2Md}, the $a^1$ and $a^2$-lines have degeneracy $3$, the $b$-lines have degeneracy $8$, and the $d$-lines have degeneracy $1$. For $i=1,2$, the slope of the $a^i_1$ and $a^i_3$ (resp. $b_1$ and $b_3$) straight lines is $2 = \sqrt{2 n_1} = \sqrt{2 n_2}$ (resp. $\sqrt{3}$). As in the previous case, modulo very small off-diagonal values in the mass matrix, the $6 = 2 \times 3$ fields in the $a^i$-lines are inherited from the $2 \times 3$ fields $A^{1, \kappa}_{\!\!\algA,\mu}$ and $A^{2, \kappa}_{\!\!\algA,\mu}$ from the two copies of $M_2$ and the $d$-lines correspond to the diagonal direction $\diag(1,1,-1,-1)/\sqrt{4} \in M_4$. The plot in Fig.~\ref{fig-2Mad_0.5} shows how the distribution of these gauge fields changes along the anti-diagonal $\lambda_1 + \lambda_2 = 0.5$. The perfect symmetry around the diagonal at $\lambda_1 = 0.25$ in Fig.~\ref{fig-2Mad_0.5} shows that the two $M_2$ blocks play an equal role, as expected. At $\lambda_1 = 0.5$, we end up on the side $\lambda_2 = 0$ in Fig.~\ref{fig-2Mlc}, where the top line (end of the $a^1$-line) has a slope $2 = \sqrt{2 n_1}$ with respect to $\lambda_1 \in [0,1]$; the middle line has a slope $\sqrt{3/2}$ (the $b$-line); and the lower line has a slope $0$ (it corresponds to the ends of the $a^2$-line and the $e$-line). Here again, at least in the region $\lambda_1 + \lambda_2 \leq 0.5$ (before the first discontinuity), we have checked numerically that the masses of the inherited fields are preserved by the map $\phi : \algA \to \algB$.
\medskip
\par
The third case $\algA = M_2 \oplus M_2$ and $\algB = M_5$, illustrated in Figs.~\ref{fig-3Vn-3Mlc}, \ref{fig-3Vd-3Md}, and \ref{fig-3Mad_0.5}, differs from the previous one by the greater number of new degrees of freedom in $\algB$. The discontinuity in Figs.~\ref{fig-3Vn} and \ref{fig-3Mlc} is larger. Its position has also moved, as can be seen also in Fig.~\ref{fig-3Md}. In this latter plot, the $a^1$ and $a^2$-lines have degeneracy $3$, the $b$-line have degeneracy $8$, the $c^1$ and $c^2$-lines have degeneracy $4$, the $d$-lines have degeneracy $1$, and the $e$-lines have degeneracy $1$. Notice that the $d$ and $e$ lines are almost always merged in the plot, except for $d_2$ and $e_2$ which are close but clearly separated. For $i=1,2$, the slope of the $a^i_1$ and $a^i_3$ (resp. $b_1$ and $b_3$,  resp. $c^i_1$ and $c^i_3$) straight lines is $2 = \sqrt{2 n_1} = \sqrt{2 n_2}$ (resp. $\sqrt{3}$, resp. $\sqrt{3/2}$). Modulo very small off-diagonal values in the mass matrix, the $6= 2 \times 3$ fields in the $a^i$-lines are inherited from the $2 \times 3$ fields $A^{1, \kappa}_{\!\!\algA,\mu}$ and $A^{2, \kappa}_{\!\!\algA,\mu}$ from the two copies of $M_2$. In accordance with the nomenclature of the labels, the $8$ fields in the $b$-lines are new degrees of freedom along directions $E^{1}_{\!\algB, \beta}$ that are contained in $M_4 \subset M_5$, where $M_4$ contains the two copies of $M_2$. The $8= 2 \times 4$ fields in the $c^i$-lines are new degrees of freedom along directions $E^{1}_{\!\algB, \beta}$ that are defined with components outside of this $M_4 \subset M_5$: the $c^1$-line (resp. $c^2$-line) corresponds to fields in the directions $E^{1}_{\!\algB, \beta}$ with non zero entries outside of $M_4 \subset M_5$ and in the same rows and same columns as the ones in $M_{n_1}$ (resp. $M_{n_2}$). In other words, the $E^{1}_{\!\algB, \beta}$ for the $c^1$-line do not commute with $\phi(M_{n_1})$ while they commute with $\phi(M_{n_2})$, and vice versa for the $c^2$-line. The $d$-lines correspond to the diagonal direction $\diag(1,1,-1,-1,0)/\sqrt{4} \in M_5$ and the $e$-lines correspond to the diagonal direction $\diag(1,1,1,1,-4)/\sqrt{20} \in M_5$. The anti-diagonal plot in Fig.~\ref{fig-3Mad_0.5} brings us more information concerning the relationship between the $a^i$ and $c^i$-lines: it seems that there is a correlation between the $a^1$-line (resp. $a^2$-line) and the $c^1$-line (resp. $c^2$-line) due to the fact that their associated directions $E^{1}_{\!\algB, \beta}$ do not commute. This non commutativity could also explain the curved $b$-line which is “constrained” by the directions in the $a^1$ and $a^2$-lines.
\medskip
\par
Finally, the fourth case $\algA = M_2 \oplus M_3$ and $\algB = M_5$, illustrated in Figs.~\ref{fig-4Vn-4Mlc}, \ref{fig-4Vd-4Md}, and \ref{fig-4Mad_0.5}, is closer to the second case than to the third case. We conjecture that this is due to the fact that the diagonal in $\algB$ is filled by $\phi$ in the second and fourth cases, while there is a remaining $0$ in the third case (which permits the existence of the directions for the $c^1$ and $c^2$-lines in Fig.~\ref{fig-3Mad_0.5}). In the mass spectrum in Fig.~\ref{fig-4Md}, the $a^1$-lines have degeneracy $3$, the $a^2$-lines have degeneracy $8$, the $b$-lines have degeneracy $12$, and the $d$-lines have degeneracy $1$. The slope of the $a^1_1$ and $a^1_3$ (resp. $a^2_1$ and $a^2_3$,  resp. $b_1$ and $b_3$) straight lines is $2 = \sqrt{2 n_1}$ (resp. $\sqrt{6} = \sqrt{2 n_2}$, resp. $\sqrt{25/6}$). Modulo very small off-diagonal values in the mass matrix, the $3$ fields in the $a^1$-lines are inherited from the $3$ fields $A^{1, \kappa}_{\!\!\algA,\mu}$ from $M_{n_1} = M_2$ and the $8$ fields in the $a^2$-lines are inherited from the $8$ fields $A^{2, \kappa}_{\!\!\algA,\mu}$ from $M_{n_2} = M_3$. The $d$-lines correspond to the diagonal direction $\diag(1,1,1,1,-4)/\sqrt{20} \in M_5$. As showed in Fig.~\ref{fig-4Mad_0.5}, the mass spectrum along the anti-diagonal is no more symmetric, as can also be seen in Fig.~\ref{fig-4Mlc} (look for instance at the singular line in the mass spectrum): this distinguishes this case from the second one and illustrates how a change in the algebra $\algA$ affects the mass spectrum.
\medskip
\par
Let us make comments on these results. The exploration of the space of configurations for the fields $B^{i, \kappa'}_{\!\!\algA, \kappa}$'s along paths parametrized by the $\lambda_i$'s already shows a rich typology concerning the possible masses for the gauge bosons  $A^{k, \beta}_{\algB,\mu}$.
\medskip
\par
As seen in Fig.~\ref{fig-1Vd-1Md} for instance, the minimum for a conflictual situation $\lambda_1 = 1$ and $\lambda_2 = 0$ (conflict between the two minimal configurations for the $B^{1}_{\!\!\algA, \kappa}$ and $B^{2}_{\!\!\algA, \kappa}$ in $M_2$) is non zero and produces a global configuration for the fields $B^{1}_{\!\algB, \beta}$ that is neither the null-configuration nor the basis-configuration. The induced masses show 3 possible values with degeneracies. Inserting this configuration as a initial data for another step into a sequence of NCGFT constructed on the sequence $\{ (\algA_n, \phi_{n,m}) \, / \,  0 \leq n < m \}$, may propagate this in-between result and produce more subtle configurations with richer possibilities for the masses of the gauge bosons.
\medskip
\par
Since the exploration of the space of configurations for the fields $B^{i, \kappa'}_{\!\!\algA, \kappa}$'s is reduced to paths parametrized by the $\lambda_i$'s, our results do not offer a general and systematic view of what could happen in our kind of models. Nevertheless, the results presented above already display a rich phenomenology from which some information can be drawn.  The first noticeable feature is that the mass spectra, constrained by the $\phi$-compatibility, reveal that the masses are grouped in specific directions, so that we have neither a full degeneracy (as in Fig.~\ref{fig-5Md}) nor a complete list of independent masses (as many masses as degrees of freedom): these specific directions are grouped according to the inherited degrees of freedom (the $a^i$-lines), according to the way the new degrees of freedom commute or not with the inherited ones (the $b$ and $c^i$-lines), and according to the possible new diagonal degrees of freedom one can introduce (the $d$ and $e$-lines). Masses for inherited gauge bosons are preserved by the $\phi$-compatibility condition quite systematically near the null-configuration. Concerning the first discontinuity on the diagonal plots before the basis-configuration, the position of this discontinuity seems to be related, by an approximate linear relationship, to the ratio of the number of new degrees of freedom over the number of inherited degrees of freedom, see Table~\ref{table discontinuity ratio}:

\newcommand{\centercell}[1]{\multicolumn{1}{c}{#1}}
\newcolumntype{d}[1]{D{.}{.}{#1}}
\begin{table}[h]
	\centering
	\begin{tabular}{rd{2.0}d{2.0}d{1.3}d{1.3}d{1.3}}
		\toprule
		\centercell{Case} & \centercell{$n_{\text{ndof}}$} & \centercell{$n_{\text{idof}}$} & \centercell{$r_{\text{dof}}$} & \centercell{$\lambda_{1, \text{first}}$} & \centercell{$\lambda_{1, \text{second}}$}\\
		\midrule
		$M_2 \oplus M_3 \to M_5$ & 13 & 11 & 1.182 & 0.475 & 2.526 \\
		$M_2 \oplus M_2 \to M_4$ & 9   & 6    & 1.5      & 0.542 & 2.456 \\
		$M_2 \to M_3$                         & 5   & 3    & 1.667 & 0.563 & 2.376 \\
		$M_2 \oplus M_2 \to M_5$ & 18 & 6   & 3          & 0.734 & 2.263 \\
		\bottomrule
	\end{tabular}
	\caption{Relationship between the positions of the first and second discontinuities and the ratio of the number of new degrees of freedom over the number of inherited degrees of freedom in the diagonal plots Figs.~\ref{fig-4Md}, \ref{fig-2Md}, \ref{fig-1Md}, and \ref{fig-3Md}: $n_{\text{ndof}}$ (resp. $n_{\text{idof}}$) is the number of new (resp. inherited) degrees of freedom, $r_{\text{dof}} = n_{\text{ndof}}/n_{\text{idof}}$ is the ratio of these degrees of freedom, $\lambda_{1, \text{first}}$ (resp. $\lambda_{1, \text{second}}$) is the value of $\lambda_1$ at the first (resp. second) discontinuity.}
	\label{table discontinuity ratio}
\end{table}

\chapter{\texorpdfstring{Spectral Triple-based Approach to NCGFT on $AF$-Algebras}{Spectral triple approach to NCGFT on AF algebras}}

\label{sec STA}
In this chapter, we propose to set up NCGFT based on $AF$-Algebras using the spectral triple-based approach. In subsection \ref{sec AF algebras}, we will see how to lift arrows of a Bratteli diagram to arrows between Krajewski diagrams. In subsection \ref{subsec normalized phiH map} we will create the “$\phi$-normalized” map for physical purposes. In subsection \ref{subsec comparison of spectral actions}, the spectral actions defining NCGFTs associated with two spectral triples related by the arrows of the Bratteli diagram are compared (tensored by a commutative spectral triple to put us in the context of Almost Commutative manifolds). This is done by using the $\phi$-normalized map and working with $\phi$-compatible operators at the level of the two successive spectral triples of the elementary step in the inductive sequence. This chapter is an account of the results given in  \cite{masson2022lifting}. Other works consisting in the development of spectral triples and non-commutative gauge theories on AF algebras in a different way can be found in \cite{christensen2006spectral,marcolli2014gauge}.

\section{Lifting one Step of the Defining Inductive Sequence}
\label{sec one step in the sequence ST}

In this section, we adapt to spectral triples the inclusion map $\phi : \algA \to \algB$. As explained in Chapt.~\ref{sec AFA}, the main idea, which is central in this thesis, is to define a notion of $\phi$-compatibility for the structures defining spectral triples $(\algA, \hsA, \DA, \JA, \gammaA)$ and $(\algB, \hsB, \DB, \JB, \gammaB)$ on top of $\algA$ and $\algB$. This construction, applied in Sect.~\ref{sec AFA} to $AF$-algebras, can be interpreted as a lift of arrows in a Bratteli diagram to arrows between Krajewski diagrams.

\subsection{General Situations}
\label{sec general situations}
In this subsection, we will explore the consequences of $\phi$-compatibility and strong $\phi$-compatibility conditions on the structure of spectral triples. 

\medskip
\par
\begin{definition}[$\phi$-compatibility of spectral triples]
	\label{def phi compatibility of spectral triples}
	Assume given a $\phi$-compatible map $\phiH : \hsA \to \hsB$.
	\medskip
	\par
	Two odd spectral triples $(\algA, \hsA, \DA)$ and $(\algB, \hsB, \DB)$ are said to be $\phi$-compatible if $\DA$ is $\phi$-compatible with $\DB$.
	\medskip
	\par
	Two real spectral triples $(\algA, \hsA, \DA, \JA)$ and $(\algB, \hsB, \DB, \JB)$ are said to be $\phi$-compatible if $\DA$ (resp. $\JA$) is $\phi$-compatible with $\DB$ (resp. $\JB$).
	\medskip
	\par
	In the even case for $\algA$, one requires that $\algB$ is also even and that the grading operators $\gammaA$ and $\gammaB$ are $\phi$-compatible.
\end{definition}

Strong $\phi$-compatibility of spectral triples can be defined in an obvious way.

\begin{remark}
	Notice that strong $\phi$-compatibility of spectral triples is similar to the condition (3) given in \cite[Def~2.1]{FlorGhor19p} where their couple $(\phi, I)$ corresponds to our couple $(\phi, \phiH)$. We depart from this paper where  inductive sequences of spectral triples are studied in the following way: we will restrict our analysis to the algebraic part of spectral triples since only $AF$-algebras will be considered later, so that the analytic part is quite trivial in our situation, and we will focus on gauge fields theories defined on top of spectral triples. For instance, conditions like (ST1) (about the $*$-subalgebra $\algA^\infty$) and (ST2) (about the compactness of the resolvent of the Dirac operator) in  \cite{FlorGhor19p} will not be considered here. Moreover, since we are interested in accumulating “new degrees of freedom” along the inductive limit, the $\phi$-compatibility condition will be more effective than the strong $\phi$-compatibility condition.
\end{remark}

Since $\JA$ and $\JB$ define $\algA^{e} = \algA \otimes \algA^\circ$ and $\algB^{e} = \algB \otimes \algB^\circ$ modules structures on $\hsA$ and $\hsB$, it is convenient to express $\phi$-compatibility in terms of this structure. The morphism $\phi$ defines a canonical morphism of algebras $\phi^\circ : \algA^\circ \to \algB^\circ$ by the relation $\phi^\circ(a^\circ) \defeq \phi(a)^\circ$. We then define $\phi^{e} : \algA^{e} \to \algB^{e}$ as $\phi^{e} \defeq \phi \otimes \phi^\circ$, \textit{\textit{i.e.}} $\phi^{e}(a_1 \otimes a_2^\circ) = \phi(a_1) \otimes \phi^\circ(a_2^\circ)$. Let $\modM$ (resp. $\modN$) be a $\algA$-bimodule (resp. $\algB$-bimodule), which is also a $\algA^{e}$-left module ( resp. $\algB^{e}$-left module) by $(a_1 \otimes a_2^\circ) e \defeq a_1 e a_2$ for any $e \in \modM$ and $a_1, a_2 \in \algA$ (and similar relations for $\algB$ and $\modN$). Then, we say that a linear map between the two bimodules $\phiMod : \modM \to \modN$ is $\phi$-compatible if it is $\phi^{e}$-compatible between the two left modules, that is $\phiMod( (a_1 \otimes a_2^\circ) e) = \phi^{e}(a_1 \otimes a_2^\circ) \phiMod(e)$, which is equivalent to $\phiMod( a_1 e  a_2) = \phi(a_1) \phiMod(e) \phi(a_2)$.

\begin{lemma}
	Suppose that $\phiH : \hsA \to \hsB$ is $\phi$-compatible as a map of left modules and that $\JA$ and $\JB$ are strong $\phi$-compatible. Then $\phiH$ is $\phi^{e}$-compatible as a map between the bimodules defined by the real operators. 
\end{lemma}

\begin{proof}
	For any $\psi \in \hsA$, $a_1, a_2 \in \algA$, by definition, one has $a_1 \psi a_2 = (a_1 \otimes a_2^\circ) \psi = a_1 \JA a_2^\ast \JA \psi$. On the one hand, since $\phiH$ is $\phi$-compatible, one has $\phiH(a_1 \psi) = \phi(a_1) \phiH(\psi)$. On the other hand, $\phiH(\psi a_2) = \phiH( \JA a_2^\ast \JA \psi) = \JB \phi(a_2)^\ast \JB \phiH(\psi) = \phiH(\psi) \phi(a_2)$.
\end{proof}

\begin{lemma}
	\label{lemma JB JA st-phi-comp}
	Suppose that $\JB$ is strong $\phi$-compatible with $\JA$:
	\begin{enumerate}
		\item $\epsilonA = \epsilonB$.
		
		\item $\JB^{-1}$ is strong $\phi$-compatible with $\JA^{-1}$
		
		\item $\JB$ is diagonal in its matrix decomposition.
		
		\item If two operators $A$ on $\hsA$ and $B$ on $\hsB$ are $\phi$-compatible, then the operators $\JA A \JA^{-1}$ and $\JB B \JB^{-1}$ are $\phi$-compatible.
	\end{enumerate}
\end{lemma}

\begin{proof}
	From $\JA^2 = \epsilonA$ and $\JB^2 = \epsilonB$, one gets $\epsilonA \phiH(\psi) = \phiH(\JA^2 \psi) = \JB^2 \phiH(\psi) = \epsilonB \phiH(\psi)$ for any $\psi \in \hsA$, so that $\epsilonB = \epsilonA$. From this we deduce that $\JB^{-1} = \epsilonB \JB$ is  strong $\phi$-compatible with $\JA^{-1} = \epsilonA \JA$.
	\medskip
	\par
	Let $\JB = \smallpmatrix{ \JB[,\phi]^{\phi} & \JB[,\phi]^{\perp} \\ \JB[,\perp]^{\phi} & \JB[,\perp]^{\perp} }$. Since $\JB$ is strong $\phi$-compatible with $\JA$, we already know that $\JB[,\perp]^{\phi} = 0$. Let $\psi_\algB \in \phiH(\hsB)$ and $\psi'_\algB \in \phiH(\hsA)^{\perp}$. Then $\JB (\psi_\algB) = \smallpmatrix{ \JB[,\phi]^{\phi}(\psi_\algB) \\ 0}$ and $\JB(\psi'_\algB) = \smallpmatrix{ \JB[,\phi]^{\perp}(\psi'_\algB) \\ \JB[,\perp]^{\perp}(\psi'_\algB) }$, so that $0 = \langle \psi'_\algB, \psi_\algB \rangle_{\hsB} = \langle \JB(\psi_\algB), \JB(\psi'_\algB) \rangle_{\hsB} = \langle \JB[,\phi]^{\phi}(\psi_\algB), \JB[,\phi]^{\perp}(\psi'_\algB) \rangle_{\hsB}$. From $\JB^{-1} = \epsilonB \JB$ and $\JB[,\perp]^{\phi} = 0$, one gets that  $\JB[,\phi]^{\phi}$ is invertible with $(\JB[,\phi]^{\phi})^{-1} = (\JB^{-1})_{\phi}^{\phi} = \epsilonB \JB[,\phi]^{\phi}$, so that $\JB[,\phi]^{\phi}(\phiH(\hsA)) = \phiH(\hsA)$, which implies that $\JB[,\phi]^{\perp}(\psi'_\algB) \in \phiH(\hsA)^{\perp}$, that is, $\JB[,\phi]^{\perp}(\psi'_\algB) = 0$ for any $\psi'_\algB \in \phiH(\hsA)^{\perp}$, and so $\JB[,\phi]^{\perp} = 0$.
	\medskip
	\par
	From $(\JB B \JB^{-1})_{\phi}^{\phi} = \JB[,\phi]^{\phi} B_{\phi}^{\phi} (\JB[,\phi]^{\phi})^{-1}$, we deduce that the operators $\JA A \JA^{-1}$ and $\JB B \JB^{-1}$ are $\phi$-compatible.
\end{proof}

\begin{lemma}
	\label{lemma gammaB gammaA phi compatibiliy diagonal}
	Let us consider the even case and suppose $\gammaB$ is $\phi$-compatible with $\gammaA$.
	\begin{enumerate}
		\item Then $\gammaB$ is diagonal in its matrix decomposition, so that strong $\phi$-compatibility and $\phi$-compatibility between $\gammaB$ and $\gammaA$ are equivalent.\label{item gammaB diagonal}
		\item Then $\phiH$ is diagonal for the matrix decomposition induced by $\hsA = \hsA^{+} \oplus \hsA^{-}$ and $\hsB = \hsB^{+} \oplus \hsB^{-}$, so that $\phiH$ restricts to maps $\hsA^{\pm} \to \hsB^{\pm}$.\label{item phiH diagonal grading}
	\end{enumerate}
\end{lemma}

\begin{proof}
	Point~\ref{item gammaB diagonal}: since $\gammaB^\dagger = \gammaB$, one has $\gammaB = \smallpmatrix{ \gammaB[,\phi]^{\phi} & \gammaB[,\phi]^{\perp} \\ \gammaB[,\phi]^{\perp \dagger} & \gammaB[,\perp]^{\perp} }$. The $\phi$-compatibility implies $(\gammaB[,\phi]^{\phi})^2 \phiH(\psi) = \phiH(\gammaA^2 \psi) = \phiH(\psi)$, so that $(\gammaB[,\phi]^{\phi})^2 = 1$. Since $\gammaB^2 = 1$, one has $(\gammaB[,\phi]^{\phi})^2 + \gammaB[,\phi]^{\perp} \gammaB[,\phi]^{\perp \dagger} = 1$, from which we get $\gammaB[,\phi]^{\perp} \gammaB[,\phi]^{\perp \dagger} =0$, which implies $\gammaB[,\phi]^{\perp} = 0$, so that $\gammaB$ is diagonal. By Prop.~\ref{prop strong and not strong phi compatibility}, this implies strong $\phi$-compatibility.
	
	Point~\ref{item phiH diagonal grading}: for every $\psi \in \hsA^{\pm}$, one has $\pm \phiH( \psi) = \phiH( \gammaA \psi) = \gammaB \phiH( \psi)$, so that $\phiH( \psi) \in \hsB^{\pm}$.
\end{proof}

\begin{proposition}\phantom{A}
	\label{prop KO dim strong phi compatibility}
	\begin{enumerate}
		\item If two (odd/even) real spectral triples are strong $\phi$-compatible, then they have the same $KO$-dimension (mod 8).
		
		\item If two (odd/even) real spectral triples are $\phi$-compatible and $\JB$ is strong $\phi$-compatible with $\JA$, then they have the same $KO$-dimension (mod 8).
	\end{enumerate}
\end{proposition}

\begin{proof}
	Let $(\algA, \hsA, \DA, \JA)$ and $(\algB, \hsB, \DB, \JB)$ be two strong $\phi$-compatible real spectral triple. In the even case, consider the gradings $\gammaA$ and $\gammaB$. Then one has $\JA^2 = \epsilonA$, $\JA \DA = \epsilonA' \DA \JA$, and $\JA \gammaA = \epsilonA'' \gammaA \JA$ (in the even case), and similar relations for $\algB$. We already know from Lemma~\ref{lemma JB JA st-phi-comp} that $\epsilonA = \epsilonB$. For any $\psi \in \hsA$, one has $\phiH(\JA \DA \psi) = \JB \DB \phiH(\psi)$ and $\phiH(\DA \JA \psi) = \DB \JB \phiH(\psi)$, so that $\epsilonB' = \epsilonA'$, and $\phiH(\JA \gammaA \psi) = \JB \gammaB \phiH(\psi)$ and $\phiH(\gammaA \JA \psi) = \gammaB \JB \phiH(\psi)$, so that $\epsilonB'' = \epsilonA''$.
	
	The second assertion follows the same line of computations, using the fact that $\JB$ and $\gammaB$ are diagonal (Lemmas~\ref{lemma JB JA st-phi-comp} and \ref{lemma gammaB gammaA phi compatibiliy diagonal}), so that in particular $(\JB \DB)_{\phi}^{\phi} = \JB[, \phi]^{\phi} \DB[, \phi]^{\phi}$ and $(\DB \JB)_{\phi}^{\phi} = \DB[, \phi]^{\phi} \JB[, \phi]^{\phi}$.
\end{proof}

The requirement that $\JB$ be strong $\phi$-compatible with $\JA$ seems to be inevitable in the generic situation to get the same $KO$-dimension. In the case of $AF$-algebras, this requirement will be a consequence of another requirement on the $\phiH$ map, see Prop.~\ref{prop JB JA strong phi compatibiliy relation on ukappa}.
\medskip
\par
Let $(\algA, \hsA, \DA, \JA)$ and $(\algAp, \hsAp, \DAp, \JAp)$ be two unitary equivalent real spectral triples for $\UA : \hsA \to \hsAp$ and $\phiA : \algA \to \algAp$ and let $(\algB, \hsB, \DB, \JB)$ and $(\algBp, \hsBp, \DBp, \JBp)$ be two unitary equivalent real spectral triples for $\UB : \hsB \to \hsBp$ and $\phiB : \algB \to \algBp$. 

\begin{proposition}
	\label{prop unitary equi triple and st phi comp}
	Suppose that $(\algA, \hsA, \DA, \JA)$ and $(\algB, \hsB, \DB, \JB)$ are strong $\phi$-compatible, and that there are a morphism of algebra $\phi' : \algA' \to \algB'$ and a linear map $\phiH' : \hsAp \to \hsBp$ such that $\phi' \circ \phiA = \phiB \circ \phi$ and $\phiH' (\UA \psi) = \UB \phiH(\psi)$ for any $\psi \in \hsA$. Then $(\algAp, \hsAp, \DAp, \JAp)$ and $(\algBp, \hsBp, \DBp, \JBp)$ are strong $\phi'$-compatible. If the spectral triples are even, the result holds also.
\end{proposition}

This result shows that strong $\phi$-compatibility is transported by unitary equivalence if one assumes some natural conditions on the maps $\phi'$ and $\phiH'$, which are the commutativity of the following diagrams:
\begin{equation*}\tikzexternaldisable
	\begin{tikzcd}[column sep=20pt, row sep=20pt]
		\algA
		\arrow[r, "\phi"]
		\arrow[d, "\phiA'"]
		& \algB
		\arrow[d, "\phiB"]
		\\
		\algAp
		\arrow[r, "\phi'"]
		&
		\algBp	
	\end{tikzcd} 
	\quad \text{ and } \quad
	\tikzsetnextfilename{fig-cd-2}
	\begin{tikzcd}[column sep=20pt, row sep=20pt]
		\hsA
		\arrow[r, "\phiH"]
		\arrow[d, "\UA'"]
		& \hsB
		\arrow[d, "\UB"]
		\\
		\hsAp
		\arrow[r, "\phiH'"]
		&
		\hsBp	
	\end{tikzcd} 
\end{equation*}\tikzexternalenable

\begin{proof}
	For any $\psi' \in \hsAp$, let $\psi \in \hsA$ be the unique vector such that $\psi' = \UA \psi$, and for any $a' \in \algAp$, let $a \in \algA$ the unique element such that $a' = \phiA(a)$. Then one has $\phiH'( \piAp(a') \psi') = \phiH'( (\piAp \circ \phiA(a)) \UA \psi) = \phiH'( \UA \piA(a) \psi) = \UB \phiH( \piA(a) \psi) = \UB (\piB \circ \phi(a)) \phiH( \UA^{-1} \psi') = \UB (\piB \circ \phi(a)) \UB^{-1} \phiH'( \psi') = (\piBp \circ \phiB \circ \phi(a)) \phiH'( \psi') = (\piBp \circ \phi'(a')) \phiH'( \psi')$, so that $\phiH'$ is $\phi'$-compatible. Let $A$ and $B$ be strong $\phi$-compatible operators on $\hsA$ and $\hsB$ and define $A' \defeq \UA A \UA^{-1}$ and $B' \defeq \UB B \UB^{-1}$ on $\hsAp$ and $\hsBp$. Then one has $\phiH'(A' \psi') = \phiH'(\UA A \psi) = \UB \phiH(A \psi) = \UB B \phiH(\psi) = B' \UB \phiH(\psi) = B' \phiH'(\UA \psi) = B' \phiH'(\psi')$, so that $A'$ and $B'$ are strong $\phi'$-compatible. Applying this result to $\DAp$ and $\DBp$ (resp. $\JAp$ and $\JBp$, resp. $\gammaAp$ and $\gammaBp$ in the even case) shows that $(\algAp, \hsAp, \DAp, \JAp)$ and $(\algBp, \hsBp, \DBp, \JBp)$ are strong $\phi'$-compatible and similarly in the even case. 
\end{proof}

In the proof, the commutativity of the first diagram is only used when the representation $\piBp$ is applied, and more specifically, when this representation acts on $\phiH'(\hsAp)$. In other words, the minimal condition in this proof is that $\piBp \circ \phi' \circ \phiA = \piBp \circ \phiB \circ \phi$ holds as operators acting on $\phiH'(\hsAp) \subset \hsBp$.
\medskip
\par
Another version using only $\phi$-compatibility can be proposed in the following way:
\begin{proposition}
	\label{prop unitary equi triple and phi comp and diag unitary}
	Suppose that $(\algA, \hsA, \DA, \JA)$ and $(\algB, \hsB, \DB, \JB)$ are $\phi$-compatible, and that there is a morphism of algebra $\phi' : \algA' \to \algB'$ and a linear map $\phiH' : \hsAp \to \hsBp$ such that $\phi' \circ \phiA = \phiB \circ \phi$ and $\phiH' (\UA \psi) = \UB \phiH(\psi)$ for any $\psi \in \hsA$, and suppose that $\UB$ is diagonal. Then $(\algAp, \hsAp, \DAp, \JAp)$ and $(\algBp, \hsBp, \DBp, \JBp)$ are $\phi'$-compatible. If the spectral triples are even, the result holds also.
\end{proposition}

\begin{proof}
	The proof follows the same line of reasoning as for the proof of Prop.~\ref{prop unitary equi triple and st phi comp}, with the key difference that, from $B' \defeq \UB B \UB^{-1}$ and the fact that  $\UB$ is diagonal, one has $B'^{\phi}_{\phi} = \UB^{\phi}_{\phi} B^{\phi}_{\phi} (\UB^{\phi}_{\phi})^{-1}$.
\end{proof}
\medskip
\par
The map $\phi$ induces a natural map of graded algebras $\phi : \calT^\grast \algA \to \calT^\grast \algB$ by the relation $\phi(a^0 \otimes \cdots \otimes a^n) = \phi(a^0) \otimes \cdots \otimes \phi(a^n)$. If $\omega \in \Omega^1_U(\algA)$, then one can check that $\phi(\omega) \in \Omega^1_U(\algB)$, so that $\phi$ restricts to a map of graded algebras $\Omega^\grast_U(\algA) \to \Omega^\grast_U(\algB)$. If $\phi(\bbboneA) = \bbboneB$, then $\phi(\ddU a) = \phi(\bbboneA \otimes a - a \otimes \bbboneA) = \bbboneB \otimes \phi(a) - \phi(a) \otimes \bbboneB = \ddU \phi(b)$. If $\phi(\bbboneA) \neq \bbboneB$, let $p_\phi \defeq \phi(\bbboneA) \in \algB$ be the induced projection. Then $\phi(\ddU a) = p_\phi \otimes \phi(a) - \phi(a) \otimes p_\phi \in \Omega^1_U(\algB)$ can be written as $\phi(\ddU a) = p_\phi \ddU \phi(a) + \phi(a) \ddU (\bbboneB - p_\phi) = p_\phi \ddU \phi(a) - \phi(a) \ddU p_\phi$. This shows that $\phi$ is a morphism of differential algebras only when it is unital. In the following, we will use the most general relation $\phi(a^0 \ddU a^1) = \phi(a^0) \ddU \phi(a^1) - \phi(a^0 a^1) \ddU p_\phi$ since $\phi(a) p_\phi = \phi(a)$. 

\begin{proposition}
	\label{prop universal forms and Dirac phi compatibility}
	Suppose that $\DB$ is $\phi$-compatible with $\DA$. 
	\begin{enumerate}
		\item For any $\omega \in \Omega^1_U(\algA)$, $\piDB\circ \phi(\omega)$ is $\phi$-compatible with $\piDA(\omega)$.
		
		\item Suppose that $\JB$ is strong $\phi$-compatible with $\JA$. For any unitaries $\uA \in \algA$ and $\uB \in \algB$ such that $\piA(\uA)$ and $\piB(\uB)$ are $\phi$-compatible and $\piB(\uB)$ is diagonal in the matrix decomposition, $\DB^{\uB}$ is $\phi$-compatible with $\DA^{\uA}$. \label{enum unitaries in algebras}
		
		\item Using the hypothesis of the previous points, $\DB[, \phi(\omega)]^{\uB}$ is $\phi$-compatible with $\DA[,\omega]^{\uA}$.
	\end{enumerate}
\end{proposition}
\medskip
\par
Condition~\ref{enum unitaries in algebras} in this Proposition implies in particular that $\piBp \circ \phi' \circ \phiA = \piBp \circ \phiB \circ \phi$ (see comment after Prop.~\ref{prop unitary equi triple and st phi comp}) with $\algA' = \algA$, $\algB' = \algB$ and $\phi' = \phi$.

\begin{proof}
	We can reduce the general case to $\omega = a^0 \ddU a^1 \in \Omega^1_U(\algA)$. Let us then consider $\piDB\circ \phi(a^0 \ddU a^1) = \phi(a^0) [\DB, \phi(a^1)] - \phi(a^0 a^1) [\DB, p_\phi]$ (with $\piB$ omitted in this relation and the following). For any $\psi \in \hsA$, one has $\phi(a^0 a^1) [\DB, p_\phi] \phiH(\psi) = \phi(a^0 a^1) \DB \phiH(\psi) - \phi(a^0 a^1) \phi(\bbboneA) \DB \phiH(\psi) = 0$, since $p_\phi \phiH(\psi) = \phiH(\psi)$, so that $\piDB\circ \phi(a^0 \ddU a^1) \phiH(\psi) = \phi(a^0) [\DB, \phi(a^1)] \phiH(\psi)$. Using the matrix decomposition $\DB = \smallpmatrix{ \DB[,\phi]^{\phi} & \DB[,\phi]^{\perp} \\ \DB[,\perp]^{\phi} & \DB[,\perp]^{\perp} }$ and Point~\ref{item piBphi(a) diagonal} in Prop.~\ref{prop strong and not strong phi compatibility}, one gets
	\begin{align*}
		\phi(a^0) [\DB, \phi(a^1)] 
		\begin{pmatrix} \phiH(\psi) \\  0 \end{pmatrix}
		&= 
		\begin{pmatrix}
			\phi(a^0)_{\phi}^{\phi} [\DB[,\phi]^{\phi}, \phi(a^1)_{\phi}^{\phi}] \phiH(\psi)
			\\ 
			\phi(a^0)_{\perp}^{\perp} (\DB[,\perp]^{\phi} \phi(a^1)_{\phi}^{\phi} - \phi(a^1)_{\perp}^{\perp} \DB[,\perp]^{\phi} ) \phiH(\psi)
		\end{pmatrix}
	\end{align*}
	From this relation we get $\piDB (\phi(a^0 \ddU a^1))_{\phi}^{\phi} = \phi(a^0)_{\phi}^{\phi} [\DB[,\phi]^{\phi}, \phi(a^1)_{\phi}^{\phi}]$ and then $\piDB (\phi(a^0 \ddU a^1))_{\phi}^{\phi} \phiH(\psi) = \phi(a^0)_{\phi}^{\phi} [\DB[,\phi]^{\phi}, \phi(a^1)_{\phi}^{\phi}] \phiH(\psi) = \phiH( a^0 [\DA, a^1] \psi) = \phiH( \piDA(a^0 \ddU a^1) \psi)$ since $\DB$ is $\phi$-compatible with $\DA$.
	\medskip
	\par
	Using the hypothesis that $\piB(\uB)$ is diagonal, a straightforward computation gives $(\piB(\uB)^\dagger [\DB, \piB(\uB)])_{\phi}^{\phi} = (\piB(\uB)^\dagger)_{\phi}^{\phi} [\DB[,\phi]^{\phi}, \piB(\uB)_{\phi}^{\phi} ]$ from which we deduce that $\piB(\uB)^\dagger [\DB, \piB(\uB)]$ is $\phi$-compatible with $\piA(\uA)^\dagger [\DA, \piA(\uA)]$. From Lemma~\ref{lemma JB JA st-phi-comp}, we deduce that $\JB \piB(\uB)^\dagger [\DB, \piB(\uB)] \JB^{-1}$ is $\phi$-compatible with $\JA \piA(\uA)^\dagger [\DA, \piA(\uA)] \JA^{-1}$, and so that $\DB^{\uB} = \DB + \piB(\uB)^\dagger [\DB, \piB(\uB)] + \epsilonB' \JB \piB(\uB)^\dagger [\DB, \piB(\uB)] \JB^{-1}$ is $\phi$-compatible with $\DA^{\uA} = \DA + \piA(\uA)^\dagger [\DA, \piA(\uA)] + \epsilonA' \JA \piA(\uA)^\dagger [\DA, \piA(\uA)] \JA^{-1}$ since $\epsilonB' = \epsilonA'$ by Prop~\ref{prop KO dim strong phi compatibility}.
	\medskip
	\par
	The last point combines the two previous results by replacing $\DA$ by $\DA[,\omega] = \DA + \piDA(\omega) + \epsilonA' \JA \piDA(\omega) \JA^{-1}$ and $\DB$ by $\DB[, \phi(\omega)] = \DB + \piDB(\phi(\omega)) + \epsilonB' \JB \piDB(\phi(\omega)) \JB^{-1}$ which are $\phi$-compatible by the first point, Lemma~\ref{lemma JB JA st-phi-comp}, and Prop~\ref{prop KO dim strong phi compatibility}.
\end{proof}

Notice that one can associate to any unitary $\uA \in \algA$ the diagonal (unitary) operator $\smallpmatrix{ \piB\circ \phi(\uA) & 0 \\ 0 & \bbbone_{\perp}^{\perp} }$ where $\piB\circ \phi(\uA) \phiH(\psi) \defeq \phiH(\piA(\uA) \psi)$ for any $\psi \in \hsA$ and $\bbbone_{\perp}^{\perp}$ is the identity operator on $\phiH(\hsA)^\perp$. But this operator is not necessarily of the form $\piB(\uB)$ for a unitary $\uB \in \algB$. In the case of $AF$-algebras, it will be possible to construct a unitary $\uB  \in \algB$ from $\uA$ such that $\piA(\uA)$ and $\piB(\uB)$ are (strong) $\phi$-compatible and $\piB(\uB)$ is diagonal, see Prop.~\ref{prop uB from uA case AF}.
\medskip
\par
A strong version of the previous proposition can be proposed, for which a proof is not necessary since it combines previous results and the same line of reasoning when computations are needed:
\begin{proposition}
	Suppose that $\DB$ is strong $\phi$-compatible with $\DA$. 
	\begin{enumerate}
		\item For any $\omega \in \Omega^1_U(\algA)$, $\piDB\circ \phi(\omega)$ is strong $\phi$-compatible with $\piDA(\omega)$.
		
		\item Suppose that $\JB$ is strong $\phi$-compatible with $\JA$. For any unitaries $\uA \in \algA$ and $\uB \in \algB$ such that $\piA(\uA)$ and $\piB(\uB)$ are strong $\phi$-compatible, $\DB^{\uB}$ is strong $\phi$-compatible with $\DA^{\uA}$.
		
		\item Using the hypothesis of the previous points, $\DB[, \phi(\omega)]^{\uB}$ is strong $\phi$-compatible with $\DA[,\omega]^{\uA}$.
	\end{enumerate}
\end{proposition}


\subsection{$AF$-Algebras in terms of Krajewski's Diagram Structures}
\label{sec AF algebras}

\medskip
\par
We will use the results of section \ref{ResMatrDirac}, in particular the diagrammatic descriptions of (odd/even) real spectral triples. Let $(\algA, \hsA, \DA, \JA)$ and $(\algB, \hsB, \DB, \JB)$ be two real spectral triples on the algebras $\algA = \toplus_{i=1}^{r} M_{n_i}$ and $\algB = \toplus_{k=1}^{s} M_{m_k}$ with $\hsA = \toplus_{v \in \GammaA^{(0)}} \hsiA[,v]$ and $\hsB = \toplus_{w \in \GammaB^{(0)}} \hsiB[,w]$. As we defined the maps $i,j$ on $\GammaA^{(0)}$, let us define the similar maps $k, \ell$ on $\GammaB^{(0)}$: for any $w \in \GammaB^{(0)}$ with $\pi_{\lambda\rho}(w) = (\bem_k, \bem_\ell)$, $k(w) \defeq k$ and $\ell(w) \defeq \ell$.
\medskip
\par
Let $\phiH : \hsA \to \hsB$ be a $\phi$-compatible linear map of bimodules ($\phi^{e}$-compatible as left modules). This map decomposes along the maps $\phiH[,w]^v : \hsiA[,v] \to \hsiB[,w]$ between irreps on both sides. For any $a = \toplus_{i=1}^{r} a_i \in \algA$, $b = \toplus_{i=1}^{r} b_i \in \algA$, and $\psi = \toplus_{v \in \GammaA^{(0)}} \psi_v$, one has $\phiH(a b ^\circ \psi) = \phi(a) \phi(b)^\circ \phiH(\psi)$ with (using \eqref{eq pi(a) decomposition along Hv})\\ $\phiH(a b ^\circ \psi) = \sum_{v \in \GammaA^{(0)}} \toplus_{w \in \GammaB^{(0)}} \phiH[,w]^{v} ( a_{i(v)} b_{j(v)}^\circ \psi_v)$ and $\phi(a) \phi(b)^\circ \phiH(\psi) = \sum_{v \in \GammaA^{(0)}} \toplus_{w \in \GammaB^{(0)}} \phi(a)_{k(w)} \phi(b)_{\ell(w)}^\circ \phiH[,w]^{v} (\psi_v)$.\\ We can select a fixed $v$ and choose $\psi$ with only one non-zero component $\psi_v$. Then one gets, for any $v \in  \GammaA^{(0)}$ and any $w \in \GammaB^{(0)}$, $\phiH[,v]^{w} ( a_{i(v)} b_{j(v)}^\circ \psi_v) = \phi(a)_{k(w)} \phi(a)_{\ell(w)}^\circ \phiH[,v]^{w} (\psi_v)$. Let us now consider a fixed index $i$ and $a$ with only non-zero component at $i(v) = i$, and the same for a fixed $j$ and $b$: with $(k, \ell) = (k(w), \ell(w))$, one has
\begin{align}
	\label{eq phiH w v a b psiv}
	\phiH[,w]^{v}(a_i b_j^\circ \psi_v) = \phi_{k}^{i}(a_i) \phi_{\ell}^{j}(b_j)^\circ \phiH[,w]^{v}(\psi_v)
\end{align}
This relation, combined with \eqref{eq phi-k-i(ai)}, suggests to decompose $\bbC^{m_k}$ in $\hsiB[,w] = \bbC^{m_k} \otimes \bbC^{m_\ell \circ}$ as $\bbC^{m_k} =  [ \toplus_{i=1}^{r} \bbC^{n_i} \otimes \bbC^{\alpha_{ki}} ] \oplus \bbC^{n_{0,k}}$ and similarly for $\bbC^{m_\ell \circ}$ with a last term $\bbC^{n_{0,\ell}}$, so that one has the orthogonal decomposition
\begin{align}
	\label{eq decomp hBw}
	\hsB[,w] = \bbC^{m_k} \otimes \bbC^{m_\ell \circ} 
	=&[ \toplus_{i,j = 1}^{r} \bbC^{n_i} \otimes \bbC^{\alpha_{ki}} \otimes \bbC^{\alpha_{\ell j}} \otimes \bbC^{n_j \circ} ] 
	\oplus [ \toplus_{i=1}^{r}  \bbC^{n_i} \otimes \bbC^{\alpha_{ki}} \otimes \bbC^{n_{0,\ell} \circ} ]& 
	\\
	&\oplus [ \toplus_{j=1}^{r} \bbC^{n_{0,k}} \otimes \bbC^{\alpha_{\ell j}} \otimes \bbC^{n_j \circ} ] 
	\oplus [ \bbC^{n_{0,k}} \otimes \bbC^{n_{0,\ell} \circ}]&.
\end{align}
For any $i,j = 1, \dots, r$ and $k, \ell = 1, \dots, s$, let us define the inclusion
\begin{align*}
	I_{k,\ell}^{i,j} : \bbC^{n_i} \otimes \bbC^{\alpha_{ki}} \otimes \bbC^{\alpha_{\ell j}} \otimes \bbC^{n_j \circ} \hookrightarrow \bbC^{m_k} \otimes \bbC^{m_\ell \circ}.
\end{align*}
Notice that $I_{k,\ell}^{i,j} = I_{k}^{i} \otimes I_{\ell}^{j \circ}$ with the inclusions $I_{k}^{i} : \bbC^{n_i} \otimes \bbC^{\alpha_{ki}} \hookrightarrow \bbC^{m_k}$ and $I_{\ell}^{j \circ} : \bbC^{\alpha_{\ell j}} \otimes \bbC^{n_j \circ} \hookrightarrow \bbC^{m_\ell \circ}$.
\medskip
\par
Let $F^{(i,k), (j, \ell)}_\algA : \bbC^{n_i} \otimes \bbC^{\alpha_{ki}} \otimes \bbC^{\alpha_{\ell j}} \otimes \bbC^{n_j \circ} \to \bbC^{n_j}  \otimes \bbC^{\alpha_{\ell j}} \otimes \bbC^{\alpha_{ki}} \otimes \bbC^{n_i \circ}$ be the involution $\xi_i \otimes \vm^k_i \otimes \vm^\ell_j \otimes \eta_j^\circ \mapsto \eta_j \otimes \vm^\ell_j \otimes \vm^k_i \otimes \xi_i^\circ$ and $F^{k\ell}_\algB : \bbC^{m_k} \otimes \bbC^{m_\ell \circ} \to \bbC^{m_\ell} \otimes \bbC^{m_k \circ}$ the involution $\varphi_k \otimes \vartheta_\ell^\circ \mapsto \vartheta_\ell \otimes \varphi_k^\circ$. Then, one can check that
\begin{align}
	\label{eq flips inclusions}
	F^{k\ell}_\algB \circ I_{k,\ell}^{i,j}
	&= I_{\ell,k}^{j,i} \circ F^{(i,k), (j, \ell)}_\algA
\end{align}
Notice that $\JA :  \hsA[,v] \to \hsA[, \JimA(v)]$ can be written as $\JA = \epsilonA(v,d_\algA) (J_0 \otimes J_0) \circ F^{i j}_\algA$ since $F^{i j}_\algA = \hJimA[,v]$ with $(i, j) = (i(v), j(v))$.
\medskip
\par
In the case of a $AF$-algebra, the inclusions $I_{k}^{i}$ (and so $I_{\ell}^{j \circ}$ and $I_{k,\ell}^{i,j}$) are defined directly from the Bratteli diagram of the algebra, that is, they depend only on the inclusion $\phi : \algA \to \algB$. We can now write $\phiH$ in terms of these inclusions.
\medskip
\par
Combining \eqref{eq phiH w v a b psiv} and \eqref{eq decomp hBw}, the map $\phiH[,w]^{v}$ first reduces to a map $\bbC^{n_i} \otimes \bbC^{n_j \circ} \to \bbC^{n_i} \otimes \bbC^{\alpha_{ki}} \otimes \bbC^{\alpha_{\ell j}} \otimes \bbC^{n_j \circ}$, and using a slight adaptation of Lemma~\ref{lemma technical module endo reduction}, it reduces to a linear map $\bbC \to \bbC^{\alpha_{ki}} \otimes \bbC^{\alpha_{\ell j}}$, that is, to the data of an element $u(v,w) \in \bbC^{\alpha_{ki}} \otimes \bbC^{\alpha_{\ell j}}$, such that $\phiH[,w]^{v}$ is the composition of $\bbC^{n_i} \otimes \bbC^{n_j \circ} \ni \xi_i \otimes \eta_j^\circ \mapsto \xi_i \otimes u(v,w) \otimes \eta_j^\circ \in \bbC^{n_i} \otimes \bbC^{\alpha_{ki}} \otimes \bbC^{\alpha_{\ell j}} \otimes \bbC^{n_j \circ}$ with the inclusion $I_{k,\ell}^{i,j}$. When $\alpha_{ki} = 0$ or $\alpha_{\ell j} = 0$, $\phiH[,w]^{v} = 0$.
\medskip
\par
Consequently, the $\phi$-compatible map $\phiH$ is completely determined by a family of matrices $u(v,w) \in M_{\alpha_{k i} \times \alpha_{\ell j}} \simeq \bbC^{\alpha_{k i}} \otimes \bbC^{\alpha_{\ell j}}$%
\footnote{\label{fn z=xyT}We use the following convention. Let $x, x' \in \bbC^n$ and $y, y' \in \bbC^m$. Define $\bbC^n \otimes \bbC^m \ni x \otimes y \simeq z = x y^\top \in M_{n \times m}$ and $z' = x' y'^\top \in M_{n \times m}$, so that $z(v) = \langle \bar{y}, v\rangle_{\bbC^m} x$ for any $v \in \bbC^m$. One then gets $\bar{y} \otimes \bar{x} \simeq z^\ast$ and $\tr(z^\ast z') = \langle x, x' \rangle_{\bbC^n} \langle y, y' \rangle_{\bbC^m}$, and by linearity, a similar relation holds for $z = \sum_{i} x_i y_i^\top$ and $z' = \sum_{i} x'_i y_i'^\top$. This relation will be used in the following.}
by the previous relation, with $(i, j) = (i(v), j(v))$ and $(k, \ell) = (k(w), \ell(w))$. Notice that for $v, v' \in \GammaA^{(0)}$ such that $\pi_{\lambda\rho}(v) = \pi_{\lambda\rho}(v')$, the ranges of $\phiH[,w]^{v}$ and $\phiH[,w]^{v'}$ are at the same place in $\hsiB[,w]$ (the range of $I_{k,\ell}^{i,j}$), and $u(v,w)$ and $u(v',w)$ define a kind of relative positioning and weight between the two ranges. If $\pi_{\lambda\rho}(v) \neq \pi_{\lambda\rho}(v')$, the ranges are orthogonal in $\hsiB[,w]$ since the ranges of $I_{k,\ell}^{i,j}$ and $I_{k,\ell}^{i',j'}$ are distinct in the orthogonal decomposition \eqref{eq decomp hBw} when $(i,j) \neq (i',j')$.
\medskip
\par
For non-real spectral triples, a similar (simpler) result can be obtained: a $\phi$-compatible map $\phiH : \hsA \to \hsB$ is completely determined by the linear maps $\phiH[,w]^{v} : \hsiA[,v] = \bbC^{n_i} \to \hsiB[,w] = \bbC^{m_k}$ for $i=i(v)$ and $k=k(w)$, and so by a family of vectors $u(v,w) \in \bbC^{\alpha_{k i}}$ such that $\phiH[,w]^{v}$ is the composition of $\bbC^{n_i} \ni \xi_i \mapsto \xi_i \otimes u(v,w) \in \bbC^{n_i} \otimes \bbC^{\alpha_{ki}}$ with the inclusion $I_{k}^{i} : \bbC^{n_i} \otimes \bbC^{\alpha_{ki}} \hookrightarrow \bbC^{m_k}$.
\medskip
\par

The following result summarizes the construction achieved so far:
\begin{lemma}
	\label{lemma matrices u(v,w)}
	There is a family of matrices $u(v,w) \in M_{\alpha_{k i} \times \alpha_{\ell j}}$ such that, for any $v \in \GammaA^{(0)}$ and $w \in \GammaB^{(0)}$, with $(i, j) = (i(v), j(v))$ and $(k, \ell) = (k(w), \ell(w))$, one has
	\begin{align}
		\label{eq phiH Iklij xi eta}
		\phiH[,w]^{v}( \xi_i \otimes \eta_j^\circ) 
		&= I_{k,\ell}^{i,j}( \xi_i \otimes u(v,w) \otimes \eta_j^\circ)
		\quad \text{for any $\xi_i \otimes \eta_j^\circ \in \hsA[, v]$.}
	\end{align}
	For any $v \in \GammaA^{(0)}$, any $w \in \GammaB^{(0)}$, and any $a = \toplus_{i=1}^{r} a_i \in \algA$, one has
	\begin{align*}
		\phi_{k}^{i}(a_i) I_{k,\ell}^{i,j} ( \xi_i \otimes u(v,w) \otimes \eta_j^\circ) 
		&= I_{k,\ell}^{i,j} ( a_i \xi_i \otimes u(v,w) \otimes \eta_j^\circ)
	\end{align*}
	with $(i, j) = (i(v), j(v))$ and $(k, \ell) = (k(w), \ell(w))$.
	
	In the even case, if $\gammaB$ is $\phi$-compatible with $\gammaA$, then $\phiH[,w]^{v}$, and so $u(v,w)$, can be non-zero only when $s(v) = s(w)$. 
\end{lemma}

\begin{proof}
	The first statement is a summary of the previous decomposition of $\phiH$. The second relation is a direct consequence of this decomposition of $\phiH$ and \eqref{eq phiH w v a b psiv}. Notice that in the LHS, one could replace $\phi_{k}^{i}(a_i)$ by $\phi_{k}(a)$ since only the entries positioned according to $i$ in the matrix $\phi_{k}(a) \in M_{m_k}$, see \eqref{eq phi-k(a)} and \eqref{eq phi-k-i(ai)}, give non-zero contributions once applied on the range of $I_{k,\ell}^{i,j}$. In the even case, the statement is a consequence of Lemma~\ref{lemma gammaB gammaA phi compatibiliy diagonal}, which implies here that $\phiH[,w]^{v} = 0$ when $s(v) \neq s(w)$.
\end{proof}
Let us stress that $u(v,w)$ is an essential object since it encodes all the data of embedding (\textit{i.e.} $\alpha_{ki}$'s and the way of embedding into these corresponding blocks) which is at the heart of the Bratteli diagram structure. Then, it allows us to link Bratteli's and Krajewski's diagrammatic structures since the way its basic elements (nodes $v$ and $w$) are now connected to Bratteli's diagram structures.

\begin{proposition}
	\label{prop st phi comp and phi comp of operators for AF}
	Two operators $A$ on $\hsA$ and $B$ on $\hsB$ are strong $\phi$-compatible if and only if
	\begin{align}
		\label{eq A B str phi comp AF}
		\tsum_{v_2 \in \GammaA^{(0)}}\phiH[,w_2]^{v_2}(A_{v_2}^{v_1} \psi_{v_1})
		&=
		\tsum_{w_1 \in \GammaB^{(0)}} B_{w_2}^{w_1} \phiH[,w_1]^{v_1}(\psi_{v_1})
	\end{align}
	for any $v_1 \in \GammaA^{(0)}$, $w_2 \in \GammaB^{(0)}$, and $\psi_{v_1} \in \hsiA[,v_1]$. They are $\phi$-compatible if and only if
	\begin{align*}
		\tsum_{v_2 \in \GammaA^{(0)}}\phiH[,w_2]^{v_2}(A_{v_2}^{v_1} \psi_{v_1})
		&=
		\tsum_{w_1 \in \GammaB^{(0)}} B_{\phi, w_2}^{\phi, w_1} \phiH[,w_1]^{v_1}(\psi_{v_1})
	\end{align*}
	for any $v_1 \in \GammaA^{(0)}$, $w_2 \in \GammaB^{(0)}$, and $\psi_{v_1} \in \hsiA[,v_1]$, where $B_{\phi, w_2}^{\phi, w_1} : \hsiB[,w_1] \cap \phiH(\hsA) \to \hsiB[,w_2] \cap \phiH(\hsA)$ is the decomposition of $B_{\phi}^{\phi}$ along the $\hsiB[,w] \cap \phiH(\hsA)$'s.
\end{proposition}

\begin{proof}
	For any $\psi = \toplus_{v_1 \in \GammaA^{(0)}} \psi_{v_1} \in \hsA$, one has $A \psi = \toplus_{v_2 \in \GammaA^{(0)}} \sum_{v_1 \in \GammaA^{(0)}} A_{v_2}^{v_1} \psi_{v_1} \psi_{v_1}$ and $\phiH(\psi) = \toplus_{w_1 \in \GammaB^{(0)}} \sum_{v_1 \in \GammaA^{(0)}} \phiH[,w_1]^{v_1}(\psi_{v_1})$. This implies $\phiH(A \psi) = \toplus_{w_2 \in \GammaB^{(0)}} \sum_{v_1, v_2 \in \GammaA^{(0)}} \phiH[,w_2]^{v_2}( A_{v_2}^{v_1}  \psi_{v_1})$ and, in a similar way, $B \phiH(\psi) = \toplus_{w_2 \in \GammaB^{(0)}} \sum_{w_1 \in \GammaB^{(0)}} \sum_{v_1 \in \GammaA^{(0)}} B_{w_1}^{w_2} \phiH[,w_2]^{v_1}(\psi_{v_1})$. The strong $\phi$-compatibility is then equivalent to $\sum_{v_1, v_2 \in \GammaA^{(0)}} \phiH[,w_2]^{v_2}( A_{v_2}^{v_1}  \psi_{v_1}) = \sum_{w_1 \in \GammaB^{(0)}} \sum_{v_1 \in \GammaA^{(0)}} B_{w_1}^{w_2} \phiH[,w_2]^{v_1}(\psi_{v_1})$ for any $w_2 \in \GammaB^{(0)}$, and, by linearity (fixing $\psi$ with one non-zero component at $v_1$), $\sum_{v_2 \in \GammaA^{(0)}} \phiH[,w_2]^{v_2}( A_{v_2}^{v_1}  \psi_{v_1}) = \sum_{w_1 \in \GammaB^{(0)}}  B_{w_1}^{w_2} \phiH[,w_2]^{v_1}(\psi_{v_1})$ for any $v_1 \in \GammaA^{(0)}$ and $w_2 \in \GammaB^{(0)}$.
	\medskip
	\par
	The $\phi$-compatibility relation follows the same computation with $B$ replaced by $B_{\phi}^{\phi}$.
\end{proof}

\begin{lemma}
	For any $a = \toplus_{i=1}^{r} a_i \in \algA$ and $b = \toplus_{k=1}^{s} b_i \in \algB$, $\piA(a)$ and $\piB(b)$ are strong $\phi$-compatible if and only if, for any $v \in \GammaA^{(0)}$, any $w \in \GammaB^{(0)}$, and any $\xi_{i(v)} \otimes \eta_{j(v)}^\circ \in \hsA[, v]$, one has
	\begin{align*}
		b_{k(w)} I_{k,\ell}^{i,j} ( \xi_{i(v)} \otimes u(v,w) \otimes \eta_{j(v)}^\circ) 
		&= I_{k,\ell}^{i,j} ( a_{i(v)} \xi_{i(v)} \otimes u(v,w) \otimes \eta_{j(v)}^\circ) 
	\end{align*}
\end{lemma}

\begin{proof}
	Inserting \eqref{eq pi(a) decomposition along Hv} into \eqref{eq A B str phi comp AF} for $A = \piA(a)$ and $B = \piB(b)$, one gets $\phiH[,w]^{v}( a_{i(v)} \psi_v) = b_{k(w)} \phiH[,w]^{v}(\psi_v)$ for any $v \in \GammaA^{(0)}$ and $w \in \GammaB^{(0)}$. Using \eqref{eq phiH Iklij xi eta} for $\phiH[,w]^{v}$ then gives the relation.
\end{proof}

\begin{proposition}
	\label{prop uB from uA case AF}
	Let $\uA \in \algA$ be a unitary element and define $\uB \defeq \phi(\uA) + p_{n_{0}} \in \algB$ (see \eqref{eq def pn0k}). Then $\uB$ is a unitary element such that $\piB(\uB)$ is diagonal (in the orthogonal decomposition defined by $\phiH$) and is strong $\phi$-compatible with $\piA(\uA)$.
\end{proposition}

\begin{proof}
	One already knows that $\piB \circ \phi(\uA)$ is strong $\phi$-compatible with $\piA(\uA)$ (see Remark~\ref{rmk piA(a) st phi comp piB(phi(a))}). By construction, the range of $\phiH[,w]$ is contained only in the first term in brackets (the double direct sum over $i,j$) in \eqref{eq decomp hBw}, while $\piB(p_{n_0})$ is non-trivial only on the last two terms (the ones with $\bbC^{n_{0,k}}$ as first factor). This implies that $\piB(p_{n_0}) \phiH(\psi) = 0 = \phiH(\uA \psi)$ for any any $\psi \in \hsA$. 
	So, one has $\piB(\uB) \phiH(\psi) = \phiH(\uA \psi)$ for any $\psi \in \hsA$, and since $\piA(\uA)$ and $\piB(\uB)$ are unitary, by Prop.~\ref{prop strong and not strong phi compatibility}, $\piB(\uB)$ is diagonal.
\end{proof}

\begin{proposition}
	\label{prop JB JA strong phi compatibiliy relation on ukappa}
	$\JB$ is strong $\phi$-compatible with $\JA$ if and only if
	\begin{align}
		\label{eq ukappa epsilon ustar}
		u(\JimA(v), \JimB(w)) &= \frac{\epsilonA(v,d_\algA)}{\epsilonB(w,d_\algB)} u(v,w)^\ast
	\end{align}
	for any $v \in \GammaA^{(0)}$ and $w \in \GammaB^{(0)}$ where $d_\algA$ (resp. $d_\algB$) is the $KO$-dimension of $\algA$ (resp. $\algB)$.
\end{proposition}

Prop.~\ref{prop KO dim AF phi compatibility} below gives a criterion on spectral triples on top of $\algA$ and $\algB$ so that $d_\algA = d_\algB$.

\begin{proof}
	For any $\psi_v = \xi_i \otimes \eta_j^\circ \in \hsiA[,v]$, one has $\phiH[,\JimB(w)]^{\JimA(v)}(\JA \psi_v) = \epsilonA(v,d_\algA) \phiH[,\JimB(w)]^{\JimA(v)}(\Beta_{j} \otimes \Bxi_{i}^\circ) = \epsilonA(v,d_\algA) I_{\ell,k}^{j,i}\big( \Beta_{j} \otimes u(\JimA(v), \JimB(w)) \otimes \Bxi_{i}^\circ \big)$ and $\JB \phiH[,w]^{v}(\psi_v) = \JB \circ  I_{k,\ell}^{i,j}(\xi_i \otimes u(v,w) \otimes \eta_j^\circ) = \epsilonB(w,d_\algB) I_{\ell,k}^{j,i} \big( \Beta_{j} \otimes u(v,w)^\ast \otimes \Bxi_{i}^\circ \big)$ when one uses \eqref{eq flips inclusions} and the identification of $M_{\alpha_{k i} \times \alpha_{\ell j}}$ with $\bbC^{\alpha_{k i}} \otimes \bbC^{\alpha_{\ell j}}$ (see Footnote~\ref{fn z=xyT}). This implies the required equivalence.
\end{proof}

\begin{corollary}
	\label{corr phiH and phiHkappa}
	If $\JB$ is strong $\phi$-compatible with $\JA$, then, for any $v \in \GammaA^{(0)}$ and $w \in \GammaB^{(0)}$, $\phiH[,\JimB(w)]^{\JimA(v)} \neq 0$ if and only if $\phiH[,w]^{v} \neq 0$.
\end{corollary}

\begin{proof}
	For any $v \in \GammaA^{(0)}$ and $w \in \GammaB^{(0)}$, with $(i, j) = (i(v), j(v))$ and $(k, \ell) = (k(w), \ell(w))$, from \eqref{eq ukappa epsilon ustar}, one has
	\begin{align*}
		\phiH[,\JimB(w)]^{\JimA(v)}(\xi_j \otimes \eta_i^\circ)
		&= I_{\ell, k}^{j,i} (\xi_j \otimes u(\JimA(v), \JimB(w)) \otimes \eta_i^\circ)= \frac{\epsilonA(v,d_\algA)}{\epsilonB(w,d_\algB)} I_{\ell, k}^{j,i} (\xi_j \otimes u(v,w)^\ast \otimes \eta_i^\circ).
	\end{align*}
	We then get the equivalence since $u(v,w)$ defines $\phiH[,w]^{v}$.
\end{proof}

\begin{figure}
	{\centering
		\subfloat[A Bratteli diagram]{\includegraphics[]{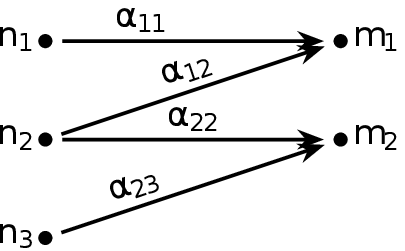}\label{fig bratteliDiagram}}\\
		\subfloat[A lift of a Bratteli diagram between two Krajewski diagrams]{\includegraphics[]{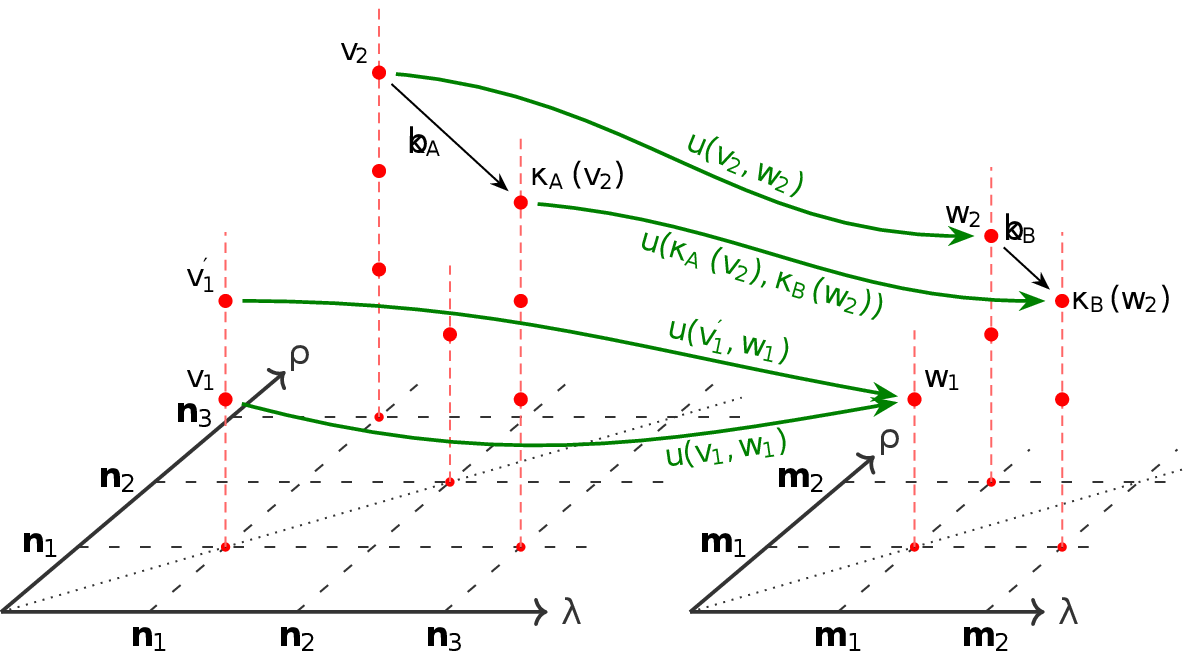}\label{fig liftingBratteliDiagram}}
		\par}
	\caption{Lift of a Bratteli diagram between two Krajewski diagrams.\\
		\protect\subref{fig bratteliDiagram}: An example of a Bratteli diagram for the inclusion $M_{n_1} \oplus M_{n_2} \oplus M_{n_3} \to M_{m_1} \oplus M_{m_2}$ with multiplicities $\alpha_{ki}$ for the inclusion of $M_{n_i}$ into $M_{m_k}$.\\
		\protect\subref{fig liftingBratteliDiagram}: Some lifts of the maps (arrows) given in the Bratteli diagram \protect\subref{fig bratteliDiagram} as maps $\phiH[,w]^{v} : \hsA[,v] \to \hsB[,w]$, here represented as (green) arrows decorated with their defining matrices $u(v,w) \in \bbC^{\alpha_{ki}} \otimes \bbC^{\alpha_{\ell j}}$, see \eqref{eq phiH Iklij xi eta}. The configuration for the arrows $v_2 \to w_2$ and $\JimA(v_2) \to \JimB(w_2)$ is the consequence of Corollary~\ref{corr phiH and phiHkappa}. In the even case, according to Lemma~\ref{lemma matrices u(v,w)}, one should have $s(v_2) = s(w_2)$ for $u(v_2,w_2)$ to be non-zero, and similarly for other arrows. The arrows $M_{n_2} \to M_{m_1}$ and $M_{n_2} \to M_{m_2}$ in \protect\subref{fig bratteliDiagram} are not lifted in order to lighten the drawing.}
\end{figure}
For any $v' = (i, p', j) \in \GammaA[,\bn_i \bn_j]^{(0)}$, we define 
\begin{align*}
	\iota_{v}^{v'} 
	: \hsiA[,v'] \to \hsiA[,v] 
	\qquad\text{as}\qquad
	\iota_{v}^{v'}(\xi_{i, \alpha}^{(1)} \otimes \vm_{i j}^{p'} \otimes \eta_{j, \alpha}^{\circ (2)}) = \xi_{i, \alpha}^{(1)} \otimes \vm_{i j}^{p} \otimes \eta_{j, \alpha}^{\circ (2)}.
\end{align*}
\begin{proposition}
	\label{prop scalar products}
	Let $v, v' \in \GammaA^{(0)}$, $w \in \GammaB^{(0)}$, and $\psi_{v} \in \hsiA[,v]$ and $\psi'_{v'} \in \hsiA[,v']$. 
	
	When $\pi_{\lambda\rho}(v) \neq \pi_{\lambda\rho}(v')$, one has $\langle \phiH[,w]^{v}(\psi_{v}), \phiH[,w]^{v'}(\psi'_{v'}) \rangle_{\hsiB[,w]} = 0$. 
	
	When $\pi_{\lambda\rho}(v) = \pi_{\lambda\rho}(v')$, one has
	\begin{align*}
		\langle \phiH[,w]^{v}(\psi_{v}), \phiH[,w]^{v'}(\psi'_{v'}) \rangle_{\hsiB[,w]}
		&=
		\langle \psi_{v}, \iota_{v}^{v'}(\psi'_{v'}) \rangle_{\hsiA[,v]} \, \tr( u(v,w)^\ast u(v',w) )
	\end{align*}
	In particular, for any $\psi_{v} \in \hsiA[,v]$ and $\psi'_{v'} \in \hsiA[,v]$, one has
	\begin{align}
		\norm{\phiH[,w]^{v}(\psi_{v})}_{\hsiB[,w]} 
		&= 
		\norm{\psi_{v}}_{\hsiA[,v]} \norm{u(v,w)}_{F} \nonumber
		\\
		\langle \phiH^{v}(\psi_{v}), \phiH^{v'}(\psi'_{v'}) \rangle_{\hsiB} 
		&= 
		\langle \psi_{v}, \iota_{v}^{v'}(\psi'_{v'}) \rangle_{\hsiA[,v]} \Big( \tsum_{w \in \GammaB^{(0)}} \tr( u(v,w)^\ast u(v',w) )  \Big) \label{eq scalar production phiH u}
	\end{align}
	where $\norm{-}_{F}$ is the Frobenius norm on matrices, defined as $\norm{A}_{F}^2 \defeq \tr(A^\ast A)$. This implies that $\phiH^{v} : \hsiA[,v] \to \hsiB$ is one-to-one if and only if $\tsum_{w \in \GammaB^{(0)}} \norm{u(v,w)}^2_{F} > 0$.
\end{proposition}

\begin{proof}
	From a previous remark, the scalar product is zero when $\pi_{\lambda\rho}(v) \neq \pi_{\lambda\rho}(v')$. So, suppose $\pi_{\lambda\rho}(v) = \pi_{\lambda\rho}(v')$. Let $i = i(v) = i(v')$ and $j = j(v) = j(v')$ and consider $\psi_{v} = \xi_{i} \otimes \eta_j^\circ$ and $\psi'_{v'} = \xi'_{i} \otimes \eta_j'^\circ$, so that $\phiH[,w]^{v}(\psi_{v}) = \xi_i \otimes u(v,w) \otimes \eta_j^\circ$ and $\phiH[,w]^{v'}(\psi'_{v'}) = \xi'_i \otimes u(v',w) \otimes \eta_j'^\circ$ both in $\bbC^{n_i} \otimes M_{\alpha_{k i} \times \alpha_{\ell j}} \otimes \bbC^{n_j} \simeq \bbC^{n_i} \otimes \bbC^{\alpha_{k i}} \otimes \bbC^{\alpha_{\ell j}} \otimes \bbC^{n_j} \subset \bbC^{m_k} \otimes \bbC^{m_\ell \circ}$. Then $\langle \phiH[,w]^{v}(\psi_{v}), \phiH[,w]^{v'}(\psi'_{v'}) \rangle_{\hsiB[,w]} = \langle \xi_i, \xi'_i \rangle_{\bbC^{n_i}} \langle \eta_j, \eta'_j \rangle_{\bbC^{n_j}} \tr(u(v,w)^\ast u(v',w))$ where the trace factor is obtained from the identification of $M_{\alpha_{k i} \times \alpha_{\ell j}}$ with $\bbC^{\alpha_{k i}} \otimes \bbC^{\alpha_{\ell j}}$ and we have used the fact that $\lambda(v) = \lambda(v') = n_i$ and $\rho(v) = \rho(v') = n_j$ to write the scalar products. This implies the formula in terms of the scalar product on $\hsiA[,v]$, from which we deduce the relations on the norm in $\hsiB[,w]$ and on the scalar product in $\hsB$. This last relation implies the norm relation $\norm{\phiH^{v}(\psi_{v})}^2_{\hsiB} = \norm{\psi_{v}}^2_{\hsiA[,v]} \Big( \tsum_{w \in \GammaB^{(0)}} \norm{u(v,w)}^2_{F}  \Big)$. Then, suppose $\tsum_{w \in \GammaB^{(0)}} \norm{u(v,w)}^2_{F} > 0$: if $\psi_{v} \in \hsiA[,v]$ is such that $\phiH^{v}(\psi_{v}) = 0$, then $\norm{\psi_{v}}^2_{\hsiA[,v]} = 0$, so that $\psi_{v} = 0$, that is, $\phiH^{v}$ is one-to-one. Suppose $\tsum_{w \in \GammaB^{(0)}} \norm{u(v,w)}^2_{F} = 0$, then $\norm{\phiH^{v}(\psi_{v})}^2_{\hsiB} = 0$ for any $\psi_{v} \in \hsiA[,v]$, so that $\phiH^{v} = 0$, that is, $\phiH^{v}$ is not one-to-one.
\end{proof}

Notice that the condition $\tsum_{w \in \GammaB^{(0)}} \norm{u(v,w)}^2_{F} > 0$ for any $v \in \GammaA^{(0)}$ does not implies that $\phiH$ is one-to-one: one can consider a situation where, for $v, v' \in \GammaA^{(0)}$ such that $\pi_{\lambda\rho}(v) = \pi_{\lambda\rho}(v')$, $\psi_{v} \in \hsiA[,v]$, and $\psi'_{v'} \in \hsiA[,v']$ , $\phiH[,w]^v(\psi_{v}) + \phiH[,w]^{v'}(\psi'_{v'}) = 0 \in \hsiB[,w]$ for some $w \in \GammaB^{(0)}$.

\medskip
From \eqref{eq scalar production phiH u}, it is natural to define, for any  $v, v' \in \GammaA^{(0)}$,
\begin{align*}
	\spm^{v_1, v_2} &\defeq
	\begin{cases}
		0 & \text{if $\pi_{\lambda\rho}(v_1) \neq \pi_{\lambda\rho}(v_2)$}\\
		\tsum_{w \in \GammaB^{(0)}} \tr( u(v_1,w)^\ast u(v_2,w) ) & \text{if $\pi_{\lambda\rho}(v_1) = \pi_{\lambda\rho}(v_2)$}
	\end{cases}
\end{align*}
so that \eqref{eq scalar production phiH u} can be written as $\langle \phiH^{v_1}(\psi_{v_1}), \phiH^{v_2}(\psi'_{v_2}) \rangle_{\hsiB} = \langle \psi_{v_1}, \iota_{v_1}^{v_2}(\psi'_{v_2}) \rangle_{\hsiA[,v_1]} \, \spm^{v_1, v_2}$. 
\medskip
\par
With $\pi_{\lambda\rho}(v_1) = \pi_{\lambda\rho}(v_2) = (\bn_i, \bn_j)$, one has $\spm^{v_2, v_1} = \tsum_{w \in \GammaB^{(0)}} \tr( u(v_2,w)^\ast u(v_1,w) ) = \tsum_{w \in \GammaB^{(0)}} \overline{\tr( u(v_1,w)^\ast u(v_2,w) )} = \overline{\spm^{v_1,v_2}}$, so that $(\spm^{v_1,v_2})_{v_1,v_2}$ is a Hermitian matrix, so that this matrix can be diagonalized. Recall that the labels $v_1, v_2$ depends on the choices of the orthonormal bases $\{ \vm_{i j}^{p} \}_{1 \leq p \leq \mu_{i j}}$ of the spaces $\bbC^{\mu_{i j}}$'s: this diagonalization (see proof of Prop.~\ref{prop diagonalization spm for KO dims}) is related to a change of these bases. This leads us to introduce the following Hypothesis.

\begin{hypothesis}
	\label{hyp phiH diagonalization}
	We suppose that $\phiH$ is one-to-one and is such that there are orthonormal bases $\{ \vm_{i j}^{p} \}_{1 \leq p \leq \mu_{i j}}$ of the spaces $\bbC^{\mu_{i j}}$ which conform to Prop.~\ref{prop basis odd case} (in the odd case) or Prop.~\ref{prop basis even case} (in the even case), and such that, for the decomposition of $\hsA$ induced by these bases, $\spm^{v_1, v_2} = \spe_{v_1} \delta^{v_1, v_2}$ when $\pi_{\lambda\rho}(v_1) = \pi_{\lambda\rho}(v_2)$, with real numbers $\spe_{v}$ such that $\spe_{\JimA(v)} = \spe_{v}$.
\end{hypothesis}

A direct consequence of these assumptions is that $\langle \phiH^{v_1}(\psi_{v_1}), \phiH^{v_2}(\psi'_{v_2}) \rangle_{\hsiB} = 0$ for any $v_1 \neq v_2$ and  $\langle \phiH^{v}(\psi_{v}), \phiH^{v}(\psi'_{v}) \rangle_{\hsiB} = \spe_{v} \, \langle \psi_{v}, \psi'_{v} \rangle_{\hsiA[,v]}$ for any $v$. The one-to-one requirement is natural in the context of $AF$-algebras, since it generalizes the one-to-one requirement on $\phi$. On the other hand, the diagonalization requirement is not mandatory for the formal developments to come, but it will be useful to compare spectral actions for $\phi$-compatible spectral triples on $\algA$ and $\algB$ in Sect.~\ref{sec spectral actions AF AC manifold}. Moreover, this requirement is satisfied for some $KO$-dimensions:

\begin{proposition}
	\label{prop diagonalization spm for KO dims}
	Suppose that $\JB$ is strong $\phi$-compatible with $\JA$, and, in the even case, that $\gammaB$ is $\phi$-compatible with $\gammaA$. Then, in $KO$-dimensions $0,1,2,6,7$, the diagonalization requirement in Hypothesis~\ref{hyp phiH diagonalization} is satisfied for any $\phiH$. 
\end{proposition}

\begin{proof}
	Let $\{ \vm_{i j}^{p} \}_{1 \leq p \leq \mu_{i j}}$ be orthonormal bases of the spaces $\bbC^{\mu_{i j}}$ which satisfy Prop.~\ref{prop basis odd case} (in the odd case) or Prop.~\ref{prop basis even case} (in the even case). Let us first complete the notations introduced before Prop.~\ref{prop scalar products}, where we have introduced the identification $v = (i,p,j)$. With this notation, we define $\JimA(v) = (j, \Bp, i)$ where $\Bp = 1, \dots, \mu_{ji} = \mu_{ij}$ and $\bar{\Bp} = p$ (obviously, this bar is not to be confused with a complex conjugation). 
	\medskip
	\par
	Let $\{ \wm_{k \ell}^{q} \}_{1 \leq q \leq \nu_{k\ell}}$ be orthonormal bases of the spaces $\bbC^{\nu_{k\ell}}$ where $\nu_{k\ell}$ are the multiplicity of the irreps $\hsB[, \bem_k \bem_\ell]$ in $\hsB$. These bases define the irreps $\hsB[,w]$ for $w = (k,q,\ell) \in \GammaB[,\bem_k \bem_\ell]^{(0)}$ as in \eqref{hsv from vmijp basis}. We have written the map $\phiH[,w]^{v} : \hsA[,v] \to \hsB[,w]$ in terms of a matrix $u(v,w)$. It is convenient to write $\phiH[,w]^{v}$ explicitly in terms of the bases $\{ \vm_{i j}^{p} \}$ and $\{ \wm_{k \ell}^{q} \}$. In order to avoid cumbersome notations, we use the identification $\bbC^{n_i} \otimes \bbC^{\mu_{i j}} \otimes \bbC^{n_j \circ} \simeq \bbC^{n_i} \otimes \bbC^{n_j \circ} \otimes \bbC^{\mu_{i j}}$ (resp. $\bbC^{m_k} \otimes \bbC^{\nu_{k\ell}} \otimes \bbC^{m_\ell \circ} \simeq \bbC^{m_k} \otimes \bbC^{m_\ell \circ} \otimes \bbC^{\nu_{k\ell}}$) so that $\vm_{i j}^{p}$ (resp. $\wm_{k \ell}^{q}$) will appear on the right in the tensor products. Then we can replace the notation $u(v,w)$ by the notation $\Mu_{k\ell,q}^{ij,p} \in M_{\alpha_{k i} \times \alpha_{\ell j}}$ which refers to the bases $\{ \vm_{i j}^{p} \}$ and $\{ \wm_{k \ell}^{q} \}$ for which, similarly to \eqref{eq phiH Iklij xi eta}, one has $\phiH[,w]^{v} (\xi_i \otimes \eta_j^\circ \otimes \vm_{i j}^{p}) = I_{k,\ell}^{i,j} \big( \xi_i \otimes \Mu_{k\ell,q}^{ij,p} \otimes \eta_j^\circ \big) \otimes \wm_{k \ell}^{q}$ (no summation). In the $p$ and $q$ variables, $\spm^{v_1, v_2}$ with $\pi_{\lambda\rho}(v_1) = \pi_{\lambda\rho}(v_2) = (\bn_i, \bn_j)$ then becomes $\hspm^{p_1, p_2}_{ij} = \tsum_{k, \ell, q} \tr( (\Mu^{ij,p_2}_{k\ell,q})^\ast \Mu^{ij,p_1}_{k\ell,q} )$ for $p_1, p_2 = 1, \dots, \mu_{ij}$. Notice the switch $1 \leftrightarrow 2$ which will be convenient later. Since we suppose that $\JB$ is strong $\phi$-compatible with $\JA$, by Prop.~\ref{prop JB JA strong phi compatibiliy relation on ukappa} one get \eqref{eq ukappa epsilon ustar} in terms of the new notations: $\Mu_{\ell k,\Bq}^{ji,\Bp} = \tfrac{\epsilonA(i,p,j,d_\algA)}{\epsilonB(k,\ell,q,d_\algB)} (\Mu_{k\ell,q}^{ij,p})^\ast$. This implies that
	\begin{align}
		\hspm^{\Bp_1, \Bp_2}_{ji} 
		&=
		\tsum_{\ell, k, \Bq} \tr( (\Mu^{ji,\Bp_2}_{\ell k,\Bq})^\ast \Mu^{ji,\Bp_1}_{\ell k,\Bq} )
		= \tsum_{k, \ell, q}  \tfrac{\epsilonA(i,p_2,j,d_\algA) \epsilonA(i,p_1,j,d_\algA)}{\epsilonB(k,\ell,q,d_\algB)^2} \tr( \Mu^{ij,p_2}_{k\ell,q} (\Mu^{ij,p_1}_{k\ell,q})^\ast ) \nonumber
		\\
		\label{eq hspmBp1Bp2ji = hspmp2p2ij}
		&= \epsilonA(i,p_1,j,d_\algA) \epsilonA(i,p_2,j,d_\algA) \hspm^{p_2, p_1}_{ij} 
	\end{align}
	
	In the following, we fix the couple $(i,j)$. Let us introduce a change of bases $\{ \vm_{i j}^{p} \}$ to $\{ \vm'^{p'}_{i j} \}$ in $\bbC^{\mu_{i j}}$, where $\vm'^{p'}_{i j} = \tsum_{p} u^{p'p} \vm_{i j}^{p}$ for a unitary matrix $U = (u^{p'p})_{p',p}$. Then a straightforward computation shows that the matrices $\Mu'^{ij,p'}_{k\ell,q}$ defined relatively to the bases $\{ \vm'^{p'}_{i j} \}$ and $\{ \wm_{k \ell}^{q} \}$ are $\Mu'^{ij,p'}_{k\ell,q} = \tsum_{p} u^{p'p} \Mu^{ij,p}_{k\ell,q}$, and the associated $\hspm'^{p'_1, p'_2}_{ij} = \tsum_{k, \ell, q} \tr( (\Mu'^{ij,p'_2}_{k\ell,q})^\ast \Mu'^{ij,p'_1}_{k\ell,q} )$ become
	\begin{align*}
		\hspm'^{p'_1, p'_2}_{ij}
		&= \tsum_{k, \ell, q} 
		\tsum_{p_1, p_2} \overline{u^{p'_2 p_2}} u^{p'_1 p_1}\, 
		\tr( (\Mu^{ij,p_2}_{k\ell,q})^\ast \Mu^{ij,p_1}_{k\ell,q} )
		= \tsum_{p_1, p_2} \overline{u^{p'_2 p_2}} u^{p'_1 p_1}\, \hspm^{p_1, p_2}_{ij},
	\end{align*}
	so that $\hspm'_{ij} = U \hspm_{ij} U^\ast$ with $U^\ast = (\overline{u^{p p'}})_{p',p}$ (here we use the switch $1 \leftrightarrow 2$ mentioned before). Since $\hspm_{ij}$ is a Hermitian matrix, there is a unitary matrix $U$ such that $\hspm'_{ij} = U \hspm_{ij} U^\ast$ is diagonal with real eigenvalues $\spe_{ij}^p = \spe_{v}$. So, for the new basis $\{ \vm'^{p'}_{i j} \}$ of $\bbC^{\mu_{i j}}$ defined by $U$, $\hspm'_{ij}$, and so $(\spm^{v_1, v_2})_{v_1,v_2}$, is diagonal.
	\medskip
	\par
	Let us now look how this diagonalization can be performed according to the constraints given in proposition \ref{prop basis odd case} (in the odd case) or Prop.~\ref{prop basis even case} (in the even case). The first constraint, common to Prop.~\ref{prop basis odd case} and \ref{prop basis even case}, is $\vm_{j i}^{\Bp} = \epsilonA(i, p, j ,d_\algA) L_{i j}(\Bvm_{i j}^{p})$ for any $p$, where $\epsilonA(i, p, j ,d_\algA) = \epsilonA(v ,d_\algA)$ is defined in \eqref{eq epsilon(v, d)}.
	\medskip
	\par
	Let us first consider the case $i<j$ (for any $KO$-dimension), for which $\epsilonA(i, p, j ,d_\algA)=1$ and $\Bp = p$, so that, from \eqref{eq hspmBp1Bp2ji = hspmp2p2ij}, one has $\hspm^{p_1, p_2}_{ji} = \hspm^{p_2, p_1}_{ij}$: $\hspm_{ji}$ is the transpose of $\hspm_{ij}$. Since this result is true in any basis of $\bbC^{\mu_{i j}}$, this implies $\hspm'_{ji} = \BU  \hspm_{ji} \BU^\ast$. On the other hand, $\vm'^{p'}_{j i} =  L_{i j}(\Bvm'^{p'}_{i j}) = \tsum_{p} \overline{u^{p'p}} L_{i j}(\Bvm_{i j}^{p}) = \tsum_{p} \overline{u^{p'p}} \vm^{p}_{j i}$, so that the change of bases from $\{ \vm^{p}_{j i} \}$ to $\{ \vm'^{p'}_{j i} \}$ in $\bbC^{\mu_{j i}}$ is performed by the unitary matrix $\BU$. From these two compatible relations, one concludes that the change of basis defined by $U$ in  $\bbC^{\mu_{i j}}$ which diagonalizes $\hspm'_{ij}$ automatically induces a change of bases $\BU$ in $\bbC^{\mu_{j i}}$ which diagonalizes $\hspm'_{ji}$. Notice then that the eigenvalues $\spe_{ji}^p$ in $\hspm'_{ji}$ are the same as the eigenvalues $\spe_{ij}^p$ in $\hspm'_{ij}$, so that $\spe_{\JimA(v)} = \spe_{v}$.
	\medskip
	\par
	Let us now consider $i=j$ in $KO$-dimensions $0,1,7$. Then, as before, $\epsilonA(i, p, i ,d_\algA)=1$ and $\Bp = p$, so that, from \eqref{eq hspmBp1Bp2ji = hspmp2p2ij}, one has  $\hspm^{p_1, p_2}_{ii} = \hspm^{p_2, p_1}_{ii}$, and we already know that $\hspm^{p_1, p_2}_{ii} = \overline{\hspm^{p_2, p_1}_{ii}}$: the matrix $\hspm_{ii}$ is a real symmetric matrix, and the diagonalizing matrix $U$ can be chosen to be an orthogonal matrix (so a real matrix). This result is compatible with the required condition $\vm_{ii}^{p} = L_{ii}(\Bvm_{ii}^{p})$ on the basis since $\vm'^{p'}_{ii} = L_{ii}(\Bvm'^{p'}_{ii}) = \tsum_{p} \overline{u^{p'p}} L_{ii}(\Bvm_{ii}^{p}) = \tsum_{p} \overline{u^{p'p}} \vm_{ii}^{p} = \tsum_{p} u^{p'p} \vm_{ii}^{p}$. Here, it is trivial that $\spe_{\JimA(v)} = \spe_{v}$ since $\JimA(v) = v$.
	\medskip
	\par
	Finally, consider $i=j$ in $KO$-dimensions $2,3,4,5,6$. In that situation, if $p=2a$ (resp. $p=2a-1$) then $\Bp = 2a -1$ (resp. $\Bp = 2a$), and  $\epsilonA(i, 2a-1, i ,d_\algA)=1$ and  $\epsilonA(i, 2a, i ,d_\algA)=\epsilonA$. The matrix $\hspm_{ii}$ is a  block matrix $\smallpmatrix{\hspm_{ii}^{e,e} & \hspm_{ii}^{e,o} \\ \hspm_{ii}^{o,e} & \hspm_{ii}^{o,o} }$ where $o$ and $e$ stand for odd and even: for instance $\hspm_{ii}^{e,e} = \left( \hspm_{ii}^{2a_1, 2a_2} \right)$ and  $\hspm_{ii}^{e,o} = \left( \hspm_{ii}^{2a_1, 2a_2-1} \right)$ with $a_1, a_2=1, \dots, \mu_{ii}/2$. Then, from \eqref{eq hspmBp1Bp2ji = hspmp2p2ij}, one has $\hspm_{ii}^{2a_1, 2a_2} = \hspm_{ii}^{2a_2-1, 2a_1-1}$, $\hspm_{ii}^{2a_1, 2a_2-1} = \epsilonA \hspm_{ii}^{2a_2, 2a_1-1}$, and $\hspm_{ii}^{2a_1-1, 2a_2} = \epsilonA \hspm_{ii}^{2a_2-1, 2a_1}$. Considering these block matrices as matrices indexed by $a_1, a_2$, this means that $\hspm_{ii}^{e,e} = {\hspm_{ii}^{o,o}}^\top$, $\hspm_{ii}^{e,o} = \epsilonA {\hspm_{ii}^{e,o}}^\top$, and $\hspm_{ii}^{o,e} = \epsilonA {\hspm_{ii}^{o,e}}^\top$. Since $\hspm_{ii}$ is Hermitian, one also has $\hspm_{ii}^{e,e} = {\hspm_{ii}^{e,e}}^\ast$ and $\hspm_{ii}^{e,o} = {\hspm_{ii}^{o,e}}^\ast$.
	\medskip
	\par
	In $KO$-dimensions $3,4,5$, one has $\epsilonA = -1$, so that $\hspm_{ii}^{e,o} = - {\hspm_{ii}^{e,o}}^\top = {\hspm_{ii}^{o,e}}^\ast$, which implies that $\hspm_{ii}^{e,o}$ and $\hspm_{ii}^{o,e}$ are antisymmetric matrices. We report the analysis for $KO$-dimensions $2,6$ after the following considerations.
	\medskip
	\par
	In the even case, since $\gammaB$ is $\phi$-compatible with $\gammaA$, from Lemma~\ref{lemma matrices u(v,w)}, $u(v,w)$ is non-zero only when $s(v) = s(w)$, so that the sum defining $\spm^{v_1,v_2}$ implies $s(w) = s(v_1) = s(v_2)$. The matrix $(\spm^{v_1, v_2})_{v_1,v_2}$ is then block diagonal along the decomposition $s(v) = \pm 1$, and its diagonalization can be done by blocks: in terms of the change of bases in $\bbC^{\mu_{i j}}$, this means that the unitary $U$ introduced above which diagonalizes $\hspm_{ij}$ can be chosen to preserve the eigenspaces defined by the maps $\ell_{ij}$ in Prop~\ref{prop basis even case}. The decomposition along $s(v) = \pm 1$ is preserved by $\JimA$ since $s_{ji}^{\Bp} = \epsilonA'' s_{ij}^{p}$: so all the previous developments are compatible with this choice for $U$.
	\medskip
	\par
	In the case  $i=j$ and $KO$-dimensions $2,6$, from Prop.~\ref{prop basis even case}, one has $s(i,2a,i) = 1$ and $s(i,2a-1,i) = -1$, so that the block decomposition $\smallpmatrix{\hspm_{ii}^{e,e} & \hspm_{ii}^{e,o} \\ \hspm_{ii}^{o,e} & \hspm_{ii}^{o,o} }$ corresponds to the block decomposition along  $s(v) = \pm 1$, and from the previous considerations, this implies that  $\hspm_{ii}^{e,o} = \hspm_{ii}^{o,e} = 0$. Since $\hspm_{ii}^{e,e}$ is Hermitean, there is a unitary matrix $\TU$ such that $\TU \hspm_{ii}^{e,e} \TU^\ast$ is diagonal, and then by transposition, $\BTU \hspm_{ii}^{o,o} \BTU^\ast$ is also diagonal with the same eigenvalues, that is, $\spe_{\JimA(v)} = \spe_{v}$. The unitary $U = \smallpmatrix{\TU & 0 \\ 0 & \BTU }$ diagonalizes $\hspm_{ii}$ and this diagonalization is compatible with the required conditions $\vm_{ii}^{2a} = L_{ii}(\Bvm_{ii}^{2a-1})$ and $\vm_{ii}^{2a-1} = \epsilonA L_{ii}(\Bvm_{ii}^{2a})$: $\TU = ( \Tu^{a',a})$ (resp. $\BTU = ( \BTu^{a',a})$) induces a change of the sub-basis $\{ \vm_{ii}^{2a} \}$ to $\{ \vm'^{2a}_{ii} \}$ (resp. $\{ \vm_{ii}^{2a-1} \}$ to $\{ \vm'^{2a-1}_{ii} \}$) with $\vm'^{2a'}_{ii} = \tsum_{a} \Tu^{a',a} \vm_{ii}^{2a}$ (resp.  $\vm'^{2a'-1}_{ii} = \tsum_{a} \BTu^{a',a} \vm_{ii}^{2a-1}$). The required condition is satisfied since then $\vm'^{2a'-1}_{ii} = \epsilonA L_{ii}(\Bvm'^{2a'}_{ii}) = \epsilonA  \tsum_{a} \BTu^{a',a} L_{ii}(\Bvm^{2a}_{ii}) = \tsum_{a} \BTu^{a',a} \vm_{ii}^{2a-1}$.
\end{proof}

We suspect that the diagonalization property proved in Prop.~\ref{prop diagonalization spm for KO dims} could be true also in $KO$-dimensions $3,4,5$. But we were unable to prove this fact. Nevertheless, the proposition fortunately covers the $KO$-dimension $6$ used in the finite part of the spectral triple for the NC version of the Standard Model of Particles Physics, see \cite{ChamConnMarc07a} and \cite{Suij15a} for instance.

\begin{proposition}
	\label{prop KO dim AF phi compatibility}
	If two (odd/even) real spectral triples are $\phi$-compatible and $\phiH$ is such that \eqref{eq ukappa epsilon ustar} holds, then they have the same $KO$-dimension (mod 8).
\end{proposition}

\begin{proof}
	Since $\phiH$ satisfies \eqref{eq ukappa epsilon ustar}, by Prop.~\ref{prop JB JA strong phi compatibiliy relation on ukappa}, $\JB$ is strong $\phi$-compatible with $\JA$ and is diagonal. By Lemma~\ref{lemma gammaB gammaA phi compatibiliy diagonal}, $\gammaB$ is strong $\phi$-compatible with $\gammaA$ and is diagonal. The difference with Prop.~\ref{prop KO dim strong phi compatibility}, is that $\DB$ is only $\phi$-compatible with $\DA$. So, we already get $\epsilonB =  \epsilonA$ and $\epsilonB'' = \epsilonA''$: it remains to consider $\epsilonA'$ and $\epsilonB'$. 
	\medskip
	\par
	Since $\JB$ is diagonal, one has $\JB \DB = \smallpmatrix{ \JB[,\phi]^{\phi} \DB[,\phi]^{\phi} & \JB[,\phi]^{\phi} \DB[,\phi]^{\perp} \\ \JB[,\perp]^{\perp} \DB[,\perp]^{\phi} & \JB[,\perp]^{\perp} \DB[,\perp]^{\perp} }$ and $\DB \JB = \smallpmatrix{ \DB[,\phi]^{\phi} \JB[,\phi]^{\phi} & \DB[,\phi]^{\perp} \JB[,\perp]^{\perp} \\ \DB[,\perp]^{\phi} \JB[,\phi]^{\phi} & \DB[,\perp]^{\perp} \JB[,\perp]^{\perp} }$, so that $\JB[,\phi]^{\phi} \DB[,\phi]^{\phi} = \epsilonB' \DB[,\phi]^{\phi} \JB[,\phi]^{\phi}$. Inserting this relation in the $\phi$-compatibility conditions on $\JB$ and $\DB$ implies $\epsilonB' = \epsilonA'$.
\end{proof}

From Prop.~\ref{prop st phi comp and phi comp of operators for AF}, the strong $\phi$-compatibility condition between $\DB$ and $\DA$ is equivalent to
\begin{align*}
	\tsum_{v_2 \in \GammaA^{(0)}}\phiH[,w_2]^{v_2}(\DA[,(v_1, v_2)] \psi_{v_1})
	&=
	\tsum_{w_1 \in \GammaB^{(0)}} \DB[,(w_1,w_2)] \phiH[,w_1]^{v_1}(\psi_{v_1})
\end{align*}
for any $v_1 \in \GammaA^{(0)}$, $w_2 \in \GammaB^{(0)}$, and $\psi_{v_1} \in \hsiA[,v_1]$, and the $\phi$-compatibility condition is equivalent to
\begin{align*}
	\tsum_{v_2 \in \GammaA^{(0)}}\phiH[,w_2]^{v_2}(\DA[,(v_1, v_2)] \psi_{v_1})
	&=
	\tsum_{w_1 \in \GammaB^{(0)}} \DB[, \phi,(w_1,w_2)]^{\phi} \phiH[,w_1]^{v_1}(\psi_{v_1})
\end{align*}
where $\DB[, \phi,(w_1,w_2)]^{\phi} : \hsiB[,w_1] \cap \phiH(\hsA) \to \hsiB[,w_2] \cap \phiH(\hsA)$. Unfortunately, from this relation, we cannot define the elementary operators $\DB[, \phi,(w_1,w_2)]^{\phi}$ in terms of the elementary operators $\DA[,(v_1, v_2)]$. Only the operators $\tsum_{w_1 \in \GammaB^{(0)}} \DB[, \phi,(w_1,w_2)]^{\phi} : \toplus_{w_1 \in \GammaB^{(0)}} \hsB[, w_1] \cap \phiH(\hsA) \to \hsB[, w_2] \cap \phiH(\hsA)$ can be recovered from the $\DA[,(v_1, v_2)]$'s.

\section{\texorpdfstring{Spectral Actions for $AF$-AC Manifolds}{Spectral actions for AF-AC Manifolds}}
\label{sec spectral actions AF AC manifold}

Given a spectral action $(\algA, \hsA, \DA, \JA, \gammaA)$ for a finite dimensional algebra $\algA$ and given a Riemannian spin manifold $(M,g)$ equipped with its canonical spectral triple $(C^\infty(M), L^2(\SpinBun), \DM, \JM, \gammaM)$, we consider the spectral triple $(\halgA \defeq C^\infty(M) \otimes \algA , \hshA \defeq L^2(\SpinBun) \otimes \hsA, \DhA \defeq \DM \otimes \bbbone + \gammaM \otimes \DA, \JhA \defeq \JM \otimes \JA, \gammahA \defeq \gammaM \otimes \gammaA)$ over the Almost Commutative algebra $\halgA$. Obviously, this makes sense only when the $KO$-dimension for $M$ and $\algA$ produces such a spectral triple, see for instance \cite{DabrDoss11a}.
\medskip
\par
Then, given two spectral triples $(\algA, \hsA, \DA, \JA, \gammaA)$ and $(\algB, \hsB, \DB, \JB, \gammaB)$ for two finite dimensional algebras $\algA$ and $\algB$, and a one-to-one morphism $\phi : \algA \to \algB$ such that the two spectral triples are $\phi$-compatible, with $\JB$ strong $\phi$-compatible with $\JA$, the aim of this section is to compare the spectral actions on $\halgA$ and $\halgB$ (for the same Riemannian spin manifold $(M,g)$). 
\medskip
\par
In order to have a good physical interpretation of the $\phi$-compatibility, in particular at the level of the fermions, we first need to introduce a “normalized” $\phiH$ map.

\subsection{Normalized $\phiH$ Map}
\label{subsec normalized phiH map}

Denote by $\phiH^0 : \hsA \to \hsB$ a given morphism as in Def.~\ref{def phi phiH}. We suppose that it satisfies Hypothesis~\ref{hyp phiH diagonalization}. Then we can choose the orthonormal bases $\{ \vm_{i j}^{p} \}_{1 \leq p \leq \mu_{i j}}$ of $\bbC^{\mu_{i j}}$ that diagonalize $(\spm^{v_1, v_2})_{v_1,v_2}$ and this implies that $\{ \phiH^0(e_{\Tv}) \}_{\Tv \in \TGammaA^{(0)}}$ is a basis of $\phiH^0(\hsA)$. For any $\Tv = (v,\alpha)$, we  can identify $\phiH^0(e_{\Tv})$ with $\phiH^{0,v}(e_{\Tv})$. Let $\Tv_1 = (v_1, \alpha_1)$ and $\Tv_2 = (v_2, \alpha_2)$. When $v_1 \neq v_2$, one has $\langle \phiH^{0,v}(e_{\Tv_1}), \phiH^{0,v}(e_{\Tv_2}) \rangle_{\hsiB} = 0$, while,  when $v = v_1 = v_2$, $\langle \phiH^{0,v}(e_{\Tv_1}), \phiH^{0,v}(e_{\Tv_2}) \rangle_{\hsiB} = \spe_{v} \, \langle \Tv_1, \Tv_2 \rangle_{\hsiA[,v]} = \spe_{v} \, \delta_{\alpha_1,\alpha_2}$. This implies that $\{ \phiH^0(e_{\Tv}) \}_{\Tv \in \TGammaA^{(0)}}$ is an orthogonal family. Since $\phiH^0$ is one-to-one and $\norm{\phiH^{0,v}(e_{\Tv})}^2 = \spe_{v}$, one has $\spe_{v} \neq 0$ for any $v \in \GammaA^{(0)}$.

\begin{definition}
	The normalized $\phiH$ map associated to the map $\phiH^0 : \hsA \to \hsB$ which satisfies Hypothesis~\ref{hyp phiH diagonalization} is the map $\phiH : \hsA \to \hsB$ defined by
	\begin{align}
		\label{NormedPhim}
		\phiH( \toplus_{v \in \GammaA^{(0)}} \psi_v) 
		\defeq \tsum_{v \in \GammaA^{(0)}} \spe_{v(\Tv)}^{-1/2} \phiH^{0,v}(\psi_v)
	\end{align}
	Using \eqref{eq phiH w v a b psiv} with $b_j = 0$, one can check that $\phiH$ satisfies Def.~\ref{def phi phiH}.
\end{definition}

The normalization has been chosen such that the family $\{ f_{\Tv} \defeq \phiH(e_{\Tv}) \}_{\Tv \in \TGammaA^{(0)}}$ is an orthonormal basis of $\phiH(\hsA)$. This basis of $\phiH(\hsA)$ can be completed with any orthonormal basis $\{ f_{\hw} \}_{\hw \in \hGammaB^{(0)}}$ of $\phiH(\hsA)^\perp$ where $\hGammaB^{(0)}$ is any index set for this basis. So, $\{ f_{\Tv} \}_{\Tv \in \TGammaA^{(0)}} \cup \{ f_{\hw} \}_{\hw \in \hGammaB^{(0)}}$ is an orthonormal basis of $\hsB$ adapted to the decomposition $\phiH(\hsA) \oplus \phiH(\hsA)^\perp$.
\medskip
\par
We now consider the normalized $\phiH$ map in place of $\phiH^0$. The relation between the scalar products in $\hsB$ and $\hsA$ reduces to the simple relation $\langle \phiH^{v}(\psi_{v}), \phiH^{v}(\psi'_{v}) \rangle_{\hsiB} = \langle \psi_{v}, \psi'_{v} \rangle_{\hsiA[,v]}$ for any $v$.
\medskip
\par
In the following, $\phi$-compatibility of operators will be relative to the normalized $\phiH$ map.
\medskip
\par
For an operator $B$ on $\hsB$ which is $\phi$-compatible with an operator $A$ on $\hsA$, the components $B_{\phi}^{\perp}$, $B_{\perp}^{\phi}$, and $B_{\perp}^{\perp}$ of $B$ in the $2\times 2$ matrix decomposition induced by $\hsB = \phiH(\hsA) \oplus \phiH(\hsA)^\perp$ will be called non-inherited, while the component $B_{\phi}^{\phi}$ will be called inherited. Let us use the acronym “\TNIC” for “Terms with Non-Inherited Components” in the following technical results, which are the main interest for the use of the normalized $\phiH$ map:
\begin{lemma}
	\label{lemma scalar product and trace phi compatible}
	For $i=1,\ldots,n$, let $B_i$ be an operator on $\hsB$ which is $\phi$-compatible with an operator $A_i$ on $\hsA$.
	\begin{enumerate}
		\item For any $\Tv_1, \Tv_2 \in \TGammaA^{(0)}$, one has\label{item scalar product of product}: $\qquad\qquad\langle f_{\Tv_1}, B_1 \cdots B_n f_{\Tv_2} \rangle_{\hsB}= \langle e_{\Tv_1}, A_1 \cdots A_n e_{\Tv_2} \rangle_{\hsA} + \text{\TNIC}$		
		\item As a consequence, one has\label{item trace of product}: $\quad\qquad\qquad\tr(B_1 \cdots B_n) = \tr(A_1 \cdots A_n) + \text{\TNIC}$
	\end{enumerate}
\end{lemma}

\begin{proof}
	First, let us prove the relation in Point~\ref{item scalar product of product} for $n=1$. We omit the index $i$. Using the matrix decomposition $A e_{\Tv} = \tsum_{\Tv' \in \TGammaA^{(0)}} A_{\Tv}^{\Tv'} e_{\Tv'}$ along the basis $\{ e_{\Tv} \}_{\Tv \in \TGammaA^{(0)}}$, the RHS is $A_{\Tv_2}^{\Tv_1}$. For the LHS, one has $\langle f_{\Tv_1}, B f_{\Tv_2} \rangle_{\hsB} = \langle \phiH^{v_1}(e_{\Tv_1}), B_{\phi}^{\phi} \phiH^{v_2}(e_{\Tv_2}) \rangle_{\hsB} = \langle \phiH^{v_1}(e_{\Tv_1}), \phiH^{v_2}(A e_{\Tv_2}) \rangle_{\hsB} = \tsum_{\Tv \in \TGammaA^{(0)}} A_{\Tv_2}^{\Tv} \langle \phiH^{v_1}(e_{\Tv_1}), \phiH^{v}(e_{\Tv}) \rangle_{\hsB}$. This expression is zero for $v_1 \neq v$, so the summation reduces to the summation over the $\Tv = (v_1, \alpha) \in  \TGammaA^{(0)}$: $\tsum_{\Tv = (v_1, \alpha)} A_{\Tv_2}^{\Tv} \langle \phiH^{v_1}(e_{\Tv_1}), \phiH^{v_1}(e_{\Tv}) \rangle_{\hsB} = \tsum_{\Tv = (v_1, \alpha)} A_{\Tv_2}^{\Tv} \langle e_{\Tv_1}, e_{\Tv} \rangle_{\hsA} = A_{\Tv_2}^{\Tv_1}$.
	\medskip
	\par
	Let us return to the general situation $n \geq 1$ in Point~\ref{item scalar product of product}. With $B_i = \smallpmatrix{ B_{i, \phi}^{\phi} & B_{i, \phi}^{\perp} \\ B_{i, \perp}^{\phi} & B_{i, \perp}^{\perp} }$, a straightforward computation shows that the only component in $B_1 \cdots B_n$ that contains only inherited components is in the block $(B_1 \cdots B_n)_{\phi}^{\phi}$ and it is $B_{1, \phi}^{\phi} \cdots B_{n, \phi}^{\phi}$, so that $\langle f_{\Tv_1}, B_1 \cdots B_n f_{\Tv_2} \rangle_{\hsB}  = \langle f_{\Tv_1}, B_{1, \phi}^{\phi} \cdots B_{n, \phi}^{\phi} f_{\Tv_2} \rangle_{\hsB} + \text{\TNIC}$. The proof that $\langle f_{\Tv_1}, B_{1, \phi}^{\phi} \cdots B_{n, \phi}^{\phi} f_{\Tv_2} \rangle_{\hsB} = \langle e_{\Tv_1}, A_1 \cdots A_n e_{\Tv_2} \rangle_{\hsA}$ is the same as before, with $B_{\phi}^{\phi} = B_{1, \phi}^{\phi} \cdots B_{n, \phi}^{\phi}$ and $A = A_1 \cdots A_n$ which satisfy $\phiH(A \psi) = B_{\phi}^{\phi} \phiH(\psi)$.
	\medskip
	\par
	Point~\ref{item trace of product} is a direct consequence of Point~\ref{item scalar product of product}. By the previous argument on the product $B_1 \cdots B_n$, one has 
	\begin{align*}
		\tr(B_1 \cdots B_n) 
		&= \tsum_{\Tv \in \TGammaA^{(0)}} \langle f_{\Tv}, B_1 \cdots B_n f_{\Tv} \rangle_{\hsB} + \tsum_{\hw \in \hGammaB^{(0)}} \langle f_{\hw}, B_1 \cdots B_n f_{\hw} \rangle_{\hsB} 
		= \tsum_{\Tv \in \TGammaA^{(0)}} \langle f_{\Tv}, B_1 \cdots B_n f_{\Tv} \rangle_{\hsB} + \text{\TNIC}
	\end{align*}  
	and $\tsum_{\Tv \in \TGammaA^{(0)}} \langle f_{\Tv}, B_1 \cdots B_n f_{\Tv} \rangle_{\hsB}  = \tsum_{\Tv \in \TGammaA^{(0)}} \langle e_{\Tv}, A_1 \cdots A_n e_{\Tv} \rangle_{\hsA} +  \text{\TNIC} = \tr(A_1 \cdots A_n) + \text{\TNIC}$ by Point~\ref{item scalar product of product}.
\end{proof}

\begin{remark}
	\label{remark non normalized}
	Point~\ref{item trace of product} can be proved directly without the assumption that $\phiH$ is normalized.
\end{remark}

The notion of $\phi$-compatibility has been developed for operators on $\hsA$ and $\hsB$. We define $\phi$-compatibility for fermions as follows:
\begin{definition}
	A vector $\psi_\algB$ is $\phi$-compatible with a vector $\psi_\algA$ if $\psi_\algB - \phiH(\psi_\algA) \in  \phiH(\hsA)^\perp$.
\end{definition}

Using this definition, we extend the acronym “\TNIC” (“Terms with Non-Inherited Components”) to terms which contains fermions.
\medskip
\par
From a physical point of view, the consequence of the normalization of $\phiH$ is that, for any $\psi \in \hsA$, one has
\begin{align*}
	\norm{\phiH(\psi)}_{\hsB} = \norm{\psi}_{\hsA}
\end{align*} 
since, with $\psi = \toplus_{v \in \GammaA^{(0)}} \psi_v$, one has $\norm{\phiH(\psi)}_{\hsB}^2 = \toplus_{v \in \GammaA^{(0)}} \langle \phiH^{v}(\psi_v), \phiH^{v}(\psi_v) \rangle_{\hsB} = \toplus_{v \in \GammaA^{(0)}} \langle \psi_v, \psi_v \rangle_{\hsA} = \norm{\psi}_{\hsA}^2$. This means that $\phiH$ respects the normalization of the state vector when it is injected from $\hsA$ into $\hsB$. This state $\psi$ can be “diluted” at different “places” in $\hsB$ (different irreps) but its norm is conserved.

\subsection{Comparison of Spectral Actions}
\label{subsec comparison of spectral actions}

The purpose of this section is to compare the spectral actions for the two $\hphi$-compatible spectral triples
\begin{align*}
	&(\halgA, \hshA, \DhA \defeq \DM \otimes \bbbone + \gammaM \otimes \DA, \JhA \defeq \JM \otimes \JA, \gammahA \defeq \gammaM \otimes \gammaA)
	\\
	&\text{and}
	\\
	&(\halgB, \hshB, \DhB \defeq \DM \otimes \bbbone + \gammaM \otimes \DB, \JhB \defeq \JM \otimes \JB, \gammahB \defeq \gammaM \otimes \gammaB)
\end{align*}
for the same Riemannian spin manifold $(M,g)$. We suppose that $\phiH : \hsA \to \hsB$ is normalized.
\medskip
\par
We extend in a natural way the map $\phi$ to a morphism of algebras $\hphi \defeq \Id \otimes \phi : \halgA \to \halgB$. In the same way, we denote by $\hphiH \defeq \Id \otimes \phiH : \hshA \to \hshB$ the natural extension of $\phiH$. The notion of (strong) $\hphi$-compatibility is then naturally defined from the notion of (strong) $\phi$-compatibility: an operator $\hB = B_M \otimes B_F$ on $\hshB$ is (strong) $\hphi$-compatible with an operator $\hA = A_M \otimes A_F$ on $\hshA$ if $\hphiH(\hA (\chi \otimes \psi)) = \hB_{\hphi}^{\hphi} \hphiH( \chi \otimes \psi)$ (resp. $\hphiH(\hA \chi \otimes \psi) = \hB \hphiH( \chi \otimes \psi)$) for any $\chi \otimes \psi \in \hshA$, that is, $\chi \otimes \phiH( A_F \psi) = \chi \otimes B_{F, \phi}^{\phi} \phiH(\psi)$ (resp. $\chi \otimes \phiH( A_F \psi) = \chi \otimes B_F \phiH(\psi)$).
\medskip
\par
Lemma~\ref{lemma scalar product and trace phi compatible} then extends naturally to:
\begin{lemma}
	\label{lemma scalar product and trace phi compatible AC}
	Let $\{ \chi_{c} \}_{c \in C}$ be an orthonormal basis of $L^2(\SpinBun)$. For $i=1,\ldots,n$, let $\hB_i$ be an operator on $\hshB$ which is $\hphi$-compatible with an operator $\hA_i$ on $\hshA$.
	\begin{enumerate}
		\item For any $\Tv_1, \Tv_2 \in \TGammaA^{(0)}$ and $c_1, c_2 \in C$, one has\label{item scalar product of product AC}
		\begin{align*}
			\langle \chi_{c_1} \otimes f_{\Tv_1}, \hB_1 \cdots \hB_n \chi_{c_2} \otimes f_{\Tv_2} \rangle_{\hshB} 
			&= \langle \chi_{c_1} \otimes e_{\Tv_1}, \hA_1 \cdots \hA_n \chi_{c_2} \otimes e_{\Tv_2} \rangle_{\hshA} + \text{\TNIC}
		\end{align*}
		
		\item As a consequence, one has\label{item trace of product AC}
		\begin{align*}
			\Tr(\hB_1 \cdots \hB_n) = \Tr(\hA_1 \cdots \hA_n) + \text{\TNIC}
		\end{align*}
	\end{enumerate}
\end{lemma}

\begin{proof}
	The proof is similar to the one of Lemma~\ref{lemma scalar product and trace phi compatible}, noticing that the geometrical part plays no role in the main steps.
\end{proof}

We follow \cite{Suij15a} to define the bosonic and the fermionic spectral actions. For any $\omega \in \Omega^1_U(\halgA)$, let us consider the operator $\DhA[,\omega] = \DhA + \omega + \epsilonhA' \JhA \omega \JhA^{-1}$ where $\omega$ is used in place of $\piDhA(\omega)$. Let $f : \bbR \to \bbR$ be a positive even function. Then the bosonic spectral action is defined by \eqref{BosAction}.
\medskip
\par
To define the fermionic spectral action, we introduce the vector space of Grassmann vectors $\ThshA$ defined from $\hshA$, and the notation $\Tpsi \in \ThshA$ for any $\psi \in \hshA$. Then, in the even case, for any $\Tpsi \in \ThshA^+$, where $\ThshA^+$ corresponds to Grassmann vectors associated to vectors $\psi \in \hshA^+ = \ker (\gammahA - 1)$ (even elements in $\hshA$). Then the fermionic spectral action is defined by \eqref{FerAction}.
\medskip
\par
From now on, we suppose that $\dim M = 4$ and, to simplify the presentation (to focus mainly on the algebraic part of the spectral actions), we suppose that $(M,g)$ is compact and flat, so that all the Riemann tensors will be trivial in the following. 
\medskip
\par
Let us use the following notations. For any $\omega\in \Omega^1_U(\halgA)$ with $\piDhA(\omega) = \gamma^\mu \otimes \tilde{A_\mu} + \gamma_M \otimes \tilde{\Phi} $, for Hermitian operators $\tilde{A_\mu}$ and $\tilde{\Phi}$ on $C^\infty(M) \otimes \hsA$, define $A_\mu \defeq \tilde{A_\mu} - \JA \tilde{A_\mu} \JA^{-1}$ and $\Phi \defeq \DA + \tilde{\Phi} + \JA \tilde{\Phi} \JA^{-1}$, so that $\DhA[, \omega] = \DM \otimes 1 + \gamma^\mu \otimes A_\mu + \gammaM \otimes \Phi$. 
\medskip
\par
Let $\nabla_\mu^\SpinBun$ be the spin connection on $\SpinBun$, and consider the vector bundle $E = \SpinBun \otimes (M \times \hsA)$ such that $L^2(E) = \hshA$, and let $\nabla_\mu^E \defeq \nabla_\mu^\SpinBun \otimes 1 + 1 \otimes (\partial_\mu + i A_\mu)$ be the natural twisted connection on $E$ defined by the spectral triple, so that $\DhA[, \omega] = -i \gamma^\mu \nabla_\mu^E + \gammaM \otimes \Phi$. Finally, let $D_\mu \defeq \partial_\mu + i \ad(A_\mu)$ and $F_{\mu\nu} \defeq \partial_\mu A_\nu - \partial_\nu A_\mu + i[A_\mu, A_\nu]$. In the same way, we introduce $\omega'$, $\tilde{A}'_\mu$, $\tilde{\Phi}'$, $A'_\mu$, $\Phi'$, $E'$, $D'_\mu$, $F'_{\mu\nu}$ for the algebra $\halgB$. 
\medskip
\par
Let $f_n \defeq \int_0^\infty f(x) x^{n-1} \dd x$ be the moments of $f$ for $n > 0$, then we have the general result \cite[Prop.~8.12]{Suij15a} that we have simplified to take into account the fact that the metric $g$ is Euclidean. In particular there is no Einstein-Hilbert Lagrangian since the purely geometric part needs not be compared from $\halgA$ to $\halgB$.  

\begin{proposition}
	\label{prop bosonic lagrangian}
	Suppose that the $KO$-dimension of $\algA$ is even, then
	\begin{align*}
		\Tr f(\DhA[,\omega] / \Lambda)
		\sim
		\int_M \lag(A_\mu, \Phi) \, \dd^4 x + \calO(\Lambda^{-1})
	\end{align*}
	with
	\begin{align*}
		\lag(A_\mu, \Phi)
		&= \lag_A(A_\mu) + \lag_{\Phi}(A_\mu, \Phi)
	\end{align*}
	where $\lag_A(A_\mu) = \frac{f(0)}{24 \pi^2} \tr( F_{\mu\nu} F^{\mu\nu})$, and, up to a boundary term,
	\begin{align*}
		\lag_{\Phi}(A_\mu, \Phi) 
		&=
		- \frac{2 f_2 \Lambda^2}{4 \pi^2} \tr(\Phi^2) 
		+ \frac{f(0)}{8 \pi^2} \tr(\Phi^4)
		+ \frac{f(0)}{8 \pi^2} \tr\big( (D_\mu \Phi) (D^\mu \Phi) \big)
	\end{align*}
\end{proposition}
We use the same function $f$ and the same cut-off $\Lambda$ for the spectral actions on $\halgA$ and $\halgB$. It corresponds to the same Lagrangian as in \eqref{LagSpectral}, without the terms coming from the metric of $\Man$.
\medskip
\par
We suppose that $\omega \in \Omega^1_U(\halgA)$ and $\omega' \in \Omega^1_U(\halgB)$ are $\hphi$-compatible in the sense that $\piDhB(\omega')$ and $\piDhA(\omega)$ are $\hphi$-compatible. Since the family of vectors $\{ \gamma^\mu, \gamma_M \}$ are linearly independent in the Clifford algebra generated by the $\gamma^\mu$'s, this implies that $\tilde{A}'_\mu$ (resp. $\tilde{\Phi}'$) is $\hphi$-compatible with $\tilde{A}_\mu$ (resp. $\tilde{\Phi}$). The strong $\phi$-compatibility between $\JB$ and $\JA$ then implies that $A'_\mu$ (resp. $\Phi'$) is $\hphi$-compatible with $A_\mu$ (resp. $\Phi$). Notice then that $\partial_\mu \Phi'$ (resp. $\partial_\mu A'_\nu$) is $\hphi$-compatible with $\partial_\mu \Phi$ (resp. $\partial_\mu A_\nu$). \footnote{Thanks to the fact that $\phiH$ does not depend on the points in $M$.} We then have:

\begin{proposition}
	Suppose that $\omega \in \Omega^1_U(\halgA)$ and $\omega' \in \Omega^1_U(\halgB)$ are $\hphi$-compatible in the previous sense. Then\GN{notation changé pour lag b}
	\begin{align*}
		\lag_{\halgB,A}(A'_\mu)
		&= \lag_{\halgA,A}(A_\mu) + \text{\TNIC}
		\\
		\lag_{\halgB, \Phi'}(A'_\mu, \Phi')
		&= \lag_{\halgA, \Phi}(A_\mu, \Phi) + \text{\TNIC}
	\end{align*}
\end{proposition}

\begin{proof}
	From Prop.~\ref{prop bosonic lagrangian}, all the terms in $\lag_{\halgB,A}(A'_\mu)$ and $\lag_{\halgB, \Phi'}(A'_\mu, \Phi')$ are traces of polynomials of $\hphi$-compatible elements. So, according to Lemma~\ref{lemma scalar product and trace phi compatible AC}, up to terms with non-inherited components, they are equal to the similar expression in terms of traces of polynomials of the corresponding elements on $\halgA$.
\end{proof}

\begin{remark}
	This Proposition can be proved without the assumption on the normalization of $\phiH$, see Remark~\ref{remark non normalized}.
\end{remark}

A slight extension of Prop.~\ref{prop unitary equi triple and phi comp and diag unitary} shows that $\omega \in \Omega^1_U(\halgA)$ and $\omega' \defeq \hphi(\omega) \in \Omega^1_U(\halgB)$ are $\hphi$-compatible. But then $\omega'$ contains only inherited degrees of freedom, and so this situation is quite trivial from a physical point of view.

\begin{proposition}
	If $\Tpsi'$ is $\hphi$-compatible with $\Tpsi$, then
	\begin{align*}
		\act_{\halgB, f}[\omega', \Tpsi'] 
		&= \langle \JhB \Tpsi', \DhB[,\omega'] \Tpsi' \rangle_{\ThshB}
		= \langle \JhA \Tpsi, \DhA[,\omega] \Tpsi \rangle_{\ThshA} + \text{\TNIC}
		= \act_{\halgA, f}[\omega, \Tpsi] + \text{\TNIC}
	\end{align*}
\end{proposition}

\begin{proof}
	Since $\piDhB(\omega')$ and $\piDhA(\omega)$ are $\hphi$-compatible, $\DhB[,\omega']$ and $\DhA[,\omega]$ are $\hphi$-compatible, and since $\JB$ and $\JA$ are strong $\phi$-compatible, then $\JhB$ and $\JhA$ are strong $\hphi$-compatible.
	\medskip
	\par
	Using previously defined notations, one can write $\Tpsi' = \tsum_{c, \Tv} \Tpsi'^{c, \Tv} \chi_{c} \otimes f_{\Tv} + \tsum_{c, \hw} \Tpsi'^{c, \Tv} \chi_{c} \otimes f_{\hw}$ and $\Tpsi = \tsum_{c, \Tv} \Tpsi^{c, \Tv} \chi_{c} \otimes e_{\Tv}$. Since $\Tpsi'$ and $\Tpsi$ are $\hphi$-compatible, one has $\Tpsi'^{c, \Tv} =\Tpsi^{c, \Tv}$ for any $c, \Tv$. So, $\langle \JhB \Tpsi', \DhB[,\omega'] \Tpsi' \rangle_{\ThshB} = \tsum_{c_1, c_2, \Tv_1, \Tv_2} \Tpsi'^{c_1, \Tv_1} \Tpsi'^{c_2, \Tv_2} \langle \JhB \chi_{c_1} \otimes f_{\Tv_1} , \DhB[,\omega'] \chi_{c_2} \otimes f_{\Tv_2} \rangle_{\ThshB} = \epsilonhB \tsum_{c_1, c_2, \Tv_1, \Tv_2} \Tpsi'^{c_1, \Tv_1} \Tpsi'^{c_2, \Tv_2} \overline{\langle \chi_{c_1} \otimes f_{\Tv_1} ,  \JhB \DhB[,\omega'] \chi_{c_2} \otimes f_{\Tv_2} \rangle_{\ThshB}}$. From Lemma~\ref{lemma scalar product and trace phi compatible AC}, one has $\langle \chi_{c_1} \otimes f_{\Tv_1} ,  \JhB \DhB[,\omega'] \chi_{c_2} \otimes f_{\Tv_2} \rangle_{\ThshB} = \langle \chi_{c_1} \otimes e_{\Tv_1} ,  \JhA \DhA[,\omega] \chi_{c_2} \otimes e_{\Tv_2} \rangle_{\ThshA} + \text{\TNIC}$ so that, since $\epsilonhB = \epsilonhA$, $\langle \JhB \Tpsi', \DhB[,\omega'] \Tpsi' \rangle_{\ThshB} = \epsilonhA \tsum_{c_1, c_2, \Tv_1, \Tv_2} \Tpsi^{c_1, \Tv_1} \Tpsi^{c_2, \Tv_2} \overline{\langle \chi_{c_1} \otimes f_{\Tv_1} ,  \JhA \DhA[,\omega] \chi_{c_2} \otimes e_{\Tv_2} \rangle_{\ThshA}} + \text{\TNIC} = \langle \JhA \Tpsi ,  \DhA[,\omega] \Tpsi \rangle_{\ThshA} + \text{\TNIC}$.
\end{proof}

\begin{remark}
	Notice that the formal proofs presented in the previous Propositions, which compare the spectral action on $\halgB$ to the spectral action on $\halgA$, do not reveal the terms which mix inherited and non-inherited components. A concrete and complete computation is necessary to compare precisely the two Lagrangians. This computation was done but is too long to be presented here.
\end{remark}

These results can be collected to construct a sequence $\{ (\halgA_n, \hs_{\halgA_n}, D_{\halgA_n}, J_{\halgA_n}, \gamma_{\halgA_n}) \}_{n \geq 0}$) of even real spectral triples (on AC manifolds) which are $\hphi_{n,n+1}$-compatible and a sequence of their corresponding spectral actions $\act_b[\omega_n] + \act_f[\omega_n, \Tpsi_n]$ with a control about their inherited and non-inherited terms. Using slight modifications of Prop.~\ref{prop universal forms and Dirac phi compatibility} and \ref{prop uB from uA case AF}, a gauge transformation on $\halgA_n$ can be transported to a gauge transformation on $\halgA_{n+1}$. So, we end up with a sequence of compatible NCGFT constructed on top of the defining sequence of an $AF$-algebra $\algA = \varinjlim \algA_n$.
\medskip
\par 

\chapter{Conclusions and Outlooks}
\label{CompMethAFNCGFT}
In this chapter, we propose to conclude this part and therefore the essential results put forward in this thesis, thus opening some perspectives for future works. 

\medskip
\par 
Both frameworks allow us to find the Lagrangian on top of $\algA$ into the one on top of $\algB$. In the case of the derivations, this depends on the convention chosen to define the integration (see \eqref{NewIntegralDer}). In the case of spectral triples, it depends on the choice of taking the $\phi$-normalized compatibility (see \eqref{NormedPhim}). These two choices are linked to how a “scalar weight” is associated with the objects used. The fact that the action of $\algA$ is recovered at the rank $\algB$ is an important indication of the naturalness of these constructions, particularly concerning the intuition (underlying the GUTs) of extending a theory by keeping some of its structures.
\medskip
\par 
The derivation framework is more constrained by the formalism than the spectral triple one. From a mathematical point of view, this may seem to be an advantage as there are fewer free choices. But from the point of view of physical applications, the resulting structure is probably poorer than with the spectral triples where additional choices to the natural structure of the $AF$-algebra appear for each of the structural operators. However, as suggested in section \ref{sec numerical exploration of the SSBM}, a potential interesting phenomenology concerning the prediction of bosons masses should be explored.
\medskip
\par 
The spectral triple framework offers the possibility to work with both notions of $\phi$-compatibility for the operators. It would be interesting to determine the results for the Lagrangian by taking the strong $\phi$-compatibility and/or any $\phi$-map (not necessarily the $\phi$-normalized). In the same way, a numerical exploration remains to be done in the case of spectral triples.
\medskip
\par 
Unfortunately, the implemented project is very large and I didn't get the time to explore all its relevant aspects. Then many questions remain to be explored, in particular, what is the tractability offered along the inductive sequence by the $\phi$-compatibility notions, by the norm or by Dirac operators? What is the meaning of the NCGFT at the end of the inductive sequence, given that it would potentially have an infinite number of degrees of freedom? Which $AF$-algebra can be interesting to make NCGFTs on it? Knowing that $AF$-algebras can approximate geometries and that our NCGFTs benefit from a geometrical interpretation, to what extent can the implemented scheme allow us to approximate geometries?
\medskip
\par 
The two frameworks thus established offer a large number of possibilities concerning the elaboration of phenomenological models. Indeed, many choices are possible, concerning the $AF$-algebra, the module (or Hilbert space), the chosen notion of $\phi$-compatibility and of the structural operators for the spectral triplets, or the way $u(v,w)$ embeds the fermionic fields, potentially allowing a “rotation" mechanism in the representation space (which could be linked to the CKM matrix mechanism). After the end of my PhD thesis, I hope to have the opportunity to work with researchers having some expertise in these fields, in order to study and explore GUT-like models beyond the NCSMPP, within this framework.
\medskip
\par 
Finally, this work goes beyond the development of NCGFTs. Indeed, a large part of the efforts made during my PhD thesis have concerned the elaboration of differential structures on $AF$-algebras, with notions of compatibility between the operators allowing to build and relate these structures at each step of the sequence. There remain many avenues to explore from this simple fact, notably on the possibility of approximating interesting, potentially commutative geometries with this procedure.

\bibliography{bibliography}

\end{document}